# Proceedings of the PHOTON-2017 Conference

Geneva, Switzerland
22 – 26 May 2017

Editors: David d'Enterria
Albert de Roeck
Michelangelo Mangano

This volume is indexed in: CERN Document Server (CDS), INSPIRE.

This volume should be cited as:

Proceedings of the PHOTON-2017 Conference, Geneva, Switzerland, 22 – 26 May 2017, edited by David d'Enterria, Albert de Roeck and Michelangelo Mangano, CERN Proceedings, Vol. 1/2018, CERN-Proceedings-2018-001 (CERN, Geneva, 2018),
https://doi.org/10.23727/CERN-Proceedings-2018-001

A contribution in this volume should be cited as:

[Author name(s)], in Proceedings of the PHOTON-2017 Conference, Geneva, Switzerland, 22 – 26 May 2017, edited by David d'Enterria, Albert de Roeck and Michelangelo Mangano, CERN Proceedings, Vol. 1/2018, CERN-Proceedings-2018-001 (CERN, Geneva, 2018), pp. [first page]–[last page],
https://doi.org/10.23727/CERN-Proceedings-2018-001.[first page]

# Abstract

This document collects the proceedings of the PHOTON 2017 conference ("International Conference on the Structure and the Interactions of the Photon", including the 22$^{\mathrm{th}}$ "International Workshop on Photon-Photon Collisions", and the "International Workshop on High Energy Photon Colliders") held at CERN (Geneva) in May, 2017. The latest experimental and theoretical developments on the topics of the PHOTON conference series are covered: (i) photon-photon processes in $e^+e^-$, proton-proton (pp) and nucleus-nucleus (AA) collisions at current and future colliders, (ii) photon-hadron interactions in $e^{\pm}p$, pp, and AA collisions, (iii) final-state photon production (including Standard Model studies and searches beyond it) in pp and AA collisions, and (iv) high-energy gamma-ray astrophysics.

# Contents

# PHOTON-2017 Conference Proceedings

*David d'Enterria*[1], *Albert de Roeck*[1] *and Michelangelo Mangano*[2] (**Conference chairs**)
[1] CERN, EP Department, CH-1211 Geneva
[2] CERN, TH Department, CH-1211 Geneva

The PHOTON 2017 conference ("International Conference on the Structure and the Interactions of the Photon", including the 22th "International Workshop on Photon-Photon Collisions, and the "International Workshop on High Energy Photon Colliders") was held at CERN (Geneva) from 22th to 26th May, 2017. The conference is part of a series that was initiated in 1973 in Paris as "International Colloquium on Photon-Photon Collisions at Electron-Positron Storage Rings". The latest Photon conferences took place in Novosibirsk (2015), Paris (2013), Spa (2011), Hamburg (2009), Paris (2007), Warsaw (2005), Frascati (2003), Ascona (2001), Freiburg (2000), Ambleside (1999), and Egmond aan Zee (1997).

The topics of the conference included (i) photon-photon processes in $\mathrm{e^+e^-}$, proton-proton (pp) and nucleus-nucleus (AA) collisions at current and future colliders, (ii) photon-hadron interactions in $\mathrm{e^\pm p}$, pp, and AA collisions, (iii) final-state photon production (including Standard Model studies and searches beyond it) in pp and AA collisions, and (iv) high-energy gamma-rays astrophysics.

The conference attracted more than 90 physicists, both experimentalists and theorists, sharing fruitful and inspiring discussions through 75 talks covering the current state-of-the-art in all relevant research directions. The meeting agenda and the presentation slides can be downloaded from the workshop website:

https://indico.cern.ch/e/photon2017

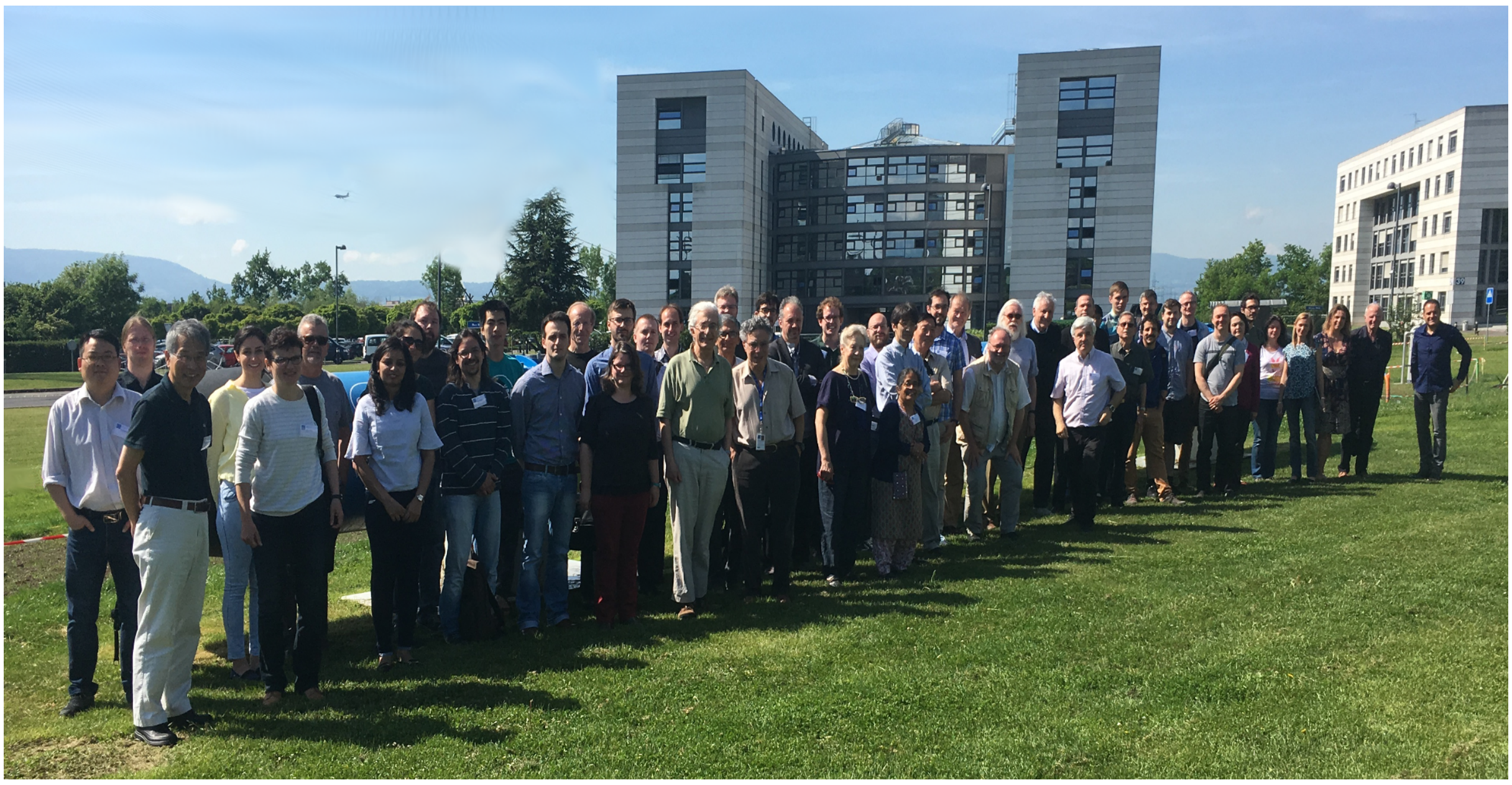

**Fig. 1:** PHOTON'17 conference picture. CERN, 25th May 2017.

The current proceedings include about 50 write-ups covering all the topics presented during the conference. Very unfortunately, during the conference (despite not being able to attend it), one of the leading scientists in the field of photon physics, Maria Krawczyk, member of the International Advisory Committee, passed away. These proceedings are dedicated to her memory. The impact of Maria's work and personality is described in the closing contribution by Rohini Godbole and Lia Pancheri.

## International Advisory Committee

Stan Brodsky (SLAC)
David d'Enterria (CERN)
Albert de Roeck (CERN)
Victor Fadin (INP)
Alex Finch (Lancaster)
Ilya Ginzburg (NSC)
Jeffrey Gronberg (LNLL)
Jack Gunion (UC Davis)
Kaoru Hagiwara (KEK)
Kapusta Frederic (LPNHE)
Uri Karshon (Weizmann Inst.)
Michael Klasen (Münster)
Maria Krawczyk (FUW)
Eric Laenen (NIKHEF)
Klaus Möning (DESY)
Giulia Pancheri (INFN)
Krzysztof Piotrzkowski (Louvain)
Mariangela Settimo (LPNHE)
Soldner Remblod (Manchester)
Yoshi Sakai (KEK)
Tohru Takahashi (Hiroshima)
Valery I. Telnov (NSK)
Sadaharu Uehara (KEK)
Mayda Velasco (Northwestern)
Shin-Shan Yu (Taiwan)
Fabian Zomer (LAL)

## Acknowledgments

This conference would not have been possible without the financial support of the CERN TH Department for a few of the participants, as well as without the valuable administrative help from Michelle Connor.

# Maria Krawczyk : friend and physicist

*R.M. Godbole* [1] *and G. Pancheri* [2,*]
[1] Center for High Energy Physics, Indian institute of Science, Bangalore, India
[2] INFN Frascati National Laboratories, I00044 Frascati, Italy

**Abstract**
With this brief note, we remember our friend Maria Krawczyk, who passed away on May 24th 2017. We briefly outline some of her physics interests and main accomplishments, and her great human and moral qualities.

On May 25th 2017, the shocking news of Maria Krawczyk's sudden passing reached all the participants of PHOTON 2017, the Workshop and Conference on Photons, whose International Advisory Committee Maria had been a member of since 2003. Maria had herself organized the 2005 Conference, in Warsaw and Kazimierz, celebrating, at the same time, 100 years since Einstein's paper on the photo-electric effect.

Our friend Maria had passed away the night before, on May 24th. All week she had been unwell, but she had thought it was a temporary illness and, until just the day before, kept on working with her long time collaborator Ilya Ginzburg. She was also expected, since Monday of that week, to attend the PLANCK 2017 meeting in Warsaw, where she was one of the Organizers.

Maria was a wonderful person to all who knew her and a loving wife, mother and grandmother of four: grand-daughter (15), and 3 grandsons (18, 11 & 8), to her family. Her demise is a painful loss for all the people who have known her. First and foremost her family and then a very large number of friends, physics colleagues and collaborators spread all over the world. Her life had not been easy. As the communist rule hardened during the 1980's, her husband Tomasz was put in jail, being a strong vocal enemy of the regime. Alone and fearing for her husband's future, she carried on with her personal and professional life, caring and supporting their two little boys. All along, she was doing physics, with great passion and a view for unconventional solutions.

She was deeply interested in a lot of different areas of particle physics and contributed to them very effectively. Her interest in QCD and resummation brought her to Frascati, and to invite one of us (G.P.) to Warsaw, and Kazimierz Workshops. High energy photons and their hadronic structure [1] occupied her for many years in her working life. She always had a 'nose' for interesting physics and pursued a given area at a time with clear vision of what she wanted to achieve there. She wrote a very useful review on 'Photon Structure Functions' [2]. It was her involvement with the photons that fed her interaction with one of us (R.G.). Her interest in photons and in the extended Higgs sector, naturally propelled her towards CP studies of the Higgs sector since photon colliders provide one of the most unambiguous probes of the CP property of the Higgs. In fact she was one of the originators of the idea of 'Workshop on CP studies and nonstandard Higgs physics' while she was a scientific associate at CERN. The idea materialised into a very active and useful workshop [3]. She also devoted a lot of her mind to the two-Higgs doublet models [4], inert doublet and implications for cosmology thereof [5].

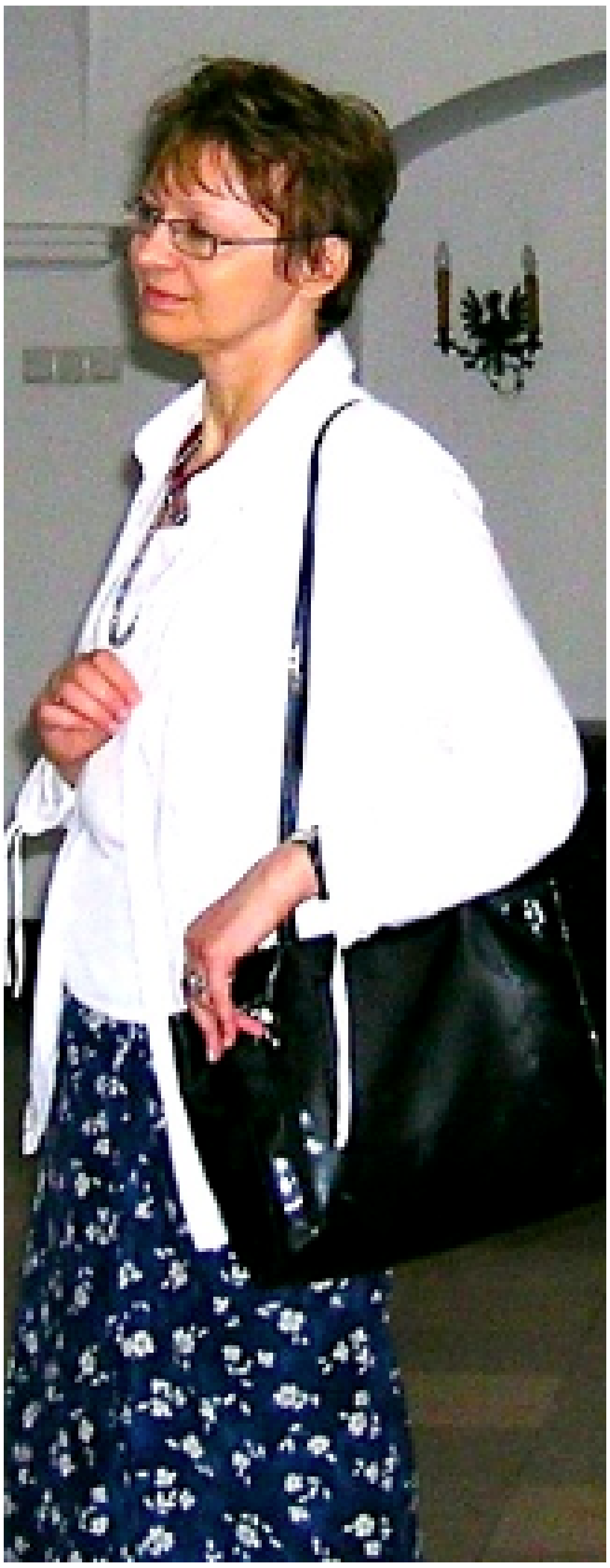

**Fig. 1:** Maria in Kazimierz (Kraków Old Town, Poland), 2006.

[*] Also Research Affiliate with MIT CTP, Cambridge, MA, USA

R.G. has known Maria since 1986, starting from a meeting at the ICHEP conference in Berkeley. At the time Maria was one of the very small number of woman particle theorists, whom one wanted to emulate. She was also the universal mother, caring about everybody with same passion that she cared about physics. She was an incredibly warm and kind person, as everybody who interacted with her knew.

She was a keen advocate of the future colliders which would be necessary to unravel the physics of the Standard Model and beyond. She had been involved very deeply in the discussions of the $e^+e^-$ Linear Collider, first the TESLA being planned in Germany and then the International Linear Collider: ILC. Even closer to her heart was the PLC: Photon Linear Collider in which she got interested during her initial collaboration with Ilya Ginzburg [6].

Her interest in the subject started in the 90's and led to a long collaboration with Ilya Ginzburg first and then with a number of students as well as some of us. Ilya Ginzburg, was in fact in Warsaw on May 18th, working with her in person until May 20th, and by phone, when she started being unwell. On May 24th she sent him a newly corrected version of the paper they are preparing together. So she was immersed in Science till the last moments of her life! Indeed she leaves behind five doctoral students who were working with her.

With Ilya and Per Osland she had found a corner of parameter space in two-Higgs doublet models where a scalar will exist which looks 'like' the SM Higgs and yet one will not be in the trivial decoupling regime [7]. Another long time collaborator of Maria, Per Osland, always admired her great optimism. One thing Per is keen remembering, and appreciated very much during the many years of working with Maria, was her good grasp of experimental facts, and their implications. Due to her interest in the two-Higgs doublet models, Maria was also involved deeply in phenomenology of charged Higgs bosons. Her interest in scalars led her to co-initiate, with Bohdan Grzadkowski, the very interesting series 'Scalars' which is held in Warsaw every two years.

Maria was an exceptionally generous soul. Later in her life, after she had become a professor at University of Warsaw, and was recognized as a very good and active particle physicist, after the regime had changed and her husband had been able to return to normal academic life, her mother fell ill and Maria cared for her until the end. In 2006 she organized a *Giulia fest* in Kazimierz, for the occasion of the EURIDICE network Final Meeting, in August 2006. It was the occasion of G.P.'s 65 years of age and the closing meeting of the Research and Training EU Network of which Maria had been the Polish node scientist-in-charge. Fig. 1 shows Maria during that meeting, in the hall of the Dom Architekta in Kazimierz. The most moving thing in this occasion is that Maria's mother had just passed away a few days earlier, but Maria courageously kept her promise to be with us and brought the conference to a very happy and warm ending.

The three of us, R.G., G.P. and Maria, shared a special bond, built on common physics interests, reciprocal respect, and an unspoken understanding of how difficult it was to charter the path of the career we had chosen. We understood each other, and we will miss her deeply. We are grateful to the friends who helped us in writing this memorial, in particular to Maria's colleagues and friends Ilya Ginzburg and Jan Kalinowski, and to Maria's family who shared with us some of their loving memories.

The short list of Maria'a papers cited in this *in memoriam* is certainly incomplete, but we hope it will be sufficient to show her versatility and the depth of physics interests which accompanied her life.

# Photon Structure Functions: past, present, future

*T. Ueda*[1], *T. Uematsu*[2] *and K. Sasaki*[3]
1. Nikhef Theory Group, Science Park 105, 1098 XG Amsterdam, The Netherlands
2. Institute for Liberal Arts and Sciences, Kyoto University, Kyoto 606-8501, Japan
3. Faculty of Engineering, Yokohama National University, Yokohama 240-8501, Japan

**Abstract**
We review the photon structure functions in the past and at present and discuss the future of this field.

## 1 Prologue

In $e^+e^-$ collider experiments, the two-photon process in which one of the virtual photons is very far off shell (large $Q^2 \equiv -q^2$) while the other is close to the mass shell (small $P^2 \equiv -p^2$) can be viewed as a deep-inelastic electron-photon scattering [1] . In the deep-inelastic scattering off a photon target, we can study the structure of photon. Two (unpolarized) structure functions $F_2^\gamma(x, Q^2)$ and $F_L^\gamma(x, Q^2)$ of the real photon ($P^2 = 0$) can be measured in the single-tag events, while in the double-tag events we observe $F_2^\gamma(x, Q^2, P^2)$ and $F_L^\gamma(x, Q^2, P^2)$ of the virtual photon.

## 2 Photon structure functions — Past

The structure functions $F_2^\gamma(x, Q^2)$ and $F_L^\gamma(x, Q^2)$ were first studied in the parton model [2] and then investigated in perturbative QCD (pQCD). The leading order (LO) [3] and the next-to-leading order (NLO) [4] QCD contributions to $F_2^\gamma$ were calculated and the moments of $F_2^\gamma$ is expressed as

$$\int_0^1 dx x^{n-2} F_2^\gamma(x, Q^2) = \frac{\alpha}{4\pi}\frac{1}{2\beta_0}\left\{\frac{4\pi}{\alpha_s(Q^2)}a_n + b_n + h_n(\alpha_s(Q^2)) + \mathcal{O}(\alpha_s(Q^2))\right\} \tag{1}$$

where $x$ is the Bjorken variable, $\beta_0$ is the one-loop QCD $\beta$ function and $\alpha$ ($\alpha_s(Q^2)$) is the QED (QCD running) coupling constant. Since $1/\alpha_s(Q^2)$ behaves as $\ln(Q^2/\Lambda^2)$ at large $Q^2$, where $\Lambda$ is the QCD scale parameter, the first term $a_n/\alpha_s(Q^2)$ dominates over the $b_n$ term and also over the hadronic term $h_n(\alpha_s(Q^2))$. The LO contributions $a_n$ were definite [3]. Meanwhile, the NLO corrections $b_n$ were calculated only for $n > 2$ [4]. For $n > 2$, the hadronic moments $h_n(\alpha_s(Q^2))$ vanish in the large-$Q^2$ limit and the $b_n$ terms give finite contributions. However, at $n = 2$, the hadronic energy-momentum tensor operator comes into play. Due to the conservation of this operator, $b_n$ shows a singularity at $n = 2$ and $h_{n=2}(\alpha_s(Q^2))$ does not vanish at large $Q^2$. Actually, $h_n(\alpha_s(Q^2))$ also develops a singularity at $n = 2$ which cancels out the one of $b_n$, and $h_n(\alpha_s(Q^2))$ and $b_n$ in combination give a finite but perturbatively incalculable contribution at $n = 2$ [5]. The fact that a definite information on the NLO second moment is missing prevents us to fully predict the shape and magnitude of the structure function of $F_2^\gamma(x, Q^2)$ up to the order $\mathcal{O}(\alpha)$.

It was then pointed out [5] that the situation changes significantly when we analyze the structure function of a virtual photon with $P^2$ much larger than $\Lambda^2$, more specifically, in the kinematical region $\Lambda^2 \ll P^2 \ll Q^2$. In this region, the hadronic component of the photon can also be dealt with *perturbatively* and thus a definite prediction of the whole structure function, its shape and magnitude, may become possible. In fact, the virtual photon structure function $F_2^\gamma(x, Q^2, P^2)$ for $\Lambda^2 \ll P^2 \ll Q^2$ was calculated in LO (the order $\alpha/\alpha_s$) and NLO (the order $\alpha$) [5] without any unknown parameters. It is notable that the pathology of singularity, which appeared at $n=2$ in the term $b_n$ of Eq. (1) for the real photon target, disappeared from the moments of $F_2^\gamma(x,Q^2,P^2)$.

Then what happens to the moments of $F_2^\gamma(x,Q^2,P^2)$ with arbitrary $P^2$ but $P^2 \ll Q^2$? If we employ the framework of the operator product expansion supplemented by the renormalization group method, we need to know the photon matrix element (PME), $\langle\gamma(p)|O_n^i(\mu^2)|\gamma(p)\rangle$, with $i=S,G,NS,\gamma$, where $|\gamma(p)\rangle$ is the "target" virtual photon state with momentum $p$, $O_n^i$ are the relevant twist-2 spin-$n$ operators and $\mu^2$ is the renormalization point. The indices $S,G,NS$ and $\gamma$ refer to singlet quark, gluon, nonsinglet quark and photon, respectively. To lowest order in the QED coupling, $\langle\gamma(p)|O_n^\gamma(\mu^2)|\gamma(p)\rangle=1$. Choosing the renormalization point at $\mu^2=Q_0^2$ with the condition $\Lambda^2 \ll Q_0^2 \ll Q^2$, we write the PME's of the hadronic operators $\vec{O}_n=(O_n^S,O_n^G,O_n^{NS})$ as $\langle\gamma(p)|\vec{O}_n(\mu)|\gamma(p)\rangle|_{\mu^2=Q_0^2}=\frac{\alpha}{4\pi}\vec{A}_n(Q_0^2;P^2)$. Now when photon state becomes far off-shell and $P^2$ approaches $Q_0^2$, its point-like nature prevails and $\vec{A}_n(Q_0^2;P^2)$ becomes calculable perturbatively. Let us put $\vec{A}_n(Q_0^2;P^2=Q_0^2)\equiv\vec{A}_n^{(1)}$ in one-loop order. For an arbitrary $P^2$ in the range $0\le P^2\le Q_0^2$, we divide $\vec{A}_n(Q_0^2;P^2)$ into two pieces such that $\vec{A}_n(Q_0^2;P^2)=\vec{\widetilde{A}}_n(Q_0^2;P^2)+\vec{A}_n^{(1)}$. Note that $\vec{\widetilde{A}}_n(Q_0^2;P^2)$ contains nonperturbative contributions (i.e., hadronic components) when $P^2$ is in the range $0\le P^2\le Q_0^2$, and satisfies the boundary condition by definition $\vec{\widetilde{A}}_n(Q_0^2;P^2=Q_0^2)=0$. Then the following formula is obtained for the moments of $F_2^\gamma$ up to NLO in QCD (the extension to NNLO is straightforward),

$$\begin{aligned}\int_0^1 dx x^{n-2}F_2^\gamma(x,Q^2,P^2)/\Big(\frac{\alpha}{4\pi}\frac{1}{2\beta_0}\Big) &= \frac{4\pi}{\alpha_s(Q^2)}\sum_{i=+,-,NS}\frac{\widetilde{\mathcal{L}}_i^n}{1+d_i^n}\Big\{1-\Big(\frac{\alpha_s(Q^2)}{\alpha_s(Q_0^2)}\Big)^{1+d_i^n}\Big\} \\ &+\sum_{i=+,-,NS}\frac{\widetilde{\mathcal{A}}_i^n}{d_i^n}\Big\{1-\Big(\frac{\alpha_s(Q^2)}{\alpha_s(Q_0^2)}\Big)^{d_i^n}\Big\}+\sum_{i=+,-,NS}\frac{\widetilde{\mathcal{B}}_i^n}{1+d_i^n}\Big\{1-\Big(\frac{\alpha_s(Q^2)}{\alpha_s(Q_0^2)}\Big)^{1+d_i^n}\Big\}+\mathcal{C}^n \\ &+2\beta_0\vec{\widetilde{A}}_n(Q_0^2;P^2)\cdot\sum_{i=+,-,NS}P_i^n\vec{C}_n(1,0)\Big(\frac{\alpha_s(Q^2)}{\alpha_s(Q_0^2)}\Big)^{d_i^n}\ ,\end{aligned} \tag{2}$$

which is applicable for an arbitrary target mass squared $P^2$ in the range $0\le P^2\le Q_0^2$. Here the coefficients $\widetilde{\mathcal{L}}_i^n$, $\widetilde{\mathcal{A}}_i^n$, $\widetilde{\mathcal{B}}_i^n$ and $\mathcal{C}^n$ are written in terms of the quantities calculable by the pQCD. Their explicit expressions are found in Ref. [5]. The exponents $d_i^n$ are given by $d_i^n=\lambda_i^n/2\beta_0$ $(i=+,-,NS)$ where $\lambda_i^n$ are the eigenvalues of the one-loop anomalous dimension matrix $\hat{\gamma}_n^{(0)}$, which is expanded as $\hat{\gamma}_n^0=\sum_i\lambda_i^nP_i^n$ with $P_i^n$ being the projection operators [4]. The terms $a_n$ and $b_n$ in Eq. (1) correspond to $\sum_i\widetilde{\mathcal{L}}_i^n/(1+d_i^n)$ and $\sum_i\widetilde{\mathcal{A}}_i^n/d_i^n+\sum_i\widetilde{\mathcal{B}}_i^n/(1+d_i^n)+\mathcal{C}^n$, respectively. Since $d_-^{n=2}=0$, $\widetilde{\mathcal{A}}_-^n/d_-^n$ and thus $b_n$ become singular at $n=2$. But we see that the product $(\widetilde{\mathcal{A}}_-^n/d_-^n)\times\Big[1-(\alpha_s(Q^2)/\alpha_s(Q_0^2))^{d_-^n}\Big]$ is finite in the limit $n\to 2$. There appear no singularities in the expression in Eq. (2). When $P^2$ approaches $Q_0^2$, the last term with $\vec{\widetilde{A}}_n(Q_0^2;P^2)$ vanishes, and we recover the result of Ref. [5]. For an arbitrary $P^2$ below $Q_0^2$, however, we need to use the experimental data once or resort to some nonperturbative methods (like lattice QCD) or employ models (like the vector meson dominance model) to estimate $\vec{\widetilde{A}}_n(Q_0^2;P^2)$.

## 3 Photon structure functions — Present

The work toward the next-to-next-to-leading order (NNLO) ($\mathcal{O}(\alpha\alpha_s)$) analysis of the real photon $F_2^\gamma$ started in Ref. [6], where the lowest six even-integer Mellin moments of the three-loop photon-parton (quark and gluon) splitting functions were calculated and the parton distributions in the real photon were analyzed. Later the virtual photon $F_2(x,Q^2,P^2)$ was investigated up to NNLO [7] using the results of the three-loop anomalous dimensions for the quark and gluon operators [8] and of the three-loop photon-parton splitting functions [9]. The NNLO result is shown in Fig. 1 together with three curves: the LO,

NLO QCD results and the box (tree) diagram contribution [5],

$$F_2^{\gamma(\mathrm{Box})}(x,Q^2,P^2)=\frac{3\alpha}{\pi}n_f\langle e^4\rangle\left\{x\left[x^2+(1-x)^2\right]\ln\frac{Q^2}{P^2}-2x\left[1-3x+3x^2+(1-2x+2x^2)\ln x\right]\right\},$$

where $n_f\langle e^4\rangle=\sum_{i=1}^{n_f}e_i^4$ with $e_i$ being the electric charge of the active quark with flavor $i$ and $n_f$ is the number of active quarks. We observe that the NNLO corrections reduce $F_2(x,Q^2,P^2)$ at large $x$.

Regarding the longitudinal structure function $F_L^\gamma$, its LO contribution which is of order $\alpha$, was calculated in QCD for the real photon ($P^2=0$) target in Ref. [3]. The analysis was made for the case of the virtual photon $F_L^\gamma(x,Q^2,P^2)$ ($\Lambda^2\ll P^2\ll Q^2$) in LO [5] and extended up to NLO ($\mathcal{O}(\alpha\alpha_s)$) [7]. The results for the virtual photon target are shown in Fig. 2 , where the box (tree) diagram contribution is expressed by $F_L^{\gamma(\mathrm{Box})}(x,Q^2,P^2)=\frac{3\alpha}{\pi}n_f\langle e^4\rangle\left\{4x^2(1-x)\right\}$.

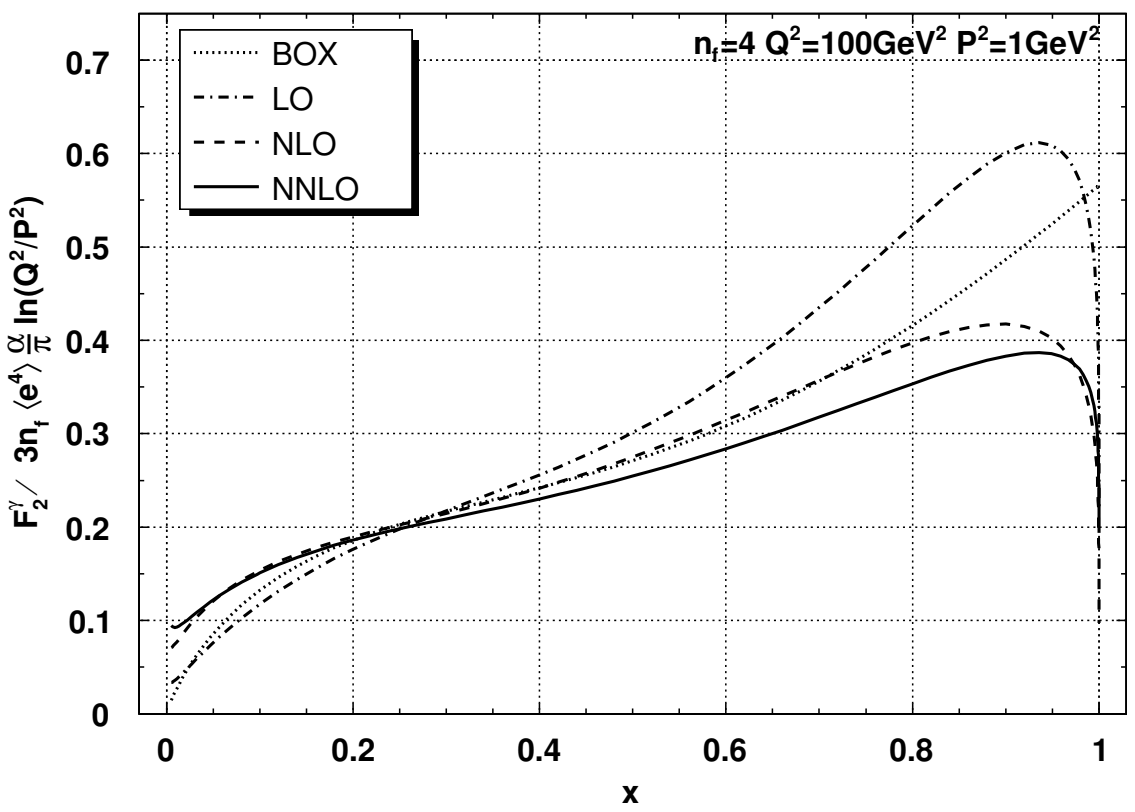

**Fig. 1:** Virtual photon structure function $F_2^\gamma(x,Q^2,P^2)$ in units of $(3\alpha n_f\langle e^4\rangle/\pi)\ln(Q^2/P^2)$ for $Q^2=100$ GeV$^2$ and $P^2=1$ GeV$^2$ with $n_f=4$ and $\Lambda=0.2$ GeV. [7]

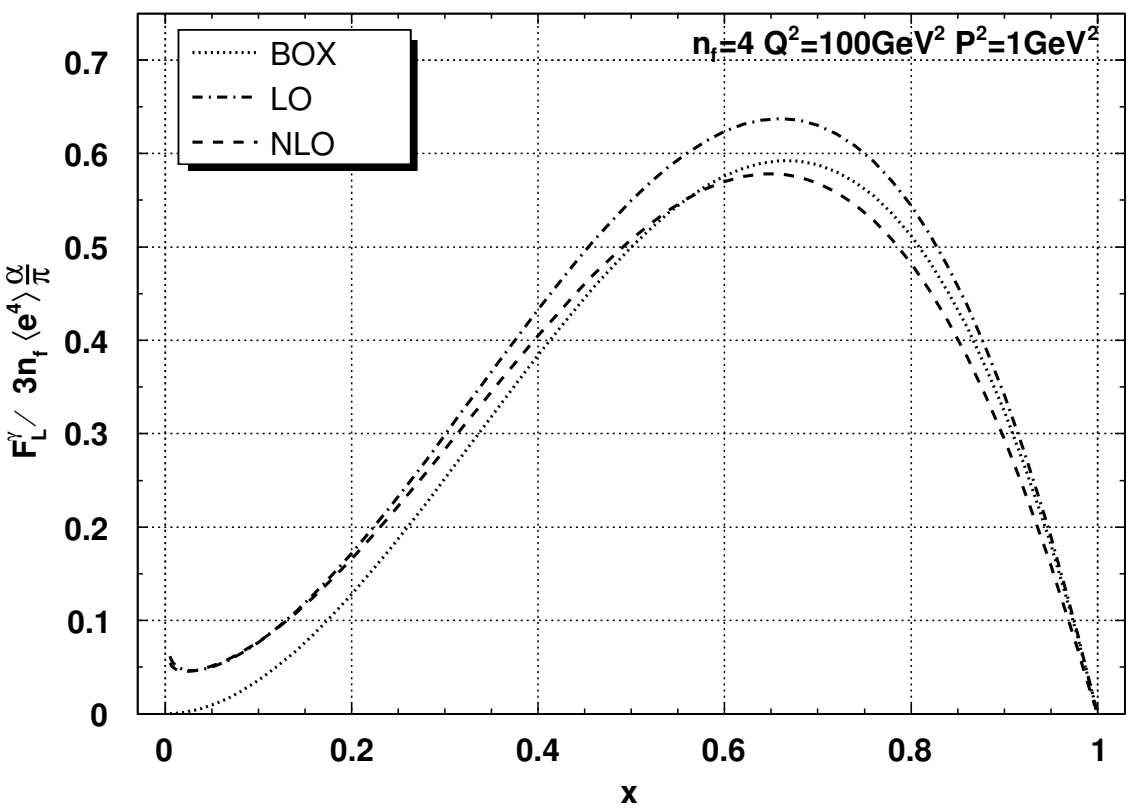

**Fig. 2:** Longitudinal photon structure function $F_L^\gamma(x,Q^2,P^2)$ in units of $(3\alpha n_f\langle e^4\rangle/\pi)$ for $Q^2=100$ GeV$^2$ and $P^2=1$ GeV$^2$ with $n_f=4$ and $\Lambda=0.2$ GeV. [7]

In the framework of the QCD improved parton model, the structure function $F_2^\gamma$ is expressed in a factorized form, i.e., as convolutions of coefficient functions and parton distributions in the photon:

$$F_2^\gamma(x,Q^2,P^2)=\sum_i C_2^i\otimes q_i^\gamma+C_2^G\otimes G^\gamma+C_2^\gamma\otimes\Gamma^\gamma \tag{3}$$

where $q_i^\gamma$, $G^\gamma$ and $\Gamma^\gamma$ are quark (with $i$-flavour), gluon and photon distributions, respectively, and $C_2^i$, $C_2^G$ and $C_2^\gamma$ are corresponding coefficient functions. In the leading order of $\alpha$, $\Gamma^\gamma$ does not evolve with $Q^2$ and we set $\Gamma^\gamma=\delta(1-x)$, which means that $C_2^\gamma$ contributes directly to $F_2^\gamma$. These parton distribution functions satisfy the well known DGLAP evolution eqs. Solving the DGLAP eqs. with the appropriate initial conditions at $P^2$ one obtains the parton distributions at $Q^2$. But the coefficient functions and parton distributions are dependent on the factorization-scheme (FS) adopted for defining these quantities. It is the standard choice to use the modified minimal subtraction ($\overline{\mathrm{MS}}$) scheme for the multi-loop calculations of the relevant quantities, namely, the coefficient functions, the splitting functions of partons and the $\beta$ function parameters. Using these results one obtains the parton distributions in the $\overline{\mathrm{MS}}$ scheme. However it was observed [6, 10] that the multi-loop photonic $\overline{\mathrm{MS}}$ contributions to $C_2^\gamma$ are negative and singular for $x\to 1$ and that, in the $\overline{\mathrm{MS}}$ scheme, these singularities have to be compensated by the quark distributions which thus have rather different behaviours at NLO and NNLO from the LO contribution. Under such circumstance a new factorization scheme, which is called DIS$_\gamma$, was introduced [10]. In this scheme the photonic coefficient function $C_2^\gamma$, which is the direct photon contribution to $F_2^\gamma$ in $\overline{\mathrm{MS}}$ scheme, is absorbed into the quark distributions, so that $C_2^\gamma|_{\mathrm{DIS}_\gamma}=0$ while the gluon distribution $G_2^\gamma$ is intact, i.e., $G_2^\gamma|_{\mathrm{DIS}_\gamma}=G_2^\gamma|_{\overline{\mathrm{MS}}}$.

Parton distributions in the real photon were investigated up to NNLO [6]. For the case of the virtual photon the analyses were made at NLO [11] and NNLO [12]. The results of the flavour-singlet-quark distribution $q_S^\gamma \equiv \sum_i q_i^\gamma$ in the virtual photon are shown in Fig. 3 ($\overline{\rm MS}$ scheme) and in Fig. 4 (DIS$_\gamma$). We observe that (i) the quark distribution shows quite different behaviours in two schemes, especially in large-$x$ region; (ii) in $\overline{\rm MS}$ scheme, the behaviours of the (LO+NLO) and (LO+NLO+NNLO) curves are quite different from the LO curve. They lie below the LO curve for $0.2 < x < 0.8$ but diverge as $x \to 1$; (iii) in DIS$_\gamma$ scheme, the three curves LO, (LO+NLO) and (LO+NLO+NNLO) rather overlap below $x = 0.6$, which means that the NLO and NNLO contributions to the quark distribution are small for moderate $x$ —- appropriate behaviours from the viewpoint of "perturbative stability" [6]; (iv) the gluon distribution in the photon is very small in absolute value except in small-$x$ region.

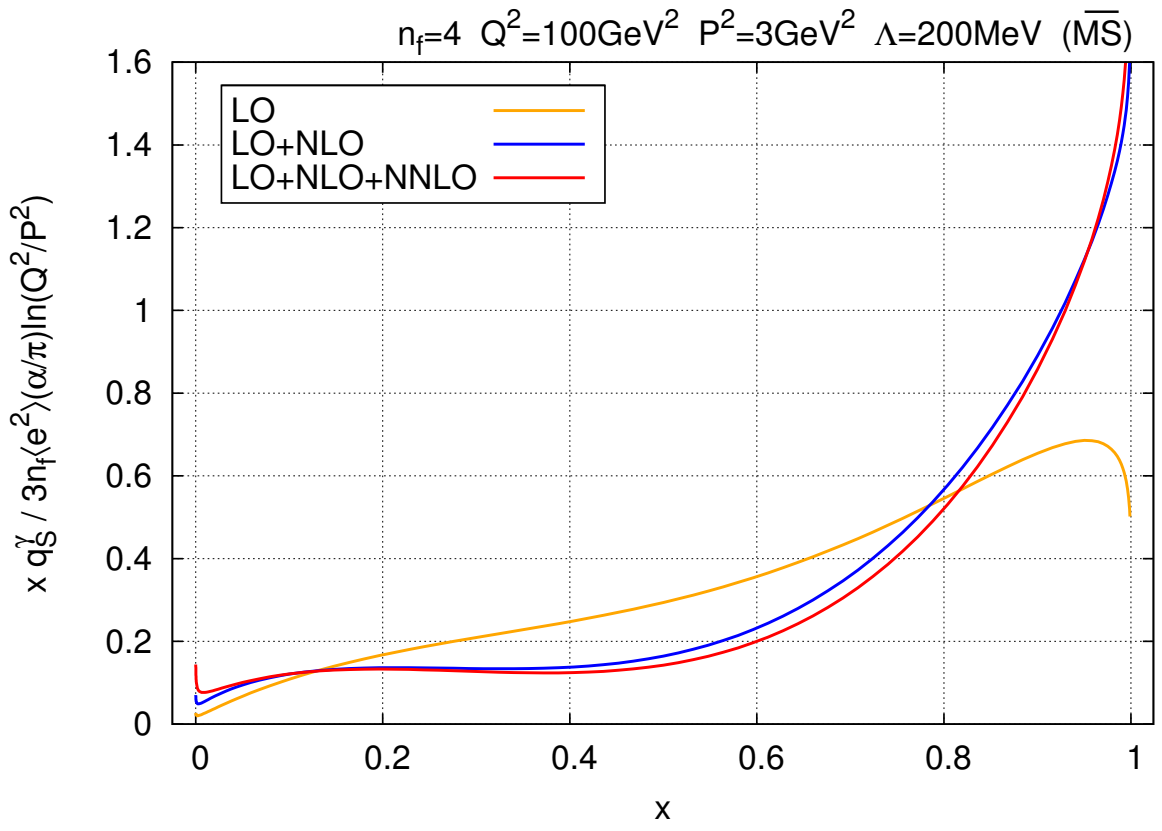

**Fig. 3:** Singlet-quark distribution $xq_S^\gamma(x, Q^2, P^2)$ in $\overline{\rm MS}$ scheme. [12]

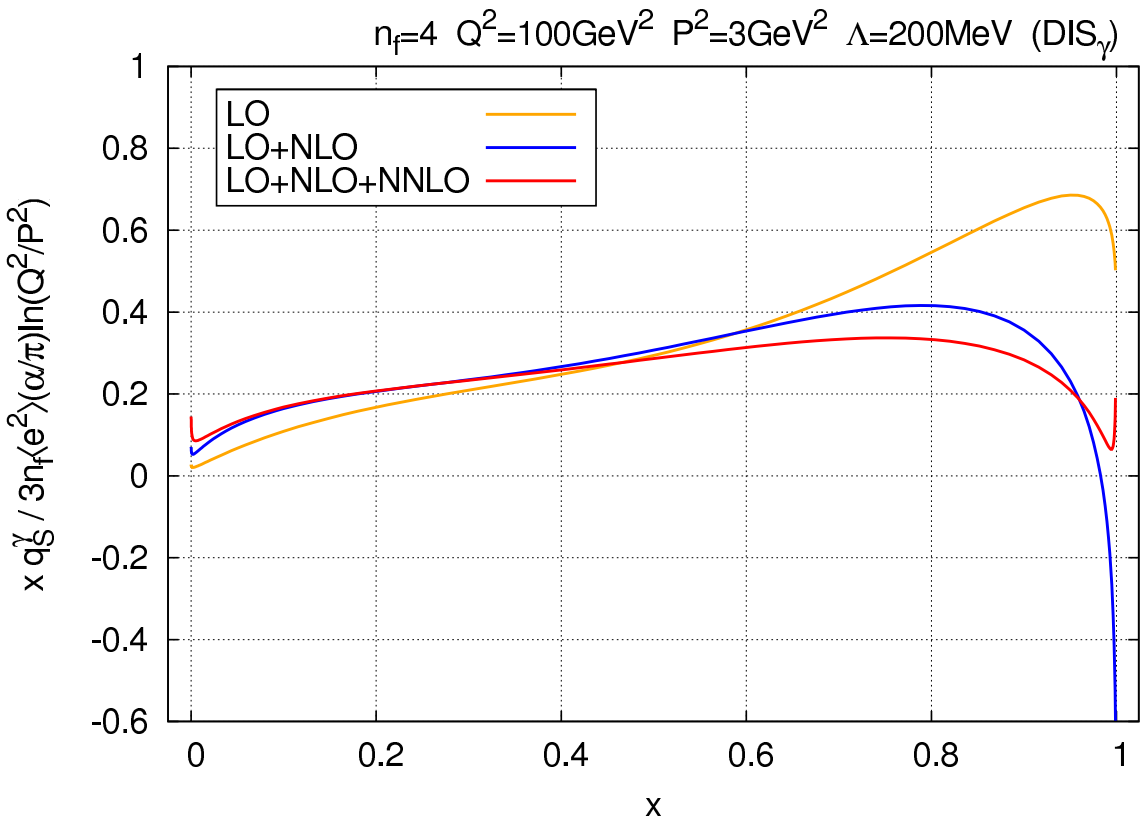

**Fig. 4:** Singlet-quark distribution $xq_S^\gamma(x, Q^2, P^2)$ in DIS$_\gamma$ scheme. [12]

A heavy quark $h$ with mass $m_h^2$ (here, charm and beauty in mind) contributes to the structure function $F_2^\gamma(x, Q^2)$ when $(p + q)^2 = Q^2(\frac{1}{x} - 1) > (2m_h)^2$. There are several approaches in which the heavy-quark mass effects are taken into account. In the fixed flavour-number scheme (FFNS), heavy quarks only appear in the final state of the process and their contributions are described by the heavy-quark coefficient functions [13, 14]. The distributions $q_h(x, Q^2)$ are set to be zero. The FFNS is not appropriate when $Q^2 \gg m_h^2$. In the zero-mass variable flavour-number scheme (ZVFNS), heavy-quark distributions appear similar to the light partons. When $Q^2$ is larger than a threshold associated with a heavy quark (usually taken as $Q^2 = m_h^2$), this quark is considered as an extra massless parton [15]. The distribution $q_h(x, Q^2)$ differs from zero when $Q^2 > m_h^2$ but otherwise $q_h(x, Q^2) = 0$. The ZVFNS is not appropriate if $Q^2 \approx m_h^2$. The third one is called as the ACOT($\chi$) scheme [16, 17], which combines both features of the FFNS and ZVFNS. In order to treat the kinematical threshold for the heavy-quark production correctly, it introduces a new variable $\chi_h \equiv x(1 + 4m_h^2/Q^2)$ and modifies the integration range of convolution as follows: $\int_x^1 \frac{dy}{y} f(y)C(\frac{x}{y}) \Rightarrow \int_{\chi_h}^1 \frac{dy}{y} f(y)C(\frac{x}{y})$. Experimental data analyses of $F_2^\gamma$ with heavy-quark mass effects taken into account have been performed by several groups [13–15, 17, 18]. But we need more data to refine models for the treatment of heavy-quark contributions.

## 4 Photon structure functions — Future

With the discovery of the Higgs boson at the LHC, plans for building the next-generation $e^+e^-$ colliders [19] are attracting growing attention. In these collider machines, we may obtain highly polarized $e^+$ and $e^-$ beams. Using these polarized beams we can study another aspect of the photon: its spin structure. For a recent review see [20]. The QCD analysis of the polarized structure function $g_1^\gamma(x, Q^2)$ for a real photon target was performed in LO [21] and in NLO [22, 23]. The polarized virtual photon structure function $g_1^\gamma(x, Q^2, P^2)$ with $\Lambda^2 \ll P^2 \ll Q^2$ was investigated up to NLO in QCD [24]. At $P^2 = 0$, the structure function $g_1^\gamma$ satisfies a remarkable sum rule, which is non-perturbative and independent of $Q^2$, due to gauge invariance [25, 26]:

$$\int_0^1 g_1^\gamma(x, Q^2)dx = 0 \,. \tag{4}$$

But when the target photon becomes off-shell, $P^2 \neq 0$, the first moment of $g_1^\gamma(x, Q^2, P^2)$ does not vanish any more. The NLO result in QCD for the case $\Lambda^2 \ll P^2 \ll Q^2$ is [24, 26],

$$\int_0^1 dx g_1^\gamma(x, Q^2, P^2) = -\frac{3\alpha}{\pi}\left[\sum_{i=1}^{n_f} e_i^4\left(1-\frac{\alpha_s(Q^2)}{\pi}\right) - \frac{2}{\beta_0}\left(\sum_{i=1}^{n_f} e_i^2\right)^2\left(\frac{\alpha_s(P^2)}{\pi}-\frac{\alpha_s(Q^2)}{\pi}\right)\right] \,. \tag{5}$$

The first term in the square brackets resulted from the QED axial anomaly while the second term from the QCD axial anomaly. The sum rule was extended up to NNLO in QCD [27].

## 5 Epilogue

For future investigation on the photon structure, we still need to understand: (i) hadronic contributions to photon; (ii) heavy-quark mass effects; (iii) transition from real to virtual photon target; (iv) behaviours of $F_2^\gamma$ and parton distributions near $x = 0$ and $x = 1$; (v) the spin structure of photon. To that end it is essential for us to have more new experimental data on the photon structure.

## Acknowledgements

We wish to thank the organizers of Photon 2017 for the hospitality at such a stimulating conference.

# Initial State Radiation and Two-photon results from Belle

*C. P. Shen and Q. Y. Guo*
School of Physics and Nuclear Energy Engineering, Beihang University, Beijing 100191, People's Republic of China

**Abstract**

We review recent results from initial state radiation and two-photon processes at the Belle experiment. The results include the measurements of $e^+e^- \to \pi^+\pi^- J/\psi$, $\pi^+\pi^-\psi(2S)$, $K^+K^-J/\psi$, and $\gamma\chi_{cJ}$ from initial state radiation processes and the measurement of $\gamma\gamma \to p\bar{p}K^+K^-$ from two-photon process. We also summarize all the published results from these two kinds of processes. Moreover, we point out some golden channels that will be studied firstly with larger statistic at Belle II in the near future.

## 1 Initial state radiation results from Belle

The idea of utilizing initial state radiation (ISR) from a high-mass state to explore electron-positron processes at all energies below that state was outlined in Ref. [1]. The possibility of exploiting such processes in high luminosity $\phi-$ and B- factories was discussed in Refs. [2–4] and motivates the hadronic cross section measurement. The traditional way of measuring of the hadronic cross section via the energy scan has one disadvantage - it needs dedicated experiments. An alternative way, the study of ISR events at B-factories provides independent and contiguous measurements of hadronic cross sections in this energy region and also contributes to the investigation of low-mass resonances spectroscopy. So states with $J^{PC} = 1^{--}$ can be studied with ISR in Belle's and BaBar's large $\Upsilon(4S)$ data samples or via direct production in $e^+e^-$ collisions at BESIII.

The study of charmoniumlike states via ISR at the B-factories has proven to be very fruitful. Many of them show properties different from the naive expectation of conventional charmonium states. The first vector charmoniumlike state $Y(4260)$ was observed and confirmed by BaBar [5], CLEO [6] and Belle experiments [7]. Besides the $Y(4260)$, Belle also observed a broad excess near 4 GeV, called $Y(4008)$ [8]. With full of BaBar data sample 454 fb$^{-1}$, the $Y(4008)$ structure was not confirmed [9]. The difference on the measured cross section from BaBar and Belle at around 4.01 GeV is large. For $Y(4260) \to \pi^+\pi^- J/\psi$ decays, Belle also observed a clear charged signal in the distributions of $M_{\rm max}(\pi^\pm J/\psi)$, the maximum of $M(\pi^+ J/\psi)$ and $M(\pi^- J/\psi)$ [10]. The measured mass and width are $(3899.0 \pm 3.6 \pm 4.9)$ MeV$/c^2$ and $(63 \pm 24 \pm 26)$ MeV/$c^2$ with a signal significance greater than $5\sigma$. This state is close to the $D\bar{D}^*$ mass threshold and is called $Z_c(3900)$.

The $Y(4360)$ was firstly found by BaBar [11], while Belle observed two resonant structures at 4.36 and 4.66 GeV/$c^2$, denoted as the $Y(4360)$ and $Y(4660)$ [12]. BaBar confirmed the existence of the $Y(4660)$ state later [13]. Besides the $Y(4360)$ and $Y(4660)$ parameters are measured with improved precision with full 980 fb$^{-1}$ data sample [14], Belle also noticed there are a number of events in the vicinity of the $Y(4260)$ mass. But the signal significance of the $Y(4260)$ is only $2.4\sigma$. Evidence for a charged charmoniumlike structure at 4.05 GeV$/c^2$, denoted as the $Z_c(4050)$, was observed in the $\pi^\pm\psi(2S)$ intermediate state in the $Y(4360)$ decays, which might be the excited state of the $Z_c(3900)$.

Using a data sample of 673 fb$^{-1}$, Belle observed abundant $e^+e^- \to K^+K^-J/\psi$ signal events [15]. There is one very broad structure in the $K^+K^-J/\psi$ mass spectrum; fits using either a single Breit-Wigner (BW) function, or the $\psi(4415)$ plus a second BW function yield resonant parameters that are very different from those of the currently tabulated excited $\psi$ states. To examine possible structures in the $K^+K^-J/\psi$, $K^+K^-$ and $K^\pm J/\psi$ systems, updated measurement of $e^+e^- \to K^+K^-J/\psi$ between threshold and 6.0 GeV/$c^2$ with an integrated luminosity of 980 fb$^{-1}$ was performed, but no clear structure is observed in any system [16].

In order to improve the understanding of the nature of vector charmoniumlike states and search for more new states, Belle studied $e^+e^- \to \gamma\chi_{cJ}$ process using ISR events with $\chi_{cJ}$ reconstructed via $\gamma J/\psi$ [17]. The integrated luminosity used in this analysis is 980 fb$^{-1}$. After all the event selections, no significant signal is observed in either $\gamma\chi_{c1}$ or $\gamma\chi_{c2}$ mode in $M(\gamma\gamma J/\psi)$ distributions for $\gamma\chi_{c1}$ and $\gamma\chi_{c2}$ candidate events as well as the sum of them. The measured upper limits on the cross sections are around a few pb to a few tens of pb.

Figure 1 shows the measured $e^+e^- \to \pi^+\pi^-J/\psi$, $\pi^+\pi^-\psi(2S)$ and $K^+K^-J/\psi$ cross sections from Belle with full of data sample. Table 1 also summarized all the published ISR results at Belle for those processes with information of used total integrated luminosity, the range of center-of-mass (C.M.) energy, the related physics topics and the corresponding reference paper.

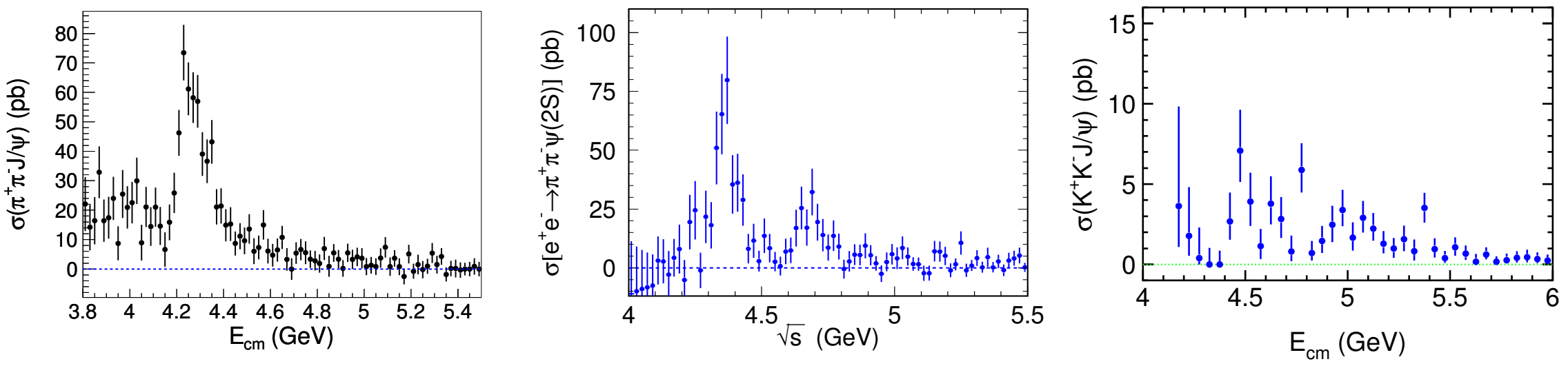

**Fig. 1:** The measured $e^+e^- \to \pi^+\pi^-J/\psi$ [10], $\pi^+\pi^-\psi(2S)$ [14] and $K^+K^-J/\psi$ [16] cross sections from Belle.

**Table 1:** Published ISR results at Belle for those processes with total integrated luminosity (Int. Lum.), the range of center-of-mass energy (C.M. energy), the related physics topics and the corresponding reference paper.

| Process | Int. Lum. | C.M. energy | Physics Covered | Reference |
|---|---|---|---|---|
| $D^{(*)\pm}D^{(*)\mp}$ | 547.8 fb$^{-1}$ | (3.9-5.0) GeV | cross sections | PRL 98, 092001(2007) |
| $DD_2^*(2460)$ | 673 fb$^{-1}$ | (4.0-5.0) GeV | $\psi(4415)$ | PRL 100, 062001 (2008) |
| $\Lambda_c^+\Lambda_c^-$ | 695 fb$^{-1}$ | (4.8-5.4) GeV | Y(4630) | PRL 101, 172001 (2008) |
| $D^0D^{*-}\pi^+$ | 695 fb$^{-1}$ | (4.1-5.2) GeV | Y(4260) | PRD80 101, 091101(R) (2009) |
| $D\bar{D}$ | 673 fb$^{-1}$ | (3.8-5.0) GeV | cross sections | PRD 77, 011103 (2008) |
| $\pi^+\pi^-J/\psi$ | 548 fb$^{-1}$ | (3.8-5.5) GeV | Y(4008),Y(4260) | PRD 99, 182004 (2007) |
| $\pi^+\pi^-\psi(2S)$ | 673 fb$^{-1}$ | (4.0-5.5) GeV | Y(4360),Y(4660) | PRD 99, 142002 (2007) |
| $K^+K^-J/\psi$ | 673 fb$^{-1}$ | (4.2-6.0) GeV | Y(4260) | PRD 77, 011105(R) (2008) |
| $\pi^+\pi^-\phi$ | 673 fb$^{-1}$ | (1.3-3.0) GeV | Y(2175),$\phi$(1680) | PRD 80, 031101 (2009) |
| $\eta J/\psi$ | 980 fb$^{-1}$ | (3.8-5.3) GeV | $\psi(4040)$, $\psi(4160)$ | PRD 87, 051101(R) (2013) |
| $\pi^+\pi^-J/\psi$ | 980 fb$^{-1}$ | (3.8-5.5) GeV | Y(4008),Y(4260), $Z_c(3900)$ | PRL 110, 252002 (2013) |
| $K^+K^-J/\psi$ | 980 fb$^{-1}$ | (4.4-5.2) GeV | Y(4260) | PRD 89, 072015 (2014) |
| $\pi^+\pi^-\psi(2S)$ | 980 fb$^{-1}$ | (4.0-5.5) GeV | Y(4260),Y(4360),Y(4660) | PRD 91, 112007 (2015) |
| $\gamma\chi_{cJ}$ | 980 fb$^{-1}$ | (3.8-5.6) GeV | cross sections | PRD 92, 012011 (2015) |

More data are necessary and better for ISR studies. Belle II will accumulate 10 ab$^{-1}$ (50 ab$^{-1}$) data at around $\Upsilon(4S)$ by 2020 (2024). Compared to the current BESIII, with ISR events the whole hadron spectrum is visible so that the line shape of the resonance and fine structures can be investigated. The disadvantage is the effective luminosity and detection efficiency are relative low although we have huge data sample. Figure 2 shows the effective luminosity from 3 to 5 GeV in the Belle II data samples.

We can see that, for 10 ab$^{-1}$ Belle II data, we have about 400–500 pb$^{-1}$ data for every 10 MeV in the range 4–5 GeV. Of course, the ISR analyses have a lower efficiency than in direct $e^+e^-$ collisions because of the extra ISR photons and the boost given to events along the beam direction. Even taking these effects into account, the full Belle II data sample, which corresponds to about 2,000–2,300 pb$^{-1}$ data for every 10 MeV from 4–5 GeV, will result in similar statistics for modes like $e^+e^- \to \pi^+\pi^- J/\psi$ at BESIII currently. Belle II has the advantage that data at different energies will be accumulated at the same time, making the analysis much simpler than at BESIII at 60 data points. In addition, Belle II can produce events above 4.6 GeV, which is currently the maximum energy of BEPCII.

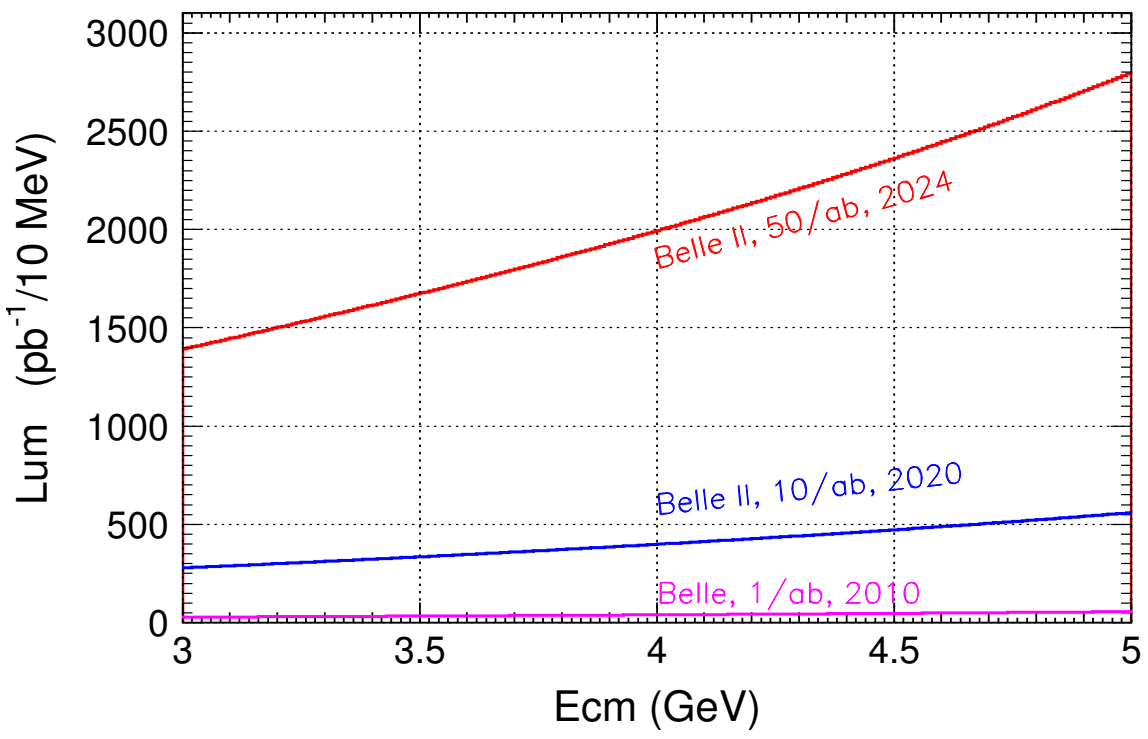

**Fig. 2:** Effective luminosity at low energy in the Belle and Belle II $\Upsilon(4S)$ data samples.

## 2 Two-photon results from Belle

At $e^+e^-$ collider, two-photon interactions are studied via the process $e^+e^- \to e^+e^-\gamma^*\gamma^* \to e^+e^-R$. Almost all of the beam energy is caught by scattered electron and positron and usually they are not detected (no-tagged events). If the scattering angle is sufficiently large, they can be detected in the forward region (tagged events). Many two-photon processes have been done by Belle. Table 2 summarizes the published results from two-photon processes with information of the analyzed mode, the used integrated luminosity (Int. Lum.), the $\gamma\gamma$ C.M. energy, the studied light mesons/baryons and charmonium states, and the corresponding reference paper.

The latest published two-photon results are from $\gamma\gamma \to p\bar{p}K^+K^-$ in order to search for exotic baryons. The LHCb observed two exotic structures, $P_c(4380)^+$ and $P_c(4450)^+$, in the $J/\psi p$ system in $\Lambda_b^0 \to J/\psi K^- p$ [18]. These two $P_c$ states must consist of at least five quarks. Several theoretical interpretations of these states have been developed, such as the diquark picture and hadronic molecules. Actually, the first strong experimental evidence for a pentaquark state, referred to as the $\Theta(1540)^+$, was reported in the reaction $\gamma n \to nK^+K^-$ in the LEPS experiment [19]. It was a candidate for a $uudd\bar{s}$ pentaquark state. However, it was not confirmed in larger-statistics data samples in the same experiment and was most probably not a genuine state [20].

To confirm the pentaquark states discovered by LHCb, further experimental searches for exotic baryons are necessary. The possibility of observing additional hypothetical exotic baryons in $\gamma\gamma$ collisions is discussed in Ref. [21]. The cross section for the reaction $\gamma\gamma \to p\bar{p}K^+K^-$ is predicted to be around 0.1 nb for $W_{\gamma\gamma} \geq 2(m_p + m_K)$ [21], where $W_{\gamma\gamma}$ is the C.M. energy of two-photon system. This presents the opportunity to search for novel exotic baryons, denoted as $\Theta(1540)^0 \to pK^-$ and $\Theta(1540)^{++} \to pK^+$ which are similar to $\Theta(1540)^+$, in intermediate processes in two-photon annihilations.

The process $\gamma\gamma \to p\bar{p}K^+K^-$ and its intermediate processes are measured for the first time using a 980 fb$^{-1}$ data sample by Belle [22]. The production of $p\bar{p}K^+K^-$ and a $\Lambda(1520)^0$ ($\bar{\Lambda}(1520)^0$) signal in the $pK^-$ ($\bar{p}K^+$) invariant mass spectrum are clearly observed. However, no evidence for an exotic baryon near 1540 MeV/$c^2$, i.e., $\Theta(1540)^0$ ($\bar{\Theta}(1540)^0$) or $\Theta(1540)^{++}$ ($\Theta(1540)^{--}$), is seen in the $pK^-$ ($\bar{p}K^+$) or $pK^+$ ($\bar{p}K^-$) invariant mass spectra. Cross sections for $\gamma\gamma \to p\bar{p}K^+K^-$, $\Lambda(1520)^0\bar{p}K^+ + c.c.$ and the products $\sigma(\gamma\gamma \to \Theta(1540)^0\bar{p}K^+ + c.c.)\mathcal{B}(\Theta(1540)^0 \to pK^-)$ and $\sigma(\gamma\gamma \to \Theta(1540)^{++}\bar{p}K^- + c.c.)\mathcal{B}(\Theta(1540)^{++} \to pK^+)$ are measured. The cross sections for $\gamma\gamma \to p\bar{p}K^+K^-$ are lower by a factor 2.5 or more than the theoretical prediction of 0.1 nb in Ref. [21].

**Table 2:** Published results from two-photon processes with information of the analyzed mode (Process), the used integrated luminosity (Int. Lum.), the $\gamma\gamma$ C.M. energy, the studied light mesons/baryons and charmonium states (Charm.), and the corresponding reference paper.

| Process | Int. Lum. | $\gamma\gamma$ C.M. energy | Light mesons/baryons | Charm. | Reference |
|---|---|---|---|---|---|
| $\gamma J/\psi$ | 32.6 fb$^{-1}$ | (3.2-3.8) GeV | — | — | PLB 540, 33(2002) |
| $K^+K^-$ | 67 fb$^{-1}$ | (1.4-2.4) GeV | $f_2'(1525)$ | — | EPJC 32, 323(2003) |
| $\pi^+\pi^-/K^+K^-$ | 87.7 fb$^{-1}$ | (2.4-4.1) GeV | — | $\chi_{c0,c2}$ | PLB 615, 39(2005) |
| $p\bar{p}$ | 89 fb$^{-1}$ | (2.03-4.0) GeV | — | $\eta_c$ | PLB 621, 41(2005) |
| $D\bar{D}$ | 395 fb$^{-1}$ | (3.7-4.3) GeV | — | $\chi_{c2}'$ | PRL 96, 082003(2006) |
| $\pi^+\pi^-$ | 85.9 fb$^{-1}$ | (0.8-1.5) GeV | $f_0(980)$, $f_2(1270)$, $\eta'$ | — | PRD 75, 051101(2007) |
| $K_s^0K_s^0$ | 397.6 fb$^{-1}$ | (2.4-4.0) GeV | — | $\chi_{c0,c2}$ | PLB 651, 15(2007) |
| four mesons | 395 fb$^{-1}$ | (1.4-3.4) GeV | — | $\chi_{c0,c2}$<br>$\eta_c(1S,2S)$ | EPJC 53, 1(2008) |
| $\pi^0\pi^0$ | 95 fb$^{-1}$ | (0.6-4.0) GeV | $f_0(980)$, $f_2(1270)$<br>$f_2'(1525)$ | $\chi_{c0,c2}$, $\eta_c$ | PRD 78, 052004(2008) |
| $\pi^0\pi^0$ | 223 fb$^{-1}$ | (0.6-4.1) GeV | $f_4(2050)$, $f_2(1950)$ | $\chi_{c0,c2}$ | PRD 79, 052009(2009) |
| $\eta\pi^0$ | 223 fb$^{-1}$ | (0.84-4.0) GeV | $a_0(980)$, $a_0(1450)$<br>$a_2(1320)$ | —<br>— | PRD 80, 032001(2009) |
| $\phi J/\psi$ | 825 fb$^{-1}$ | (4.2-5.0) GeV | — | $X(4350)$ | PRL 104, 112004(2010) |
| $\omega J/\psi$ | 694 fb$^{-1}$ | (3.9-4.2) GeV | — | $X(3915)$ | PRL 104, 092001(2010) |
| $\eta\eta$ | 393 fb$^{-1}$ | (1.096-3.8) GeV | $f_2(1270)$, $f_2'(1525)$ | $\chi_{c0,c2}$ | PRD 82, 114031(2010) |
| $\omega\omega$, $\omega\phi$, $\phi\phi$ | 870 fb$^{-1}$ | $< 4.0$ GeV | — | $\chi_{c0,c2}$, $\eta_c$ | PRL 108, 232001(2012) |
| $\gamma\gamma^* \to \pi^{0a}$ | 759 fb$^{-1}$ | $4 < Q^2 < 40$ GeV$^{2b}$ | — | — | PRD 86, 092007(2012) |
| $\eta'\pi^+\pi^-$ | 673 fb$^{-1}$ | (1.4-3.4) GeV | $\eta(1760)$, $X(1835)$ | $\eta_c$ | PRD 86, 052002(2010) |
| $K_s^0K_s^0$ | 972 fb$^{-1}$ | (1.04-4.0) GeV | $f_2(1270)$, $a_2(1320)$<br>$f_2'(1525)$, $f_0(1710)$ | $\chi_{c0,c2}$<br>$\eta_c(2S)$ | PTEP 2013, 123C01 (2013) |
| $\gamma\gamma^* \to \pi^0\pi^{0a}$ | 759 fb$^{-1}$ | $Q^2 < 30$ GeV$^{2b}$ | $f_0(980)$, $f_2(1270)$ | — | PRD 93, 032003(2016) |
| $p\bar{p}K^+K^-$ | 980 fb$^{-1}$ | (3.2-5.6) GeV | $\Lambda(1520)$, $\Theta(1540)$ | $\chi_{c0,c2}$ | PRD 93, 112017(2016) |

[a] $\gamma^*$ denotes a virtual photon.
[b] -$Q^2$ is the invariant-mass squared of a virtual (spacelike) photon.

Experimental studies for two-photon physics at Belle II have merits since all the data at any energy point can be used to investigate the lower invariant mass region. Physics at higher invariant mass region, $W_{\gamma\gamma} > 5$ GeV, is not suitable because the luminosity fucntion for two-photon collisions steeply decreases with increasing $W_{\gamma\gamma}$ and the backgrounds from single photon annihilation processes are considerable.

With the total integrated luminosity larger than 10 ab$^{-1}$ at Belle II, the below two-photon processes are our priorities: (1) $\gamma\gamma \to D\bar{D}$ to study $Z(3930)$ and search for $\chi_{c0}(2P)$. We expect obvious contributions from $\gamma\gamma \to D\bar{D}^* \to D\bar{D}\pi$ and $\gamma\gamma \to D\bar{D}(n)\gamma$. All of these processes are cross contaminated. Their cross sections can be measured by doing independent analysis and using iteration method. (2) $\gamma\gamma \to \phi J/\psi$ to confirm $X(4350)$ and search for other $X$ states.

## 3 Summary and discussion

Although dramatic progresses have been made on the study of the ISR and two-photon processes, which has improved the understanding of the light mesons, light baryons, $XYZ$ states and the conventional charmonium states greatly, there are more questions to be answered with the currently available data. All of above mentioned processes will be updated to give improved precision and more new reactions will be measured on the cross sections and search for more resonances or decay modes with larger statistic at Belle II.

## Acknowledgements

Supported in part by National Natural Science Foundation of China (NSFC) under contract No. 11575017; the Ministry of Science and Technology of China under Contract No. 2015CB856701; and the CAS Center for Excellence in Particle Physics (CCEPP).

# $F_2^\gamma$ at the ILC, CLIC and FCC-ee

*B. Krupa*, T. Wojtoń and L. Zawiejski, on behalf of the FCAL Collaboration*
Institute of Nuclear Physics PAN, ul. Radzikowskiego 152, 31-342 Cracow, Poland

**Abstract**
Despite many studies of the photon structure done in the past, still lots of issues need to be investigated. Continued research in this field would greatly benefit from measurements performed at future $e^+e^-$ colliders. The potential of this field of research at the ILC, CLIC and FCC-ee is studied. In order to check the possibility of measuring the photon structure functions at future experiments, simulations of two-photon processes are performed. This paper presents the obtained results.

## 1 Motivation

The earliest photons probably appeared about fourteen billion years ago, during the Big Bang. Unlike electrons and quarks, photons have no mass ($m < 1 \cdot 10^{-18}$ eV), so they can travel in vacuum at the speed of light. They are chargeless ($q < 1 \cdot 10^{-35} e$) [1] gauge bosons of quantum electrodynamics (QED) having no internal structure in the common sense. However, according to quantum field theory, the existence of interactions carried by the gauge boson means that this boson can develop some structure. A photon, for example, can fluctuate for a short amount of time to a lepton-antilepton or quark-antiquark pair. Photons can therefore interact with other particles directly as a whole or through particles produced by their quantum fluctuations. The diversity of the photons' behaviour allows us to investigate its leptonic or hadronic nature. The quantities used to describe the structure of a photon are photon structure functions.

The experiments measuring these functions have until now only been carried out at electron-positron colliders or electron-proton storage rings [2], where the lepton beams serve as the source of high energy photons. The first measurement was performed using the detector PLUTO at the DESY storage ring PETRA (1981) [3]. Following this pioneering work many experiments have been conducted, yet there are still a lot of problems to solve. Thus it is essential to continue the studies. However, the last papers containing experimental data on the photon structure functions were published in 2005. They were the papers of the two-photon working group from the L3 experiment at LEP [4]. New experimental data can be anticipated from a new linear electron-positron collider: ILC, CLIC or FCC-ee. As the beam energy at the ILC / CLIC will be higher than at LEP, it is expected that it will be possible to measure the $Q^2$ evolution of the structure function $F_2^\gamma$ in a wider range. On the other hand, although the beam energy at FCC-ee will be lower than at CLIC, thanks of its high luminosity [5] this machine will offer large statistics. Therefore, one can expect valuable results on many issues related to the photon structure function. It would be moreover interesting to study the structure function for highly virtual photons, because the interaction of two virtual photons is the so-called 'golden' process to distinguish between DGLAP and BFKL parton dynamics [6]. For this purpose, the ability to tag both scattered electrons (the so-called double-tagged events) is needed. It would allow us also to determine the invariant mass squared of the $\gamma\gamma$ system $W^2$ independently of the hadronic final state and thus to increase the precision of the measurement of the photon structure function [7]. Additionally, new light on the photon structure would be shed by spin-dependent structure functions (more details in [8,9]), which have not been measured so far.

*e-mail: beata.krupa@ifj.edu.pl

This would be possible in the polarized $e^+e^-$ collisions at the future colliders. Furthermore, two-photon processes are a background for physical analyses in which signals from new physics (physics beyond the Standard Model) are being searched for, therefore the knowledge of their nature is important [10].

## 2 Expected measurement capabilities

By analyzing the possibilities of measuring the structure function $F_2^\gamma$ in the experiments at the future $e^+e^-$ colliders, the expected numbers of events of deep inelastic electron-photon scattering was estimated. The estimates assumed the predicted values of luminosities for the various stages of these experiments [5, 11, 12], whereas the cross sections were determined using the Monte Carlo generator PYTHIA 6.4 [13]. The projections for the ten-month period of data collection are presented in Table 1. The number of events included in the last column of the table have been estimated assuming the detection of scattered electron at the LumiCal detector. This detector will be the luminometer of the future experiment what will be run on ILC/CLIC or FCC-ee using Bhabha scattering as the gauge process [11, 14–16].

| | sqrt(s) [GeV] | Luminosity $[10^{34}cm^{-2}s^{-1}]$ | $\sigma(e^+e^- \to e^+e^-\gamma\gamma \to e^+e^-hadrons)$ [pb] | Acceptance | N(events) |
|---|---|---|---|---|---|
| FCC-ee | 90 | 68.0 | 287.8 | 0.170 | $8.6 \times 10^8$ |
| | 160 | 19.0 | 419.4 | 0.076 | $1.6 \times 10^8$ |
| | 240 | 4.9 | 540.8 | 0.042 | $2.9 \times 10^7$ |
| | 350 | 1.3 | 674.1 | 0.024 | $5.4 \times 10^6$ |
| ILC | 500 | 1.8 | 823.5 | 0.040 | $15.4 \times 10^6$ |
| | 1000 | 3.6 | 1035.5 | 0.040 | $38.8 \times 10^6$ |
| CLIC | 1500 | 3.7 | 1212.6 | 0.015 | $4.6 \times 10^6$ |
| | 3000 | 5.9 | 1425.1 | 0.015 | $8.7 \times 10^7$ |

**Table 1:** Estimated number of events of deep inelastic electron-photon scattering per 10 months of data collection at FCC-ee, ILC, and CLIC assuming tagging of scattered electrons at the LumiCal detector.

The distributions of polar angles at which these electrons are scattered are shown in Figures 1 and 2. They show clearly that with increasing beam energy, the more electrons are scattered at small angles. Therefore, they become more difficult to register in the detectors and, consequently, we lose a large number of events. Thus despite the fact that the cross section increases with the energy (cf. Fig. 3), the measurement capabilities decrease.

The kinematic variable dependent on the polar angle $\Theta$ of the scattered electron is the virtuality of the photon:

$$Q^2 = 4E_bE\sin^2\left(\frac{\theta}{2}\right), \tag{1}$$

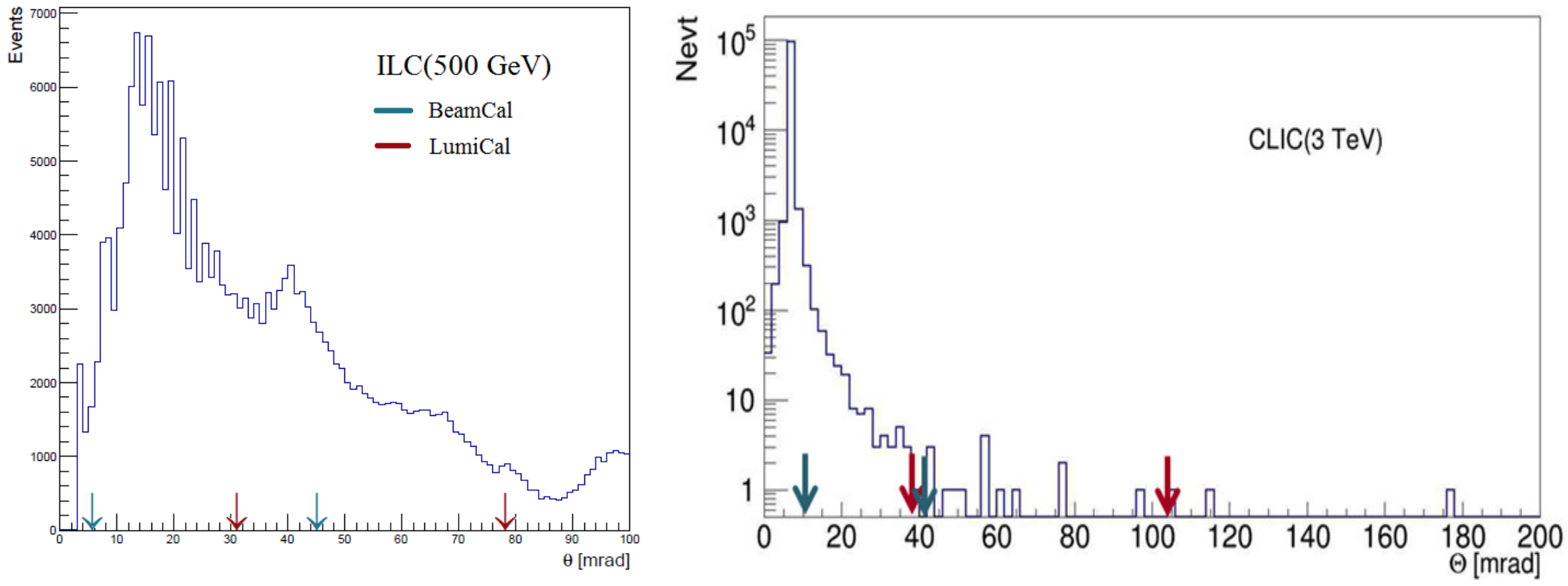

**Fig. 1:** The distributions of polar angles of scattered electrons expected in the energy range of the linear collider: ILC (left), and CLIC (right). Arrows indicate the angular acceptances for the LumiCal (red arrow) and BeamCal (blue arrow) detectors. The BeamCal will be used to measure remnants of beamstrahlung, to assist beam tuning, as well as to detect high energy single electrons [14].

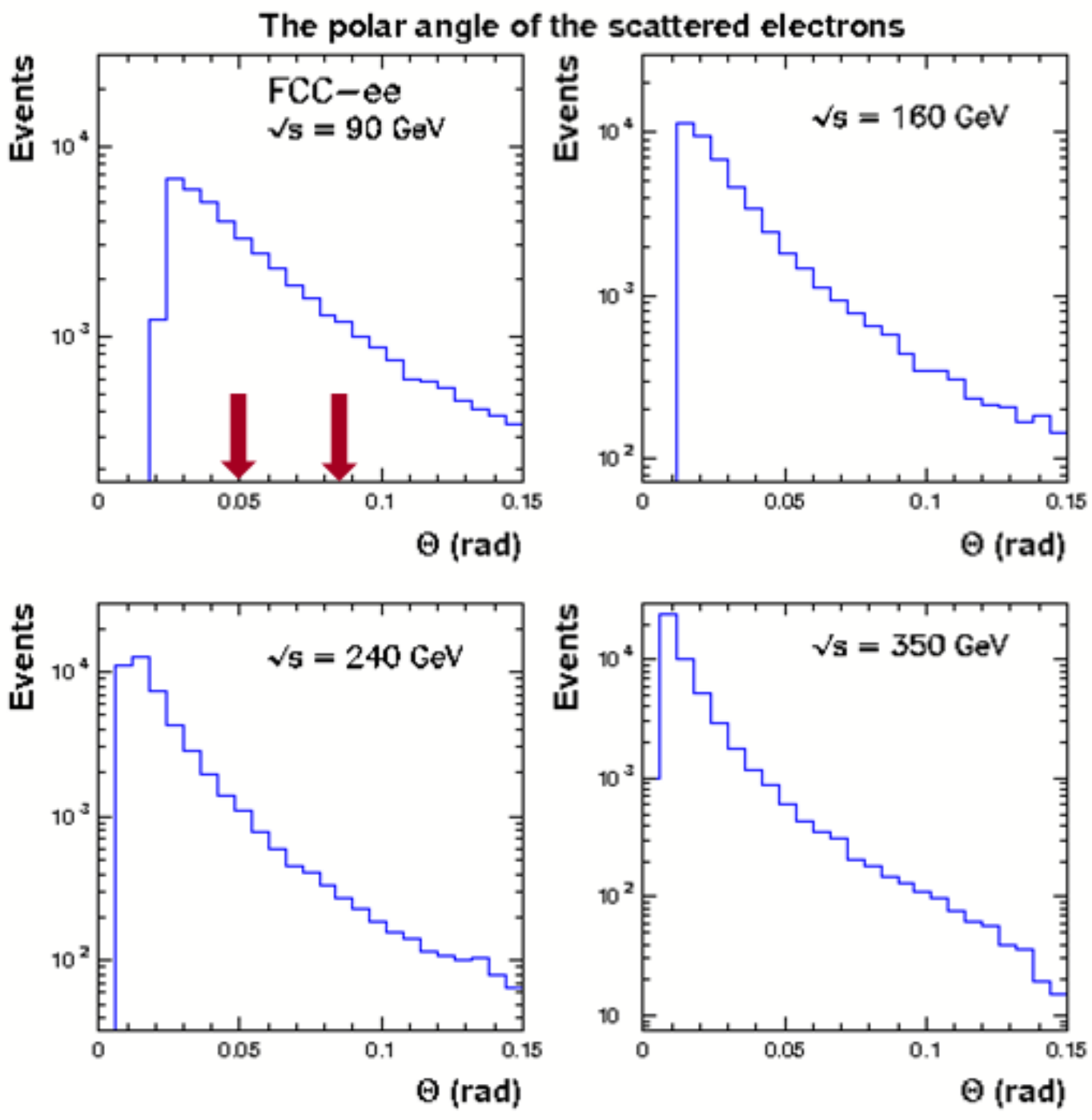

**Fig. 2:** The polar angle of the scattered electrons for energies proposed for FCC-ee project. The arrows indicate the indicative acceptance range for the LumiCal detector.

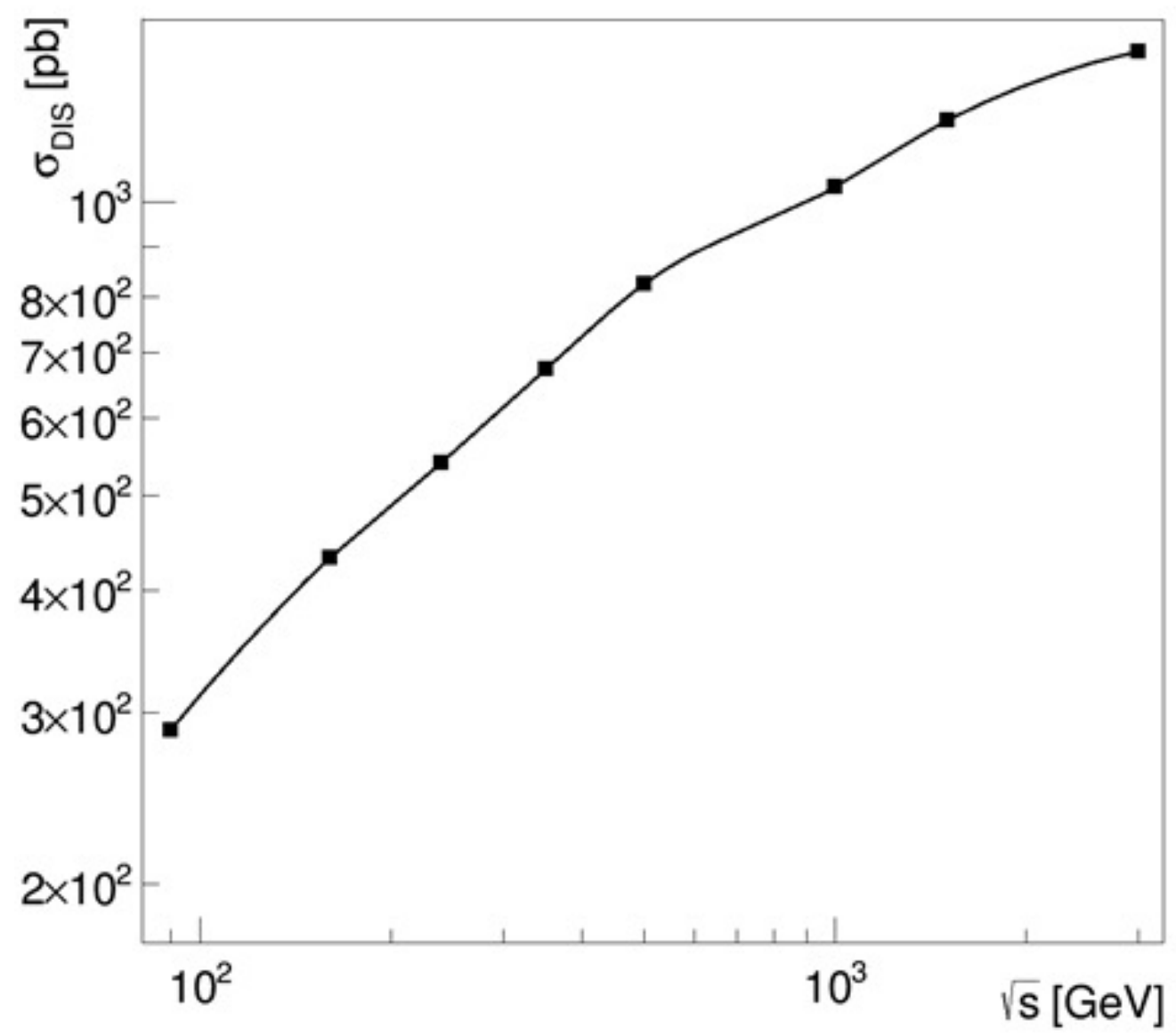

**Fig. 3:** Dependence of the cross section on the process $e^+e^- \xrightarrow{\gamma\gamma} e^-e^+$ hadrons on the center-of-mass collision energy. The results were obtained using the PYTHIA 6.4 generator.

where $E_b$ is the energy of the beam electron and E refer to the energy of the scattered electron. The $Q^2$ should take higher values for the higher beam energy, as it was confirmed by the simulations. Figure 4 shows the distributions of variable $Q^2$ obtained using three different Monte Carlo generators (PYTHIA 6.4 [13], TWOGAM (2.04) [17], HERWIG 6.5 [18]) for ILC (500 GeV) and CLIC (3000 GeV). They show a clear shift towards higher values of $Q^2$ with increasing beam energy. Also differences in the results obtained from different generators are visible. This result provides additional motivation to design future experiments to investigate photon properties, since only the experimental data will validate the models adopted in each generator.

Another kinematic variable describing the process of deep inelastic electron-photon scattering is the Bjorken variable $x$, which is determined experimentally from the formula:

$$x = \frac{Q^2}{(Q^2 + W^2 + P^2)}, \tag{2}$$

where $W^2$ is the invariant mass squared of $\gamma\gamma$ system, $P^2$ is the virtuality of the second photon and is usually assumed to be zero. The $W^2$ is determined from the energy and momenta of the particles in the final state. It is therefore fraught with high uncertainty due to the limited detection capabilities of the produced particles (angular acceptances of detectors). The effect of the available angular ranges in which the produced hadrons are detected on the distribution of the variable $x$ illustrate the histograms shown in Figure 5. The blue colour indicates the distribution in the ideal case when all produced hadrons are seen in the detector. The red colour corresponds to the results obtained with the adoption of certain (indicated on the individual histograms) conditions for these polar angles of the hadrons. In addition, the lines in green correspond to situations where hadrons are not detected within the LHCAL angular acceptance range. LHCAL is the hadronic calorimeter proposed for ILC detectors which will measure hadrons at small produced at small polar angles. The results show that the ability to detect hadrons produced at low polar angles (e.g. using a LHCAL detector) can improve the accuracy of the measurement of the effective mass W and thus increase the precision of the determination of the variable $x$.

Determined $x$ and $Q^2$ variables were used to calculate the photon structure function. Example results of PYTHIA 6.4 Monte Carlo studies, as well as those obtained using the reconstructed kinematic variables, are presented in Figures 6 and 7.

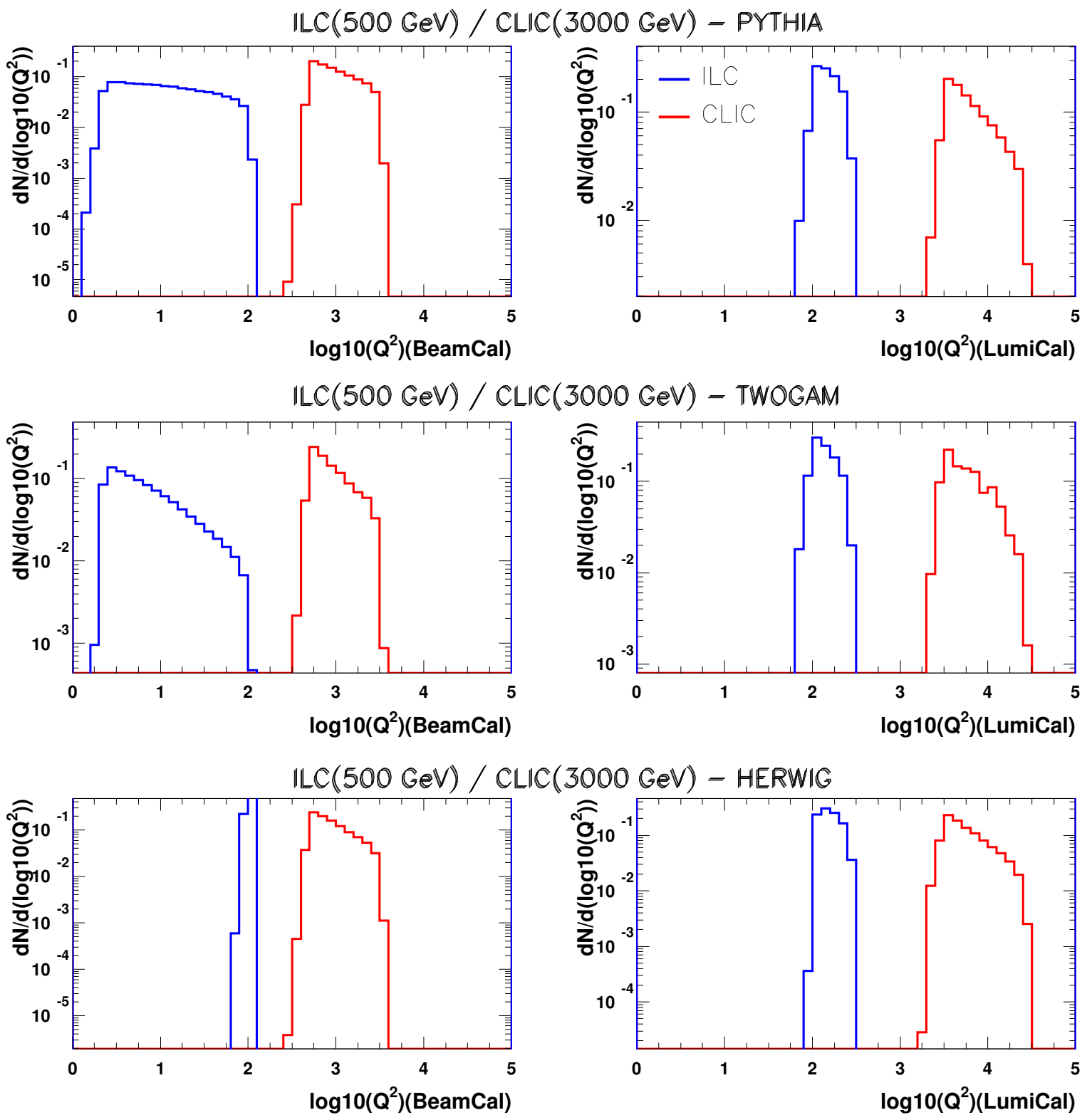

**Fig. 4:** Distributions of the variable $Q^2$ obtained using three Monte Carlo generators (PYTHIA 6.4, TWOGAM 2.04, HERWIG 6.5) for photon structure studies at the ILC(500 GeV) and CLIC(3 TeV).

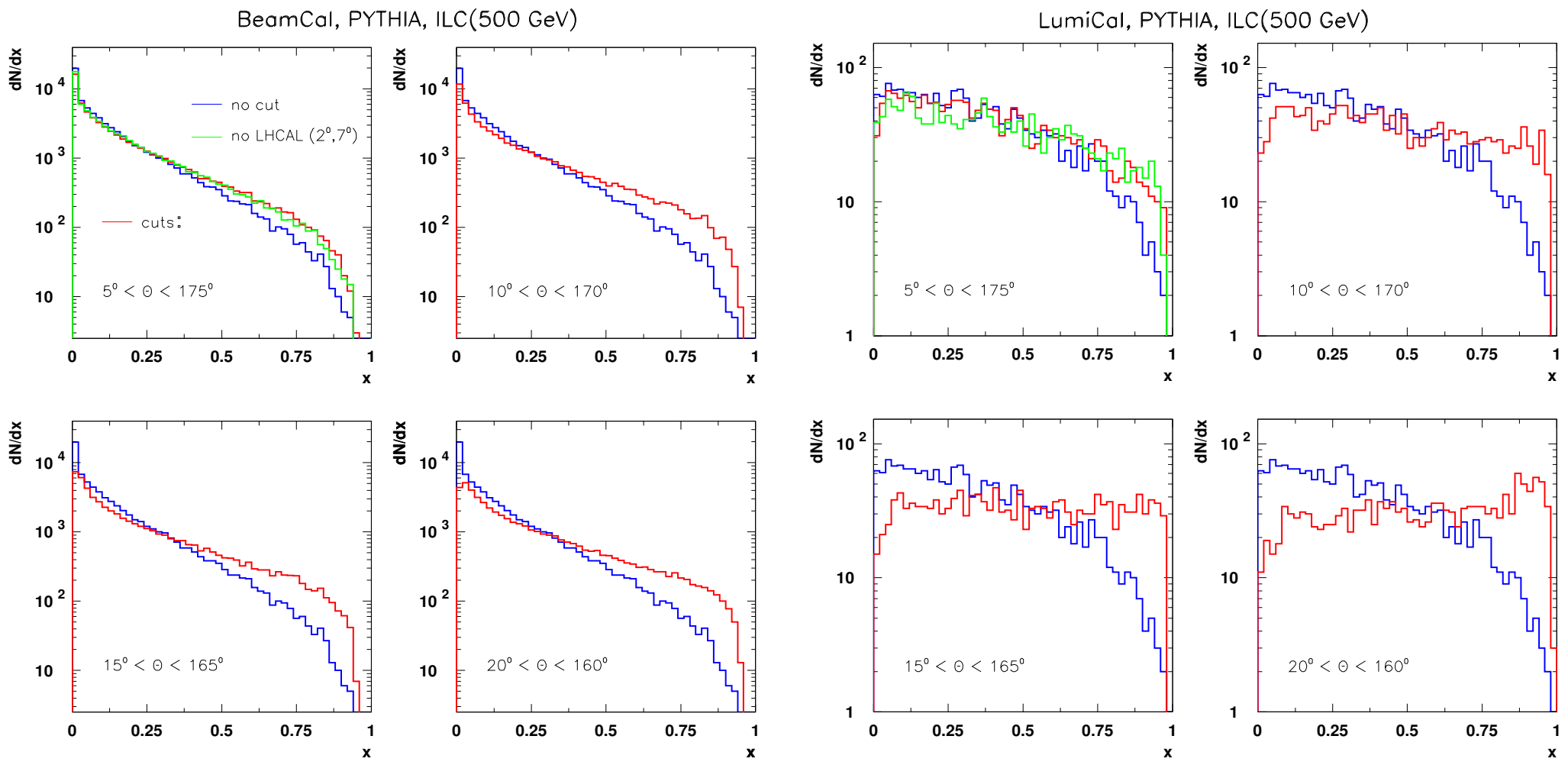

**Fig. 5:** The distribution of the $x$ variable in the case when scattered electrons are detected only in the BeamCal detector (left) and only in the LumiCal detector (right) depending on the available angular ranges within which it is possible to register hadrons produced in the process $e^+e^- \xrightarrow{\gamma\gamma} e^-e^+$ hadrons. The range 2–7 degrees was adopted as polar angle acceptance for the LHCAL detector.

The obtained results indicate that the forward detectors will allow the photon structure function in $e\gamma$ DIS processes to be measured. Available angular ranges will make it possible to obtain the functions in a wide range of $Q^2$. Since the results obtained at the reconstruction level differ from the results obtained at the generation level, it is necessary to introduce corrections due to the detector effects.

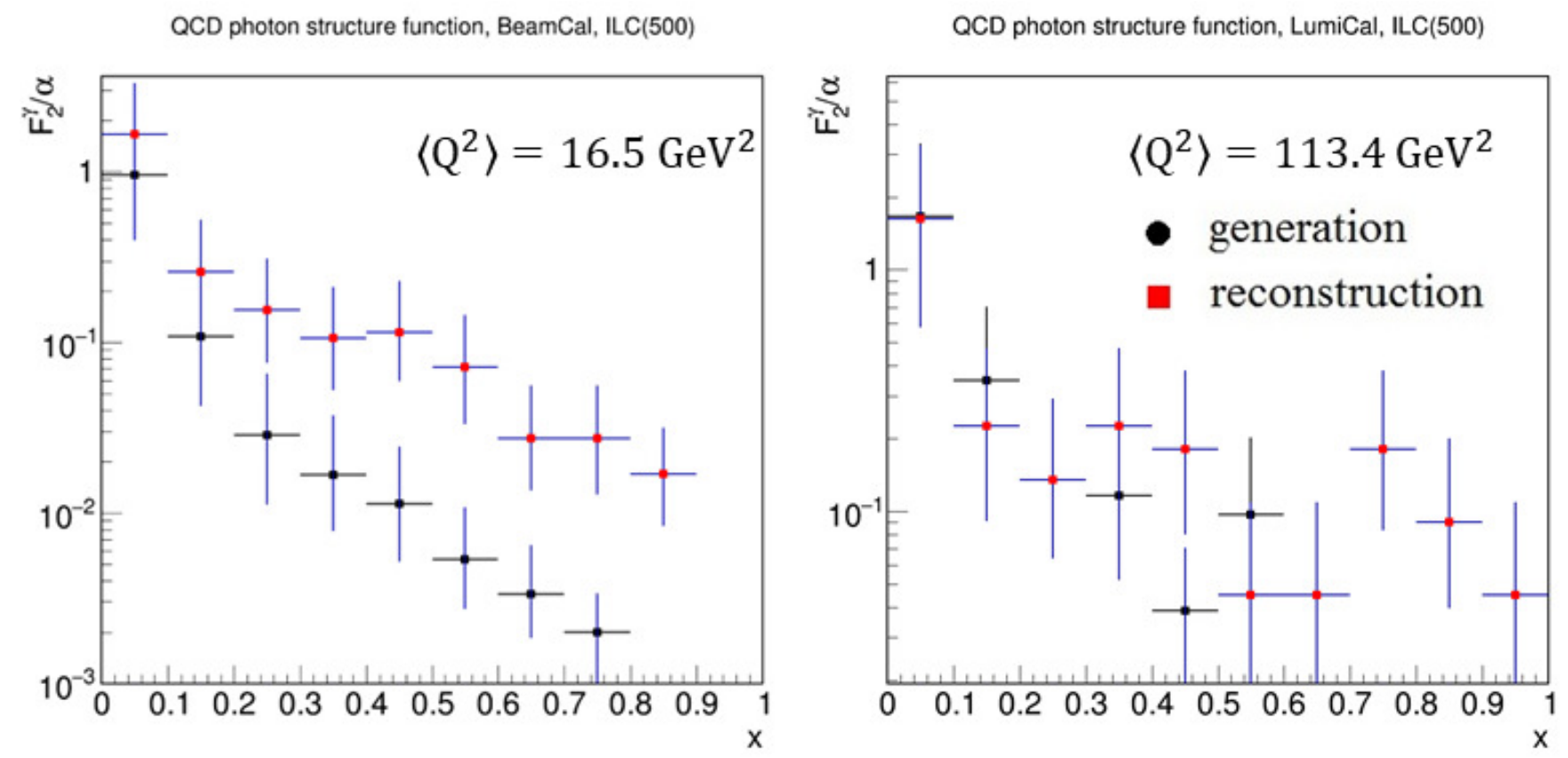

**Fig. 6:** The hadronic photon structure functions divided by the fine structure constant as a function of $x$ variable in the case of tagging the scattered electrons at BeamCal (left) and LumiCal (right) detectors planned for ILC (500 GeV). Distributions at generator (dots) and reconstruction (squares) levels are shown. The uncertainties marked on these plots are statistical only.

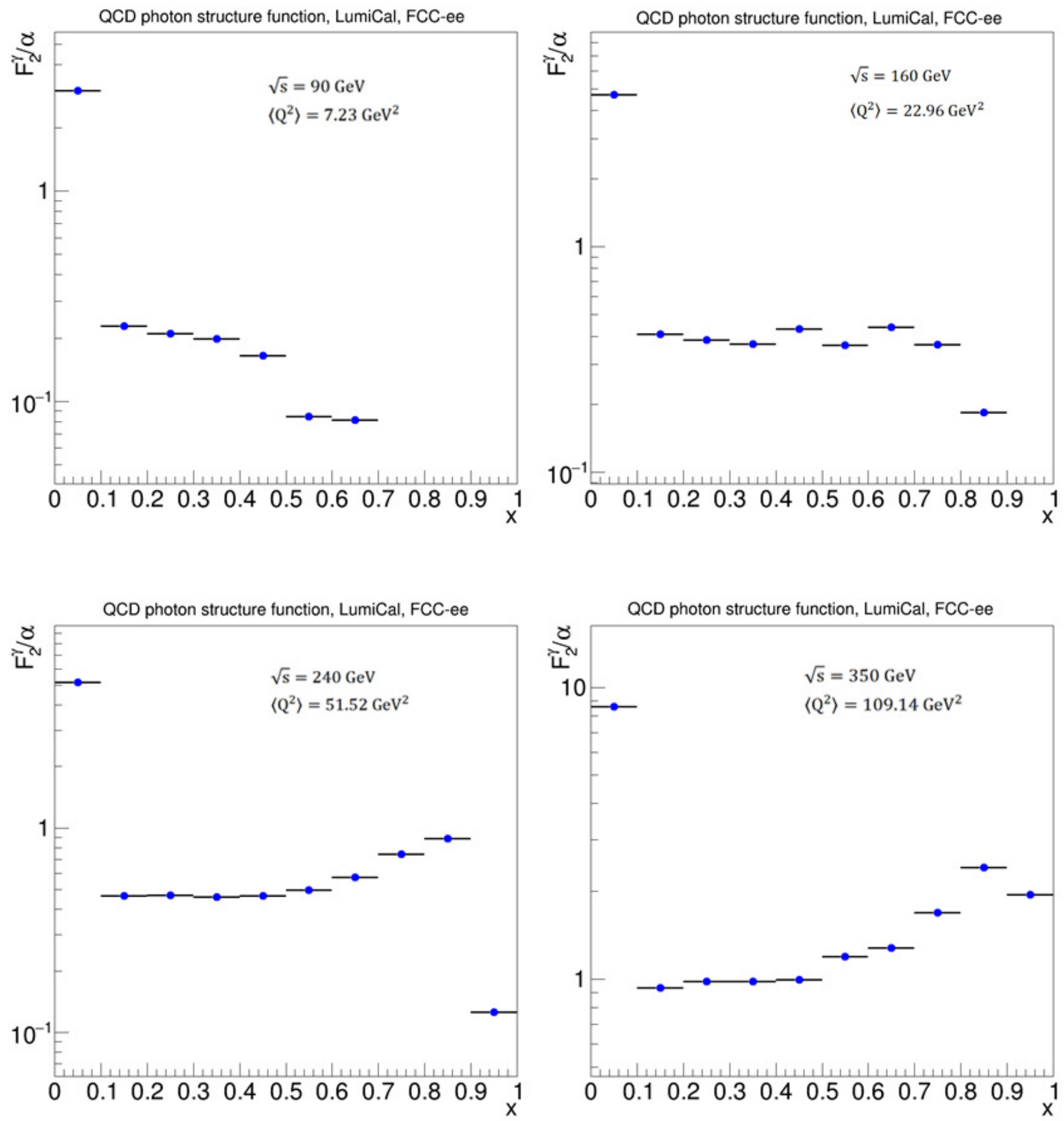

**Fig. 7:** The hadronic photon structure functions divided by the fine structure constant as a function of $x$ variable in the case of tagging the scattered electrons at LumiCal detector planned for FCC-ee. Generator-level distribution with negligible statistical uncertainties.

# Higgs boson production at an electron-photon collider

*N. Watanabe, Y. Kurihara*[1], *T. Uematsu*[2], *K. Sasaki*[3]
1. High Energy Accelerator Research Organization (KEK), Tsukuba, Ibaraki 305-0801, Japan
2. Institute for Liberal Arts and Sciences, Kyoto University, Kyoto 606-8501, Japan
3. Faculty of Engineering, Yokohama National University, Yokohama 240-8501, Japan

**Abstract**
We investigate the Standard Model Higgs boson production in $e^-\gamma$ collisions. The electroweak one-loop contributions to the scattering amplitude for $e^-\gamma \to e^-H$ are calculated and expressed in analytical form. We analyze the cross section for the Higgs boson production in $e^-\gamma$ collisions for each combination of polarizations of the initial electron and photon beams. The feasibility of observing the Higgs boson in $e^- + \gamma \to e^- + b + \bar{b}$ channel is examined.

## 1 Introduction

After a Higgs boson with mass 125 GeV was discovered by ATLAS and CMS at LHC [1], there has been growing interest in building a new accelerator facility [2], a linear $e^+e^-$ collider, which offers much cleaner experimental collisions. Along with $e^+e^-$ collider, other options such as $e^-e^-$, $e^-\gamma$ and $\gamma\gamma$ colliders have also been discussed [3, 4]. An $e^-e^-$ collider is easier to build than an $e^+e^-$ collider and may stand as a potential candidate before positron sources with high intensity are available. The $e^-\gamma$ and $\gamma\gamma$ options are based on $e^-e^-$ collisions, where one or two of the electron beams are converted to the photon beams. In [5] we analyzed the Standard Model (SM) Higgs boson production in a $e^-$ and real $\gamma$ collision experiment,

$$e^-(k_1) + \gamma(k_2) \to e^-(k_1') + H(p_h) \ . \tag{1}$$

In this talk we summarize the results of [5].

## 2 Higgs boson production in $e^-\gamma$ collisions

The relevant Feynman diagrams for the process (1) start at the one-loop level in the electroweak interaction. We calculate the relevant one-loop diagrams in unitary gauge using dimensional regularization which respects electromagnetic gauge invariance. The one-loop diagrams which contribute to the reaction (1) are classified into four groups: $\gamma^*\gamma$ fusion diagrams (Fig. 1), $Z^*\gamma$ fusion diagrams, "$W\nu_e$" diagrams (Fig. 2) and "$Ze$" diagrams (Fig. 3).

Since $k_2$ is the momentum of a real photon, we have $k_2^2 = 0$ and $k_2^\beta \epsilon_\beta(k_2) = 0$, where $\epsilon_\beta(k_2)$ is the photon polarization vector. We set $q = k_1 - k_1'$. Assuming that electrons are massless so that $k_1^2 = {k_1'}^2 = 0$, we introduce the following Mandelstam variables: $s = (k_1 + k_2)^2$, $t = (k_1 - k_1')^2 = q^2$, $u = (k_1 - p_h)^2 = m_h^2 - s - t$, where $p_h^2 = m_h^2$ with $m_h$ being the SM Higgs boson mass.

Charged fermions and $W$ boson contribute to the one-loop $\gamma^*\gamma$ fusion diagrams. Since the couplings of the Higgs boson to fermions are proportional to the fermion masses, we only consider the top quark for the charged fermion loop diagrams. The $\gamma^*\gamma$ fusion diagrams we calculate are shown in Fig. 1. We obtain the contribution from the one-loop $\gamma^*\gamma$ fusion diagrams to the gauge-invariant scattering amplitude as follows:

$$A_{\gamma\gamma} = \left(\frac{e^3 g}{16\pi^2}\right)\left[\overline{u}(k_1')\gamma_\mu u(k_1)\right]\frac{1}{t}\left(g^{\mu\beta} - \frac{2k_2^\mu q^\beta}{m_h^2 - t}\right)\epsilon_\beta(k_2)\ F_{\gamma\gamma}\ , \tag{2}$$

with

$$F_{\gamma\gamma} = \frac{2m_t^2}{m_W} N_c Q_t^2 \, S^{\gamma\gamma}_{(T)}(t, m_t^2, m_h^2) - m_W S^{\gamma\gamma}_{(W)}(t, m_W^2, m_h^2) \, , \tag{3}$$

where $e$, $g$, $m_t$ and $m_W$ are the electromagnetic and weak gauge couplings, the top-quark and $W$-boson masses, respectively, and $N_c = 3$ and $Q_t = \frac{2}{3}$. $S^{\gamma\gamma}_{(T)}$ and $S^{\gamma\gamma}_{(W)}$ are contributions from top loops and $W$ loops, respectively, and are expressed in terms of the Passarino-Veltman two-point integrals $B_0$'s and three-point integrals $C_0$'s [6]. For their explicit expressions, see Eqs.(2.6) and (2.7) of Ref. [5]. Although the two-point integrals $B_0$'s have ultraviolet divergences, these divergences are cancelled out when they are added and, therefore, $S^{\gamma\gamma}_{(T)}$ and $S^{\gamma\gamma}_{(W)}$ give finite results.

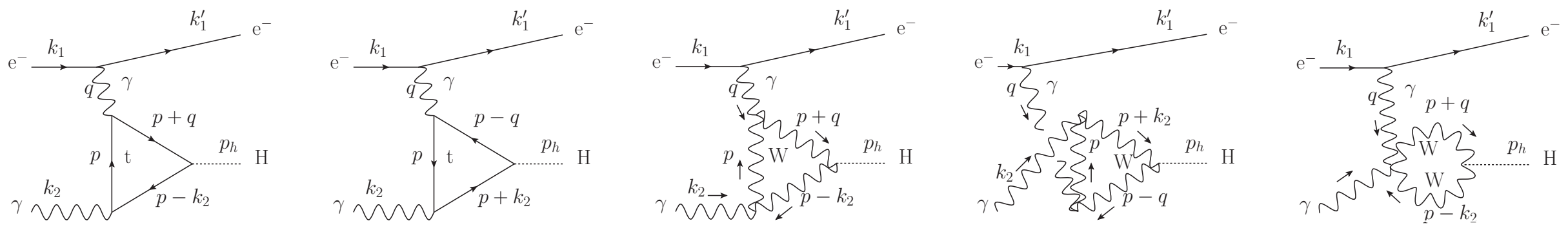

**Fig. 1:** $\gamma^*\gamma$ fusion diagrams: top-quark and $W$-boson loop contributions.

The one-loop $Z^*\gamma$ fusion diagrams for the Higgs boson production are obtained from the one-loop $\gamma^*\gamma$ fusion diagrams given in Fig. 1 by replacing the photon propagator with that of the $Z$ boson with mass $m_Z$. Charged fermions and $W$ boson contribute to the one-loop $Z^*\gamma$ fusion diagrams. Again we only consider the top quark for the charged fermion loop diagrams. We calculate the contribution from the $Z^*\gamma$ fusion diagrams and obtain,

$$A_{Z\gamma} = \left(\frac{eg^3}{16\pi^2}\right) \left[\overline{u}(k_1')\gamma_\mu \left(f_{Ze} + \gamma_5\right) u(k_1)\right] \frac{1}{t - m_Z^2} \left(g^{\mu\beta} - \frac{2k_2^\mu q^\beta}{m_h^2 - t}\right) \epsilon_\beta(k_2) \, F_{Z\gamma} \, , \tag{4}$$

with

$$F_{Z\gamma} = -\frac{m_t^2}{8 m_W \cos^2\theta_W} N_c Q_t f_{Zt} S^{Z\gamma}_{(T)}(t, m_t^2, m_h^2) + \frac{m_W}{4} S^{Z\gamma}_{(W)}(t, m_W^2, m_h^2) \, , \tag{5}$$

where $f_{Ze}$ and $f_{Zt}$ are the strength of the vector part of the $Z$-boson coupling to the electron and top quark, respectively, and are given by $f_{Ze} = -1 + 4\sin^2\theta_W$ and $f_{Zt} = 1 - \frac{8}{3}\sin^2\theta_W$ with $\theta_W$ being the Weinberg angle. The axial-vector part of the $Z$-boson coupling to the top quark has a null effect and we find $S^{Z\gamma}_{(T)}(t, m_t^2, m_h^2) = S^{\gamma\gamma}_{(T)}(t, m_t^2, m_h^2)$ and $S^{Z\gamma}_{(W)}(t, m_W^2, m_h^2) = S^{\gamma\gamma}_{(W)}(t, m_W^2, m_h^2)$.

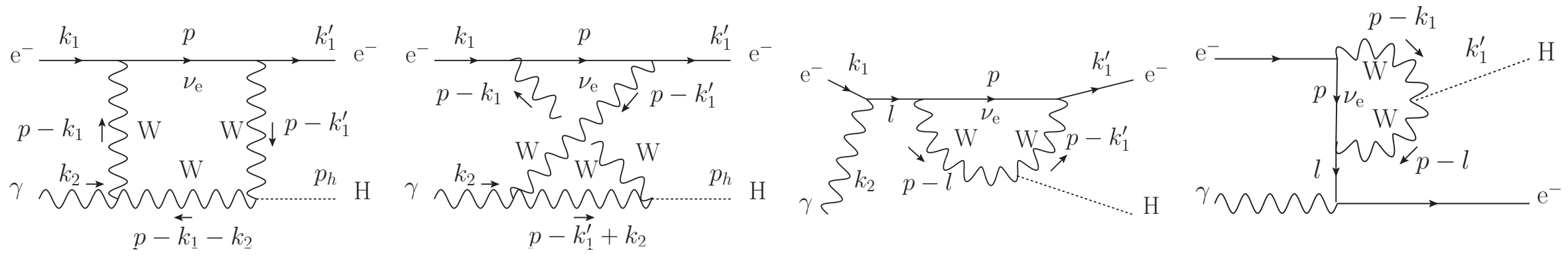

**Fig. 2:** "$W\nu_e$" diagrams.

The Feynman diagrams involving the $W$ boson and electron neutrino, which are shown in Fig. 2, also contribute to the Higgs boson production in $e^-\gamma$ collisions. They yield the "$W\nu_e$" amplitude which is written in the following form,

$$A_{W\nu_e} = \left(\frac{eg^3}{16\pi^2}\right) \frac{m_W}{4} \left[\overline{u}(k_1') \, F_{(W\nu_e)\beta} \, (1 - \gamma_5) u(k_1)\right] \epsilon(k_2)^\beta \, , \tag{6}$$

where the factor $(1 - \gamma_5)$ is due to the $e$-$\nu$-$W$ vertex. Thus, when the electron beams are right-handedly polarized, these "$W\nu_e$" diagrams do not contribute. The factor $F_{(W\nu_e)\beta}$ is written in a gauge-invariant

form as

$$F_{(W\nu_e)\beta} = \Big(\frac{2k_{1\beta}\not{k}_2}{s} - \gamma_\beta\Big) S^{W\nu_e}_{(k_1)}(s,t,m_h^2,m_W^2) + \Big(\frac{2k'_{1\beta}\not{k}_2}{u} + \gamma_\beta\Big) S^{W\nu_e}_{(k'_1)}(s,t,m_h^2,m_W^2)\ , \tag{7}$$

where $S^{W\nu_e}_{(k_1)}$ and $S^{W\nu_e}_{(k'_1)}$ are expressed in terms of the scalar integrals $B_0$'s, $C_0$'s and the scalar four-point integrals $D_0$'s. See Eqs.(2.15) and (2.16) of Ref. [5]. Again the ultraviolet divergences of $B_0$'s cancel out and, in the end, $S^{W\nu_e}_{(k_1)}$ and $S^{W\nu_e}_{(k'_1)}$ are finite. Finally we note that $S^{W\nu_e}_{(k'_1)}$ vanishes at $u = 0$, which is anticipated from the expression of the second term in Eq. (7).

The last one-loop contributions to the Higgs boson production in $e^-\gamma$ collisions come from the Feynman diagrams shown in Fig. 3. These "$Ze$" diagrams give the following amplitude,

$$A_{Ze} = \Big(\frac{eg^3}{16\pi^2}\Big)\Big(-\frac{m_Z}{16\cos^3\theta_W}\Big) \times \Big[\overline{u}(k'_1)\ F_{(Ze)\beta}\ (f_{Ze}+\gamma_5)^2 u(k_1)\Big]\epsilon(k_2)^\beta\ , \tag{8}$$

where the factor $(f_{Ze}+\gamma_5)^2$ arises from the $Z$-boson coupling to electrons. The factor $F_{(Ze)\beta}$ is written in a gauge-invariant form as

$$F_{(Ze)\beta} = \Big(\frac{2k_{1\beta}\not{k}_2}{s} - \gamma_\beta\Big) S^{Ze}_{(k_1)}(s,t,m_h^2,m_Z^2) + \Big(\frac{2k'_{1\beta}\not{k}_2}{u} + \gamma_\beta\Big) S^{Ze}_{(k'_1)}(s,t,m_h^2,m_Z^2)\ , \tag{9}$$

where $S^{Ze}_{(k_1)}$ and $S^{Ze}_{(k'_1)}$ are expressed in terms of the scalar integrals $B_0$'s, $C_0$'s and $D_0$'s. See Eqs.(2.19) and (2.20) of Ref. [5]. Collinear singularities appear in some of $C_0$'s and $D_0$'s. These collinear divergences are handled by dimensional regularization. Actually scalar integrals with collinear divergences appear in combination. When they are added their divergences cancel out. Thus $S^{Ze}_{(k_1)}$ and $S^{Ze}_{(k'_1)}$ are both finite. Note also that $S^{Ze}_{(k'_1)}$ vanishes at $u = 0$.

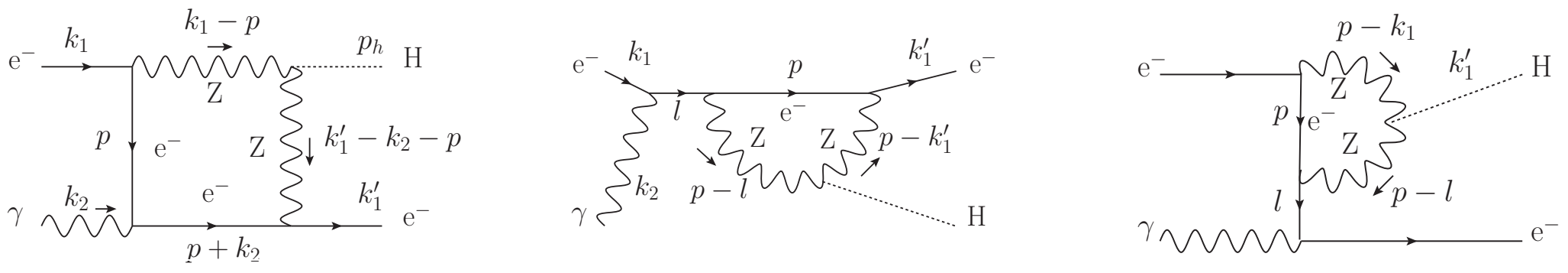

**Fig. 3:** "$Ze$" diagrams.

## 3 Numerical analysis

One of the advantages of linear colliders is that we can acquire highly polarized colliding beams. Let us consider the Higgs boson production reaction (1) when both the initial electron and photon beams are fully polarized. We denote the polarizations of the electron and photon as $P_e = \pm 1$ and $P_\gamma = \pm 1$, respectively. When an initial electron is polarized with polarization $P_e$, we modify $u(k_1)$ as $u(k_1) \to \frac{1+P_e\gamma_5}{2}\ u(k_1)$. And when a photon with momentum $k_2$ is moving in the $+z$ direction in the center-of-mass (CM) frame, the circular polarization ($P_\gamma = \pm 1$) of the photon is taken to be $\epsilon(k_2, \pm 1)_\beta = \frac{1}{\sqrt{2}}(0, \mp 1, -i, 0)$. Then the Higgs boson production cross section is given by

$$\sigma_{(e\gamma\to eH)}(s, P_e, P_\gamma) = \frac{1}{16\pi s^2}\int_{\rm cut} dt\ \Big\{\sum_{\text{final electron spin}} |A(P_e,P_\gamma)|^2\Big\}\ . \tag{10}$$

where $A(P_e, P_\gamma)$ is written at the one-loop level as

$$A(P_e,P_\gamma) = A_{\gamma\gamma}(P_e,P_\gamma) + A_{Z\gamma}(P_e,P_\gamma) + A_{W\nu_e}(P_e,P_\gamma) + A_{Ze}(P_e,P_\gamma)\ . \tag{11}$$

Since the forward and backward directions in an $e^-\gamma$ collider are blind spots for the detection of scattered particles, we introduce kinematical cuts for the scattered electron in $e^-\gamma$ collisions. Denoting $\theta$ as the angle between the initial and scattered electrons in the CM frame, we choose the allowed region of $\theta$ as $10^\circ \le \theta \le 170^\circ$, which leads to the integration range of $t$ in Eq. (10) as $(-s + m_h^2 - t_{\rm cut}) \le t \le t_{\rm cut}$ with $t_{\rm cut} = -\frac{1}{2}(s - m_h^2)(1 - \cos 10^\circ)$. Note that $A_{W\nu_e} = 0$ when $P_e = +1$.

For numerical analysis we choose the mass parameters and the coupling constants as follows: $m_h = 125$ GeV, $m_t = 173$ GeV, $m_Z = 91$ GeV, $m_W = 80$ GeV, $\cos\theta_W = \frac{m_W}{m_Z}$, $e^2 = 4\pi\alpha_{em} = \frac{4\pi}{128}$ and $g = \frac{e}{\sin\theta_W}$. The electromagnetic constant $e^2$ is chosen to be the value at the scale of $m_Z$. We plot $\sigma_{(e\gamma\to eH)}(s, P_e, P_\gamma)$ in Fig. 4 as a function of $\sqrt{s}$ ($\sqrt{s} \ge 130$ GeV) for each case of polarizations of the electron and photon beams. For the case $P_eP_\gamma = -1$, the cross section $\sigma_{(e\gamma\to eH)}$ is very small at $\sqrt{s} = 130$ GeV, since the integration range of $t$ is small and the differential cross section vanishes as $t \to t_{\rm min}$. The cross section $\sigma_{(e\gamma\to eH)}(s, P_e = -1, P_\gamma = +1)$ rises gradually up to about 2 fb, while $\sigma_{(e\gamma\to eH)}(s, P_e = +1, P_\gamma = -1)$ increases rather slowly up to 0.4 fb. This is due to the interference between $A_{\gamma\gamma}$ and $A_{Z\gamma}$, which acts constructively for ($P_e = -1, P_\gamma = +1$) but destructively for ($P_e = +1, P_\gamma = -1$). For the case $P_eP_\gamma = +1$, the cross section $\sigma_{(e\gamma\to eH)}$ is about 2 fb at $\sqrt{s} = 130$ GeV. The cross section $\sigma_{(e\gamma\to eH)}(s, P_e = -1, P_\gamma = -1)$ rises above 3 fb around $\sqrt{s} = 200$ GeV and then gradually decreases as $\sqrt{s}$ increases. This is due to the destructive interference both between $A_{W\nu_e}$ and $A_{\gamma\gamma}$ and between $A_{W\nu_e}$ and $A_{Z\gamma}$ in the range of large $-t$. Again the destructive interference between $A_{\gamma\gamma}$ and $A_{Z\gamma}$ is responsible for the decrease of $\sigma_{(e\gamma\to eH)}(s, P_e = +1, P_\gamma = +1)$ as $\sqrt{s}$ increases.

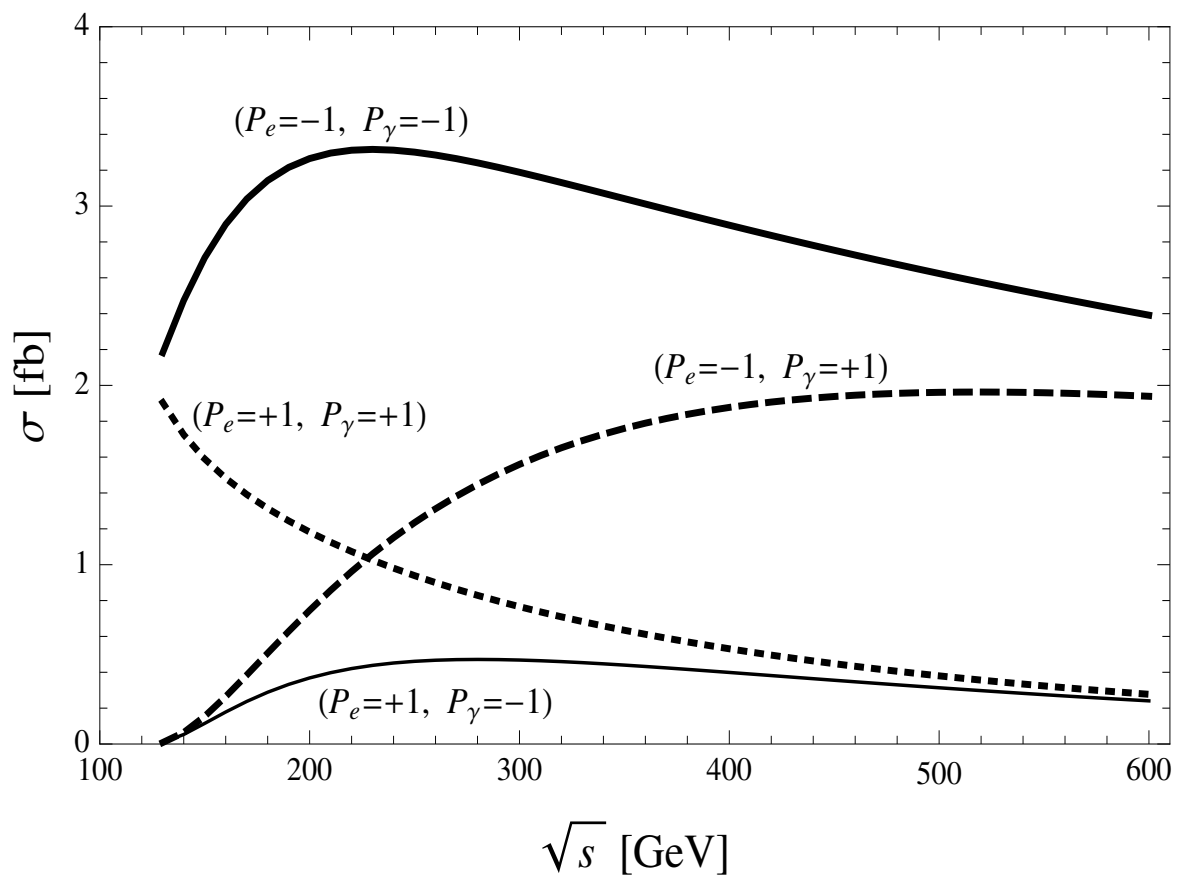

**Fig. 4:** Higgs boson production cross section $\sigma_{(e\gamma\to eH)}(s, P_e, P_\gamma)$. The kinematical cut is chosen such that the allowed angle $\theta$ of the scattered electron in the center-of-mass frame is $10^\circ \le \theta \le 170^\circ$.

A high-intensity photon beam can be produced by laser light backward scattering off a high-energy electron beam, $e^-\gamma_{\rm Laser} \to e^-\gamma$, where the backward-scattered photon receives a major fraction of the incoming electron energy [7]. Its energy distribution depends on the polarizations of the initial electron ($P_{e2} = \pm$) and laser photon ($P_{\rm Laser} = \pm$). Suppose we have a highly polarized $e^-e^-$ collider machine. Converting one of the electron beams to photon beam by means of backward Compton scattering of a polarized laser beam, we obtain an $e^-\gamma$ collider with high polarization. The Higgs boson production cross section for $e^-\gamma \to e^-H$ in an $e^-e^-$ collider, whose beam energies are $E_{e1}$ and $E_{e2}$ and polarizations are $P_{e1}$ and $P_{e2}$, is expressed as,

$$\sigma_{e\gamma\ \rm collision}(s_{ee}, E_{\rm Laser}, P_{e1}, P_{e2}, P_{\rm Laser}) = \sum_{P_\gamma} \int dy\ N(y, E_{e2}, E_{\rm Laser}, P_{e2}, P_{\rm Laser}, P_\gamma)\sigma_{(e\gamma\to eH)}(s, P_{e1}, P_\gamma), \quad (12)$$

where $\sigma_{(e\gamma\to eH)}(s, P_{e1}, P_\gamma)$ is given in Eq. (10) with $P_e$ replaced by $P_{e1}$, and $s_{ee}$ is the CM energy squared of the two initial electron beams and related to $s$ as $s = ys_{ee}$. The photon energy is $yE_{e2}$ and its spectrum is given by $N(y, E_{e2}, E_{\text{Laser}}, P_{e2}, P_{\text{Laser}}, P_\gamma)$ [8,9] with $P_{e2}P_{\text{Laser}} = -1$. We take the case $P_{e2}P_{\text{Laser}} = -1$ so that the spectrum has a peak at the highest energy.

We analyze the cross section of the Higgs boson production in an $e^-e^-$ collider through $b\bar{b}$ decay channel, $e + \gamma \to e + H \to e + b + \bar{b}$, since it has a large branching ratio. We consider the case: $E_{\text{Laser}} = 1.17$eV (the YAG laser with wavelength 1064nm) and $E_{e2}$ =250GeV ($\sqrt{s_{ee}} = 500$GeV). The integration range of $y$ in Eq. (12) is given by $y_{\text{min}} \leq y \leq y_{\text{max}}$ with $y_{\text{min}} = 0.0625$ and $y_{\text{max}} = 0.82$. The Monte Carlo method is used. A $b$-quark mass is chosen to be 4.3 GeV. The angle cuts of the scattered electron and $b(\bar{b})$ quarks are chosen such that the allowed regions are $10^\circ \leq \theta_{e^-} \leq 170^\circ$ and $10^\circ \leq \theta_{b(\bar{b})} \leq 170^\circ$, respectively, and the energy cuts of these particles are set to be 3GeV. The Monte Carlo statistical error is about $0.1\%$ when sampling number is taken to be 200,000. The result is shown in Table 1 for each combination of polarizations $P_{e1}$ and $P_{\text{Laser}}$.

We also analyze the significance $S/\sqrt{B}$ of the Higgs boson production. The $b\bar{b}$ decay channel of the Higgs boson in $e^-\gamma$ collisions has a substantial background. Especially a huge background appears at the $Z$-boson pole. However, it is expected that when we measure the invariant mass $m_{b\bar{b}}$ of $b$ and $\bar{b}$ quarks, the background is small in the region $m_{b\bar{b}}$ >120GeV compared with the signals of the Higgs boson production. We use GRACE [9] to write down all the tree Feynman diagrams for $e+\gamma \to e+b+\bar{b}$ and to evaluate their contributions to the background cross section. We assume that the integrated luminosity is 250fb$^{-1}$. The significance is calculated by taking samples in the region 120GeV $\leq m_{b\bar{b}} \leq$ 130GeV at the parton level. The results are given in Table 1. Large values of significance are obtained for the cases of $(P_{e1}, P_{\text{Laser}}) = (-1, \mp 1)$ with $\sqrt{s_{ee}} = 500$GeV.

| $\sqrt{s_{ee}}$ GeV | $P_{e1}$ | $P_{\text{Laser}}$ | $\sigma_{\text{cut}}$ fb | $S/\sqrt{B}$ |
|---|---|---|---|---|
| 500 | 1 | -1 | 0.11 | 2.93 |
| | 1 | 1 | 0.19 | 1.31 |
| | -1 | -1 | 1.22 | 10.6 |
| | -1 | 1 | 1.01 | 6.8 |

**Table 1:** Higgs boson production cross section and significance in $e^-\gamma$ collision in an $e^-e^-$ collider for the case $E_{\text{Laser}} = 1.17$eV, $E_{e2}$ =250GeV, $\sqrt{s_{ee}}$ =500GeV, and for each combination of polarizations $P_{e1}$ and $P_{\text{Laser}}$. $P_{e2}$ is chosen to be $-P_{\text{Laser}}$.

## 4 Summary

We have investigated the SM Higgs boson production in $e^-\gamma$ collisions. The electroweak one-loop contributions to the scattering amplitude for $e^-\gamma \to e^-H$ were calculated and they were expressed in analytical form.ãĂĂ We analyzed the cross section of the Higgs boson production through the $b\bar{b}$ decay channel, $e + \gamma \to e + H \to e + b + \bar{b}$, in an $e^-\gamma$ collision in an $e^-e^-$ collider. A high-energy photon beam was assumed to be produced by laser light backward scattering off one of the high-energy electron beams of the $e^-e^-$ collider. We obtained large values of the significance $\sqrt{S}/B$ for the Higgs boson production for both $\sqrt{s_{ee}} = 250$GeV and $\sqrt{s_{ee}} = 500$GeV. We therefore conclude that the Higgs boson will be clearly observed in $e^-\gamma$ collision experiments.

## Acknowledgements

We wish to thank the organizers of Photon 2017 for the hospitality at such a stimulating conference.

# Prospects for $\gamma\gamma \to$ Higgs observation in ultraperipheral ion collisions at the Future Circular Collider

*David d'Enterria*[1], *Daniel E. Martins*[2], *and Patricia Rebello Teles*[3]
[1]CERN, EP Department, 1211 Geneva.
[2]UFRJ, Univ. Federal do Rio de Janeiro, 21941-901, Rio de Janeiro, RJ.
[3]CBPF, Centro Brasileiro de Pesquisas Físicas, 22290-180, Rio de Janeiro, RJ.

**Abstract**
We study the two-photon production of the Higgs boson, $\gamma\gamma \to$ H, at the Future Circular Collider (FCC) in ultraperipheral PbPb and pPb collisions at $\sqrt{s_{\rm NN}} = 39$ and 63 TeV. Signal and background events are generated with MADGRAPH 5, including $\gamma$ fluxes from the proton and lead ions in the equivalent photon approximation, yielding $\sigma(\gamma\gamma \to \mathrm{H})$ = 1.75 nb and 1.5 pb in PbPb and pPb collisions respectively. We analyse the H$\to b\bar{b}$ decay mode including realistic reconstruction efficiencies for the final-state $b$-jets, showered and hadronized with PYTHIA 8, as well as appropriate selection criteria to reduce the $\gamma\gamma \to b\bar{b}, c\bar{c}$ continuum backgrounds. Observation of PbPb$\xrightarrow{\gamma\gamma}$(Pb)H(Pb) is achievable in the first year with the expected FCC integrated luminosities.

## 1 Introduction

The observation of the predicted Higgs boson [1] in proton-proton collisions at the Large Hadron Collider [2, 3] has represented a breakthrough in our scientific understanding of the particles and forces in nature. A complete study of the properties of the scalar boson, including its couplings to all known particles, and searches of possible deviations indicative of physics beyond the Standard Model (SM), require a new collider facility with much higher center-of-mass (c.m.) energies [4]. The Future Circular Collider (FCC) is a post-LHC project at CERN, aiming at pp collisions up to at a c.m. energy of $\sqrt{s}$ = 100 TeV in a new 80–100 km tunnel with 16–20 T dipoles [5]. The FCC running plans with hadron beams (FCC-hh) includes also heavy-ion operation at nucleon-nucleon c.m. energies of $\sqrt{100}$ TeV$\cdot\sqrt{Z_1Z_2/(A_1A_2)} = 39$ TeV, 63 TeV for PbPb, pPb with (monthly) integrated luminosities of 110 nb$^{-1}$ and 29 pb$^{-1}$ [6]. Such high collision energies and luminosities, factors of 7 and 30 times higher respectively than those reachable at the LHC, open up the possibility to study the production of the Higgs boson in nuclear collisions, both in central hadronic [7] as well as in ultraperipheral (electromagnetic) [8] interactions. The observation of the latter $\gamma\gamma \to$ H process provides an independent measurement of the H-$\gamma$ coupling not based on Higgs decays but on its $s$-channel production mode.

The measurement of exclusive $\gamma\gamma \to$ H in ultraperipheral collisions (UPCs) [9, 10] of pPb and PbPb beams was studied in detail for LHC energies[1] in [8], although its observation there is unfeasible with the nominal luminosities (Fig. 1, left). We extend such studies for FCC energies, where such an observation is warranted. All charges accelerated at high energies generate electromagnetic fields which, in the equivalent photon approximation (EPA) [12], can be considered as quasireal photon beams[2] [13]. The highest available photon energies are of the order of the inverse Lorentz-contracted radius $R$ of the source charge, $\omega_{\rm max} \approx \gamma/R$, which at the FCC yield photon-photon collisions above 1 TeV (Table 1). In addition, since the photon flux scales as the squared charge of the beam, $Z^2$, two-photon cross sections are enhanced millions of times for ions ($Z_{\rm Pb}^4$ = 5·10$^7$ for PbPb) compared to proton or electron beams, thereby featuring the largest $\gamma\,\gamma$ luminosities among all colliding systems (Fig. 1, left).

[1]A few older papers had already previously discussed the possibility to produce the Higgs boson in heavy-ion UPCs [11].
[2]The emitted photons are almost on mass shell, with virtuality $-Q^2 < 1/R^2$, where $R$ is the radius of the charge, i.e. $Q \approx 0.28$ GeV for protons ($R \approx 0.7$ fm) and $Q < 0.06$ GeV for nuclei ($R_A \approx 1.2\,A^{1/3}$ fm) with mass number $A > 16$.

**Table 1:** Relevant parameters for photon-photon processes in ultraperipheral pPb and PbPb collisions at the FCC: (i) nucleon-nucleon c.m. energy, $\sqrt{s_{\rm NN}}$, (ii) integrated luminosity per year, $\mathcal{L}_{\rm int}$, (iii) beam energies, E$_{\rm beam}$, (iv) Lorentz factor, $\gamma = \sqrt{s_{\rm NN}}/(2\,m_N)$, (v) effective (Pb) radius, $R_A$, (vi) photon "cutoff energy" in the c.m. frame, $\omega_{\max}$, and (vii) "maximum" photon-photon c.m. energy, $\sqrt{s_{\gamma\gamma}^{\max}}$. The last column lists the $\gamma\,\gamma \to$ H cross sections.

| System | $\sqrt{s_{\rm NN}}$ | $\mathcal{L}_{\rm int}$ | $E_{\rm beam1} + E_{\rm beam2}$ | $\gamma$ | $R_A$ | $\omega_{\max}$ | $\sqrt{s_{\gamma\gamma}^{\max}}$ | $\sigma(\gamma\gamma \to {\rm H})$ |
|---|---|---|---|---|---|---|---|---|
| pPb | 63 TeV | 29 pb$^{-1}$ | 50. + 19.5 TeV | 33 580 | 7.1 fm | 950 GeV | 1.9 TeV | 1.5 pb |
| PbPb | 39 TeV | 110 nb$^{-1}$ | 19.5 + 19.5 TeV | 20 790 | 7.1 fm | 600 GeV | 1.2 TeV | 1.75 nb |

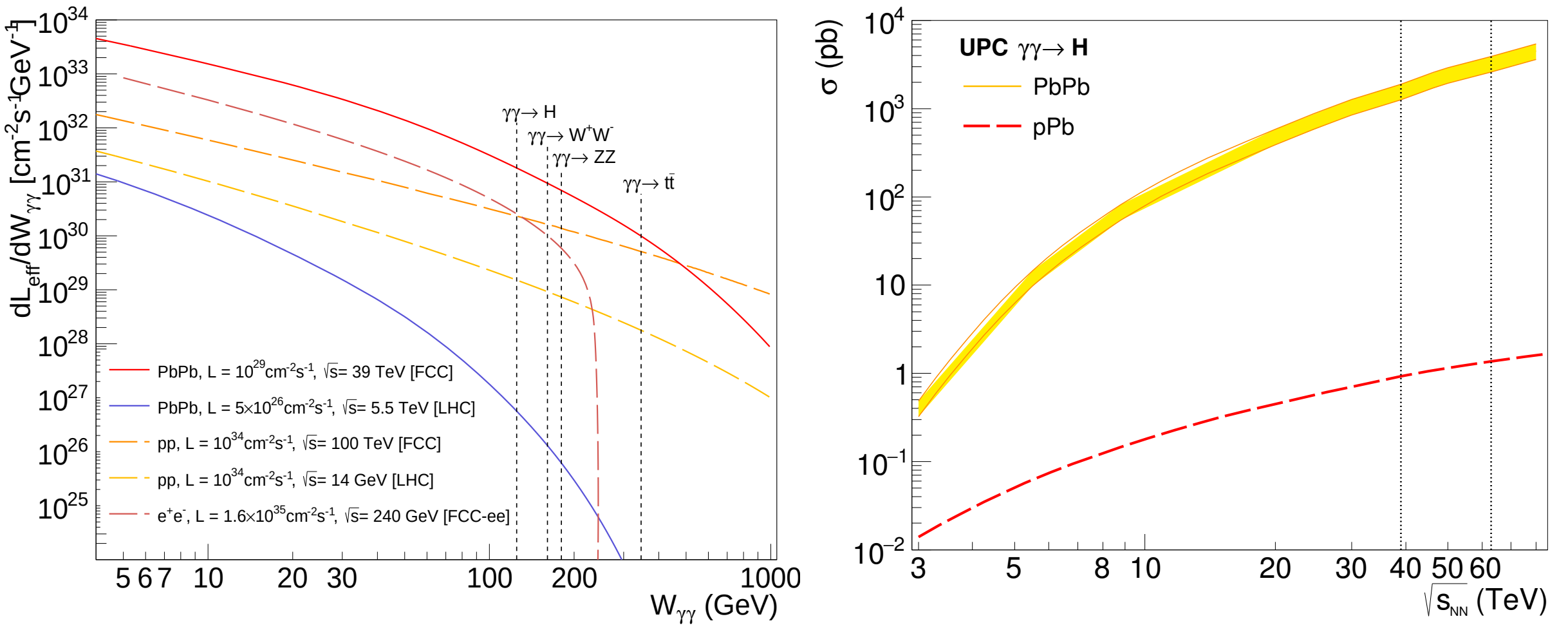

**Fig. 1:** Left: Two-photon effective luminosities as a function of $\gamma\,\gamma$ c.m. energy over $W_{\gamma\,\gamma} \approx$ 5–1000 GeV in PbPb, pp, and $e^+e^-$ collisions at the FCC [5, 6, 14], and in PbPb and pp collisions at the LHC. Right: Two-photon fusion Higgs boson cross section versus nucleon-nucleon c.m. energy in ultraperipheral PbPb (top) and pPb (bottom curve) collisions. The vertical lines indicate the expected FCC running energies at $\sqrt{s_{\rm NN}}$ = 39 and 63 TeV.

## 2 Theoretical setup

The MADGRAPH 5 (v.2.5.4) [15] Monte Carlo (MC) event generator is used to compute the relevant cross sections from the convolution of the Weizsäcker-Williams EPA photon fluxes [12] for the proton and lead ion, and the H-$\gamma$ coupling parametrized in the Higgs effective field theory [16], following the implementation discussed in [8] with a more accurate treatment of the non hadronic-overlap correction. The proton $\gamma$ flux is given by the energy spectrum $f_{\gamma/p}(x)$ where $x = \omega/E$ is the fraction of the beam energy carried by the photon [17]:

$$f_{\gamma/p}(x) = \frac{\alpha}{\pi}\,\frac{1 - x + 1/2x^2}{x}\int_{Q^2_{\rm min}}^{\infty}\frac{Q^2 - Q^2_{\rm min}}{Q^4}|F(Q^2)|^2 dQ^2\;, \qquad (1)$$

with $\alpha = 1/137$, $F(Q^2)$ the proton electromagnetic form factor, and the minimum momentum transfer $Q_{\rm min}$ is a function of $x$ and the proton mass $m_p$, $Q^2_{\rm min} \approx (xm_p)^2/(1-x)$. The photon energy spectrum of the lead ion ($Z = 82$), integrated over impact parameter $b$ from $b_{\rm min}$ to infinity, is given by [18]:

$$f_{\gamma/Pb}(x) = \frac{\alpha Z^2}{\pi}\,\frac{1}{x}\left[2x_i K_0(x_i)K_1(x_i) - x_i^2(K_1^2(x_i) - K_0^2(x_i))\right]\;, \qquad (2)$$

where $x_i = x\, m_N\, b_{\rm min}$, and $K_0$, $K_1$ are the modified Bessel functions of the second kind of zero and first order, related respectively to the emission of longitudinally and transversely polarized photons. The latter dominating for ultrarelativistic charges ($\gamma \gg 1$). The dominant Higgs decay mode is $\rm H \to b\bar{b}$, with a branching fraction of 58% as computed with HDECAY [19]. The PYTHIA8.2 [20] MC generator was employed to shower and hadronize the two final-state $b$-jets, which are then reconstructed with the Durham $k_t$ algorithm [21] (exclusive 2-jets final-state) using FASTJET 3.0 [22]. The same setup is used to generate the exclusive two-photon production of $b\bar{b}$ and (possibly misidentified) $c\bar{c}$ and light-quark ($q\bar{q}$) jet pairs, which constitute the most important physical background for the measurement of the H$\to b\bar{b}$ channel.

## 3 Results

The total elastic Higgs boson cross sections in ultraperipheral PbPb and pPb collisions as a function of $\sqrt{s}$ are shown in Fig. 1 (right). We have assigned a conservative 20% uncertainty to the predicted cross sections to cover different charge form factors. At LHC energies, we find a slightly reduced cross section, $\sigma(\rm PbPb \to \gamma\gamma \to H) = 15 \pm 3$ pb, compared to the results of [8] due a more accurate treatment of the non hadronic-overlap correction based on [23]. The predicted total Higgs boson cross sections are $\sigma(\gamma\gamma \to \rm H)$ = 1.75 nb and 1.5 pb in PbPb and pPb collisions at $\sqrt{s_{_{\rm NN}}}$ = 39 and 63 TeV which, for the nominal $\mathcal{L}_{\rm int}$ = 110 nb$^{-1}$ and 29 pb$^{-1}$ luminosities per "year" (1-month run), imply ~200 and 45 Higgs bosons produced (corresponding to 110 and 25 bosons in the $b\bar{b}$ decay mode, respectively). The main backgrounds are pairs from the $\gamma\gamma \to b\bar{b}, c\bar{c}, q\bar{q}$ continuum, where charm and light ($q = uds$) quarks are misidentified as $b$-quarks. The irreducible $\gamma\gamma \to b\bar{b}$ background over the mass range $100 < W_{\gamma\gamma} <$ 150 GeV is ~20 times larger than the signal, but can be suppressed (as well as that from misidentified $c\bar{c}$ and $q\bar{q}$ pairs) via various kinematical cuts. The data analysis follows closely the similar LHC study [8], with the following reconstruction performances assumed: jet reconstruction over $|\eta| < 5$, 7% $b$-jet energy resolution (resulting in a dijet mass resolution of ~6 GeV at the Higgs peak), 70% $b$-jet tagging efficiency, and 5% (1.5%) $b$-jet mistagging probability for a $c$ (light-flavour $q$) quark. For the double $b$-jet final-state of interest, these lead to a ~50% efficiency for the MC-generated signal (S), and a total reduction of the misidentified $c\bar{c}$ and $q\bar{q}$ continuum backgrounds (B) by factors of ~400 and ~400 000.

**Table 2:** Summary of the cross sections and expected number of events per run after event selection criteria (see text) for signal and backgrounds in the $\gamma\gamma \to \rm H(b\bar{b})$ analysis, obtained from events generated with MADGRAPH 5+PYTHIA 8 for PbPb and pPb collisions at FCC energies.

| PbPb at $\sqrt{s_{_{\rm NN}}}$ = 39 TeV | cross section (b-jet (mis)tag efficiency) | visible cross section after $p_T^j, \cos\theta_{jj}, m_{jj}$ cuts | $N_{\rm evts}$ ($\mathcal{L}_{\rm int}$ = 110 nb$^{-1}$) |
|---|---|---|---|
| $\gamma\gamma \to \rm H \to b\bar{b}$ | 1.02 nb (0.50 nb) | 0.19 nb | 21.1 |
| $\gamma\gamma \to b\bar{b}$ [$\rm m_{b\bar{b}}$=100−150 GeV] | 24.3 nb (11.9 nb) | 0.23 nb | 25.7 |
| $\gamma\gamma \to c\bar{c}$ [$\rm m_{c\bar{c}}$=100−150 GeV] | 525 nb (1.31 nb) | 0.02 nb | 2.3 |
| $\gamma\gamma \to q\bar{q}$ [$\rm m_{q\bar{q}}$=100−150 GeV] | 590 nb (0.13 nb) | 0.002 nb | 0.25 |
| pPb at $\sqrt{s_{_{\rm NN}}}$ = 63 TeV | | | $N_{\rm evts}$ ($\mathcal{L}_{\rm int}$ = 29 pb$^{-1}$) |
| $\gamma\gamma \to \rm H \to b\bar{b}$ | 0.87 pb (0.42 pb) | 0.16 pb | 4.8 |
| $\gamma\gamma \to b\bar{b}$ [$\rm m_{b\bar{b}}$=100−150 GeV] | 21.8 pb (10.7 pb) | 0.22 pb | 6.3 |
| $\gamma\gamma \to c\bar{c}$ [$\rm m_{c\bar{c}}$=100−150 GeV] | 410. pb (1.03 pb) | 0.011 pb | 0.3 |
| $\gamma\gamma \to q\bar{q}$ [$\rm m_{q\bar{q}}$=100−150 GeV] | 510. pb (0.114 pb) | 0.001 pb | 0.04 |

As proposed in [8], various simple kinematical cuts can be applied to enhance the S/B ratio. Since the transverse momenta of the Higgs decay $b$-jets peak at $p_T^j \approx m_{\rm H}/2$, selecting events with at least one jet within $p_T$ = 55–62.5 GeV suppresses $\sim$ 96% of the continuum backgrounds, while removing only half of the signal. Also, one can exploit the fact that the angular distribution of the Higgs decay $b$-jets in the helicity frame is isotropically distributed in $|\cos\theta_{j_1j_2}|$, i.e., each jet is independently emitted either in the same direction as the $b\bar{b}$ pair or opposite to it, while the continuum (with quarks propagating in the $t$- or $u$- channels) is peaked in the forward–backward directions. Thus, requiring $|\cos\theta_{j_1j_2}| < 0.5$ further suppresses the continuum contaminations by another $\sim$ 20% while leaving untouched the remaining signal. The significance of the signal can then be computed from the remaining number of counts within 1.4$\sigma$ around the Gaussian Higgs peak (i.e., $117 < m_{b\bar{b}} < 133$ GeV) over the underlying dijet continuum. Table 2 summarizes the visible cross sections and the number of events after cuts for the nominal luminosities of each system.

In PbPb $\sqrt{s}$ = 39 GeV for the nominal integrated luminosity of $\mathcal{L}_{\rm int} = 110$ nb$^{-1}$ per run, we expect about $\sim$21 signal counts over $\sim$28 for the sum of backgrounds in a window $m_{b\bar{b}}$ = 117–133 GeV around the Higgs peak. Reaching a statistical significance close to 5$\sigma$ (Fig. 2, left) would require to combine two different experiments (or doubling the luminosity in a single one). Similar estimates for pPb at 63 TeV (29 pb$^{-1}$) yield about 5 signal events after cuts, over a background of 6.7 continuum events. Reaching a 5$\sigma$ significance for the observation of $\gamma\gamma \to$ H production (Fig. 2, right) would require in this case to run for about 8 months (instead of the nominal 1-month run per year), or running 4 months and combining two experiments. All the derived number of events and significances are based on the aforementioned set of kinematical cuts, and can be likely improved by using a more advanced multivariate analysis.

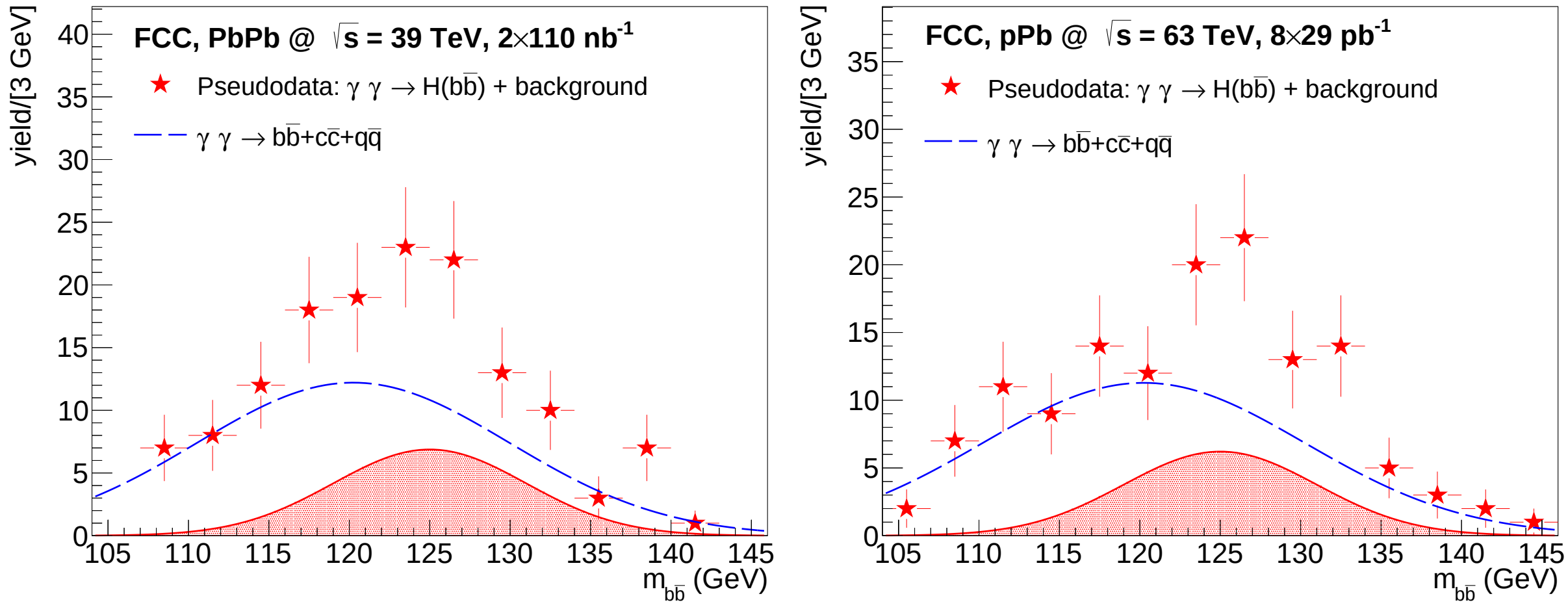

**Fig. 2:** Expected dijet invariant mass distributions for the combination of photon-fusion Higgs signal (hatched Gaussian) and $b\bar{b} + c\bar{c} + q\bar{q}$ continuum (dashed line) in ultraperipheral PbPb ($\sqrt{s_{\rm NN}}$ = 39 TeV, left) and pPb ($\sqrt{s_{\rm NN}}$ = 63 TeV, right) collisions, after event selection criteria with the quoted integrated luminosities (see text).

## 4 Conclusion

We have presented prospect studies for the measurement of the two-photon production of the Higgs boson in the $b\bar{b}$ decay channel in ultraperipheral PbPb and pPb collisions at the FCC. Cross sections have been obtained with MADGRAPH 5, using the Pb (and proton) equivalent photon fluxes and requiring no hadronic overlap of the colliding particles, at nucleon-nucleon c.m. energies of $\sqrt{s_{\rm NN}}$ = 39, and 63 TeV. The $b$-quarks have been showered and hadronized with PYTHIA 8, and reconstructed in a exclusive two-jet final-state with the $k_T$ algorithm. By assuming realistic jet reconstruction performances and (mis)tagging efficiencies, and applying appropriate kinematical cuts on the jet $p_T$ and dijet mass and angles in the helicity frame, we can reconstruct the H($b\bar{b}$) signal on top of the dominant $\gamma\gamma \to b\bar{b}$ continuum background. The measurement of $\gamma\gamma \to$ H $\to b\bar{b}$ would yield 21 (5) signal counts over 28

(7) continuum dijet pairs around the Higgs peak, in PbPb (pPb) collisions for their nominal integrated luminosities per run. Observation of the photon-fusion Higgs production at the $5\sigma$-level is achievable in the first year by combining the measurements of two experiments (or doubling the luminosity in a single one) in PbPb, and by running for about 8 months (or running 4 months and combining two experiments) in the pPb case. The feasibility studies presented here confirm the interesting Higgs physics potential open to study in $\gamma\gamma$ ultraperipheral ion collisions at the FCC, providing an independent measurement of the H-$\gamma$ coupling not based on Higgs decays but on a $s$-channel production mode.

**Acknowledgments –** P. R. T. acknowledges financial support from the CERN TH Department and from the FCC project.

# $\gamma\gamma$ Collider – A Brief History and Recent Developments

*W. Chou*
IHEP, Beijing, China

**Abstract**
There is renewed interest in constructing a $\gamma\gamma$ collider following the discovery of the Higgs boson and the formation of an ICFA (*International Committee for Future Accelerators*) – ICUIL (*International Committee on Ultrahigh Intensity Lasers*) collaboration. ICUIL brings state-of-the-art laser technology to build new types of accelerators such as a $\gamma\gamma$ collider. A recent ICFA mini-workshop on $\gamma\gamma$ colliders investigated the possibility and physics value of a low energy $\gamma\gamma$ collider. This paper uses a 1 MeV (c.m.) $\gamma\gamma$ collider currently under design at IHEP as an example to discuss various aspects of this collider, including the physics case, requirements on the laser beam, the electron beam, the accelerator and detector.

## 1 A Brief History

$\gamma\gamma$ colliders were first suggested in the early 1980s as a possible extension of two proposed linear colliders: VLEPP and SLC [1,2]. BINP played an important role in the early development of fundamental accelerator physics of $\gamma\gamma$ colliders [3,4]. In the 1990s, there were a number of high-energy linear collider (LC) proposals including the NLC from SLAC, the JLC from KEK and TESLA from DESY. In each of these three LC design reports, there was an appendix on a $\gamma\gamma$ collider as a potential add-on. In the 2000s, under the leadership of ICFA, the world HEP community activities towards building a linear collider converged to a single effort – the International Linear Collider Global Design Effort (ILC GDE) led by B. Barish. In 2007 the GDE released an initial ILC cost estimate (US$ 6.74 billion plus 24 million person-hours), which exceeded the previous TESLA cost estimate for a similar machine by a factor of two. This figure was further inflated by the US accounting system and the US Department of Energy quickly withdrew from bidding for hosting the ILC in the United States. Another potential host country, Japan, also expressed serious concern about the high cost. To lower the upfront cost, H. Sugawara in 2008 proposed to the ICFA to build a pair of 90 GeV $e^-$ linacs and use them as a $\gamma\gamma$ collider as an initial phase towards an eventual full-scale 250×250 GeV ILC. ICFA commissioned a working group headed by M. Peskin to study this proposal. However, at a meeting in February 2009, ICFA decided to reject this proposal for three reasons: (1) the physics of a $\gamma\gamma$ collider was not as strong as that from an $e^+e^-$ collider; (2) the cost saving (US$ 3 billion and 11 million person-hours) was insufficient; (3) it would require a major laser R&D program, which ICFA was not prepared to initiate. After this decision, the $\gamma\gamma$ collider activity stalled.

A change came in 2012 when the Higgs boson was discovered at the LHC. Because the Higgs mass is low (125 GeV), there are many options for a Higgs factory as documented in a report from the 2012 ICFA workshop on "*Accelerators for a Higgs Factory: Linear vs. Circular*" [5], which included the option of a $\gamma\gamma$ collider. The advantage of a $\gamma\gamma$ collider is that the cross section for $\gamma\gamma \rightarrow H$ is large and comparable to $e^+e^- \rightarrow ZH$ (~200 fb) but the required energy is much lower (63 GeV for a photon beam, corresponding to 80 GeV for an electron beam, compared to 120 GeV per electron beam in an $e^+e^-$ collider). This makes a $\gamma\gamma$ collider an attractive option for either a low energy linear collider (80

GeV per electron beam) or a low energy circular collider (80 GeV per beam). Furthermore, for a $\gamma\gamma$ collider there is no need for positrons and only one damping ring is required.

In the meantime, another important event gave new momentum to the $\gamma\gamma$ collider, namely, the formation of an ICFA-ICUIL collaboration. ICFA is the leading body of the world HEP community, whereas ICUIL the leading body of the world laser community. The "marriage" between the two has had a profound impact on both communities. A White Paper of the ICFA-ICUIL Joint Task Force was published in December 2011 [6], in which there was detailed discussion about applying high power laser technology to a $\gamma\gamma$ collider.

In addition to an add-on to a linear collider such as ILC and CLIC, a $\gamma\gamma$ collider for a Higgs factory can also be an add-on to a circular collider such as FCC-ee and CEPC, or a dedicated recirculating linac such as SAPPHiRE and HFiTT, as reported at the 2017 ICFA mini-workshop on $\gamma\gamma$ colliders [7]. There is also a recent review article discussing various aspects of $\gamma\gamma$ colliders [8].

## 2 Low Energy $\gamma\gamma$ Colliders

Among the many $\gamma\gamma$ collider proposals listed above, a common feature is that they all will explore the energy frontier, i.e., at the energy scale of a Higgs factory or higher. However, construction of such a high energy $\gamma\gamma$ collider would require at least 20 years. Therefore, an important topic at the ICFA mini-workshop was to identify the shortest path for designing and constructing a low energy $\gamma\gamma$ collider.

Several low energy $\gamma\gamma$ colliders were proposed. V. Telnov suggested using the European XFEL (17.5 GeV) as the linac for a $\gamma\gamma$ collider to study the physics in the c-quark and b-quark energy region of 3-12 GeV. M. Velasco proposed a $\gamma\gamma$ collider at the $\tau\tau$ threshold (3.6-8 GeV) to study $\tau$ physics including $\tau$(g-2). In these energy ranges, one can contemplate using existing electron linacs instead of waiting for the construction of major new linacs (ILC or CLIC).

Further discussions revealed that even a $\gamma\gamma$ collider at 1 MeV can provide important physics such as light-by-light scattering and Breit-Wheeler pair production. Both processes were predicted in the 1930s but have never been observed or measured in the laboratory [9-13]. This opened an avenue to explore the feasibility of designing and constructing a $\gamma\gamma$ collider in the near future because low energy electron linacs are widely available.

In the following we will use the BEPC linac at the IHEP, China as an example for a low energy $\gamma\gamma$ collider.

## 3 $\gamma\gamma$ Collider Study at IHEP

The BEPC linac is the injector to the $e^+e^-$ collider at IHEP, which serves as a tau-charm factory. The maximum energy of the linac is 2.6 GeV; it is continuously tunable down to ~200 MeV. The linac has the capability of generating two consecutive electron bunches in each pulse. The spacing between the two bunches can vary from 7 ns to 4 μs. The pulse rate is 50 Hz. Currently it uses a thermionic gun, which gives rise to large beam emittance (2 μm at 2 nC, unnormalized). To use the linac for a $\gamma\gamma$ collider, the gun must be replaced by a photo-injector. In order to minimize the disruption to collider operation, the plan is to add a photo-injector to the last 200 MeV section of the linac.

Figure 1 shows the layout of a $\gamma\gamma$ collider with a center-of-mass energy around 1 MeV for direct observation of $\gamma\gamma \to \gamma\gamma$ scattering and $\gamma\gamma \to e^+e^-$ pair production [9,10]. The size of the building is based on an existing experimental hall, which is a candidate site to host this $\gamma\gamma$ collider.

Two consecutive electron bunches, spaced by tens of ns (several meters), come from the BEPC linac, each of 200 MeV and 2 nC. A kicker gives a kick on the second bunch and sends the two bunches to two separate transport lines. Each bunch passes through a final focusing system (FFS) and is focused to a few μm in the transverse dimension. An intense laser beam hits each of the electron bunches at the conversion point (CP) and generates a back-scattered high energy (~1 MeV) γ-ray. The two γ-rays then collide at the interaction point (IP). The used electrons remain at about 200 MeV and go downstream to a beam dump near the bending magnet. There is also a dump for the used laser beam (not shown in Fig. 1). The FFS is a permanent magnet quadrupole (PMQ) triplet chosen due to the requirements of high gradient and small size. The details of the interaction region (IR) including the triplet and laser path are under design. Table 1 lists a preliminary set of parameters.

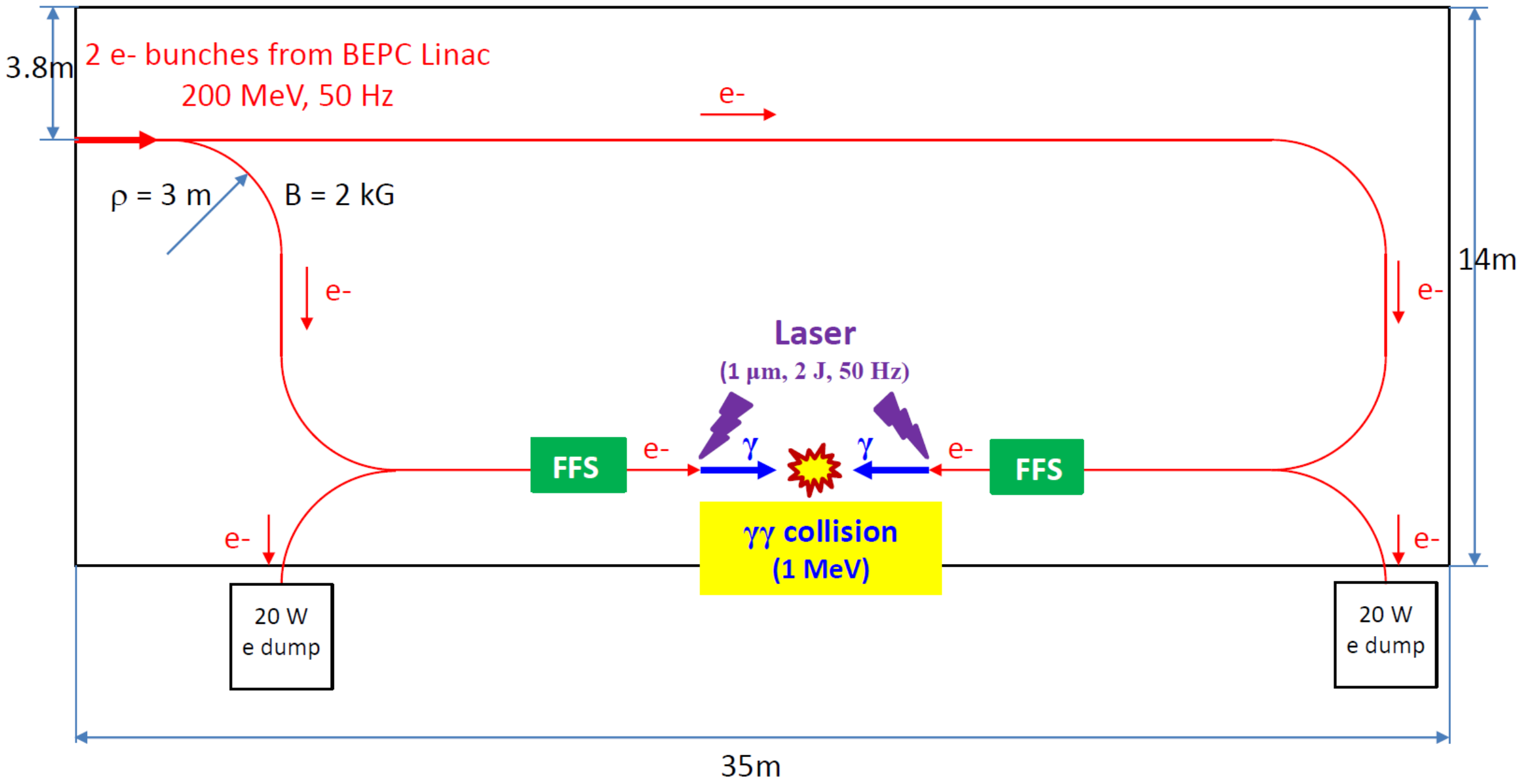

**Fig. 1:** Layout of a 1 MeV γγ collider in an experimental hall at IHEP.

**Table 1:** Preliminary parameters of a 1 MeV (c.m.) γγ collider

| **Laser beam** | | **Electron beam** | |
|---|---|---|---|
| Wavelength | 1.064 μm | Energy | 200 MeV |
| Waist size (RMS) | 5 μm | Bunch charge | 2 nC |
| Rayleigh range | 298 μm | Size at CP (RMS) | 2 μm |
| Pulse energy | 2 J | Emittance | 6.4E–3 μm |
| Pulse length | 0.33 ps | β* | 626 μm |
| Repetition rate | 50 Hz | Bunch length (RMS) | 2 ps |
| Cross angle | 167 mrad | Repetition rate | 50 Hz |
| IP – CP distance | 313 μm | Crossing angle | 0 |
| Nonlinear parameter $a_0$ | 0.45 | Geometric luminosity of $e^-e^-$ | 1.6E28 |

Based on the parameters in Table 1, CAIN simulation gives a γγ collision luminosity of $3.3\times10^{27}$ $cm^{-2}s^{-1}$ for all energies and $1.1\times10^{27}$ $cm^{-2}s^{-1}$ for energies above 0.9 MeV, as shown in Figure 2 [14]. At 1 MeV, the cross section of $\gamma\gamma \rightarrow \gamma\gamma$ is peaked at about 3 μb, $\gamma\gamma \rightarrow e^+e^-$ about 100 mb. So, the expected event rate is, respectively, several per hour for light-by-light scattering and ~100 per second for pair production. (It is interesting to note the light-by-light scattering event rate is comparable to the Higgs event rate from the CEPC, in which the luminosity is higher by 7 orders of magnitude but the cross section for $e^+e^- \to ZH$ is smaller by the same order of magnitude [15].)

The required laser (2 J, 50 Hz, 100 W, 0.33 ps) is challenging but achievable according to several laser experts and vendors. It is a TW system at 50 Hz. Compared to HAPLS, a PW/10 Hz laser made for ELI-Beamlines, it should be simpler, easier and less expensive. The synchronization between the two laser pulses and between the laser and electron is critical. Modern technology can give a time jitter at 100 fs or less, which should be adequate for our purpose [16].

The detector is another challenge because of large background from $e^-e^-$ and $e^-\gamma$ collisions in addition to background from the environment and cosmic rays. A preliminary design uses a multilayer cylinder consisting of plastic scintillators, CsI crystals and photo multiplier tubes (PMT). Physics simulations using CAIN, PYTHIA and other generators and detector simulations using Geant4 are underway.

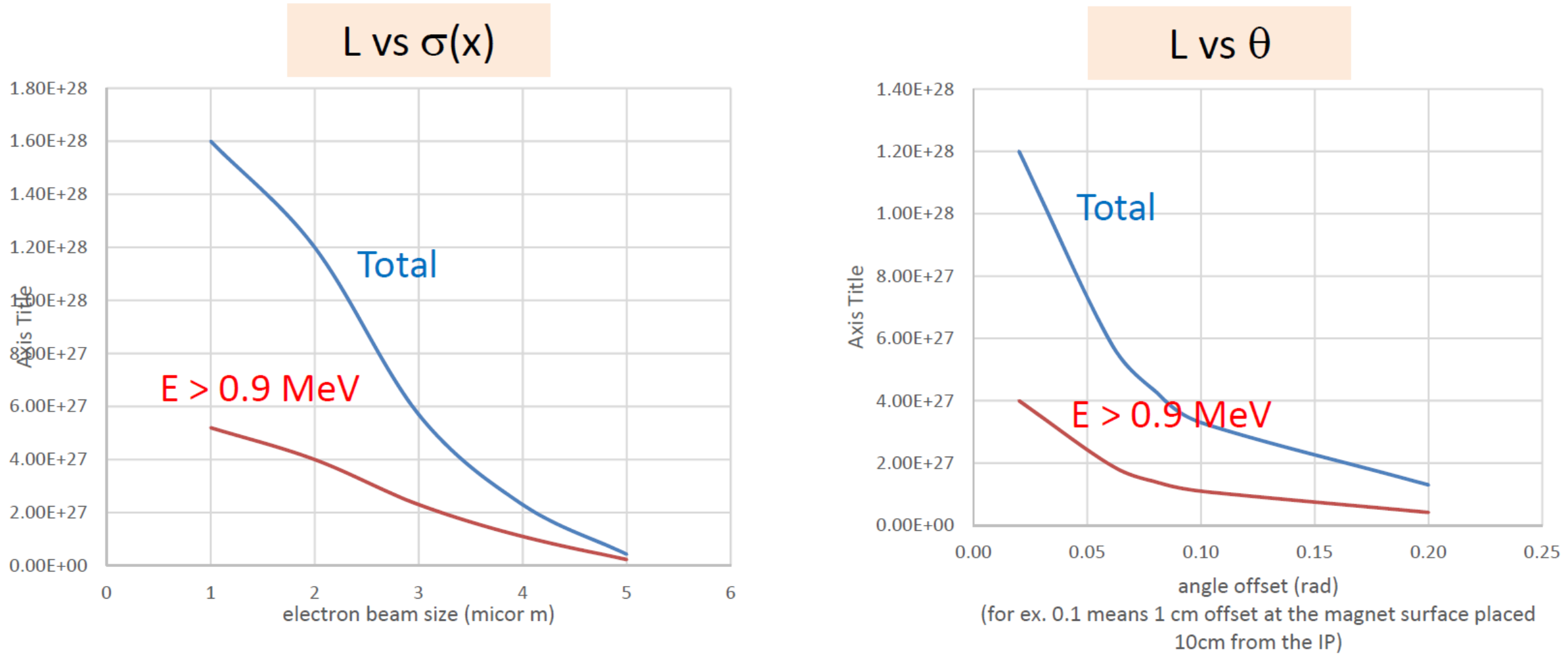

**Fig. 2:** Left: luminosity vs. bunch length. Right: luminosity vs. laser crossing angle. (courtesy T. Takahashi)

## Acknowledgement

The author thanks the participants of the 2017 ICFA mini-workshop on $\gamma\gamma$ colliders and Chair Professor Wei Lu for stimulating discussions. He also thanks the IHEP – Tsinghua University – Beihang University team for the progress in the study of a low energy $\gamma\gamma$ collider. Discussions with Professor T. Takahashi are particularly enlightening and greatly appreciated.

# Photon-Photon Measurements in ATLAS

*Chav Chhiv Chau on behalf of the ATLAS Collaboration*
Carleton University, Ottawa, Canada

**Abstract**
At the Large Hadron Collider, two-photon production is a key process for studying photon interactions at high energies. Measurements of two-photon processes provide a direct access to the photon distribution function in protons and nuclei, and can probe the couplings between four vector bosons ($WW\gamma\gamma$) to find any deviation from the Standard Model. Two-photon measurements reported by the ATLAS Collaboration are presented. These measurements are performed using the proton-proton collisions at the centre-of-mass energy of 7 TeV ($pp \rightarrow pp\ell^+\ell^-$) and 8 TeV ($pp \rightarrow ppW^+W^- \rightarrow ppe^\pm\mu^\mp$), and using the lead-lead collisions at the centre-of-mass energy of 5.02 TeV (Pb + Pb $\rightarrow$ Pb + Pb + $\mu^+\mu^-$).

## 1 Introduction

Two-photon induced production is a purely electroweak process and has a relatively clean signature compared to most interactions recorded in the ATLAS experiment [1]. Indeed, the incoming protons or lead nuclei can create nearly-real photons coherently and escape the interaction with only a slight deviation with respect to their initial trajectories. Then, exclusive dileptons or $W$-boson pairs are produced via photon-photon scattering ($\gamma\gamma \rightarrow \ell^+\ell$, $\gamma\gamma \rightarrow W^+W^-$).

The ATLAS Collaboration has measured the exclusive production of dileptons ($pp \rightarrow pp\ell^+\ell^-$, where $\ell$ is an electron or a muon) and $W$-boson pairs ($pp \rightarrow ppW^+W^- \rightarrow ppe^\pm\mu^\mp$) in proton-proton ($pp$) collisions and exclusive dimuon production (Pb + Pb $\rightarrow$ Pb + Pb + $\mu^+\mu^-$) in lead-lead collisions. The measurements cover a wide range of the two-photon mass spectrum thanks to the LHC high centre-of-mass energy; for example, above the dilepton pair continuum of $m_{\mu\mu} = 10$ GeV for Pb + Pb $\rightarrow$ Pb + Pb + $\mu^+\mu^-$ production (a phase space not covered by previous measurements). They provide a way to determine the distribution of photons emitted from a proton and a nucleus. Furthermore, exclusive $W^+W^-$ production is sensitive to the $WW\gamma\gamma$ quartic gauge couplings. The strength of the couplings can be extracted in order to compare to Standard Model prediction, since any deviation could indicate the existence of new physics.

The exclusive signal is modeled using the equivalent photon approximation (EPA) [2], which describes the fast moving proton and lead ion as structureless charged particle. As shown in Fig. 1, the proton scattering can be elastic where both protons remain intact, single-dissociative (SD) where one of them dissociates and double-dissociative (DD) where both fragment. The elastic process is characterized by the production of back-to-back leptons, i.e. the transverse momentum of the dilepton system $p_{\mathrm{T}}^{\ell\ell} \sim 0$, providing a way to separate the elastic from the dissociative production. For the heavy ion analysis, the dissociative interactions are not considered.

The following three measurements are presented here. The exclusive dilepton measurement [2] is performed using 4.6 fb$^{-1}$ of $pp$ collisions at the centre-of-mass energy of $\sqrt{s}$ = 7 TeV recorded in 2011. The exclusive $W^+W^-$ analysis [3] is also performed with $pp$ collisions, but using the dataset recorded in 2012 at $\sqrt{s}$ = 8 TeV corresponding to 20.2 fb$^{-1}$. Reported more recently, the Pb + Pb $\rightarrow$ Pb + Pb + $\mu^+\mu^-$ measurement [4] is done using a 515 $\mu$b$^{-1}$ data sample collected at $\sqrt{s}$ = 5.02 TeV in 2015.

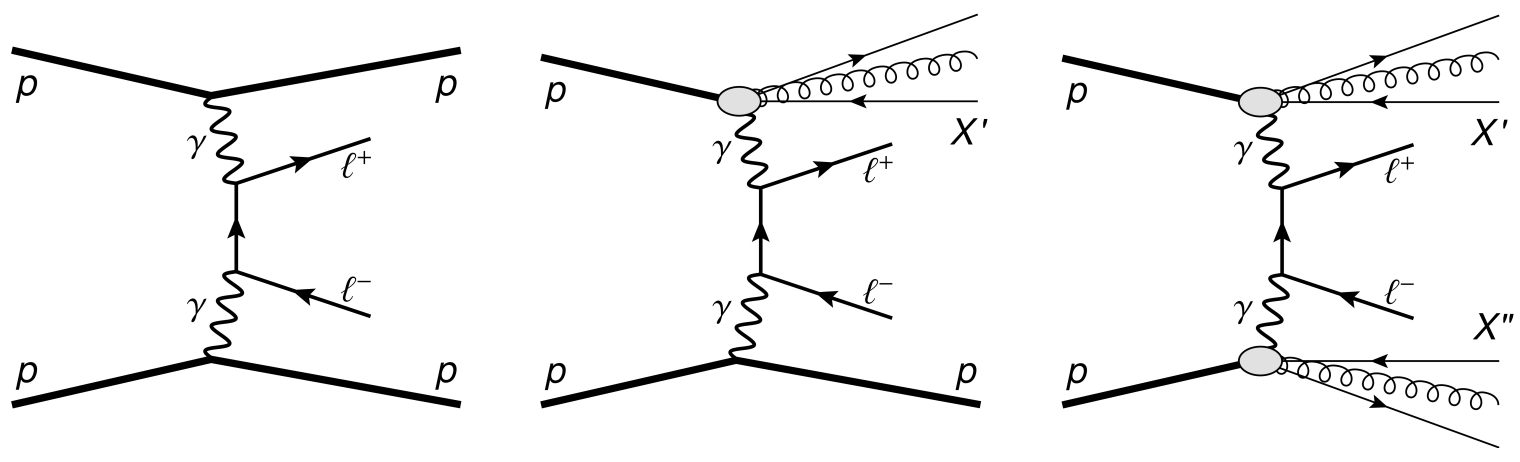

**Fig. 1:** Diagrams for the exclusive dilepton production representing the (left) elastic process, (centre) single-dissociation where one initial proton dissociates (SD) and (right) double-dissociation where both protons fragment (DD). The symbols $X'$ and $X''$ denote any additional final state created.

## 2 Exclusive Dilepton Production

The exclusive dilepton analysis considers the $e^+e^-$ and $\mu^+\mu^-$ final states. The data sample corresponds to 4.6 fb$^{-1}$ of $pp$ collision at $\sqrt{s}$ = 7 TeV recorded by the ATLAS experiment during 2011. The signal is the elastic dilepton production, while the single- and double-dissociative processes are considered as background. During the 2011 operation, many extra interactions occur during a same bunch crossing, so the signal events are selected only if they have a hard scattering vertex with exactly two lepton tracks isolated from other tracks and vertices.

### 2.1 Event Selection and Triggers

The events are first required to satisfy the following trigger logic: a muon with $p_{\mathrm{T}}^{\mu} > 18$ GeV or a dimuon with $p_{\mathrm{T}}^{\mu} > 10$ GeV for the muon channel; an electron with $p_{\mathrm{T}}^{e} > 20$ GeV or a di-electron with $p_{\mathrm{T}}^{e} > 12$ GeV for the electron channel. The exclusive events are selected by requiring a vertex with exactly two charged-particle tracks with $p_{\mathrm{T}}^{\mathrm{track}} > 0.4$ GeV and no extra tracks or vertices within a 3 mm longitudinal isolation distance from that dilepton vertex ($\Delta z_{\mathrm{Vtx}}^{\mathrm{iso}} > 3$ mm). Figure 2 illustrates the efficiency of the criteria in the muon channel. The exactly-two-tracks criterion suppresses Drell-Yan production and eliminates completely the diboson, multi-jet and top-quark backgrounds. The $\Delta z_{\mathrm{Vtx}}^{\mathrm{iso}} >$ 3 mm requirement further suppresses Drell-Yan background, while retaining most of the signal. A large fraction of the Drell-Yan production that remains is rejected by vetoing the region 70 GeV $< m_{\ell^+\ell^-} <$ 105 GeV. Finally, a requirement on the dimuon transverse momentum $p_{\mathrm{T}}^{\ell^+\ell^-} < 1.5$ GeV is applied to reduce the dissociative background. This requirement suppresses single-dissociation by a factor of 3.

### 2.2 Measured cross section

At the LHC, the exclusive processes are suppressed due to the finite size of protons. The suppression is related directly to the probability of inelastic collisions [5] that rises with the centre-of-mass energy. The suppression factors are extracted from fits of the acoplanarity variable ($1 - |\Delta\phi_{\ell^+\ell^-}|/\pi$, where $\Delta\phi_{\ell^+\ell^-}$ is the azimuthal opening angle between the leptons.) An example of the fit is shown in Fig. 3 for the electron channel. As expected, the elastic dileptons have small acoplanarity compared to the single-dissociative events. The double-dissociative and Drell-Yan contamination is insignificant and is fixed to the Monte Carlo (MC) predictions. The acoplanarity fits estimate the elastic and SD fractions in data and determine the correction factors required to bring the simulation to agree with data. For the elastic signal, these factors are found to be $R^{\mathrm{excl}}_{\gamma\gamma\to e^+e^-} = 0.86 \pm 0.07$ and $R^{\mathrm{excl}}_{\gamma\gamma\to\mu^+\mu^-} = 0.70 \pm 0.04$ for the electron and muon channel, respectively. The SD factors are 0.76 for both channels.

The cross sections are measured in the fiducial regions defined by the following criteria: $p_{\mathrm{T}}^{e} >$ 12 GeV, $|\eta^e| < 2.4$ and $m_{ee} > 24$ GeV for the electron channel; $p_{\mathrm{T}}^{\mu} > 10$ GeV, $|\eta^\mu| < 2.4$ and $m_{\mu\mu} > 20$ GeV for the muon channel. They are evaluated using the following relation:

$$\sigma^{\mathrm{excl}}_{\gamma\gamma\to\ell^+\ell^-} = R^{\mathrm{excl}}_{\gamma\gamma\to\ell^+\ell^-} \cdot \sigma^{\mathrm{EPA}}_{\gamma\gamma\to\ell^+\ell^-}, \tag{1}$$

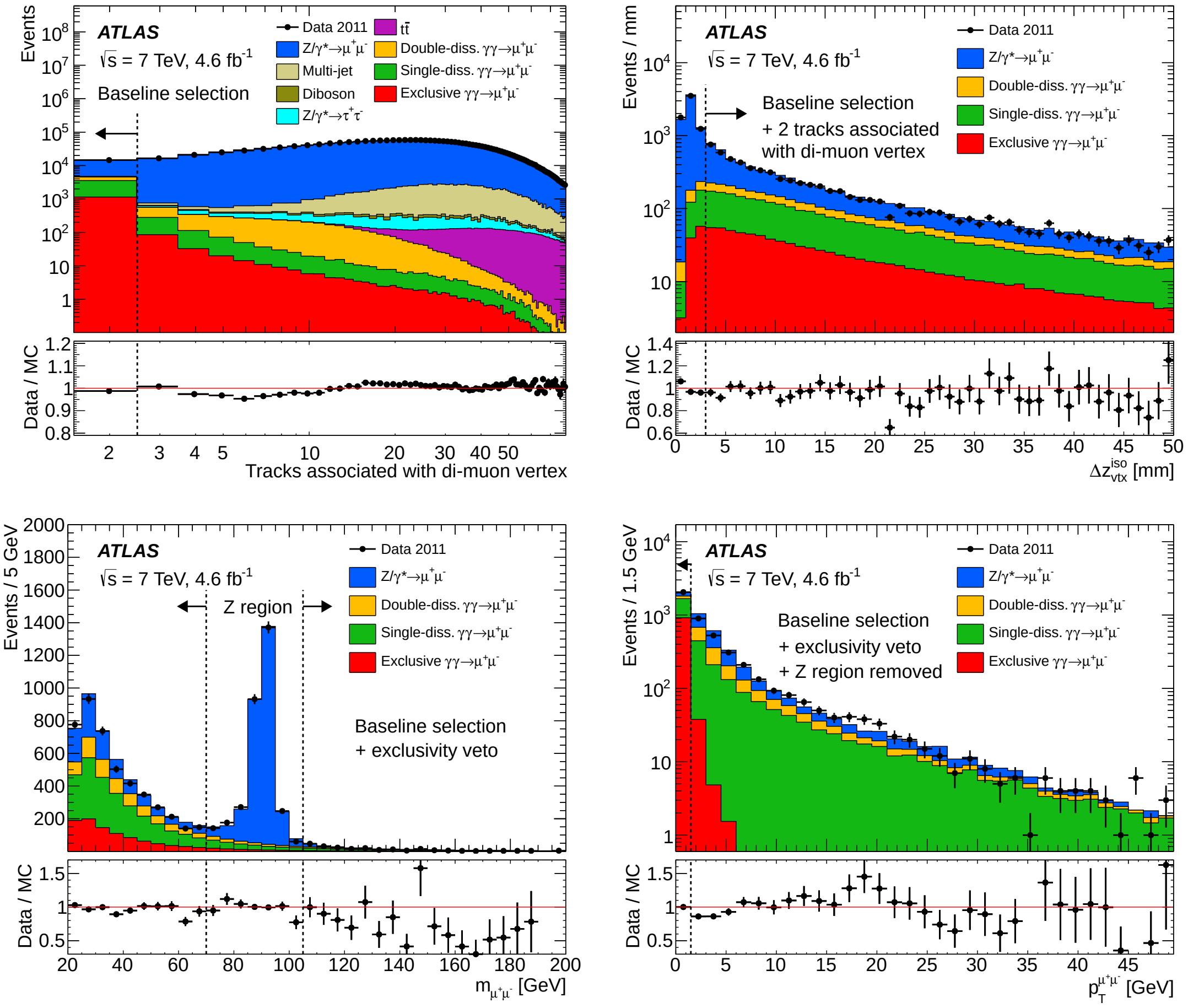

**Fig. 2:** The distributions of (top left) the number of tracks associated with the dimuon vertex, (top right) the longitudinal distance between the dimuon vertex and any other tracks or vertices, (bottom left) the invariant mass and (bottom right) the transverse momentum of the dimuon system. The black dashed lines and arrow indicate the requirements. More details can be found in Ref. [2]

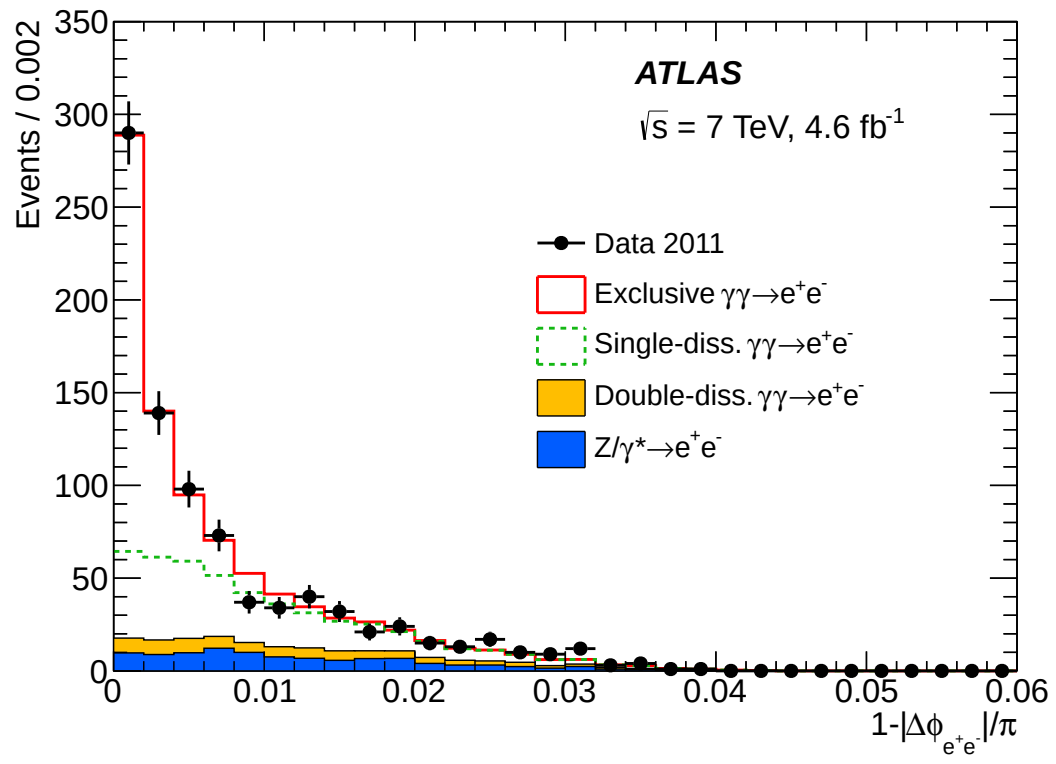

**Fig. 3:** Acoplanarity distribution of electron pairs passing all requirements. The elastic and SD yields are determined from the fit described in the text. More details can be found in Ref. [2].

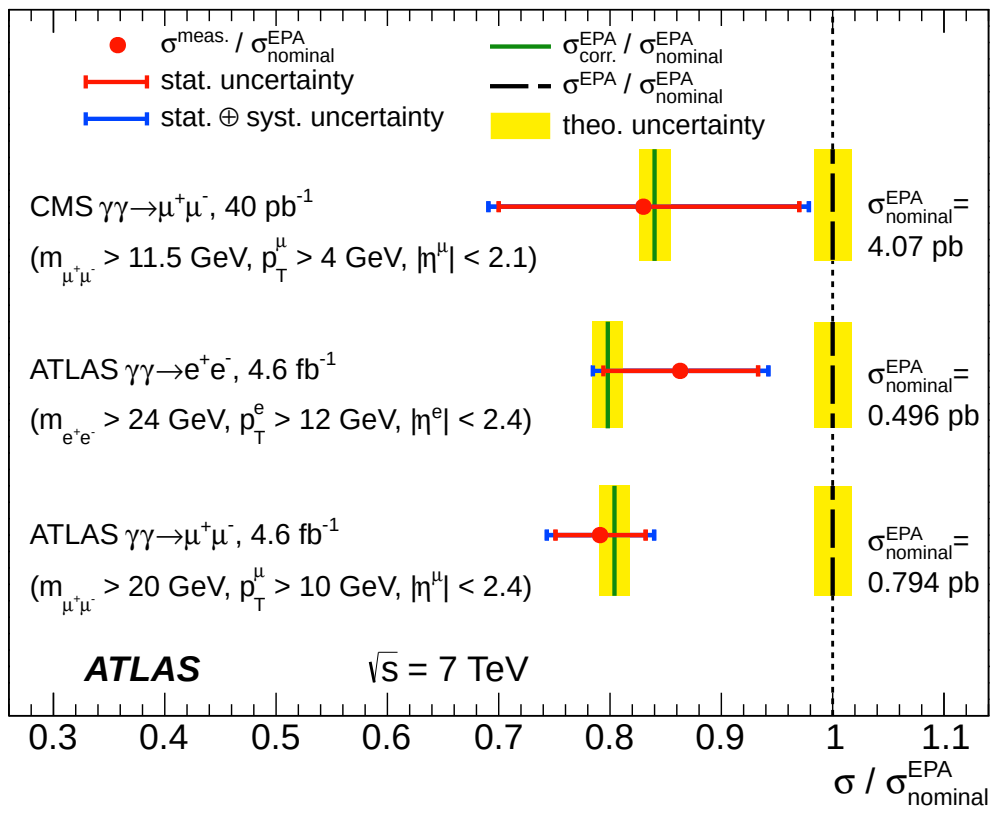

**Fig. 4:** Comparison of the ratios of measured (red points) and ratios of predicted (dashed line) cross section to the uncorrected EPA prediction [2]. A similar CMS measurement is also shown.

where $\sigma^{\mathrm{EPA}}_{\gamma\gamma\to\ell^+\ell^-}$ is the cross section calculated with the HERWIG++ generator [6] for the corresponding fiducial region.

The cross sections are determined to be 0.428 ± 0.035 (stat.) ± 0.018 (syst.) pb for the electron channel, and 0.628 ± 0.032 (stat.) ± 0.021 (syst.) pb for the muon channel. They agree within experimental uncertainties with the predictions [5]: $0.398 \pm 0.007$ (theor.) for the electron channel and $0.638 \pm 0.011$ (theor.) for the muon channel. More predictions [7] are available recently. Figure 4 shows the ratios of measured cross section to the prediction corrected for proton absorptive effects and to the uncorrected (nominal) prediction. A similar CMS measurement [8] is also included in the comparison. The measured cross sections are in agreement with the predicted values corrected for the proton absorptive effects [5].

## 3 Exclusive $W^+W^-$ Production

The exclusive $W^+W^-$ analysis considers the opposite-charge different-flavour $e^{\pm}\mu^{\mp}$ final states. The measurement is performed using 20.2 fb$^{-1}$ of $pp$ collisions at $\sqrt{s}$ = 8 TeV recorded by the ATLAS experiment during 2012. The exclusive $W^+W^-$ process has a clean signature, since no extra activity is produced in addition to the two leptons. Also, it is a key process to probe the $WW\gamma\gamma$ quartic gauge couplings. Any enhancement of the coupling strength could indicate the existence of new physics.

Without tagging the initial state protons combined with having neutrinos in the final states, it is not possible to separate the elastic from the dissociative events. Therefore, the elastic, SD and DD $W^+W^-$ processes are together considered as signal. The dissociative contribution is estimated from data, using an exclusive dimuon sample with $m_{\mu\mu} > 160$ GeV (twice the mass of $W$ boson). The data-driven scale factor $f_\gamma$ defined in Eq. (2) is given by the ratio of the number of data after subtracting off the number of background events ($\mathrm{N_{Observed}}$ - $\mathrm{N_{Background}}$) to the number of elastic events ($\mathrm{N_{Elastic}}$) predicted by HERWIG++. So the total expected signal for analysis is the prediction for the elastic $W^+W^-$ production scaled by $f_\gamma$,

$$f_\gamma = \frac{\text{Elastic+SD+DD}}{\text{Elastic from MC}} = \left.\frac{\mathrm{N_{Observed}} - \mathrm{N_{Background}}}{\mathrm{N_{Elastic}}}\right|_{m_{\mu\mu}>160\ \mathrm{GeV}}. \tag{2}$$

### 3.1 Exclusivity Selection

The events selected for analysis satisfy any of the following triggers: a single-muon trigger with $p_{\mathrm{T}}^{\mu} > 24$ GeV, a single-electron trigger with $p_{\mathrm{T}}^{e} > 24$ GeV or an electron-muon trigger with $p_{\mathrm{T}}^{e} > 12$ GeV and $p_{\mathrm{T}}^{\mu} > 8$ GeV. The pileup conditions in 2012 are higher than in 2011. Therefore, a new track-based strategy for selecting exclusive events is developed for the analysis. The strategy starts by selecting two leptons with $p_{\mathrm{T}}^{\ell} > 20$ GeV and $m_{\ell\ell} > 20$ GeV. The event vertex is then reconstructed as the average of the longitudinal impact parameters ($z_0$) of the leptons: $(z_0^{\ell 1} + z_0^{\ell 2})/2$. Also, the two leptons are required to be within 1 mm in $z$ of each other. At the next step, the longitudinal distance between the lepton vertex and other extra tracks ($\Delta z_0$) is computed. Finally, the exclusivity selection keeps only events with zero extra track within a 1 mm longitudinal distance of the lepton vertex ($\Delta z_0^{\mathrm{iso}}$ = 1 mm).

Cross-checks are done in the muon channel to validate the new strategy. Two of them are discussed here. The first cross-check extracts the suppression factor ($f_{\mathrm{EL}}$) related to the proton absorptive effects and compares it to the expected value deduced from the exclusive dilepton measurement (Section 2.2). Similar fits of the acoplanarity variable are done, but with slightly different selection criteria: $p_{\mathrm{T}}^{\mu} > 20$ GeV, $m_{\mu\mu} > 45$ GeV, $|m_{\mu\mu} - m_Z| > 15$ GeV, $p_{\mathrm{T}}^{\mu\mu} < 3$ GeV and $\Delta z_0^{\mathrm{iso}}$ = 1 mm. The correction factor obtained is $f_{\mathrm{EL}}$ = 0.76 ± 0.04(stat.) ± 0.10(syst.). Figure 5 shows the acoplanarity distributions of dimuons passing all the requirements. The elastic and dissociative contributions are corrected using the $f_{\mathrm{EL}}$ factor and the background is fixed to the Monte Carlo prediction.

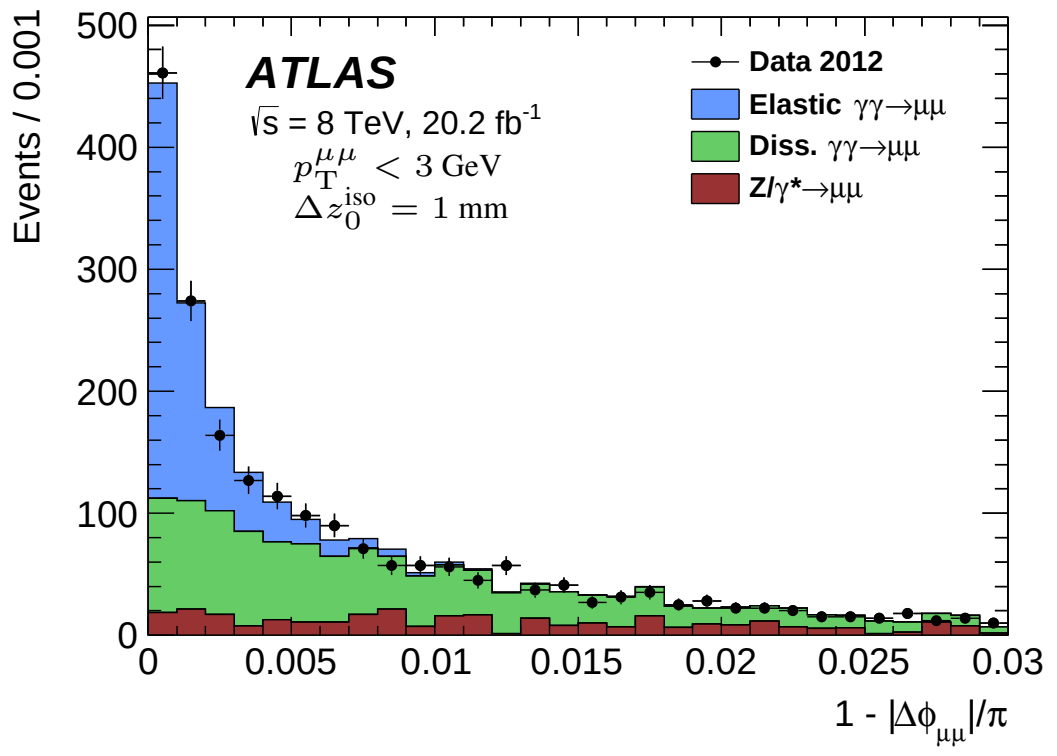

**Fig. 5:** Acoplanarity distribution of dimuons after requiring $p_{\mathrm{T}}^{\mu\mu} < 3$ GeV and $\Delta z_0^{\mathrm{iso}} = 1$ mm compared to data [3]. The elastic and dissociative yields are determined from the fit described in Section 2.2.

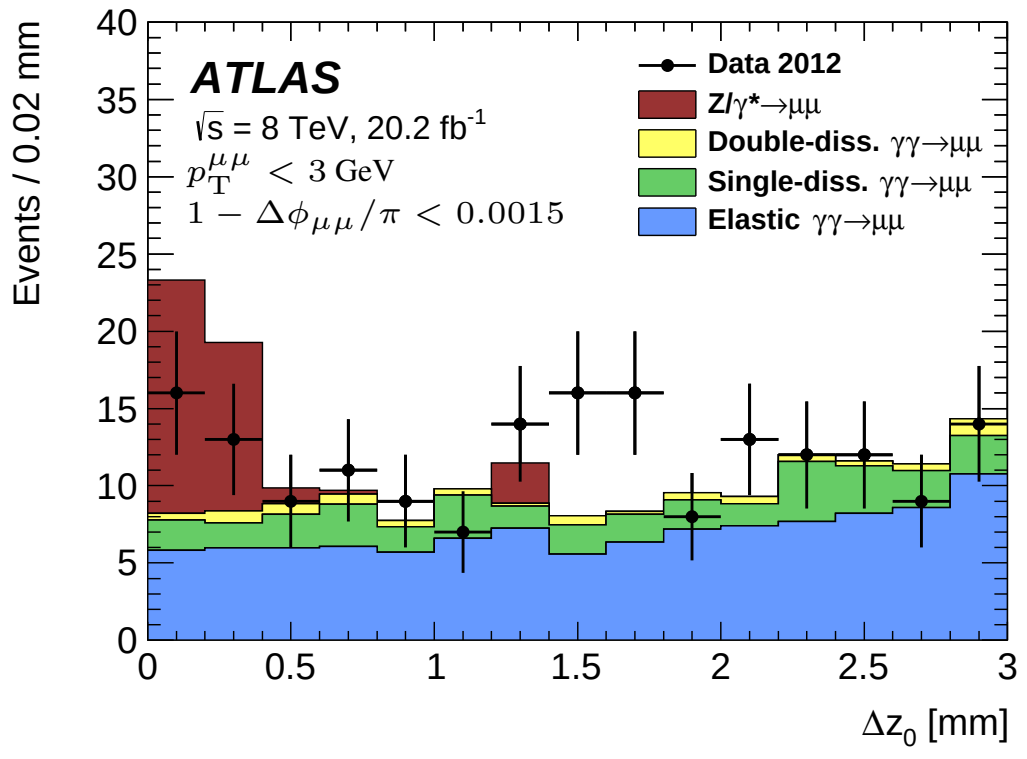

**Fig. 6:** Longitudinal distance between the lepton vertex and the closest extra track for dimuon events passing the criteria mentioned in the text [3].

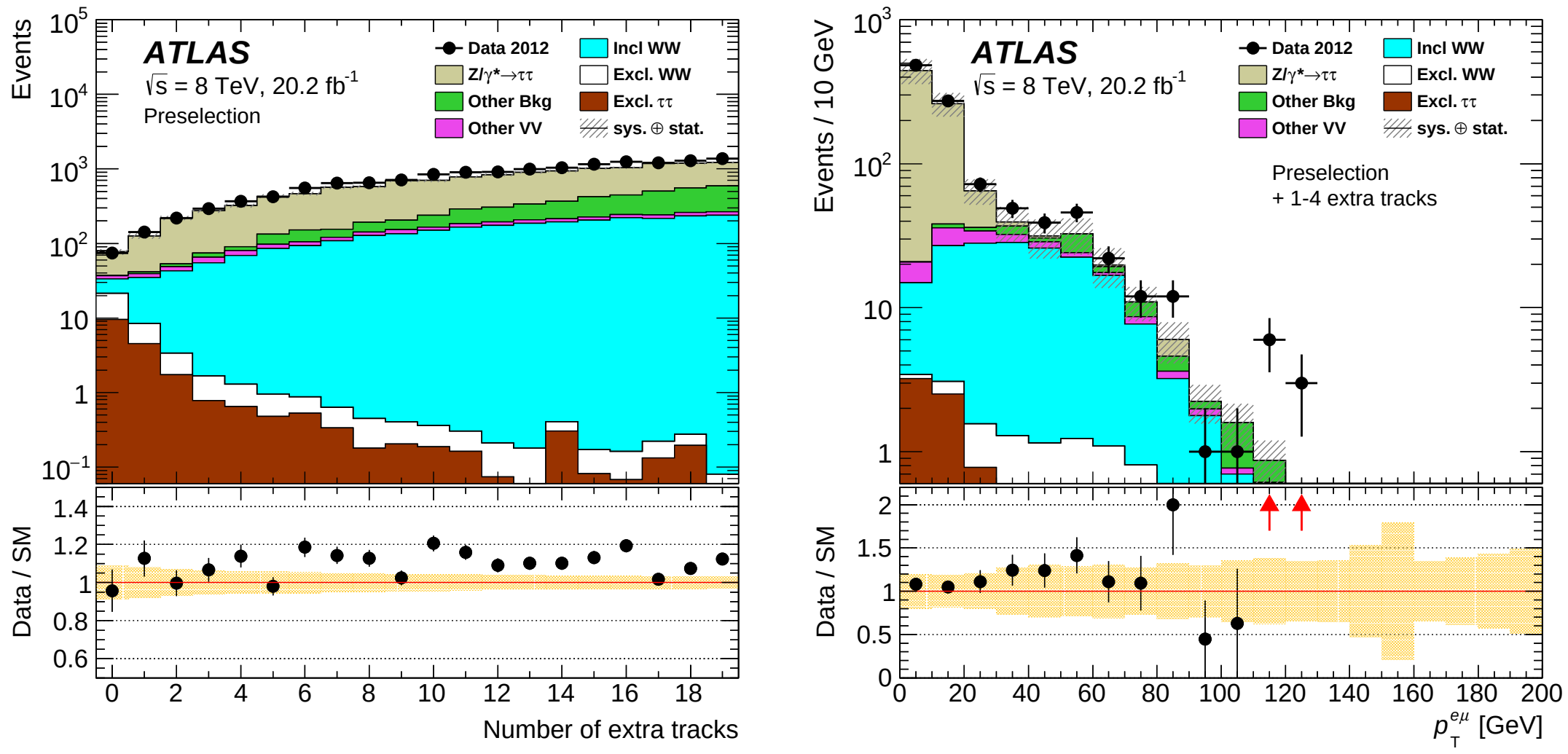

**Fig. 7:** Distribution of track multiplicities after requiring the exclusive $W^+W^-$ preselection (Table 1) with no number of track dependent correction (left), and the $p_{\mathrm{T}}^{e\mu}$ distribution of candidates that have 1–4 extra tracks (right). The enriched inclusive $W^+W^-$ control region is the 1–4 extra-track region above $p_{\mathrm{T}}^{e\mu} > 30$ GeV. The band around the Data/SM ratio of unity illustrates the systemic uncertainties. More details can be found in Ref. [3].

A second cross-check studies the pileup modelling and the effects of pileup on the efficiency of the exclusivity selection. Figure 6 compares the $\Delta z_0$ of simulated dimuon events to data. The events pass the same criteria as the previous cross-check, except that the acoplanarity < 0.0015 requirement replaces the exclusivity selection. The background passing all the criteria is Drell-Yan production and its $\Delta z_0$ distribution peaks at zero due to the underlying event (accompanying the hard interaction) producing extra tracks. For exclusive production, the extra track is a pileup track, so the $\Delta z_0$ distribution is flat.

### 3.2 Control Regions

Figure 7 shows the distribution of the number of extra tracks and the $p_{\mathrm{T}}$ spectrum of the events with 1 to 4 extra tracks. Selecting events with zero extra track keeps most of the signal, while eliminating a large fraction of background. Therefore, the events with 1–4 extra tracks constitute a good sample to control the background.

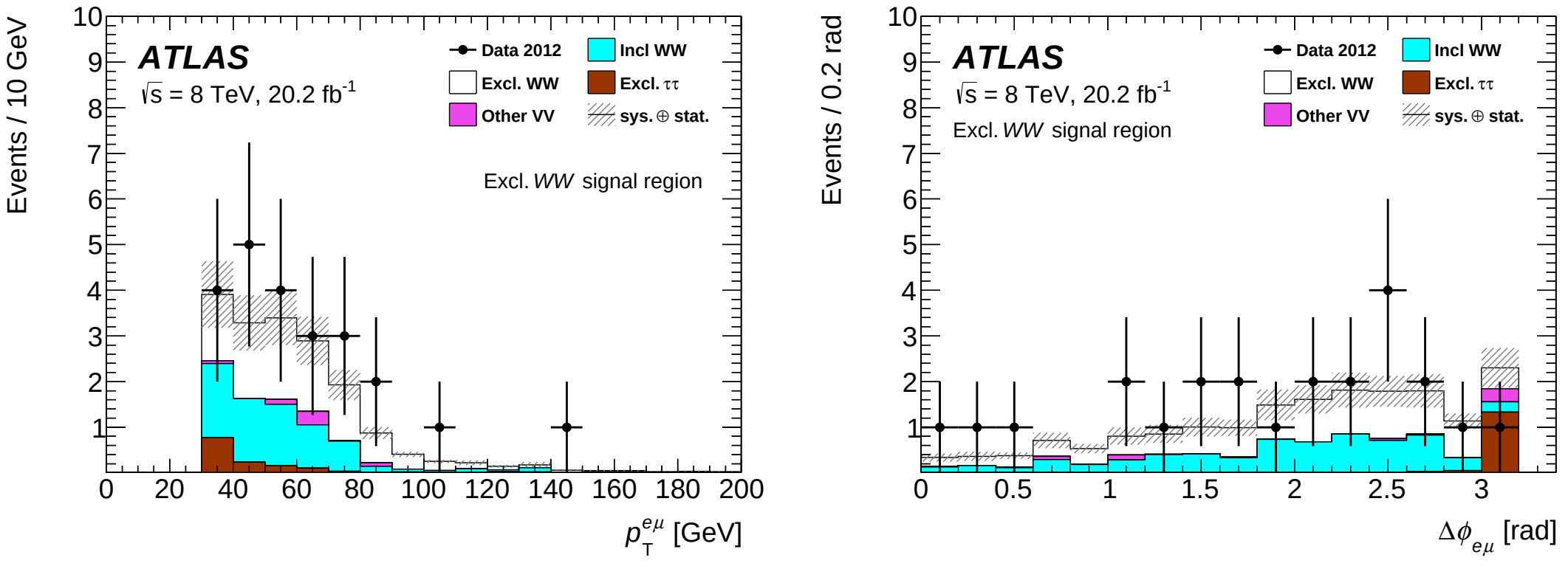

**Fig. 8:** Kinematic distributions in the exclusive $W^+W^-$ signal region comparing the simulation to data [3].

### 3.3 Measurement

The event yields can be found in Table 1. The Preselection keeps good $e^\pm\mu^\mp$ events consistent with $W$-boson pair production. The $p_T^{e\mu} > 30$ GeV requirement further reduce the Drell-Yan production and dissociative backgrounds. The exclusivity selection suppresses the inclusive $W^+W^-$ background by three orders of magnitude. So, in the exclusive $W^+W^-$ signal region, 23 candidates are observed in data, while 9.3 signal and 8.3 background events are expected. The observed significance over the background-only hypothesis is $3\sigma$, constituting evidence of exclusive $W^+W^-$ production in $pp$ collisions. Figure 8 shows the $p_T^{e\mu}$ and $\Delta\phi_{e\mu}$ distributions in the signal region. The simulation describes well the data.

The cross section times branching ratio is determined to be 6.9 ± 2.7 fb. It is consistent with the prediction. Since no significant excess over the Standard Model is observed, a 95% C.L. limit on anomalous quartic gauge couplings (aQGCs) is set.

| | Signal | Data | Total Bkg | Incl $WW$ | Excl. $\tau\tau$ | non-$WW$ | Other Bkg | MC/Data |
|---|---|---|---|---|---|---|---|---|
| Preselection | 22.6 ± 1.9 | 99424 | 97877 | 11443 | 21.4 | 1385 | 85029 | 0.98 |
| $p_T^{e\mu} > 30$ GeV | 17.6 ± 1.5 | 63329 | 63023 | 8072 | 4.30 | 896.3 | 54051 | 1.00 |
| Exclusivity selection | 9.3 ± 1.2 | 23 | 8.3 ± 2.6 | 6.6 ± 2.5 | 1.4 ± 0.3 | 0.3 ± 0.2 | − − − | 0.77 |
| aQGC signal region | | | | | | | | |
| $p_T^{e\mu} > 120$ GeV | 0.37 ± 0.04 | 1 | 0.37 ± 0.13 | 0.32 ± 0.12 | 0.05 ± 0.03 | 0 | − − − | 0.74 |

**Table 1:** The event yields at different stages of exclusive $W^+W^-$ event selection [3]. The Preselection consists of the following requirements: exactly two $e^\pm\mu^\mp$ leptons, $p_T^{\ell 1} > 25$ GeV, $p_T^{\ell 2} > 20$ GeV and $m_{e\mu} > 20$ GeV.

The aQGC signal region has one additional requirement: $p_T^{e\mu} > 120$ GeV. Figure 9 shows the $p_T^{e\mu}$ of events passing all selection criteria apart for the one on $p_T^{e\mu}$ itself. Several predictions for the aQGC parameters are shown as well, indicating that aQGCs would enhance the high $p_T^{e\mu}$ region. The aQGC limits are evaluated using the one data candidate as constraint. The 95% confidence level contour and 1-dimensional limits are shown in Fig. 10. The limits obtained are consistent with the CMS combined 7 and 8 TeV measurement [9].

## 4 Dimuons in Lead-Lead Collisions

The heavy-ion measurement is of the cross section for exclusive dimuon production in lead nuclei collisions for invariant mass $m_{\mu\mu} > 10$ GeV. The data sample corresponds to 515 $\mu$b$^{-1}$ of Pb+Pb collisions recorded by the ATLAS experiment in 2015. At high energy, the nuclei can produce quasi-real photons coherently with an enhancement of $Z^2$ over the incoherent production. In ultra-peripheral collisions, the two-photon production would be enhanced by $Z^4$. Therefore a significant rate of such QED processes is expected, although they have small cross section.

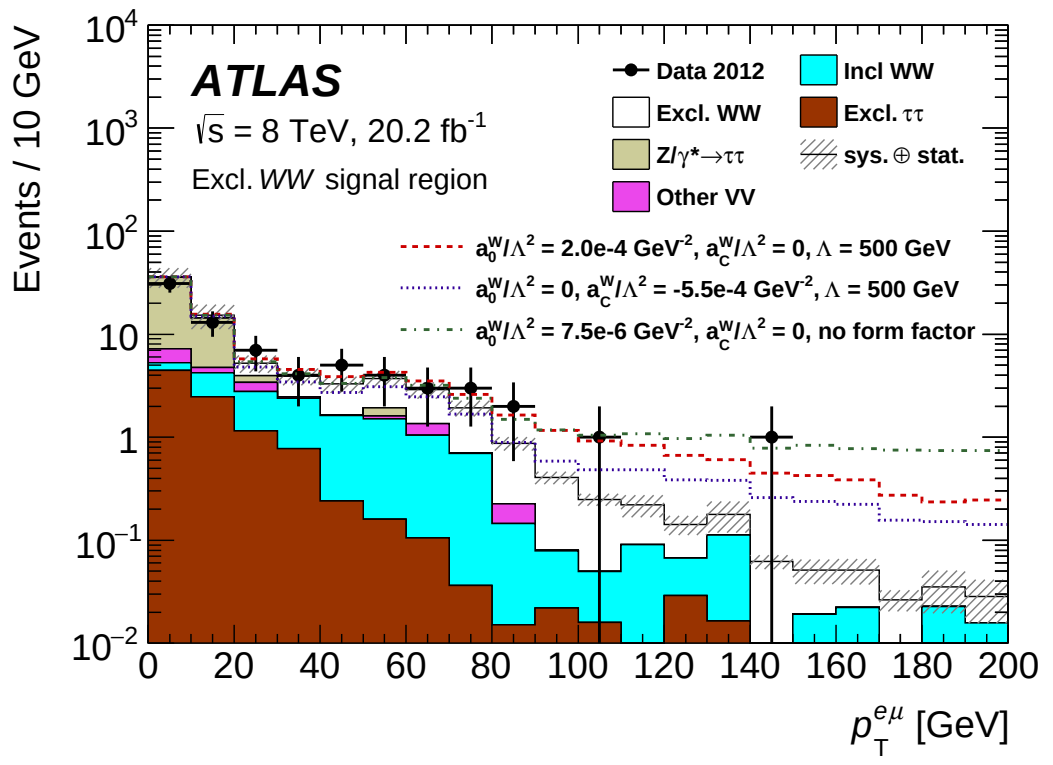

**Fig. 9:** Distribution of events passing all exclusive $W^+W^-$ selection, except the requirement on $p_{\mathrm{T}}^{e\mu}$ [3]. Several aQGC scenarios are overlaid (dashed lines) as examples.

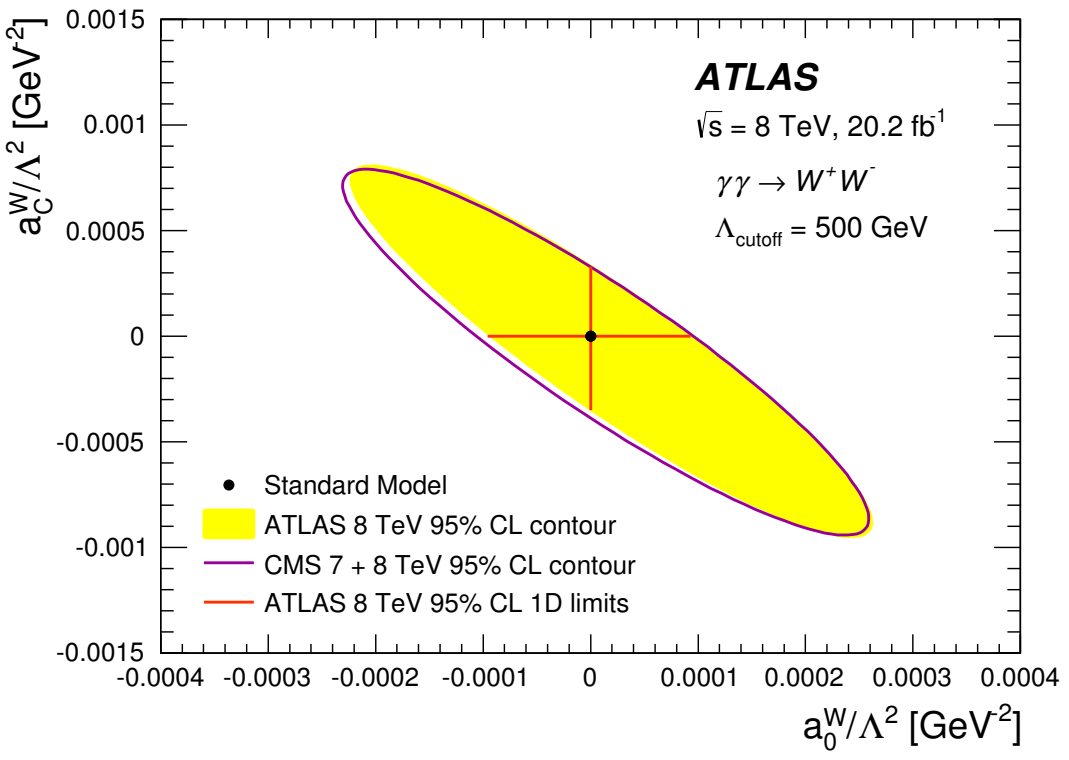

**Fig. 10:** The observed 95% confidence level contour and 1-dimensional limits [3]. The CMS combined 7 and 8 TeV result is shown for comparison.

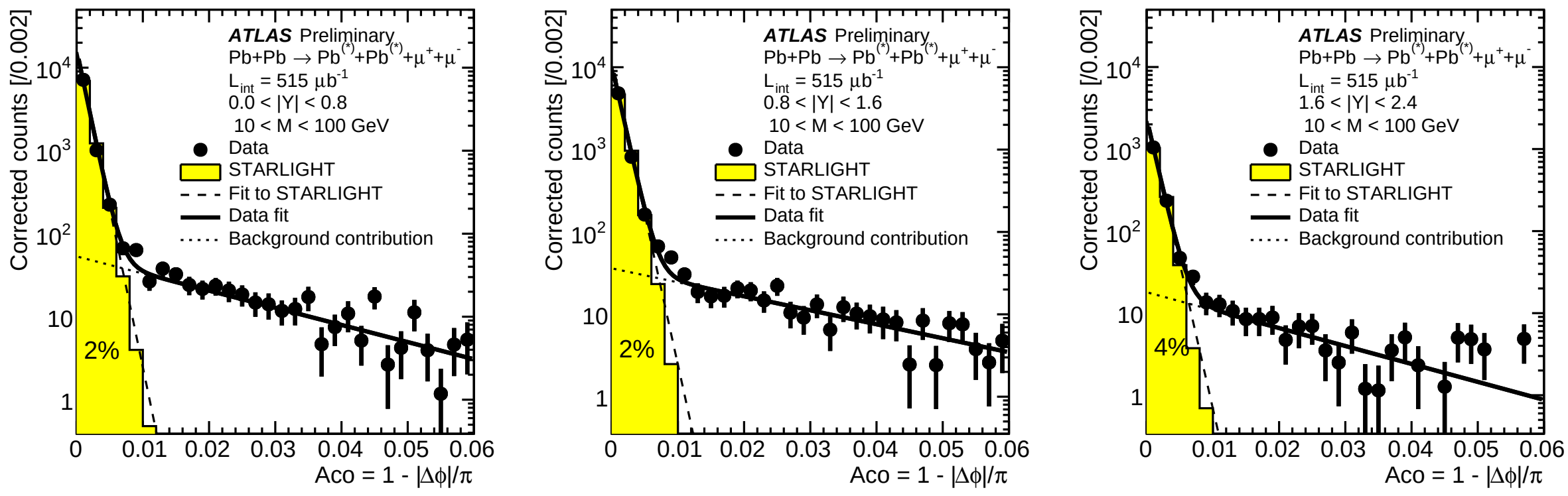

**Fig. 11:** Acoplanarity distributions for 10 GeV $< m_{\mu\mu} <$ 100 GeV in three dimuon rapidity bins: (left) $|Y_{\mu\mu}| <$ 0.8, (centre) 0.8 $< |Y_{\mu\mu}| <$ 1.6 and (right) 1.6 $< |Y_{\mu\mu}| >$ 2.4. More details can be found in Ref. [4].

### 4.1 Measurement

The events are selected if they have a muon and little additional activity in the detector. The selection of exclusive dimuons is similar to the dilepton analysis in $pp$ collisions, i.e. events with a vertex with exactly two good opposite charged muons are selected. The signal is modelled using STARLIGHT [10]. For the run conditions of this analysis, it predicts kinematics coverage of dimuon mass up to 100 GeV.

Figure 11 shows the acoplanarity fits in the region 10 GeV $< m_{\mu\mu} <$ 100 GeV in three rapidity bins: $|Y_{\mu\mu}| <$ 0.8, 0.8 $< |Y_{\mu\mu}| <$ 1.6 and 1.6 $< |Y_{\mu\mu}| >$ 2.4. Most signal events have very small acoplanarity, less than 0.008, as expected. Some activity is observed in the high acoplanarity tail. However, the composition of the events in the tail is unclear: it could be signal, background or a combination of the two. Therefore, two extreme cases are considered: all events in the high acoplanarity tail are signal; all of these events are background. The background estimate is then assumed to be the average of the two cases.

The cross section is determined in the following region: $p_{\mathrm{T}}^{\mu} >$ 4 GeV, $|\eta^{\mu}| <$ 2.4, $m_{\mu\mu} >$ 10 GeV. The cross section for the fiducial region is 32.2 $\pm$ 0.3(stat.) $^{+4.0}_{-3.4}$(syst.) $\mu$b, consistent with the STARLIGHT prediction of 31.64 $\pm$ 0.04(stat.) $\mu$b. Figure 12 shows the cross section as function of $m_{\mu\mu}$ and $Y_{\mu\mu}$ for a few overlapping regions. The simulation describes the data well over the full kinematic range.

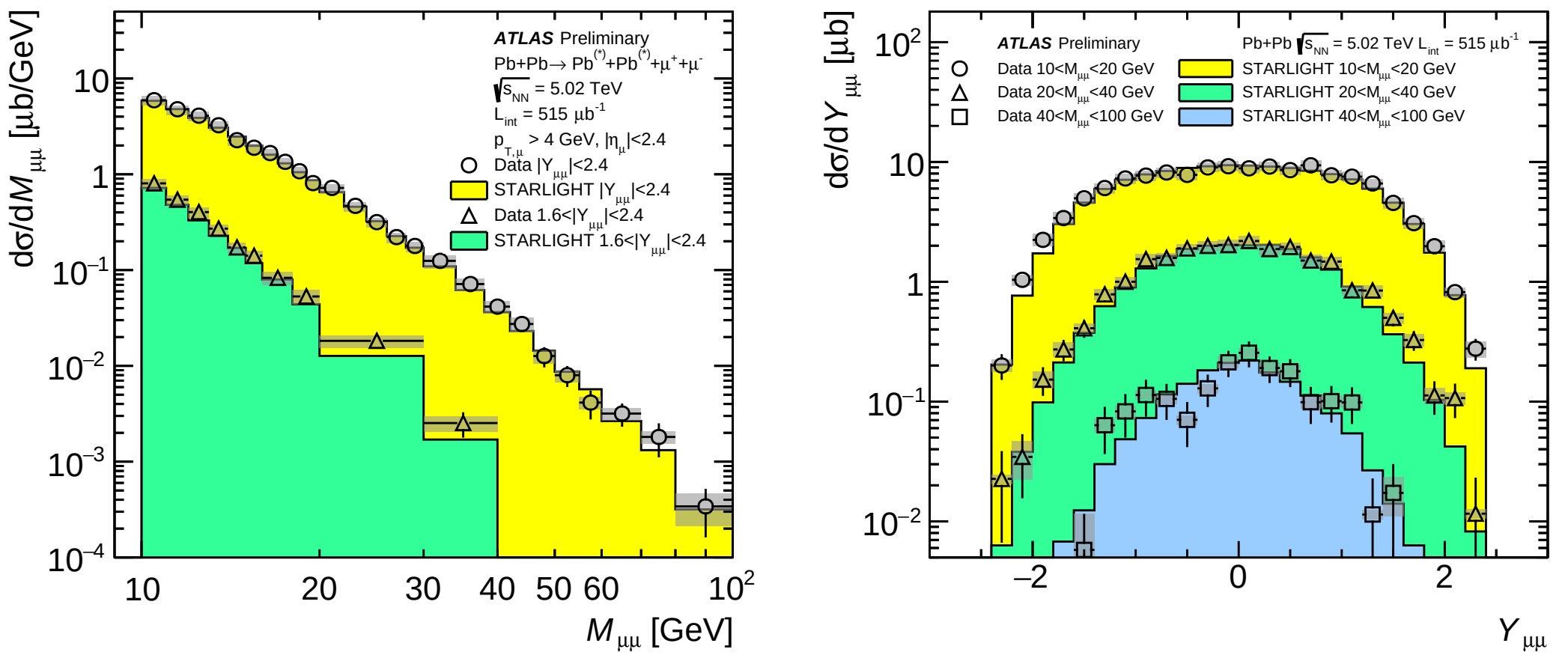

**Fig. 12:** Cross section for exclusive dimuon production in lead-lead collisions as a function of dimuon mass (left) and rapidity (right) [4]. Error bars indicate statistical uncertainties, while grey bands indicate the total systematic uncertainties.

## 5 Conclusion

Two-photon production of lepton pairs and $W$-boson pairs have been measured with the ATLAS detector. The proceedings present two measurements done using $pp$ collision data, $pp \to pp\ell^+\ell^-$ at $\sqrt{s}$ = 7 TeV and $pp \to ppW^+W^- \to ppe^\pm\mu^\mp$ at $\sqrt{s}$ = 8 TeV, and a measurement of Pb + Pb $\to$ Pb + Pb + $\mu^+\mu^-$ production at $\sqrt{s}$ = 5.02 TeV. The ATLAS experiment has also reported 95% C.L. limits on the $WW\gamma\gamma$ anomalous quartic gauge couplings, extracted from the exclusive production of $W$-boson pairs. The measurements provide direct access to the elastic photon distributions in the proton and lead nucleus. Performed using a larger data sample compared to previous measurements resulting in an improved statistical precision, they provide a better understanding of photon interactions in hadron colliders.

# Photon-photon measurements in CMS

*Ruchi Chudasama on behalf of the CMS collaboration*
Bhabha Atomic Research Center, Mumbai, India

**Abstract**
We discuss the measurements of exclusive photon-photon processes ($\gamma\gamma \rightarrow \ell^+\ell^-, \mathrm{W^+W^-}, \gamma\gamma$ ) using data collected by the CMS experiment in pp collisions at $\sqrt{s} = 7$ and 8 TeV and in PbPb collisions at $\sqrt{s_{\mathrm{NN}}}$ =5.02 TeV.

## 1 Introduction

Photon and photon (or "two photon") fusion processes have long been studied at $e^+e^-$, ep and hadron colliders, at the LHC they can be studied at much higher energies than available before [1]. Two photon processes provides a wide range of opportunities from testing fundamental Quantum Electro Dynamics (QED) to searches for physics beyond the Standard Model (SM). The results presented in this report are based on the data collected by the CMS experiment.

The central feature of the CMS detector is a superconducting solenoid that provides a magnetic field of 3.8 T, required to bend the charged particle's trajectory and measure it's momentum accurately. Within the solenoid volume are a silicon pixel and strip tracker, electromagnetic calorimeter (ECAL), hadron calorimeter (HCAL), each composed of a barrel and two endcap sections. Muons are measured in gas-ionization detectors embedded in the steel flux-return yoke outside the solenoid over the range $|\eta| < 2.4$. Two Hadron Forward (HF) calorimeters cover $2.9 < |\eta| < 5.2$, and two zero degree calorimeters (ZDC) are sensitive to neutrons and photons with $|\eta| > 8.3$. A more detailed description of the CMS detector can be found in Ref [2].

## 2 Exclusive two-photon production of lepton pairs in pp collisions at $\sqrt{s} = 7$ TeV

The higher energies and luminosities available at CMS allow for significant improvements in the measurement of exclusive two-photon production of lepton pairs. This process can be reliably calculated within QED, within uncertainties of less than 1% associated with the proton form factor. Exclusive two-photon production of lepton pairs provides an excellent control sample for photon fluxes and cross-sections for other exclusive processes.

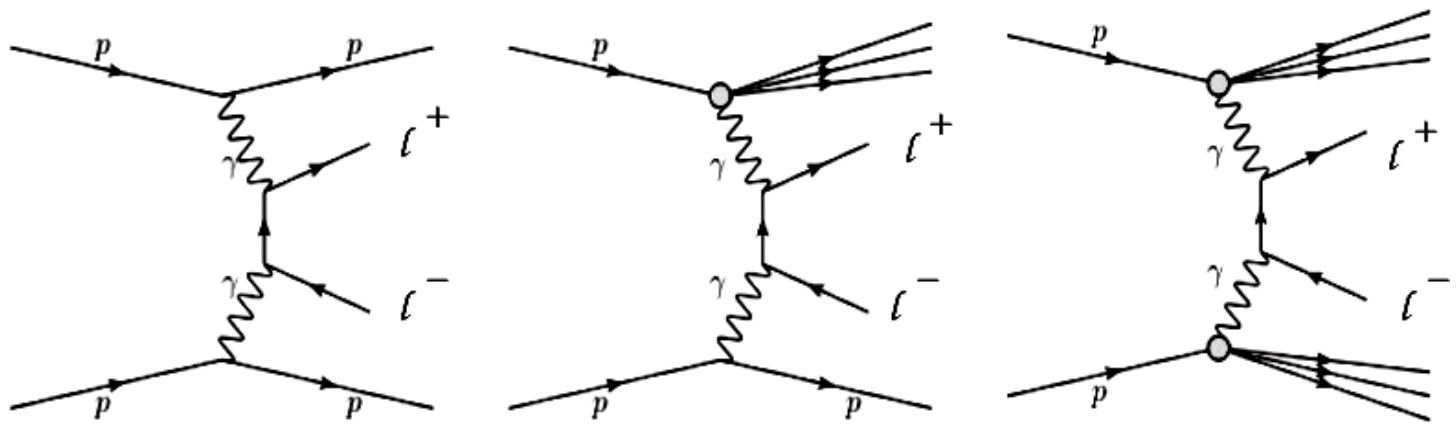

**Fig. 1:** Schematic diagrams for the elastic (left), single dissociative (center), and double dissociative (right) two-photon production of dilepton pairs.

Measurements of exclusive two-photon production of lepton pairs in pp collisions at $\sqrt{s} = 7$ TeV have been presented in [3, 4], collected by the CMS experiment. The measurement of $\gamma\gamma \rightarrow \mu^+\mu^-$ corresponds to an integrated luminosity of 40 $\mathrm{pb}^{-1}$ [3], and that of $\gamma\gamma \rightarrow e^+e^-$ to 36 $\mathrm{pb}^{-1}$ [4]. The LPAIR 4.0 [5] MC event generator was used to generate exclusive and semi-exclusive dilepton pair events. In semi-exclusive production, one or both the protons dissociate into a low mass system (Fig.1). The dimuon events are selected with a dedicated trigger that required the presence of two muons with minimum $p_T > 3$ GeV. In order to minimize the uncertainties related to the knowledge of the low $p_T$ and larger $\eta$ muon efficiencies, muons with $p_T > 4$ GeV and $|\eta| < 2.1$ are selected. In order to reduce the contamination from dimuon decays of the Upsilon meson, the invariant mass of the dimuon pair is required to be above 11.5 GeV. To suppress the proton dissociation background contribution, the muon pair is required to be back-to-back in azimuthal angle $(1-|\Delta\phi/\pi| < 0.1)$ and $\Delta p_T < 1.0$ GeV. The elastic pp $\to$ p$\mu^+\mu^-$p contribution is extracted by performing a binned maximum-likelihood fit to the measured $p_T$ distribution (Fig.2) extracting a cross section of $\sigma = 3.38^{+0.58}_{-0.55}$ (stat.) $\pm 0.16$ (syst.) $\pm 0.14$ (lumi.) pb. The measured data-theory signal ratio is $0.83^{+0.14}_{-0.13}$ (stat.) $\pm 0.04$ (syst.) $\pm 0.03$ (lumi.) [3]. The measured cross-section is consistent with the predicted value from LPAIR event generator.

The electron pair events are selected with a dedicated trigger, which selects at least two electrons with $E_T > 5$GeV and $\Delta\phi$ between the two electrons greater than 2.5 rad. At the offline level, electrons with $E_T > 5.5$ GeV and $|\eta| < 2.5$ are required. In order to reduce the contamination from Upsilon meson decays, the invariant mass of electron pair is required to be above 11 GeV. The exclusive events are selected by requiring no additional tracks in the tracker and no additional tower above noise threshold in the calorimeters. Seventeen exclusive or semi-exclusive $e^+e^-$ candidates are observed (Fig. 2), with an expected background of $0.85 \pm 0.28$ (stat.) events, consistent with the theoretical prediction for the combined elastic and inelastic yield of $16.3 \pm 1.3$ (syst.) events [4].

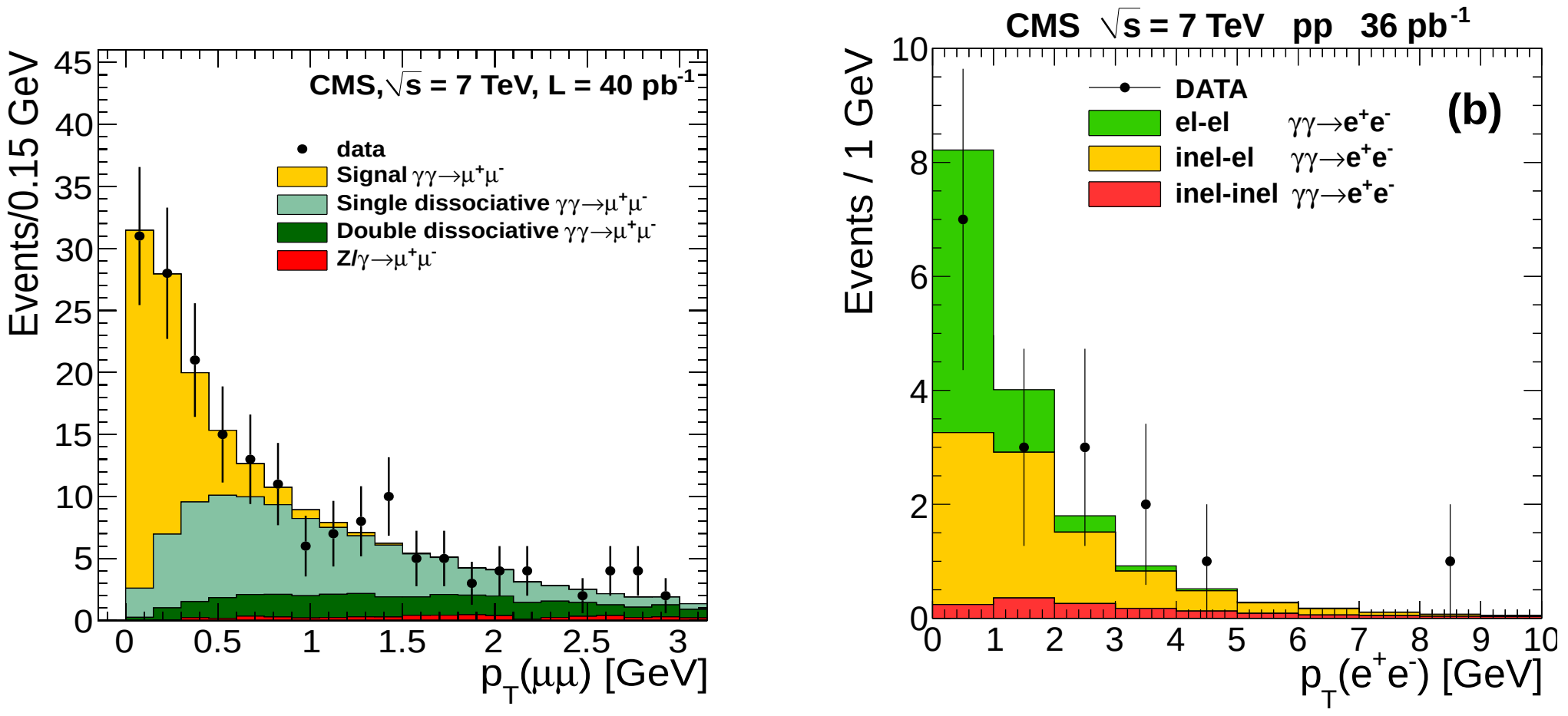

**Fig. 2:** Transverse momentum distributions for exclusive two-photon production of $\mu^+\mu^-$ (left) and $e^+e^-$ (right) events. Histograms show the predictions for the exclusive, single and double-dissociative processes [3, 4].

## 3 Measurement of $W^+W^-$ pair production in pp collisions at $\sqrt{s} = 7$ and 8 TeV

High energy photon interactions at the LHC provide a unique opportunity to study exclusive production of W pairs. At leading order, quartic, t-channel and u-channel processes contribute to $\gamma\gamma \to W^+W^-$ production (Fig.3). The measurement of the quartic WW$\gamma\gamma$ coupling is particularly sensitive to deviations from the Standard Model (SM) and searches for new physics. A genuine anomalous quartic gauge coupling (AQGC) is introduced via an effective Lagrangian with two additional dimension-6 terms containing the parameters $a_0^W$ and $a_C^W$. With the discovery of a light Higgs boson [6–8] a linear realization of the SU(2)$\times$U(1) symmetry of the SM, spontaneously broken by the Higgs mechanism, is possible. Thus, the lowest order operators, where new physics may cause deviations in the quartic gauge boson couplings alone, are of dimension 8. By assuming the anomalous WWZ$\gamma$ vertex vanishes, a direct relationship between the dimension-8 and dimension-6 couplings can be recovered [9]. In both dimension-6 and dimension-8 scenarios, the $\gamma\gamma \to W^+W^-$ cross section increases quadratically with energy, therefore a dipole form factor is introduced to preserve unitarity.

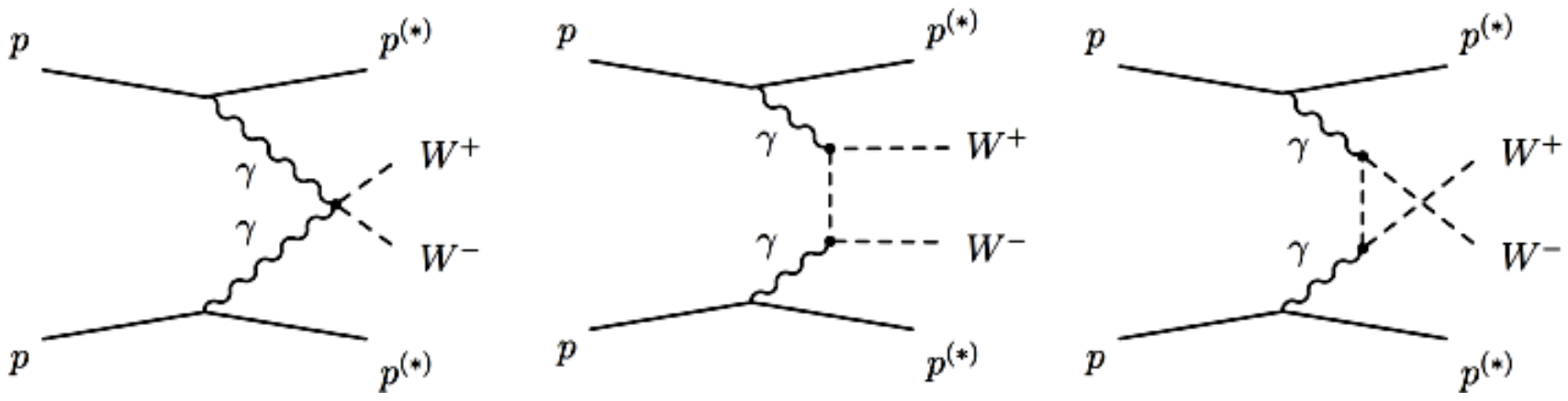

**Fig. 3:** Quartic (left), t-channel (center), and u-channel (right) diagrams contributing to the $\gamma\gamma \to W^+W^-$ process at leading order in the SM [10].

CMS has carried out a measurement of exclusive and quasi-exclusive $\gamma\gamma \to W^+W^-$ production, via $pp \to p^{(*)}W^+W^-p^{(*)} \to p^{(*)}\mu^\pm e^\mp p^{(*)}$ at $\sqrt{s} = 7$ and 8 TeV , corresponding to luminosities of 5.5 fb$^{-1}$ and 19.7 fb$^{-1}$, respectively [10, 11]. The production of W pairs was measured in $\mu^\pm e^\mp$ final state, since $W^+W^- \to \mu^+\mu^-$ or $W^+W^- \to e^+e^-$ would be dominated by Drell-Yan events and $\gamma\gamma \to l^+l^-$ production. The $\mu^\pm e^\mp$ events were extracted from the data with a dedicated trigger that selects two leptons with transverse momentum $p_T > 17(8)$ GeV for the leading (subleading) lepton. Offline, the events with an opposite-charge electron-muon pair originating from a common primary vertex that has no additional tracks associated with it were selected to remove the underlying event activity. The events with transverse momentum of the pair p$_T(\mu^\pm e^\mp) >$ 30 GeV were selected to suppress backgrounds from $\tau^+\tau^-$ production, including the exclusive and quasi-exclusive $\gamma\gamma \to \tau^+\tau^-$ processes.

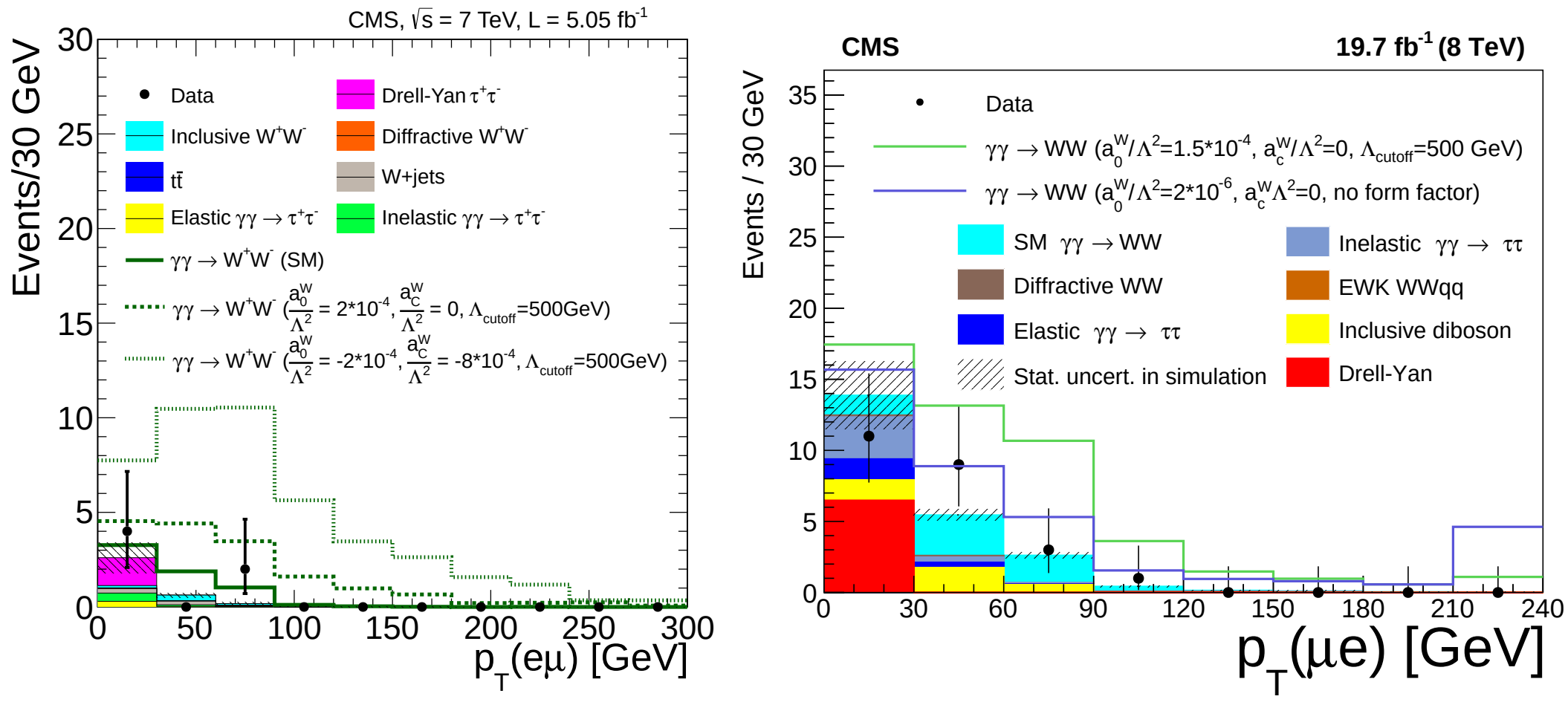

**Fig. 4:** The p$_T(\mu^\pm e^\mp)$ distribution for events with zero extra tracks at 7 TeV (left) [11] and at 8 TeV(right) [10].

Simulations show that in high-mass $\gamma\gamma$ interactions one or both of the protons dissociate, which may result in events being rejected by the veto on extra tracks. To estimate this effect from data, a sample is selected where the dilepton invariant mass is greater than 160 GeV so that W$^+$W$^-$ pairs can be produced on shell. The ratio of the observed number of events to the calculated number of elastic pp $\to$ p$l^+l^-$p events is used as a scale factor to calculate, from the predicted elastic pp $\to$ pW$^+$W$^-$p events, the total number of pp $\to$ p$^*$ W$^+$W$^-$ p$^*$ to be expected when including proton dissociation. The numerical value of the scale factor thus obtained is F = 4.10±0.43. Fig. 4 shows the p$_T(\mu^\pm e^\mp)$ distribution for events passing all other selection requirements. In the signal region with no additional tracks and p$_T(\mu^\pm e^\mp) >$ 30 GeV, two events are observed at 7 TeV compared to the expectation of 2.2±0.4 signal events and 0.84±0.15 background events, corresponding to an observed (expected) significance of 0.8$\sigma$ (1.8$\sigma$). While 13 events are observed at 8 TeV compared to the ex-pectation of 5.3±0.7 signal events and 3.9±0.6 background events which corresponds to a mean expected signal significance of 2.1 $\sigma$. When combining the 7 and 8 TeV results, treating all systematic uncertainties as fully uncorrelated between the two measurements, the resulting observed (expected) significance for the 7 and 8 TeV combination is 3.4$\sigma$ (2.8$\sigma$), constituting evidence for $\gamma\gamma \rightarrow W^+W^-$ production in proton-proton collisions at the LHC. Interpreting the 8 TeV results as a cross section multiplied by the branching fraction to $\mu^\pm e^\mp$ final states, corrected for all experimental efficiencies and extrapolated to the full space, yields : $\sigma(\mathrm{pp} \rightarrow \mathrm{p}^{(*)}\mathrm{W}^+\mathrm{W}^-\mathrm{p}^{(*)} \rightarrow \mathrm{p}^{(*)}\mu^\pm \mathrm{e}^\mp \mathrm{p}^{(*)}) = 11.9^{+5.6}_{-4.5}$ fb.

The corresponding 95% confidence level (CL) upper limit obtained from the 7 TeV data was <10.6 fb, with a central value of $2.2^{+3.3}_{-2.0}$ fb. The transverse momentum $\mathrm{p}_T(\mu^\pm e^\mp)$ distribution was also used to search for signals of anomalous quartic gauge couplings. A selection of $\mathrm{p}_T(\mu^\pm e^\mp) > 100$ GeV is used at 7 TeV, while two bins, with boundaries $\mathrm{p}_T(\mu^\pm e^\mp) = 30$–130 GeV and $\mathrm{p}_T(\mu^\pm e^\mp) > 130$ GeV, are used to set the limit on aQGC. Fig. 5 shows the excluded values of the anomalous coupling parameters $\mathrm{a}_0^W/\Lambda^2$ and $\mathrm{a}_C^W/\Lambda^2$ with $\Lambda_{cutoff} = 500$ GeV. The exclusion regions are shown at 7 TeV (outer contour), 8 TeV (middle contour), and the 7+8 TeV combination (innermost contour). The areas outside the solid contours are excluded by each measurement at 95% CL. The cross indicates the one-dimensional limits obtained for each parameter from the 7 and 8 TeV combination, with the other parameter fixed to zero, more details can be found in [10].

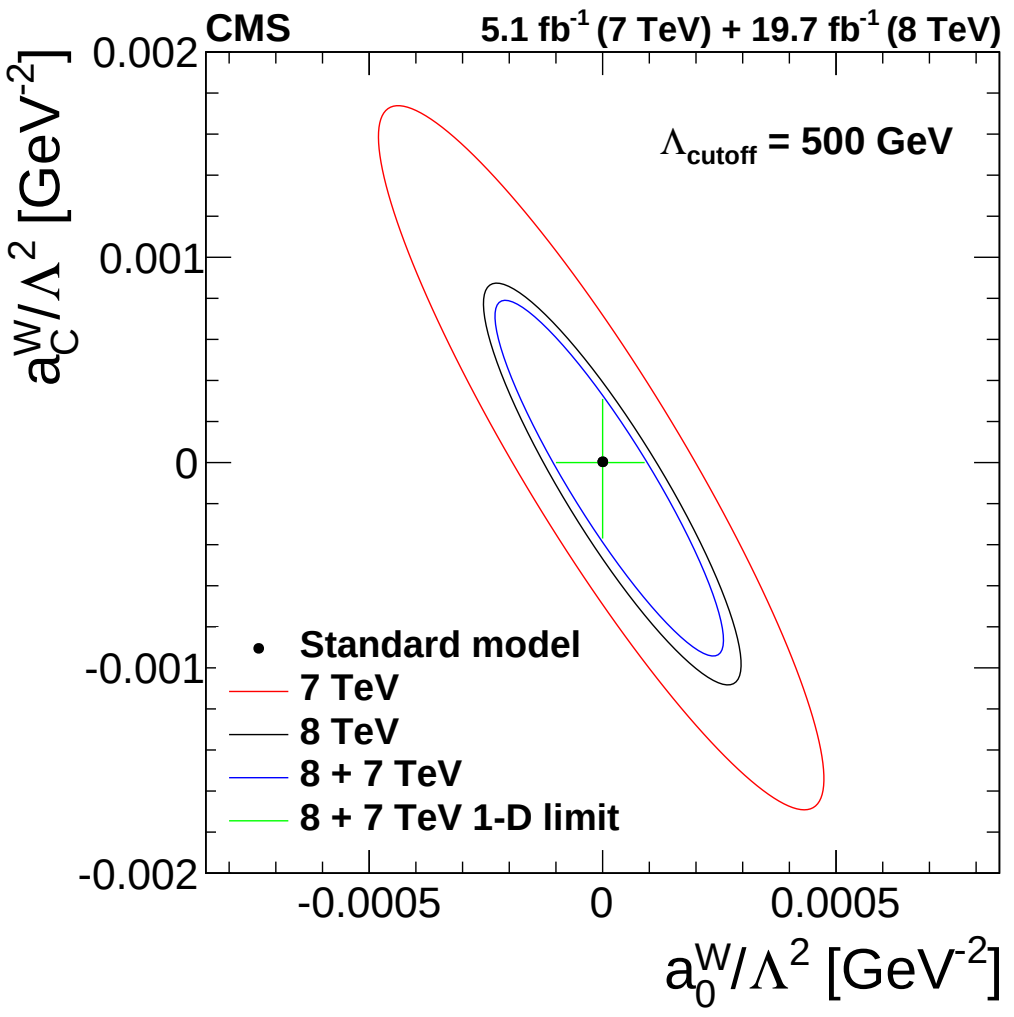

**Fig. 5:** Excluded values of the anomalous coupling parameters [10].

## 4 Search for Light-by-light scattering in PbPb collisions at $\sqrt{s_{\mathrm{NN}}}$ =5.02 TeV

The elastic light-by-light (LbyL) scattering, $\gamma\gamma \rightarrow \gamma\gamma$, is a pure quantum mechanical process that proceeds at leading order in the fine structure constant, $\mathcal{O}(\alpha^4)$, via virtual box diagrams containing charged particles. In the standard model (SM), the box diagram of Fig. 6 involves charged fermions (leptons and quarks) and boson ($\mathrm{W}^\pm$) loops. Despite its simplicity, LbyL scattering was unobserved before LHC because of its tiny cross section $\sigma_{\gamma\gamma} \propto \mathcal{O}(\alpha^4) \approx 3 \times 10^{-9}$. The feasibility to study this process at LHC was provided in Ref. [12] and evidence for its observation has been claimed by the ATLAS collaboration [13] in ultra-peripheral PbPb collisions at $\sqrt{s_{\mathrm{NN}}}$ =5.02 TeV.

**Fig. 6:** Schematic diagram of elastic $\gamma\gamma \rightarrow \gamma\gamma$ collisions in electromagnetic proton or ion interactions at the LHC.

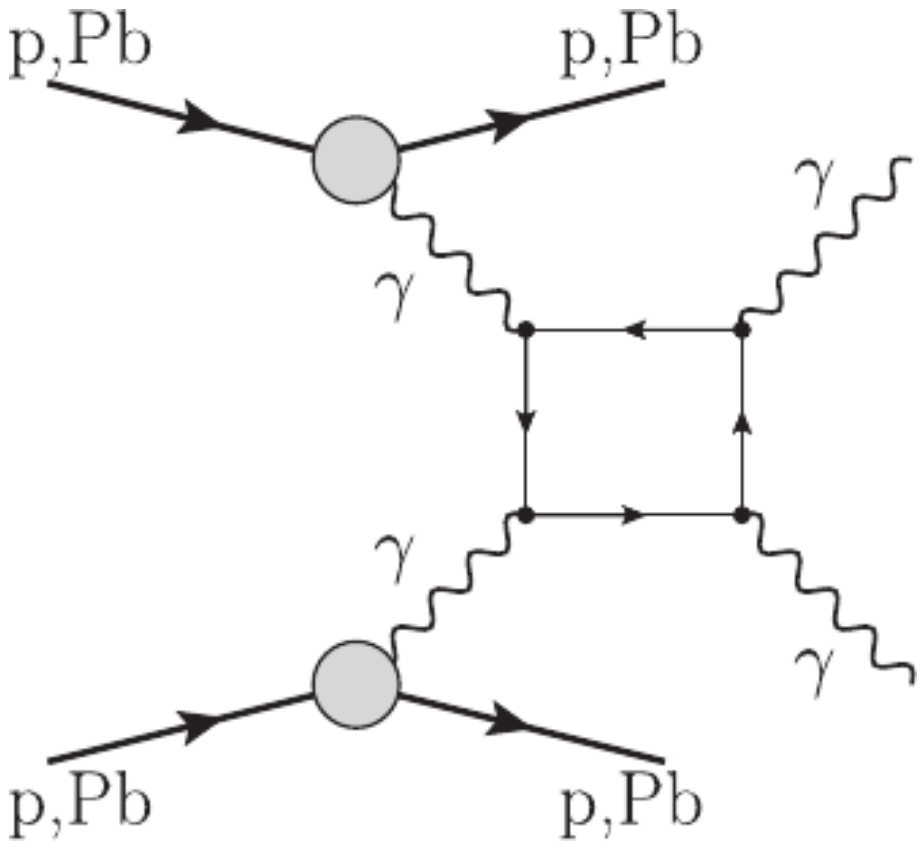

The final-state signature of interest here is the exclusive production of two photons, Pb-Pb $\rightarrow$ Pb$\gamma\gamma$ Pb where the diphoton final-state is measured in the central detector, and the incoming Pb ions survive the electromagnetic interaction and are scattered at very low angles with respect to the beam. Hence, it is expected to detect two low-energy photons and no further activity in the detector, in particular no reconstructed charged-particle tracks coming from the interaction region.

In this report, we introduce the methodology of an ongoing study of LbyL scattering, $\gamma\gamma \rightarrow \gamma\gamma$ us-ing Pb-Pb collisions, recorded by the CMS experiment in 2015 at $\sqrt{s_{\rm NN}}$ =5.02 TeV, corresponding to an integrated luminosity of 388 $\mu b^{-1}$. The MADGRAPH V.5 MC event generator [14], modified as discussed in [12], was used to simulate the leading-order exclusive diphoton cross section including all quark and lepton loops. The exclusive QED, $\gamma\gamma \rightarrow e^+e^-$, background where electrons can be misidentified as photons, was generated with STARLIGHT [15]. The central-exclusive diphoton production gg$\rightarrow \gamma\gamma$ (CEP) is simulated using SUPERCHIC 2.0 [16], in which the proton-proton cross section has been scaled by $A^2R_g^4$, where A = 208 and $R_g \approx 0.7$ is a gluon shadowing correction, and further normalized in a region of the PbPb data where such background is dominant. The events were selected by applying a dedicated UPC trigger, which requires at least two e/$\gamma$ objects with $E_T > 2$ GeV and at least one HF empty of hadronic activity. The photon reconstruction algorithm at CMS are optimized to reconstruct the photons with $E_T > 10$ GeV, while most photons in this analysis have $E_T$ between 2-10 GeV. Therefore, the thresholds of photon $E_T$ , electron $p_T$ and supercluster seed $E_T$ were reduced to 1 GeV from the default of 10 GeV. Exclusive $\gamma\gamma \rightarrow \gamma\gamma$ events are selected by requiring exactly two photons with $E_T > 2$ GeV, no charged particle with $p_T$ greater than 0.1 GeV and no additional tower above noise threshold in the calorimeter. In order to reduce the QED background, events with at least one hit in the pixel detector are vetoed. To suppress the CEP background, the photon pair is required to be back-to-back in azimuthal angle $(1-|\Delta\phi/\pi| < 0.01)$ and $\Delta E_T < 2.0$ GeV. Since, gg$\rightarrow \gamma\gamma$ process has a large theoretical uncertainty, $\mathcal{O}(50\%)$, mostly related to the modelling of the rapidity gap survival probability, the absolute prediction of this process is estimated by normalizing the MC prediction to data for acoplanarity $(1-|\Delta\phi/\pi|) >$ 0.05. After applying all event selection criteria in the used Monte Carlo simulations, we can clearly observe the light by light signal over background and demonstrate the feasibility for a measurement of this process with the CMS experiment

## 5 Conclusions

Exclusive photon-fusion production of pairs of leptons and $W^+W^-$, and of pairs of photons ("light-by-light" scattering) have been measured by the CMS experiment in "ultraperipheral" pp at $\sqrt{s} = 7$ and 8 TeV and in PbPb at $\sqrt{s_{\rm NN}}$ =5.02 TeV collisions, respectively. Such measurements provide novel access to electroweak physics in processes and/or at energies never studied before in the laboratory.

# Photon–Photon Collisions with SuperChic

*L. A. Harland–Lang*[a], *V. A. Khoze*[b,c], *M. G. Ryskin*[c]
[a] Department of Physics and Astronomy, University College London, WC1E 6BT, UK
[b] Institute for Particle Physics Phenomenology, Durham University, DH1 3LE, UK
[c] Petersburg Nuclear Physics Inst., NRC Kurchatov Institute, Gatchina, St. Petersburg, 188300, Russia

**Abstract**
The `SuperChic` Monte Carlo generator provides a common platform for QCD–mediated, photoproduction and photon–induced Central Exclusive Production (CEP), with a fully differential treatment of soft survival effects. In these proceedings we summarise the processes generated, before discussing in more detail those due to photon–photon collisions, paying special attention to the correct treatment of the survival factor. We briefly consider the light–by–light scattering process as an example, before discussing planned extensions and refinements for the generator.

## 1 Introduction

Central exclusive production (CEP) is the process

$$pp \to p + X + p\,, \tag{1}$$

where the '+' signs indicate the presence of large rapidity gaps between the outgoing protons and the central system. That is, the protons remain intact after the collision, with just the system $X$ and nothing else (at least in the absence of pile–up) produced in the detector. The experimental signal for this process is highly favourable, with the crucial advantage that the outgoing protons can be measured by 'tagging' detectors, providing a unique insight into the properties of the central state. This has become particularly topical in light of the installation of the AFP and CT–PPS tagging detectors, which are now taking data in association with the ATLAS and CMS detectors, respectively. In addition, a novel approach based on combining the LHC Beam Loss Monitoring system with the LHC experiments may provide an alternative way to select such events [1].

A CEP event may be produced in three ways, through a purely QCD–mediated interaction, the collision of two photons emitted from each proton, or the photoproduction process where both of these mechanisms operate. The first case requires the development of a completely distinct theoretical framework which covers both the perturbative and non–perturbative QCD regimes, and is particularly sensitive to the nature of the produced state (see [2] for a review). In the second case, the QED initial state is particularly well understood, being simply given in terms of the known electromagnetic proton form factors, while it can be shown that the impact of non–perturbative QCD effects is small. This can therefore serve as a unique laboratory with which to observe QED mediated particle production, including of electromagnetically coupled BSM states, at the LHC. In effect, we can turn the LHC into a photon–photon collider.

Thus, there is a rich phenomenological CEP programme at the LHC. To fully exploit these possibilities, the `SuperChic` Monte Carlo event generator has been developed over a number of years, see [3] for details. This simulates a broad range of exclusive processes, including QCD–mediated, those due to $\gamma\gamma$ collisions and photoproduction, and is the most complete and up-to-date generator of its kind. In these proceedings, we will discuss in detail the simulation of CEP via photon–photon collisions in SuperChic, demonstrating how this is achieved in the MC, describing the simulated processes, discussing one such process, namely light–by–light scattering, in more detail, before finally considering the planned extensions in the future.

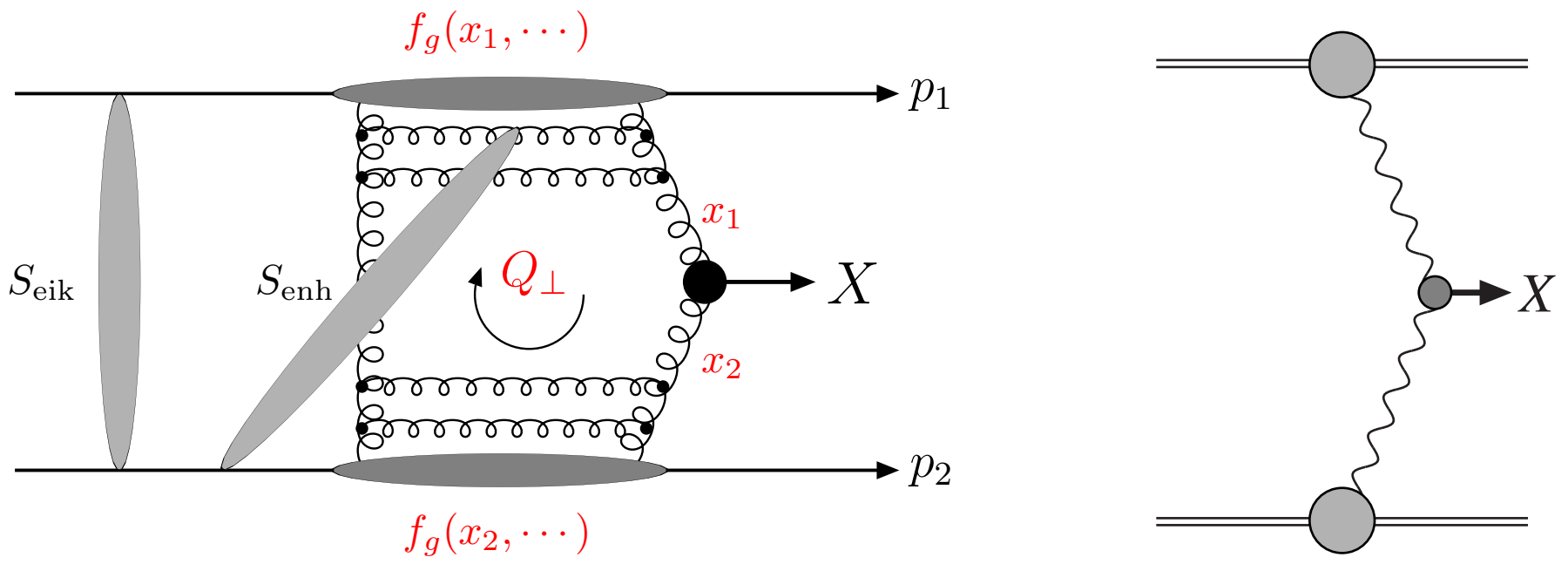

**Fig. 1:** The mechanisms for QCD (left) and (right) photon–initiated CEP.

## 2 The SuperChic MC – overview

The SuperChic MC is a Fortran–based generator for CEP, providing a common platform for QCD–mediated, photoproduction and $\gamma$–induced reactions, with a fully differential treatment of soft survival effects. Arbitrary user–defined histograms may be produced, with arbitrary cuts, as well as unweighted events in Les Houches and HEPEVT formats. The code is available on the Hepforge website [4].

### 2.1 QCD–initiated processes

A representative diagram of the QCD–initiated mechanism for CEP is shown in Fig. 1 (left). The pQCD–based Durham model is used to calculate the basic cross section, while the model of [5] is used to include survival effects, that is the probability that additional soft particle production will not spoil the exclusivity of the event (the same model is applied to the photon–initiated and photoproduction processes listed in Sections 2.2 and 2.3). The processes currently generated are the production of:

- Standard Model Higgs boson via the $b\bar{b}$ decay.
- 2– and 3–jet events.
- Light meson pairs ($\pi\pi$, $\eta(')\eta(')$, $KK$, $\phi\phi$, $\rho\rho$).
- Quarkonium pairs ($J/\psi$, $\psi(2S)$)
- $\chi_{c,b(J)}$ quarkonia, via 2– and 3–body decays, and $\eta_{c,b}$.
- Photon pairs, $\gamma\gamma$.

### 2.2 Photon–initiated processes

The basic CEP process is shown in Fig. 1 (right) and is discussed in more detail in the following sections. Here we simply list the generated processes, which are available for proton and lepton beams:

- Standard Model Higgs boson via the $b\bar{b}$ decay.
- $W^+W^-$ via leptonic decays, including full spin correlations.
- Lepton pairs, $l^+l^-$.
- Light–by–light scattering, $\gamma\gamma$.
- Monopolium and Monopole pairs[1].

[1] Available on request but not in official release at the time of these proceedings.

### 2.3 Photoproduction

These are simulated in the MC according to a fit to the available HERA data. The generated final–states are:

- $\rho(\to \pi^+\pi^-)$.
- $\phi(\to K^+K^-)$.
- $J/\psi(\to \mu^+\mu^-)$.
- $\Upsilon(\to \mu^+\mu^-)$.
- $\psi(2S)(\to \mu^+\mu^-, J/\psi\pi^+\pi^-)$.

## 3 Modelling $\gamma\gamma$ collisions

Exclusive photon–exchange processes in $pp$ collisions are described in terms of the equivalent photon approximation (EPA) [6]. The quasi–real photons are emitted by the incoming proton $i = 1, 2$ with a flux given by

$$n(x_i) = \frac{\alpha}{\pi x_i} \int \frac{\mathrm{d}^2 p_{i_\perp}}{p_{i_\perp}^2 + x_i^2 m_p^2} \left( \frac{p_{i_\perp}^2}{p_{i_\perp}^2 + x_i^2 m_p^2} (1 - x_i) F_E(Q_i^2) + \frac{x_i^2}{2} F_M(Q_i^2) \right) , \tag{2}$$

where $x_i$ is photon momentum fraction, and $Q^2$ is the photon virtuality, given by

$$Q_i^2 = \frac{p_{i_\perp}^2 + x_i^2 m_p^2}{1 - x_i} . \tag{3}$$

The functions $F_E$ and $F_M$ are given in terms of the proton electric and magnetic form factors, and are known in the phenomenologically relevant region to sub–percent level precision; we take the ‘double–dipole’ parameterisation as measured by the A1 collaboration [7]. The CEP cross section is then given in terms of the photon flux by

$$\frac{\mathrm{d}\sigma^{pp\to pXp}}{\mathrm{d}M_X^2\, \mathrm{d}y_X} \sim \frac{1}{s} n(x_1) n(x_2) \cdot \hat{\sigma}(\gamma\gamma \to X) , \tag{4}$$

where $M_X, Y_X$ are the mass and rapidity of the produced object $X$. According to a naive application of the EPA this would in fact be an exact equality. However, this omits the fact that we are asking for an exclusive final–state, that is the production of $X$ accompanied by no additional particles. In addition to the photon–initiated interaction above, the protons may interact independently, producing soft particles and spoiling the exclusivity of the final state. In other words, for an exclusive process we must include the probability of no multi–parton interactions, known as the ‘survival factor’.

The inclusion of the survival factor in a MC environment requires some care, as this is not a simple multiplicative factor, but rather it depends on the final–state kinematics. To see why this is the case we note that the probability for additional particle production must depend on physical grounds on the impact parameter of the colliding protons; most simply, if the protons collide at larger impact parameter they will be less likely to interact independently and produce additional particles. More concretely, the survival factor obeys

$$\frac{\mathrm{d}S^2}{\mathrm{d}^2\mathbf{b}_{1t}\mathrm{d}^2\mathbf{b}_{2t}} \sim |T(s, \mathbf{b}_{1t}, \mathbf{b}_{2t})|^2 \exp(-\Omega(s, \mathbf{b}_t)) , \tag{5}$$

where $\mathbf{b}_{it}$ are the transverse positions of the colliding protons, and $\mathbf{b}_t = \mathbf{b}_{1t} - \mathbf{b}_{2t}$ is the impact parameter of the collision. $T$ is the amplitude corresponding to the cross section (4) excluding survival effects, and $\Omega$ is the proton opacity, which relates to the non–perturbative structure of the proton, and can be extracted from such hadronic observables as the elastic and total cross sections. Thus $\exp(-\Omega)$ corresponds to the Poissonian probability for no additional particle production.

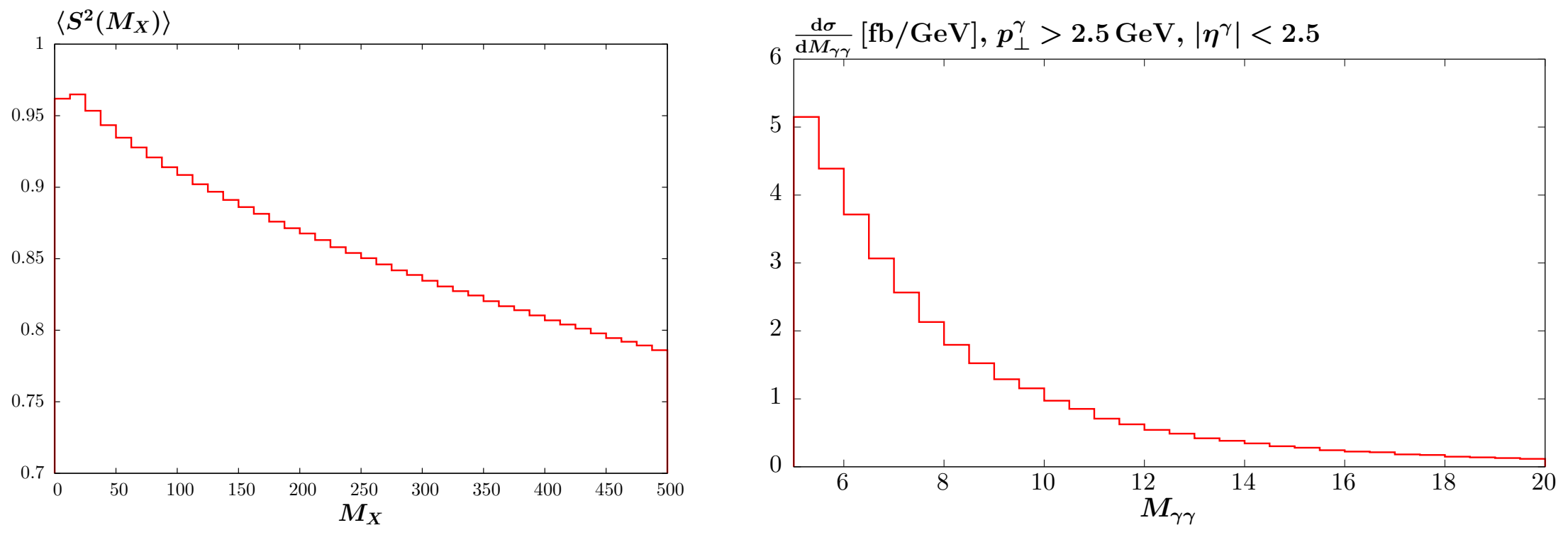

**Fig. 2:** (Left) Average survival factor for lepton pair production at the 14 TeV LHC, taken from [3]. The leptons are required to have $p_\perp > 2.5$ GeV and $|\eta| < 2.5$. (Right) $\gamma\gamma$ invariant mass distribution for light–by–light scattering in $pp$ collisions at the 13 TeV LHC. The photons are required to have $p_\perp > 2.5$ GeV and $|\eta| < 2.5$. Both distributions calculated using `SuperChic`.

In the MC we do not work explicitly in impact parameter space, but rather (5) translates into a dependence on the transverse momenta $\mathbf{p}_{i\perp}$ of the outgoing protons, which are the Fourier conjugates of the $\mathbf{b}_{it}$ variables. By considering the photon flux (2) this allows us to make some immediate conclusions about the impact of the survival factor. In particular, due to the form factors $F_E$, $F_M$, which are steeply falling with photon virtuality, the average $Q_i^2 \sim p_{i\perp}^2 \sim 0.05$ GeV$^2$ is very low in photon–initiated CEP. This corresponds to large impact parameters $\gtrsim 1$ fm and therefore $S^2 \sim 1$. In other words, the impact of non–perturbative QCD effects is low, and to good approximation we are dealing with a purely QED initial state. This supports the use of such processes as tools to search for electromagnetically charged BSM states.

A further implication of this derives from (3), from which we can see that the average photon virtuality will increase with increasing momentum fraction $x$, and therefore we will expect the survival factor to decrease as the invariant mass and/or rapidity of the produced object increases. This trend is clear in Fig. 2 (left), which shows the dependence of the average survival factor on the invariant mass $M_X$ of produced system for the case of lepton pair production at the 14 TeV LHC. In `SuperChic`, a complete differential treatment of the survival factor is provided, so that all such kinematic effects are automatically accounted for. In fact, while (4) and (5) are written at the cross section level, a proper treatment of survival effects requires that we work at the amplitude level. In this way, the survival factor is sensitive to the helicity structure of the underlying $\gamma\gamma \to X$ subprocess, see [3] for further details.

It is worth emphasising that the impact parameter dependence of both the opacity $\Omega$ and the $\gamma\gamma \to X$ amplitude $T$ in (5) must be accounted for, and if this is omitted it will give misleading results. This is the case in for example [8], which has been compared to the recent ATLAS measurement [9] of exclusive muon pair production. The principle cause for the difference between these results and the `SuperChic` prediction is not the choice of model for the opacity $\Omega$ (which may have some genuine model variation) but rather the fact that the impact parameter dependence of the $\gamma\gamma \to \mu^+\mu^-$ amplitude is omitted in [8]. This has been checked explicitly in [3].

## 4 Example process: light–by–light scattering

The full list of generated photon–initiated CEP processes is given in Section 2.2. As an example, we will consider the case of light–by–light scattering, $\gamma\gamma \to \gamma\gamma$, where in the SM the continuum process proceeds via an intermediate lepton, quark and $W$ boson box, see [10, 11] for a detailed study. Until recently, this process had not been observed directly, and it is also sensitive to BSM effects, see e.g. [12, 13]. In addition, it is particularly topical in light of the first direct evidence for this process by the ATLAS collaboration [14], in Pb–Pb collisions. The invariant mass distribution for the 13 TeV in $pp$ collisions is shown in Fig. 2 (right).

While in the official SuperChic release only lepton and proton beams are available, a version with the heavy ion flux implemented using code provided by the authors of [10] is available on request. Work is currently ongoing to include heavy ion beams in the MC, including an exact treatment of the initial–state kinematics and a proper evaluation of survival effects.

In addition to the light–by–light signal, it is in general possible for the exclusive $\gamma\gamma$ final–state to be produced via the QCD interaction $gg \to \gamma\gamma$ as in Fig. 1 (left). In the ATLAS analysis, to estimate the $Pb$–$Pb$ cross section the SuperChic prediction for $pp$ collisions is corrected by a factor of $A^2 R_g^2$, taken from [10], where $A = 208$ is the lead mass number and $R_g \approx 0.7$ accounts for nuclear shadowing effects. In other words, up to the shadowing correction the predicted cross section in $pp$ collisions is simply scaled by the number of participating nucleons in the collision. However, this argument is not justified. In particular, as the range of QCD $R_{\rm QCD} \ll R_A$ only those nucleons which are situated on the ion periphery may interact while leaving the ions intact. A detailed calculation is therefore required, with the survival factor evaluated by correctly accounting for the geometry of the heavy ion collision. In this way, we find that the CEP cross section in heavy ion collisions will instead scale like $\sim A^{1/3}$ [15]. We will therefore expect the QCD–initiated contribution to be lower than a simple $A^2$ scaling would suggest, although a precise numerical prediction is required to confirm the level of suppression[2].

## 5 Conclusion and outlook

The SuperChic Monte Carlo generator provides a common platform for QCD–mediated, photoproduction and photon–induced reactions, with a fully differential treatment of soft survival effects. The latest version is available on the Hepforge website [4]. In these proceedings we have concentrated on the case of photon–photon collisions, but the full list of the generated processes has been given in Section 2.

A number of extensions to the MC are planned for the future. There are two general possibilities to pursue, either including new beam types or adding new processes, and extensions in both directions are foreseen. In the former case, as discussed in Section 4, work is ongoing to include a complete treatment of heavy ion beams for photon–induced processes, as well as a precise evaluation of the QCD–initiated cross section. In the latter case, a range of additions are anticipated for the next MC version, including axion–like particles, monopolium and monopole pairs and the inclusion of $W$ boson loops in the case of light–by–light scattering (which can be important at higher masses, and is so far omitted). By continuing to develop and extend this tool, we hope to exploit as fully as possible the exciting potential to use the LHC as photon–photon collider at unprecedented energies.

## Acknowledgements

MGR thanks the IPPP at Durham University for hospitality and VAK thanks the Leverhulme Trust for an Emeritus Fellowship. The research of MGR was supported by the RSCF grant 14-22-00281. LHL thanks the Science and Technology Facilities Council (STFC) for support via the grant award ST/L000377/1.

[2] In fact in [14] the normalization of the SuperChic result is determined by the data, giving a value of $f_g = 0.5 \pm 0.3$ relative to the prediction. This however is driven by a limited number of events in the tail of the photon acoplanarity distribution which may also be due to dissociation of the ions; as the ZDCs were not used in the ATLAS analysis, this cannot be excluded.

# LHC limits on axion-like particles from heavy-ion collisions

*Simon Knapen*[1,2], *Tongyan Lin*[1,2,3], *Hou Keong Lou*[1,2] *and Tom Melia*[1,2,4]
[1] Department of Physics, University of California, Berkeley, California 94720, USA
[2] Theoretical Physics Group, Lawrence Berkeley National Laboratory, Berkeley, California 94720, USA
[3] Department of Physics, University of California, San Diego, California 92093, USA
[4] Kavli Institute for the Physics and Mathematics of the Universe (WPI), University of Tokyo Institutes for Advanced Study, University of Tokyo, Kashiwa 277-8583, Japan

**Abstract**
In these proceedings we use recent LHC heavy-ion data to set a limit on axion-like particles coupling to electromagnetism with mass in the range 10-100 GeV. We recast ATLAS data as per the strategy proposed in [1], and find results in-line with the projections given there.

## 1 Introduction

The LHC has completed its highest luminosity heavy-ion collision run (Pb-Pb), with ATLAS, CMS and ALICE all recording data at a centre-of-mass energy per nucleon of $\sqrt{s_{NN}} = 5.02$ TeV. In previous work [1] we showed that the large charge of the lead ions ($Z = 82$) results in a huge $Z^4$ coherent enhancement in the exclusive production of axion-like particles (ALPs) that couple to electromagnetism, which can lead to competitive limits for ALPs. This proceeding is an update to our previous work; we recast the analysis of the ATLAS $480\,\mu b^{-1}$ data set [2] to provide limits on ALPs in the mass region $10\,\text{GeV} < m_a < 100\,\text{GeV}$. In line with the projections in [1], we find that the LHC heavy-ion data provides the strongest limits to date in this parameter range. While the physics potential of exclusive heavy ion collisions has been known for decades [3–5], to our knowledge this represents the first time LHC heavy-ion data sets the most stringent limit on a specific beyond the Standard Model physics scenario.

Ultra-peripheral collisions (UPCs) are quasi-elastic processes where the impact parameter is much greater than the ion radius (see *e.g.* Refs. [6–8]). The ions remain (largely) intact, and there is a large rapidity gap between any produced particle and the beam-line with very little detector activity. This clean environment, along with the $Z^4$ enhanced signal rate, provides a low background ALP search channel that can perform better than searches using the p-p run.

The production of an ALP in a UPC proceeds via photon fusion—see Fig. 1—where we consider a Lagrangian of the form

$$\mathcal{L}_a = \frac{1}{2}(\partial a)^2 - \frac{1}{2}m_a^2 a^2 - \frac{1}{4}\frac{a}{\Lambda}F\widetilde{F}\,, \tag{1}$$

where $\widetilde{F}^{\mu\nu} \equiv \epsilon^{\mu\nu\rho\sigma}F_{\rho\sigma}/2$, and with $a$ being the pseudoscalar ALP of mass $m_a$ which couples to electromagnetism via the dimensionful coupling $1/\Lambda$. Such a coupling can be obtained through the $SU(2)_L$ invariant operator $-aB\widetilde{B}/(4\cos^2\theta_W\Lambda)$ where $B$ is the hypercharge field strength. Polarization effects of the incoming photons can lead to different scalar and pseudoscalar production rates, but the effects are relatively small when integrating over all impact parameters [9, 10]. Our limits therefore apply for scalar particles through the replacement $\tilde{F}(\tilde{B}) \to F(B)$ in Eq. (1).

Here we recap some of the details of our treatment in [1]. The ALP parameter space is already substantially constrained by cosmological and astrophysical observations, as well as by a broad range of collider and intensity frontier experiments (see *e.g.* [11, 12] for reviews and recent results). In the regime of interest for UPCs (1 GeV $\lesssim m_a \lesssim$ 100 GeV), the existing constraints however come from LEP and LHC [13–15]. In Fig. 2, we show the expected sensitivity from performing a bump hunt in $m_{\gamma\gamma}$ for UPCs, assuming a luminosity for the current ($1\,\text{nb}^{-1}$) and the high luminosity ($10\,\text{nb}^{-1}$) Pb-Pb runs.[1] For each mass point we computed the expected Poisson limit [16]. The dominant backgrounds are estimated to be light-by-light scattering [3] and fake photons from electrons, and become negligible for $m_{\gamma\gamma} \gtrsim 20$ GeV. In the region which there is background, we assume the entire signal falls into a bin of width 1 GeV. The signal selection criteria in this case are $E_T > 2$ GeV and $|\eta| < 2.5$ for the two photons and $|\phi_{\gamma\gamma} - \pi| < 0.04$. The analogous limit from the exclusive p-p analysis performed by CMS [17] is also shown, which is very weak due to low photon luminosities. For the $F\tilde{F}$ operator the heavy-ion limits are significantly stronger, whereas for the $B\tilde{B}$ operator, traditional p-p collider limits are enhanced due to additional production channels through the $Z$ coupling.

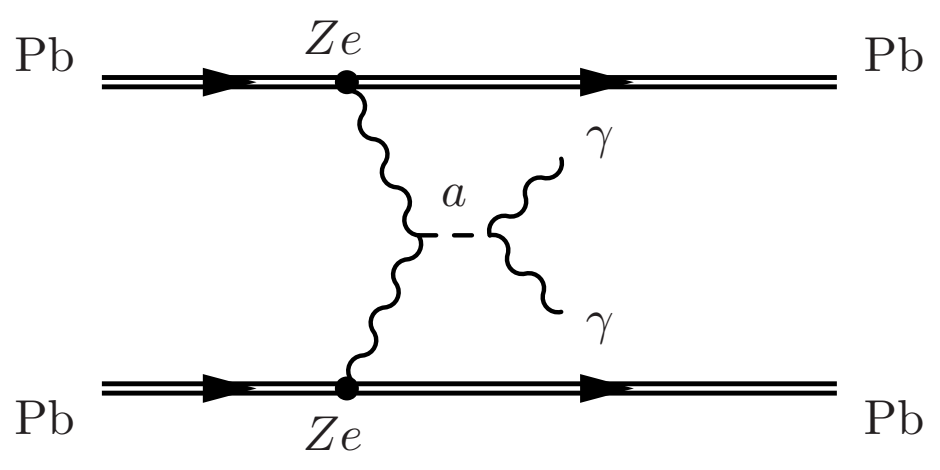

**Fig. 1:** Exclusive ALP production in ultra-peripheral Pb-Pb collisions.

Light-by-light scattering has been measured by the ATLAS collaboration [2], and the results were consistent with our estimates and those in earlier computations [18–20]. Using the observed $m_{\gamma\gamma}$ spectrum, we then derive an observed limit on ALPs for $F\widetilde{F}$ and $B\widetilde{B}$ couplings, which are shown in black in Fig. 2. In detail, we generated Monte Carlo samples for the ALP signal using a modified version of the `STARlight` code [21],[2] which assigns a small virtuality to the photons and as such leads to a typical $p_T^{\gamma\gamma} \lesssim 100\,\text{MeV}$ for the recoil of the $\gamma\gamma$-system. We then follow the ATLAS analysis and apply the following selection cuts on the signal:

1. Require exactly two photons with $E_T > 3$ GeV and $|\eta| < 2.4$
2. Demand $|\phi_{\gamma\gamma} - \pi| < 0.03$, where $\phi_{\gamma\gamma}$ is the azimuthal angle between the two photons

The signal efficiency is ∼70% near threshold and becomes fully efficient if the sum of the photon energies exceeds 9 GeV. The selection criteria are slightly different from our previous theoretical analysis, however we note that only the larger $E_T$ cut leads to noticeable changes for the efficiencies. Given that we do not model photon identification at the detector level, we apply an extra total reconstruction efficiency of 90%, which roughly takes into account the per-photon ID efficiency of 95% measured by ATLAS.

The $m_{\gamma\gamma}$ spectrum measured by ATLAS is plotted in bin-widths of 3 GeV, starting at $m_{\gamma\gamma} = 6$ GeV. For our exclusion, we generated samples with $m_{\gamma\gamma} = 7, 10, 13, 16, ...$ GeV, and assume that all the events are contained in their respective bins after final selection. We further assume that ATLAS did not observe any events with $m_{\gamma\gamma} \gtrsim 30$ GeV. The 95% exclusion limits on the coupling $1/\Lambda$ are obtained assuming only statistical uncertainties. A more detailed CL$_s$ analysis that includes a proper treatment of systematics would yield slightly more conservative limits, and we encourage the experimental community to include such an analysis as it is beyond the scope of our simulation framework.

In summary, we have found that heavy-ion collisions at the LHC can provide the best limits on ALP-photon couplings for $7\,\text{GeV} < m_a < 100\,\text{GeV}$, confirming our previous estimates. The very large photon flux and extremely clean event environment in heavy-ion UPCs provides a rather unique opportunity to search for BSM physics.

[1] Limits from the p-Pb runs are not competitive despite their higher luminosity, because of the less advantageous $Z^2$ scaling of the production rate. Collisions with lighter elements, *e.g.* Ar-Ar, may set relevant limits if the luminosity could be enhanced by two to three orders of magnitude, as compared to current Pb-Pb run.

[2] Our patch for ALP production is now included in the latest `STARlight` release.

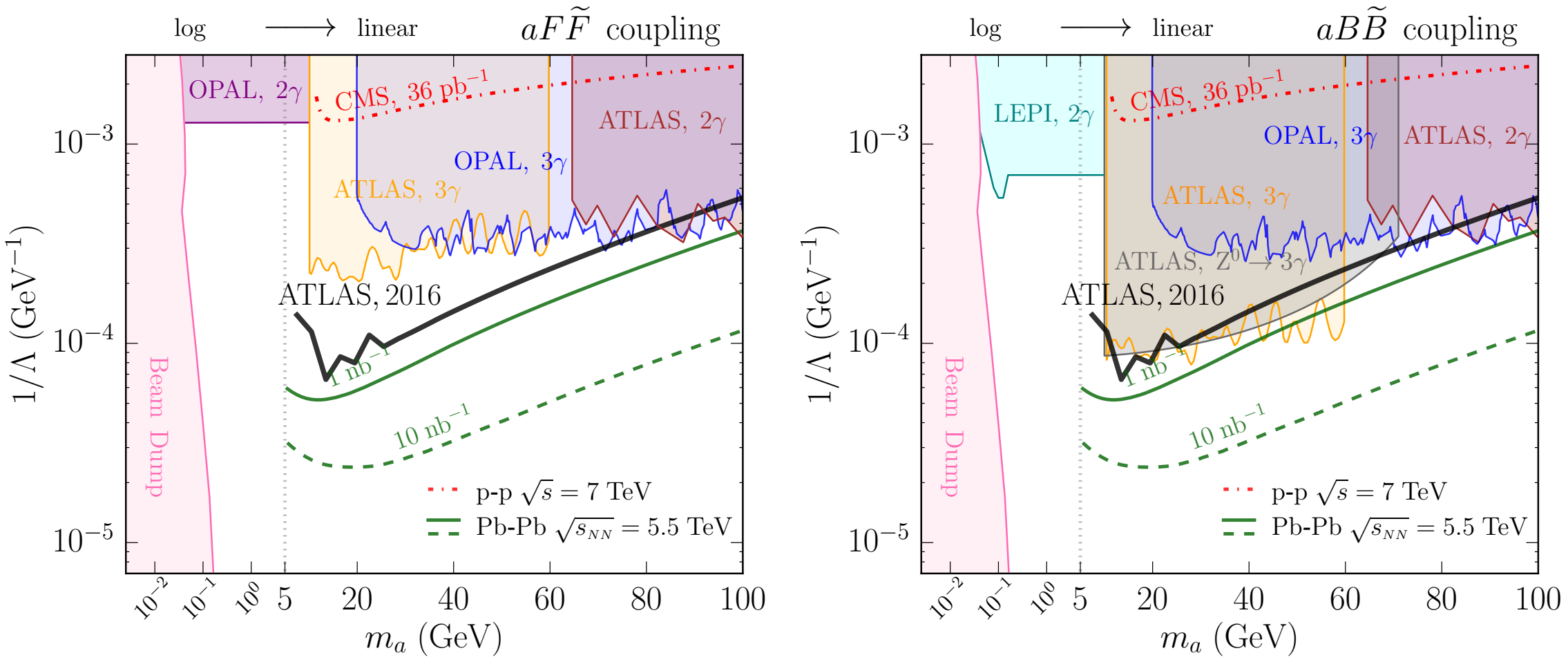

**Fig. 2:** *Left:* We show 95% exclusion limits on the operator $\frac{1}{4}\frac{1}{\Lambda}aF\tilde{F}$ using recent ATLAS results on heavy-ion UPCs [2] (solid black line). The expected sensitivity assuming a luminosity of 1 nb$^{-1}$ (10 nb$^{-1}$) is shown in solid (dashed) green. For comparison, we also give the analogous limit from 36 pb$^{-1}$ of exclusive p-p collisions [17] (red dot-dash). Remaining exclusion limits are recast from LEP II (OPAL $2\gamma$, $3\gamma$) [22] and from the LHC (ATLAS $2\gamma$, $3\gamma$) [23, 24] (see [1] for details). *Right*: The corresponding results for the operator $\frac{1}{4\cos^2\theta_W}\frac{1}{\Lambda}aB\tilde{B}$. The LEP I, $2\gamma$ (teal shaded) limit was obtained from [14].

## Acknowledgements

We thank Bob Cahn, Lucian Harland-Lang, Yonit Hochberg, Joerg Jaeckel, Spencer Klein, Hitoshi Murayama, Michele Papucci, Dean Robinson, Sevil Salur, and Daniel Tapia Takaki for useful conversations. SK and TL also thank the participants of the 3rd NPKI workshop in Seoul for useful comments and discussions. This work was performed under DOE contract DE-AC02-05CH11231 and NSF grant PHY-1316783.

# Constraint on Born-Infeld Theory from Light-by-Light Scattering at the LHC

*John Ellis*[1,2], *Nick E. Mavromatos*[1], *Tevong You*[3*]
[1]Theoretical Particle Physics and Cosmology Group, Physics Department, King's College London, London WC2R 2LS, UK
[2]Theoretical Physics Department, CERN, CH-1211 Geneva 23, Switzerland
[3]DAMTP, University of Cambridge, Wilberforce Road, Cambridge, CB3 0WA, UK and
Cavendish Laboratory, University of Cambridge, J.J. Thomson Avenue, Cambridge, CB3 0HE, UK

**Abstract**
The recent measurement by ATLAS of light-by-light scattering in LHC Pb-Pb collisions is the first direct evidence for this basic process. We find that it excludes a range of the mass scale of a nonlinear Born-Infeld extension of QED that is $\lesssim$ 100 GeV. In the case of a Born-Infeld extension of the Standard Model in which the $\mathrm{U(1)_Y}$ hypercharge gauge symmetry is realized nonlinearly, the limit on the corresponding mass reach is $\sim$ 90 GeV, which in turn imposes a lower limit of $\gtrsim$ 11 TeV on the magnetic monopole mass in such a $\mathrm{U(1)_Y}$ Born-Infeld theory.

## 1 Introduction

Over 80 years ago, soon after Dirac proposed his relativistic theory of the electron [1] and his interpretation of 'hole' states as positrons [2], Halpern [3] in 1933 and Heisenberg [4] in 1934 realized that quantum effects would induce light-by-light scattering, which was first calculated in the low-frequency limit by Euler and Kockel [5] in 1935. Subsequently, Heisenberg and Euler [6] derived in 1936 a more general expression for the quantum nonlinearities in the Lagrangian of Quantum Electrodynamics (QED), and a complete calculation of light-by-light scattering in QED was published by Karplus and Neuman [7] in 1951. However, measurement of light-by-light scattering has remained elusive until very recently. In 2013 d'Enterria and Silveira [8] proposed looking for light-by-light scattering in ultraperipheral heavy-ion collisions at the LHC, and evidence for this process was recently presented by the ATLAS Collaboration [9], at a level consistent with the QED predictions in [8] and [10].

In parallel with the early work on light-by-light scattering in QED, and motivated by a 'unitarian' idea that there should be an upper limit on the strength of the electromagnetic field just as the speed of light is an upper limit, Born and Infeld [11] proposed in 1934 a conceptually distinct nonlinear modification of the Lagrangian of QED:

$$\begin{aligned}\mathcal{L}_{\mathrm{QED}} &= -\frac{1}{4}F_{\mu\nu}F^{\mu\nu} \;\rightarrow \\ \mathcal{L}_{\mathrm{BI}} &= \beta^2\Big(1-\sqrt{1+\frac{1}{2\beta^2}F_{\mu\nu}F^{\mu\nu}-\frac{1}{16\beta^4}(F_{\mu\nu}\tilde{F}^{\mu\nu})^2}\;\Big),\end{aligned} \tag{1}$$

where $\beta$ is an *a priori* unknown parameter with the dimension of [Mass]$^2$ that we write as $\beta \equiv M^2$, and $\tilde{F}_{\mu\nu}$ is the dual of the field strength tensor $F_{\mu\nu}$. Interest in Born-Infeld theory was revived in 1985 when

*Based on Phys. Rev. Lett. **118** (2017) no.26, 261802, arXiv:1703.08450

Fradkin and Tseytlin [12] discovered that it appears when an Abelian vector field in four dimensions is coupled to an open string, as occurs in models inspired by M theory in which particles are localized on lower-dimensional 'branes' separated by a distance $\simeq 1/\sqrt{\beta} = 1/M$ in some extra dimension [1]. Depending on the specific brane scenario considered, $M$ might have any value between a few hundred GeV and the Planck scale $\sim 10^{19}$ GeV.

When considering phenomena at energies $\ll M$ as in this paper, the most relevant terms are those of fourth order in the gauge field strengths in (1). Until now, there has been no strong lower limit on the Born-Infeld scale $\beta$ or, equivalently, the brane mass scale $M$ and the brane separation $1/M$. A constraint corresponding to $M \gtrsim 100$ MeV was derived in [14] from electronic and muonic atom spectra, though the derivation has been criticized in [15]. Measurements of photon splitting in atomic fields [16] were considered in [17], where it was concluded that they provided no limit on the Born-Infeld scale, and it was suggested that measurements of the surface magnetic field of neutron stars [18] might be sensitive to $M = \sqrt{\beta} \sim 1.4 \times 10^{-5}$ GeV. More recently, measurements of nonlinearities in light by the PVLAS Collaboration [19] are somewhat more sensitive to the individual nonlinear terms in (1), but are insensitive to the particular combination appearing in the Born-Infeld theory, as discussed in [20] where more references can be found. Fig. 1 is taken from that paper: the left panel shows the prediction of Heisenberg and Euler for the coefficients of the $(F_{\mu\nu}F^{\mu\nu})^2$ and $(F_{\mu\nu}\tilde{F}^{\mu\nu})^2$ terms in $\mathcal{L}_{\rm BI}$ (1) (black dot), denoted by $c_{2,0}$ and $c_{0,2}$, respectively, and the right panel displays the experimental constraints available until now. The dashed lines are possible values in Born-Infeld theory [11], and the only constraint on the Born-Infeld combination of $c_{2,0}$ and $c_{0,2}$ comes from Lamb shift measurements (region on the right side of the right panel with close diagonal lines), which yield $M \gtrsim \mathcal{O}(100)$ MeV.

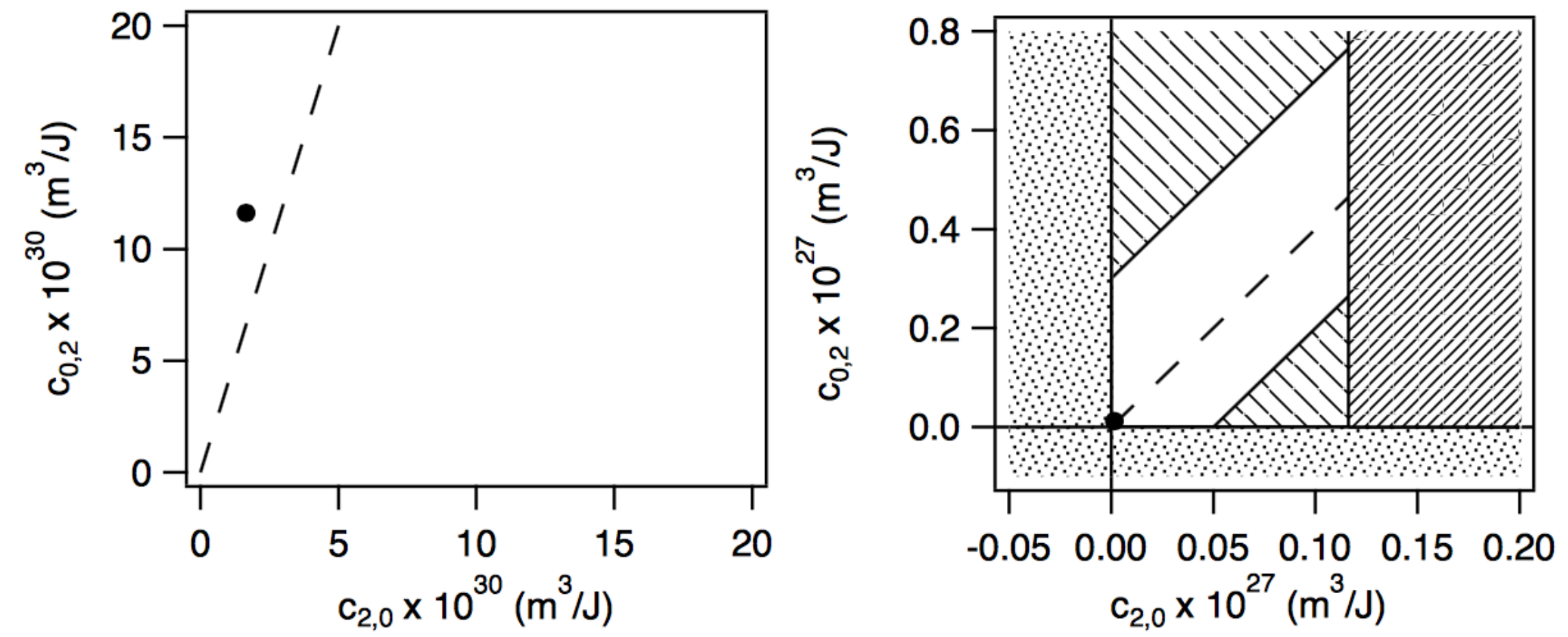

**Fig. 1:** *Left panel: The prediction of Heisenberg and Euler [6] (black dot) for the coefficients of the* $(F_{\mu\nu}F^{\mu\nu})^2$ *and* $(F_{\mu\nu}\tilde{F}^{\mu\nu})^2$ *terms in* $\mathcal{L}_{\rm BI}$ *(1), denoted by* $c_{2,0}$ *and* $c_{0,2}$*, respectively. Right panel: Previous experimental constraints on these coefficients. Born-Infeld theory predicts [11] values of* $c_{2,0}$ *and* $c_{0,2}$ *along the dashed lines. Figure adapted from [20].*

Here we show that the agreement of the recent ATLAS measurement of light-by-light scattering with the standard QED prediction provides the first limit on $M$ in the multi-GeV range, excluding a significant range extending to

$$M \gtrsim 100\,\text{GeV}\,, \tag{2}$$

and entering the range of interest to brane theories. This limit is obtained under quite conservative assumptions, and plausible stronger assumptions discussed later would strengthen this lower bound to $M \gtrsim 200$ GeV.

[1]Remarkably, the maximum field strength is related to the fact that the brane velocity is limited by the velocity of light [13], confirming the insight of Born and Infeld [11].

One may also consider a string-motivated Born-Infeld extension of the Standard Model in which the hypercharge $\mathrm{U(1)_Y}$ gauge symmetry is realised non-linearly, in which case the limit (2) is relaxed to

$$M_Y = \cos\theta_W M \gtrsim 90 \text{ GeV}, \tag{3}$$

where we have used $B^\mu_Y = \cos\theta_W\, A^\mu_{\mathrm{EM}} - \sin\theta_W\, Z^\mu$ and $\sin^2\theta_W \simeq 0.23$, with $\theta_W$ the weak mixing angle. As a corollary of this lower limit on the $\mathrm{U(1)_Y}$ brane scale, we recall that Arunasalam and Kobakhidze recently pointed out [21] that the Standard Model modified by a Born-Infeld $\mathrm{U(1)_Y}$ theory has a finite-energy electroweak monopole [22, 23] solution $\mathcal{M}$, whose mass they estimated as $M_{\mathcal{M}} \simeq 4$ TeV $+\ 72.8\, M_Y$. Such a monopole is less constrained by Higgs measurements than electroweak monopoles in other extensions of the Standard Model [24], and hence of interest for potential detection by the ATLAS [25], CMS and MoEDAL experiments at the LHC [26]. However, our lower limit $M_Y \gtrsim$ 90 GeV (2) corresponds to a 95% CL lower limit on the mass of this monopole $M_{\mathcal{M}} \gtrsim 11$ TeV, excluding its production at the LHC.

## 2 Light-by-Light Scattering in LHC Pb-Pb Collisions

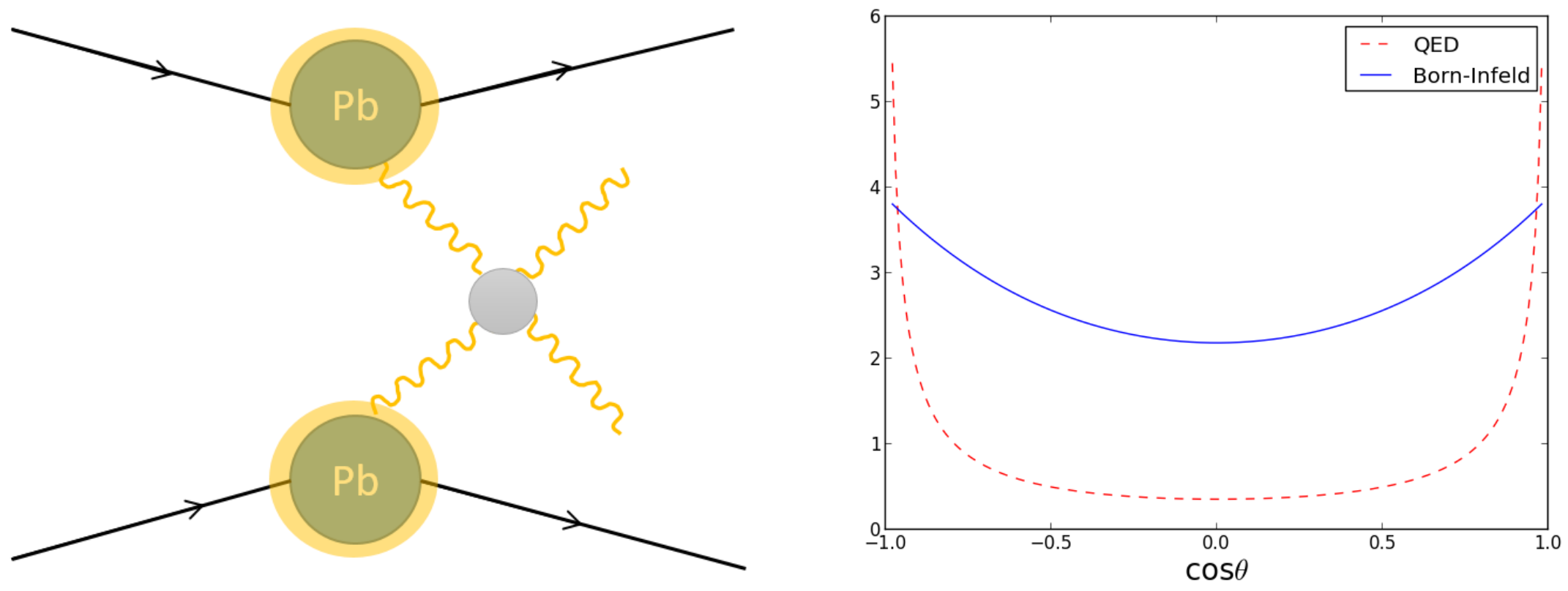

**Fig. 2:** *Left panel: Cartoon of light-by-light scattering through photon-photon collisions in ultra-peripheral Pb-Pb collisions. Right panel: Comparison between the angular distributions (with arbitrary normalisations) as functions of* $\cos\theta$ *in the centre-of-mass frame (where* $\theta$ *is the polar angle) for the leading-order differential cross-sections in U(1)*$_{\mathrm{EM}}$ *Born-Infeld theory and QED, plotted as solid blue and dashed red lines, respectively.*

Following the suggestion of [8], we consider ultra-peripheral heavy-ion collisions in which the nuclei scatter quasi-elastically via photon exchange: Pb + Pb $(\gamma\gamma)$→Pb$^{(*)}$ + Pb$^{(*)}$+ X, as depicted in the left panel of Fig. 2, effectively acting via the equivalent photon approximation (EPA) [27] as a photon-photon collider. The EPA allows the electromagnetic field surrounding a highly-relativistic charged particle to be treated as equivalent to a flux of on-shell photons. Since the photon flux is proportional to $Z^2$ for each nucleus, the coherent enhancement in the exclusive $\gamma\gamma$ cross-section scales as $Z^4$, where $Z =$ 82 for the lead (Pb) ions used at the LHC. This is why heavy-ion collisions have an advantage over proton-proton or proton-lead collisions for probing physics in electromagnetic processes [8]. Photon fusion in ultra-peripheral heavy-ion collisions has been suggested as a way of detecting the Higgs boson [28, 29] and, more recently, the possibility of constraining new physics beyond the Standard Model (BSM) in this process was studied in [30, 31].

As already mentioned, the possibility of directly observing light-by-light scattering at the LHC was proposed in [8], and this long-standing prediction of QED was finally measured earlier this year with $4.4\sigma$ significance by the ATLAS Collaboration [9] at a level in good agreement with calculations in [8, 10]. The compatibility with the Standard Model constrains any possible contributions from BSM physics. Born-Infeld theory is particularly interesting in this regard, in the absence of constraints from low-energy optical and atomic experiments [19, 20].

The leading-order cross-section for unpolarised light-by-light scattering in Born-Infeld theory in the $\gamma\gamma$ centre-of-mass frame is given by [17, 32]:

$$\sigma_{\rm BI}(\gamma\gamma \to \gamma\gamma) = \frac{1}{2}\int d\Omega \frac{d\sigma_{\rm BI}}{d\Omega} = \frac{7}{1280\pi}\frac{m_{\gamma\gamma}^6}{\beta^4}\,, \tag{4}$$

where $m_{\gamma\gamma}$ is the diphoton invariant mass and the differential cross-section is

$$\frac{d\sigma_{\rm BI}}{d\Omega} = \frac{1}{4096\pi^2}\frac{m_{\gamma\gamma}^6}{\beta^4}\left(3+\cos\theta\right)^2\,. \tag{5}$$

We recall that the parameter $\beta = M^2$ enters as a dimensionful parameter in the Born-Infeld theory of non-linear QED defined by the Lagrangian (1). If this originates from a Born-Infeld theory of hypercharge then the corresponding mass scale is $M_Y = \cos\theta_W M$.

We plot in the right panel of Fig. 2 the angular distributions as functions of $\cos\theta$ in the centre-of-mass frame (where $\theta$ is the polar angle) for the leading-order differential cross-sections in both Born-Infeld theory and QED (with arbitrary normalisations), as solid blue and dashed red lines, respectively. We see that the Born-Infeld distribution is less forward peaked than that for QED. For the latter, we used the leading-order amplitudes for the quark and lepton box loops in the ultra-relativistic limit from [33], omitting the percent-level effects of higher-order QCD and QED corrections, as well the $W^{\pm}$ contribution that is negligible for typical diphoton centre-of-mass masses at the LHC.

The total exclusive diphoton cross-section from Pb+Pb collisions is obtained by convoluting the $\gamma\gamma \to \gamma\gamma$ cross-section with a luminosity function $dL/d\tau$ [34],

$$\sigma_{\rm excl.} = \int_{\tau_0}^{1} d\tau \frac{dL}{d\tau}\sigma_{\gamma\gamma\to\gamma\gamma}(\tau)\,. \tag{6}$$

We have introduced here a dimensionless measure of the diphoton invariant mass, $\tau \equiv m_{\gamma\gamma}^2/s_{NN}$, where $\sqrt{s_{NN}} = 5.02$ TeV is the centre-of-mass energy per nucleon pair in the ATLAS measurement. The luminosity function, derived for example in [34], can be written as an integral over the number distribution of photons carrying a fraction $x$ of the total Pb momentum:

$$\frac{dL}{d\tau} = \int_{\tau}^{1} dx_1 dx_2 f(x_1) f(x_2)\delta(\tau - x_1 x_2)\,, \tag{7}$$

where the distribution function $f(x)$ depends on a nuclear form factor. We follow [34] in adopting the form factor proposed in [29], while noting that variations in the choice leads to $\sim 20\%$ uncertainties in the final cross-sections [8]. A contribution with a non-factorisable distribution function should also be subtracted to account for the exclusion of nuclear overlaps, but this is not a significant effect for the relevant kinematic range, causing a difference within the $20\%$ uncertainty [31] from the photon luminosity evaluated numerically using the `STARlight` code [35]. For $\sqrt{s_{NN}} = 5.5$ TeV and $m_{\gamma\gamma} > 5$ GeV we obtain a QED cross-section of $\sigma_{\rm excl.}^{\rm QED} = 385 \pm 77$ nb, in good agreement with [8]. The ATLAS measurement is performed at $\sqrt{s_{NN}} = 5.02$ TeV and for $m_{\gamma\gamma} > 6$ GeV, for which we find $\sigma_{\rm excl.}^{\rm QED} = 220 \pm 44$ nb.

This total $\gamma\gamma \to \gamma\gamma$ cross-section is reduced by the fiducial cuts of the ATLAS analysis, which restrict the phase space to a photon pseudorapidity region $|\eta| < 2.4$, and require photon transverse energies $E_T > 3$ GeV and the diphoton system to have an invariant mass $m_{\gamma\gamma} > 6$ GeV with a transverse momentum $p_T^{\gamma\gamma} < 2$ GeV and an acoplanarity $A{\rm co} = 1 - \Delta\phi/\pi < 0.01$. We simulate the event selection using Monte-Carlo sampling, implementing the cuts with a $15\%$ Gaussian smearing in the photon transverse energy resolution at low energies and 0.7% at higher energies [9, 36] above 100 GeV. Since the differential cross-section does not depend on $\phi$ we implement the acoplanarity cut as a fixed 85% efficiency in the number of signal events after the $p_T^{\gamma\gamma}$ selection, following the ATLAS analysis [9]. The total reduction in yield for the QED case is a factor $\epsilon \sim 0.30$, which results in a fiducial cross-section $\sigma_{\text{fid.}}^{\text{QED}} = 53 \pm 11$ nb for $\sqrt{s}_{NN} = 5.02$ TeV, in good agreement with the two predictions of 45 and 49 nb quoted by ATLAS [9].

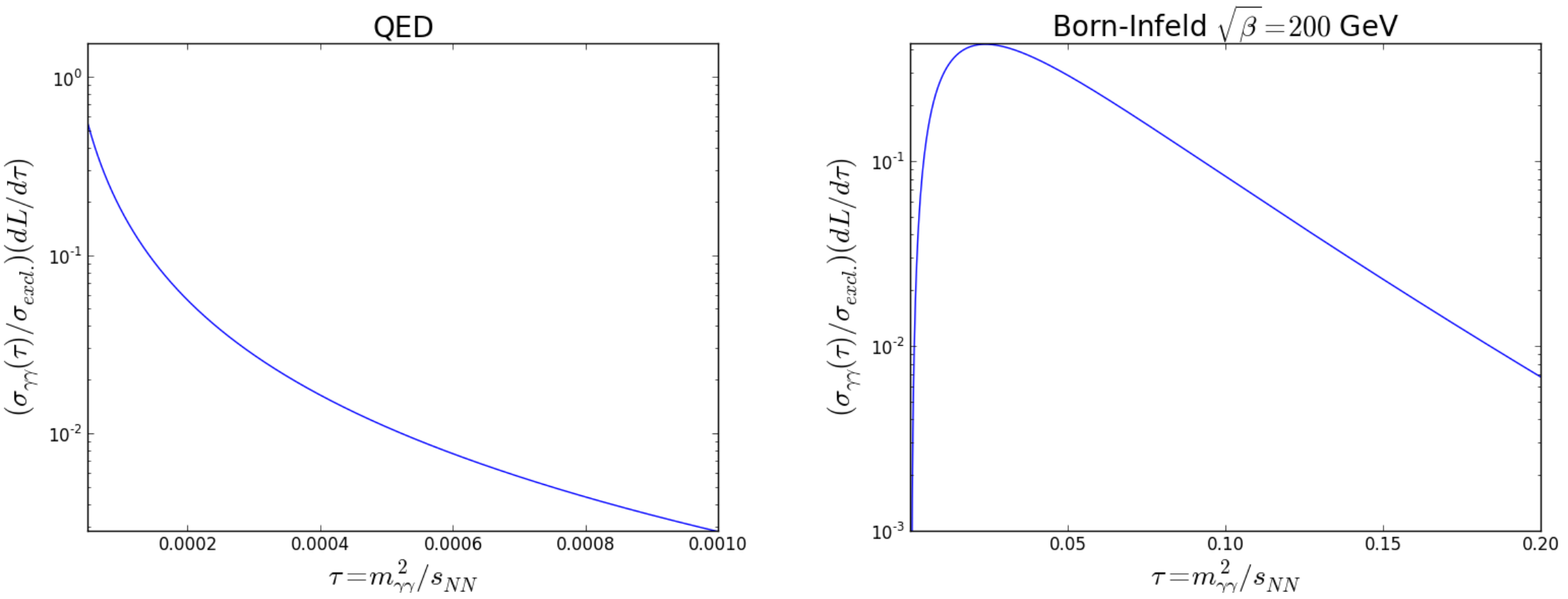

**Fig. 3:** *The distributions in the scaled diphoton invariant mass $\tau \equiv m_{\gamma\gamma}^2/s_{NN}$, normalised by the total $\gamma\gamma \to \gamma\gamma$ cross-section, for the QED case in the upper panel and for U(1)$_{\text{EM}}$ Born-Infeld theory with $M = \sqrt{\beta} = 200$ GeV in the lower panel.*

Following this validation for the QED case, we repeat the procedure for the Born-Infeld cross-section. Since the Born-Infeld $\gamma\gamma \to \gamma\gamma$ cross-section grows with energy, the dominant contribution to the cross-section comes from the $\tau \lesssim 0.2$ part of the integral, compared with $\tau \lesssim 10^{-4}$ for the QED case. We show in Fig. 3 the distributions of the $\sigma(\gamma\gamma \to \gamma\gamma)$ cross-section multiplied by the photon flux luminosity factor – normalised by the total exclusive cross-section – as functions of the invariant diphoton mass distribution, for the QED case in the left panel and in Born-Infeld theory with $M = \sqrt{\beta} = 200$ GeV in the right panel.

We see that the invariant-mass distribution in the Born-Infeld case extends to $m_{\gamma\gamma} > M$, where the validity of the tree-level Born-Infeld Lagrangian may be questioned because the Taylor expansion of the square root in the non-polynomial Born-Infeld Lagrangian (1) could break down. With this in mind, we use two approaches to place plausible limits on $M = \sqrt{\beta}$. In the first and most conservative method we consider $\gamma\gamma$ scattering only for $m_{\gamma\gamma} \leq M$, while in the second approach we integrate the $\gamma\gamma$ cross-section (4) up to the diphoton invariant mass where the unitarity limit $\sigma_{\text{BI}} \sim 1/m_{\gamma\gamma}^2$ is saturated, beyond which we assume that the cross-section saturates the unitarity limit and falls as $\sim 1/m_{\gamma\gamma}^2$.

We find fiducial efficiencies for the cut-off and unitarization approaches to be $\epsilon \sim 0.39$ and 0.14, respectively. Whilst the $E_T$ and $\eta$ cuts have much less effect than for QED, as expected from the difference in the angular distributions visible in the right panel of Fig. 2, the larger invariant masses appearing in the Born-Infeld case are much more affected by the $p_T^{\gamma\gamma}$ requirement.

## 3 Constraint on Born-Infeld Extension of QED

Our calculations of the corresponding U(1)$_{\text{EM}}$ Born-Infeld fiducial cross-sections are plotted in the left panel of Fig. 4 as a function of $M = \sqrt{\beta}$: the green curve is for the more conservative cut-off approach, and the blue curve assumes that unitarity is saturated. These calculations are confronted with the ATLAS measurement of $\sigma_{\text{fid.}} = 70 \pm 24$ (stat.) $\pm\, 17$ (sys.) nb [9], assuming that these errors are Gaussian and adding them in quadrature with a theory uncertainty of $\pm 10$ nb. We perform a $\chi^2$ fit to obtain the 95% CL upper limit on a Born-Infeld signal additional to the 49 nb Standard Model prediction [2]. This corresponds to the excluded range shaded in pink above $\sigma^{95\%\mathrm{CL}}_{\mathrm{fid.}} \sim 65$ nb in the left panel of Fig. 4, which translates into the limit $M = \sqrt{\beta} \gtrsim 100\,(190)$ GeV in the cut-off (unitarized) approach, as indicated by the green (blue) vertical dashed line in Fig. 4, respectively.

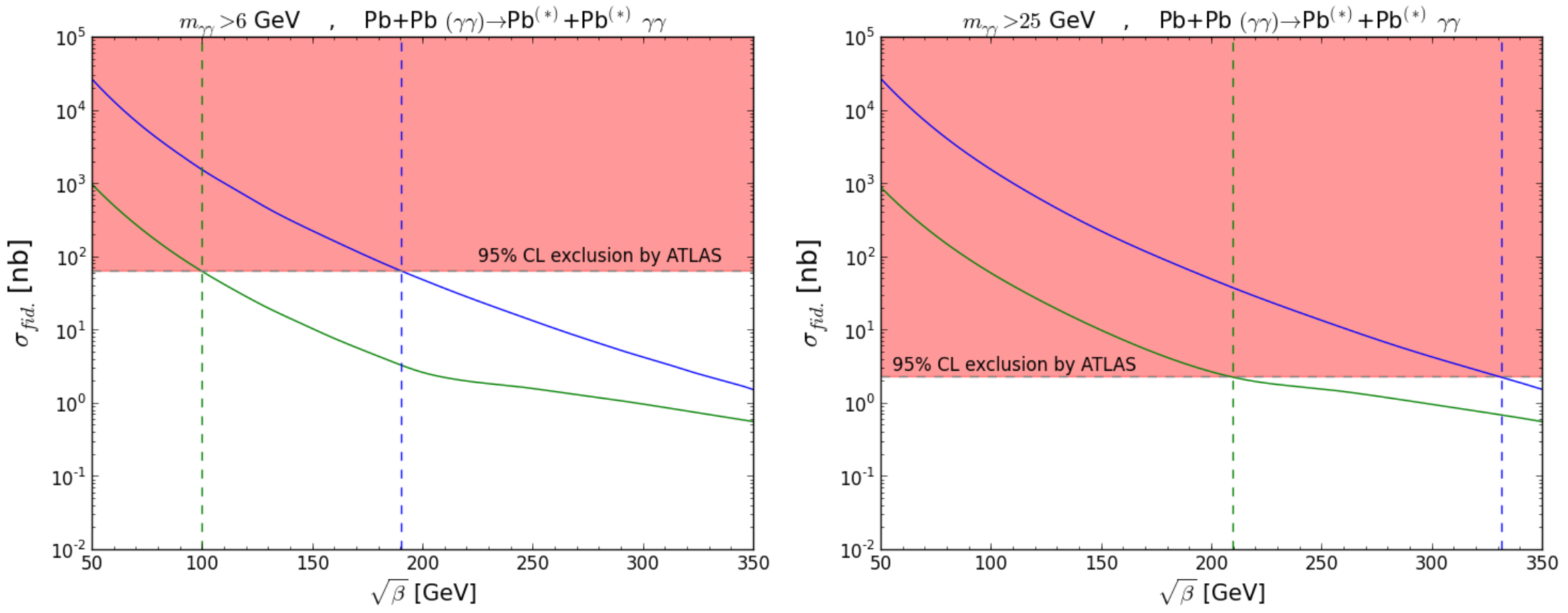

**Fig. 4:** *The fiducial cross section for light-by-light scattering in relativistic heavy-ion collisions, $\sigma(\mathrm{Pb{+}Pb}(\gamma\gamma) \to \mathrm{Pb}^{(*)}{+}\mathrm{Pb}^{(*)}\gamma\gamma)$ as a function of $M = \sqrt{\beta}$ in the U(1)$_{\rm EM}$ Born-Infeld theory is shown as a solid green (blue) line for a hard cut-off (unitarized) approach, respectively, as discussed in the text. The lower diphoton invariant mass cut-off is set at 6 GeV (25 GeV) on the upper (lower) plot. This is compared with the 95% CL upper limit obtained from the ATLAS measurement [9] by combining the statistical and systematic errors in quadrature as well as a 10 nb theoretical uncertainty in the cross section predicted in QED [8, 10] (horizontal dashed line), which excludes the higher range shaded pink. The corresponding 95% CL lower limits $M \gtrsim 100\,(190)$ GeV for $m_{\gamma\gamma} > 6$ GeV, and $M \gtrsim 210\,(330)$ GeV for $m_{\gamma\gamma} > 25$ GeV, are shown as vertical dashed lines in green (blue).*

These limits could be strengthened further by considering the $m_{\gamma\gamma}$ distribution shown in Fig. 3(b) of [9], where we see that all the observed events had $m_{\gamma\gamma} < 25$ GeV, in line with expectations in QED, whereas in the Born-Infeld theory most events would have $m_{\gamma\gamma} > 25$ GeV. Calculating a ratio of the total exclusive cross-section of QED for $m_{\gamma\gamma} > 6$ GeV and $> 25$ GeV as $\sigma^{m_{\gamma\gamma}>25\,\mathrm{GeV}}_{\mathrm{excl.}}/\sigma^{m_{\gamma\gamma}>6\,\mathrm{GeV}}_{\mathrm{excl.}} \sim 0.02$, we estimate a 95% CL upper limit of $\sim 2$ nb for $m_{\gamma\gamma} > 25$ GeV. The corresponding exclusion plot is shown in the right panel of Fig. 4, where we see a stronger limit $M = \sqrt{\beta} \gtrsim 210\,(330)$ GeV in the cut-off (unitarized) approach with the same colour coding as previously.

Our lower limit on the QED Born-Infeld scale $M = \sqrt{\beta} \gtrsim 100$ GeV is at least 3 orders of magnitude stronger than the sensitivities to $M = \sqrt{\beta}$ of previous measurements of nonlinearities in light [14–17, 19, 20]. Because of the kinematic cuts made in the ATLAS analysis, our limit does not apply to a range of values of $M \lesssim 10$ GeV for which the nonlinearities in (1) should be taken into account. However, our limit is the first to approach the range of potential interest for string/M theory constructions, since models with (stacks of) branes separated by distances $1/M : M = \mathcal{O}(1)$ TeV have been proposed in that context [37]. Our analysis could clearly be refined with more sophisticated detector simulations and the uncertainties reduced. However, in view of the strong power-law dependence of the Born-Infeld cross-section on $M = \sqrt{\beta}$ visible in (4), the scope for significant improvement in our constraint is limited unless experiments can probe substantially larger $m_{\gamma\gamma}$ ranges. In this regard, it would be interesting to explore the sensitivities of high-energy $e^+e^-$ machines considered as $\gamma\gamma$ colliders.

[2] We neglect possible interference effects that are expected to be small due to the different invariant mass and angular distributions involved.

## 4 Born-Infeld Extension of the Standard Model and the Mass of a Magnetic Monopole

As mentioned in the Introduction, Arunasalam and Kobakhidze have recently pointed out [21] that the Standard Model modified by a Born-Infeld theory of the hypercharge $\mathrm{U(1)_Y}$ contains a finite-energy monopole solution with mass $M_{\mathcal{M}} = E_0 + E_1$, where $E_0$ is the contribution associated with the Born-Infeld $\mathrm{U(1)_Y}$ hypercharge, and $E_1$ is sssociated with the remainder of the Lagrangian. Arunasalam and Kobakhidze have estimated [21] that $E_0 \simeq 72.8\, M_Y$, where $M_Y = \cos\theta_W M$, and Cho, Kim and Yoon had previously estimated [23] that $E_1 \simeq 4$ TeV [3]. Combining these calculations and using our lower limit $M \gtrsim 100$ GeV (2), we obtain a lower limit $M_{\mathcal{M}} \gtrsim 11$ TeV on the $\mathrm{U(1)_Y}$ Born-Infeld monopole mass [4]. Unfortunately, this is beyond the reach of MoEDAL [26] or any other experiment at the LHC [25], but may be accessible at a future 100-TeV $pp$ collider [38] or in a cosmic-ray experiment.

## Acknowledgements

The work of JE and NEM was supported partly by the STFC Grant ST/L000326/1. The work of TY was supported by a Junior Research Fellowship from Gonville and Caius College, Cambridge. We thank Vasiliki Mitsou for drawing our attention to [21], and her and Albert De Roeck, Igor Ostrovskiy and Jim Pinfold of the MoEDAL Collaboration for their interest and relevant discussions. TY is grateful for the hospitality of King's College London where part of this work was completed.

[3] Both these estimates are at the classical level, and quantum corrections have yet to be explored.

[4] For completeness, we recall that it was argued in [21] that nucleosynthesis constraints on the abundance of relic monopoles require $M_{\mathcal{M}} \lesssim 23$ PeV.

# Anomalous gauge interactions in photon collisions at the LHC and the FCC

*S. Fichet*[a*], *C. Baldenegro*[b]
[a] ICTP-SAIFR & IFT-UNESP, R. Dr. Bento Teobaldo Ferraz 271, São Paulo, Brazil
[b] University of Kansas, Lawrence, Kansas, U.S.

**Abstract**

The forward proton detectors recently installed and operating at the LHC open the possibility to observe photon collisions with high precision, providing a novel window on physics beyond the Standard Model. We review recent simulations and theoretical developments about the measurement of anomalous $\gamma\gamma\gamma\gamma$ and $Z\gamma\gamma\gamma$ interactions. The searches for these anomalous gauge interactions are expected to set bounds on a wide range of particles including generic electroweak particles, neutral particles with dimension-5 coupling to gauge bosons, polarizable dark particles, and are typically complementary from new physics searches in other channels.

## 1 Forward proton detectors and photon collisions

The new forward detectors have been installed at both ATLAS (ATLAS Forward Proton detector [1]) and CMS (CT-PPS detector [2]) and have started to take data since the 2017 run. The purpose of these detectors is to measure intact protons arising from diffractive processes at small angle, giving access to the so-called central exclusive processes

$$pp \to p \oplus X \oplus p\,, \tag{1}$$

where the $\oplus$ denote gaps with no hadronic or electromagnetic activity between the central system $X$ and the outgoing protons, see Fig. 1. These central exclusive processes can potentially provide a new window on physics beyond the Standard Model (SM) at the LHC. The special role of the forward detectors is to characterize the outgoing intact protons, hereby giving access to the complete kinematics of the event. Such information can then be used to drastically reduce the backgrounds.

The forward detectors are installed at $\sim 200$ m on both sides of CMS and ATLAS and host tracking stations. Their acceptance in the fractional momentum loss of the intact protons $\xi$ is approximately $0.015 < \xi < 0.15$ at the nominal accelerator magnetic lattice and beam conditions, which corresponds to an acceptance of 300 to 1900 GeV in the invariant mass of the central system for the nominal LHC beam optics when both protons stay intact. The CMS and TOTEM collaborations presented the first results of said system at the LHC by measuring the central semi-exclusive production of high-mass muon pairs at 13 TeV with an integrated luminosity of 10 fb$^{-}$1 collected in high-luminosity fills [3], which proves the feasibility of the search for New Physics in the exclusive channel. Timing detectors are scheduled to be installed in both CT-PPS and ATLAS to measure the protons time-of-flight of protons with an expected precision of $\sim 15$ ps, which would allow to determine the primary vertex with a $\sim 1$ mm precision [4]. Timing detectors will not be used in the estimations presented today, however they

*Speaker

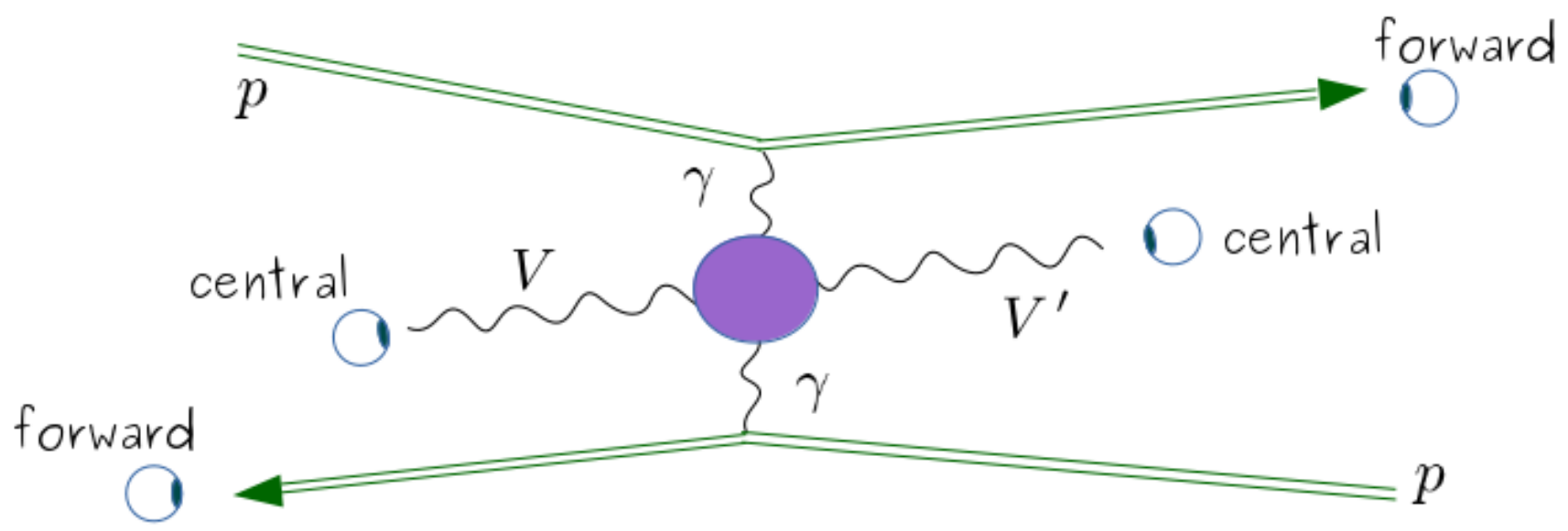

**Fig. 1:** Central exclusive photon collision with gauge boson final states. The protons remain intact and are seen by the forward detectors. The outgoing gauge bosons have high $p_T$ and are seen in the central detectors.

are absolutely necessary for final states with missing transverse energy (for example, $W^+W^-$ in the leptonic channel).

Central exclusive processes with intermediate photons – *i.e.* photon collisions – are especially interesting as the equivalent photon flux from the protons is large and well understood. In terms of an effective theory description of the new physics effects, one can for instance test operators like $|H|^2 F_{\mu\nu}F^{\mu\nu}/\Lambda^2$ which induce anomalous single or double Higgs production (for the MSSM case, see [5,6]). Our focus is on quartic gauge couplings, which leads to diboson final states as shown in Fig. 1. Such self-interactions of neutral gauge bosons are particularly appealing to search for new physics, because the SM irreducible background is small. The neutral quartic gauge interactions constitute smoking gun observables for new physics. Studies using proton-tagging at the LHC for new physics searches can be found in [7–24].

## 2 Searching for anomalous $\gamma\gamma\gamma\gamma$ and $Z\gamma\gamma\gamma$ interactions

Given the promising possibilities of forward detectors, realistic simulations of the search for $\gamma\gamma \to \gamma\gamma$, $\gamma\gamma \to Z\gamma$ have been carried out in [15,21,25]. The search for light-by-light scattering at the LHC without proton tagging has been first thoroughly analyzed in [26]. Let us review the setup, the backgrounds and the event selection for $\gamma\gamma$ and $Z\gamma$ final states.

The Forward Physics Monte Carlo generator (FPMC, [27]) is designed to produce within a same framework the double pomeron exchange (DPE), single diffractive, exclusive diffractive and photon-induced processes. The emission of photons by protons is correctly described by the Budnev flux [28,29], which takes into account the proton electromagnetic structure. The SM $\gamma\gamma \to \gamma\gamma$ process induced by loops of SM fermions and $W$ boson, the exact contributions from new particles with arbitrary charge and mass, and the anomalous vertices described by the effective operators of Eq. (3) have been implemented into FPMC.

The main background for the signals $\gamma\gamma \to \gamma\gamma, Z(ll, jj)\gamma$ is the detection of central final states (including misidentified jet or electrons) originating from a non-exclusive process, occurring *simultaneously* with the forward detection of two protons from pile-up. The probability to detect at least one proton in each of the forward detectors is estimated to be 32%, 66% and 93% for 50, 100 and 200 additional interactions respectively. Other backgrounds include double Pomeron exchange and central exclusive QCD production.

The knowledge of the full event kinematics is a powerful constraint to reject the huge background from pile-up. The key requirements consist in matching the missing momentum (rapidity difference) of the di-proton system with the invariant mass (rapidity difference) of the central system ($XY \equiv \gamma\gamma, (jj)\gamma, (\ell\bar{\ell})\gamma$), with $m_{XY}/m_{pp} < 5-10\%$ and $|y_{XY} - y_{pp}| < 3-10\%$ depending on the channel. Extra cuts rely on the event topology, using the fact that the $XY$ states are typically back-to-back with similar $p_T$.

**Table 1:** $5\,\sigma$ sensitivity to the effective quartic gauge couplings of Eq. (3) in GeV$^{-4}$.

| Assumptions | $\gamma\gamma$ final state | $\gamma jj$, $\gamma ll$ final state |
|---|---|---|
| 13 TeV, 300 fb$^{-1}$, $\mu = 50$ | $\zeta^\gamma(\tilde{\zeta}^\gamma) < 9 \cdot 10^{-15}$ | $\zeta^{\gamma Z}(\tilde{\zeta}^{\gamma Z}) < 1.9 \cdot 10^{-13}$ $[\gamma jj + \gamma ll]$<br>$\zeta^{\gamma Z}(\tilde{\zeta}^{\gamma Z}) < 2.8 \cdot 10^{-13}$ $[\gamma ll]$<br>$\zeta^{\gamma Z}(\tilde{\zeta}^{\gamma Z}) < 2.3 \cdot 10^{-13}$ $[\gamma jj]$ |
| 13 TeV, 3000 fb$^{-1}$, $\mu = 200$ | $\zeta^\gamma(\tilde{\zeta}^\gamma) < 1 \cdot 10^{-14}$ | $\zeta^{\gamma Z}(\tilde{\zeta}^{\gamma Z}) < 1.8 \cdot 10^{-13}$ $[\gamma ll]$ |
| 100 TeV, 3000 fb$^{-1}$, $\mu = 200$ | $\zeta^\gamma(\tilde{\zeta}^\gamma) < 1.1 \cdot 10^{-16}$ | |
| 100 TeV, 3000 fb$^{-1}$, $\mu = 1000$ | $\zeta^\gamma(\tilde{\zeta}^\gamma) < 2 \cdot 10^{-16}$ | |

This translates to cuts of the form $|\Delta\phi_{X,Y} - \pi| < 0.02$, $p_{TX}/p_{TY} > 0.90 - 0.95$. The signal is harder than the background hence one cuts on the invariant mass typically as $m_{XY} > 600 - 700$ GeV. Further background reduction could be possible by measuring the protons time-of-flight, which would allow to constrain the event vertex.

## 3 Sensitivity to heavy new particles at LHC and FCC

In a scenario where new particles are too heavy to be produced on-shell at the collider, the presence of these new states is best studied using effective field theory methods. The low-energy effects of the new particles are parametrized by higher dimensional operators made of SM fields. For the quartic gauge interactions of our interest we use the basis of operators

$$\begin{aligned} \mathcal{L}_{4\gamma} &= \zeta^\gamma F_{\mu\nu}F^{\mu\nu}F_{\rho\sigma}F^{\rho\sigma} + \tilde{\zeta}^\gamma F_{\mu\nu}\tilde{F}^{\mu\nu}F_{\rho\sigma}\tilde{F}^{\rho\sigma} \\ \mathcal{L}_{3\gamma Z} &= \zeta^{\gamma Z} F_{\mu\nu}F^{\mu\nu}F_{\rho\sigma}Z^{\rho\sigma} + \tilde{\zeta}^{\gamma Z} F_{\mu\nu}\tilde{F}^{\mu\nu}F_{\rho\sigma}\tilde{Z}^{\rho\sigma}\,. \end{aligned} \tag{2}$$

Other operators like $\mathcal{O}_2^\gamma = F_{\mu\nu}F^{\nu\rho}F_{\rho\sigma}F^{\sigma\mu}$ are sometimes used. They are linearly dependent of the ones above, with $4\mathcal{O}_2 = 2\mathcal{O} + \tilde{\mathcal{O}}$.

These expected sensitivities are given in Table 1 for different scenarios corresponding to medium luminosity (300 fb$^{-1}$) and to high luminosity (3000 fb$^{-1}$) at the LHC. The background being small, statistical significance grows quickly with the event number. Typically $O(10)$ events are enough to reach $5\sigma$ significance. We also provide an estimation for FCC with proton collisions, assuming 3000 fb$^{-1}$ and average pile-up of $\mu = 200 - 1000$. We assume the forward proton detectors properties at FCC to be similar to those at the LHC, including the acceptance range.

The ATLAS Collaboration has set a bound on the $Z \to \gamma\gamma\gamma$ decay of $\mathcal{B}(Z \to \gamma\gamma\gamma) < 2.2 \cdot 10^{-6}$ [30], beating the ones from LEP. This bound translates as a limit [25]

$$\sqrt{\zeta^2 + \tilde{\zeta}^2 - \frac{\zeta\tilde{\zeta}}{2}} < 1.3 \cdot 10^{-9}\ \mathrm{GeV}^{-4}\ \ (95\%\mathrm{CL})\,. \tag{3}$$

Imagining the same search is done at 13 TeV data with 300 fb$^{-1}$ in the same conditions, we expect very roughly an improvement by an order of magnitude of the bound of Eq. 3. In addition, the current number of pile-up interactions at 13 TeV sets a challenge to the measurement of $3\gamma$ final states. This remains far away from the expected sensitivities obtained from photon collisions at the same luminosity by roughly three orders of magnitudes.

## 4 Sensitivity to new electroweak particles

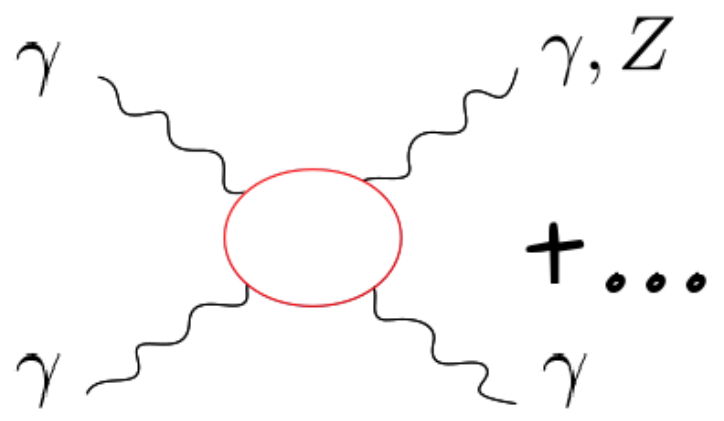

**Fig. 2:** A loop of electroweak particles inducing a quartic gauge interaction.

New EW charged particles contribute to anomalous gauge couplings at one-loop (Fig. 2). Because of gauge invariance, these contributions can be parametrized in terms of the mass and quantum numbers of the new particle [14]. Using the heat kernel calculation of [14], we obtain the effective couplings

$$\begin{aligned}\left(\zeta^{\gamma Z},\tilde{\zeta}^{\gamma Z}\right) &= \left(c_s,\tilde{c}_s\right)\frac{\alpha_{\rm em}^2}{s_w c_w\, m^4}N\,d\left(c_w^2\frac{3d^4-10d^2+7}{240}+(c_w^2-s_w^2)\frac{(d^2-1)Y^2}{4}-s_w^2Y^4\right)\\ &\equiv \left(c_s,\tilde{c}_s\right)\frac{\alpha_{\rm em}^2}{s_w c_w\, m^4}\frac{Q_{\rm eff}^4}{4}\,,\end{aligned} \tag{4}$$

$$\left(\zeta^{\gamma},\tilde{\zeta}^{\gamma}\right)=\left(c_s,\tilde{c}_s\right)\frac{\alpha_{\rm em}^2}{m^4}N\,d\left(\frac{3d^4-10d^2+7}{960}+\frac{(d^2-1)Y^2}{8}+\frac{Y^4}{4}\right)\,. \tag{5}$$

We have evaluated the sensitivities using the exact amplitudes. For example, the sensitivity for the case of a charged vector is shown in Fig. 3 for a luminosity of 300 fb$^{-1}$ and $\mu = 50$.

Certain candidates like the typical vector-like quarks arising in Composite Higgs models are already excluded by stronger bounds. On the other hand, our expected bounds from photon collisions are completely model-independent. They apply to any quantum number, are independent of the amount of mixing with the SM quarks, and even apply to vector-like leptons.

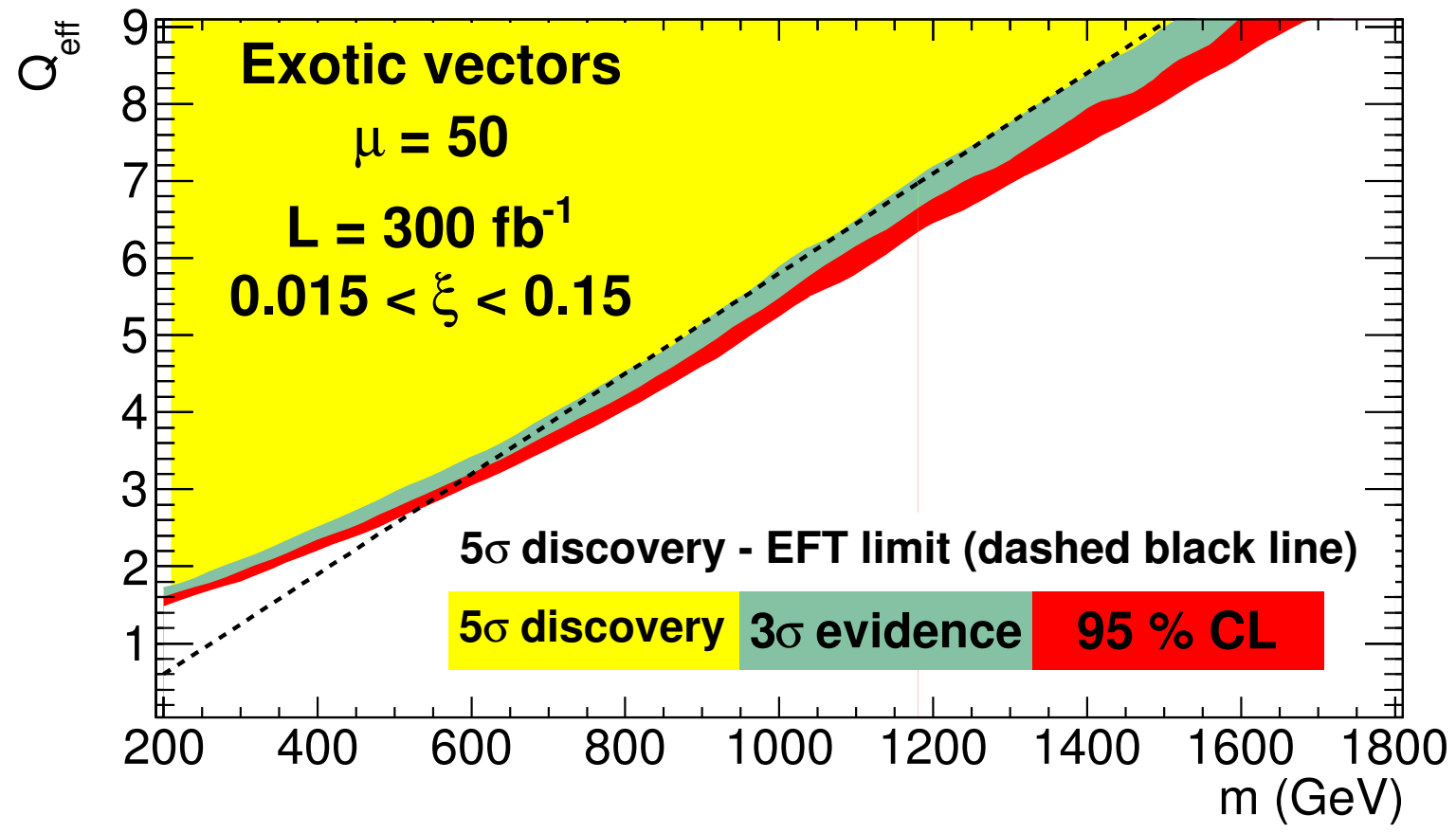

**Fig. 3:** Expected sensitivity on a spin-1 particle as a function of its mass and of its effective charge $Q_{\rm eff}$.

## 5 Sensitivity to new neutral particles

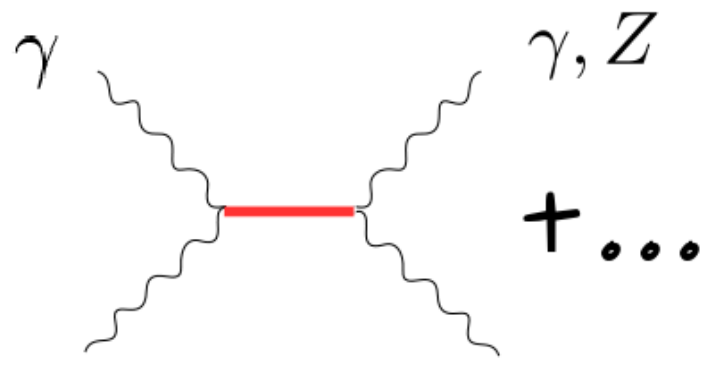

**Fig. 4:** The exchange of a neutral particle inducing a quartic gauge interaction.

Non-renormalizable interactions of neutral particles are also present in common extensions of the SM. Such theories can contain scalar, pseudo-scalar and spin-2 resonances, respectively denoted by $\varphi$, $\tilde{\varphi}$ and $h^{\mu\nu}$ [15], that can potentially be strongly-coupled to the SM. The full effective theory for such neutral resonances is given in [31].

$$\begin{aligned}\mathcal{L}_{\gamma\gamma} =& \varphi \left[ \frac{1}{f_{0+}^{\gamma\gamma}} (F_{\mu\nu})^2 + \frac{1}{f_{0+}^{\gamma Z}} F_{\mu\nu} Z_{\mu\nu} \right] + \tilde{\varphi} \left[ \frac{1}{f_{0-}^{\gamma\gamma}} F_{\mu\nu} \tilde{F}_{\mu\nu} + \frac{1}{f_{0-}^{\gamma Z}} F_{\mu\nu} \tilde{Z}_{\mu\nu} \right] \\ & + h^{\mu\nu} \left[ \frac{1}{f_2^{\gamma\gamma}} (-F_{\mu\rho} F_\nu{}^\rho + \eta_{\mu\nu} (F_{\rho\lambda})^2/4) + \frac{1}{f_2^{\gamma Z}} (-F_{\mu\rho} Z_\nu{}^\rho + \eta_{\mu\nu} F_{\rho\lambda} Z_{\rho\lambda}/4) \right] ,\end{aligned} \tag{6}$$

where the $f_S$ have mass dimension one. Such neutral particle induces quartic gauge couplings (Fig.4) given by

$$(\zeta^\gamma, \tilde{\zeta}^\gamma) = \frac{1}{(f_s^{\gamma\gamma})^2 m^2} (d_s, \tilde{d}_s) , \quad (\zeta^{\gamma Z}, \tilde{\zeta}^{\gamma Z}) = \frac{1}{f_s^{\gamma\gamma} f_s^{\gamma Z} m^2} (d_s, \tilde{d}_s) , \quad (d_s, \tilde{d}_s) = \begin{cases} 1, 0 & s = 0^+ \\ 0, 1 & s = 0^- \\ \frac{1}{4}, \frac{1}{4} & s = 2 . \end{cases} \tag{7}$$

For example, at 13 TeV, 300 fb$^{-1}$, $\mu = 50$, the expected $5\sigma$ sensitivities on the scalar are

$$m < 4.5\,\text{TeV} \cdot \left( \frac{1\,\text{TeV}}{f^{\gamma\gamma}} \right) , \quad m < 2.3\,\text{TeV} \cdot \left( \frac{1\,\text{TeV}}{\sqrt{f^{\gamma\gamma} f^{\gamma Z}}} \right) . \tag{8}$$

Following [15, 21], the bounds on a dilaton mass can reach 4260 GeV ($5\sigma$) and those on a KK graviton with IR gauge fields can reach 5670 GeV ($5\sigma$). Interestingly, the $Z\gamma$ coupling vanishes( $f_{Z\gamma} \to \infty$) when the particle couples universally to the $SU(2) \times U(1)_Y$ field strengths $(B^{\mu\nu})^2$, $(W^{I,\mu\nu})^2$, hence the $\gamma Z$ channel provides a powerful piece of information.

## 6 Sensitivity to a polarizable dark sector

It is very plausible that a new particle be *almost dark* in the sense that it interacts with light only via higher-dimensional operators. Let us focus on self-conjugate particles, which have no dipole – our approach applies very similarly if the dark particle is polarized. For concreteness we focus on a real scalar. Its interactions with light have the form $\frac{1}{\Lambda^2}\phi^2(F^{\mu\nu})^2$, $\frac{1}{\Lambda^4}\partial_\mu\phi\partial_\nu\phi F^{\mu\rho}F_\rho^{\ \nu}$, .... The property of polarizability can be either induced by mediators, or arise from the inner structure of the particle (intrinsic polarizability) [24]. The latter happens in particular if the dark particle is a composite made of electrically charged constituents. As a matter of fact, many models of dark sectors can feature this kind of composite dark particles (see for instance vectorlike confinement, stealth DM [32]).

The polarizable dark particle can induce quartic gauge interactions at one-loop as shown in Fig. 5. This contribution is potentially dominant in the scenario of intrinsic polarizability. The photon collision then quite literally sheds light on the dark particle. In a simulation at 13 TeV, 300 fb$^{-1}$ $\mu = 50$, the sensitivities in mass and in $\Lambda$ go typically up to the TeV scale – details depend on which polarizability operator is considered.

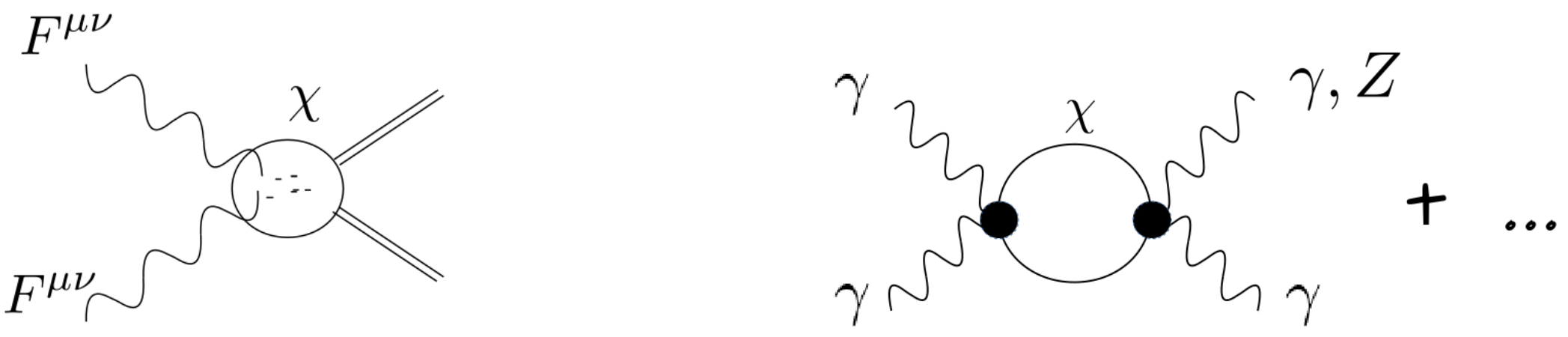

**Fig. 5:** *Left:* Sketch of a composite polarizable dark particle. *Right:* Sketch of the quartic gauge interaction induced by the polarizable dark particle.

Searching for dark sectors via their virtual effects is a recent trend [33, 34], and many interesting developments are yet to be done. An advantage of such searches is they do not rely on the hypothesis of stability of the dark particle. When the dark particle is stable it is a dark matter candidate, in which case the search in photon collisions turns out to be complementary to monojet+ missing energy searches. The search in photon collisions gets favored by larger multiplicities in the loop ($\propto N^2$ versus $\propto N$) and for large dark particle mass as it is not constrained by the kinematic threshold of DM pair production.

## Acknowledgements

I would like to thank the organizers of the Photon 2017 conference for the kind invitation and the São Paulo Research Foundation (FAPESP) for the support under grants #2011/11973 and #2014/21477-2.

# The LHC Ring as a Photon-Photon Search Machine

*Risto Orava*[1]* *for the LHC ring proto-collaboration*
[1]University of Helsinki, Helsinki Institute of Physics and CERN-EP, CH-1211 Geneva 23, Switzerland

**Abstract**
The LHC Collider Ring can be turned into a photon-photon search machine based on Central Exclusive Process (CEP): $pp \to p + X + p$. The present extensive Beam Loss Monitoring (BLM) system of the LHC precisely registers the exit points of the final state CEP protons according to their longitudinal momentum losses. The BLM system is being continuously extended, and equipped with fast timing and data acquisition systems, will enable efficient new physics event tagging together with the LHC experiments. The LHC Ring can be used to facilitate an on-line automatic search machine for the physics of tomorrow.

## 1 Introduction

In the following, the LHC Ring is described as a new photon-photon physics search facility based on existing instrumentation of the LHC ring and the LHC experiments. The approach presented here is novel, and uses the LHC Beam Loss Monitoring (BLM) and other LHC beam instrumentation devices for tagging the new physics event candidates in a model-independent way. The physics potential of the proposed facility is huge, and highly complementary to the present experimental installations at the LHC (ALICE, ATLAS/ALPHA, CMS/TOTEM, LHCb/MoEDAL experiments).

A few selected Central Exclusive Production (CEP) processes are discussed together with high mass Single Diffractive (SD) scattering. The CEP processes provide an ideal test ground for the proposed approach - here a pair of coincident final state protons, exiting the LHC beam vacuum chamber, are used to tag the event candidates. The fractional momenta of the final state protons are directly related to the invariant mass of the centrally produced system. The proposed approach [1, 2] is independent of the particular decay modes of a centrally produced system, and substantially enhances the potential of observing new heavy particle states at the LHC. Performance of the customary Roman Pot technology is limited by the location of the pots, and the allowed transverse access to the beam.

The collaborators represent the key areas of this proposal: in accelerator physics and LHC instrumentation (S. Redaelli et al., CERN Beams Division), accelerator theory (Werner Herr, CERN Beams Division), theoretical high energy physics (Lucian Harland-Lang, University College, London, K. Huitu, Division of Particle Physics and Astrophysics, University of Helsinki; Valery Khoze, University of Durham University; M.G. Ryskin Petersburg Nuclear Physics Institute, Gatchina, St. Petersburg; V. Vento, University of Valencia and CSIC) and experimental high energy physics (A. De Roeck, CERN EP; M. Kalliokoski, CERN Beams Division; Beomkyu Kim, University of Jyväskylä; Jerry W. Lämsä, Iowa State University, Ames; C. Mesropian, Rockefeller University; Mikael Mieskolainen, University of Helsinki; Toni Mäkelä, Aalto University, Espoo; Risto Orava, University of Helsinki, Helsinki Institute of Physics and CERN; J. Pinfold, FRSC, Centre for Particle Physics Research, Physics Department, University of Alberta; Sampo Saarinen, University of Helsinki; M. Tasevsky, Institute of Physics of Academy of Sciences, Czech Republic.

*e-mail:risto.orava@cern.ch

The project has break-through potential in a number of physics processes beyond the examples discussed here. The basic infrastructure, the LHC Ring with its beam instrumentation and experiments, already exists, and only minor extensions are proposed for relatively inexpensive additional detectors and for facilitating triggering and automatic event selection. Preliminary analyses of the BLM signals validate the basic approach adopted by the authors, and include exiting candidate events in different physics categories listed below.

## 2 Scanning for new physics

The Central Exclusive Production (CEP) of particle state, $X$, is described by the following three processes:

$$pp \to p + (\gamma\gamma \to X) + p, \tag{1}$$

$$pp \to p + (\gamma + gg \to X) + p, \tag{2}$$

$$pp \to p + (gg \to X) + p, \tag{3}$$

where the $+$ signs indicate rapidity gaps. The CEP sub-processes are facilitated by the photons (photon-photon interaction) (1), photons and gluons ("photo-production" or "photon-pomeron" interaction) (2), and gluons ("diffractive" or "double pomeron exchange") (3). In Figure 2, the corresponding Feynman diagrams for the processes (1-3) are shown.

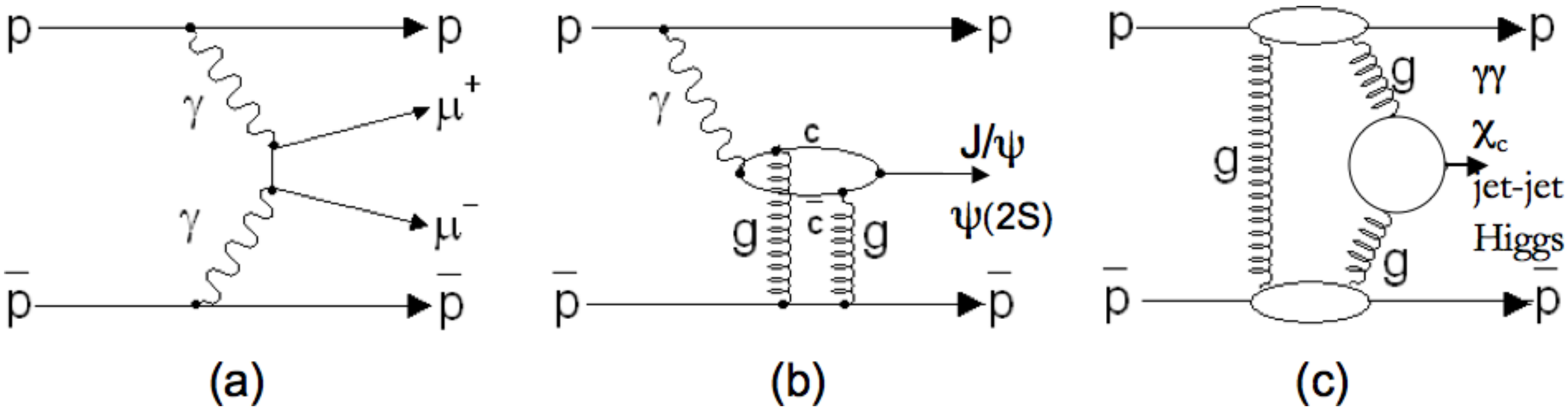

**Fig. 1:** The Central Exclusive Production (CEP) processes facilitated by: (a) photon-photon sub interaction $pp(\gamma\gamma) \to p + (l^+l^-, W^+W^-) + p$, (b) photon-gluon sub-interaction $pp(\gamma + gg) \to p + (J/\psi, \psi(2S)) + p$, and (c) gluon-gluon sub-interaction $pp(gg) \to p$ + ($\chi_c$, jet-jet, Higgs) + $p$.

The respective cross sections for the processes ((1-3)) are calculated as the convolutions of the effective luminosities $L(\gamma\gamma)$, $L(\gamma + gg)$, or $L(gg)$ (Figure 2) and the square of the matrix element of the corresponding sub-process [3]. In the Central Exclusive Production (CEP), a number of advantageous properties exist compared to inclusive (or semi-inclusive) production: The mass and width of the centrally produced state, $X$, is correlated with the fractional (longitudinal) momentum losses, $\xi_i = 1 - p'_{z_i}/p_z$, of the final state protons $p'_z$ and the initial beam proton $p_z$, as:

$$M^2 \simeq \xi_1\xi_2 s \tag{4}$$

where $s$ is the centre-of-mass energy squared. A measurement of the invariant mass of the decay products would be required to match the missing mass condition available by the measurement of the pair of final state proton fractional momentum losses. At higher central masses, $M_X \gtrsim 200$ GeV, the photon-photon process (1) dominates.

The following example processes are considered: Magnetic monopolium: Numerous experimental searches for magnetic monopoles have been carried out but all have met with failure. These experiments have led to a lower mass limit in the range from 350 . . . 500 GeV. A way out of this impasse is the above old idea of Dirac [4], namely, monopoles are not seen freely because they are confined by their strong magnetic forces forming a bound state called monopolium [5]. Here the CEP produced monopolium states [6] are searched for by registering the pair of final state protons exiting the beam vacuum chamber at the distance of some $\sim$ 230 meters from IP8 (MoEDAL/LHCb experiment); $W^+W^-$ pairs and anomalous couplings: Central exclusive production (CEP) of $W^+W^-$ $(Z^0Z^0)$ pairs can be used both as a luminosity monitoring process [7] and as a process for studying basic physics questions beyond the Standard Model, such as anomalous vector boson couplings.

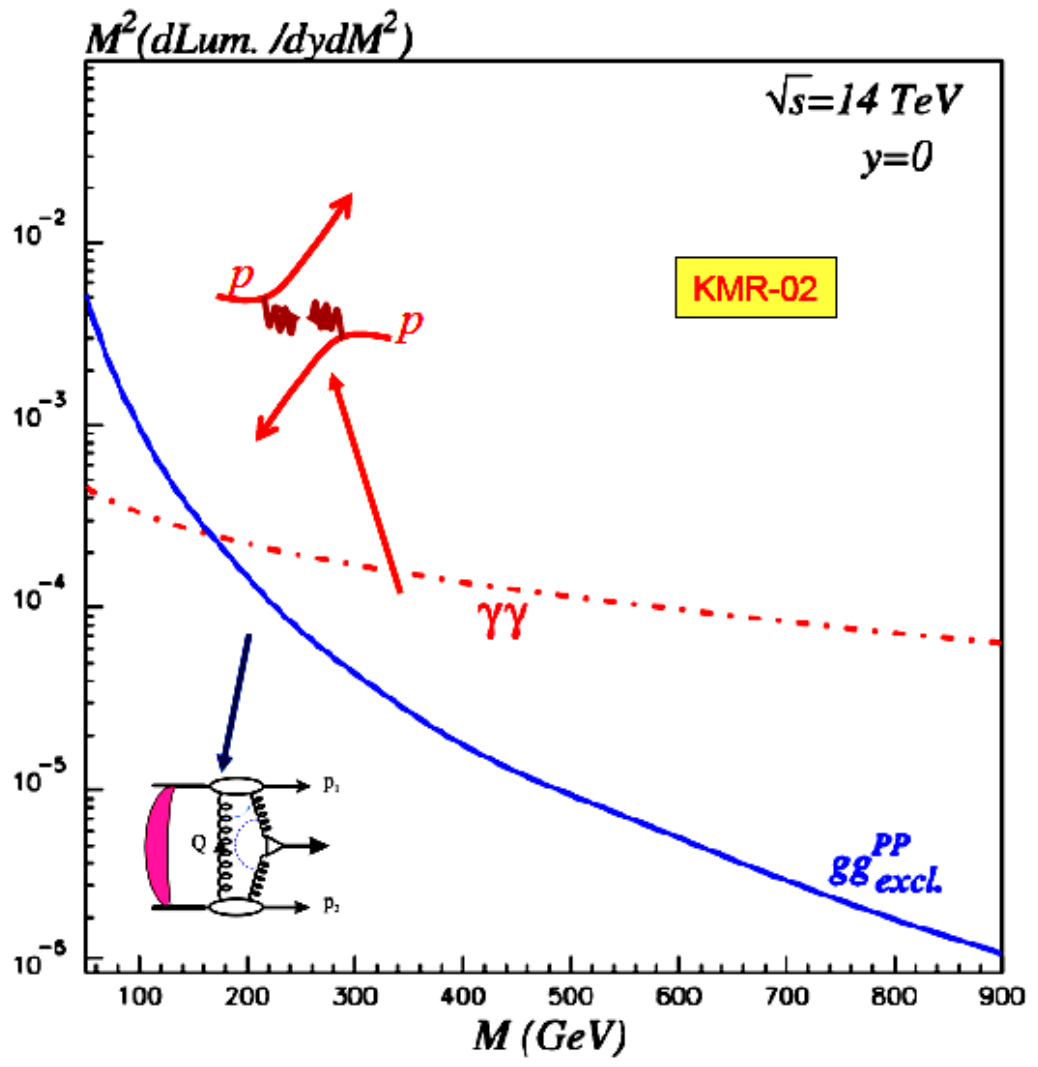

**Fig. 2:** Gluon (solid curve) and photon (dotted-dashed curve) luminosities as a function of the central mass in Central Exclusive Production (CEP) [3].

In a recent publication the contribution of the $W^+W^-$ $(Z^0Z^0)$ mechanism is compared to the gluon induced CEP process $gg \to W^+W^-$ [8]. The phase space integrated gluon induced CEP cross section is found to be considerably smaller (less than 1 fb), while the photon induced CEP is calculated to have a cross section of $115$ fb. The photo-production process dominates at small four-momentum transfers for a wide range of $W^+W^-$ $(Z^0Z^0)$ invariant masses, and allows efficient analyses of anomalous triple-boson $(WW/ZZ)$ and quartic-boson $(WW/ZZ)$ couplings together with tests of the models beyond the Standard Model. The $\gamma\gamma \to W^+W^-$ cross section peaks at $M_X \simeq 200$ GeV yielding (in a symmetric case) a pair of protons exiting the beam vacuum chamber at $\sim$ 330 meters from the interaction point.

All four LHC experiments have sufficient luminosity for studying the processes; The Standard Model (SM) and BSM Higgs bosons: The Higgs boson observations at the LHC are almost exclusively based on the $\gamma\gamma$ and $ZZ \to 4l$ decay mode [9]. Measurement of the Higgs boson production in Central Exclusive Process (CEP) was first analysed by some of the authors, and the process $pp \to p + (gg \to h^0) + p$ provides important complementary information concerning the spin-parity state of Higgs since $J^{PC} = 0^{++}$ state is strongly favoured in CEP. By tagging the Higgs event candidates independently of the Higgs decay products enables detailed analysis of the production mechanism and Higgs couplings. A measurement of the azimuthal angle between the final state protons can be used to discriminate between different Higgs production scenarios [10]. The Standard Model Higgs boson, when produced in the CEP process (3), has a proton pair exiting the LHC beam vacuum chamber at a distance of $\sim$ 427 meters from the IP. ATLAS, CMS and LHCb experiments are here relevant counterparts due to their sufficiently high integrated luminosities; Single Diffractive (SD), where the exiting proton at the longest distance from a given LHC Interaction Point, combined with a hadron shower at the experiment, efficiently identifies high mass SD event candidates.

## 3 CEP protons exiting the LHC ring

By tracing CEP protons of different $z$-values through the LHC accelerator lattice, a relation between the CEP proton exit points and the $\xi$-values of the final state protons is established. For the background studies of this proposal, both $\xi$ and the transverse momentum $p_T$ of the final state protons are considered in mapping out the exit points around the LHC ring.

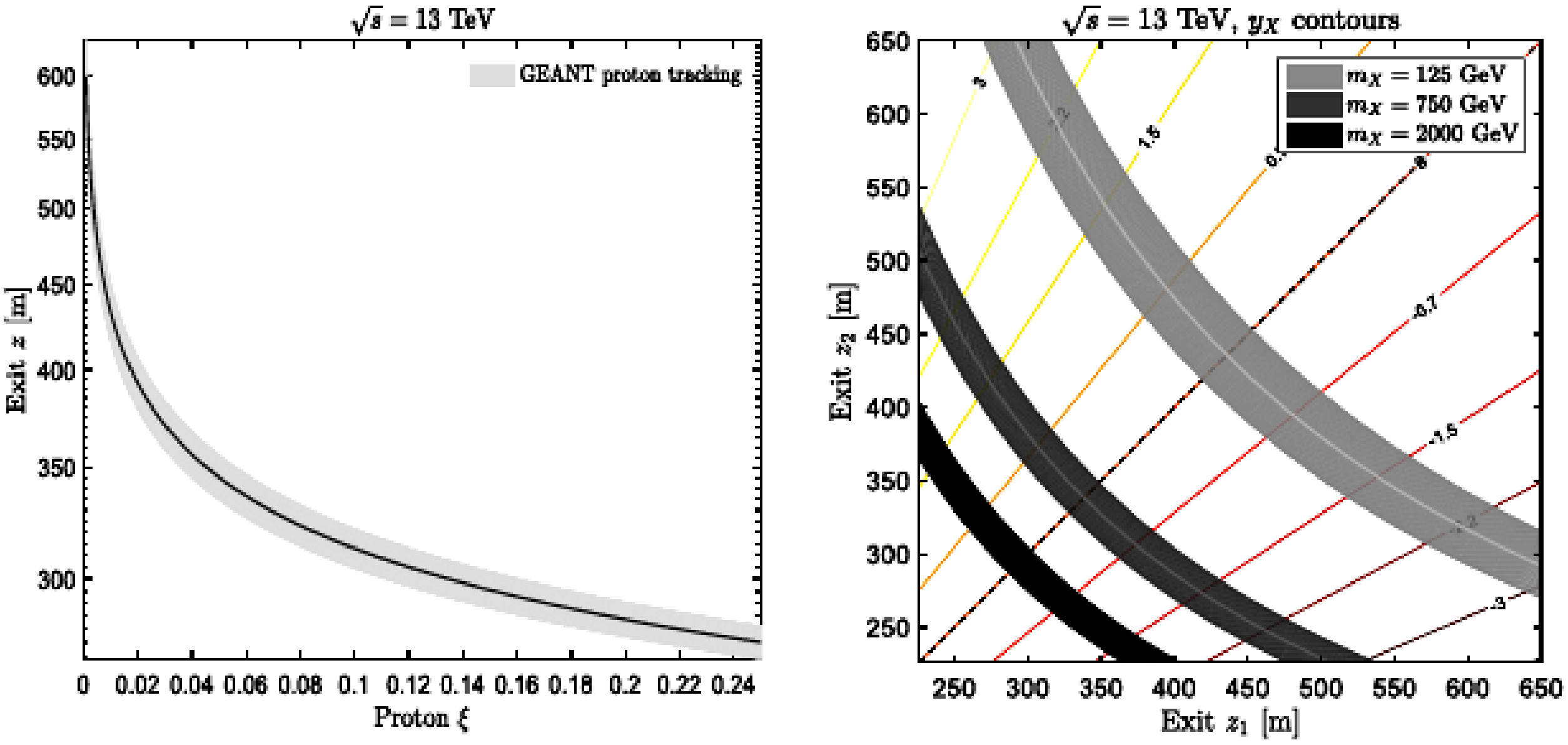

**Fig. 3:** Left: Proton exit point, $z$, in CEP: $pp \to p + X + p$, as a function of the fractional momentum loss, $\xi$ (solid line). The exit points of the leading protons out from the beam vacuum chamber are given in meters from IP5, the shaded band reflects smearing in proton transverse momentum. Right: The proton exit point combinations in CEP: $pp \to p + X + p$, as a function of central mass $M_X$ (grey bands). The exit points of the leading protons out from the beam vacuum chamber are given in meters from the Interaction Point 5 (IP5), the symmetric cases ($\xi_1 \simeq \xi_2$) have $z_1 \simeq z_2$ (dashed diagonal line). Rapidity of the centrally produced state is given as $y_X = 0.5 \ln(\xi_1/\xi_2)$ [1].

In Figure 3 (left panel), the proton exit points, shown as a function of their fractional momentum loss, $\xi_i$, are produced by the proton tracing codes. Through Equation 4, the measured proton exit locations can then be used for an $M_X$ mass scan of the centrally produced systems (Figure 3, right panel). The band widths reflect smearing in proton transverse momentum, $p_T$. The following steps are taken in tagging the CEP event candidates for each IP (IP1/ATLAS, IP2/ALICE, IP5/CMS, and IP8/LHCb): (i) The candidate CEP events are scanned by locating pairs of coincident proton exits on the opposite sides of the interaction point (IP) in question (Figure 3, right panel), (ii) The tagged events are correlated with the LHC Beam Cross Overs (BCOs) within the time window for the chosen IP, (iii) The tagged LHC BCOs are analysed as candidates for the CEP events with central masses, $M_X$, corresponding to a registered pair of exit points (Figure 3, right panel).

In Figure 4, the registered proton exit points are plotted together with the reconstructed ones obtained by fits. In Figure 5, the reconstructed diffractive masses in high mass SD scattering are shown for $M_X = 500, 1000$ and 1500 GeV. A resolution of $\Delta M_X/M_X \sim 10\%$ is obtained.

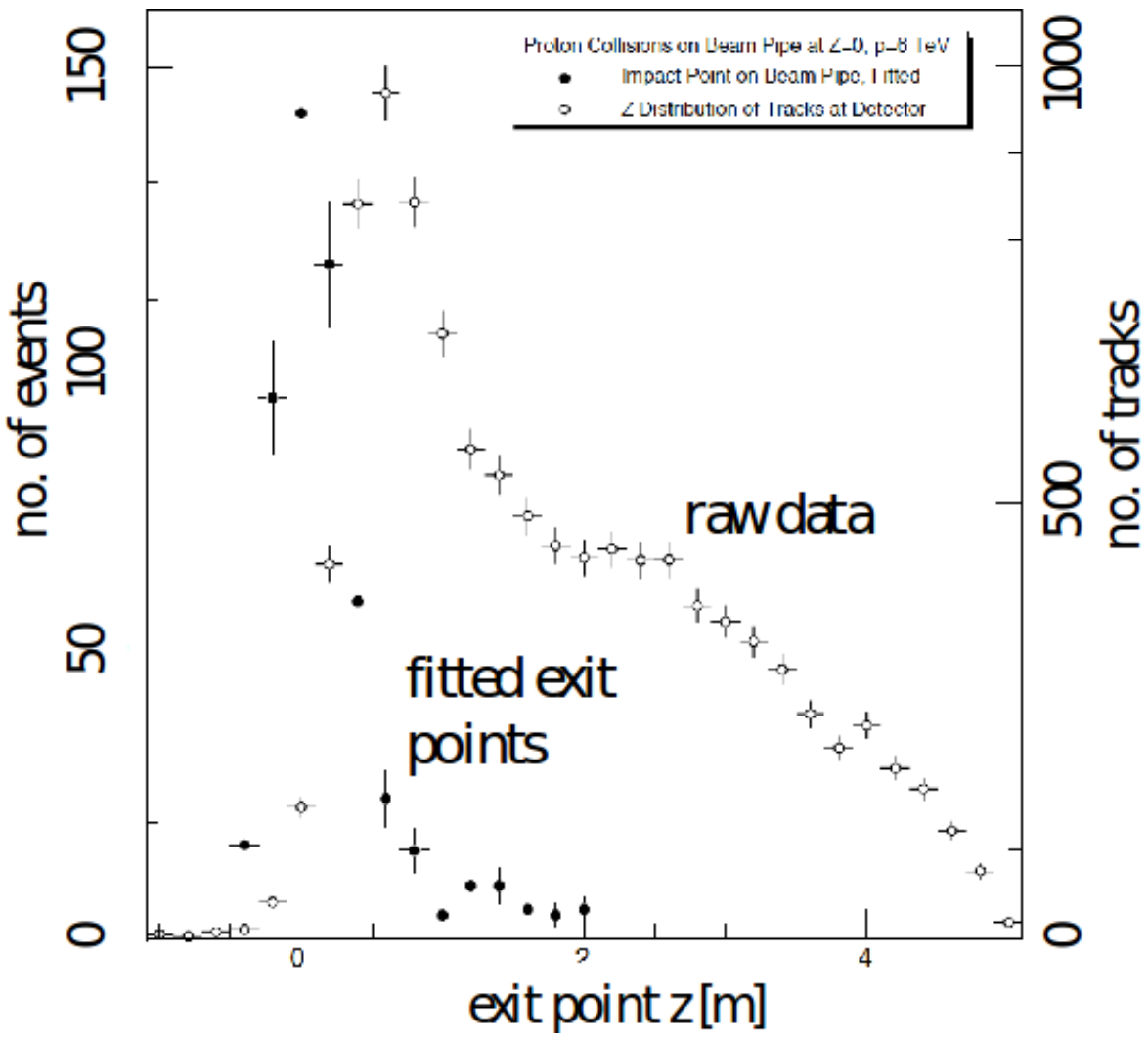

**Fig. 4:** The registered proton exit points, $z$, for raw data (open circles) and after reconstruction based on the detected hits in scintillators installed around the gaps of LHC magnets.

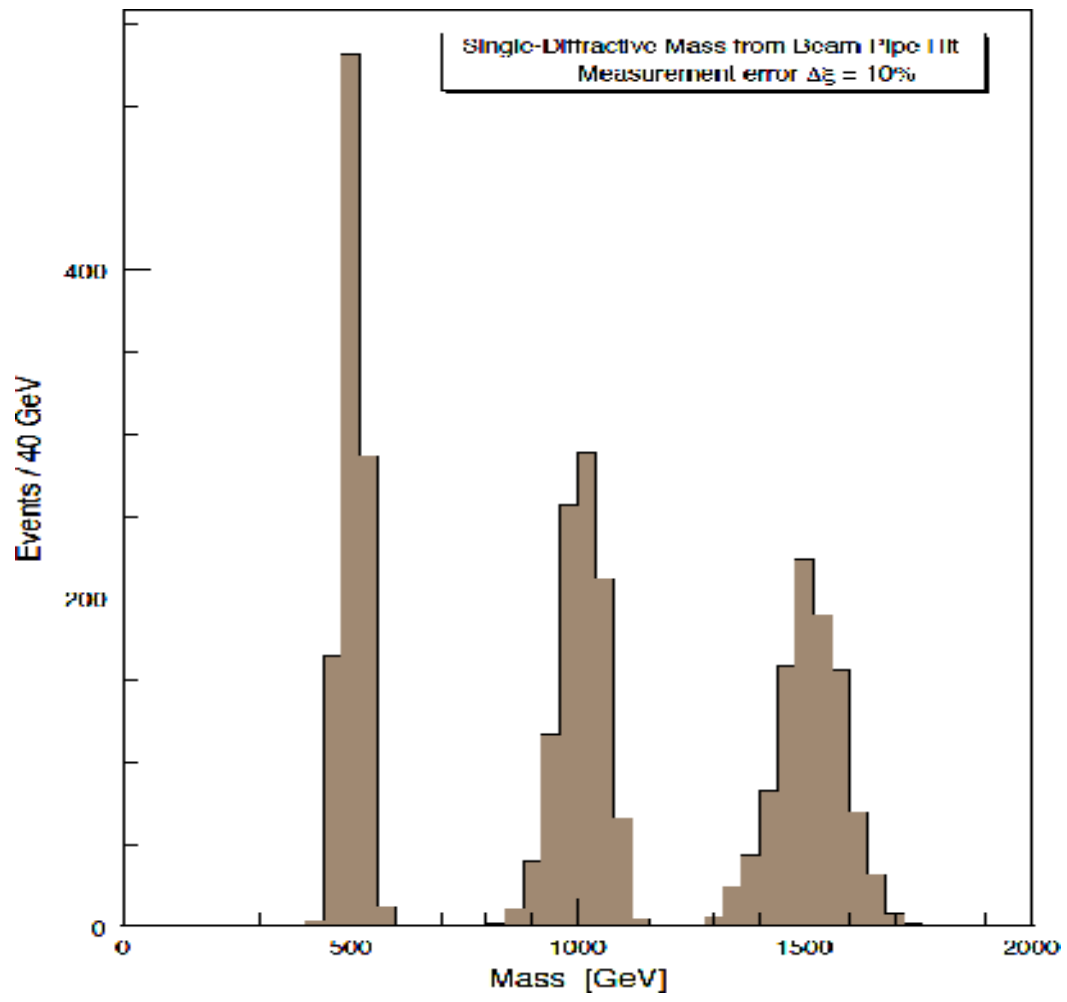

**Fig. 5:** The reconstructed central masses, $M_X$, in Single Diffractive process (SD): $pp \to p + X$, for $M_X = 500, 1000$ and 1500 GeV (Phojet simulation by J.W.Lämsä). The proton showers detected by scintillators assembled around the gaps between the LHC dipole magnets.

# The photon PDF from high-mass Drell-Yan data at the LHC

*V. Bertone*[*]
Department of Physics and Astronomy, VU University, NL-1081 HV Amsterdam,
and Nikhef Theory Group Science Park 105, 1098 XG Amsterdam, The Netherlands

**Abstract**
I present a determination of the photon PDF from a fit to the recent ATLAS measurements of high-mass Drell-Yan lepton-pair production at $\sqrt{s} = 8$ TeV. This analysis is based on the `xFitter` framework interfaced to the `APFEL` program, that accounts for NLO QED effects, and to the `aMCfast` code to account for the photon-initiated contributions within `MadGraph5_aMC@NLO`. The result is compared with other recent determinations of the photon PDF finding a general good agreement. This writeup is based on the results presented in Ref. [1].

## 1 Introduction and motivation

In order to achieve accurate predictions for the LHC phenomenology, QCD corrections, where NNLO is becoming the standard, have to be supplemented with electroweak (EW) effects. One of the direct consequences of these corrections is the introduction of the photon PDF.

An number of determinations of the photon PDF based on a variety of different approaches has been achieved in the past [3–8, 19]. The aim of this particular work is to obtain a model-independent determination of the photon PDF exploiting the recent high-mass Drell-Yan measurements at $\sqrt{s} = 8$ TeV from ATLAS [9], that have proven to provide a significant constraint on this distribution.

The constraining power of the Drell-Yan process on the photon PDF can be easily understood in terms of Feynman diagrams. Indeed, in the presence of EW corrections, the partonic channel $\gamma\gamma \to \ell^+\ell^-$ contributes to the leading order (LO) cross section for lepton-pair production in $pp$ scattering. This is illustrated in Fig. 1 where the LO diagrams contributing to this process are shown.

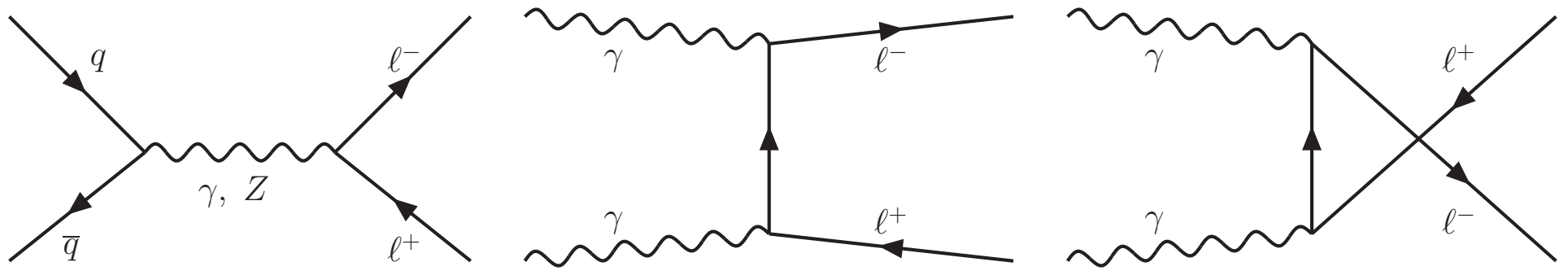

**Fig. 1:** LO diagrams that contribute to lepton-pair production at hadron colliders.

The high invariant-mass distribution of the lepton pair is particularly relevant because this observable is such that the $\gamma\gamma$ contribution, despite the relatively small size of the photon PDF, becomes comparable to that induced by the $q\overline{q}$ channel. As an illustration, the LO prediction for the lepton-pair invariant mass distribution in $e^+e$ production at the 13 TeV LHC is shown in Fig. 2 [14]. This plot indicates that the $\gamma\gamma$ channel becomes increasingly important at large values of the invariant mass and eventually dominates the distribution. Based on simple kinematic considerations, one can show that the high invariant-mass distribution in lepton-pair production probes the photon PDF at relatively large values of Bjorken-$x$, indicatively $x \gtrsim 0.02$.

[*]On behalf of the `xFitter` developer's team.

This observation constitutes a compelling motivation to exploit the precise experimental data produced by the LHC, such as the recent ATLAS data at 8 TeV published in Ref. [9], to constrain the photon PDF in this region. A crucial aspect of this analysis is the consistent inclusion of the relevant EW corrections. As it was shown in Ref. [15], the Drell-Yan process receives sizeable pure weak corrections that almost balance the corrections induced by the photon-initiated channels. Therefore, the inclusion of the NLO EW corrections to the computation of the Drell-Yan cross sections is extremely important to achieve a reliable determination of the photon PDF. This study was carried out within the open-source `xFitter` framework [10] that provides a unique environment to extract PDFs from experimental data.

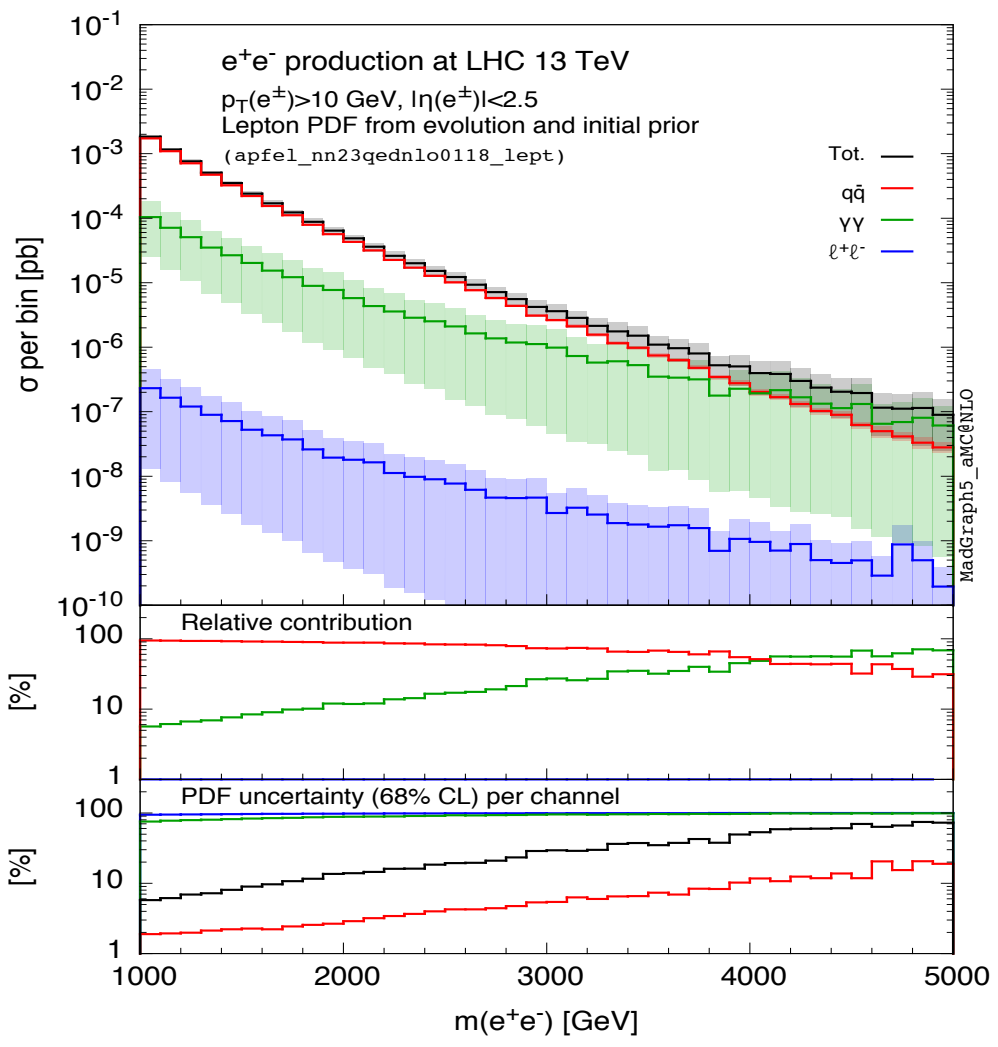

**Fig. 2:** Predictions at LO for the lepton-pair invariant mass distribution in $e^+e$ production at the LHC at 13 TeV [14].

## 2 The dataset

As mentioned above, our determination of the photon PDF relies on the recent ATLAS 8 TeV high-mass Drell-Yan data [9]. Measurements are delivered in three different formats:

1. as single-differential cross-section distributions in the lepton-pair invariant mass $m_{ll}$,
2. as double-differential cross-section distributions in $m_{ll}$ and the rapidity of the lepton pair $|y_{ll}|$,
3. and as double-differential cross-section distributions in $m_{ll}$ and the difference in pseudo-rapidity between the two leptons $\Delta\eta_{ll}$.

In our analysis we have chosen to use the second format that counts 48 data points distributed in 5 $m_{ll}$ bins: [116-150], [150-200], [200-300], [300-500], [500-1500] GeV. The first three (last two) $m_{ll}$ bins are divided into 12 (6) bins in $|y_{ll}|$ extending up to 2.4. The relevant analysis cuts on the data are: $m_{ll} \geq 116$ GeV, $|\eta_{ll}| \leq 2.5$, and $p_T^l \geq 40$ GeV (30) GeV for the leading (sub-leading) lepton.

The ATLAS data alone would clearly be insufficient to carry out an analysis aimed at the extraction of a reliable set of PDFs. Therefore, this data is supplemented by the combined inclusive deep-inelastic scattering (DIS) cross-section data from HERA [16], on which we imposed the cut $Q^2 \geq Q^2_{\rm min} = 7.5$ GeV$^2$. While the ATLAS data is directly sensitive to the photon PDF, the HERA data carries detailed information on the quark and gluon content of the proton. The union of these two datasets allows us to perform a solid determination of a the proton PDFs.

## 3 Electroweak corrections

A central aspect of this analysis is the inclusion of the EW effects. More in particular, we employed predictions accurate to NNLO in QCD and consistently included NLO EW/QED corrections. In the present analysis, this concerns three main sectors which I discuss in turn: the QED corrections to the evolution of PDFs, the QED corrections to the DIS structure functions, and the full EW corrections to the Drell-Yan cross sections.

### 3.1 Evolution

The evolution of PDFs is governed by the DGLAP equations. The DGLAP splitting functions are known up to $\mathcal{O}(\alpha_s^3)$, *i.e.* NNLO in QCD, since long [17,18]. The QED corrections are instead much more recent. The $\mathcal{O}(\alpha)$ corrections, where $\alpha$ is the QED coupling, were derived in Ref. [19], while the $\mathcal{O}(\alpha\alpha_s)$ and $\mathcal{O}(\alpha^2)$, which represent the NLO QED corrections, where computed in Refs. [20, 21].

The implementation of the full NLO QED corrections to the DGLAP evolution was achieved very recently in the APFEL program [11] following the approach of Ref. [14] and documented in Ref. [1]. A cross-check of the implementation was performed using the independent QEDEVOL code [12] based on the QCDNUM evolution program [13].

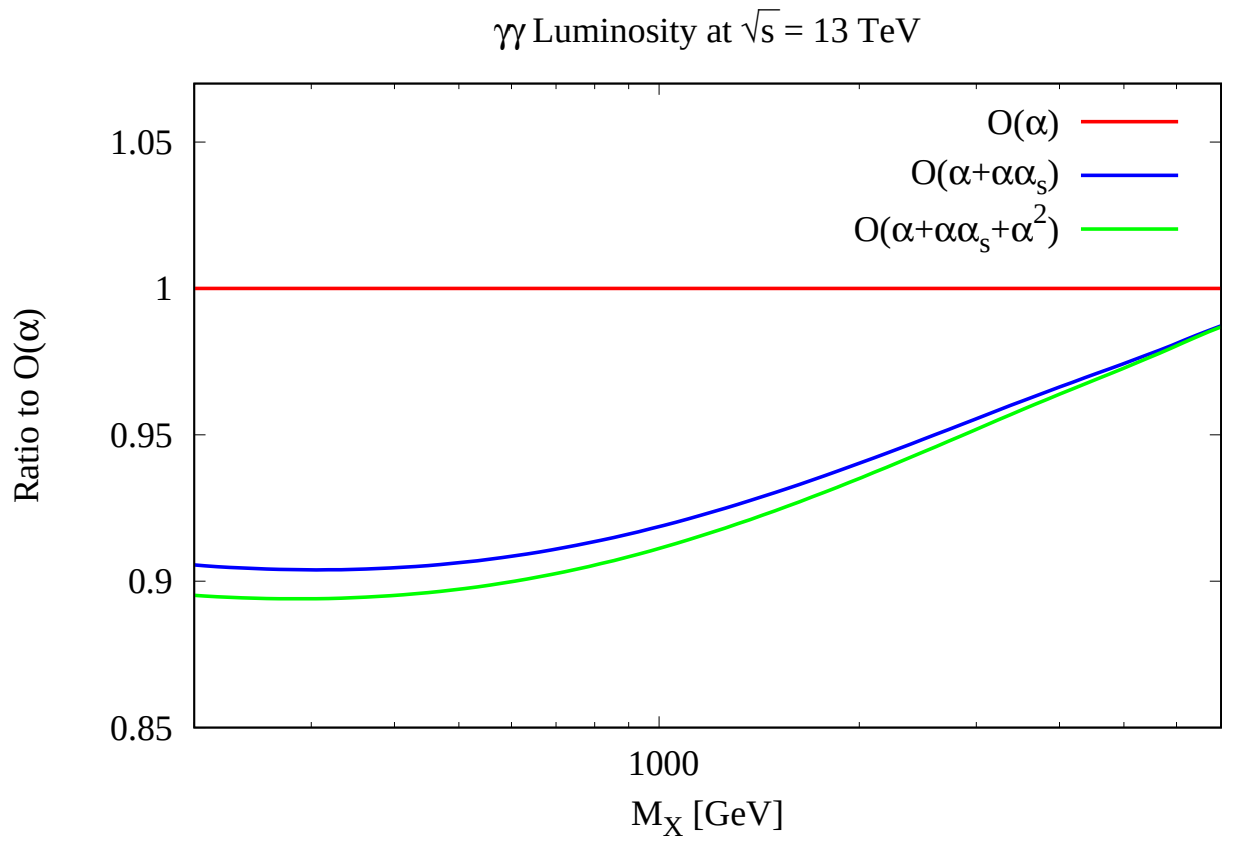

**Fig. 3:** $\gamma\gamma$ luminosity at $\sqrt{s} = 13$ TeV as a function of the final state invariant mass $M_X$ obtained with NNPDF3.0QED NNLO. The curves, taking into account the $\mathcal{O}(\alpha)$, the $\mathcal{O}(\alpha+\alpha_s\alpha)$, and the complete $\mathcal{O}(\alpha+\alpha_s\alpha+\alpha^2)$ effects in the DGLAP evolution, are presented as ratios to the $\mathcal{O}(\alpha)$ result. set.

The effect of the NLO QED corrections on the $\gamma\gamma$ luminosity, relevant to the computation of the Drell-Yan invariant mass distribution, is shown in Fig. 3 as a function of the final state invariant mass $M_X$. The photon PDF taken from the NNPDF3.0QED set is evolved including in the DGLAP equation the $\mathcal{O}(\alpha)$, the $\mathcal{O}(\alpha + \alpha_s\alpha)$, and the complete $\mathcal{O}(\alpha + \alpha_s\alpha + \alpha^2)$ corrections. Results are shown as ratios to the $\mathcal{O}(\alpha)$ curve. While the effect of the $\mathcal{O}(\alpha^2)$ is very mild, the impact of the $\mathcal{O}(\alpha_s\alpha)$ at relatively small values of $M_X$ can be as big as 10%. This reduces to around 3-5% at large invariant masses: this is the region of relevance in our study. This is still a significant effect due to the experimental uncertainty of the ATLAS data and thus it is important to take it into account.

In principle, the NLO QED corrections influence also the running of the QCD coupling $\alpha_s$ and the QED coupling $\alpha$. In fact, they introduce additional mixing terms in the respective $\beta$-functions that couple the evolution of the two couplings. However, it turns out that the impact of the mixing terms is tiny on both couplings and thus we decided not to included them as this would uselessly complicated the implementation.

### 3.2 DIS structure functions

When considering NLO QED corrections to the DIS structure functions, it is necessary to include into the hard cross sections all the $\mathcal{O}(\alpha)$ diagrams. The coefficient functions of these diagrams, being of purely QED origin, can be easily derived from the QCD expressions by properly adjusting the colour factors. This correspondence holds irrespective of whether mass effects are included. This allowed for an easy implementation of the QED corrections to the FONLL general-mass scheme [22]. Specifically, in this work we have used the variant C of the FONLL scheme, accurate to NNLO in QCD, supplemented by the NLO QED corrections.

An intereresting feature of the NLO QED corrections to the DIS structure function is that they introduce photon-initiated diagrams providing a direct handle on the photon PDF. Fig. 4 displays the effect of the NLO QED corrections on the neutral-current structure functions $F_2$, $F_L$, and $xF_3$. The predictions have been obtained including the NLO QED corrections also to the DGLAP evolution and are shown normalised to the pure QCD results. The impact of the NLO QED corrections is very moderate especially at low $x$ but becomes more significant at large $x$, where it is of the order of 2%. The same behaviour is observed also for the charged-current structure functions.

Although the net effect of the NLO QED corrections on the DIS structure functions is small, it is significant when compared to the typical size of the uncertainties of the HERA combined data. This implies that the DIS data, despite very moderately, contributes to constrain the photon PDF in the large-$x$ region.

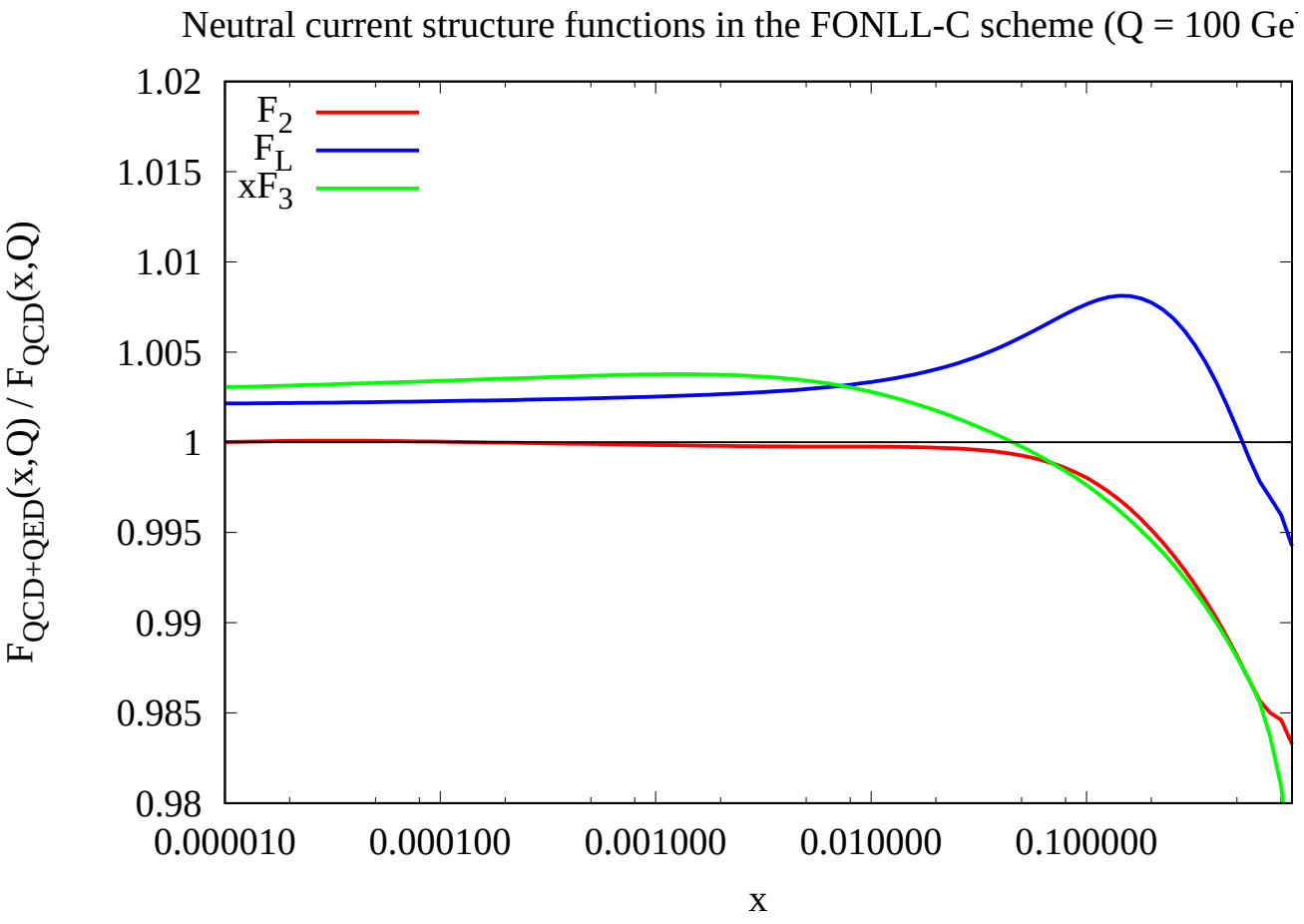

**Fig. 4:** Effects of the NLO QED corrections on the neutral-current DIS structure functions $F_2, F_L$ and $xF_3$ at $Q = 100$ GeV as functions of $x$, normalised to the pure QCD results, obtained in the FONLL-C scheme using NNPDF3.0QED NNLO evolved including the QED corrections discussed above.

### 3.3 Drell-Yan cross sections

For the calculation of the Drell-Yan cross sections at NLO in QCD, we have used the `MadGraph5_aMC@NLO` [23] program interfaced to `APPLgrid` [24] through `aMCfast` [25]. The computation also includes the contribution from the photon-initiated diagrams shown in Fig. 1. Finite mass effects of charm and bottom quarks in the matrix elements are neglected, as appropriate for a high-scale process.

The NLO calculations are supplemented by $K$-factors obtained with the `FEWZ` code [26] to account for the NNLO QCD and the NLO EW corrections. The $K$-factors are defined as:

$$K(m_{ll}, |y_{ll}|) \equiv \frac{\text{NNLO QCD} + \text{NLO EW}}{\text{NLO QCD} + \text{LO EW}}, \tag{1}$$

and computed using the MMHT2014 NNLO [27] PDF set both in the numerator and in the denominator.

Fig. 5 shows the $K$-factors of Eq. (1) as a function of the lepton-pair rapidity $|y_{ll}|$ for each $m_{ll}$ bin. The points correspond to the kinematics of double-differential distributions in $(m_{ll}, |y_{ll}|)$ of the ATLAS high-mass Drell-Yan data included in our analysis.

The $K$-factors vary between 0.98 and 1.04, indicating that higher-order corrections are generally moderate. The trend follows the expectation: the $K$-factors are particularly small at low invariant masses and in the central region, and tend to grow at larger values of $m_{ll}$ and in the forward region where they can be as large ar 4%.

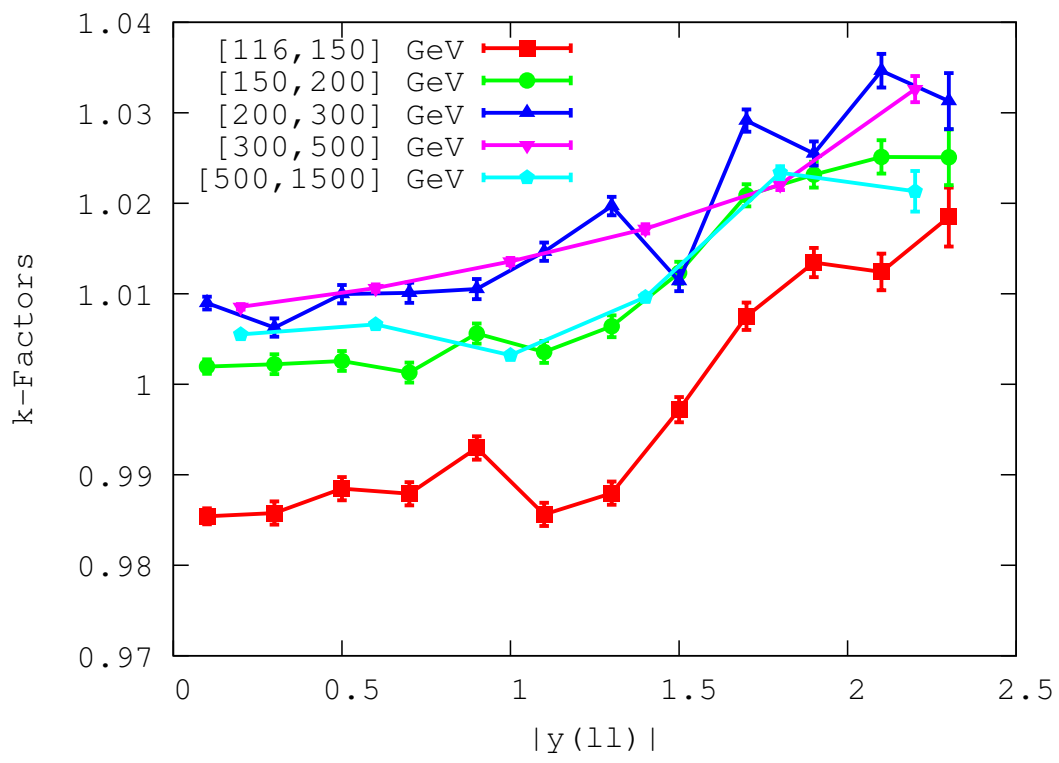

**Fig. 5:** The $K$-factors, defined in Eq. (1), as a function of the lepton-pair rapidity $|y_{ll}|$ for each $m_{ll}$ bin.

## 4 Fit settings

Our determination of the photon PDF, along with quark and gluon PDFs, was carried out in the `xFitter` framework interfaced to the `APFEL` code. The dataset included in our fit was discussed in Sect. 2 and the theory setup presented in Sect. 3. In this section we discuss the fit settings.

We parametrise the following six independent distributions at the initial scale $Q_0$:

$$\begin{aligned} xu(x,Q_0) - x\overline{u}(x,Q_0) \equiv xu_v(x,Q_0) &= A_{u_v} x^{B_{u_v}} (1-x)^{C_{u_v}} (1+E_{u_v}x^2)\,, \\ xd(x,Q_0) - x\overline{d}(x,Q_0) \equiv xd_v(x,Q_0) &= A_{d_v} x^{B_{d_v}} (1-x)^{C_{d_v}}\,, \\ x\overline{u}(x,Q_0) \equiv x\bar{U}(x,Q_0) &= A_{\bar{U}} x^{B_{\bar{U}}} (1-x)^{C_{\bar{U}}}\,, \\ x\overline{d}(x,Q_0) + x\overline{s}(x,Q_0) \equiv x\bar{D}(x,Q_0) &= A_{\bar{D}} x^{B_{\bar{D}}} (1-x)^{C_{\bar{D}}}\,, \\ xg(x,Q_0) &= A_g x^{B_g} (1-x)^{C_g} (1+E_g x^2)\,, \\ x\gamma(x,Q_0) &= A_\gamma x^{B_\gamma} (1-x)^{C_\gamma} (1+D_\gamma x + E_\gamma x^2)\,. \end{aligned} \tag{2}$$

The parameters $B_{\bar{U}}$ and $B_{\bar{D}}$ are set equal so that the quark sea distributions have the same small-$x$ behaviour. Moreover, we assume $x\bar{s} = r_s x\bar{d}$, with $r_s = 1$ [28], and $A_{\bar{U}} = A_{\bar{D}}/2$, such that $x\bar{u} \to x\bar{d}$ for $x \to 0$.

The numerical values of the heavy-quark masses are taken to be $m_c = 1.47$ GeV and $m_b = 4.5$ GeV. The reference values of the QCD and QED couplings are chosen to be $\alpha_s(M_Z) = 0.118$ and $\alpha(m_\tau = 1.777\ \text{GeV}) = 1/133.4$. As for the initial scale, we choose $Q_0 > m_c = \sqrt{7.5}$ GeV such that it is below the scale of all data points included in our fit. This particular value of the initial scale $Q_0$ is peculiar as compared to the typical choice $Q_0 < m_c \simeq 1$ GeV. The reason for choosing a somewhat larger scale is that it helps stabilise the photon PDF. However, in order to still be able to generate the charm PDFs perturbatively without the need to parameterise them, we exploited one of the functionalities of `APFEL` to set charm threshold $\mu_c = Q_0 > m_c$ [29].

## 5 Results

I finally turn to discuss the results of our fit. The partial $\chi^2$'s normalised to the number of data points for the HERA combined data and for the ATLAS high-mass Drell-yan data (bin by bin in $m_{ll}$ and total), as well as the total $\chi^2$ normalised to the number of degrees of freedom, are reported in Tab. 1. The overall fit quality is acceptably good. On the one hand, the description of the HERA data is comparable to that achieved in the HERA-PDF2.0 analysis [16]. On the other hand, despite the small experimental uncertainties, the ATLAS Drell-Yan data is perfectly fitted with a $\chi^2/N_{\text{dat}}$ equal to $48/48$. Remarkably, the single $m_{ll}$ bins of this dataset have all a good $\chi^2$.

| Dataset | $\chi^2/N_{\text{dat}}$ |
|---|---|
| HERA combined DIS data | 1236/1056 |
| ATLAS DY data [116 GeV $\leq m_{ll} \leq$ 150 GeV] | 9/12 |
| ATLAS DY data [150 GeV $\leq m_{ll} \leq$ 200 GeV] | 15/12 |
| ATLAS DY data [200 GeV $\leq m_{ll} \leq$ 300 GeV] | 14/12 |
| ATLAS DY data [300 GeV $\leq m_{ll} \leq$ 500 GeV] | 5/6 |
| ATLAS DY data [500 GeV $\leq m_{ll} \leq$ 1500 GeV] | 4/6 |
| Total ATLAS DY data $\chi^2/N_{\text{dat}}$ | 48/48 |
| Combined HERA I+II and high-mass DY $\chi^2/N_{\text{dof}}$ | 1284/1083 |

**Table 1:** $\chi^2/N_{\text{dat}}$ of the fit to the combined HERA data and the five $m_{ll}$ bins of the ATLAS Drell-Yan data. The global $\chi^2/N_{\text{dof}}$ is also reported, where $N_{\text{dof}}$ is the number of degrees of freedom in the fit.

The photon PDF at $Q^2 = 10^4$ GeV$^2$ obtained from our analysis, that we dubbed `xFitter_epHMDY`, is shown in Fig. 6 and compared to the LUXqed [7], the HKR16 [8], and the NNPDF3.0QED [5] results. The absolute distributions are shown in the lef plot, while they are displayed as ratios to the central value of `xFitter_epHMDY` in the right plot. The uncertainty bands represent the 68% confidence level for all distributions but for HKR16 for which only the central value is made available by the authors. The $x$-range shown Fig. 6 is limited to the region $0.02 \leq x \leq 0.9$ where the ATLAS Drell-Yan data are expected to constrain the photon PDF.

Fig. 6 shows that for $x \geq 0.1$ the four determinations are consistent within PDF uncertainties. For smaller values of $x$, the photon PDFs from LUXqed and HKR16 are lower than `xFitter_epHMDY` but the agreement remains at the 2-$\sigma$ level. A better agreement with the NNPDF3.0QED photon PDF is observed all over the considered range also due to the larger uncertainties. Interestingly, Fig. 6 also shows that for $0.04 \leq x \leq 0.2$ the present analysis exhibits smaller PDF uncertainties as compared to those of NNPDF3.0QED. We conclude that this is the effect of the constraining power of the ATLAS

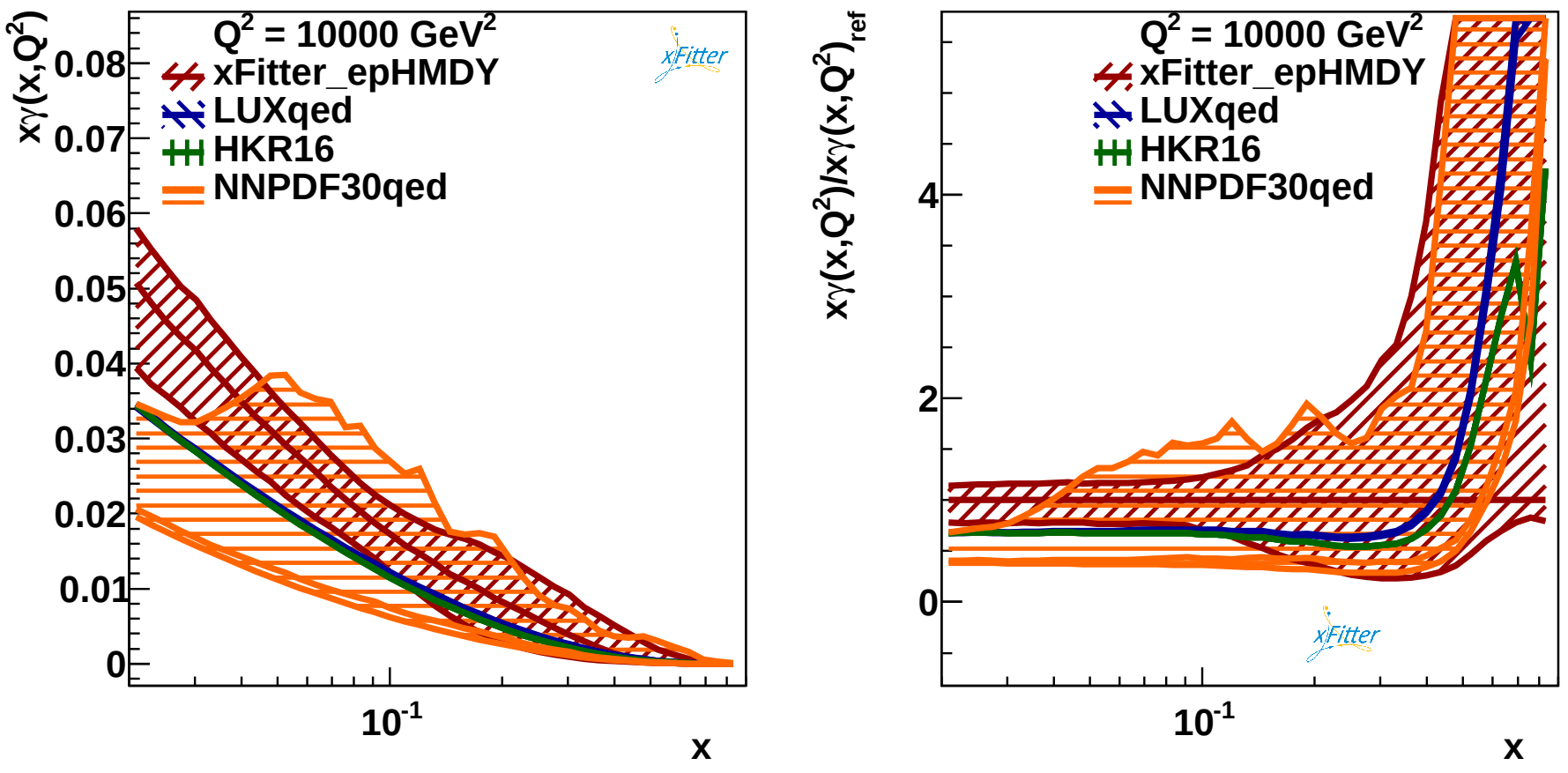

**Fig. 6:** Left plot: comparison between the photon PDF at $Q^2 = 10^4$ GeV$^2$ from the present analysis (`xFitter_epHMDY`) and the results from LUXqed, HKR16, and NNPDF3.0QED. Right plot: same as the left plot but with the distributions normalised to the central value of `xFitter_epHMDY`. The uncertainty bands represent the 68% confidence level. For HKR16 only the central value is available.

Drell-Yan data used in this analysis but not in NNPDF3.0QED. We also observe that this dataset has very little impact of the other PDFs.

Finally, in order to assess the robustness of our fit, we have performed a number of variations with respect to the default settings. Specifically, we considered variations of: the values of the input physical parameters, such as $\alpha_s$, the heavy-quark masses $m_c$ and $m_b$, and the strangeness fraction $r_r$; the PDF parametrisation and the input scale $Q_0$; the cut $Q^2_{\rm min}$ on the scale of the data included in the fit. In all cases, the resulting distributions of a given variation were in agreement, typically well within one standard deviation, with the result obtained with the default settings.

The `xFitter_epHMDY` presented in this work is available in the `LHAPDF6` format [30] upon request from the authors.

## Acknowledgements

I would like to heartily thank Ringaile Plačakytė and Voica Radescu for their invaluable contribution to this work and for their outstanding dedication as conveners of the `HERAFitter-xFitter` project from 2012 until May 2017.

# Ridge Production in High-Multiplicity Hadronic Ultra-Peripheral Proton-Proton Collisions

*Stanley J. Brodsky*
SLAC National Accelerator Laboratory, Stanford University
*Stanislaw D. Glazek*
Faculty of Physics, University of Warsaw
*Alfred S. Goldhaber*
C. N. Yang Institute for Theoretical Physics, Stony Brook University
*Robert W. Brown*
Case Western Reserve University

**Abstract**

An unexpected result at the RHIC and the LHC is the observation that high-multiplicity hadronic events in heavy-ion and proton-proton collisions are distributed as two "ridges", approximately flat in rapidity and opposite in azimuthal angle. We propose that the origin of these events is due to the inelastic collisions of aligned gluonic flux tubes that underly the color confinement of the quarks in each proton. We predict that high-multiplicity hadronic ridges will also be produced in the high energy photon-photon collisions accessible at the LHC in ultra-peripheral proton-proton collisions or at a high energy electron-positron collider. We also note the orientation of the flux tubes between the $q\bar{q}$ of each high energy photon will be correlated with the plane of the scattered proton or lepton. Thus hadron production and ridge formation can be controlled in a novel way at the LHC by observing the azimuthal correlations of the scattering planes of the ultra-peripheral protons with the orientation of the produced ridges. Photon-photon collisions can thus illuminate the fundamental physics underlying the ridge effect and the physics of color confinement in QCD.

## 1 Introduction

One of the striking features of proton-proton collisions at RHIC [1, 2] and the LHC [3–5] is the observation that high multiplicity events are distributed as "ridges" which are approximately flat in rapidity. Two ridges appear, with opposite azimuthal angles, simultaneously reflecting collective multiple particle flow and transverse momentum conservation. This statement follows from the analyses in the quoted references, although it does not appear there explicitly.

Experimental results from PHENIX [1, 2] are illustrated in Fig. 1. Since ridges appear in proton-proton collisions [6] as well as heavy ion collisions, this phenomenon evidently does not require the formation of a quark-gluon plasma. In addition, the high-multiplicity events show an unexpectedly high strangeness content [5].

In a previous publication with J. D. Bjorken, we suggested [7] that the "ridge" correlations reflect the rare events generated by the collision of aligned flux tubes that connect the quark to the diquark in the wave function of the colliding protons. The "spray" of particles resulting from the approximate line source produced in such inelastic collisions then gives rise to events with strong correlations over a large range of both positive and negative rapidity.

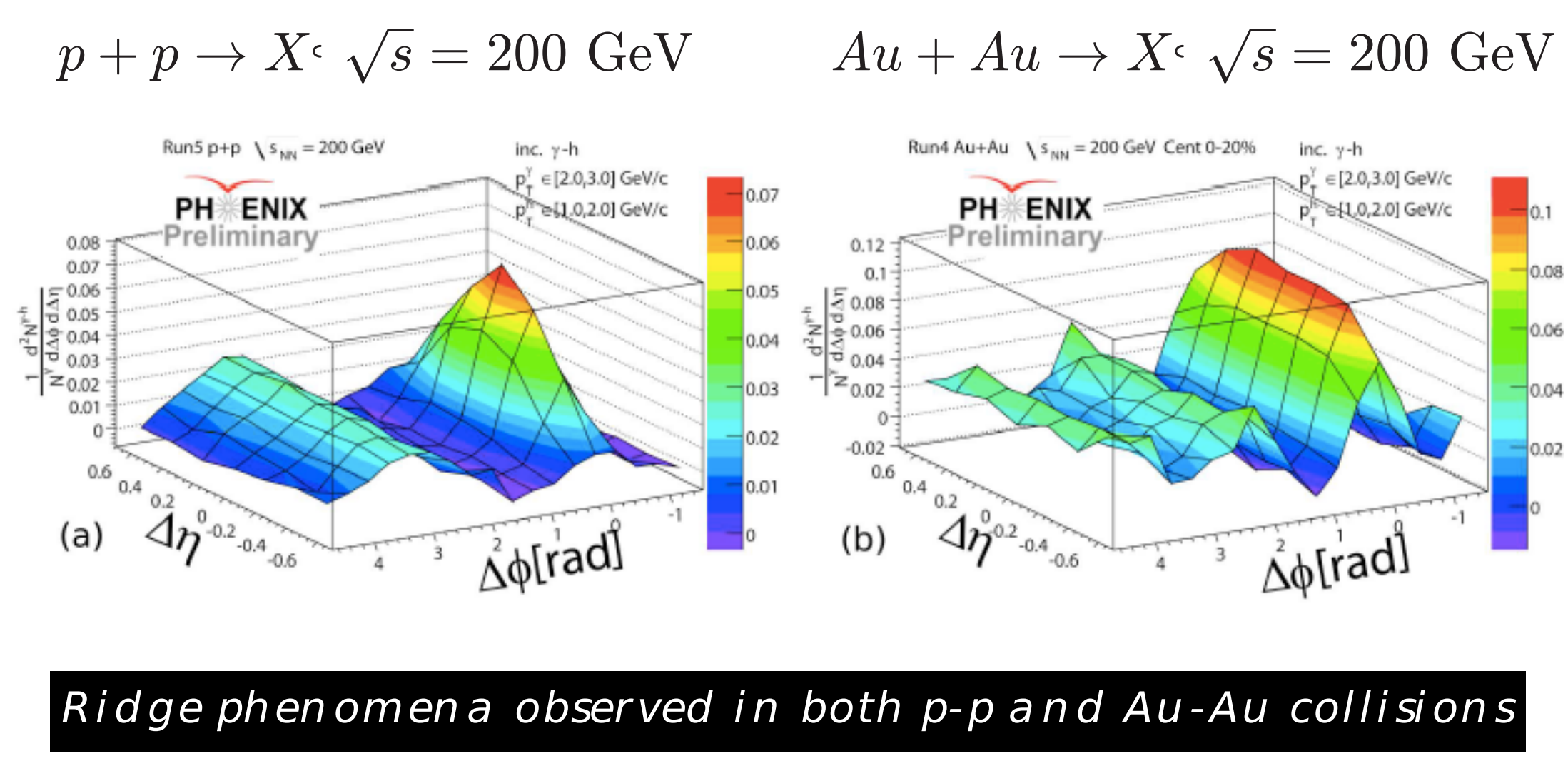

**Fig. 1:** Ridge formation in proton-proton and nucleus-nucleus collisions

Ultra-peripheral proton-proton and heavy ion collisions (UPC) would allow the study of multi-TeV photon-photon interactions at the LHC [8, 9]. Possible photon-photon studies include light-by-light scattering [10], top-pair production via $\gamma\gamma \to t\bar{t}$ processes, electroweak tests such as $W-$ pair production in $\gamma\gamma \to W^+W^-$ events [11], QCD studies, such as hard inclusive and exclusive hadronic reactions, and measurements sensitive to the photon structure function [12–14]. In this report we show that the photon-photon collisions provided by ultra-peripheral proton-proton collisions at the LHC can illuminate the physical QCD mechanisms which underly high mutiplicity hadronic events and ridge formation, including the role of color confinement and gluonic string formation.

In ultra-peripheral proton-proton collisions, each of the virtual photons can couple to a virtual $q\bar{q}$ pair. The quark and antiquark are connected by a flux tube, reflecting color-confining QCD interactions, as illustrated in Fig. 2. One can identify the flux tubes with the string-like network of gluonic interactions which confine color. Such gluonic flux tubes were originally postulated by Isgur and Paton in ref. [15]. The high-energy inelastic collisions of the two flux tubes when they are maximally aligned will then lead to high-multiplicity hadronic events distributed across the rapidity plateau. Moreover, one expects the planes of the ridges to be correlated with the planes of the flux tubes. Thus in a $\gamma - \gamma$ UPC collision, the two overlapping flux tubes can collide and interact (by multi-gluon exchange) to produce the final-state hadrons. The final-state interactions put the system on-shell so that four-momentum is conserved. This is illustrated in Fig. 3.

Photon-photon collisions with aligned flux tubes can also be studied at a high energy electron-ion collider (EIC) or in photon-proton collisions at the proposed LHeC collider, as well as with UPC proton-proton collisions at the LHC.

We come now to an interesting puzzle about the process of forming two coordinated ridges in $pp$ collisions, where both protons suffer momentum transfer along the same axis.

Here are two different scenarios: In each, we will assume that the momentum transfer to each proton produces a quark-antiquark pair aligned with that momentum transfer, with a tube of color-electric flux connecting quark to antiquark.

1. It may be that the eventually produced particles also exhibit transverse momentum along that same direction, resulting from violent motions along both of the parallel tubes.

2. The second possibility is that collision of the two parallel flux tubes causes them to generate particles that emerge out of the long sides of the relatively quiescent tubes, and therefore come out in ridges with a range of transverse momenta extending out of the plane defined by the beam and the momentum transfer.

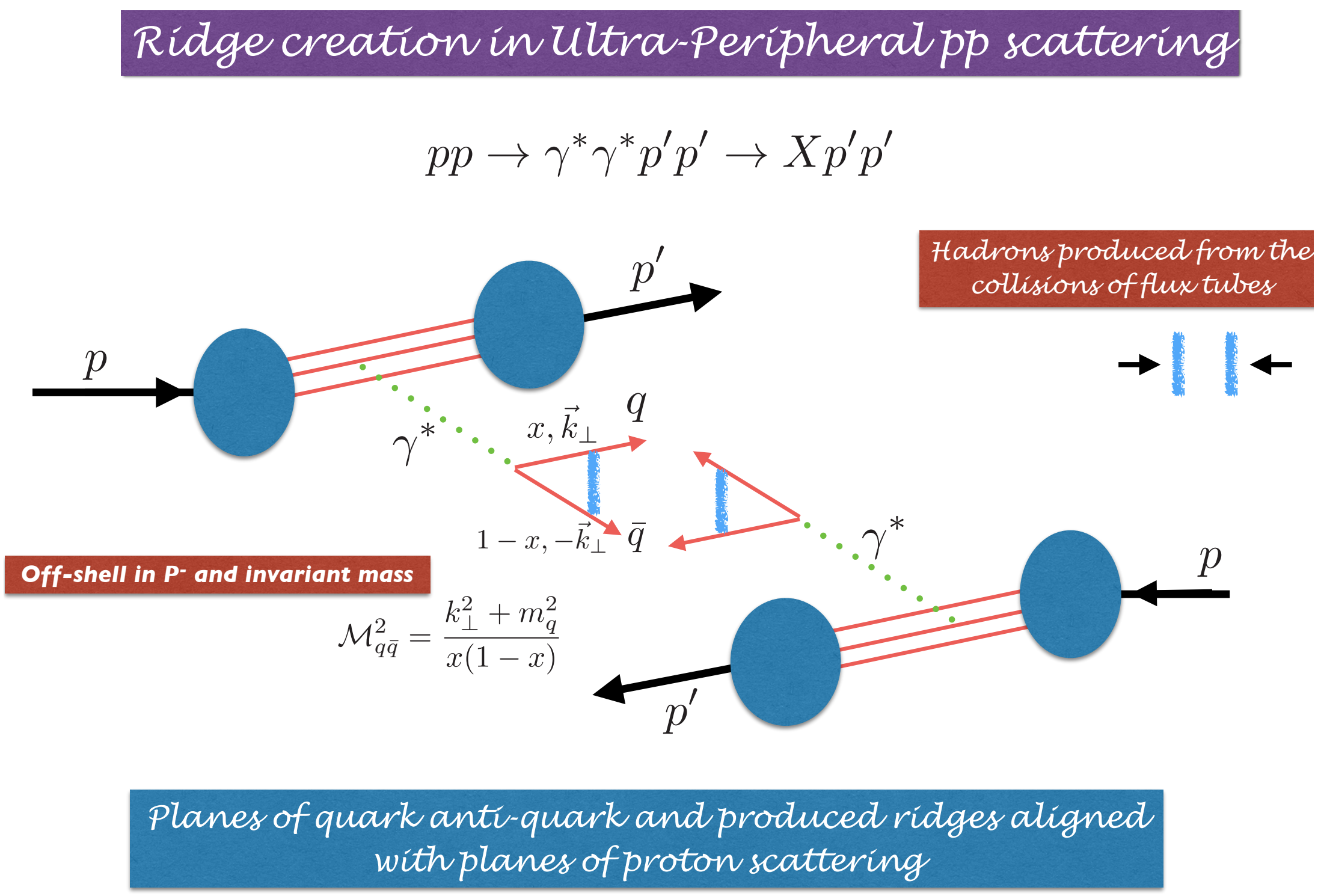

**Fig. 2:** Hadron production from aligned flux tubes in UPC collisions

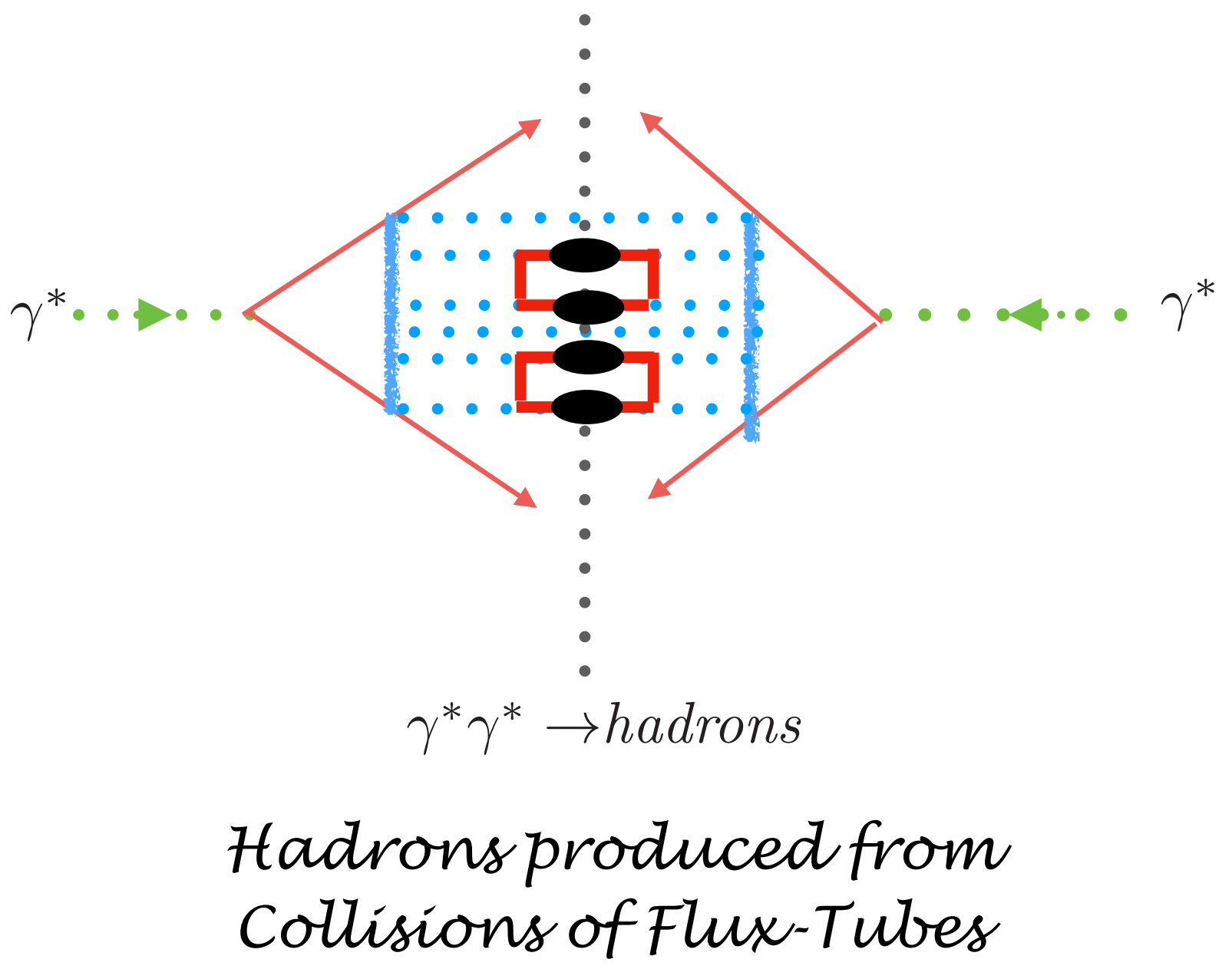

**Fig. 3:** Hadron formation from the collision of flux tubes in two-photon reactions

Thus the first possibility could be called violent and classical, while the second is quiet and coherent, whether hydrodynamical or quantum-mechanical. These two very different possibilities make the search for correlated ridges in $pp$ scattering an extremely interesting experiment.

A light-front wavefunction (LFWF) of a hadron $\psi_H(x_i, \vec{k}_{\perp i}, \lambda_i) =< n|\psi_H >$ for an $n$-parton Fock state is the hadronic eigensolution $|\psi_H >$ of the QCD light-front Hamiltonian $H_{LF}|\psi >= M^2_H|\psi >$ projected on the free parton basis. Here $x_i = k^+_i/P^+ = (k^0 + k^3)/(P^0 + P^3)$ is the boost-invariant LF momentum fraction of constituent $i$, with $\sum^n_{i=1} x_i = 1$. The squares of the LFWFs integrated over transverse momentum underly the hadronic structure functions, and the overlaps of the LFWFs generate the hadronic form factors. A light-front wavefunction is defined at a fixed LF time $\tau = t + z$; it thus can be arbitrarily off-shell in $P^-$ and in invariant mass $\mathcal{M}^2 = P^+P^- - P^2_\perp = \sum_i(\frac{k^2_\perp+m^2}{x})_i$. For example, the pointlike-coupling of a photon in perturbative QED to an intermediate lepton pair $\ell^+\ell^-$ has the form $\psi_{\gamma\to\ell\bar\ell} \propto \sqrt{\alpha}\frac{\vec\epsilon\cdot\vec k_\perp}{\mathcal{M}^2}$, where $\int d^2k_\perp dx|\psi^2| \sim \alpha$. One can study analogous double-lepton-pair formation in UPC collisions $pp \to p'p' + [\ell^+\ell^-] + [\ell^+\ell^-]$ as a check on the basic formalism. For related calculations, see ref. [16]. The coupling of the photon to quark pairs in QCD has both soft and hard contributions. The same couplings contribute to the structure and evolution of the photon structure function [12–14].

We have found that it can be useful to analyze high energy collisions in the "Fool's ISR" frame, where the two incident projectiles both have positive $P^+ = P^0 + P^z$ and nonzero transverse momenta $\pm\vec r^\perp$. The CM energy squared $s = (p_A+p_B)^2 = 4r^2_\perp$ is then carried by the nonzero transverse momenta. For an example, see ref. [17]. This frame choice simplifies factorization analyses for pQCD in the front form since it allows a single light-cone gauge $A^+ = 0$ for both projectiles.

## 2 Origin of Flux Tubes in UPC and $\gamma\gamma$ collisions

We will assume that QCD color confinement creates a gluonic string between the $q$ and the $\bar q$ of the photon. This can be motivated using AdS/QCD, together with LF holography. This formalism has been successful in predicting virtually the entire hadronic spectrum, as well as dynamics such as hadron form factors, and structure functions at an initial nonperturbative scale, as well as the QCD running coupling $\alpha_s(Q^2)$ at all scales [18, 19]. An example of the predicted meson and baryon Regge spectroscopy using superconformal algebra [18] is shown in Fig. 4.

The LF wavefunction $\psi_{q\bar q}(x, \vec k_\perp)$ is the off-shell amplitude connecting the photon to the $q\bar q$ at invariant mass $\mathcal{M}^2 = \frac{k^2_\perp+m^2_q}{x(1-x)}$, where $x = \frac{k^+}{P^+}$ at fixed LF time $\tau$. The $q\bar q$ color-confining frame-independent potential for light quarks derived from AdS/QCD and light-front holography has the form $U(\zeta)^2 = \kappa^4\zeta^2 = \kappa^4 b^2_\perp x(1-x)$ in the light-front Hamiltonian [18]. The color-confining potential that acts between the $q\bar q$ pair for the virtual photon then leads to Gaussian fall-off for the photon's LFWF $\psi_{q\bar q}(x, \vec k_\perp)$ with increasing invariant mass as well as Gaussian fall-off in transverse coordinate space: $\sim e^{-\kappa^2\zeta^2} = e^{-\kappa^2 b^2_\perp x(1-x)}$ as shown in Fig. 5 The same color-confining dynamics implies a string-like flux tube of gluons appearing between the $q$ and $\bar q$. The gluonic flux tube (illustrated as a thick blue line) shown in Fig. 3 represents the network of gluons that connects the quark to the antiquark. In effect, the transverse width of the flux tube is characterized by $b^2_\perp \propto \frac{1}{\kappa^2 x(1-x)}$, where $\kappa \sim 1/2$ GeV is the characteristic mass scale of QCD, and $x$ and $1-x$ are the LF momentum fractions of the $q$ and the $\bar q$. The width of the stringlike flux tube is thus smallest for $x \sim 1/2$ and largest at $x \to 0, 1$; i.e., at large $\mathcal{M}^2_{q\bar q}$.

Note that one is looking at the virtual $q\bar q$ state and its gluonic string at fixed LF time $\tau$. The longitudinal spatial coordinate is $x^- = x^3-x^0$, which is conjugate to the LF momentum $k^+ = k^0+k^3 = xP^+$. Thus the domain $x \to 0$, and large invariant pair mass corresponds to large $x^-$; i.e., large spatial separation between the $q$ and $\bar q$. One can identify the rapidity $y$ of a parton with respect to the parent state as $y = \log x$. The rapidity difference $y_q - y_{\bar q}$ between the quark and antiquark thus grows as $\log x$. The inelastic collision of two flux tubes when they are maximally aligned will then lead to high-multiplicity hadronic events distributed across the rapidity plateau, where the plane of the primary ridge is aligned in azimuthal angle parallel to the aligned flux tubes.

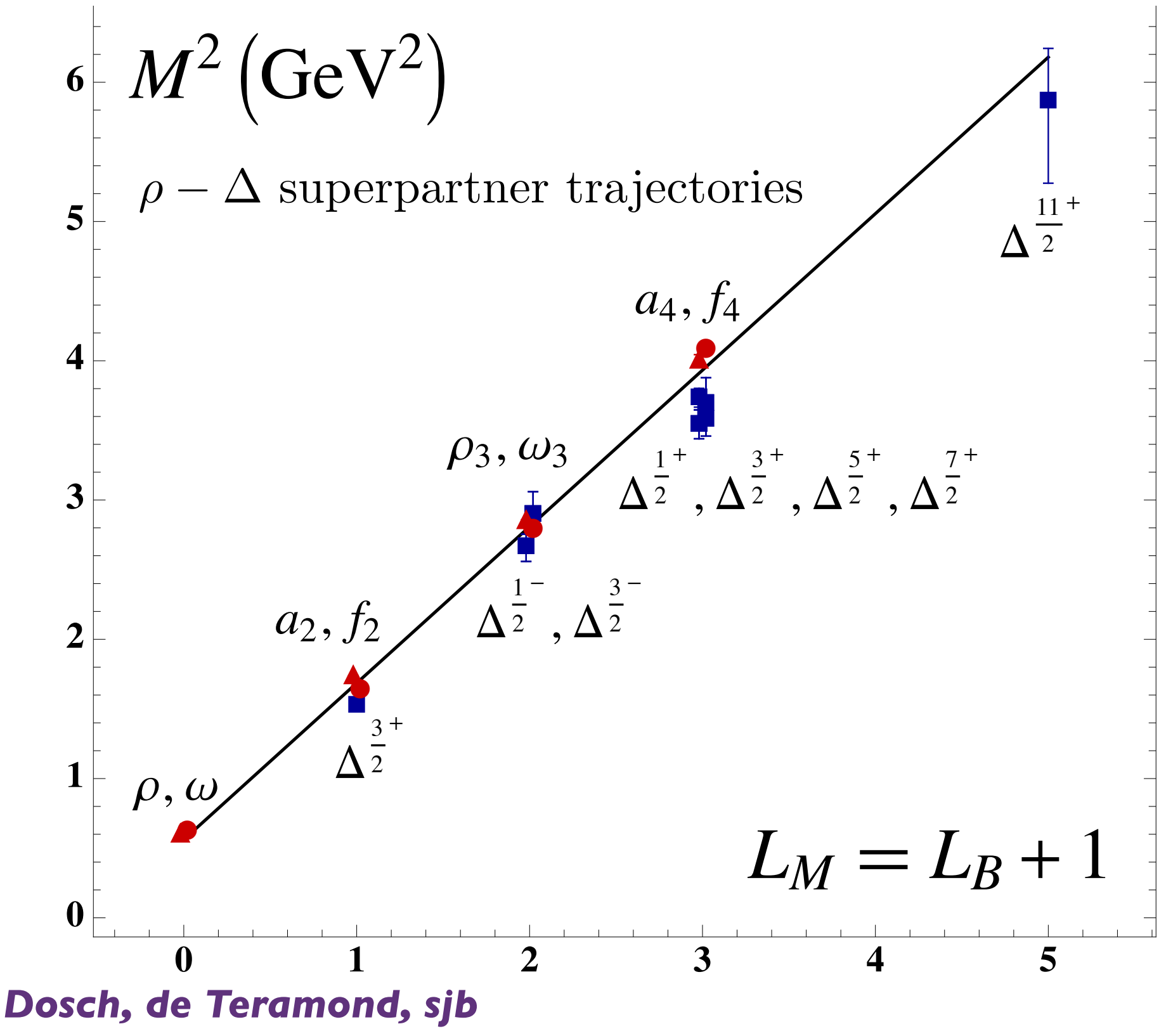

**Fig. 4:** Prediction of meson and baryon Regge spectroscopy from AdS/QCD, light-front holography, and superconformal algebra. The predictions for the meson and baryon mass spectra have the form $M_M^2 = 4\kappa^2(n+L_M)$ for mesons and $M_B^2 = 4\kappa^2(n + L_B + 1)$ for baryons; i.e., universal Regge slopes in the principal quantum number $n$ and orbital angular momentum $L$ for both mesons and baryons. The baryons have a quark plus scalar diquark structure with relative orbital angular momentum $L_B$. Superconformal algebra, together with LF holography, predicts the equality of meson and baryon masses for $L_M = L_B + 1$.

It is thus clear that the maximum number of hadrons will be created when each virtual $q\bar{q}$ pair has maximum $\mathcal{M}^2$ and the gluonic strings are long and maximally aligned; i.e., the collision of long flux tubes. The hadrons in such high multiplicity events will be produced nearly uniformly in rapidity and thus appear as ridges. This description of $\gamma\gamma$ collisions also provides a model for Pomeron exchange between the colliding $q\bar{q}$ systems. It would be interesting to relate this physical picture to the Pomeron and string-based analyses such as that given in Ref. [20]

An important aspect of the UPC events is that the plane of each produced $q\bar{q}$ (and thus the orientation of the flux tube) is correlated with the scattering plane of the parent proton since the virtual photons are transversely polarized to the fermion scattering planes. The correlation to first approximation is proportional to $\cos^2 \Delta\phi$ where $\Delta\phi = \phi_1 - \phi_2$. Since the flux tubes are aligned with the proton scattering planes, one can enhance the probability for high multiplicity hadron production by selecting events where the planes of the scattered UPC protons are parallel. Conversely, one will produce mininum hadron multiplicity if the scattering planes are orthogonal $\Delta\phi = \phi_1 - \phi_2 \simeq \pi/2$. The coupling of the highly virtual photons to strange and charm $q\bar{q}$ pairs as well as the composition of the flux tubes themselves can lead to enhanced charm and strange hadron multiplicity. We also note that in addition to the hadrons produced by the collision of the flux tubes, the $q\bar{q}$ pairs can also interact with each other by gluon exchange and produce up to four near-forward quark jets of various flavors and combinations. Such jet patterns [21] could provide additional information on the physics arising from the collisions of gluonic strings spanned between quarks.

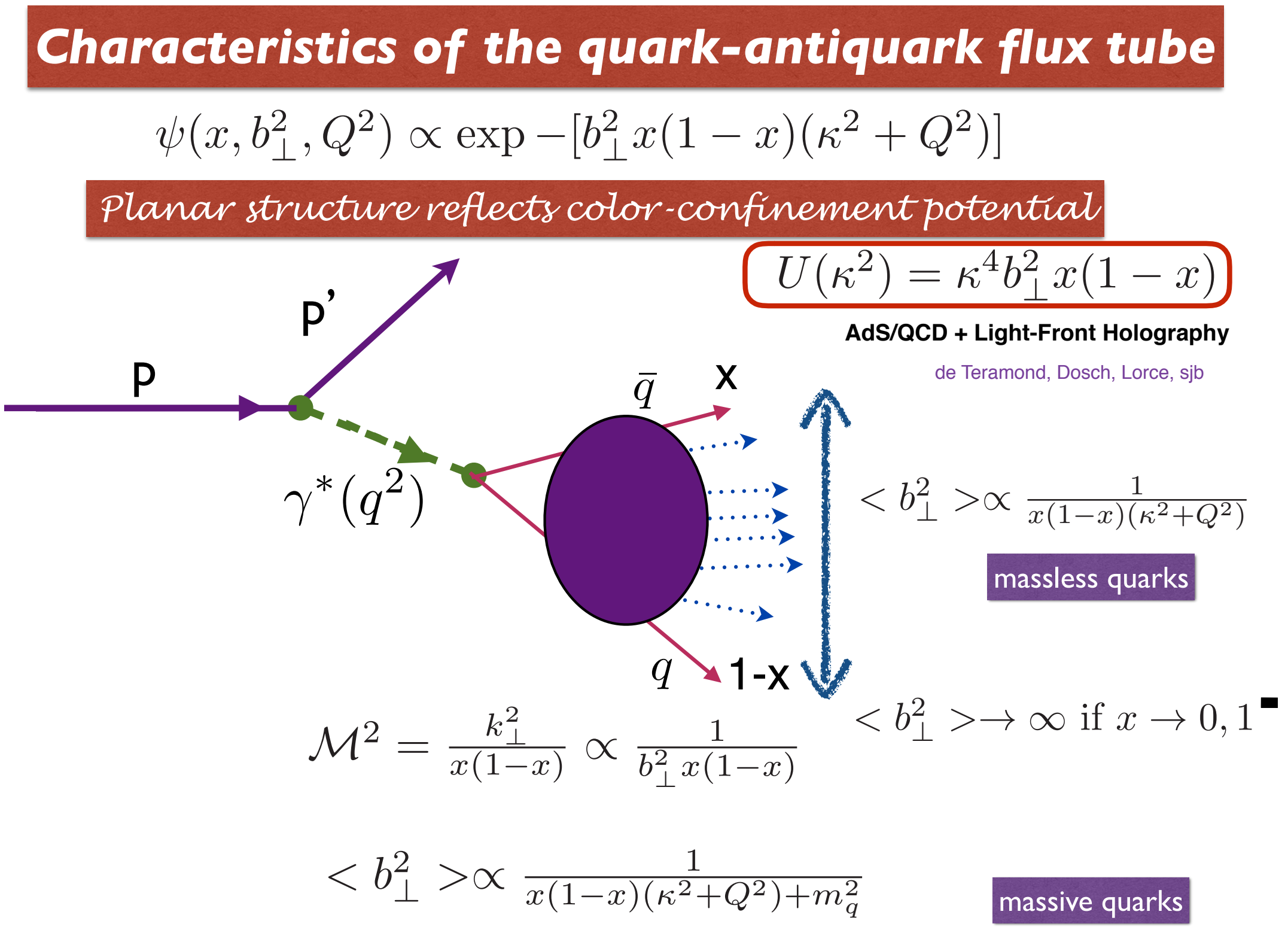

**Fig. 5:** Origin of the gluonic flux-tube based on the color-confining light-front potential derived from AdS/QCD and light-front holography.

One complication which we are currently investigating is whether the collision of the flux tubes itself can affect the orientation of the incoming $q\bar{q}$ planes and thus dilute the predicted alignment between the colliding gluonic strings. One expects that the rotation of the $q\bar{q}$ plane will be important when the total mass of the produced hadronic system is comparable to the $q\bar{q}$ invariant mass. However, the prediction that minimal hadron multiplicity will be produced when the scattering planes of the UPC protons are perpendicular would not be affected.

One can think of the initial configuration shown in Fig. 2 as similar to the configuration one has for initial distributions in pQCD factorization, such as Drell-Yan lepton pair production. The initial configuration can however be modified by the collision itself. This is analogous to the initial-state scattering in lepton-pair production that produces the Sivers single-spin correlation [22] or the double- -Mulders effect [23].

In the case of the proton, AdS/QCD predicts a color flux tube which combines two quarks into a $\bar{3}_C$ diquark system, plus a flux tube that connects the remaining $3_C$ quark to the flux tube of the diquark system. See Fig. 6. The configuration of flux tubes in the proton is a special case of the $Y$ configuration discussed in ref. [24]. The activation of both the $q[qq]$ and $[qq]$ flux tubes in a proton-proton collision could thus lead to both $v_2$ and $v_3$ correlations in the distributions of the final-state hadrons. In contrast, the UPC photon-photon collisions would only lead to a $v_2$ correlation from the activation of the $[q\bar{q}]$ flux tubes. The dependence of the distributions of high multiplicity events on proton structure is discussed in ref. [24]. One could also study the interactions of flux tubes in $\gamma p$ collisions using a single UPC proton at the LHC. See Fig. 7. The oriented flux tube of the photon generated by the single UPC proton can interact with either of the two flux tubes within the proton quark-diquark LFWF to produce high multiplicity hadronic events. The hadrons will tend to be distributed with $v_2$ or $v_3$ moments depending on the details of the collision. The produced ridges of hadrons will in this case tend to be oriented with the scattering plane of the UPC proton.

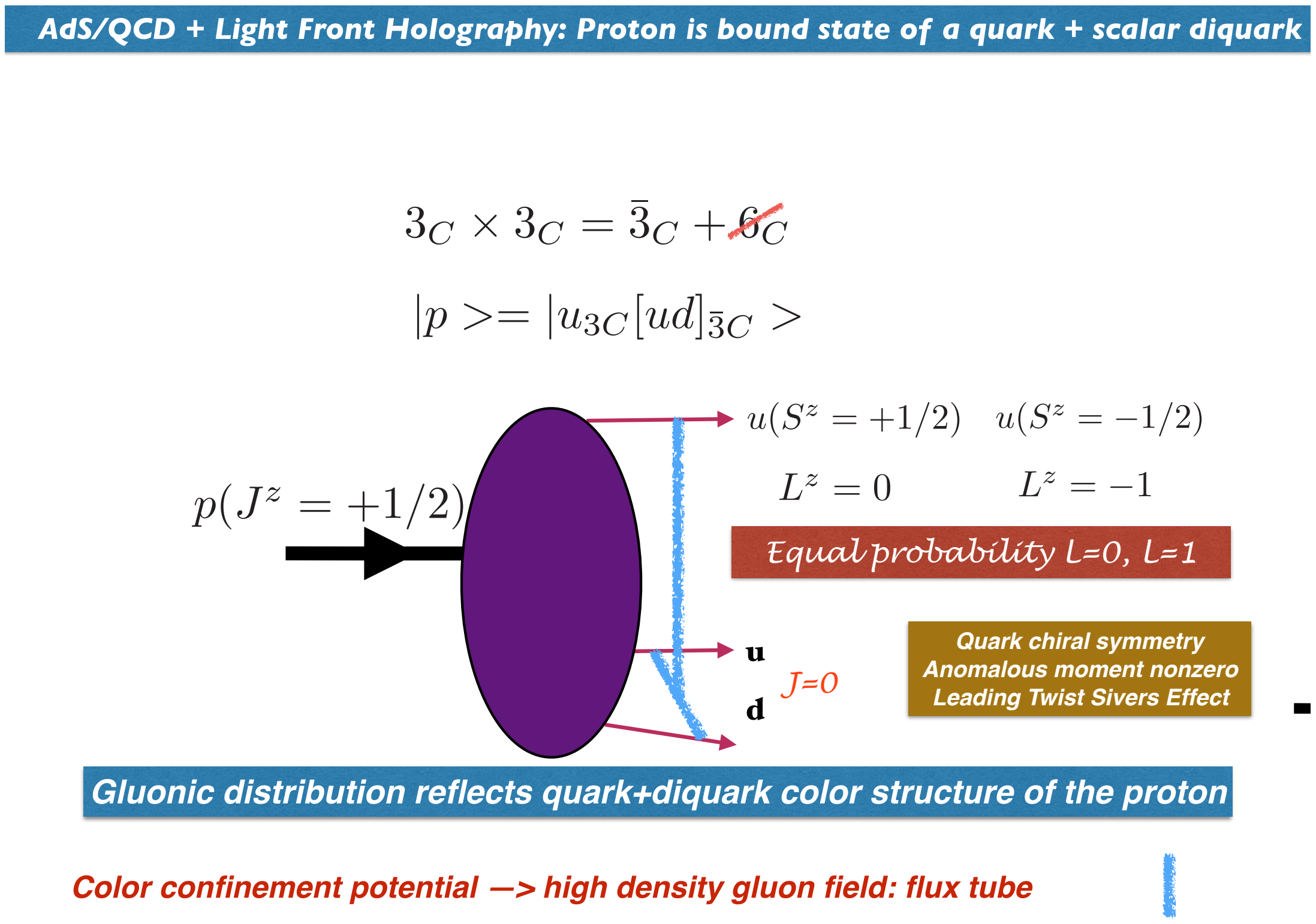

**Fig. 6:** Quark-diquark configuration of baryons predicted by AdS/QCD, light-front holography, and superconformal algebra. One predicts a color flux tube connecting the $3_C$ quark to the $\bar{3}_C$ diquark and a second flux tube within the spin-zero diquark.

## 3 Acknowledgements

This research was supported by the U. S. Department of Energy, contract DE–AC02–76SF00515. SLAC-PUB-17106.

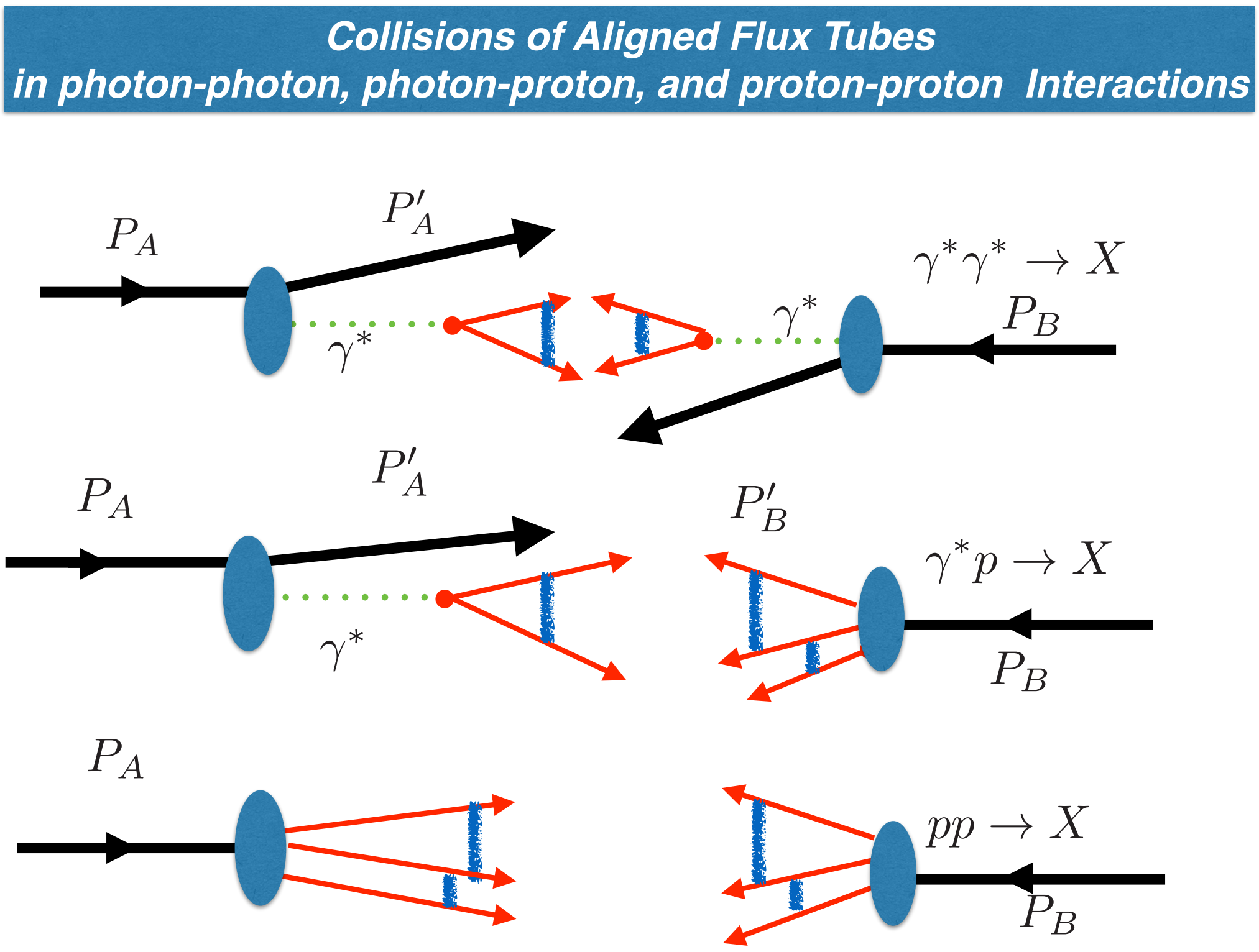

**Fig. 7:** Collisions of flux tubes in $\gamma - \gamma, \gamma - p$ and $p - p$ reactions at the LHC

# Photon-proton and photon-nucleus measurements in CMS

*Ruchi Chudasama and Dipanwita Dutta on behalf of the CMS collaboration*
Bhabha Atomic Research Centre, Mumbai, India

**Abstract**
The CMS measurements of exclusive photoproduction of $\Upsilon$ in pPb and $J/\psi$ in PbPb collisions, that probe the low-$x$ gluon density in the proton and the lead nucleus respectively, are discussed.

## 1 Introduction

The exclusive photoproduction of quarkonia in $\gamma$-proton and $\gamma$-nucleus interactions from ultraperipheral proton-nucleus and nucleus-nucleus collisions at high-energies [1], where one of the incoming hadrons emits a photon that interacts with the "target" proton/nucleus via a colour-singlet gluon exchange and materializes into a $\Upsilon$ or $J/\psi$ meson, provides a powerful tool to directly probe the gluon density inside the proton/nucleus [2].

The central feature of the CMS detector is a superconducting solenoid that provides a magnetic field of 3.8 T, required to bend the charged particle's trajectory and measure its momentum accurately. Within the solenoid volume are a silicon pixel and strip tracker, electromagnetic calorimeter (ECAL), hadron calorimeter (HCAL), each composed of a barrel and two endcap sections. Muons are measured in gas-ionization detectors embedded in the steel flux-return yoke outside the solenoid over the range $|\eta| < 2.4$. Two Hadron Forward (HF) calorimeters cover $2.9 < |\eta| < 5.2$, and two zero degree calorimeters (ZDC) are sensitive to neutrons and photons with $|\eta| > 8.3$. The beam scintillator counters (BSCs) are plastic scintillators that partially cover the face of the HF calorimeters over the range $3.9 < |\eta| < 4.4$. A more detailed description of the CMS detector can be found in Ref. [3].

## 2 Exclusive photoproduction of $\Upsilon$ in pPb collisions at $\sqrt{s_{NN}} = 5.02$ TeV

In Ultra-Peripheral Collisions (UPCs), the interactions are mainly photon-induced and strong interactions are largely suppressed. Since the photon flux scales with the square of the charge ($Z^2$) of the emitting particle, UPC are strongly enhanced for Pb compared to proton. In exclusive quarkonia photoproduction processes, the photon emitted by one of the accelerated charges (electron, proton or ion) fluctuates into $q\bar{q}$ bound pair (vector meson) and interacts with the other "target" proton or ion through a color-singlet gluon exchange. The corresponding photoproduction cross-section is thereby proportional to the square of the gluon density inside the "target".

Exclusive $\Upsilon$ photoproduction was first observed at HERA [4–6] and has recently been studied at the LHC by the LHCb experiment [7] in pp UPCs at $\sqrt{s} = 7$ and 8 TeV. CMS has carried out a similar measurement in p-Pb collisions, including the $\Upsilon$(1S) cross section as a function of the photon-proton center-of-mass energy, $W_{\gamma p}$, in the interval 91–826 GeV, corresponding to the $\Upsilon$ rapidity, y $< |2.2|$, and Bjorken-$x$ values of the order $x \sim 10^{-4}$ to $x \sim 1.3 \cdot 10^{-2}$. The differential cross-section for $\Upsilon$(nS) states as a function of transverse momentum squared, $p_T^2$, has been measured, where $p_T^2 \approx |t|$ is the four-momentum transfer at the proton vertex. At low values of $|t|$, the cross-section can be parameterized as $e^{-b|t|}$, where $b$ provides also information on the transverse density profile of the proton.

The measurement of exclusive $\Upsilon$ photoproduction in ultra-peripheral p-Pb collisions at $\sqrt{s_{NN}} = 5.02$ TeV has been presented in [8] corresponding to an integrated luminosity of $L_{\text{int}} = 32.6$ nb$^{-1}$ collected by the CMS experiment. The STARLIGHT [9] MC event generator was used to simulate exclusive $\Upsilon$(nS) photoproduction events, Fig. 1 (left), and the elastic QED background, Fig. 1 (right). The $\Upsilon$(nS) states are studied in their dimuon decay channel. The UPC events were selected with a dedicated trigger, which requires at least one muon in each event and at least one to six tracks. In order to reduce the muon inefficiencies at low $p_{\mathrm{T}}$ , muons with $p_{\mathrm{T}} > 3.3$ GeV and pseudo-rapidity $|\eta| < 2.2$ are selected. Dimuons are selected in the invariant mass range 9.1–10.6 GeV. Exclusive $\Upsilon$ candidates are selected by requiring only one vertex and no extra charged particles with $p_{\mathrm{T}} > 0.1$ GeV in the event. The $p_{\mathrm{T}}$ of the dimuon is restricted from 0.1 to 1 GeV, to reduce the contamination from elastic QED and inelastic background contributions.

The dominant background contribution to the exclusive $\Upsilon$ signal originates from QED, $\gamma\gamma \rightarrow \mu^+\mu^-$, which was estimated by STARLIGHT. The absolute prediction of QED was checked by comparing the data between invariant mass regions 8–9.12 and 10.64–12 GeV for dimuon $p_{\mathrm{T}} < 0.15$ GeV to the simulation. The contribution of non-exclusive background (inclusive $\Upsilon$, Drell-Yan and proton dissociation) was estimated by a data-driven method by loosening the exclusivity cuts. A background template is build with events with more than 2 tracks. This template was normalized to the exclusive sample in the region of dimuon $p_T > 1.5$ GeV to estimate the data-driven background. A small additional background originates from exclusive $\gamma$Pb $\rightarrow \Upsilon$ Pb events. The fraction of these events in the total number of exclusive $\Upsilon$ events was estimated using a reweighted STARLIGHT $\Upsilon$ MC sample. These backgrounds were subtracted from data to extract the exclusive signal. The background subtracted $|t|$ and y distributions were used to measure the $b$ parameter, and to estimate the exclusive $\Upsilon$ photoproduction cross-section as a function of $W_{\gamma p}$, respectively. The distributions were first unfolded in the region $0.01 < |t| < 1$ GeV$^2$, $|y| < 2.2$, and muon $p_{\mathrm{T}}^{\mu} > 3.3$ GeV, using the iterative Bayesian unfolding technique, and were further extrapolated to transverse momenta of zero using acceptance correction factors.

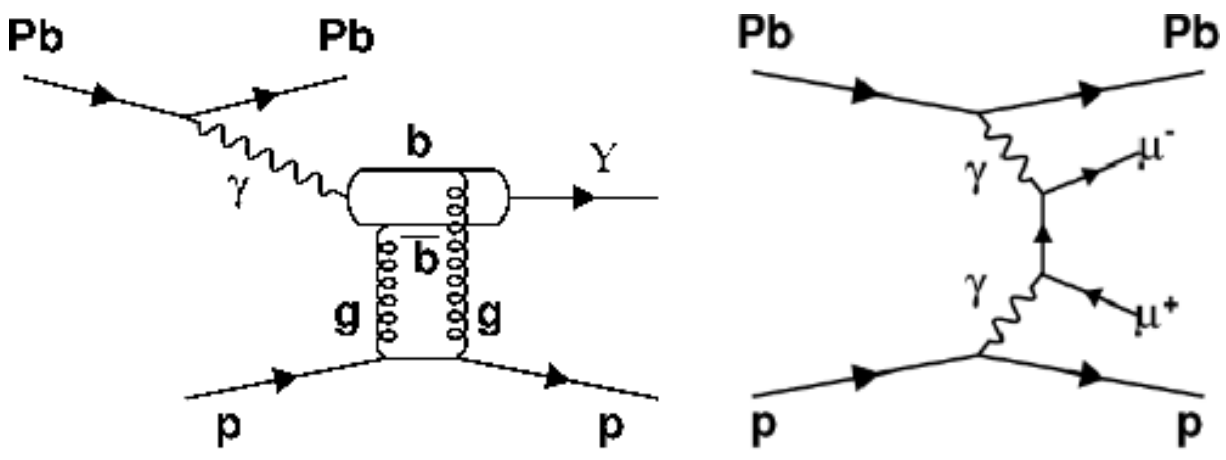

**Fig. 1:** Diagrams representing exclusive $\Upsilon$ photoproduction (left), and exclusive dimuon QED continuum (right) in pPb collisions [8].

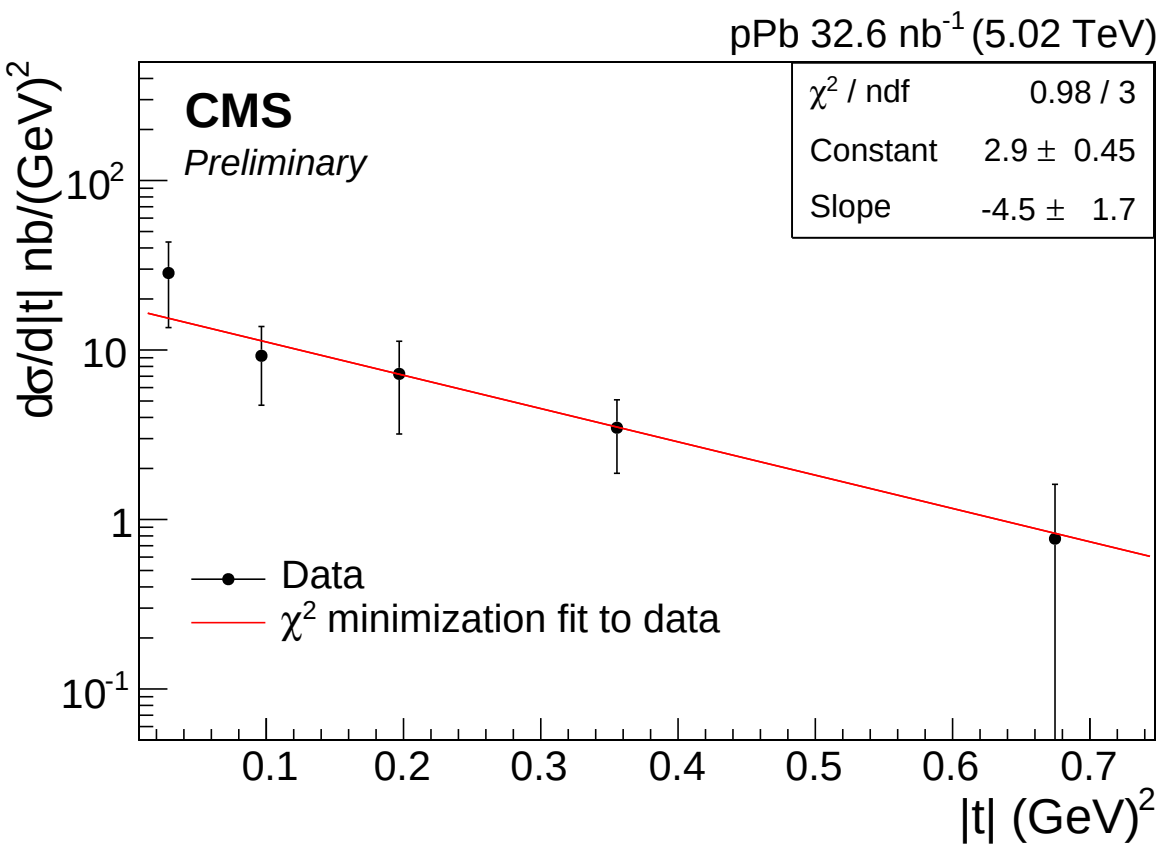

**Fig. 2:** Differential $\Upsilon$(nS) photoproduction cross section as a function of $|t|$. The solid line represents the result of a fit with an exponential function $Ne^{-b|t|}$ [8].

The differential $d\sigma/dt$ cross section is shown in Fig. 2 and extracted for the combined three $\Upsilon$(nS) states according to the following formula:

$$\frac{\mathrm{d}\sigma_{\Upsilon}}{\mathrm{d}|t|} = \frac{N^{\Upsilon(\mathrm{nS})}}{\mathcal{L} \times \Delta|t|}, \qquad (1)$$

where $|t|$ is approximated by the dimuon transverse momentum squared $p_{\mathrm{T}}^2$, $N^{\Upsilon(\mathrm{nS})}$ denotes the yield of background-subtracted, unfolded and acceptance-corrected signal events in each $|t|$ bin, $\mathcal{L}$ is the integrated luminosity, and $\Delta|t|$ is the width of each $|t|$ bin. The cross section is fitted with an exponential function $N\, e^{-b|t|}$ in the region $0.01 < |t| < 1.0$ GeV$^2$, using an unbinned $\chi^2$ minimization method. A value of $b = 4.5 \pm 1.7$ (stat) $\pm$ 0.6 (syst) GeV$^{-2}$ is extracted from the fit. This result is in agreement with the value $b = 4.3^{+2.0}_{-1.3}$ (stat) measured by the ZEUS experiment [10] for the photon-proton center-of-mass energy $60 < W_{\gamma\mathrm{p}} < 220$ GeV, and with the predictions based on pQCD models [11].

The differential $\Upsilon$(1S) photoproduction cross section, $\mathrm{d}\sigma/\mathrm{d}y$, is extracted in four bins of dimuon rapidity according to:

$$\frac{d\sigma_{\Upsilon(1S)}}{dy} = \frac{f_{\Upsilon(1S)}}{\mathcal{B}(1+f_{FD})} \frac{N^{\Upsilon(nS)}}{\mathcal{L} \times \Delta y}, \quad (2)$$

where $N^{\Upsilon(nS)}$ denotes the background-subtracted, unfolded and acceptance-corrected number of signal events in each rapidity bin. The factor $f_{\Upsilon(1S)}$ describes the ratio of $\Upsilon(1S)$ to $\Upsilon(nS)$ events, $f_{FD}$ is the feed-down contribution to the $\Upsilon(1S)$ events originating from the $\Upsilon(2S) \to \Upsilon(1S) + X$ decays (where $X = \pi^+\pi^-$ or $\pi^0\pi^0$), $\mathcal{B} = (2.48 \pm 0.05)\%$ is the branching ratio for muonic $\Upsilon(1S)$ decays, and $\Delta y$ is the width of the $y$ bin. The $f_{\Upsilon(1S)}$ fraction is used from the results of the inclusive $\Upsilon$ analysis [12] at CMS. The feed-down contribution of $\Upsilon(2S)$ decaying to $\Upsilon(1S)+\pi^+\pi^-$ and $\Upsilon(1S)+\pi^0\pi^0$ was estimated as 15% from STARLIGHT. The contribution of feed-down from exclusive $\chi_b$ states was neglected, as these double-pomeron processes are expected to be comparatively much suppressed in proton-nucleus collisions [13, 14].

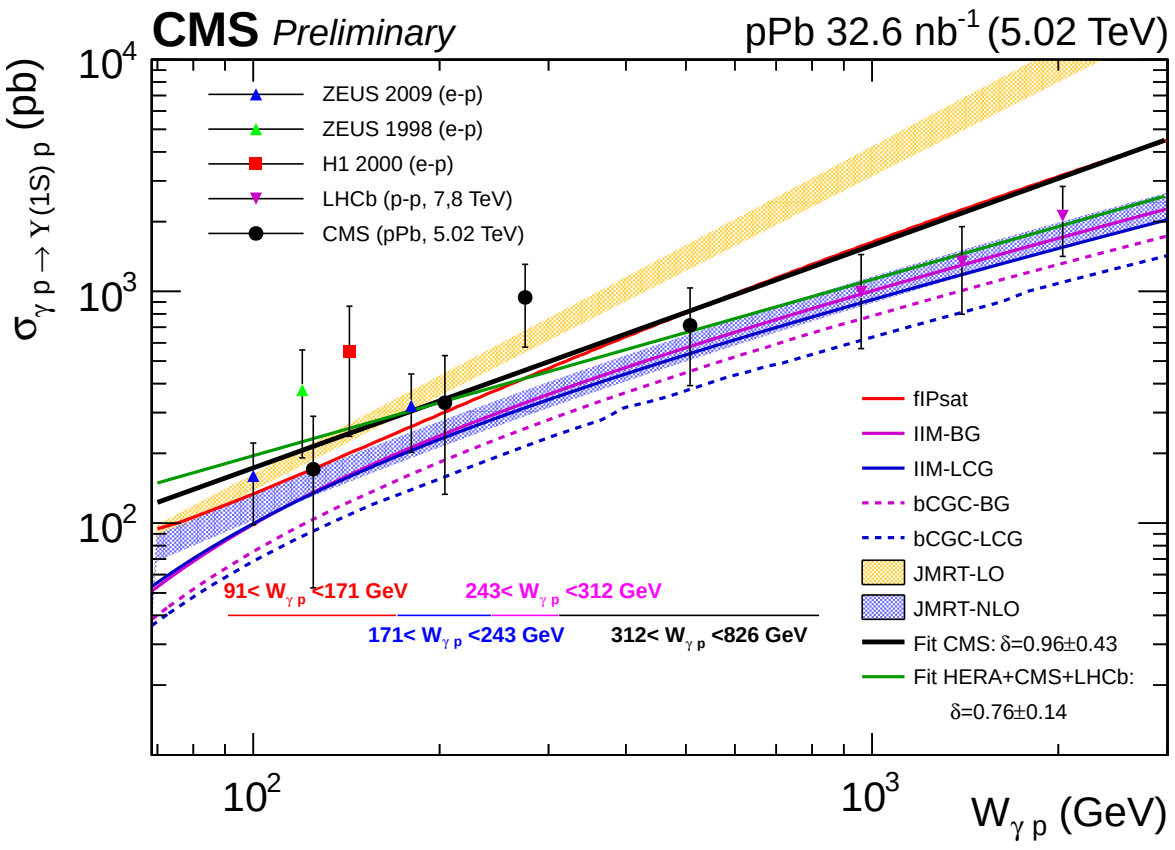

**Fig. 3:** Cross section for exclusive $\Upsilon(1S)$ photoproduction, $\gamma p \to \Upsilon(1S)p$ as a function of photon-proton center-of-mass energy, $W_{\gamma p}$ [8].

The exclusive $\Upsilon(1S)$ photoproduction cross section as a function of $W_{\gamma p}$ shown in Fig. 3, is obtained by using,

$$\sigma_{\gamma p \to \Upsilon(1S)p}(W^2_{\gamma p}) = \frac{1}{\Phi} \frac{d\sigma_{\Upsilon(1S)}}{dy}, \quad (3)$$

where $\Phi$ is the photon flux evaluated at the mean of the rapidity bin, estimated from STARLIGHT. The CMS data are plotted together with the previous measurements from H1 [4], ZEUS [5, 6] and LHCb [7] data. It is also compared with different theoretical predictions of the JMRT model [11], factorized IPsat model [15, 16], IIM [17, 18] and bCGC model [19]. As $\sigma(W_{\gamma p})$ is proportional to the square of the gluon PDF of the proton and the gluon distribution at low Bjorken $x$ is well described by a power law, the cross section will also follow a power law. Any deviation from such trend would indicate a different behavior of the gluon density function. We fit a power law $A \times (W/400)^{\delta}$ with CMS data alone that gives $\delta = 0.96 \pm 0.43$ and $A = 655 \pm 196$, and is shown by the black solid line. The extracted $\delta$ value is comparable to the value $\delta = 1.2 \pm 0.8$, obtained by ZEUS [5]. Our data are compatible with a power-law dependence of $\sigma(W_{\gamma p})$ and disfavor a faster increase with energy as predicted by LO pQCD approaches.

## 3 Coherent photoproduction of $J/\psi$ in PbPb collisions at $\sqrt{s_{NN}} = 2.76$ TeV

CMS has also carried out a measurement of the coherent $J/\psi$ photoproduction in ultra-peripheral PbPb collisions at $\sqrt{s_{NN}} = 2.76$ TeV [20], in a data sample corresponding to an integrated luminosity of $L_{\text{int}} = 159\ \mu b^{-1}$. In PbPb UPC events, vector mesons are produced in $\gamma$Pb interactions off one of the nuclei and thus, the gluon distribution inside the Pb ion can be probed for low values of Bjorken x, of the order of $x \sim 10^{-5}$ to $x \sim 2 \cdot 10^{-2}$. The STAR and PHENIX collaborations at RHIC have studied $\rho^0$ and $J/\psi$ photoproduction in ultra-peripheral AuAu collisions at $\sqrt{s_{NN}} = 200$ GeV [21, 22]. The measurement of coherent photoproduction of the $J/\psi$ meson has been performed at the LHC by the ALICE collaboration [23] in ultra-peripheral PbPb collisions at $\sqrt{s_{NN}} = 2.76$ TeV. The results provided by the ALICE collaboration have been used to compute the nuclear suppression factor, and it provides the evidence that the nuclear gluon density is below that expected for a simple superposition of protons and neutrons in the nucleus [23].

The exclusive $J/\psi$ photoproduction can be classified as coherent if the photon interacts with the whole nucleus, leaving the nucleus intact. In incoherent interactions, the photon interacts with a single nucleon, and the nucleus breaks apart. The STARLIGHT [9] MC event generator was used to simulate coherent and incoherent $J/\psi$ photoproduction events and the elastic QED background. The $J/\psi$ candidates are reconstructed through the dimuon decay channel in the rapidity interval $1.8 < |y| < 2.3$ . The UPC events were selected with a dedicated trigger, which requires an energy deposit consistent with at least one neutron in either of the ZDCs; a low signal in at least one of the BSC+ or BSC- scintillators; the presence of at least one single muon without a $p_{\mathrm{T}}$ threshold requirement, and at least one track in the pixel detector. The coherent $J/\psi$ cross section is measured for the case when the $J/\psi$ mesons are accompanied by at least one neutron on one side of the interaction point and no neutron activity on the other side ($\mathrm{X_n0_n}$), to reject the non-UPC events. In addition to the ZDC requirements, the exclusive $J/\psi$ events are selected by requiring exactly two muon tracks within the phase space region $1.2 < |\eta| < 2.4$ and $1.2 < p_{\mathrm{T}} < 1.8$ GeV and no activity above noise threshold, 3.85 GeV, in the HF detectors. Dimuon candidates with $p_{\mathrm{T}} < 1.0$ GeV within the rapidity interval $1.8 < |y| < 2.3$ and with invariant mass between 2.6–3.5 GeV are considered.

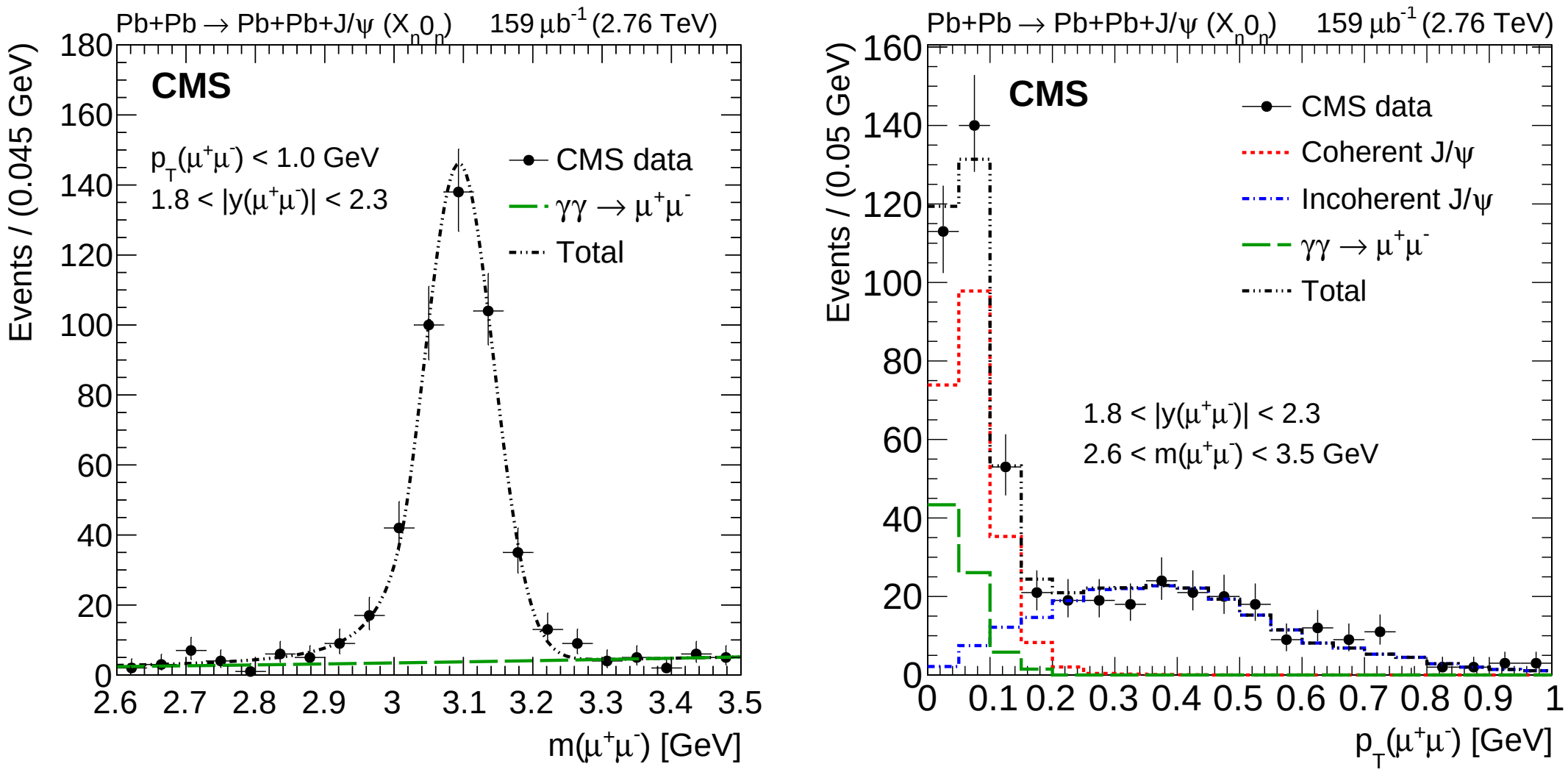

**Fig. 4:** Results from the simultaneous fit to the dimuon invariant mass (left) and $p_{\mathrm{T}}$ (right) for the $\mathrm{X_n0_n}$ break-up mode, after all selections are applied [20].

The dimuon invariant mass (Fig. 4, left) and $p_{\mathrm{T}}$ (Fig. 4, right) distributions satisfying the selection criteria are simultaneously fitted in order to extract the yield of coherent $J/\psi$ , incoherent $J/\psi$ , and $\gamma\gamma \to \mu^+\mu^-$ events. The fit uses a maximum likelihood algorithm that takes unbinned projections of the data in invariant mass and $p_{\mathrm{T}}$ as inputs. The shapes of the $p_{\mathrm{T}}$ distributions for these three processes are determined from a STARLIGHT simulation. The fit yields $207 \pm 18$ (stat) for the coherent $J/\psi$ candidates, $75 \pm 13$ (stat) for the incoherent $J/\psi$ candidates, and $75 \pm 13$ (stat) for $\gamma\gamma$ events with dimuon $p_{\mathrm{T}} < 0.15$ GeV in the rapidity interval $1.8 < |y| < 2.3$.

The coherent $J/\psi$ photoproduction cross-section in the $\mathrm{X_n0_n}$ break-up mode is given by the following formula,

$$\frac{d\sigma^{coh}_{X_\mathrm{n}0_\mathrm{n}}}{dy} = \frac{N^{coh}_{X_\mathrm{n}0_\mathrm{n}}}{\mathcal{B}(1+f_{\mathrm{FD}})\mathcal{L} \times \Delta y (A\epsilon)^{J/\psi}} \quad (4)$$

where $\mathcal{B} = 5.96 \pm 0.03$ (syst)% is the branching fraction of $J/\psi$ to dimuons, $\mathrm{N}^{coh}_{X_\mathrm{n}0_\mathrm{n}}$ is the coherent $J/\psi$ yield for $p_{\mathrm{T}} < 0.15$ GeV, $\mathcal{L} = 159 \pm 8$ (syst) $\mu\mathrm{b}^{-1}$ is the integrated luminosity, $\Delta$ y = 1 is the rapidity bin width, and $(A\epsilon)^{J/\psi}$ = 5.9 ± 0.5 (syst)% is the combined acceptance times efficiency correction factor.

The coherent $J/\psi$ yield is given by,

$$N^{coh}_{X_n0_n} = \frac{N_{yield}}{1 + f_{FD}} \quad (5)$$

where $N_{yield}$ is the coherent $J/\psi$ yield as extracted from the fit, and $f_{FD}$ is the fraction of $J/\psi$ mesons coming from coherent $\psi(2S) \rightarrow J/\psi$ + anything, estimated from STARLIGHT to be 0.018 ± 0.011 (theo). The resulting $J/\psi$ yield is, $N^{coh}_{X_n0_n} = 203 \pm 18$ (stat). Thus, the coherent $J/\psi$ photoproduction cross section for prompt $J/\psi$ mesons in the $X_n0_n$ break-up mode is $d\sigma^{coh}_{X_n0_n} = 0.36 \pm 0.04$ (stat) ± 0.04 (syst) mb. The $d\sigma^{coh}_{X_n0_n}$ cross-section was measured for the $X_n0_n$ break-up mode, and is scaled to the total cross-section by correcting it with a scale factor 5.1 ± 0.5 (theo), estimated with STARLIGHT [9]. After applying this scaling factor we obtain the total coherent $J/\psi$ photoproduction cross section $d\sigma^{coh} = 1.82$ ± 0.22 (stat) ± 0.20 (syst) ± 0.19 (theo) mb.

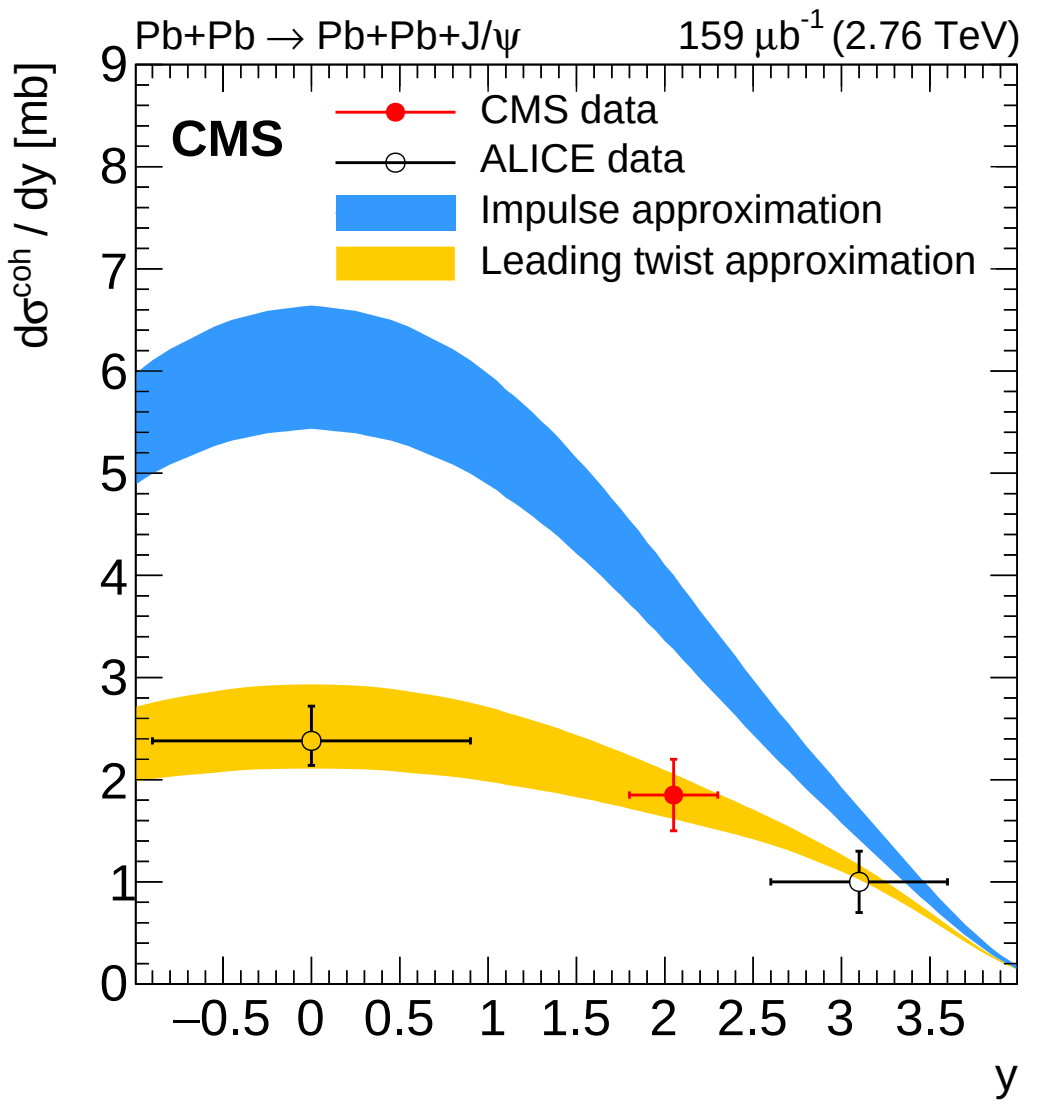

**Fig. 5:** Differential cross section as a function of rapidity for coherent $J/\psi$ photoproduction [20].

The coherent $J/\psi$ photoproduction cross section is shown in Fig. 5, compared to recent ALICE measurements [23] and to the theoretical predictions by Guzey *et al.* [24, 25] based on the impulse and leading twist approximations. The data from ALICE and CMS show a steady decrease with rapidity. The leading twist approximation prediction obtained from Ref. [25] is in good agreement with the data, while an impulse approximation model prediction is strongly disfavored, indicating that nuclear effects expected to be present at low $x$ and $Q^2$ values are needed to describe the data. The prediction given by the leading twist approximation, which includes nuclear gluon shadowing, is consistent with the data.

## 4 Conclusions

Exclusive photoproduction of $\Upsilon$ in pPb at 5.02 TeV and $J/\psi$ in PbPb at 2.76 TeV collisions have been measured by the CMS experiment at the LHC. Both measurements provide new constraints on the poorly-known gluon density at low $x$ in the proton and the lead nucleus respectively.

# Prospects for Photon-Photon and Photon-Proton Measurements with Forward Proton Taggers in ATLAS

*M. Trzebiński, on behalf of the ATLAS Collaboration*
Institute of Nuclear Physics Polish Academy of Sciences, Cracow, Poland

**Abstract**
ATLAS forward detectors, ALFA and AFP, are described with a focus on their geometric acceptance. The main results of the total, elastic and inelastic cross sections measurements performed with the ATLAS/ALFA detectors are presented. Possibility of diffractive bremsstrahlung and exclusive pion-pair photoproduction studies with the data from ATLAS/ALFA are discussed. Finally, the advantage of using a proton tagging technique for a high-statistics data recorded by the AFP detectors in the precision measurements of anomalous gauge couplings and invisible objects is discussed.

## 1 Introduction

In the majority of events of photon-photon and photon-proton scatterings at the LHC one or both outgoing protons stay intact. Since photon is a colourless object, such an exchange results in a presence of the rapidity gap between the centrally produced system and scattered protons. Thus, such events are of diffractive nature.

Diffractive studies are one of the important parts of the physics programme for the LHC experiments. This is also true for ATLAS, where a large community works on both phenomenological and experimental aspects of diffraction. In this report, the results of the total and elastic cross-sections measurements from the ATLAS experiment are presented. Moreover, predictions for new diffractive measurements, such as exclusive di-pion production, bremsstrahlung and anomalous gauge couplings, are summarized.

## 2 Forward Detectors

Diffractive production could be recognised by a search for a rapidity gap in the forward direction or by measuring forward protons. In this report, the proton tagging technique will be discussed.

Diffractive protons are usually scattered at very small angles (hundreds of microradians). In order to measure them, special devices that allow for the detector movement (so-called Roman pots, RP) are commonly used. In ATLAS [1] two systems of such detectors were installed: ALFA [2, 3] and AFP [4].

The ALFA (Absolute Luminosity For ATLAS) set-up consists of four detector stations placed symmetrically with respect to the ATLAS Interaction Point (IP) at 237 m and 245 m. Each ALFA station contains two Roman pot devices allowing vertical movement of the detectors. The spatial resolution of the ALFA detectors is of about 30 $\mu$m in the horizontal ($x$) and vertical ($y$) direction.

Since 2016 a new set of forward detectors is also installed far away form the ATLAS Interaction Point – the ATLAS Forward Proton (AFP). These detectors are placed symmetrically with respect to the ATLAS IP at 204 m and 217 m. Stations located closer to the IP contain the tracking detectors, whereas the further ones are equipped with tracking and timing devices. The reconstruction resolution of tracking detectors is estimated to be of 10 and 30 $\mu$m in $x$ and $y$, respectively. The precision of Time-of-Flight measurement is expected to be of about 20 ps.

## 3 Geometric Acceptance

There are several LHC machine running configurations at which the ALFA and AFP detectors could take data. In the simplest possible way they could be characterized by the value of the betatron function at the Interaction Point, $\beta^*$. For simplicity, one can put them into two categories: a standard ($\beta^* < 1$ m) and a special high-$\beta^*$ ($\beta^* \gg 1$ m) optics. The details of these optics are described in Ref. [5], while here only the key features are presented.

The standard optics is a typical setting for all LHC high-luminosity runs – the beam is strongly focused at the IP and the non-zero value of the crossing angle is introduced in order to avoid proton collisions outside the IP region. The high-$\beta^*$ (90, 1000 and 2500 m) optics was developed in order to measure the properties of the elastic scattering. Due to the high value of the betatron function, the beam angular divergence is very small and the beam is not as strongly focused as in the case of the standard optics. In these settings the value of the crossing angle could be either zero or non-zero.

Not all scattered protons can be measured in the forward detectors. A proton can be too close to the beam to be detected or it can hit the LHC accelerator components (collimator, beam pipe, magnet) upstream the AFP or ALFA station. The geometric acceptance is defined as the ratio of the number of protons of a given relative energy loss ($\xi = 1 - E_{proton}/E_{beam}$) and transverse momentum ($p_{\mathrm{T}}$) that reached the detector station to the total number of scattered protons with the same $\xi$ and $p_{\mathrm{T}}$. Since AFP (ALFA) is intended to operate mostly during standard (special) optics settings, only these cases are shown in Fig. 1. In the calculations, the beam properties at the IP, the beam pipe geometry, the LHC lattice magnetic properties and the distance between the beam centre and the detector edge were taken into account. The distance from the beam centre was set to 15 $\sigma$ for the standard optics, and to 10 $\sigma$ for the high-$\beta^*$ ones, where $\sigma$ is the beam size at the location of the detector station.

In order to account for the dead material of the Roman pot window, a 0.3 mm distance was added in all cases. The acceptance of ALFA and AFP detectors at various optics are complementary. For the AFP run with the standard optics the region of high acceptance (black area, >80%) is limited to $p_{\mathrm{T}} < 3$ GeV and $0.02 < \xi < 0.12$). For the ALFA detectors and high-$\beta^*$ optics the acceptance starts from $\xi = 0$ as these settings are optimised for the elastic scattering measurement.

## 4 Elastic Scattering and Total Cross Section Measurement

The elastic scattering process has the simplest signature that can be imagined: two protons exchange their momentum and are scattered at small angles.

The measurements described in this section were done using the following data samples collected by ATLAS: 80 $\mu\mathrm{b}^{-1}$ at $\sqrt{s} = 7$ TeV and 500 $\mu\mathrm{b}^{-1}$ at $\sqrt{s} = 8$ TeV, both taken with $\beta^* = 90$ m. The detailed description of these analyses can be found in [6] (7 TeV) and [7] (8 TeV) and here only the main results are presented.

The measured elastic cross-sections and the nuclear slope parameters, $B$, are:

$$\sigma_{\mathrm{el}}^{\mathrm{ALFA}}(7\ \mathrm{TeV}) = 24.00 \pm 0.19\ (\mathrm{stat.}) \pm 0.57\ (\mathrm{syst.})\ \mathrm{mb},$$

$$\sigma_{\mathrm{el}}^{\mathrm{ALFA}}(8\ \mathrm{TeV}) = 24.33 \pm 0.04\ (\mathrm{stat.}) \pm 0.39\ (\mathrm{syst.})\ \mathrm{mb},$$

$$B_{\mathrm{nucl}}^{\mathrm{ALFA}}(7\ \mathrm{TeV}) = 19.73 \pm 0.14\ (\mathrm{stat.}) \pm 0.26\ (\mathrm{syst.})\ \mathrm{GeV}^{-2},$$

$$B_{\mathrm{nucl}}^{\mathrm{ALFA}}(8\ \mathrm{TeV}) = 19.74 \pm 0.05\ (\mathrm{stat.}) \pm 0.23\ (\mathrm{syst.})\ \mathrm{GeV}^{-2},$$

where the first error is statistical and the second accounts for all experimental systematic uncertainties, from which the largest one is due to the luminosity uncertainty.

By using the optical theorem [8], the total cross section was determined:

$$\sigma_{\mathrm{tot}}^{\mathrm{ALFA}}(7\ \mathrm{TeV}) = 95.35 \pm 0.38\ (\mathrm{stat.}) \pm 1.25\ (\mathrm{exp.}) \pm 0.37\ (\mathrm{extr.})\ \mathrm{mb},$$

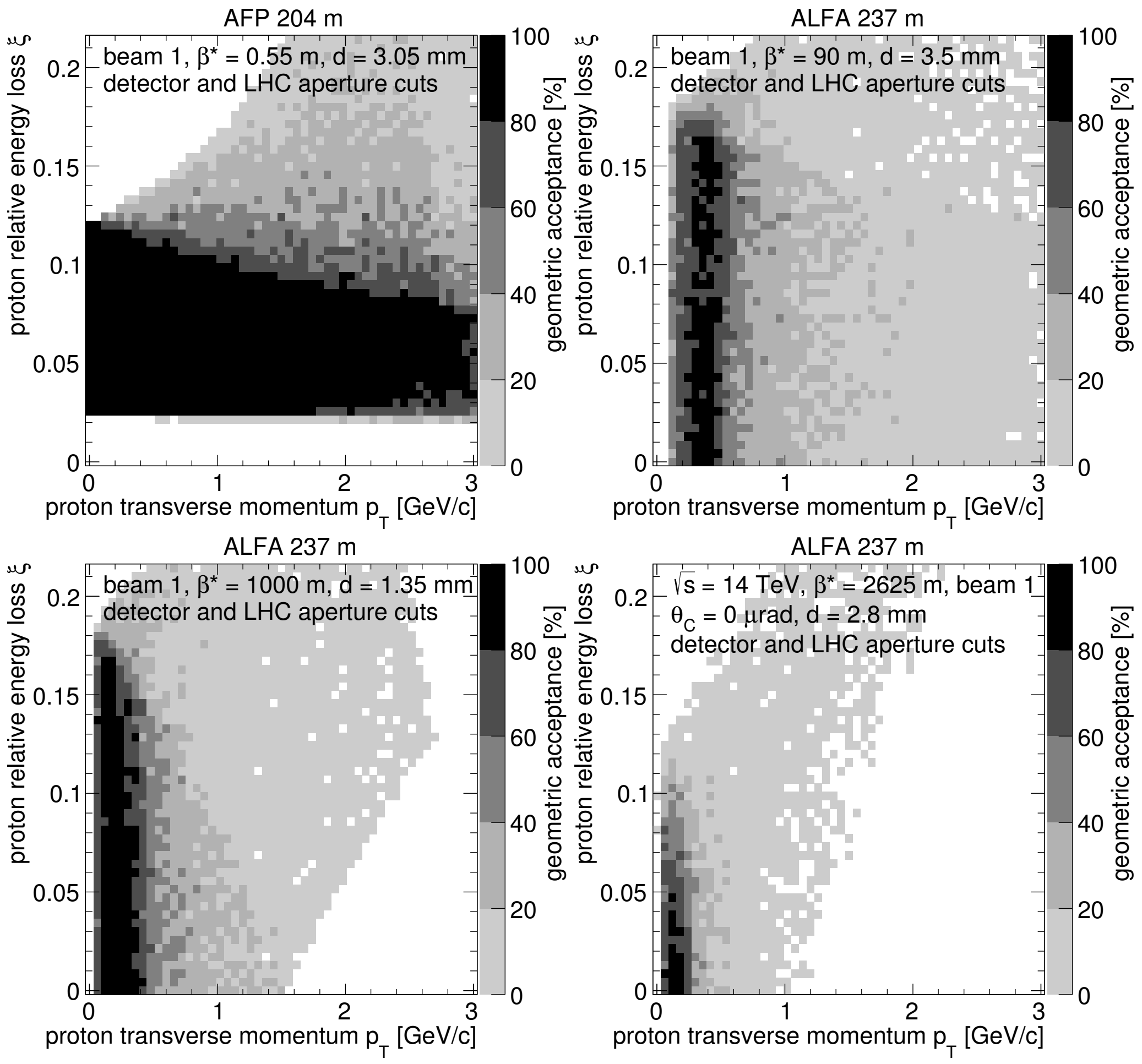

**Fig. 1:** Geometric acceptance for: AFP with $\beta^* = 0.55$ m and ALFA with $\beta^* = 90,\ 1000$ and 2625 m.

$$\sigma_{\text{tot}}^{\text{ALFA}}(8\ \text{TeV}) = 96.07 \pm 0.18\ (\text{stat.}) \pm 0.85\ (\text{exp.}) \pm 0.31\ (\text{extr.})\ \text{mb},$$

where the last error is related to uncertainties on the extrapolation to the zero four-momentum transfer ($|\text{t}| \to 0$).

Finally, by subtracting the elastic cross section from the total cross section, the inelastic cross section was calculated:

$$\sigma_{\text{inel}}^{\text{ALFA}}(7\ \text{TeV}) = 71.34 \pm 0.36\ (\text{stat.}) \pm 0.83\ (\text{syst.})\ \text{mb},$$

$$\sigma_{\text{inel}}^{\text{ALFA}}(8\ \text{TeV}) = 71.73 \pm 0.15\ (\text{stat.}) \pm 0.69\ (\text{syst.})\ \text{mb}.$$

All these results are in agreement with Monte Carlo predictions and the expected global trend [6,7]. It is also worth noting that the results from data taken at 8 TeV with $\beta^* = 1000$ m and at 13 TeV with $\beta^* = 2500$ m are on the way. One could expect that with these new data taken with a higher value of betatron function, Coulomb-nuclei interference and even Coulomb regions will be accessible [8].

## 5 Diffractive Bremsstrahlung

Diffractive bremsstrahlung is typically an electromagnetic process (see Fig. 2 (left)). However, as postulated in [9], high energy photons can be radiated in the elastic proton-proton scattering – see Fig. 2 (right). This idea was extended in [10] by introducing the proton form-factor into the calculations and by considering also other mechanisms leading to the $pp\gamma$ final state, such as a virtual photon re-scattering.

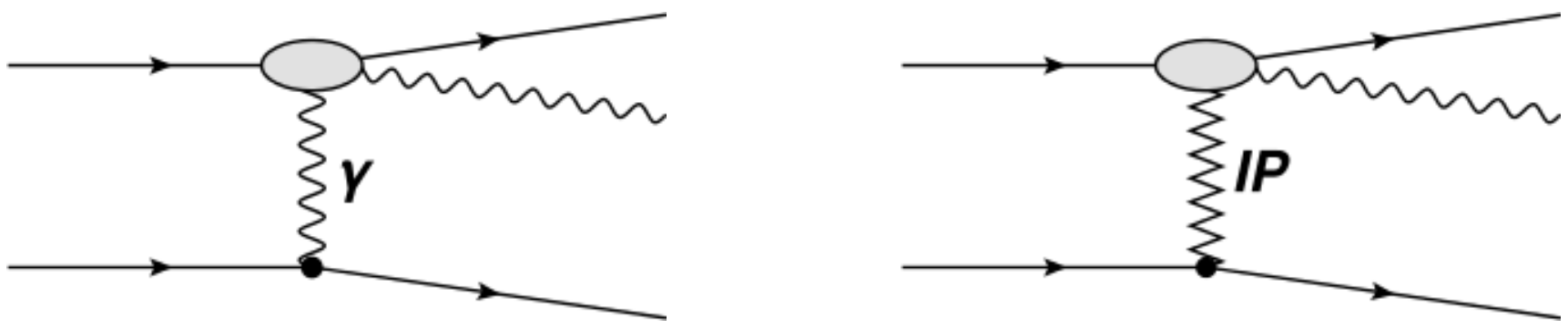

**Fig. 2:** Diagrams of electromagnetic (left) and diffractive (right) bremsstrahlung. The blobs represent various mechanisms of photon emission.

Recently, diffractive bremsstrahlung was implemented into the GenEx Monte Carlo generator [11, 12], and feasibility studies were preformed assuming $\sqrt{s} = 13$ TeV, $\beta^* = 90$ m and ALFA (ATLAS Zero Degree Calorimeter, ZDC) detectors as proton (photon) taggers [13]. It was shown that a measurement should be possible in ATLAS assuming about 10 h of data taking.

## 6 Exclusive Pion Pair Production

Exclusive pion pair production is a $2 \to 4$ process. The dominant diagram is a Pomeron-induced continuum, see Fig. 3 (left). Possibility of measurement of such process with ATLAS/ALFA detectors and high-$\beta^*$ optics was discussed in [14]. Recently, the production of a photon-induced continuum (middle) and a resonant $\rho^0$ photoproduction (right) was calculated [15]. These processes are being currently implemented to the GenEx MC generator. It is worth mentioning that exclusive pion measurements at 7 and 8 TeV with ATLAS/ALFA are under way.

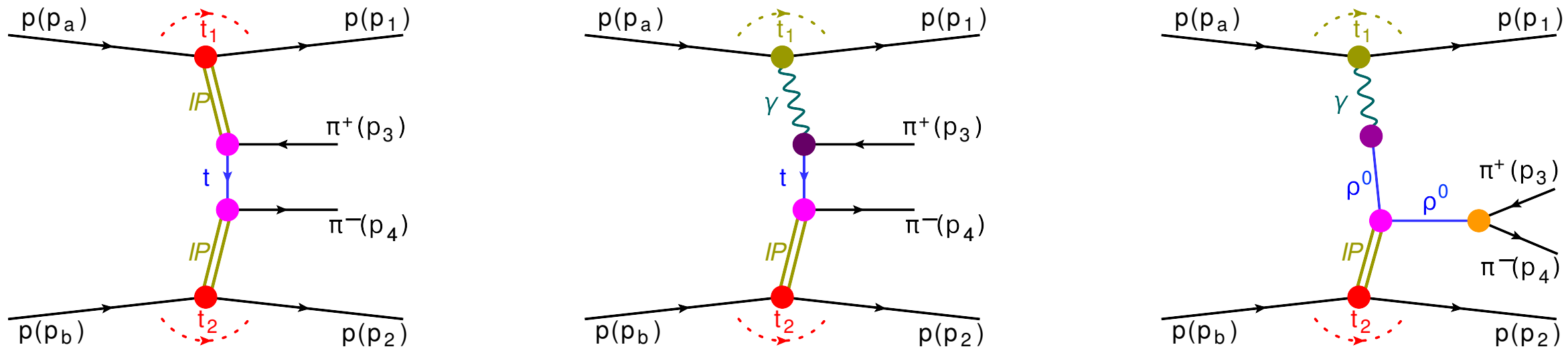

**Fig. 3:** Diagrams of exclusive non-resonant Pomeron-induced (left), non-resonant photon-induced (middle) and resonant $\rho^0$ (right) production.

## 7 Anomalous Gauge Couplings and New Physics Searches

Measurement of $W$ and $Z$ boson pair production via the exchange of two photons (see Fig. 4 (left)) allows to perform a stringent test of the electroweak symmetry breaking [16]. Standard Model predicts the existence of $\gamma\gamma WW$ quartic couplings while there is no $\gamma\gamma ZZ$ coupling. As was shown in [17, 18], collecting 30 – 300 fb$^{-1}$ of data with the ATLAS detector and using protons tagged in AFP should result in a gain in the sensitivity of about two orders of magnitude over a standard ATLAS analysis.

Proton tagging may also serve as a powerful technique for new physics searches as the backgrounds can be significantly reduced by the kinematic constraints coming from the AFP proton measurements. The general idea of background reduction was presented in [19–21] on a basis of the exclusive jet measurement.

Proton tagging technique might be also used for the invisible object searches. As an example, the case of magnetic monopoles produced by the photon exchange can be considered. From a diagram (*cf.* Fig. 4 (right)) one can conclude that, even if the centrally produced system escapes detection (or is not measurable) in ATLAS, one can measure scattered protons in AFP. In general, any production of new objects (with mass up to 2 TeV) via photon or gluon exchanges should be possible to be observed.

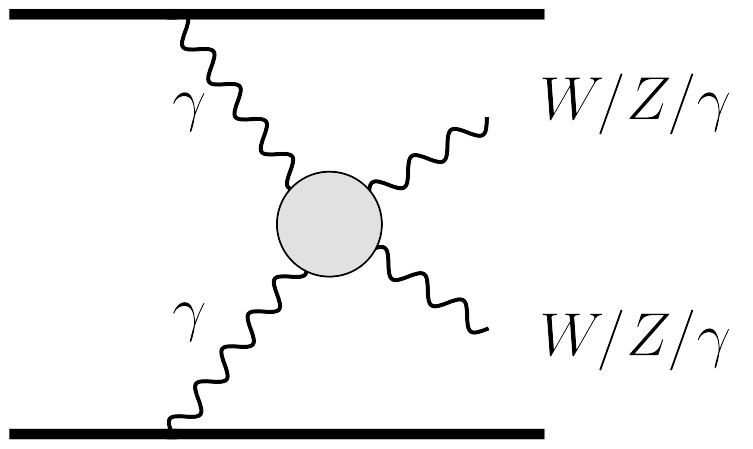

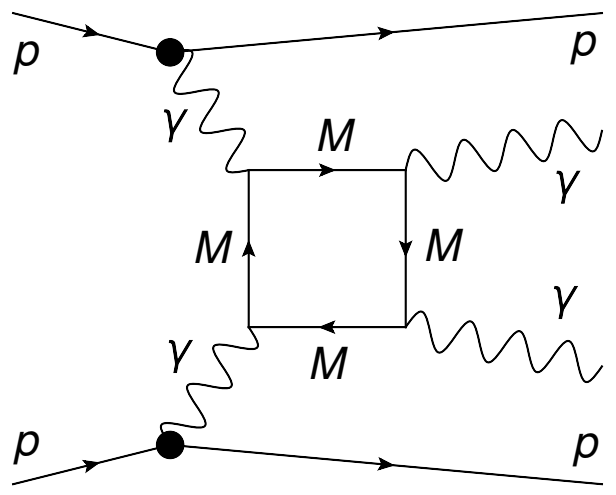

**Fig. 4:** Diagrams of anomalous gauge coupling (left) and magnetic monopole (right) production.

## 8 Summary

Two forward detectors systems are currently installed in ATLAS: ALFA (vertical RPs located ~240 m from IP) and AFP (horizontal RPs located around 210 m from IP). ALFA successfully took data at: 7 TeV with $\beta^* = 90$ m, 8 TeV with $\beta^* = 90$ m, 8 TeV with $\beta^* = 1000$ m and 13 TeV with $\beta^* = 2500$ m. AFP so far collected data at $\sqrt{s} = 13$ TeV in few special, low-luminosity runs with $\beta^* = 0.4$ m and plans to join ATLAS data-taking on a regular basis.

Properties of the elastic scattering were measured by ALFA for both runs with $\beta^* = 90$ m. Results from data taken at 8 TeV with $\beta^* = 1000$ m and 13 TeV with $\beta^* = 2500$ m (Coulomb-Nuclear Interference and potentially Coulomb regions) are on the way. When ALFA is used together with ZDC it would be possible to measure diffractive bremsstrahlung. It should be also possible to measure a photon-induced exclusive pion-pair production continuum with a resonant $\rho^0$ photoproduction on top of it.

AFP took data in two special low luminosity runs in 2016 (with detectors on one side). In 2017 a fully equipped AFP detector will take data with a proton tag on both sides during special and standard LHC runs. It is planned to measure anomalous gauge couplings ($W$ , $Z$ and photon pairs) with the AFP data. A gain in sensitivity of about two orders of magnitude over a standard analysis, which uses the data from the central detector, is expected. In principle one can try to search for any production of a new object produced via photon or gluon exchanges (magnetic monopoles, invisible particles, ...). Forward proton measurements can be used for a significant background reduction.

## Acknowledgements

The work of M.T. was supported in part by Polish National Science Centre under grant number UMO-2015/17/D/ST2/03530 and Polish Ministry of Science and Higher Education under the Mobility Plus programme (1285/MOB/IV/2015/0).

# Photon-photon and photon-hadron processes in PYTHIA 8

*I. Helenius*
Institute for Theoretical Physics, Tübingen University, Auf der Morgenstelle 14, 72076 Tübingen, Germany

**Abstract**
We present a new implementation of photoproduction processes in $e^+e^-$ and ep collisions into PYTHIA 8 Monte-Carlo event-generator. In particular we discuss how the parton showers and multiparton interactions are generated with a resolved photon beam and what is the relative contribution from direct processes in different kinematical regions. As an application we show comparisons to data for charged-particle production in $e^+e^-$ and ep collisions at LEP and HERA. We consider also photoproduction of dijets comparing to data for ep collisions at HERA and discuss about possibility to further constrain nuclear PDFs with ultra-peripheral heavy-ion collisions at the LHC.

## 1 Introduction

Photon-initiated processes can be studied in many different collider setups. In the future $e^+e^-$ colliders photon-photon processes can generate additional QCD background for many processes and, for example, provide an additional channel to produce a Higgs boson. In ep colliders one can study photon-hadron interactions which are sensitive to the structure of the resolved photon and the target hadron, and contribution of the multiparton interactions (MPIs) for particle production with photon beams. Furthermore, different nuclear modifications can be probed in ultra-peripheral heavy-ion collisions. Here there are no hadronic interactions but one of the nuclei emit a photon that interact with the other nucleus. In particular these collisions can provide important future constraints for nuclear PDFs (nPDFs).

PYTHIA 8 [1] is a general purpose Monte-Carlo event generator that is capable of simulating all particles created in an event. Event generation starts from the hard process of interest. The next step is to generate initial- (ISR) and final-state radiation (FSR) and the MPIs, evolving from the hard-process scale down to a scale below which physics becomes non-perturbative, see Ref. [2] for details. The event is then hadronized using Lund string model and unstable hadrons are decayed into stable ones measured in the detector. Main emphasis has been on pp collisions at the LHC but extensions to other collision systems have been developed. Here we discuss about recent developments for photon-photon and photon-hadron interactions in $e^+e^-$ and ep collisions [3].

## 2 Framework

Probability for the MPIs in PYTHIA 8 is given by the $2 \to 2$ QCD processes [4]. The divergence in the $p_{\mathrm{T}} \to 0$ GeV/$c$ limit is regulated with a screening parameter $p_{\mathrm{T0}}$ such that

$$\frac{\mathrm{d}\sigma}{\mathrm{d}p_{\mathrm{T}}^2} \propto \frac{\alpha_{\mathrm{S}}(p_{\mathrm{T}}^2)}{p_{\mathrm{T}}^4} \to \frac{\alpha_{\mathrm{S}}(p_{\mathrm{T}}^2 + p_{\mathrm{T0}}^2)}{(p_{\mathrm{T}}^2 + p_{\mathrm{T0}}^2)^2}. \qquad (1)$$

The parameter is taken to be energy dependent and is parameterized as $p_{\mathrm{T0}}(\sqrt{s}) = p_{\mathrm{T0}}^{\mathrm{ref}}(\sqrt{s}/7\ \mathrm{TeV})^{\alpha}$, where the default Monash-tune provides values $p_{\mathrm{T0}}^{\mathrm{ref}} = 2.28$ GeV/$c$ and $\alpha = 0.215$ for (anti)proton beams. Since the structure of a resolved photon is evidently different than the structure of a proton, the value of the screening parameter should be revised. This is one of the outcomes of the presented work.

### 2.1 Photon beam

Photons may interact as an unresolved particle (direct photon), or fluctuate into a hadronic state with equal quantum numbers (resolved photon). In the former case the photon itself act as an initiator of the hard process whereas in the latter case the constituent partons are the initiators. As with hadrons, the distribution of the partons can be described with PDFs, $f_i^{\gamma}(x,Q^2)$, which scale evolution are given by the DGLAP equations. In addition to the usual splittings of partons, for a resolved photon one needs to take into account also $\gamma \rightarrow q\bar{q}$ splittings of the beam photon, giving [5]

$$\frac{\partial f_i^{\gamma}(x,Q^2)}{\partial \log(Q^2)} = \frac{\alpha_{\mathrm{EM}}}{2\pi} e_i^2 P_{i\gamma}(x) + \frac{\alpha_{\mathrm{S}}(Q^2)}{2\pi} \sum_j \int_x^1 \frac{\mathrm{d}z}{z} P_{ij}(z)\, f_j^{\gamma}(x/z,Q^2), \tag{2}$$

where $P_{ij}(z)$'s are the usual DGLAP splitting kernels for a given $j \rightarrow ik$ splittings. The additional $\gamma \rightarrow q\bar{q}$ splittings provide more quarks at higher values of $x$ than with hadron beams. In this work we use photon PDFs from CJKL analysis [6]. For the parton shower generation the additional term corresponds to a probability to end up to the original beam photon when tracing back the ISR splittings that have taken place for the hard-process initiators. If this happens, there are no further ISR emissions or MPIs below the scale where this happens, or need for any beam remnants.

Since the interacting photons can be either resolved or direct, there are three different types of processes that needs to be taken into account for a photon-photon interaction: resolved-resolved, resolved-direct, and direct-direct. For the resolved-resolved contribution the full parton-level evolution needs to be generated including ISR and FSR and also possible MPIs. For direct-resolved case no MPIs are present nor ISR for the direct side. For direct-direct case only FSR is relevant. The relative contribution of each process type depends on the kinematics, typically direct (resolved) processes dominate when $x$ is large (small). In case of photon-hadron interaction the photons can be either resolved or direct.

### 2.2 Photon flux from leptons

The photon flux from lepton $l$ can be modelled with equivalent photon approximation (EPA). Integrating the flux over allowed photon virtuality yields

$$f_{\gamma}^{l}(x_{\gamma},Q^2_{\mathrm{max}}) = \frac{\alpha_{\mathrm{EM}}}{2\pi} \frac{1+(1-x_{\gamma})^2}{x_{\gamma}} \log\left(\frac{Q^2_{\mathrm{max}}}{Q^2_{\mathrm{min}}(x_{\gamma})}\right). \tag{3}$$

The virtuality of the photon, $Q^2$, is related to the lepton scattering angle so the lower $Q^2$ limit for the splitting can be derived from the kinematics. The appropriate upper limit depends experimental configuration. Here we have considered only quasi-real photons so $Q^2_{\mathrm{max}} \lesssim 1\ \mathrm{GeV}^2$.

For the direct contribution the spectrum of photons can be obtained directly from the flux. The distribution of partons in resolved photons that is coming from the lepton beam can be obtained by convoluting the photon flux $f_{\gamma}^{l}$ with the parton-inside-photon PDFs $f_i^{\gamma}$

$$x f_i^{l}(x,Q^2) = \int_x^1 \frac{\mathrm{d}x_{\gamma}}{x_{\gamma}} x_{\gamma} f_{\gamma}^{l}(x_{\gamma},Q^2_{\mathrm{max}}) x' f_i^{\gamma}(x',Q^2), \tag{4}$$

where $Q^2$ is now the factorisation scale and $x' = x/x_{\gamma}$. The partonic evolution is then performed for the photon-photon(hadron) sub-collision constructed according to sampled $x_{\gamma}$ values.

## 3 Results

### 3.1 Charged-hadron photoproduction in $e^+e^-$

Photon-photon interactions in $e^+e^-$ collider have been studied in LEP. A suitable observable to compare our new framework is the charged hadron production for which measurement from OPAL experiment exists [7]. The measurement used anti-tagged events where the angle of scattered leptons is beyond the acceptance so that they are not seen in the detector. With the OPAL kinematics this translates into a virtuality cut $Q^2 < 1$ GeV$^2$.

Figure 1 shows the ratio between the OPAL data and the result of simulations combining direct and resolved contributions without any MPIs with different invariant mass $W$ bins. Also the individual contributions from direct-direct, direct-resolved and resolved-resolved are plotted separately to quantify the contribution from each of these. The data is then compared to the results with MPIs using different values for the parameter $p_{\mathrm{T0}}^{\mathrm{ref}}$. The comparison shows that the default $p_{\mathrm{T0}}^{\mathrm{ref}} = 2.28$ GeV/$c$ generates too many charged particles around $p_{\mathrm{T}} \sim 2$ GeV/$c$. Increasing the value of this parameter reduces the number of charged particles in this region and a good agreement in all $W$ bins is obtained with $p_{\mathrm{T0}}^{\mathrm{ref}} = 3.30$ GeV/$c$. The systematic increase of the number of charged particles from the MPIs with increasing $W$ is supported by the data.

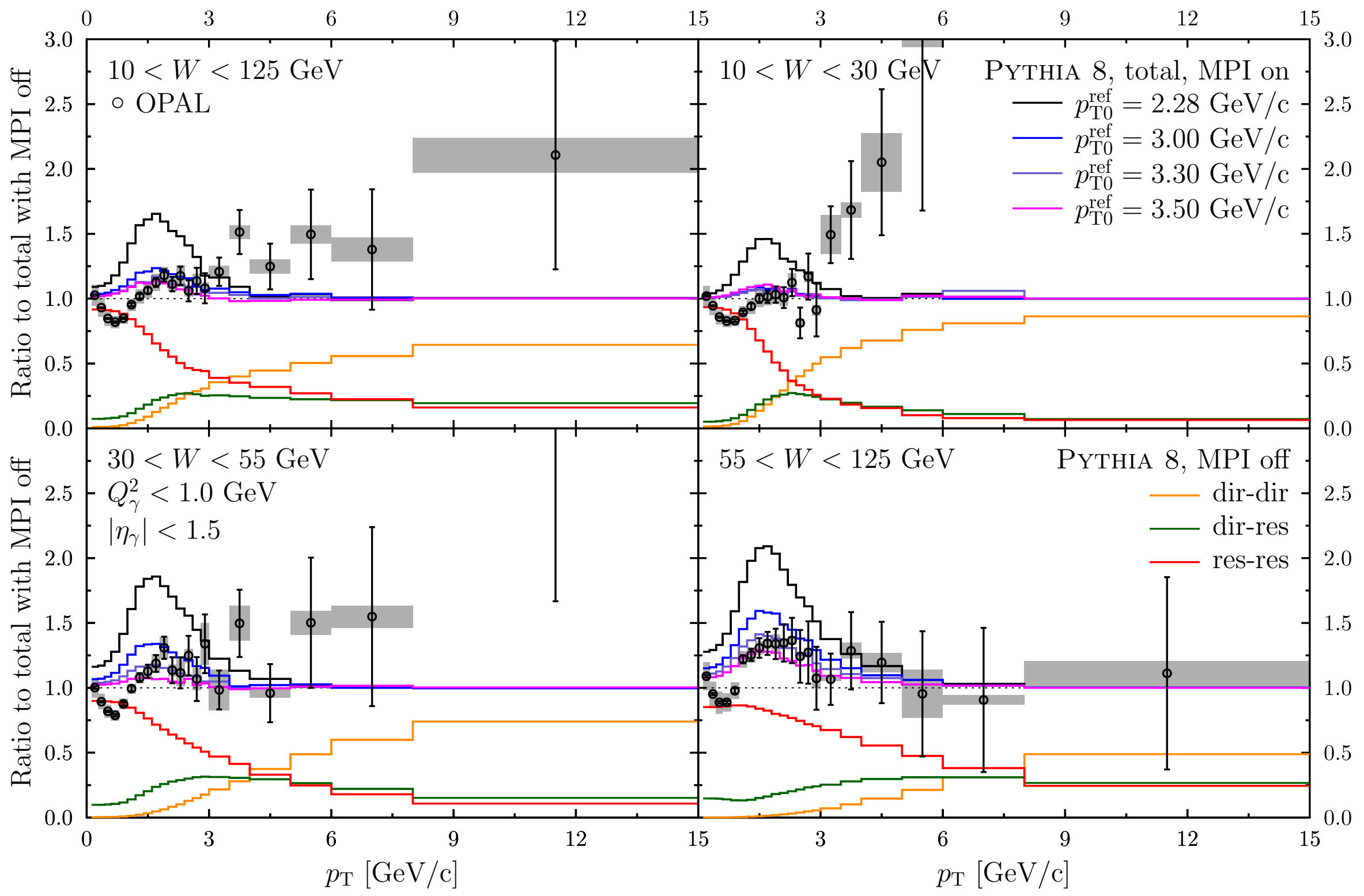

**Fig. 1:** Ratio between measured [7] and simulated charged-hadron cross sections in $\gamma\gamma$ interactions in $\mathrm{e^+e^-}$ collisions at $\sqrt{s} = 166$ GeV for different invariant mass bins, $10 < W < 125$ GeV (top left), $10 < W < 30$ GeV (top right), $30 < W < 55$ GeV (bottom left) and $55 < W < 125$ GeV (bottom right). Contribution from direct-direct (orange), direct-resolved (green) and resolved-resolved without MPIs (red) are shown separately and the sum of these are shown for different values of $p_{\mathrm{T0}}^{\mathrm{ref}}$.

### 3.2 Charged-hadron and dijet photoproduction in ep

There are plenty of data available for the photoproduction in ep collisions from HERA collider. Again a useful observable to study the effect from MPIs is the charged-hadron production for which data exists from H1 [8] and ZEUS [9] experiments. The kinematical cuts in the H1 measurements corresponds to average invariant mass of photon-proton system of $\langle W_{\gamma\mathrm{p}} \rangle = 200$ GeV. The simulations are compared to this data in Fig. 2 as a function of $p_{\mathrm{T}}$ and $\eta$. The data is best described with $p_{\mathrm{T0}}^{\mathrm{ref}} = 3.0$ GeV/$c$ which conveniently lie between the values optimal for $\gamma\gamma$ and pp. This difference in $p_{\mathrm{T0}}^{\mathrm{ref}}$ values could reflect that the photon is a cleaner state than the proton, but also that a more sophisticated energy scaling of $p_{\mathrm{T0}}(\sqrt{s})$ may be required, also for protons.

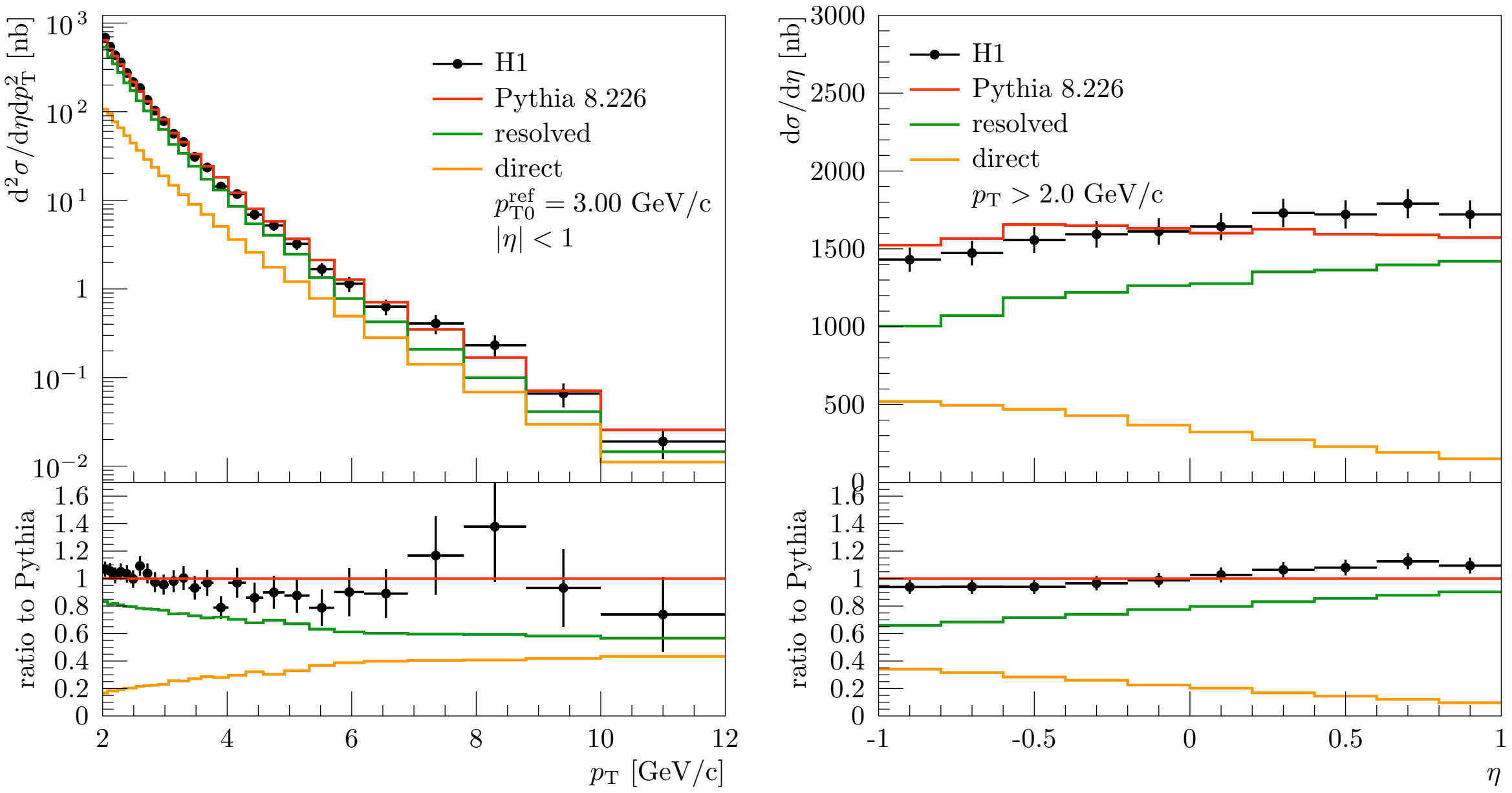

**Fig. 2:** Cross section for charged particle photoproduction as a function of $p_{\mathrm{T}}$ (left panel) and $\eta$ (right panel) in ep collisions at HERA at mid-rapidity. Data from H1 [8] are compared to PYTHIA simulations with $p_{\mathrm{T0}}^{\mathrm{ref}} = 3.0$ GeV/$c$ (red), decomposed to direct (orange) and resolved (green) contributions.

Another useful observable to study photon-hadron interactions is photoproduction of dijets. For this we can use data from ZEUS [10] where the photon virtuality is restricted to $Q^2 < 1.0$ GeV$^2$ and $134 < W_{\gamma\mathrm{p}} < 277$ GeV. The transverse energy cuts in the data are $E_{\mathrm{T}}^{\mathrm{jet1}} > 14$ GeV and $E_{\mathrm{T}}^{\mathrm{jet2}} > 11$ GeV, where jet 1 (2) is chosen to be the jet with the (second) highest $E_{\mathrm{T}}$ within pseudorapidities $-1 < \eta < 2.4$. Still it is not possible to separate the direct and resolved processes apart, but by defining

$$x_{\gamma}^{\mathrm{obs}} = \frac{E_{\mathrm{T}}^{\mathrm{jet1}} \mathrm{e}^{\eta^{\mathrm{jet1}}} + E_{\mathrm{T}}^{\mathrm{jet2}} \mathrm{e}^{\eta^{\mathrm{jet2}}}}{2yE_{\mathrm{e}}} \tag{5}$$

some sensitivity for different contributions can be obtained. Here $y$ is the inelasticity of the event and $E_{\mathrm{e}}$ is the energy of the positron beam. Figure 3 shows a comparison of data and PYTHIA simulations for the dijet cross section as a function of $x_{\gamma}^{\mathrm{obs}}$, where again the direct and resolved contributions are shown separately. The simulations are performed with $p_{\mathrm{T0}}^{\mathrm{ref}} = 3.0$ GeV/$c$ tuned to charged particle production data above. In general the agreement is decent and indeed the events from direct processes tend to sit at higher values of $x_{\gamma}^{\mathrm{obs}}$ as expected. However, the resolved processes do provide some contribution also at $x_{\gamma}^{\mathrm{obs}} > 0.8$. A possible explanation of the slight overshoot of the data might result from differences in the applied jet algorithms—this will be studied in more detail later on.

Recently it has been argued that dijet production in ultra-peripheral heavy-ion collisions at the LHC could provide further constraints for nuclear modification of the PDFs [11]. Since the expected (per-nucleon) invariant mass of photon-ion system is not that far from the $W_{\gamma\mathrm{p}}$ in ep at HERA, a qualitative study for the argument can be done just by using nPDFs for the target and quantifying the uncertainty using a realistic nPDF set. Figure 3 shows the result of this exercise using EPS09LO nPDFs [12] for the dijet cross section as a function of $\eta^{\mathrm{jet2}}$ where $0 < \eta^{\mathrm{jet1}} < 1$ and $x_{\gamma}^{\mathrm{obs}} > 0.75$. The latter condition reduces the contribution from resolved processes and therefore minimizes the uncertainty from the photon PDFs. With the given kinematics $\sim 10$ % nPDF-originating uncertainty is found demonstrating the experimental accuracy required to further constrain the nPDFs. In addition, such a measurement would provide an unique test for the factorisation of the nuclear modifications. A more detailed study using the accurate photon flux from a nucleus and LHC kinematics is in the works.

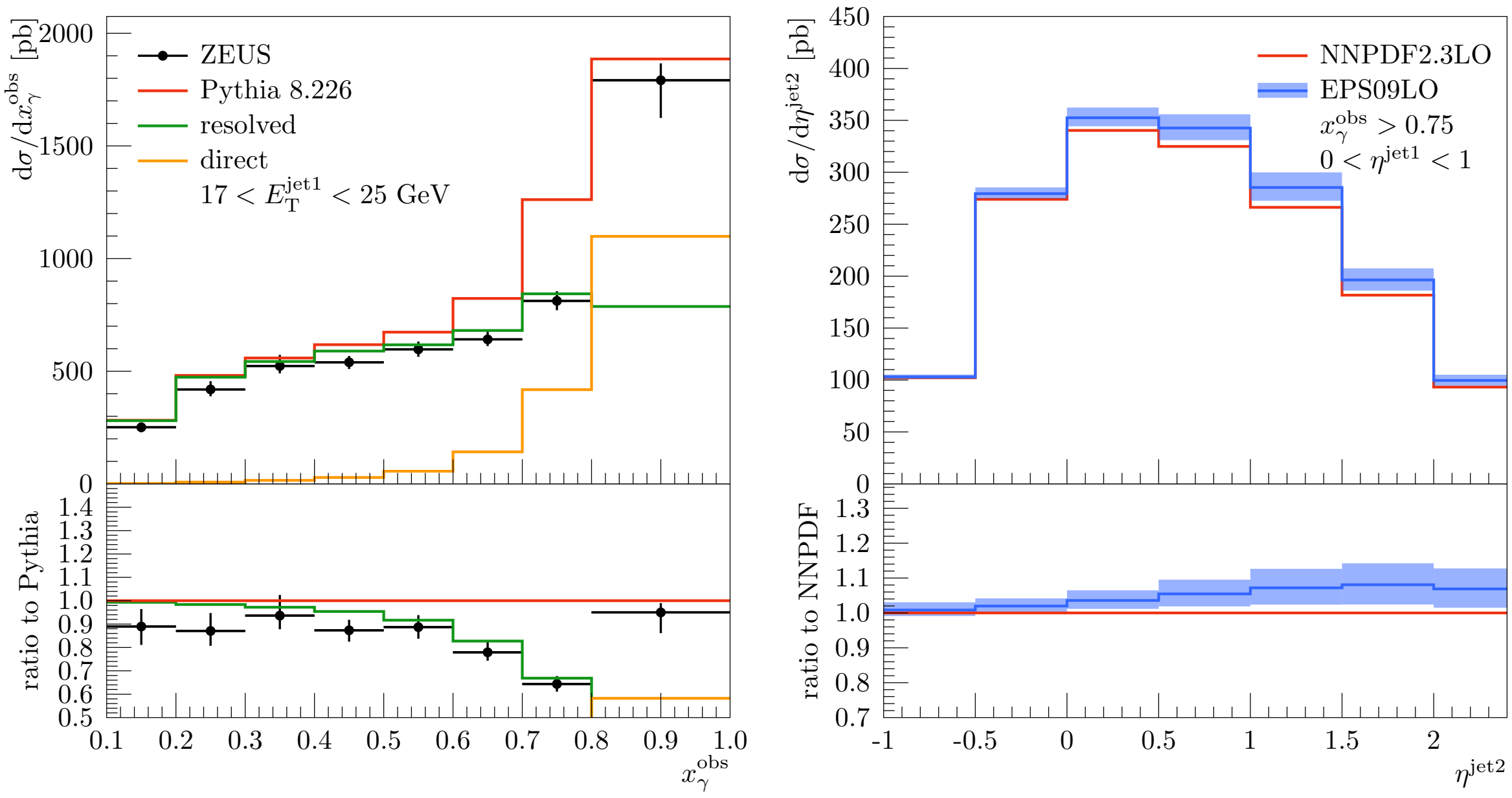

**Fig. 3: Left:** Cross section for dijet photoproduction in HERA as a function of $x_\gamma^{obs}$. Data from ZEUS [10] are compared to PYTHIA simulations (red), decomposed to direct (orange) and resolved (green) contributions. **Right:** Cross section for dijet photoproduction as a function of $\eta^{jet2}$ for events with $x_\gamma^{obs} > 0.75$ using a proton PDFs only (red) and with nuclear modifications from EPS09 (blue) including the nPDF uncertainties (blue band).

## 4 Summary

We have included a framework to simulate different photoproduction processes for different collision systems for PYTHIA 8 event generator including direct and resolved processes. The framework is validated by comparing charged-particle photoproduction cross sections in $e^+e^-$ and ep collisions to experimental data from LEP and HERA. These data are also used to constrain the role of MPIs for the resolved photon processes for which studies have been few. The data favoured $\sim 45$ $(30)$ % larger value for $p_{T0}^{ref}$ for $\gamma\gamma$ ($\gamma$p) than what were found optimal for pp, which translates into a smaller MPI cross-section. Also comparisons to data for photoproduction of dijets at HERA showed a reasonable agreement with the simulations. In future we will extend the photoproduction framework to ultra-peripheral heavy-ion collisions which can further constrain nPDFs.

## Acknowledgements

Work have been supported by the MCnetITN FP7 Marie Curie Initial Training Network, contract PITN-GA-2012-315877 and has received funding from the European Research Council (ERC) under the European Union's Horizon 2020 research and innovation programme (grant agreement No 668679).

# $\gamma\gamma$ and $\gamma p$ measurements with forward proton taggers in CMS+TOTEM

*J. Hollar, on behalf of the CMS and TOTEM Collaborations*
LIP, Lisbon, Portugal

**Abstract**
The CMS+TOTEM Precision Proton Spectrometer operated for the first time in 2016, collecting data during $pp$ collisions at $\sqrt{s} = 13$ TeV at the CERN Large Hadron Collider. Procedures for the detector alignment, optics corrections, and background estimations were developed, and applied to an analysis of the process $pp \rightarrow p\mu^+\mu^-p^{(*)}$ with dimuon masses larger than 110 GeV. A total of 12 candidate events are observed, corresponding to an excess of $> 4\sigma$ over the background prediction. This constitutes the first evidence for this process at such masses, and demonstrates the good performance of CT-PPS.

## 1 Introduction

The CMS+TOTEM Precision Proton Spectrometer (CT-PPS) [1] is a joint program of the CMS [2] and TOTEM [3] collaborations, to operate near-beam forward proton detectors in high luminosity proton-proton running at the Large Hadron Collider (LHC). The detectors consist of silicon tracking and fast timing detectors, installed in Roman Pot stations $\sim$ 210-220m from the P5 interaction region. The detectors are designed to detect the intact protons from "exclusive" production ($pp \rightarrow pXp$), primarily via either $\gamma\gamma$ fusion, or gluon-gluon interactions (with a second screening gluon exchanged to cancel the color flow).

At the LHC, a special class of collisions involves the exchange of quasi-real photons, with the incident beam particles remaining intact. In high energy proton-proton collisions, the spectrum of these $\gamma\gamma$ interactions can extend to the TeV scale, well beyond the range probed at previous colliders. This provides a unique opportunity to study photon interactions in a new energy regime. Detection of the outgoing protons provides strong background suppression and kinematic constraints, making this topology an excellent way to search for new particles and deviations from the Standard Model in low cross section processes (for some recent examples see [4–8]).

### 1.1 The CT-PPS detectors and 2016 operations

The initial design of CT-PPS called for the installation and commissioning of detectors in 2016, with physics data-taking to start in 2017. However, by making use of the existing TOTEM silicon strip detectors in RP stations at 206m and 214m from CMS, it was possible to begin data taking for physics already in 2016. This required validation of safe Roman Pot insertions at high beam intensities, as well as integration of the data acquisition and reconstruction software between CMS and TOTEM. This was accomplished at the beginning of the 2016 LHC run, and from May the Roman Pots were regularly inserted to $15\sigma$ from the beam, with the Si-strip detectors read out through the central CMS DAQ. During the summer of 2016 diamond fast timing detectors were installed in new cylindrical Roman Pots in the 220m region, and began taking data by the end of the LHC proton-proton run.

## 2 Alignment and optics corrections

### 2.1 Alignment

The alignment is performed in two steps [10]. First, special low-luminosity fills are used to determine the absolute alignment. This is achieved by inserting the RPs to within $5\sigma$ of the beam. By also inserting vertical RPs, a sample of elastic scattering $pp \to pp$ events can be collected, allowing alignment with respect to the beam based on the azimuthal symmetry of this process. The horizontal RPs can then be aligned with respect to the vertical RPs, using a subsample of tracks passing through both.

In a second step, the absolute alignment determined in the special fills must be transferred to the case of normal high-intensity fills, where only the horizontal RPs are inserted to a distance $\sim 15\sigma$ from the beam. This is based on matching the measured x distribution of track impact points, under the assumption that the same physics processes contribute to all fills. This results in a set of fill by fill alignment corrections, determined separately for each RP. The total uncertainty of the horizontal alignment procedure is on the order of 150 $\mu$m. The procedure is illustrated in Fig. 1.

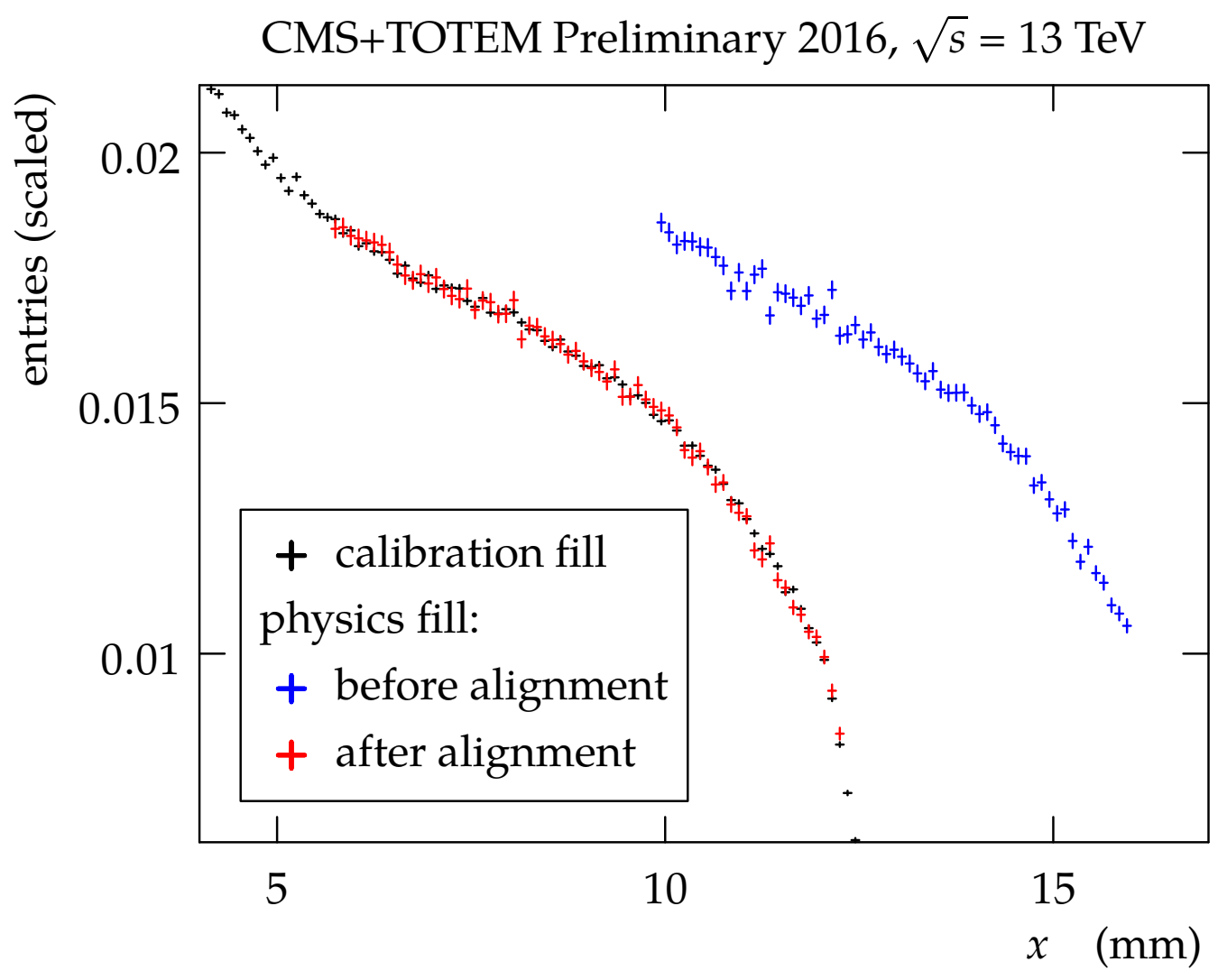

**Fig. 1:** Distribution of the track impact points as a function of the horizontal coordinate for the alignment fill (black points), a physics fill before alignment (blue points), and after alignment (red points).

### 2.2 Optics

Having aligned the RPs, a precise determination of the LHC beam optics is required to derive the proton's fractional energy loss $\xi$ from the measured x coordinates of the tracks. The procedure developed [11] relies on using real data to constrain the elements of the single pass transport matrix $T(s, \xi)$, whose elements are the optical functions of the beam line.

The leading horizontal term in the transport matrix is x = $D_x(\xi)\ \xi$, where $D_x$ is the dispersion, which has a mild dependence on $\xi$. The leading term in the vertical plane is $y \approx L_y(\xi)\Theta_y^*$, where $L_y(\xi)$ is the vertical effective length and $\Theta_y^*$ is the vertical angle of the proton at the interaction point. The value of $L_y(\xi)$ will go to zero at a particular value of $\xi_0$, leading to a "pinch" in the vertical distribution of tracks reconstructed in the RP. By determining the horizontal position of this "pinch", the value of the dispersion can be solved for as $x_0 \approx D_x \xi_0$ (Fig. 2) where higher-order terms are neglected and included in the systematic uncertainties.

A second independent method is also used to determine the difference in the dispersions in the two LHC beams, by comparing the measured physics proton distribution in the RPs.

### 2.3 Uncertainties

Given the alignment and optics corrections described in the previous sections, the proton $\xi$ can be reconstructed from the measured x position of the tracks in the strip detectors of the horizontal RPs. For large values of $\xi$ the dominant uncertainty in this determination is $5.5\%$, arising from the dispersion $D_x$. Additional uncertainties come from the alignment ($\sigma(x) \approx 150\mu$m), and the neglected terms in the horizontal terms of the transport matrix ($\sigma(x) \approx 100\mu$m).

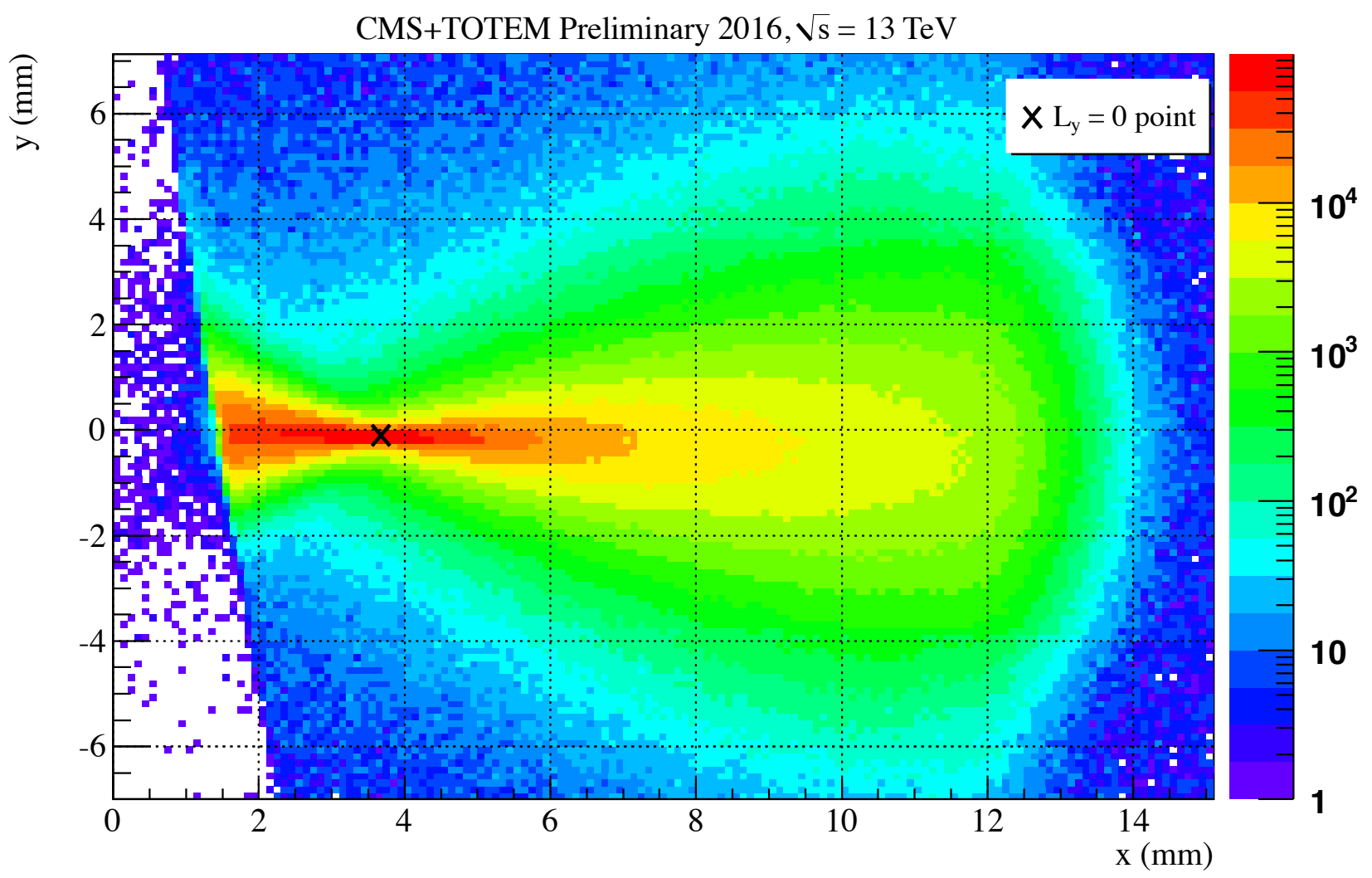

**Fig. 2:** Distribution of the track impact points measured in RP 210F, in sector 45, in the alignment fill. The point where $L_y = 0$ and its effect in the impact point distribution are shown. The beam center is at $x = y = 0$.

## 3 Analysis of $\gamma\gamma \to \mu^+\mu^-$ production

The alignment and optics procedures are then applied to an analysis of $\gamma\gamma \to \mu^+\mu^-$ production [12], using 10 fb$^{-1}$ of data collected during 2016. In order to increase the acceptance at lower masses, only one detected proton is required. The signal therefore contains a mix of both $pp \to p\mu^+\mu^-p$ events, and $pp \to p\mu^+\mu^-p^*$ events, in which one of the protons dissociates into an undetected system $p^*$.

### 3.1 Event selection and proton-dimuon matching

An event sample enriched in $\gamma\gamma$ interactions is selected in the central CMS detectors following a procedure similar to that used in earlier studies, in which no detection of the protons was possible. Events are required to have a dimuon vertex with no additional tracks within a veto region of 0.5 mm. The muons are required to have a transverse momentum $p_T > 40$ GeV, and invariant mass $m(\mu\mu) > 110$ GeV. In addition, the "acoplanarity" $(1 - |\Delta\phi(\mu\mu)|/\pi)$ of the muons is required to be less than 0.009. The selection criteria are chosen such that the expected signal to background ratio after the central detector requirements is $> 1$. Because of the high rate of multiple collisions within the same bunch crossing ("pileup"), the selection is based on information from reconstructed charged tracks and muons, without using information from the calorimeters.

In the case of events in which both protons stay intact ($pp \to p\mu^+\mu^-p$), the kinematics of the muons and the protons can be precisely related via the expression:

$$\xi(\mu\mu) = \frac{1}{\sqrt{s}}(p_T(\mu_1)e^{\pm\eta(\mu_1)} + p_T(\mu_2)e^{\pm\eta(\mu_2)}).$$

When only one of the two protons remains intact ($pp \to p\mu^+\mu^-p^*$), the same expression approximately holds when the mass of the dissociating system $M_X$ is small; the deviation becomes comparable to the experimental dimuon resolution only for masses $M_X \geq 400$ GeV, corresponding to a small fraction of events surviving the zero extra tracks requirement. The signal region is defined to include events where $\xi(\mu\mu)$ and $\xi$(RP) match within $2\sigma$ of their combined experimental resolution.

### 3.2 Backgrounds and systematics

After the central CMS detector selection, the dominant backgrounds are expected to arise from Drell-Yan $\mu^+\mu^-$ production, and from $\gamma\gamma \to \mu^+\mu^-$ production with both protons dissociating. These processes can mimic the signal when they occur in the same bunch crossing as a proton from a pileup collision, or a Roman Pot track arising from beam-related backgrounds. A control sample of $Z \to \mu^+\mu^-$ events is used to estimate the probability of a high mass $\mu^+\mu^-$ event overlapping with an uncorrelated RP track. The $\xi(\mu^+\mu^-)$ distribution in the control region is reweighted to match the distribution predicted by the Drell-Yan simulation for events entering the signal region. Simulation is then used to extrapolate to the number of such events passing the central detector selection on the track multiplicity and acoplanarity. In the case of double dissociation backgrounds, simulated events are normalized to the predicted number passing the central detector selection from simulation, and randomly mixed with protons from the $Z \to \mu^+\mu^-$ data control sample.

Systematic uncertainties on the background yield include those arising from the statistical uncertainty in the $Z \to \mu^+\mu^-$ sample used to estimate the backgrounds. For the Drell-Yan background, the modeling of the track multiplicity in the simulation, and the effect of reweighting the $\xi(\mu^+\mu^-)$ distribution, are also considered as systematic uncertainties. For the double dissociation background, uncertainties in the integrated luminosity, and in the theoretical predictions of the survival probability [9], are included as systematic uncertainties. The dominant uncertainties are due to the effect of reweighting the $\xi(\mu^+\mu^-)$ distribution (25%, taken as the full difference between the results with and without reweighting), and the modeling of the track multiplicity distribution (28%, taken as the full difference between data and simulation in the region with 1-5 extra tracks at the dimuon vertex).

The event-by-event uncertainty on $\xi(\mu\mu)$ is estimated to be 3.3%, based on simulation and data-simulation corrections derived from $Z \to \mu^+\mu^-$ events. The uncertainty on $\xi$(RP) is taken to be 5.5%, as described in section 2.

The total background estimate, including systematic uncertainties, is $1.47 \pm 0.06$ (stat.) $\pm 0.52$ (syst.) events, dominated by the Drell-Yan backgrounds.

### 3.3 Results

The correlation between the predicted $\xi(\mu\mu)$ compared and the measured $\xi$(RP) is shown in Fig. 3, for the two arms separately. In the region of low $\xi(\mu\mu)$, any signal protons should be outside of the RP acceptance, and only random background correlations are expected. In the region compatible with the RP acceptance, 17 events are observed. Of these, 12 have $\xi(\mu\mu)$ and $\xi$(RP) compatible within $2\sigma$ of the resolution, compared to $1.47 \pm 0.06$ (stat.) $\pm 0.52$ (syst.) such events expected from the backgrounds only. The significance for observing 12 matching events, including systematic uncertainties, is estimated to be $4.3\sigma$.

In Fig. 4, the signal candidate events are overlaid with an approximate CT-PPS acceptance (including the assumption that the 4-momentum transfer squared $t = 0$) in the dimuon mass-rapidity plane. The events are consistent with the acceptance for detecting one of the two protons in CT-PPS. No events with two protons are seen in the data. The highest mass candidate has m($\mu\mu$) = 341 GeV, below the region of acceptance for detecting both protons.

## 4 Conclusions

The CT-PPS project has a broad physics program, including exploration of new physics in $\gamma\gamma$ interactions at very high energies. The detectors operated for the first time in the 2016 LHC proton-proton run, collecting $\sim$ 15 fb$^{-1}$ of data. The Roman Pot insertions were validated, the detectors were commissioned, and the data acquisition and reconstruction software were fully integrated between CMS and TOTEM.

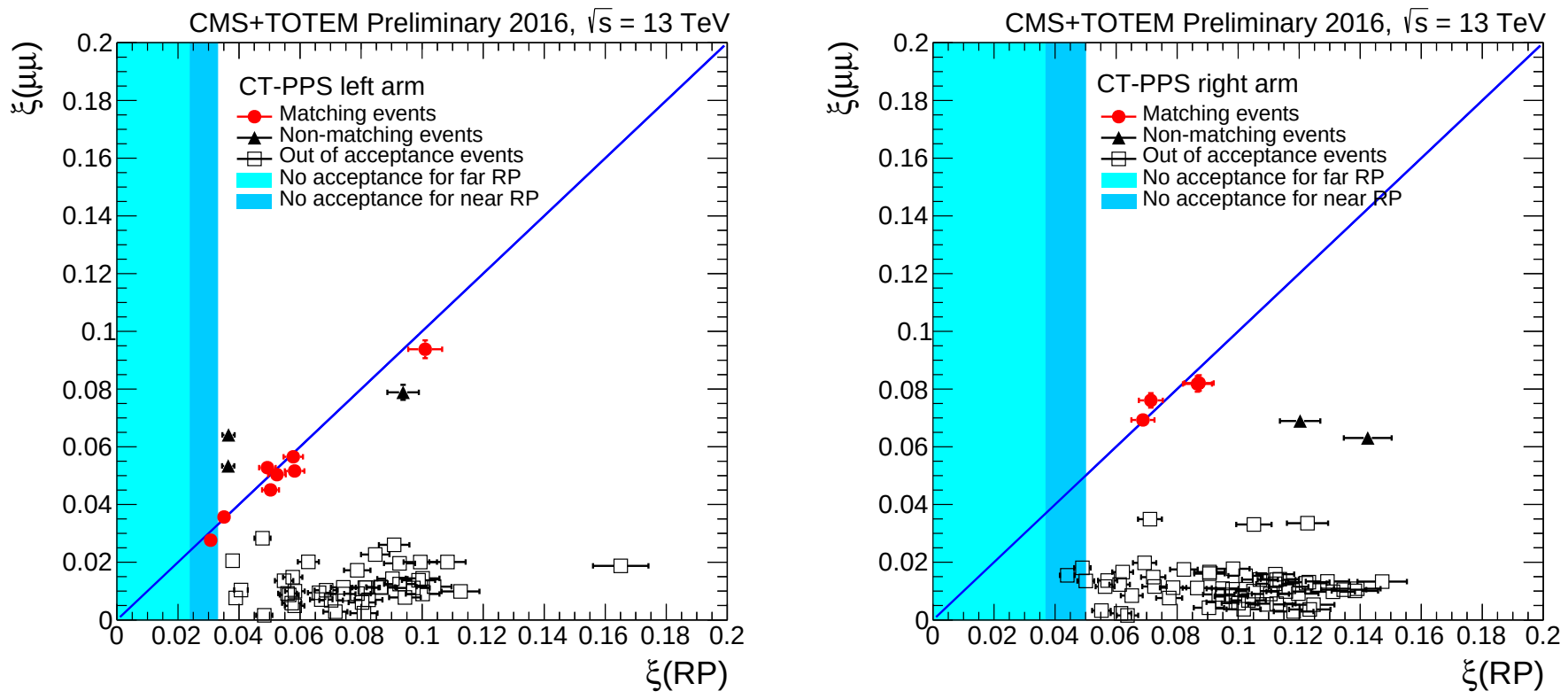

**Fig. 3:** Correlation between $\xi(\mu\mu)$ and $\xi$ measured in the Roman Pots, for both Roman Pots in each arm combined. The 45 (left) and 56 (right) arms are shown. The light shaded region corresponds to the kinematic region outside the acceptance of both the near and far RPs, while the darker shaded region corresponds to the region outside the acceptance of the near RP. For the events in which a track is detected in both, the $\xi$ value measured at the near RP is plotted.

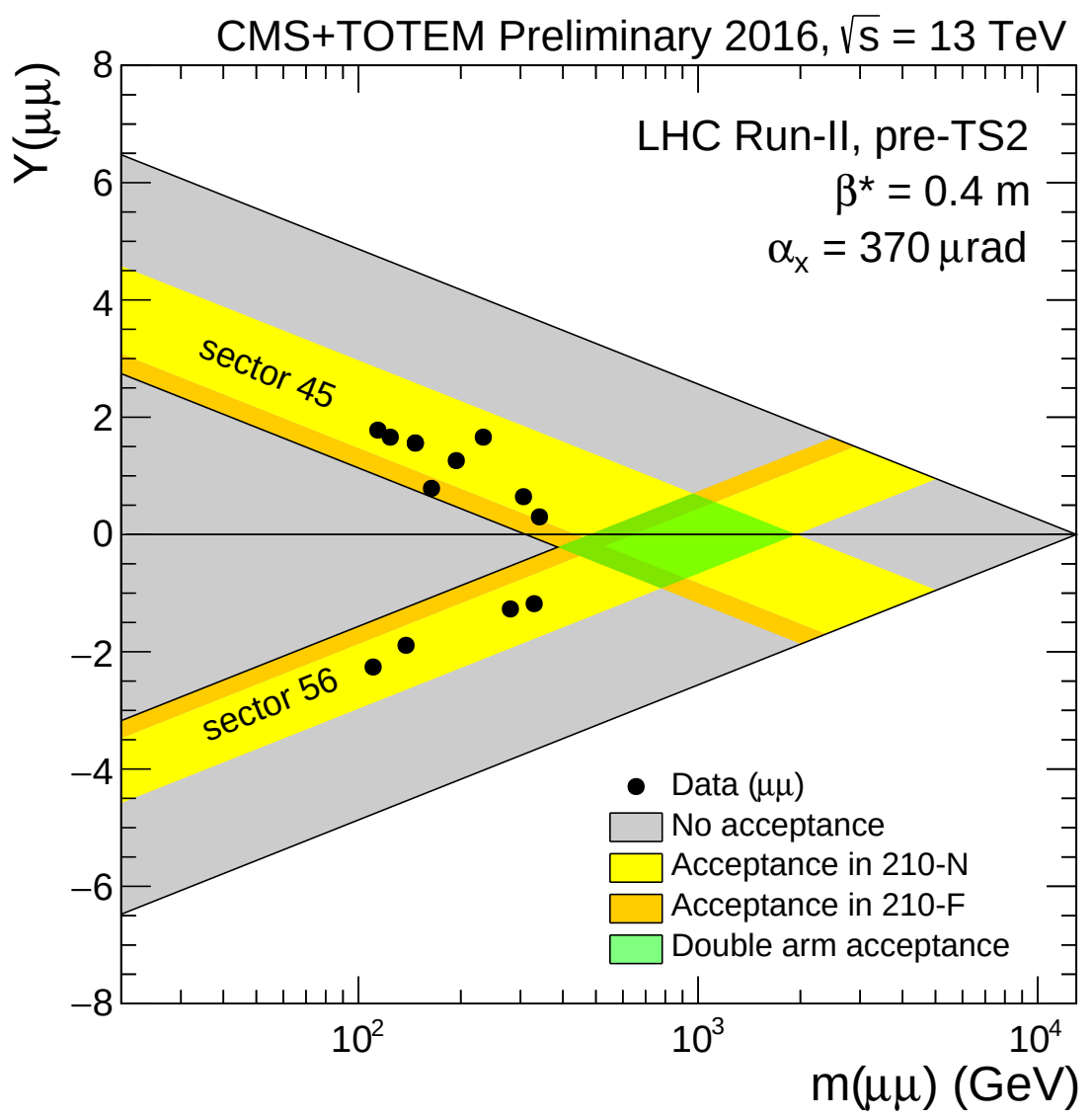

**Fig. 4:** Expected approximate coverage in the rapidity vs invariant mass plane, overlaid with the observed dimuon signal candidate events.

Data-driven procedures for the alignment and optics corrections were developed using a combination of special alignment runs, and standard high luminosity data taking. These were applied to an analysis of $\gamma\gamma \rightarrow \mu^+\mu^-$ production with single proton tags. A 4.3$\sigma$ excess of events with correlated proton and $\mu^+\mu^-$ kinematics was observed with m($\mu^+\mu^-$) = 110-341 GeV, representing evidence for tagged $\gamma\gamma$ collisions at the electroweak scale. The present data demonstrate the excellent performance of CT-PPS and its potential. With its 2016 operation, CT-PPS has proven for the first time the feasibility of operating a near-beam proton spectrometer at a high luminosity hadron collider on a regular basis.

# Photoproduction of vector mesons in ultra-peripheral Pb-Pb interactions with ALICE

*Christopher D. Anson, for the ALICE Collaboration*
Creighton University, Omaha, NE, USA

**Abstract**
Cross section measurements for vector meson production in ultra-peripheral collisions (UPCs) provide important insight into nuclear gluon distributions, the production mechanisms of vector mesons and a better understanding of the early stages of heavy ion collisions. With data collected during LHC Run 1, ALICE measured vector meson production in ultra-peripheral Pb-Pb collisions at $\sqrt{s_{NN}} = 2.76$ TeV. Detector upgrades and improved triggers before LHC Run 2 will be discussed. These improvements have allowed a much higher statistics data set to be collected for UPCs at $\sqrt{s_{NN}} = 5.02$ TeV. The new data allow for more differential studies to be performed and to probe nuclear gluon distributions at lower values of Bjorken $x$. The most recent ALICE measurements of vector meson production for $\rho^0$, J/$\psi$, and $\psi$(2S) mesons will be presented along with comparisons to the latest available models.

## 1 Introduction

Ultra-peripheral collisions (UPC) occur when two colliding particles pass by one another with an impact parameter, $b$, larger than the sum of the radii of the two particles: $b > R_1 + R_2$. Hadronic interactions are greatly suppressed in these collisions whereas long-range electromagnetic interactions between the nuclei may still proceed. These electromagnetic interactions are enhanced in Pb-Pb collisions compared to p-p collisions at the same energy because of the much larger charge of the Pb ions. The reason is that at ultra-relativistic energies, the electromagnetic field of each ion can be replaced by an equivalent flux of quasi-real photons. When a UPC event occurs, a photon from either nuclei may interact with either a photon from the other nucleus, the other nucleus as a whole or a single nucleon in the other nucleus, producing new particles. These interactions are referred to, respectively, as photon-photon interactions, photo-nuclear (coherent) interactions or photon-nucleon (incoherent) interactions. Vector mesons, such as $\rho^0$, J/$\psi$ and $\psi$(2S) are commonly created in these processes. The equivalent flux of photons that may participate in UPC events is enhanced by a factor of 6724 for Pb-Pb collisions compared to p-p collisions due to the larger electric charge of the heavy ions. The energy spectrum of these photons reaches a maximum value which is proportional to the Lorentz factor of the colliding particles. Measurements of heavy ion interactions at the energies available at the LHC provide an ideal environment to study interactions in UPC events.

Measurements of ultra-peripheral collisions provide insight into both the production mechanisms for heavy vector mesons and the initial distribution of gluons inside nuclei. Cross sections for coherent photo-production of vector mesons are proportional to the square of the nuclear gluon distribution. Measurements at different rapidities and energies, both with and without accompanying photo-nuclear dissociation of the nuclei, provide experimental constraints for these distributions over a range of values of Bjorken $x$. The studies that can be performed at ALICE with Run 2 data will provide results at lower values of Bjorken $x$, where theoretical uncertainties in the nuclear gluon distribution are largest and experimental guidance is especially needed. Because different models may include different amounts of gluon shadowing and may incorporate different assumptions, measurements of differential cross sections, $d\sigma/dy$, can help distinguish between models.

The results presented in this paper focus on the most recent ALICE measurements of UPC events in Pb-Pb collisions. An overview of the detector configuration and triggers after the upgrade for Run 2 is provided in Sec. 2. In Sec. 3 new results and comparison to models are discussed for coherent vector meson production of $\rho^0$, and $\psi$(2S) at mid-rapidity and J/$\psi$ at both forward and mid-rapidity.

## 2 ALICE detector and UPC triggers

The ALICE detector is optimized for measurements of heavy ion collisions produced at the LHC. Detailed overviews of the detector and its performance can be found in Refs. [1] and [2], respectively. The detectors currently used for UPC measurements include the Time Projection Chamber (TPC), Time-of-Flight (TOF) and Silicon Pixel Detector (SPD) at mid rapidities and the Muon Arm at backward rapidity. Additionally, the V0, ALICE Diffractive (AD) and Zero Degree Calorimeters (ZDC), which have components in both the forward and backward rapidity regions, are utilized. Each of these detectors are used either for tracking and particle identification, as part of the trigger for selecting UPC events, or in some cases both. The addition of the AD detector for Run 2 enhances the ability to suppress events that may contain hadronic interactions. A summary of the subdetectors, their rapidity coverage, function in UPC measurements and specific requirements are given in Table 1.

**Table 1:** Summary of ALICE UPC triggers for Run 2

| Detector | Rapidity | Function | Requirements |
|---|---|---|---|
| | Central trigger | | |
| TPC | $\|\eta\| < 0.9$ | Tracking/PID | |
| TOF | $\|\eta\| < 0.9$ | Trigger | $\geq$ 2 back-to-back hits |
| SPD | $\|\eta\| < 1.4$ (outer), $\|\eta\| < 2.0$ (inner) | Trigger | $\geq$ 2 back-to-back hits |
| V0[a] | $-3.7 < \eta < -1.7, 2.8 < \eta < 5.1$ | Trigger | V0A and V0C empty |
| AD[a] | $-7.0 < \eta < -4.9, 4.8 < \eta < 6.3$ | Trigger | ADA and ADC empty |
| | Forward trigger | | |
| Muon Arm[b] | $-4.0 < \eta < -2.5$ | Tracking/PID/Trigger | 2 muons, $p_T > 1GeV$ |
| V0[a] | $-3.7 < \eta < -1.7, 2.8 < \eta < 5.1$ | Trigger | Only V0A empty |
| AD[a] | $-7.0 < \eta < -4.9, 4.8 < \eta < 6.3$ | Trigger | ADA and ADC empty |
| SPD | $\|\eta\| < 1.4$ (outer), $\|\eta\| < 2.0$ (inner) | Trigger | No SPD hits |

[a] V0 and AD have components on both the "A" side (V0A, ADA) and "C" side (V0C, ADC).

[b] The Muon Arm is located on the "C" side of ALICE in the backward rapidity region.

For UPC measurements at mid-rapidity, the TPC provides tracking and particle identification. The central UPC trigger requires back-to-back hits in the SPD near the beam pipe and back-to-back hits in the TOF, just outside of the TPC. This signal selects events with at least two back-to-back tracks. In central UPC events there should be no activity in the forward or backward regions of the detector. The V0 and AD detectors are therefore required to have no signal, effectively suppressing events with hadronic interactions. The $\rho^0$ analysis used a separate trigger class without the TOF requirement.

For UPC measurements at backward rapidity, the Muon Arm provides tracking and particle identification. In addition, it is used as part of the trigger by requiring only two muons with $p_T > 1.0$ GeV/c be identified. In this case the SPD is required to have no signal, indicating the central region of the detector is empty. The V0 opposite the Muon Arm and the AD on both sides are required to have no signal, providing a veto to suppress events with hadronic interactions. The other V0 lies in front of the Muon Arm so the muons may pass through it; it is excluded from the veto in the case of the forward UPC trigger. The ZDCs lie at very forward and backward rapidities and may be used to study UPC events in which one or more of the colliding nuclei may photo-dissociate and emit neutrons. They will allow measurements at lower values of Bjorken $x$, providing tighter constraints on gluon distribution functions.

## 3 Latest results

Using Run 1 data, ALICE published measurements of $\rho^0$ [3] and $\psi(2S)$ [4] at mid-rapidity and of J/$\psi$ at both forward [5] and mid-rapidity [6] for Pb-Pb collisions at $\sqrt{s_{NN}} = 2.76$ TeV. The latest results, using Run 2 data, benefit from significantly higher statistics that were obtained because of higher luminosity, the addition of the AD detector to enhance the UPC triggers, and the higher energy of the collisions. The higher statistics allow for more differential measurements and smaller uncertainties, thereby providing improved constraints for models.

### 3.1 Coherent $\rho^0$

The coherent $\rho^0$ measurement is made in the $\pi^+\pi^-$ channel. The two pion invariant mass spectrum is shown in Fig. 1 (left). To extract the $\rho^0$ yield, a Söding fit of the form

$$\frac{d\sigma}{dm_{\pi\pi}} = |A \cdot BW + B + C \cdot e^{i\phi} \cdot BW|^2 + N \cdot Pol6 \qquad (1)$$

was performed both with and without the $\rho^0$-$\omega$ interference term (third term on the right). The mass and width for $\rho^0$ were fixed to the PDG values. Background was estimated using a $6^{th}$ order polynomial (Pol6). The $\rho^0$ contribution is represented by a Breit-Wigner distribution (BW). To obtain the $\rho^0$ yield, the Breit-Wigner component obtained from the fit was integrated between the limits $2m_\pi$ to $M_\rho + 5\Gamma_\rho$.

The coherent cross section at mid-rapidity ($|y| < 0.5$) is $d\sigma/dy = 448 \pm 2$ (stat)$^{+38}_{-75}$ (sys) mb. Comparing to available model predictions, the Run 2 measurement is consistent with STARLIGHT [7], as was the Run 1 measurement. The GKZ model [8,9] overpredicts the data. The GM model [10], which agreed with the Run 1 measurement within uncertainties, overpredicts the $\rho^0$ cross section measured with Run 2 data.

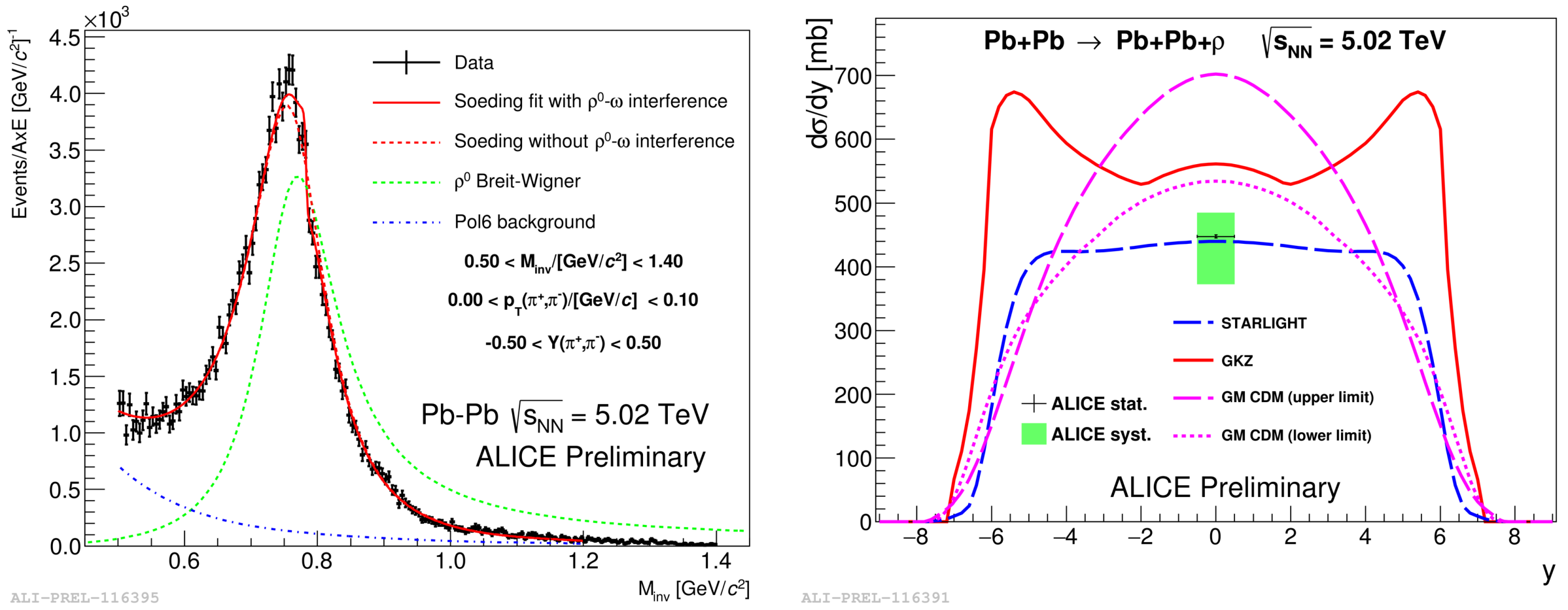

**Fig. 1:** (left) Run 2 $\pi^+\pi^-$ invariant mass spectrum and fits for the coherent $\rho^0$ analysis and (right) coherent $\rho^0$ cross section and available model predictions.

### 3.2 Coherent J/$\psi$

Current results for J/$\psi$ in Run 2, at forward rapidity ($-4.0 < |y| < -2.5$) are shown in Fig. 2 for the di-muon channel. The $p_T$ distribution is shown on the left along with Monte Carlo templates used to reproduce the distribution. The templates reproduce the data quite well. On the right, the di-muon invariant mass spectrum is shown. The J/$\psi$ and $\psi$(2S) peaks are fit with Crystal Ball functions. The $\psi$(2S) peak has a $3\sigma$ significance.

While Fig. 2 shows results for all the data at forward rapidity, the much higher statistics in Run 2 allow for more differential measurements. Results obtained for three regions of forward rapidity are shown in Fig. 3 (left). STARLIGHT [7] and models using the impulse approximation overpredict the data. As in Run 1, the cross sections measured in Run 2 are consistent with models that incorporate moderate nuclear gluon shadowing. The CGC model, LM IPsat [11], appears consistent with the data although the range reported for that model only reaches to one of the data points. The data are also consistent with the EPS09 leading order prediction [8] within uncertainties, as was the Run 1 data.

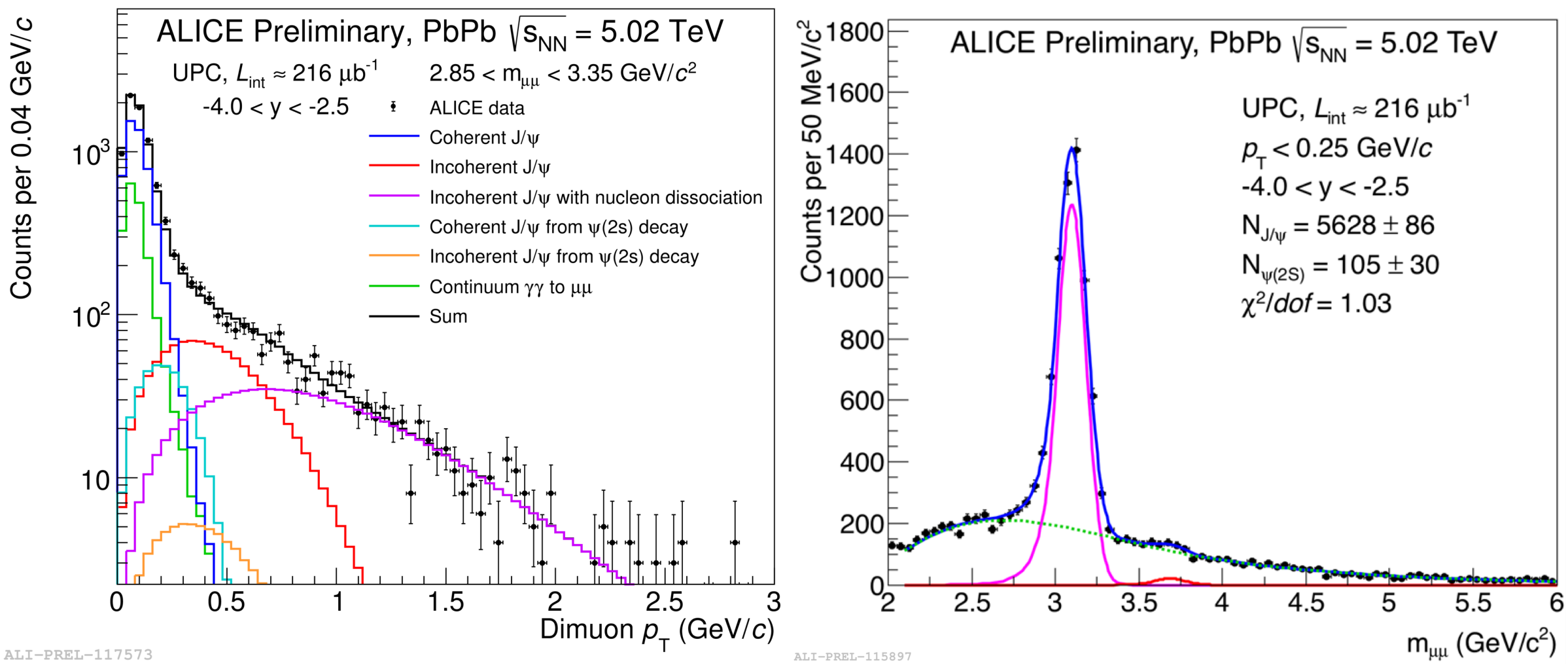

**Fig. 2:** (left) Run 2 coherent J/$\psi$ $p_T$ distribution and Monte Carlo templates that reproduce it well. (right) Di-muon invariant mass spectrum for J/$\psi$ and $\psi$(2S) and fits using Run 2 data.

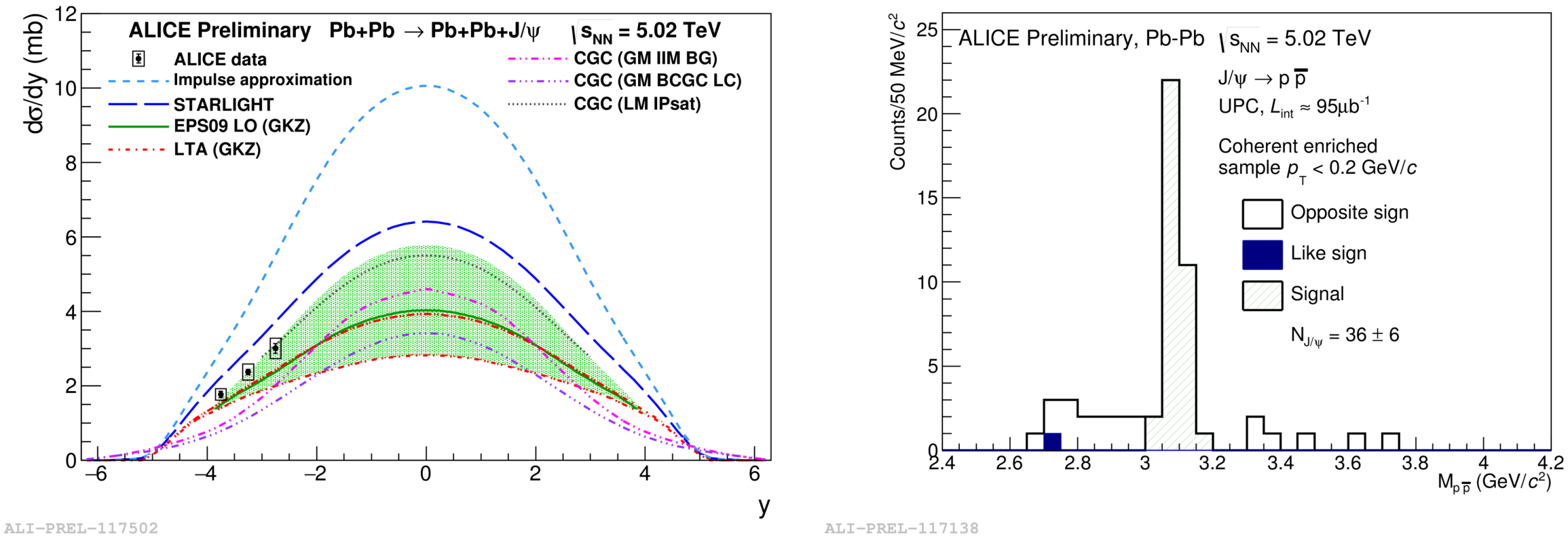

**Fig. 3:** (left) Cross section measurement, from Run 2, for coherent J/$\psi$ production at forward rapidities with comparison to models. (right) Using Run 2 data, the first observed signal for J/$\psi \rightarrow p\bar{p}$ in UPC events.

Similar measurements for J/$\psi$ at mid-rapidity are in progress. The mid-rapidity study will benefit from larger statistics compared to Run 1 and should provide access to $x \approx 5 \cdot 10^{-4}$. One particularly interesting measurement at mid-rapidity is shown in Fig. 3 (right). This is the first observation of J/$\psi \rightarrow p\bar{p}$ in ultra-peripheral collisions. There are 36$\pm$6 candidates although this signal may also contain entries from $\gamma\gamma \rightarrow p\bar{p}$.

### 3.3 Coherent $\psi$(2S)

The $\psi$(2S) signal, extracted from 95 $\mu$b$^{-1}$ of data obtained in Run 2, are shown in Fig. 4 for $\psi$(2S)$\rightarrow$ $\mu^+\mu^-\pi^+\pi^-$ (left) and $\psi$(2S)$\rightarrow$ $e^+e^-\pi^+\pi^-$ (right). These results are for measurements at mid-rapidity.

During Run 1, the ratio of the mid-rapidity cross sections for J/$\psi$ to $\psi$(2S) was found to be quite large, $0.34^{+0.08}_{-0.07}$ (stat+syst), compared to expectations from $\gamma p$ measurements and nearly all model predictions [4]. With the current results for Run 2, using J/$\psi$ and $\psi$(2S) cross sections at forward rapidity, the ratio is found to be similar to that measured by HERA in $\gamma p$ interactions. The HERA result was 0.166±0.011 [12].

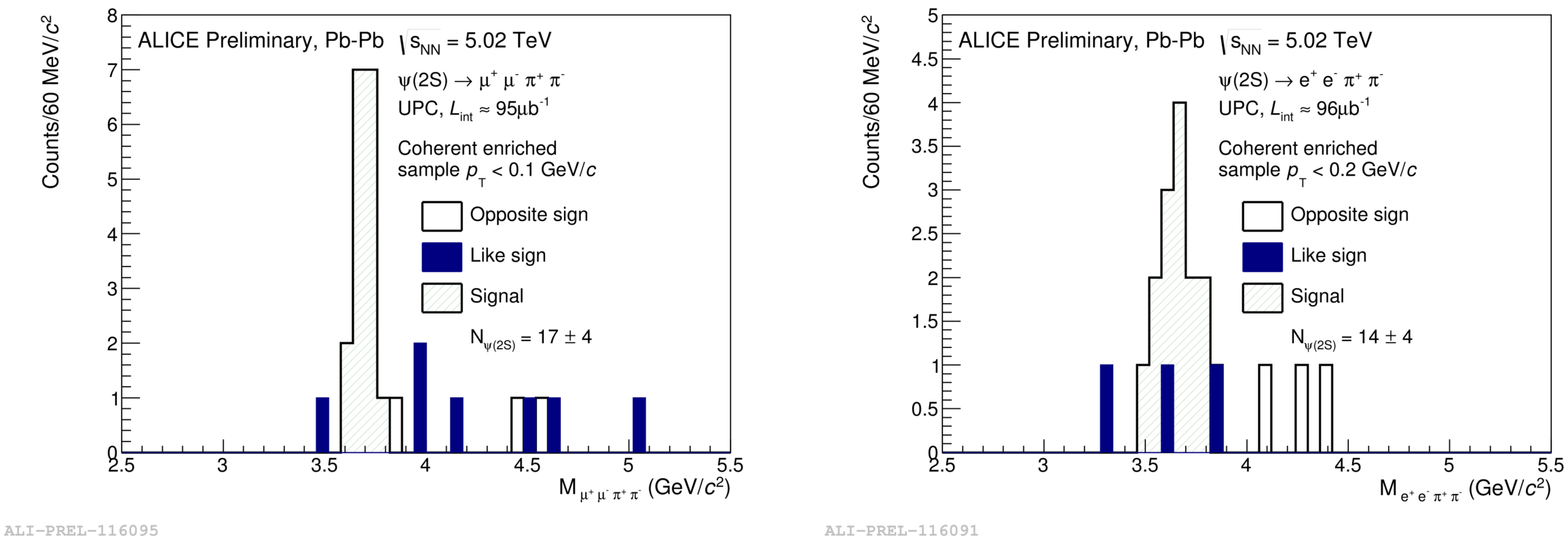

**Fig. 4:** Observed signals, in Run 2, for the $\psi$(2S) invariant mass spectra in the $\mu^+\mu^-\pi^+\pi^-$ channel (left) and $e^+e^-\pi^+\pi^-$ channel (right) at mid-rapidity.

## 4 Outlook and Summary

The latest UPC measurements from ALICE using data from Run 2 have significantly higher statistics compared to Run 1 allowing more differential measurements and smaller uncertainties, as well as opportunities for new measurements. The J/$\psi$ cross section was measured in three forward rapidity ranges and the first observation of $J/\psi \rightarrow p\bar{p}$ in UPC events has been obtained. The $\rho^0$ cross section measurement places new constraints on the available models at mid-rapidity. The ratio of cross sections for J/$\psi$ and $\psi$(2S) at forward rapidity is found to be consistent with results from HERA $\gamma$p measurements and expectations from theory, much lower than the mid-rapidity measurement from Run 1.

Future plans for ALICE UPC measurements include studying events with different amounts of nuclear dissociation by counting very forward emitted neutrons in the ZDCs. In addition to ongoing measurements of $\rho^0$, J/$\psi$ and $\psi$(2S), efforts are underway to measure additional particles.

## Acknowledgements

This work has been supported in part by the U.S. Department of Energy, Office of Science.

# Photon interactions in ultra-peripheral heavy-ion collisions in the ATLAS detector at the LHC

*Samuel Webb on behalf of the ATLAS Collaboration*
Johannes Gutenberg University Mainz, Mainz, Germany

**Abstract**

Two analyses involving photon interactions in ultra-peripheral lead-lead collisions in the ATLAS detector at the LHC are described, namely the study of dijet production in photo-nuclear ultra-peripheral collisions and a measurement of light-by-light scattering. The first, is an important way to probe and constrain nuclear parton distribution functions, which are known to exhibit suppression at low Bjorken-$x$ with respect to proton PDFs, as well as enhancement at higher Bjorken-$x$. Light-by-light scattering is forbidden in classical electrodynamics as it violates the super-position principle and is a fundamental prediction of quantum mechanics. The first direct evidence for this interaction with two quasi-real initial state photons is presented.

## 1 Introduction

Ultra-peripheral collisions (UPC) of heavy ions occur when the impact parameter is greater than twice the nuclear radius. In these events the strong interaction plays a limited role [1] and as such they provide an ideal environment for photon interaction studies. Additionally the high nuclear charge compared to proton-proton collisions gives an enhanced photon flux (by a factor of the nuclear charge, $Z$, squared), which compensates for the smaller integrated luminosity typical for such datasets.

A distinctive signature of such UPC events (which is employed in both analyses described) is that the ion emitting an initial state photon does not generally break-up, leaving a rapidity gap in the detector in the flight direction of the ion. Two ATLAS forward sub-detectors are particularly suited to identifying this characteristic signature, the Zero Degree Calorimeters (ZDC) and Minimum Bias Trigger Scintillators (MBTS). The ZDC are located $\pm$140m either side of the nominal interaction point [2], covering an absolute pseudorapidity range $|\eta| > 8.3$. Figure 1 shows an event display of an event which was obtained using a trigger that required one or more neutrons in the ZDC on one side and none on the other side [3]. The rapidity gap can be seen in the left half of the detector and note that the ZDC itself is not pictured. This trigger was used in the dijet production measurements described in Section 2. The octagonal sub-detectors shaded yellow are the MBTS. These are positioned between the Inner Detector and end-cap calorimeters with a coverage in absolute pseudorapidity between 2.07 and 3.86. The light-by-light scattering analysis (described in Section 3) employs a trigger which rejects events if more than one hit was found in the inner ring of the MBTS.

## 2 Photo-nuclear dijet production in ultra-peripheral Pb+Pb collisions

This section summarises the analysis described in Ref. [4], namely the study of several event-level observables in events with two or more jets produced via photo-nuclear interactions in UPC lead-lead (Pb+Pb) collisions in the ATLAS detector. The data was taken in 2015 at a centre of mass energy per nucleon pair of 5.02 TeV.

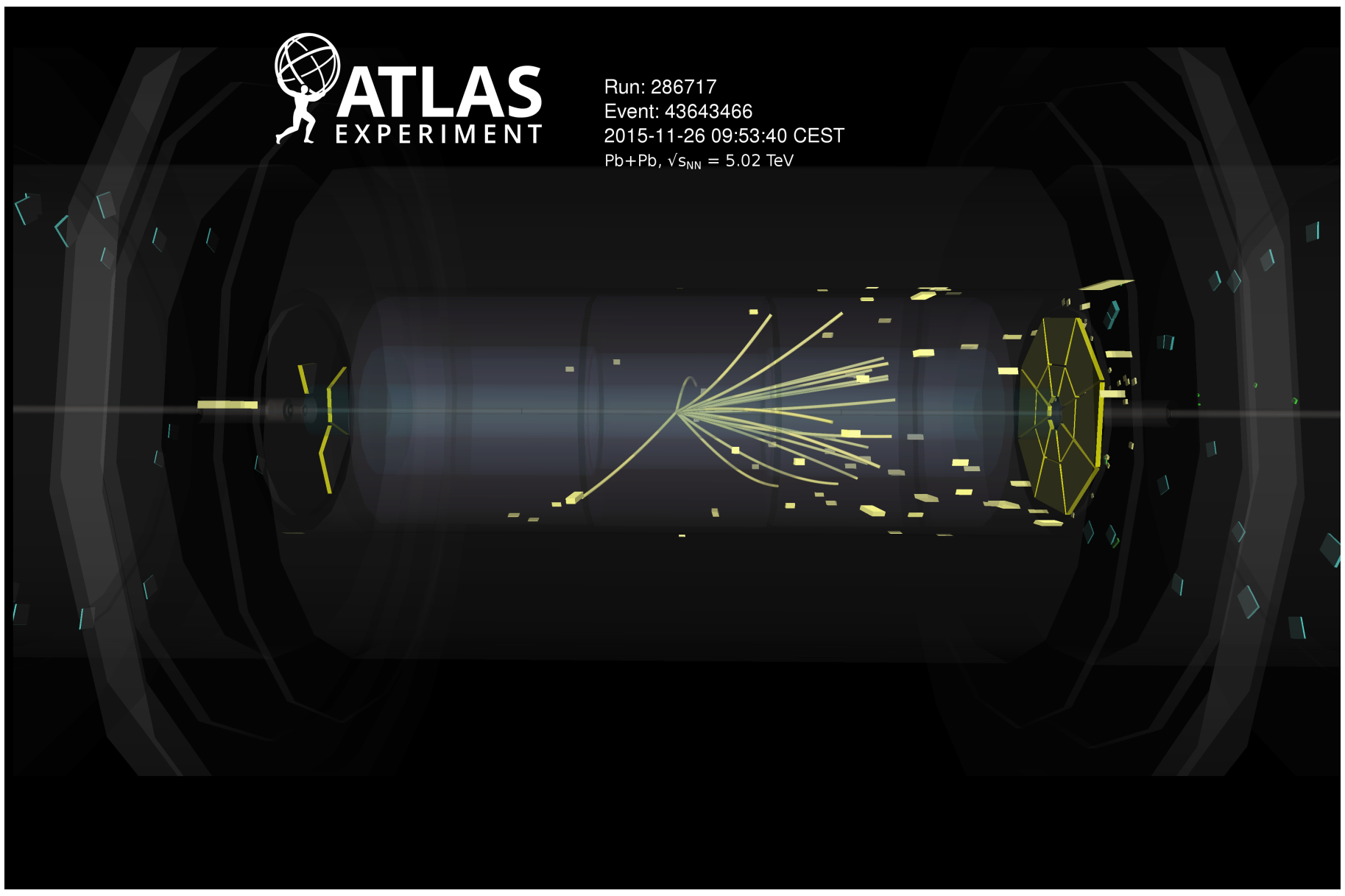

**Fig. 1:** Display of an event with large rapidity gap taken with the ZDC_XOR trigger, firing on more than one spectator neutrons on one side and no neutrons on the other side. The rapidity gap is on the side with no neutrons in the ZDC. The MBTS sub-detectors are also shown, shaded in yellow [3].

### 2.1 Motivation

The motivation for such measurements is to probe the nuclear parton distribution functions (PDFs), which are known to differ from proton PDFs [5]. At parton momentum fraction values below $x \approx 0.03$ the nuclear cross-section is suppressed (known as 'shadowing'), whilst at higher values of $x$ between about 0.03 and 0.3 an enhancement (or 'anti-shadowing') is observed. Knowledge of the nuclear PDFs at low-$x$ is however limited by the lack of experimental data [4] - an issue that this analysis addresses.

In addition to the factor of $Z^2$ enhancement with respect to the same process in proton-proton collisions, the cross-section is further increased by a Lorentz boost factor giving a total enhancement of $1.5 \times 10^6$ extending to initial state photon energies of 50 GeV [4].

### 2.2 Observables and measurements

The process of interest in this measurement can be separated into two scenarios. The first is known as direct photo-production and is shown in the left panel of Figure 2 [4]. In this case the nucleus emitting the initial state photon remains intact forming a clear rapidity gap in the flight direction of the nucleon. The other nucleus breaks up and no rapidity gap is seen. The other scenario is known as resolved photo-production, shown in the right panel of Figure 2. In this case the initial state photon serves as a source of partons which go on to participate in the hard interaction. The rapidity gap is then partially filled.

Events of interest are selected for study using a set of three triggers. Each of these require zero neutrons to be detected in one of the ZDCs (defined to be the photon-going direction), and one or more neutrons in the ZDC on the opposite side, as well as the sum of transverse energy measured in the calorimeters to be between 5 and 200 GeV. Two of the triggers have additional requirements on the transverse momentum, $p_T$, and pseudorapidity of jets. The total luminosity sampled by these triggers is 380 $\mu$b$^{-1}$.

Background events are rejected by employing rapidity gap requirements on each side. Firstly clusters and charged particle tracks are ordered in $\eta$ and intervals between adjacent tracks or clusters with separation $\Delta\eta > 0.5$ are recorded. The rapidity gap sum, $\Sigma\Delta\eta$, is required to be greater than 2 on the photon-going side, and less than 3 on the opposite side (in order to reduce contamination from $\gamma + \gamma$ collisions and non-photo-nuclear UPC processes) [4]. Events are then required to have at least two jets with $p_{\mathrm{T}} > 20$ GeV and $|\eta| < 4.4$. Jets are reconstructed using the anti-$k_t$ algorithm with $R = 0.4$ [6] and with a heavy ion subtraction technique described in Ref. [7].

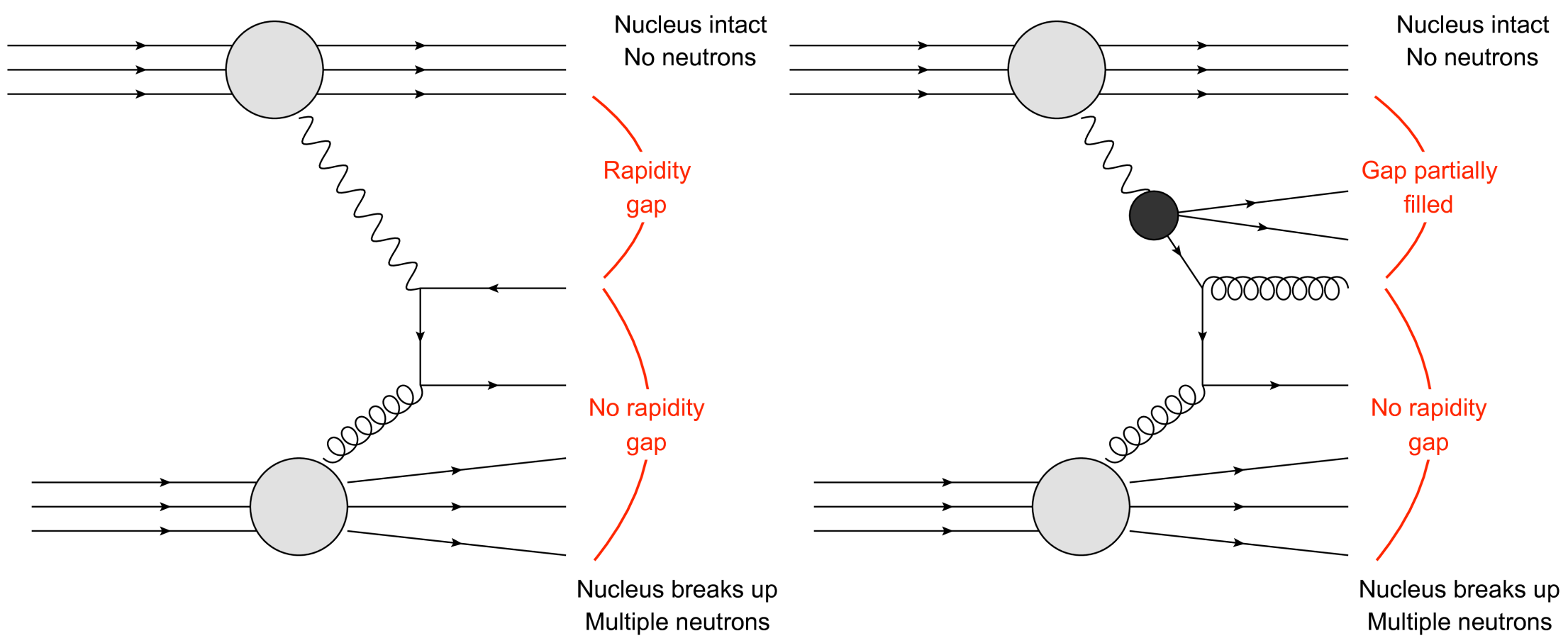

**Fig. 2:** Diagrams representing different types of leading-order contributions to dijet production in high-energy photo-nuclear collisions. The left diagram represents the direct contribution in which the photon itself participates in the hard scattering. The right diagram represents the 'resolved' contribution in which virtual excitations of the photon, into a state involving at least a $q\bar{q}$ pair and possibly multiple gluons, participates in the hard scattering in the target nucleus [4].

Three event-level variables are formed from the selected jets in an event and are defined in Equation 1, where $i$ runs over the measured jets in an event, $E$ is the jet energy, $\vec{p}$ is the jet momentum, and $p_z$ is the longitudinal component of that momentum.

$$H_{\mathrm{T}} \equiv \sum_i p_{\mathrm{T}i}, \quad m_{\mathrm{jets}} \equiv \left[ \left( \sum_i E_i \right)^2 - \left| \sum_i \vec{p_i} \right|^2 \right]^{1/2}, \quad y_{\mathrm{jets}} \equiv \frac{1}{2} \ln \left( \frac{\sum_i E_i + \sum_i p_{z,i}}{\sum_i E_i - \sum_i p_{z,i}} \right) \tag{1}$$

Two derived quantities can then be formed using $m_{\mathrm{jets}}$ (the jet system mass) and $y_{\mathrm{jets}}$ (the jet system rapidity). These are defined in Equation 2, where $x_{\mathrm{A}}$ corresponds to the ratio of the energy of the struck parton in the nucleus to the (per nucleon) beam energy and $z_\gamma$ is the equivalent quantity for the photon (multiplied by the fraction of the photon's energy carried by the resolved parton, in the resolved photo-production scenario).

$$x_{\mathrm{A}} \equiv \frac{m_{\mathrm{jets}}}{\sqrt{s}} e^{-y_{\mathrm{jets}}}, \quad z_\gamma \equiv \frac{m_{\mathrm{jets}}}{\sqrt{s}} e^{+y_{\mathrm{jets}}} \tag{2}$$

A triple-differential cross section is defined in Equation 3, where $\Delta N$ is the number of events measured in a particular bin of width $\Delta H_{\mathrm{T}}$, $\Delta x_{\mathrm{A}}$ and $\Delta z_\gamma$ in the variables $H_{\mathrm{T}}$, $x_{\mathrm{A}}$ and $z_\gamma$ respectively. The integrated luminosity is labelled $\mathcal{L}$ and $\epsilon_{\mathrm{sel}}$ and $\epsilon_{\mathrm{trig}}$ are the selection and trigger efficiencies. Note that no correction is made for the detector response, which is indicated by the tilde on the cross-section $\sigma$. This cross-section is presented in various 2D and 1D slices.

$$\frac{d^3\tilde{\sigma}}{dH_{\mathrm{T}} dx_{\mathrm{A}} dz_\gamma} = \frac{1}{\mathcal{L}} \frac{\Delta N}{\Delta H_{\mathrm{T}} \Delta x_{\mathrm{A}} \Delta z_\gamma} \frac{1}{\epsilon_{\mathrm{trig}} \epsilon_{\mathrm{sel}}} \tag{3}$$

### 2.3 Results and comparison to theory

PYTHIA version 6.41 [8] is used to simulate the photo-nuclear events using the equivalent photon flux from a muon beam as the source of photons. This provides the correct mixture of direct and resolved processes, however the photon energy spectrum in this Pythia version is not appropriate for nuclear collisions. The predicted spectrum is therefore reweighted to that from STARLIGHT [9], a MC model and event generator that has been used to simulate two-photon and photon-pomeron scattering in heavy ion collisions. STARLIGHT has been tested using data from both RHIC and LHC including the ATLAS measurement of $\gamma+\gamma \rightarrow \mu^+\mu^-$ in 5.02 TeV Pb+Pb collisions [10]. STARLIGHT was found in that particular measurement to reproduce the shape and normalisation of the di-muon spectrum well, indicating the photon flux is under control.

The necessity of reweighting the energy spectrum is demonstrated in Figure 3. Here the distributions of $z_\gamma$ and the rapidity gap on the photon-going side, $\Sigma_\gamma \Delta\eta$, are shown for data (black points), the nominal PYTHIA model (blue points), and for PYTHIA reweighted to STARLIGHT (red points). The reweighted model is in much better agreement with data than the nominal model.

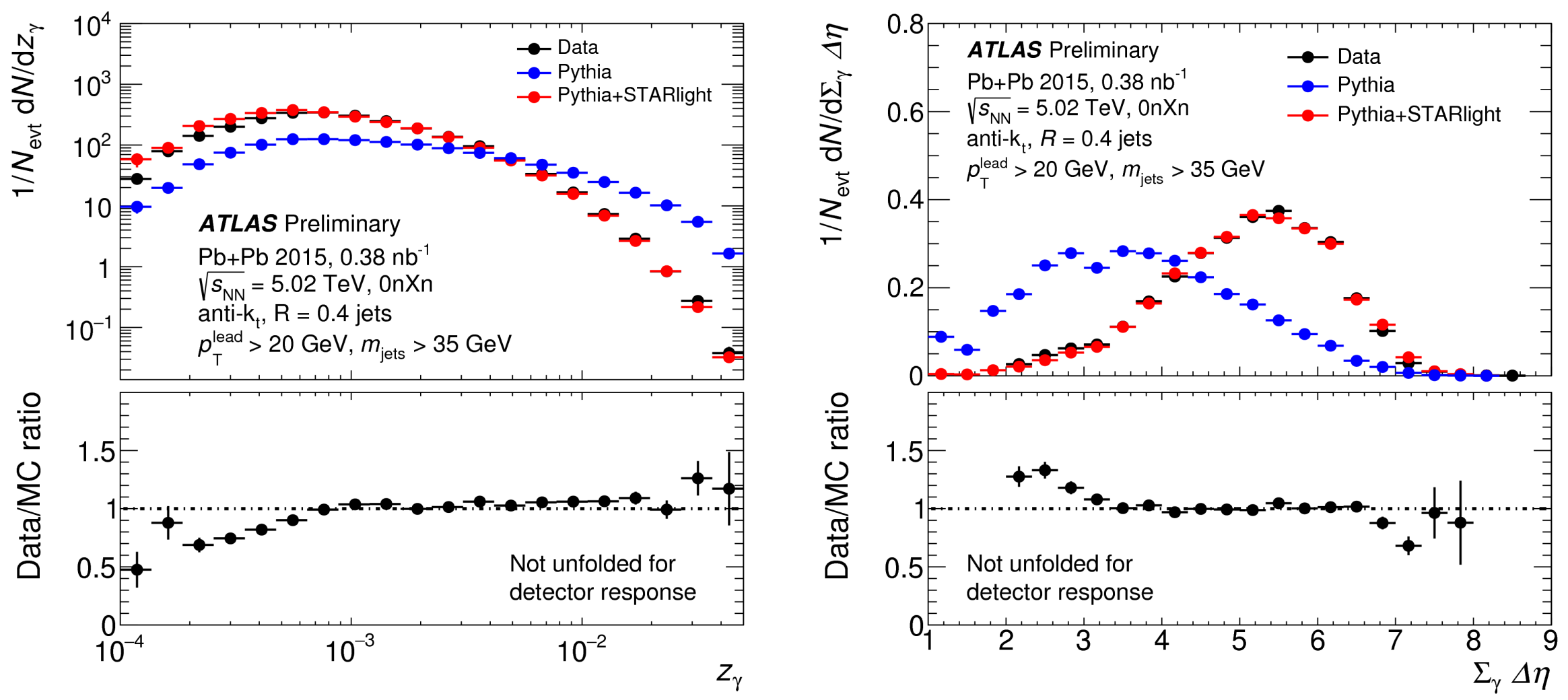

**Fig. 3:** Comparison of data (black points) and PYTHIA MC before (blue points) and after (red points) reweighting to STARLIGHT. The left figure shows the $z_\gamma$ distribution, and the right figure shows the $\Sigma_\gamma \Delta\eta$ distribution. The bottom panels shows the ratio of data to the reweighted model. The error bars represent the statistical uncertainties only. [4].

Two examples of the final cross-section distributions are shown in Figure 4. The left plot shows the double differential cross-section $d^2\tilde{\sigma}/dH_\mathrm{T}dz_\gamma$ as a function of $z_\gamma$ for different intervals of $H_\mathrm{T}$. The right plot shows $d^2\tilde{\sigma}/dz_\gamma dx_\mathrm{A}$ as a function of $x_\mathrm{A}$ for different intervals of $z_\gamma$. Data is shown in comparison to PYTHIA MC reweighted to STARLIGHT. The statistical uncertainty on the data is shown as error bars and the systematic uncertainty is shown as shaded bands. Not shown is the uncertainty on the cross-section normalisation (6.2%) which comes mostly from the integrated luminosity uncertainty (6.1%). The largest systematic uncertainties are due to the event selection efficiency, specifically the requirements on $\Sigma\Delta\eta$ which are between 10-25% depending on $x_\mathrm{A}$, and the jet energy scale and resolution, which can be up to 40% depending on the $H_\mathrm{T}$, $x_\mathrm{A}$ and $z_\gamma$. Generally the model is in good agreement with data, except at the lowest values of $z_\gamma$ and $x_\mathrm{A}$, where the cross-sections are particularly sensitive to the kinematic selections.

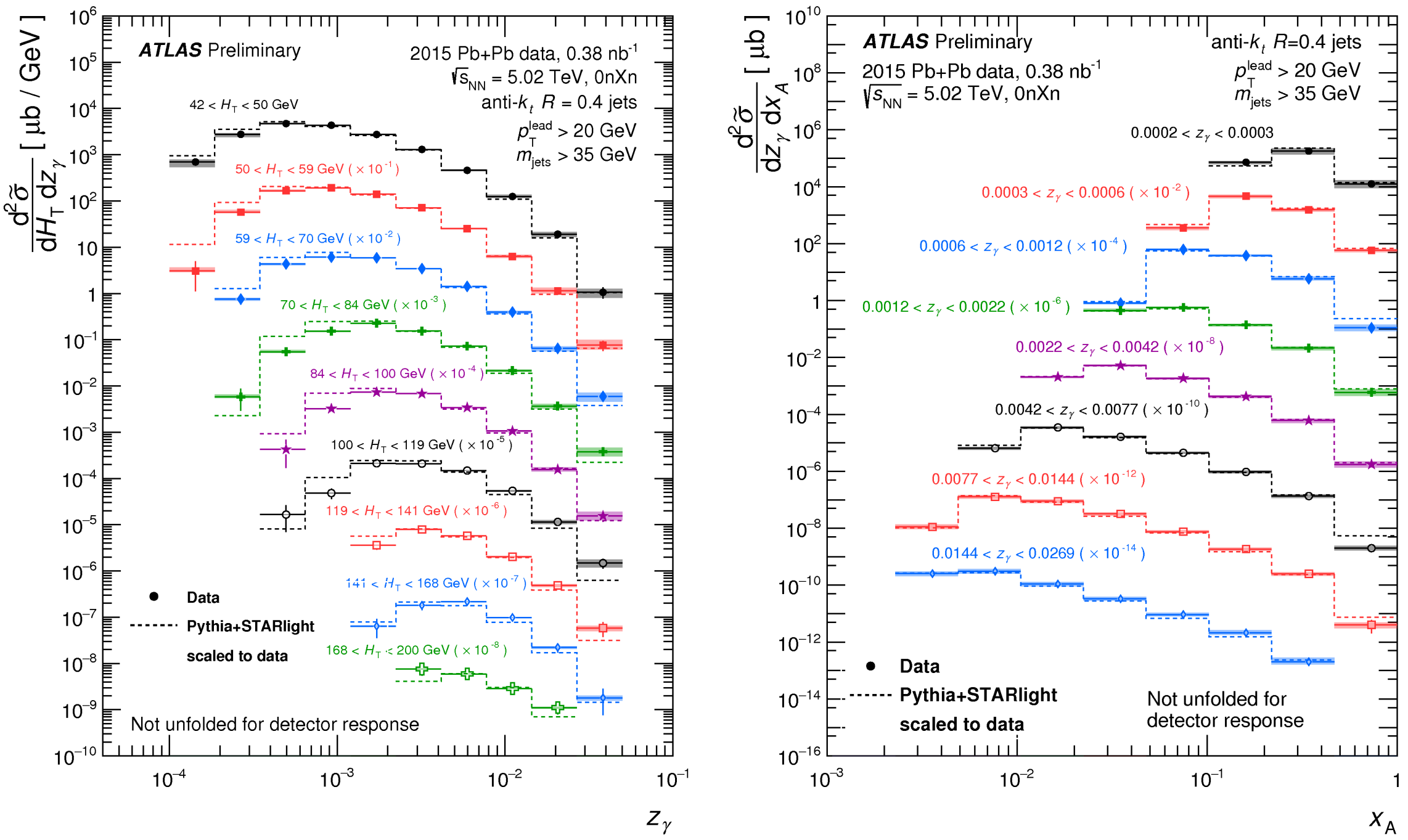

**Fig. 4:** The left plot shows the double differential cross-section $d^2\tilde{\sigma}/dH_\mathrm{T}dz_\gamma$ as a function of $z_\gamma$ for different intervals of $H_\mathrm{T}$. The right plot shows $d^2\tilde{\sigma}/dz_\gamma dx_\mathrm{A}$ as a function of $x_\mathrm{A}$ for different intervals of $z_\gamma$ [4].

## 3 Evidence for light-by-light scattering in heavy-ion collisions with the ATLAS detector at the LHC

This section summarises the analysis described in Ref. [1], which provides evidence for the process $\gamma\gamma \to \gamma\gamma$ or 'light-by-light scattering' in UPC Pb+Pb collisions in the ATLAS detector. This analysis uses the same 2015 dataset as the dijet analysis.

### 3.1 Introduction

The light-by-light scattering process is a purely quantum mechanical effect forbidden in classical electrodynamics as it violates the superposition principle [1]. In QED the reaction proceeds at lowest order via box diagrams involving fermions, which is an $\mathcal{O}\left(\alpha_\mathrm{EM}^4\right)$ process. Feynman diagrams for light-by-light scattering as well as two related processes with higher photon virtuality (Delbrück scattering of a photon in the Coulomb field of a nucleus, and photon-splitting via interaction with external fields) are shown in Figure 5. Delbrück scattering and photon-splitting have previously both been directly observed but evidence for low virtuality (or 'quasi-real') light-by-light scattering exists only indirectly from measurements of the anomalous magnetic moments of the electron and muon.

The signature for such a process in ATLAS is two low-energy photons and no further activity in the central detector. Typically this measurement could not be performed with proton-proton collisions where the average number of interactions per bunch crossing (pileup) is higher. A further advantage of using the Pb+Pb dataset is that the cross-section is enhanced by the high nuclear charge (as for the dijet measurement) [11]. This more than compensates the fact that the integrated luminosity is much smaller than the proton-proton datasets.

The signal light-by-light Monte Carlo (MC) samples are generated using calculations from Ref. [12]. The calculations are convolved with the Pb+Pb equivalent photon approximation spectrum from STARLIGHT. The resulting predictions are cross-checked with those from Ref. [11] and found to be in good agreement.

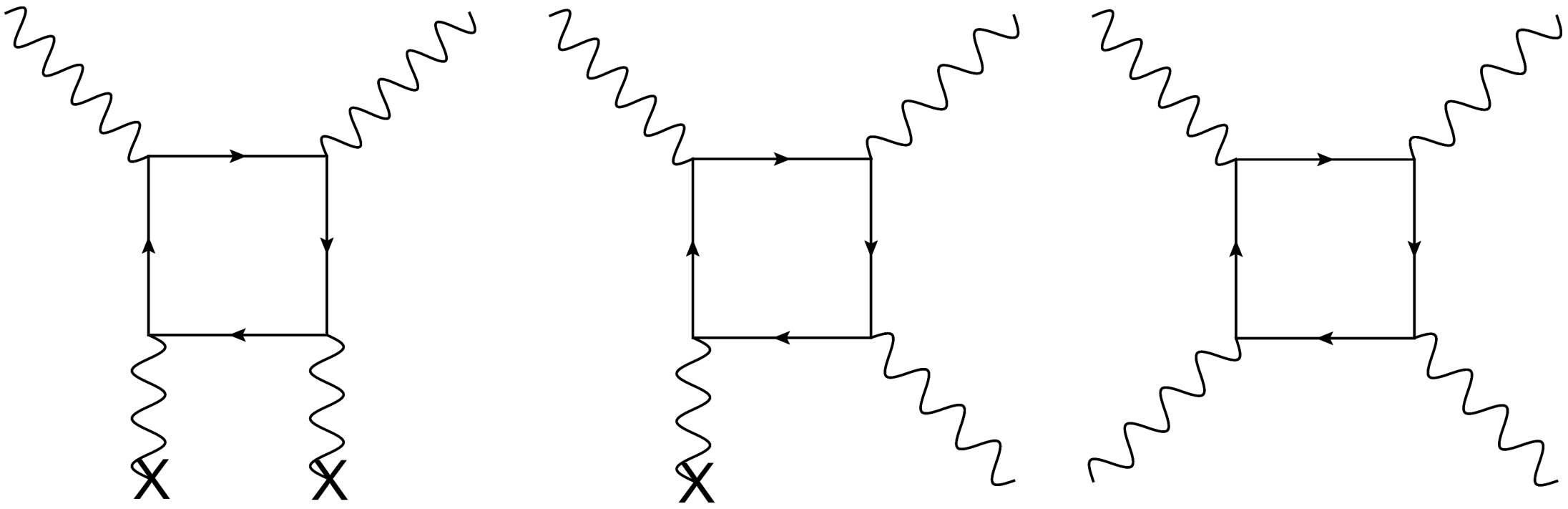

**Fig. 5:** Diagrams illustrating the QED light-by-light interaction processes: Delbrück scattering (left), photon splitting (middle) and elastic light-by-light scattering (right) [1].

The theoretical uncertainties on the model are mainly due to limited knowledge of the electromagnetic form factors and the initial photon fluxes (conservatively estimated to be 20%) and of higher order corrections (a few percent).

### 3.2 Photon identification and event selection

The final state photons in the process of interest typically have a transverse energy, $E_{\mathrm{T}}$, between 3 and 20 GeV, which is lower than in most other ATLAS analyses. Therefore a dedicated photon identification technique is used and detailed studies of the photon reconstruction and identification efficiencies are performed.

Photons are reconstructed from electromagnetic clusters in the calorimeter and with tracking information from the Inner Detector in order to identify photon conversions. The reconstruction efficiency is measured using $\gamma\gamma \to e^+e^-$ events which have emitted a hard bremsstrahlung photon. The data and model are found to have similar efficiencies as a function of the photon $E_{\mathrm{T}}$ and an uncertainty of between $5\%$ and $10\%$ is applied to cover any residual differences.

Photons are identified using three shower shape variables (a sub-set of the variables used for standard ATLAS photon identification). The identification is optimised using a multi-variate technique in order to maintain a constant efficiency of 95% as a function of $\eta$. The identification efficiency is measured using $\gamma\gamma \to \ell^+\ell^-$ events with a final state radiation photon. The efficiencies of the data and of the model are found to agree within the statistical precision and an uncertainty of up to $10\%$ is applied.

The trigger used to select candidate events requires the sum of transverse energy measured in the calorimeters to be between 5 and 200 GeV, there to be 1 hit or less in the MBTS inner ring, and 10 hits or less in the pixel detector. A total luminosity of 480 $\mu$b$^{-1}$ is sampled by this trigger, which has a relative uncertainty of $6\%$. Events are then required to have two photons with an $E_{\mathrm{T}} > 3$ GeV and $|\eta| < 2.37$ excluding the calorimeter transition region $1.37 < |\eta| < 1.52$. The two-photon system is also required to have an invariant mass, $m_{\gamma\gamma}$ above 6 GeV and a transverse momentum less than 2 GeV to reduce backgrounds. Events are vetoed if a charged track is reconstructed in the pixel detector to reduce the background from $\gamma\gamma \to e^+e^-$ events and finally it is demanded that the photons are back to back, i.e. the acoplanarity $\mathrm{Aco} = 1 - \Delta\phi/\pi < 0.01$, to reduce the central exclusive production (CEP) $gg \to \gamma\gamma$ background. 13 data events pass all of the cut criteria. Figure 6 shows the distributions of diphoton invariant mass and absolute rapidity for data and the model. Good agreement is seen between the two.

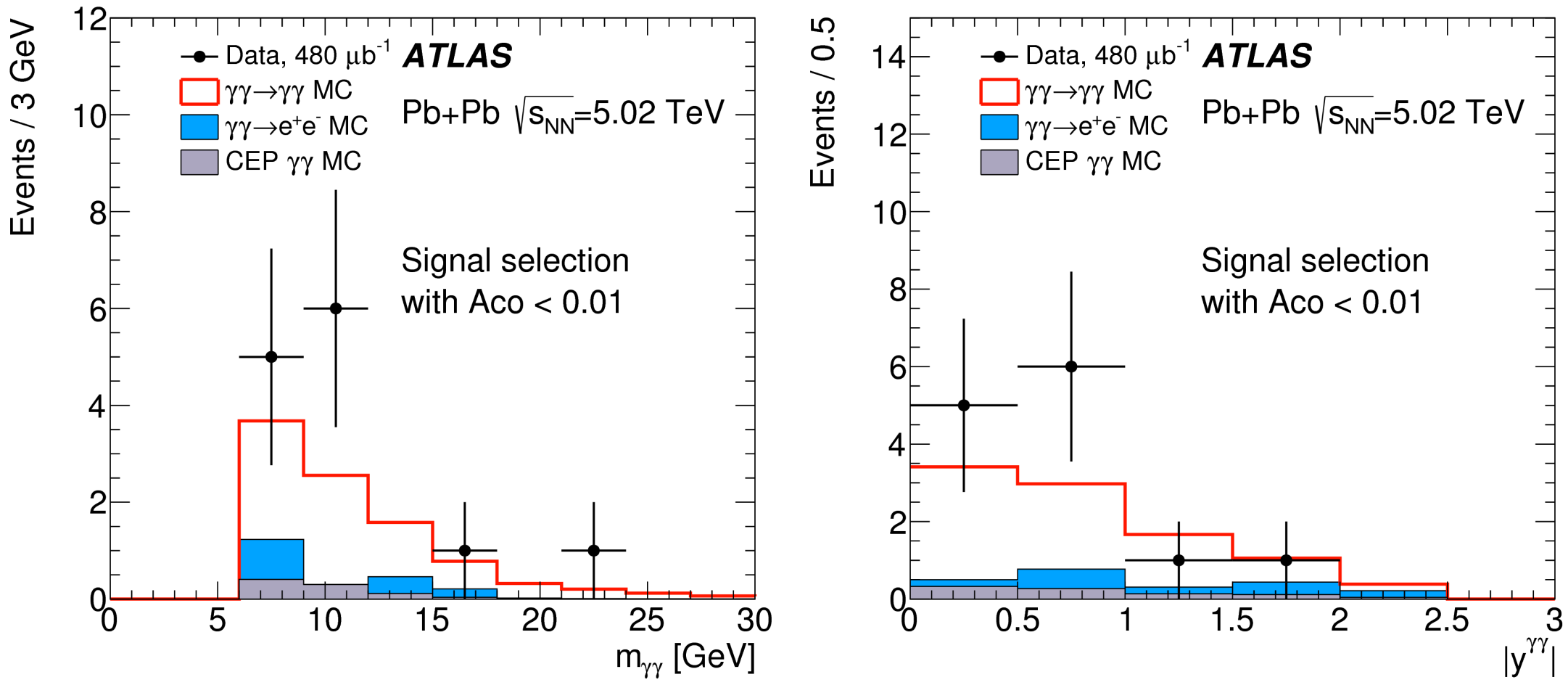

**Fig. 6:** Kinematic distributions for $\gamma\gamma \to \gamma\gamma$ event candidates: the diphoton invariant mass (left), the diphoton absolute rapidity (right). Data (points) are compared to MC expectations (histograms) [1].

### 3.3 Backgrounds

The dominant background comes from the process $\gamma\gamma \to e^+e^-$, due to its high rate as well as the fact that it has a peak at low acoplanarity like the signal. This is estimated with MC and then verified in control regions which require exactly one, or exactly two tracks to be identified. The process contributes 1.3 events in the signal region and is assigned an uncertainty of 25% based on the limited data statistics in the control regions. The other important background is CEP $gg \to \gamma\gamma$, which is again estimated with MC. This has a large theoretical uncertainty so a data-driven normalisation is performed in the acoplanarity distribution. The estimated contribution of this background in the signal region is $0.9 \pm 0.5$ events. Smaller and negligible backgrounds also considered are $\gamma\gamma \to q\bar{q} \to$ multiple $\pi^0$ mesons, exclusive neutral two-meson production and fake events, for example cosmic-ray muons.

### 3.4 Results

Figure 7 shows the final $\gamma\gamma$ acoplanarity distribution (without the Aco cut) [1]. In the region Aco $< 0.01$ 13 data events are observed where 7.3 signal events and 2.6 background events were expected. This gives a significance of $4.4\sigma$ over the background-only hypothesis.

The data is then corrected for selection inefficiencies and the impact of the photon energy and angular resolution to obtain a fiducial cross section of $70 \pm 24$ (stat.) $\pm$ 17 (syst.) nb. The dominant contributions to the total systematic uncertainty come from the photon reconstruction and identification efficiency determination, as well as the photon energy resolution estimate.

## 4 Summary

Ultra-peripheral lead-lead collisions are an ideal environment to study photo-nucleon and photon-photon collisions due to the high photon flux and little additional activity originating from the ion emitting the photon. Two results using events from UPC lead-lead collisions have been presented: dijet production in photo-nuclear interactions, and evidence for light-by-light scattering.

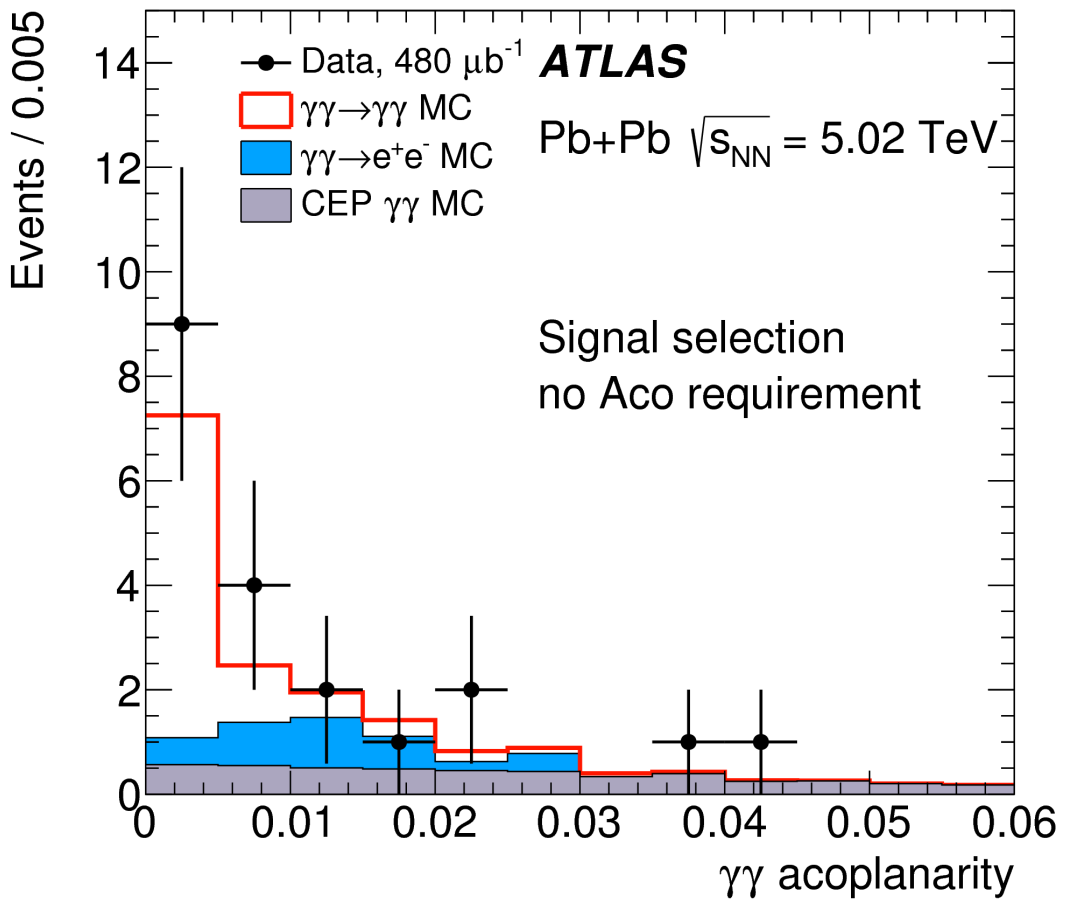

**Fig. 7:** The diphoton acoplanarity before applying Aco<0.01 requirement. Data (points) are compared to MC predictions (histograms). The statistical uncertainties on the data are shown as vertical bars. [1].

# Estimating gluon saturation in dijet photoproduction in UPC

*P. Kotko**
Department of Physics, The Pennsylvania State University, University Park, PA 16802, USA

**Abstract**
We study dijet production in the ultra-peripheral heavy ion collisions (UPC) at LHC within the saturation formalism. More precisely, we use an approach which is the large-$p_T$ approximation to the Color Glass Condensate on one hand, and the small-$x$ limit of the transverse momentum dependent (TMD) factorization on the other. The direct component of the dijet production in UPC at small $x$ probes the so-called Weizsäcker-Williams (WW) TMD gluon distribution, which is not accessible in more inclusive processes, where rather the dipole TMD gluon distribution is probed. Although the WW TMD gluon distribution is not known from data, it can be calculated from the data-restricted dipole TMD gluon distribution using the mean field approximation. Using such approximated WW distribution we calculate various dijet observables in UPC and estimate the saturation effects.

## 1 Introduction

The phenomenon of gluon saturation is most often described within the Color Glass Condensate (CGC) effective theory [1]. Within this picture, a scattering is described by an interaction of a (color) dipole with a shock-wave corresponding to the color field of a nucleus. Depending on the color flow in the particular process, the interaction with the shock-wave involves color averages of various number of the Wilson line operators: average of two Wilson lines (a dipole) appears in simplest inclusive processes, while also quadrupoles and more complicated correlators are possible. In CGC these correlators can be in principle calculated in the classical McLerran-Venugopalan (MV) model [2] with some free parameters, but truly, they contain a non-perturbative information.

Indeed, in certain limit, the CGC correlators can be related [3] to the small $x$ limit of transverse momentum dependent (TMD) gluon distributions, known from the semi-inclusive collinear factorization (see e.g. [4]). It turns out that the TMD approach gives a transparent and universal interpretation to the two fundamental objects appearing in CGC [5]: the correlator of the dipole operator (two Wilson lines), and the correlator of the gluon number operator. The first object appears in the description of inclusive particle production and can be reformulated as an unintegrated gluon distribution (UGD), dubbed thus the dipole UGD. The second object, the true gluon number distribution (called the Weizsäcker-Williams (WW) UGD) does not explicitly appear in formulae for any process within CGC. It is however related to the quadrupole correlator which appears in two particle production processes. In particular, for two particle production in $\gamma A$ collision in the near back-to-back configuration, the WW UGD is the only gluon distribution probed [3]. Within the TMD approach, the two gluon distributions are represented by the hadronic matrix element of bilocal gluon operator with fields displaced in the light-cone and transverse directions. To ensure the gauge invariance, the Wilson links have to be inserted, but their structure turns out to be different for the dipole UGD and the WW UGD. Here, let us only mention that for the latter they can be removed by a choice of gauge so that the gluon number interpretation is apparent.

*In collaboration with K. Kutak, S. Sapeta, A. Stasto and M. Strikman

The dipole UGD is well studied and there are several fits to inclusive DIS data. This is not the case for the WW UGD. In the view of the above discussion, it is a fundamental quantity and needs to be constrained from data as well. The future Electron Ion Collider will definitely provide a good source of data. Before that, however, it is interesting to investigate the ultra-peripheral heavy ion collisions (UPC) at LHC to see what possibilities it provides (for a review of UPC see [6]). In our study presented in [7], which we summarize in the following report, we took the following strategy. First, we have calculated the WW UGD using the so-called Gaussian approximation from the realistic dipole UGD. Next, using such approximated WW UGD we calculate various observables for the direct component of UPC at LHC within a framework similar to [8]. We concentrate mainly on the saturation effects as these are what make the dipole and WW UGDs different.

## 2 Framework

We use the standard approach to set up the UPC collisions: we consider a photon flux from a heavy ion in the equivalent photon approximation (see e.g. [6]). What is truly interesting, is the $\gamma A$ hard collision which we calculate as described in the following section.

Since we are interested in jet production at LHC, we assume that the typical transverse momentum $P_T$ of produced particles is rather large, definitely larger than the saturation scale $Q_s$, $P_T \gg Q_s$. Another requirement is that we want to probe the nucleus at as small $x$ as possible to justify the usage of the saturation formalism. Let us note, that although we deal with rather large $P_T$, we still can be sensitive to the saturation effects. This is because we study a dijet system, and, unlike in the inclusive jet production, the $p_T$ of jets does not translate into the transverse momentum entering the gluon distribution. Rather, it is the dijet imbalance what enters.

The factorization formula within the framework of [8] but adjusted to the present process reads

$$d\sigma_{\gamma A\to 2\,\mathrm{jet}+X} = \sum_{\{q,\bar{q}\}} \int \frac{dx_A}{x_A} \int d^2k_T\, x_A G_1\left(x_A, k_T\right)\, d\sigma_{\gamma g^*\to q\bar{q}}\left(x_A, k_T\right)\,, \tag{1}$$

where $xG_1$ is the Weizsäcker-Williams UGD. The partonic cross section $d\sigma_{\gamma g^*\to q\bar{q}}$ is calculated using the LO amplitude for the process $\gamma g^* \to q\bar{q}$, where $g^*$ denotes the off-shell gluon. It is calculated in the high energy approximation, where the momentum of the gluon has only one longitudinal component, parallel to the parent hadron. That is, taking the momentum of the nucleus to be $p_A$, the momentum of $g^*$ is $k_A^\mu = x_A p_A^\mu + k_T^\mu$, where $p_A^\mu = (1,0,0,-1)\sqrt{s}/2$ and $k_T^\mu = \left(0, k_T^1, k_T^2, 0\right)$. The off-shell gluon couples eikonally to the rest of the process, that is via the vector $p_A$. The amplitude $\gamma g^* \to q\bar{q}$ constructed in such a way is gauge invariant and is essentially the same as used in the high energy factorization (HEF) [9]. In practical Monte Carlo calculations we used the helicity amplitudes calculated using the program described in [10] but extended to quarks. The two-particle phase space is constructed with account of the initial state transverse momentum $k_T$. The factorization formula (1) has two limits which are well settled QCD results: when $k_T \sim P_T \gg Q_s$ the formula recovers the HEF result, because in that dilute limit any saturation effects are gone, in particular then the WW UGD and dipole UGD become equal. Second, in the limit $k_T \sim Q_s \ll P_T$ it reproduces the leading power limit of CGC formula [3].

The formula (1) is not actually completely correct as there is no hard scale dependence in the gluon distribution $xG_1$. Since we aim at rather large transverse momenta of jets $\sim P_T$, the hard scale adequate to the process is of the same order: $\mu \sim P_T$. Thus, in the saturation region $k_T \sim Q_s$ we have $\mu \gg k_T$ which gives rise to Sudakov-type logs which should be resummed. The general formalism to do so was developed in [11] and is rather complicated. Here, instead, we use a physical interpretation of the Sudakov form factor as a probability not to emit partons between scales $k_T$ and $P_T$ and apply it to Monte Carlo generated events. The procedure was described in [12] and has a similar effect on linear evolution as the hard scale dependence in widely used Kimber-Martin-Ryskin (KMR) model [13].

As mentioned in the Introduction, the WW UGD appearing in (1) was calculated using the Gaussian approximation following the methodology of [14] (another possible approach was presented in [15]).

In the Gaussian approximation the relation between the WW UGD $xG_1$ and the dipole UGD $xG_2$ reads:

$$\nabla^2_{k_T} G_1(x,k_T) = \frac{4\pi^2}{N_c S_\perp(x)} \int \frac{d^2 q_T}{q_T^2} \frac{\alpha_s(k_T^2)}{\left(\vec{k}_T - \vec{q}_T\right)^2} xG_2(x,q_T)\, G_2\left(x, \left|\vec{k}_T - \vec{q}_T\right|\right), \quad (2)$$

where $S_\perp(x)$ is the effective transverse area of the target. Instead of using the pure Balitsky-Kovchegov (BK) evolution equation [16,17] for the dipole UGD, we used a more involved Kwiecinski-Martin-Stasto equation [18] with the nonlinear term [19], which includes subleading effects that may be important at non-asymptotically small $x$. This evolution equation reads: (below we set $xG_2(x,k_T) \equiv \mathcal{F}(x,k_T^2)$ for more compact expression):

$$\mathcal{F}(x,k_T^2) = \mathcal{F}_0(x,k_T^2) + \frac{\alpha_s(k_T^2)N_c}{\pi} \int_x^1 \frac{dz}{z} \int_{k_{T0}^2}^\infty \frac{dq_T^2}{q_T^2} \left\{ \frac{q_T^2 \mathcal{F}\left(\frac{x}{z},q_T^2\right)\theta\left(\frac{k_T^2}{z} - q_T^2\right) - k_T^2 \mathcal{F}\left(\frac{x}{z},k_T^2\right)}{|q_T^2 - k_T^2|} \right.$$
$$\left. + \frac{k_T^2 \mathcal{F}\left(\frac{x}{z},k_T^2\right)}{\sqrt{4q_T^4 + k_T^4}} \right\} + \frac{\alpha_s(k_T^2)}{2\pi k_T^2} \int_x^1 dz \left\{ \left(P_{gg}(z) - \frac{2N_c}{z}\right) \int_{k_{T0}^2}^{k_T^2} dq_T^2 \mathcal{F}\left(\frac{x}{z},q_T^2\right) + zP_{gq}(z)\,\Sigma\left(\frac{x}{z},k_T^2\right) \right\}$$
$$- d\frac{2\alpha_s^2(k_T^2)}{R^2} \left\{ \left[ \int_{k_T^2}^\infty \frac{dq_T^2}{q_T^2} \mathcal{F}(x,q_T^2) \right]^2 + \mathcal{F}(x,k_T^2) \int_{k_T^2}^\infty \frac{dq_T^2}{q_T^2} \ln\left(\frac{q_T^2}{k_T^2}\right) \mathcal{F}(x,q_T^2) \right\}. \quad (3)$$

Above $\Sigma(x,k_T)$ is the accompanying singlet sea quark distribution and $R$ is the target radius appearing from the integration of the impact parameter dependent gluon distribution assuming the uniform distribution of matter. The parameter $d$, $0 < d \leq 1$ is set to $d = 1$ for proton and can be varied for nucleus to study theoretical uncertainty. The initial condition $\mathcal{F}_0$ was fitted to the inclusive DIS HERA in [20] with $R \approx 2.4\,\mathrm{GeV}^{-1}$. In what follows we shall name this set KS (Kutak-Sapeta) UGD. The Pb nucleus was modelled using the Woods-Saxon formula $R_A = A^{1/3}R$ where $A$ is the mass number. In the present work we use $d = 0.5$ value for the Pb ion.

## 3 Results

The results for the WW UGD are presented in Fig. 1. Let us notice, in particular, that for large $k_T$ the $xG_1$ and $xG_2$ gluons become equal, as required by the dilute limit of Eq. (1). The cuts for numerical studies are presented in Table 1. We note, that the main issue for saturation studies is that in order to have small $x$ on the nucleus side, the photon flux should be probed at rather larger $x$, for which, however, the flux becomes small. This forces us to go to rather small $p_T$ of jets to see significant effects.

**Table 1:** The kinematic cuts used in calculations of the dijet cross section in the ultra-peripheral Pb-Pb collisions.

| | |
|---|---|
| CM energy | $\sqrt{s} = 5.1\,\mathrm{TeV}$ |
| rapidity | $0 < y_1, y_2 < 5$ |
| transverse momentum | $p_{T1}, p_{T2} > p_{T0},\ p_{T0} = 25,\ 10,\ 6\,\mathrm{GeV}$ |

In the present report we shall concentrate on the results for nuclear modification ratios defined as $R_{\gamma A} = d\sigma_{AA}^{UPC}/Ad\sigma_{Ap}^{UPC}$ that is, the photon flux in both numerator and denominator originates from a nucleus. More results are given in our original work [7]. In Fig. 2 we present the results for $R_{\gamma A}$ as a function of the azimuthal angle between the dijets. Again, the maximal suppression for 6 GeV jets in the back-to-back region is about 20%. The Sudakov resummation model widens the suppression towards smaller $\Delta\phi$. In Fig. 3 we show $R_{\gamma A}$ as a function of the $p_T$ of jets (for the leading and subleading jets). We see that the maximal suppression is about 20% for $p_T$ of jets as low as $\sim$ 6 GeV. Interestingly, the leading twist nuclear shadowing model [21] predicts similar overall suppression, but the slope is different. The Sudakov resummation model changes the spectra only slightly, especially for the subleading jet.

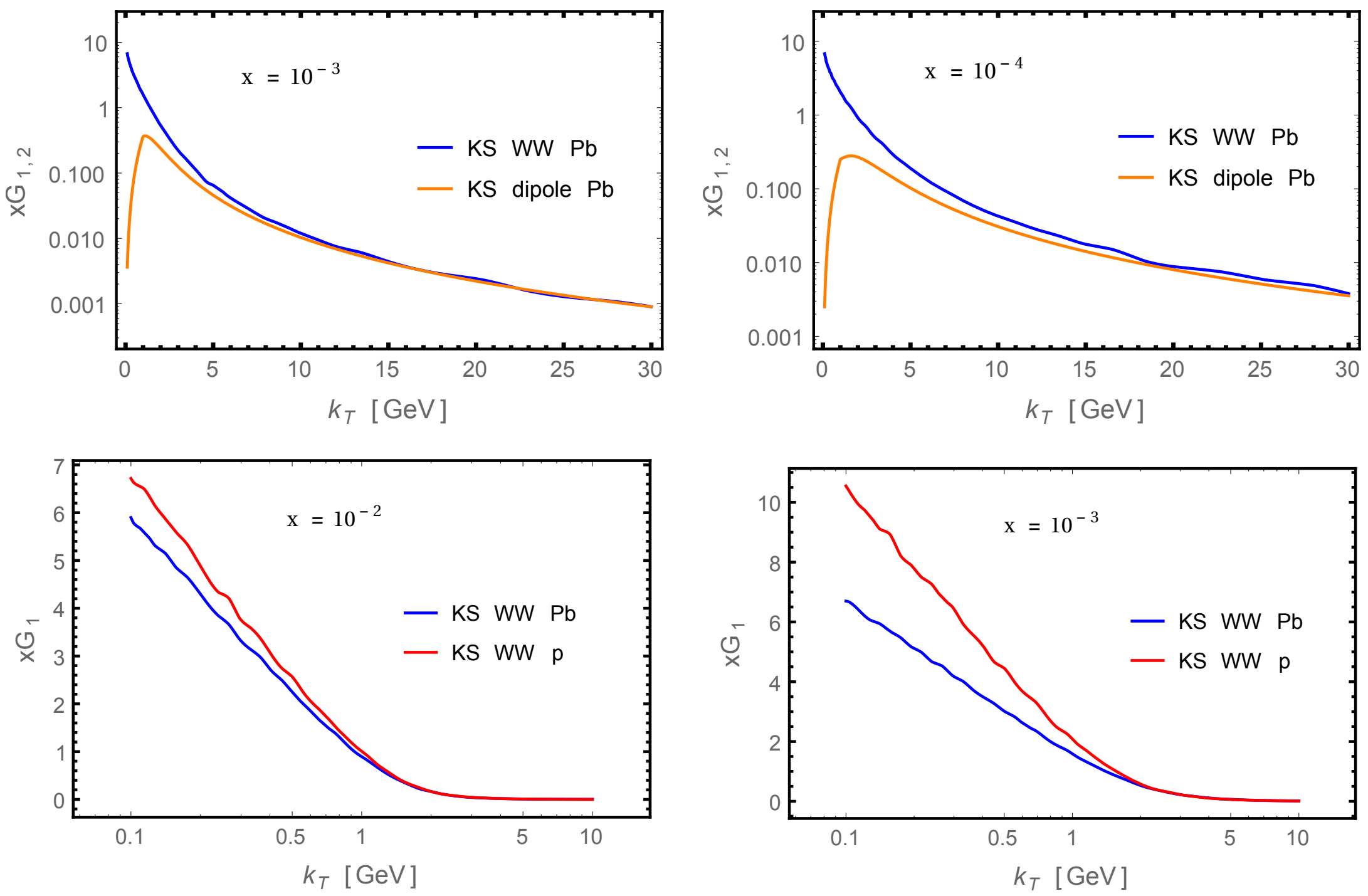

**Fig. 1:** The Weizsäcker-Williams (WW) unintegrated gluon distributions for proton and lead obtained from the KS dipole distributions [20]. The top row compares the WW UGD for Pb with the dipole UGD for Pb for two values of $x$. The bottom row shows the WW UGD for proton and lead as a function of $k_T$ for two values of $x$.

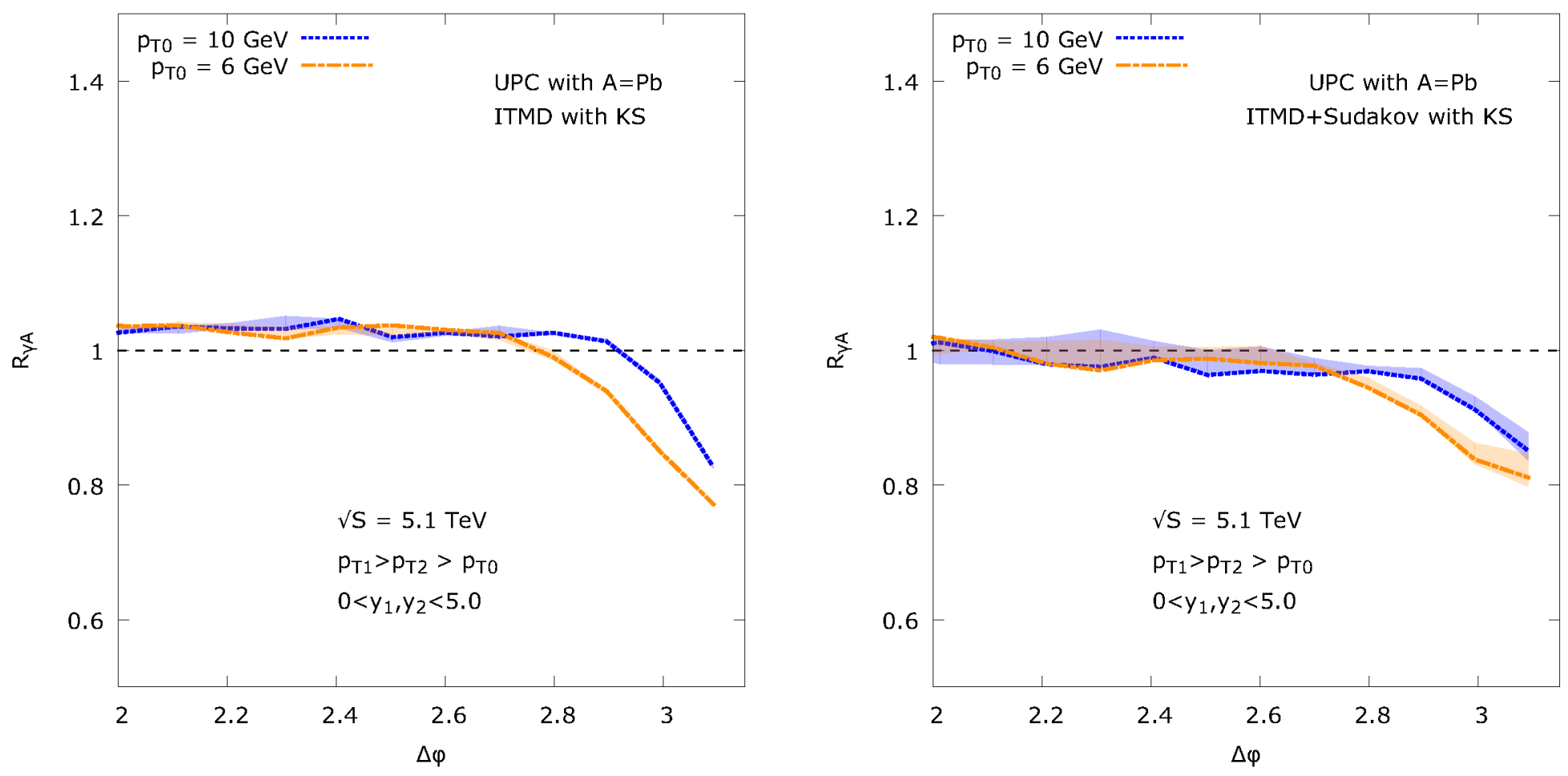

**Fig. 2:** Nuclear modification ratio as a function of the azimuthal angle between the jets, with (right) and without (left) the Sudakov resummation model.

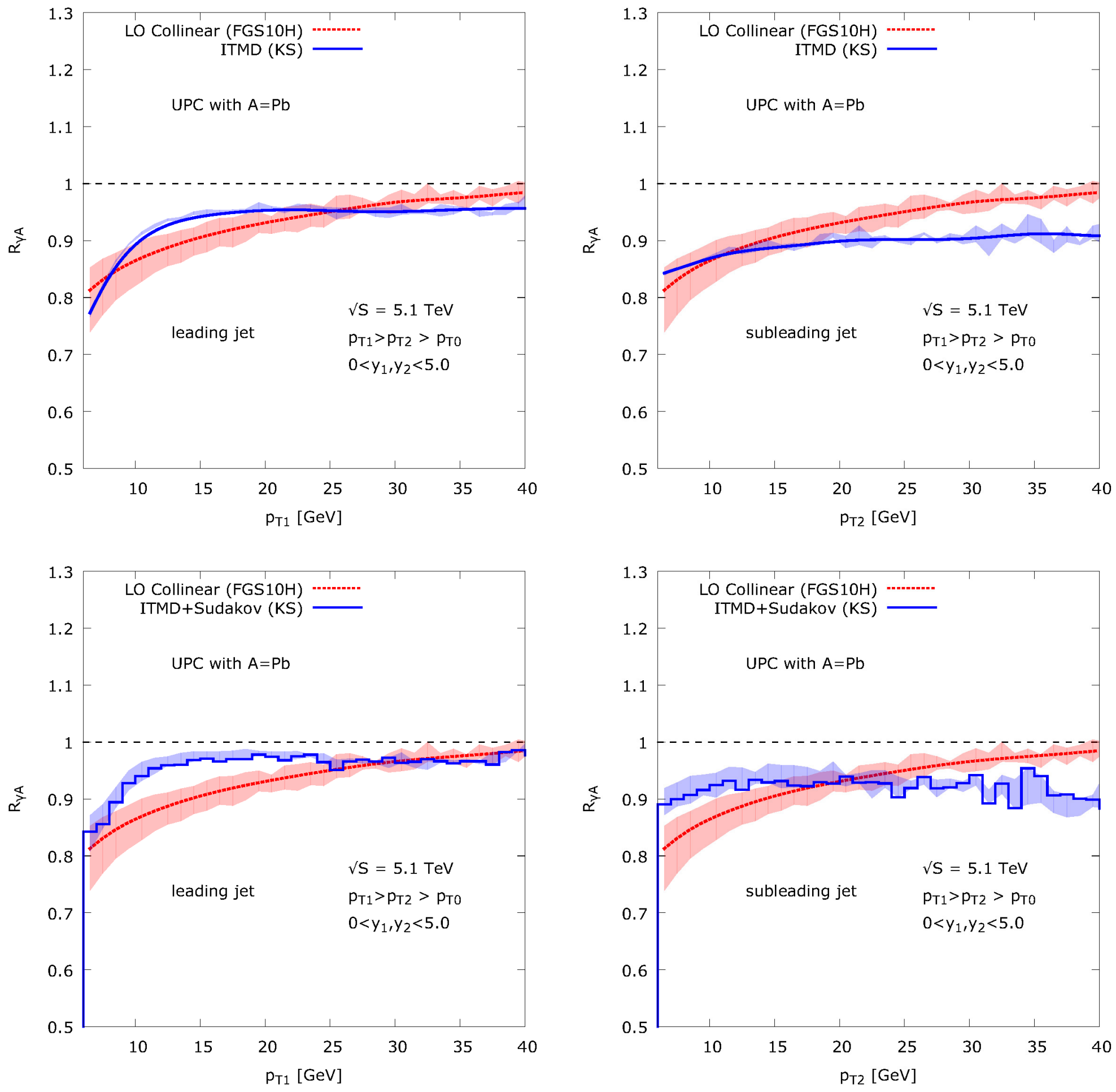

**Fig. 3:** Nuclear modification ratios as a function of the transverse momenta for leading (left column) and subleading (right column) jets. The bottom row shows the effect of the Sudakov resummation model applied to the generated events. For comparison we show the results from the LO collinear factorization using nuclear PDFs with the leading twist nuclear shadowing.

## 4 Conclusions

We have calculated the direct component of the dijet production cross section in ultra-peripheral heavy ion collision at LHC. Such process is sensitive to the Weizsäcker-Williams (WW) unintegrated gluon distribution (UGD) of nucleus, which is basically unknown from the data. According to our calculations with the approximate WW UGD obtained from the data-restricted dipole UGD using the Gaussian approximation, the saturation effects are visible, though moderate. Due to the saturation, we observe a suppression of $\gamma A$ differential cross sections comparing to $\gamma p$ of about 20% in the approximately back-to-back region for the jets with $p_T$ 6-10 GeV. Similar effects are however observed for a different mechanism then saturation, namely the leading twist nuclear shadowing.

## Acknowledgements

I wish to thank my collaborators K. Kutak, S. Sapeta, A. Stasto and M. Strikman for a common work on this project. I also thank A. van Hameren, C. Marquet and E. Petreska for motivating discussions. The work was supported by the DEO grants No. DE-SC-0002145, DE-FG02-93ER40771.

# Color fluctuation phenomena in $\gamma A$ collisions at the LHC

*M. Alvioli* [1], *L. Frankfurt* [2,3], *V. Guzey* [4], *M. Strikman* [3], *M.Zhalov* [4]
[1] Consiglio Nazionale delle Ricerche, Istituto di Ricerca per la Protezione Idrogeologica, via Madonna Alta 126, I-06128 Perugia, Italy
[2] Particle Physics Dept., School of Physics & Astronomy, Tel Aviv University, 69978 Tel Aviv, Israel
[3] Department of Physics, the Pennsylvania State University, State College, PA 16802, USA
[4] National Research Center "Kurchatov Institute", Petersburg Nuclear Physics Institute (PNPI), Gatchina, 188300, Russia

**Abstract**
We explain that color fluctuations (CFs) in the light-cone photon wave function lead to much stronger shadowing in the coherent production both in the soft regime ($\rho$ -meson photoproduction) and in the hard regime ($J/\psi$ photoproduction). We make CF based predictions for the distribution over the number of wounded nucleons $\nu$ in the inelastic photon–nucleus scattering. We show that CFs lead to a dramatic enhancement of this distribution at $\nu = 1$ and large $\nu > 10$. Our predictions can be tested in proton–nucleus and nucleus–nucleus ultraperipheral collisions.

## 1 Introduction

It is instructive to consider hadron (photon) high energy collisions in the target rest frame where the wave function of a projectile is the superposition of coherent (so-called frozen) configurations [1, 2], as a consequence of the uncertainty principle and Lorentz slowing down of the interaction time. In QCD coherence of high energy processes is well understood theoretically and established experimentally, for a review, see, e.g. [3, 4]. A distinctive feature of the QCD dynamics is that the interaction strength of different configurations of quarks and gluons, which are QCD constituents of projectile hadrons, photons, etc., varies. We refer to this phenomenon as color fluctuations (CFs). In the literature one alternatively uses the term cross section fluctuations, which refer predominantly to soft hadron (photon) interactions at high energies.

This space time picture is qualitatively different from the Glauber model where only planar diagrams for the total cross section of a projectile –nucleus collision are considered since this contribution tends to zero with an increase of the collision energy, similar to the case of hadron-hadron interactions [5, 6]. This theoretical puzzle was solved by Gribov in Refs. [1] where contribution of non-planar diagrams was calculated and duality between non-planar diagrams and a sum of the elastic contribution and the diffractive intermediate states (duality between $s$ and $t$ channels) was used to rewrite formulae in the form rather similar to the Glauber approximation [1] with a Glauber like elastic term and inelastic diffraction term.

The relative importance of the inelastic term grows with decrease of the strength of interaction of average configurations in the projectile. So deviations from prediction of the Glauber model are expected to be small for the proton projectile, strongly increase for the pion case and photoproduction of $\rho$ mesons and be even larger for the $J/\psi$ photoproduction. The same effect is present for the nuclear shadowing for the DIS cross section. For example, in the case $\sigma_{tot}(\gamma_L A)$ elastic (dipole) term gives only a higher twist contribution to the shadowing, while multiparton states give the leading twist contribution. Hence the inelastic term dominates in this case.

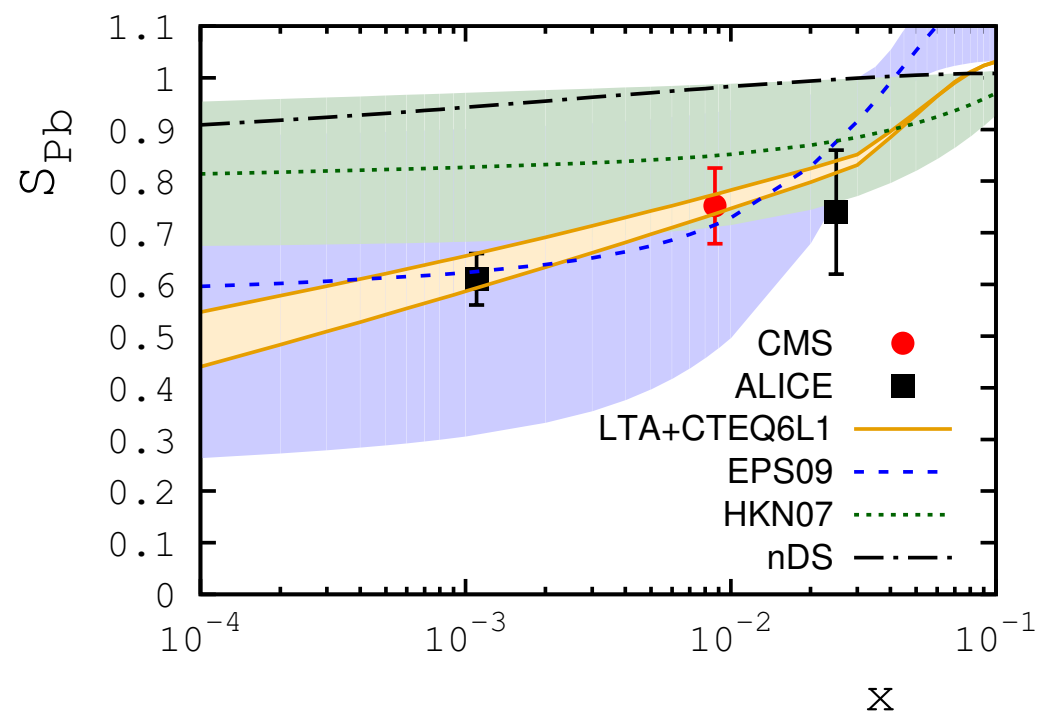

**Fig. 1:** Comparison of the suppression factor $S$ for the $J/\psi$ production extracted [8] from ALICE [9] and CMS [10] data. Prediction of the leading twist gluon shadowing approximation [7] is the yellow band. The range of expectations for the gluon shadowing in a number of models based on the fits to existing DIS data is also shown.

Ultraperipheral collisions at the LHC opened a new avenue for studies of CF since photon wave function contains components of very different size which interact with nucleon with very different strengths. In this talk we consider exclusive coherent vector meson production and effects of CFs for inclusive $\gamma A$ scattering.

## 2 Coherent production of vector mesons off nuclei

Recently coherent production of $\rho$ mesons and $J/\psi$ was studied at the LHC in the ultraperipheral heavy ion collisions. It was observed that the $\rho$-meson cross section is reduced as compared to the impulse approximation by a factor of ten, while in the $J/\psi$ reduction is by a factor of three. Both reductions are much larger than naive expectations. The standard Glauber model predicts a factor of two smaller reduction for $\rho$ production, while eikonal dipole models of $J/\psi$ production predict only a 20% reduction.

It appears that the reason for an underestimate of the reductions is neglect by inelastic intermediate states. In the $J/\psi$ case the effects of the inelastic states can be taken into account by absorbing these effects into the nuclear gluon density, $g_A(x, Q^2)$. The nuclear shadowing $(g_A/g_N)$ can be calculated using gluon diffractive PDFs measured at HERA in $\gamma^* + p \to X + p$ and few other channels (for a review and references see [7]). One finds

$$S_A = \frac{\sigma(\gamma A \to J/\psi A)}{\sigma_{imp.approx.}(\gamma A \to J/\psi A)} = \frac{g_A(x, Q^2)}{g_N(x, Q^2}, \qquad (1)$$

where $x = m^2_{J/\psi}/W^2, Q^2 \approx 3$ GeV$^2$, Our predictions agree well with the LHC data, see Fig.1. Note that elementary amplitudes of $J/\psi$ production are expressed through non-diagonal generalized parton densities. However in $J/\psi$ case light cone fractions of gluons attached to $c\bar{c} - x_1$ and $x_2$ are comparable: $x_1 \sim 1.5x, x_2 \sim x/2$ so that $(x_1 + x_2)/2$ is close to x.

In the case of $\rho$ production the challenge is that the coherent cross section is described very well by the Glauber model for moderate energies $\sim 10$ GeV [11], so a factor of two larger shadowing observed by ALICE [12] signals presence of new physics. The Gribov theory of inelastic shadowing provides the framework to take into account a different picture of high energy scattering. The configurations which are present in the intermediate states are frozen and cannot go back to $\rho$ during the passage of the nucleus.

In the hadronic basis one needs to include not only transitions $\gamma \to \rho \to \rho$ but also transitions $\gamma \to M_X \to \rho$ (for scattering off two nucleons). It has been suggested that the interaction matrix of the initial hadron or diffractively produced hadronic states with target nucleons, which arises within Gribov–Glauber approach, can be diagonalized [13, 14]. In the particular case, when diffractive intermediate states are resonances, this diagonalisation has been performed in Ref. [15]. The method of CFs developed in [16] and discussed below is the further generalization of the Gribov–Glauber approximation, which allows one to account for the fluctuations of the interaction strength and other implications of QCD.

For the soft dynamics we need to introduce $P(\sigma)$ – probability that the frozen configuration of the projectile interacting with the nucleus has interaction cross section $\sigma$. In the case of $\rho$ coherent photoproduction this amounts to the presence of the addition factor $P(\sigma)$ in the Glauber expression:

$$\sigma_{\gamma A\to\rho A} = \left(\frac{e}{f_\rho}\right)^2 \int d^2\vec{b} \left| \int d\sigma P(\sigma) \left(1 - e^{-\frac{\sigma}{2}T_A(b)}\right) \right|^2 , \tag{2}$$

Based on the similarity between the pion and $\rho$ meson wave functions suggested by the additive quark model, it is natural to assume that $P(\sigma)$ for the $\rho N$ interaction should be similar to the pion $P_\pi(\sigma)$, which we additionally multiply by the factor of $1/(1+(\sigma/\sigma_0)^2)$ to take into account the enhanced contribution of small $\sigma$ in the photoproduction due to singular behavior of the photon wave function at small quark-antiquark separations leading to

$$P(\sigma) = C\frac{1}{1+(\sigma/\sigma_0)^2} e^{-(\sigma/\sigma_0-1)^2/\Omega^2} . \tag{3}$$

The parameterization of Eq. (3) satisfies the basic QCD constraint of $P(\sigma=0) \neq 0$ and also $P(\sigma \to \infty) \to 0$. The free parameters $C$, $\sigma_0$ and $\Omega$ are found from the following constraints:

$$\int d\sigma P(\sigma), \int d\sigma P(\sigma)\sigma\langle\sigma\rangle , \int d\sigma P(\sigma)\sigma^2 = \langle\sigma\rangle^2(1+\omega_\sigma) , \tag{4}$$

where $\langle\sigma\rangle = \hat{\sigma}_{\rho N}$ in the modified VMD model, and $\omega_\sigma$ is equal to the ratio of inelastic and elastic diffraction at t=0, see discussion in [17]. The results of calculation are in a reasonable agreement with the data obtained at the LHC [12] and at RHIC [18–20].

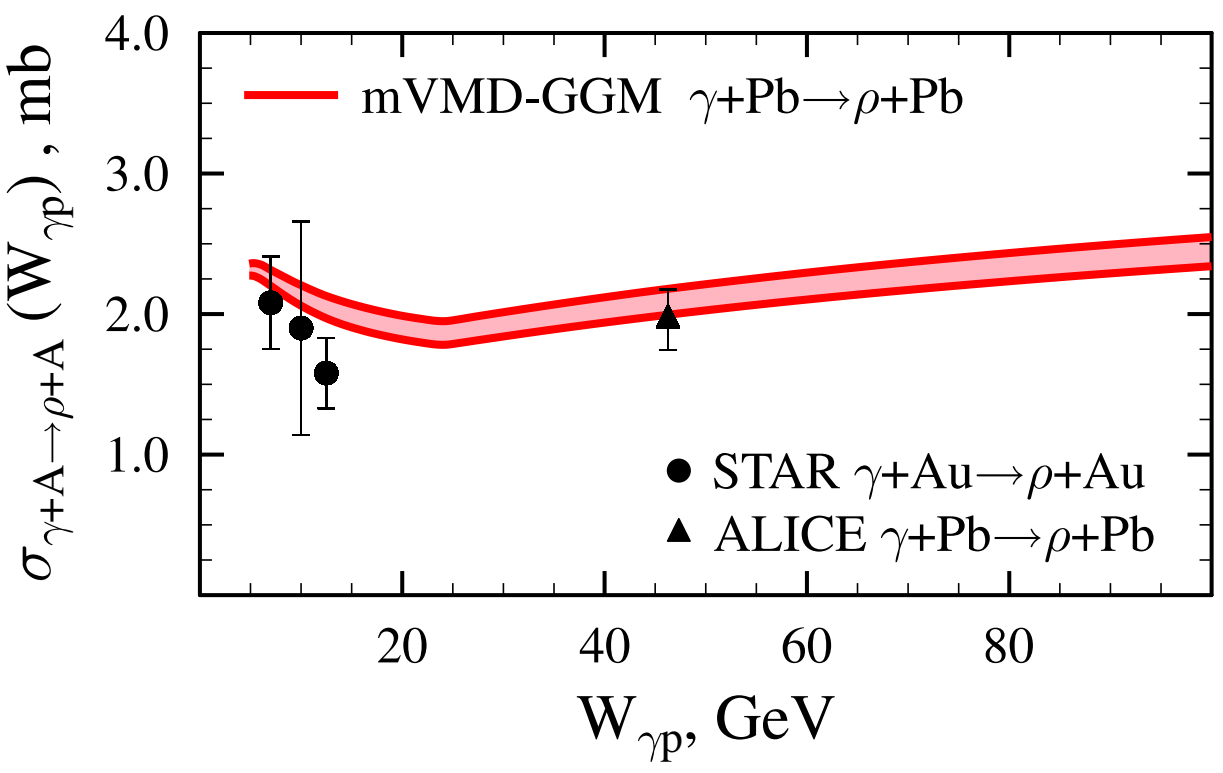

**Fig. 2:** The $\sigma_{\gamma A\to\rho A}$ cross section as a function of $W_{\gamma p}$. The theoretical predictions of Ref. [17] using the modified VMD model for the $\gamma p \to \rho p$ cross section and the Gribov–Glauber model with cross section fluctuations for the $\gamma A \to \rho A$ amplitude are compared to the STAR (circle) [18–20] and ALICE (triangle) data [12]. The shaded area reflects the theoretical uncertainty associated with the estimate of the strength of cross section fluctuations [17].

## 3 Color fluctuations in inelastic $\gamma A$ scattering [21]

Coherent production of vector mesons considered above give examples of processes which select configuration in the photon interacting with very different strength. We combine the information about interaction of small dipoles and soft, light vector meson like configurations probed in coherent $\rho$ photoproduction to build the probability distribution $P_\gamma(\sigma, W)$ for the interactions of the photon. Specifically we use the dipole approximation for $\sigma \leq 10$ mb, Eq. 3 adjusted for the contributions of $\omega, \phi$ mesons for $\sigma \geq 20$ mb and interpolate in between:

$$P_\gamma(\sigma, W) = \begin{cases} P_\gamma^{\text{dipole}}(\sigma, W)\,, & \sigma \leq 10\text{ mb}\,,\\ P_{\text{int}}(\sigma, W)\,, & 10\text{ mb} \leq \sigma \leq 20\text{ mb}\,,\\ P_{(\rho+\omega+\phi)/\gamma}(\sigma, W)\,, & \sigma \geq 20\text{ mb}\,. \end{cases} \tag{5}$$

where $P_{\text{int}}(\sigma, W)$ is a smooth interpolating function which matches dipole expression especially well for $m_q = 300$ MeV. The resulting $P_\gamma(\sigma, W)$ is shown by the red solid curve in Fig. 3. The derived $P_\gamma(\sigma, W)$ can be used to calculate distribution over the number of inelastic interaction, $\nu$ in the color fluctuation model. Also, one can include shadowing effects for the interaction of small dipoles which is smaller than for soft configurations but still significant as we have seen on the example of the $J/\psi$ production. The results of the calculation are presented in Fig. 4. One can see that probability of interactions with $\nu = 1$ corresponding to the $\gamma p$ scattering is very sensitive to CF and LT nuclear shadowing. Same holds for the $\nu \geq 10$ tail.

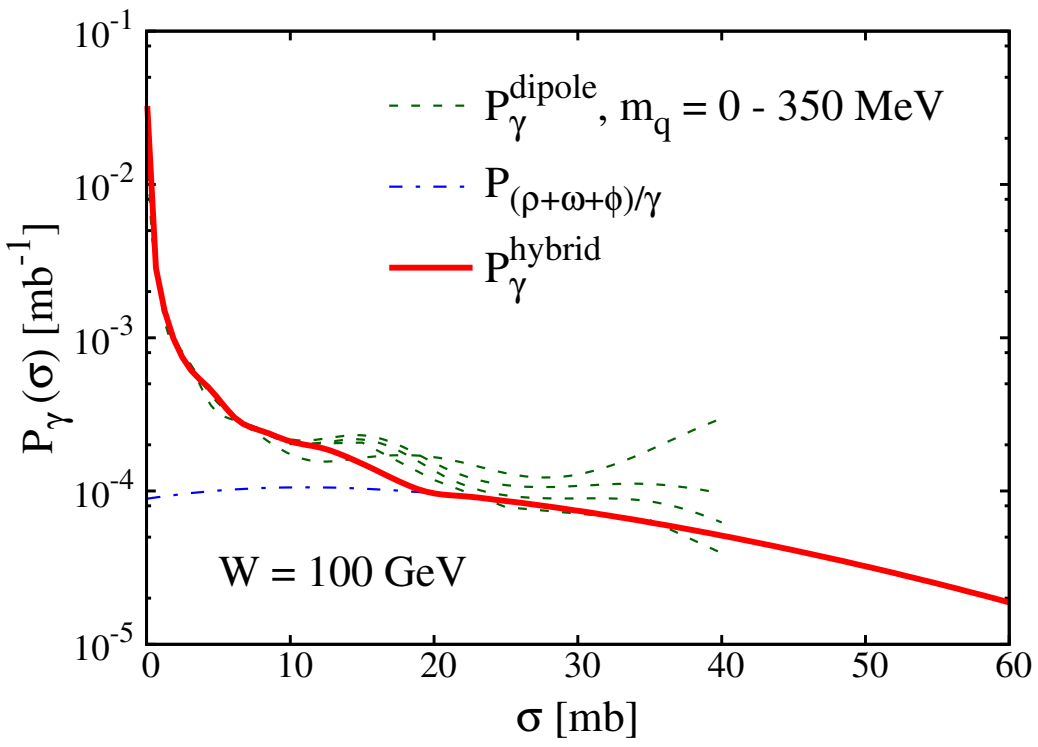

**Fig. 3:** The distributions $P_\gamma(\sigma, W)$ for the photon at $W = 100$ GeV. The red solid curve shows the full result of the hybrid model, see Eq. (5). The green dashed and blue dot-dashed curves show separately the dipole model and the vector meson contributions.

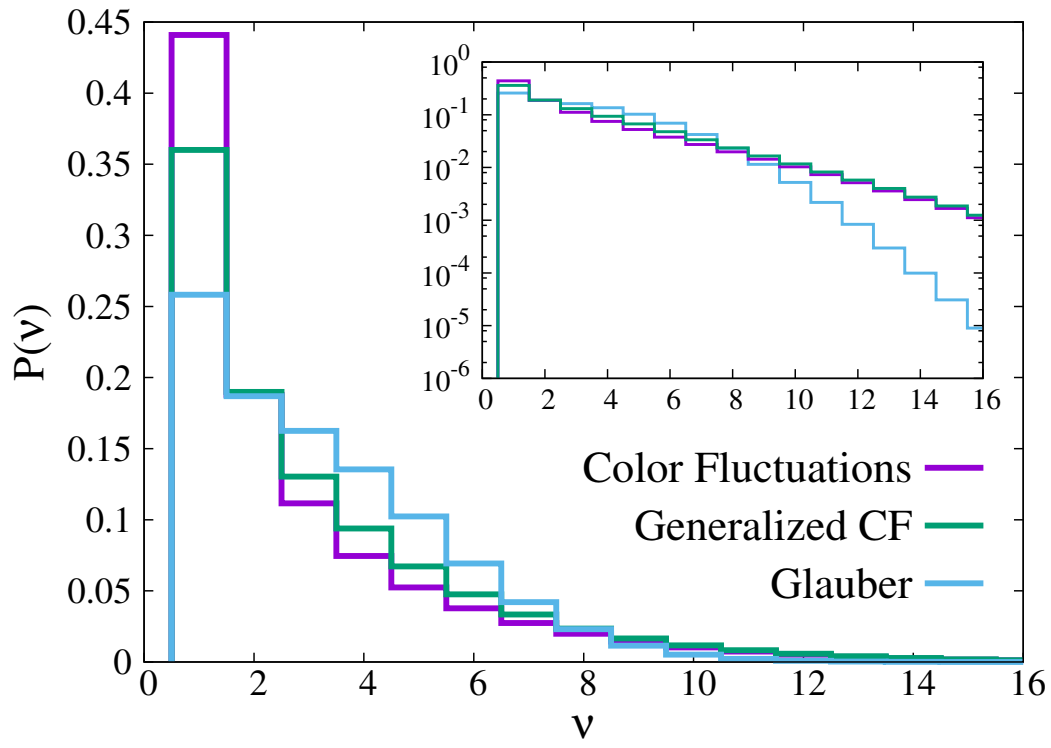

**Fig. 4:** The probability distributions $P(\nu, W)$ of the number of inelastic collisions $\nu$. Predictions based on the color fluctuation model with $P_\gamma(\sigma, W)$ given by Eq.5 are labeled "Color Fluctuations", the predictions including leading twist gluon shadowing are labeled "Generalized CF".For comparison, the CF model calculation with $\sigma = 25$ mb, which neglects the effect of CFs, is shown by the curve labeled "Glauber".

## 4 Color fluctuations and the distribution over transverse energy [21]

It is impossible to directly measure the number of inelastic interactions $\nu$ for $\gamma(h)A$ collisions. Modeling the distribution over the hadron multiplicity is also difficult due to the lack of the relevant data from $\gamma p$ scattering and issues with implementing energy–momentum conservation. However, the analysis of [22] suggests that the distribution over the total transverse energy, $\Sigma E_T$, sufficiently far away from the projectile fragmentation region (at sufficiently large negative pseudorapidities) is weakly influenced by energy conservation effects (due to the approximate Feynman scaling in this region) and is also weakly correlated with the activity in the rapidity-separated forward region. This expectation is validated by a recent measurement of $\Sigma E_T$ as a function of rapidity of a dijet in $pp$ collisions at the LHC [23].

Due to the weak sensitivity to the projectile fragmentation region, we expect that the $\Sigma E_T$ distributions in $pA$ and $\gamma A$ scattering at similar energies should have similar shapes for the same $\nu$. In Ref. [22], a model was developed for the distribution over $\Sigma E_T$ as a function of centrality (number, $\nu$, of wounded nucleons) in $pA$ scattering at large negative pseudorapidities (in the Pb-going direction) and $\sqrt{s} = 5.02$ TeV. We denote this distribution $f_\nu(\Sigma E_T) = 1/N_{\rm evt} dN/d\Sigma E_T$. In the spirit of the KNO scaling, it is natural to expect that the distribution over the $\Sigma E_T$ total transverse energy in $\gamma A$ scattering, when normalized to the average energy release in $pp$ scattering $\langle \Sigma E_T(NN) \rangle$, weakly depends on the incident collision energy. That is, the distribution over $y = \Sigma E_T(\gamma N)/\langle \Sigma E_T(\gamma N)\rangle$ has approximately the same shape at different energies. Hence we model the distribution over $y$ for photon–nucleus collisions using $F_\nu(y) = \langle \Sigma E_T(NN)\rangle f_\nu(y)$, where the factor of $\langle \Sigma E_T(NN)\rangle$ is a Jacobian to keep normalization of $\int F_\nu(y)dy = P(\nu)$.

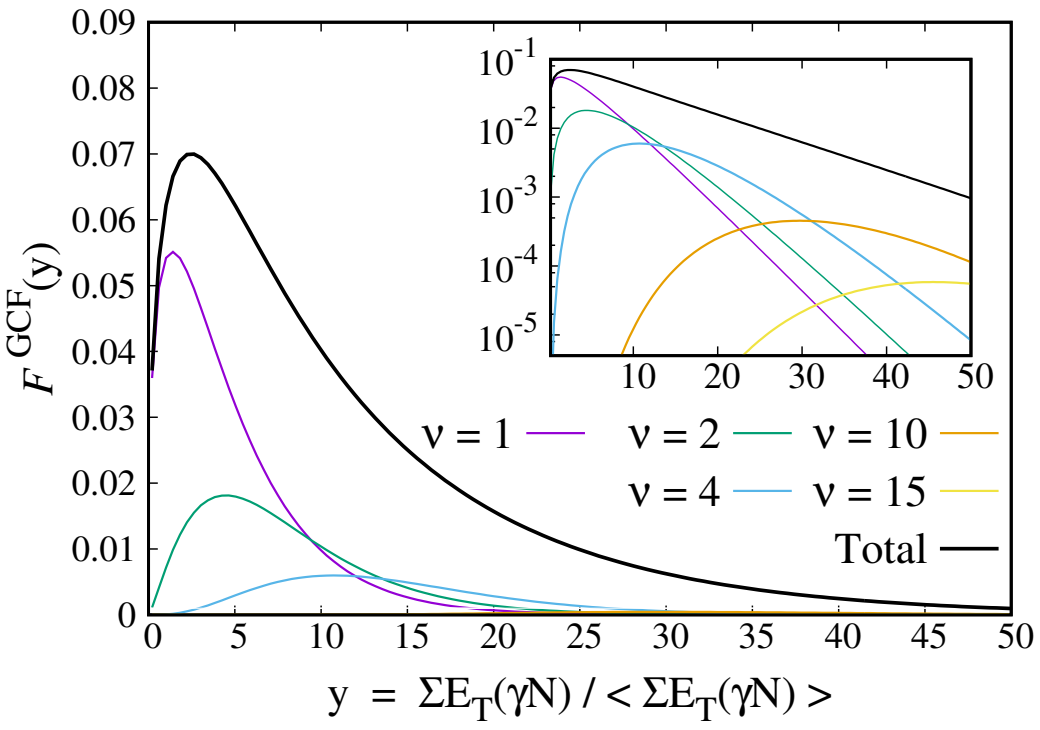

**Fig. 5:** The net probability distribution $\sum_\nu F_\nu(y)$ as a function of $y$ for different models including (curves labeled "Generalized CFs" and "Color Fluctuations") and neglecting (the curve labeled "Glauber") CFs in the photon.

The results of the calculation of $F_\nu(y)$ are presented in Fig. 5 for the Generalized Color Fluctuations (GCF) model which includes leading twist nuclear shadowing. One can see that the net distribution is predicted to be much broader than that for the $\nu = 1$ case corresponding to the $\gamma p$ scattering. Also, our results indicate that for $y = \Sigma E_T(\gamma N)/\langle \Sigma E_T(\gamma N)\rangle \leq 1$, the contribution of the interactions with one nucleon dominates. On the other hand, the distribution over $y$ in $\gamma p$ scattering can be measured in $pA$ UPCs. A first step would be to test that the $y$ distribution in $\gamma p$ and in the $\gamma A$ process with $\nu = 1$ [e.g., in the interaction of the direct photon ($x_\gamma = 1$) with a gluon with $x_A \geq 0.01$] is the same. Among other things this would give valuable information on the rapidity range affected by cascade interactions of slow (in the nucleus rest frame) hadrons which maybe formed inside the nucleus. It would be also interesting to measure separately the $y$ distribution for processes with production of leading charm with moderate $p_t$. In this case the y-distribution should be broader than for dijet large $x$ trigger, but more narrow than in the case min bias trigger. Another very interesting channel is hard resolved photon collisions where one expect a broadening of the $y$ distribution with decrease of $x_\gamma$. This effect is analogous to observed centrality dependence of the forward jet production in $pA$ scattering with increase of $x_p \geq 0.2$ which can be explained by decrease of the strength of interaction of configurations in protons for such $x$ [24].

## 5 Conclusions

In conclusion, studies of the ultraperipheral collisions at the LHC would allow to map in great detail photon wave function and investigate interplay of soft and hard physics in the photon-nucleus interactions. Selection of different final states in the photon fragmentation region would serve as an effective "strengthonometer" of the different components of the photon wave function.

**Acknowledgments –** L.F.'s and M.S.'s research was supported by the US Department of Energy Office of Science, Office of Nuclear Physics under Award No. DE-FG02-93ER40771.

# Ultra-peripheral collisions in STAR

*Jaroslav Adam for the STAR collaboration*
Creighton University, Omaha, United States of America

**Abstract**
A collision occurring at impact parameter larger than the sum of nuclear radii is denoted as an Ultra-Peripheral Collision (UPC). Such a collision is mediated by electromagnetic forces, because hadronic interactions are strongly suppressed. Ultra-relativistic heavy ions allow to study of photon-nucleus and photon-photon collisions, providing a sensitive probe of phenomena of gluon shadowing, nuclear diffraction and fundamental quantum electrodynamics. Here we report on the STAR measurements of coherent photoproduction of $\rho^0$ and $J/\psi$ mesons and of exclusive photoproduction of $e^+e^-$ pairs in Au+Au collisions at $\sqrt{s_{\rm NN}}$ = 200 GeV

## 1 Introduction

Electromagnetic fields of ultra-relativistic heavy ions are described as a flux of virtual photons. The flux of such photons is proportional to the square of the electric charge. Photoproduction of vector mesons in UPC is described by (i) photon fluctuation to a virtual quark-antiquark pair and (ii) interaction of the pair on the nuclear target. The $\rho^0$ and $J/\psi$ mesons provide photoproduction reactions at different scales, given by their masses. Reviews of UPC physics can be found in Refs. [1, 2].

The diagram for $\rho^0$ photoproduction is shown in Fig. 1. In addition to the photoproduction process, one or both nuclei may excite to Giant Dipole Resonances (GDRs) [3] or higher resonances, as also shown in Fig. 1. Those resonances decay by emitting one more neutrons, which allows to tag the UPC events experimentally.

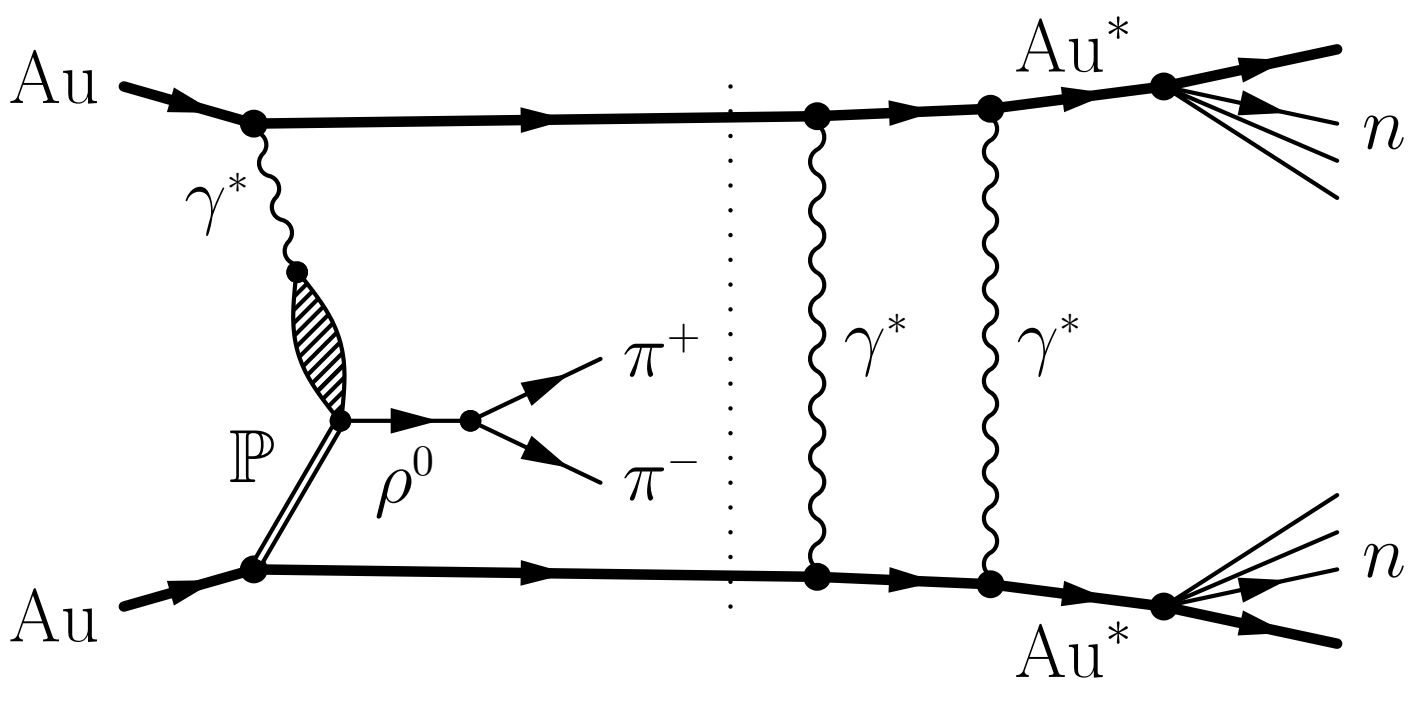

**Fig. 1:** Photoproduction of $\rho^0$ meson accompanied by additional photon exchange.

## 2 The STAR experiment at RHIC

UPC measurements at STAR [4] use tracking and particle identification in the Time Projection Chamber (TPC), covering central rapidities $|\eta| < 1$, the Time of Flight Detector (TOF) to select tracks belonging to the current event, the Beam-Beam Counters (BBC) to veto non-UPC activity at forward (backward) rapidities $2.1 < |\eta| < 5.2$ and very forward Zero-Degree Calorimeters (ZDC), $|\eta| > 6.6$, to detect neutrons from additional photon excitations. The trigger for UPC is based on inputs from BBC, TOF and ZDC.

## 3 Coherent photoproduction of $\rho^0$ mesons

Coherent photoproduction occurs when the photon interacts with the whole nucleus. Recent results on coherent $\rho^0$ photoproduction are given in Ref. [4]. Theoretical calculations of coherent photoproduction cross section can be done with a quantum Glauber model [5].

Selection criteria are aimed to select events with just two tracks from the decay $\rho^0 \to \pi^+\pi^-$. The invariant mass distribution of $\pi^+\pi^-$ pairs is shown in Fig. 2, with a fit by a modified Söding parameterization [6]. The fit includes Breit-Wigner resonances for $\rho^0$ and for $\omega$ mesons, where the contribution of the $\omega$ is needed to get an acceptable fit.

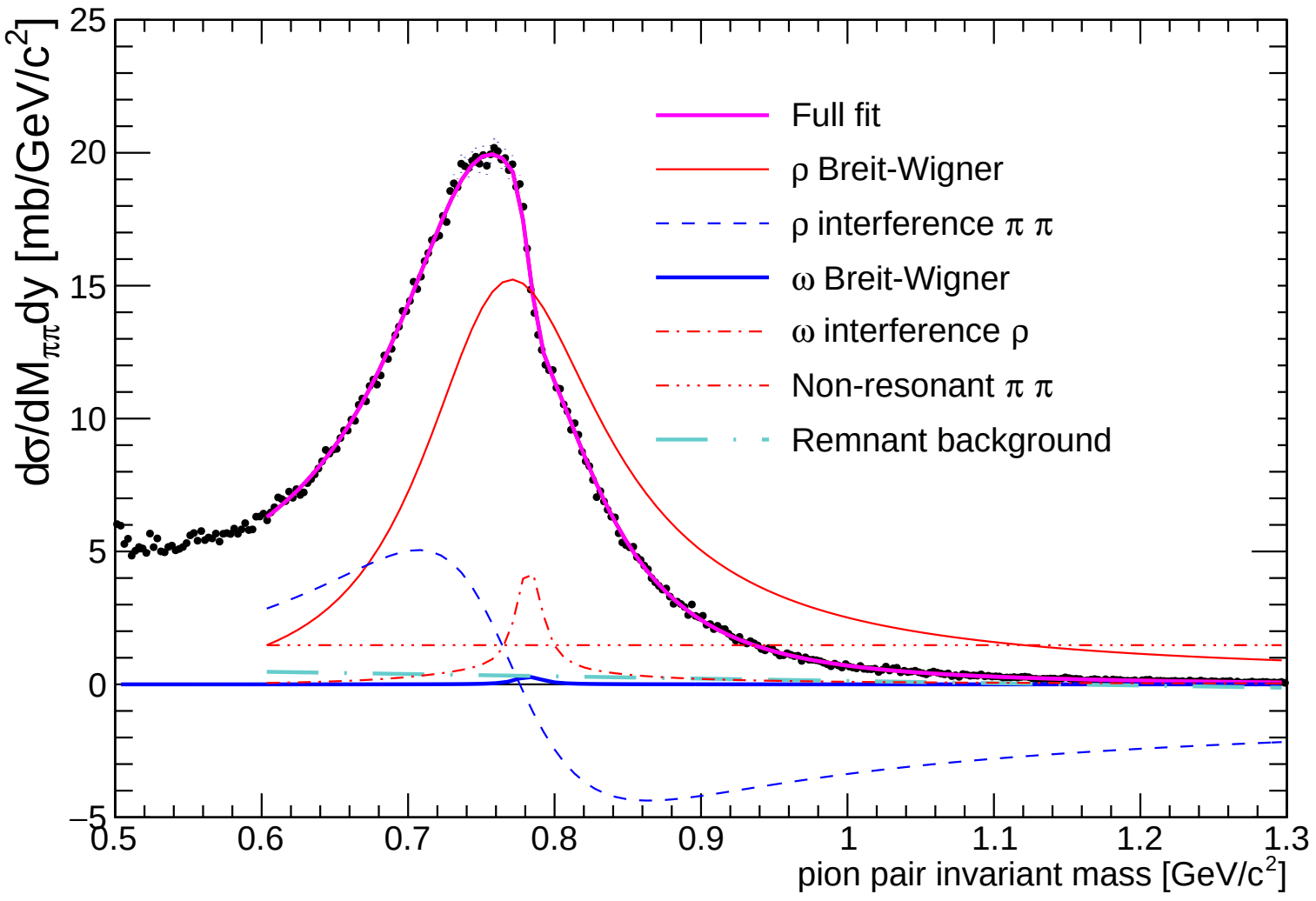

**Fig. 2:** Invariant mass of selected $\pi^+\pi^-$ candidates with transverse momentum $p_T < 100$ MeV/c [4].

Relative amplitudes of non-resonant $\pi^+\pi^-$ and $\rho^0$ ($B/A$), and of $\omega$ and $\rho^0$ ($C/A$) are shown in Fig. 3 as a function of $\pi^+\pi^-$ rapidity. The rapidity is kinematically related to the photon energy; the rapidity distribution therefore provides energy dependence of the ratios. The only previous data on $C/A$ come from the DESY-MIT experiment [7] at lower photon energies corresponding to rapidity $y = -2.5$. The ratio $C/A$ is well described by STARlight [8].

The coherent $\rho^0$ cross section as a function of $-t$ in Fig. 4 shows characteristic diffractive dips. Two categories of neutron emission are present, 1n1n for one neutron emitted on each side and XnXn for at least one neutron on each side. The positions of diffractive dips at $-t = 0.018 \pm 0.005$ GeV$^2$ and $-t = 0.043 \pm 0.01$ GeV$^2$ are correctly predicted by a quantum Glauber calculation without nuclear shadowing correction [5]. A fit by an exponential function is performed below the first peak; the slope of the fit is proportional to the nuclear size, and the result is consistent with ALICE data [9] at $\sqrt{s_{\rm NN}}$ = 2.76 TeV. The inset in Fig. 4 for very small $-t$ shows the effect of destructive interference between production on the two nuclei.

## 4 Photoproduction of $J/\psi$

Photoproduction of $J/\psi$ occurs at a harder scale given by the mass of the $J/\psi$. The process is described by two-gluon exchange, where the photon fluctuates to a dipole of a $c\bar{c}$ pair and the dipole interacts with the target nucleus. Coherent $J/\psi$ photoproduction has recently been measured at the LHC [10, 11].

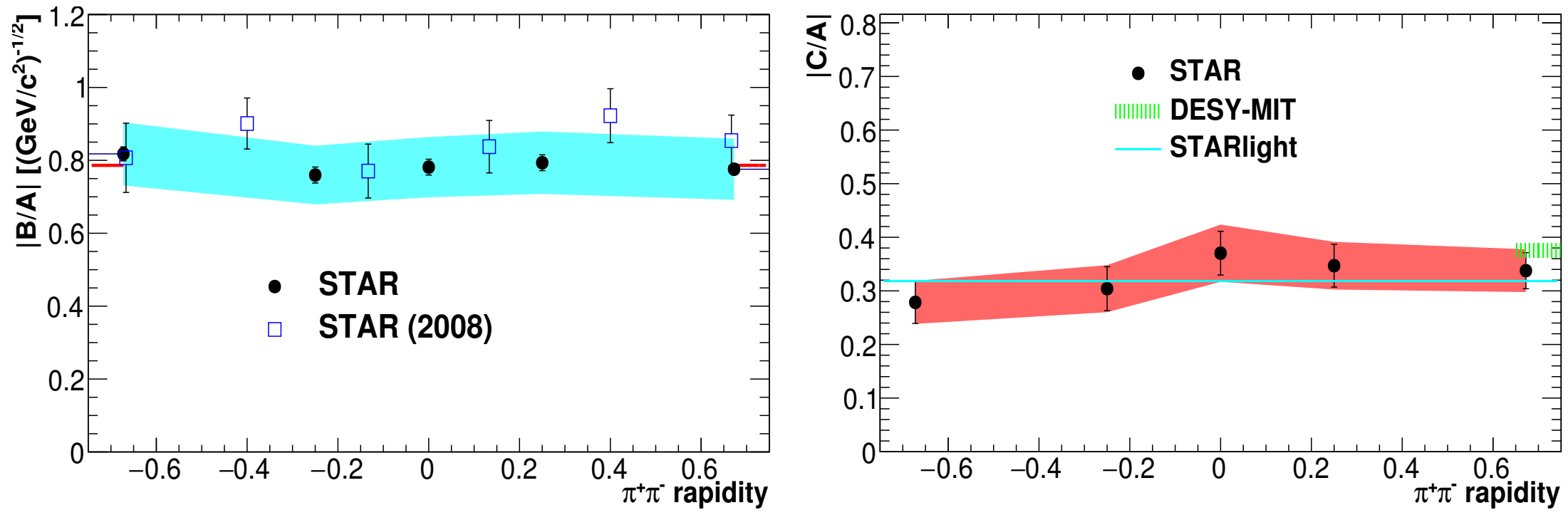

**Fig. 3:** Ratios of amplitudes of non-resonant $\pi^+\pi^-$ and $\rho^0$ (top) and $\omega$ and $\rho^0$ (bottom) [4]. The red line is the rapidity-averaged result.

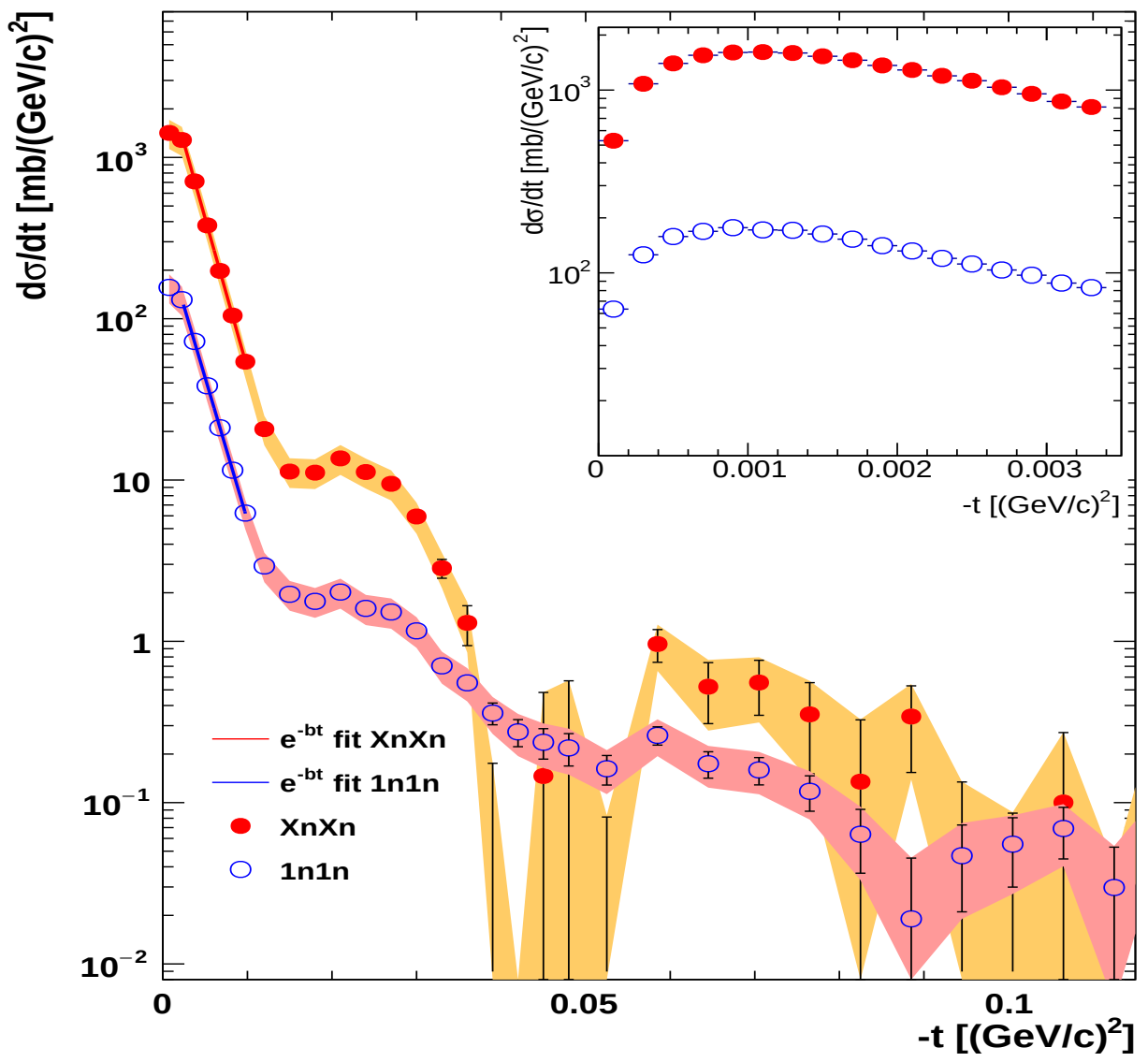

**Fig. 4:** Cross section for coherent $\rho^0$ vs. $-t$ for 1n1n and XnXn events [4].

Nuclear gluon shadowing at low Bjorken-$x$ can be studied using the data on coherent $J/\psi$ photoproduction by extracting the nuclear suppression factor $S$, defined as a ratio of measured photon-nucleus cross section to a calculation in impulse approximation. The effect of shadowing is observed as a partial depletion of the nuclear (w.r.t. nucleon) gluon density. Previous experimental and theoretical results on $S_{\mathrm{Pb}}$ are given in Fig. 5. STAR probes the gluon distribution at $x \approx 0.015$.

Selection criteria for $J/\psi \to e^+e^-/\mu^+\mu^-$ are similar to those for $\rho^0$, to get events with just two tracks corresponding to dilepton pairs. In this case the very central rapidity region $|y| < 0.02$ is excluded to prevent cosmic background. The data provide a clean $J/\psi$ signal and expected background from two-photon production of dilepton pairs.

The cross section of $J/\psi$ as a function of transverse momentum is shown in Fig. 6. A coherent peak at low $p_T$ is followed by a tail from incoherent production at higher values of $p_T$. Shape of the coherent peak in the region $p_T < 0.15$ GeV/c is consistent with STARlight estimation.

## 5 Two-photon production of electron-positron pairs

The photoproduction reaction $\gamma\gamma \to e^+e^-$ is a probe of quantum electrodynamics and the description of fields by a flux of virtual photons. The cross section of $\gamma\gamma \to e^+e^-$ is given by Breit-Wheeler for-mula [13], and the cross section for $e^+e^-$ pairs in nucleus-nucleus collisions is obtained by convolution of elemental cross section with photon fluxes of the nuclei [8]. Selection criteria for $e^+e^-$ pairs are similar to those of the $\rho^0$ mesons, to get two-track events in an otherwise empty detector (except very forward neutrons). About 13000 of $e^+e^-$ candidates were selected and only $\sim$10 like-sign candidates are present in the data, manifesting thus a very clean signal.

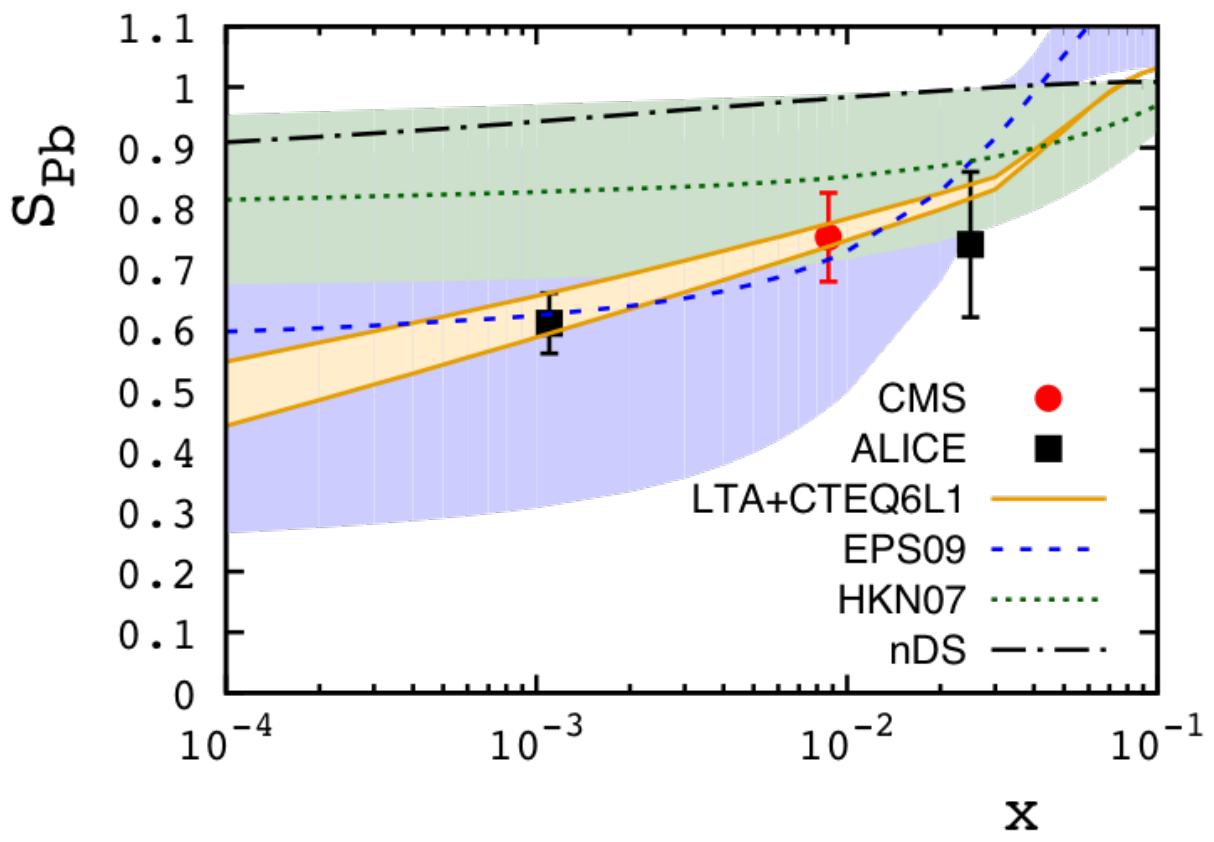

**Fig. 5:** Nuclear shadowing obtained from the LHC data and comparison to theoretical models [12].

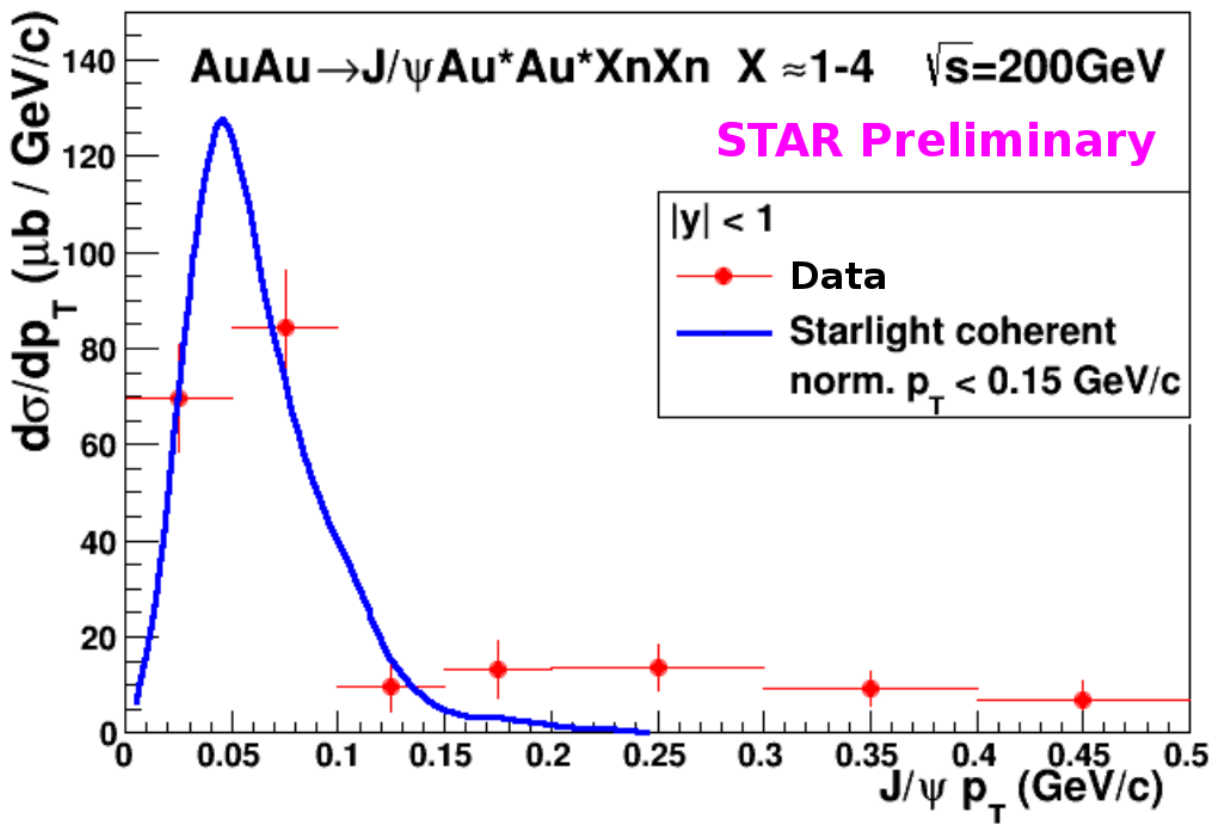

**Fig. 6:** $J/\psi$ cross section as a function of $p_T$. Error bars represent statistical uncertainties.

The distributions of invariant masses and $p_T$ of selected $e^+e^-$ candidates are shown in Fig. 7. There is exponential decrease of the yield with increasing mass of the $e^+e^-$ pairs, given by the nature of the cross section. Also a signal of $J/\psi \to e^+e^-$ is present. The $p_T$ distribution is consistent with an expected peak at low-$p_T$.

## 6 Conclusions

The STAR experiment has measured coherent photoproduction of $\rho^0$ mesons using a high statistics sample of $\rho^0 \to \pi^+\pi^-$ events [4]. Ratios of amplitudes of non-resonant $\pi^+\pi^-$ and $\rho^0$ and of $\omega$ and $\rho^0$ have been measured as a function of $\pi^+\pi^-$ rapidity. In the case of $\omega$ to $\rho^0$ amplitude, the only previous data were obtained at much lower photon energies. The cross section of coherent $\rho^0$ as a function of $-t$ shows diffractive dips, which were compared to a theoretical Glauber prediction.

Currently work is in progress in obtaining the cross sections of $J/\psi$ and $e^+e^-$ pairs, with newly reconstructed data samples. The results will help in the understanding of nuclear effects on gluon density in the transition region from shadowing to non-shadowing in the case of the $J/\psi$, and the mechanism of photon fluxes in the case of the $e^+e^-$ pairs.

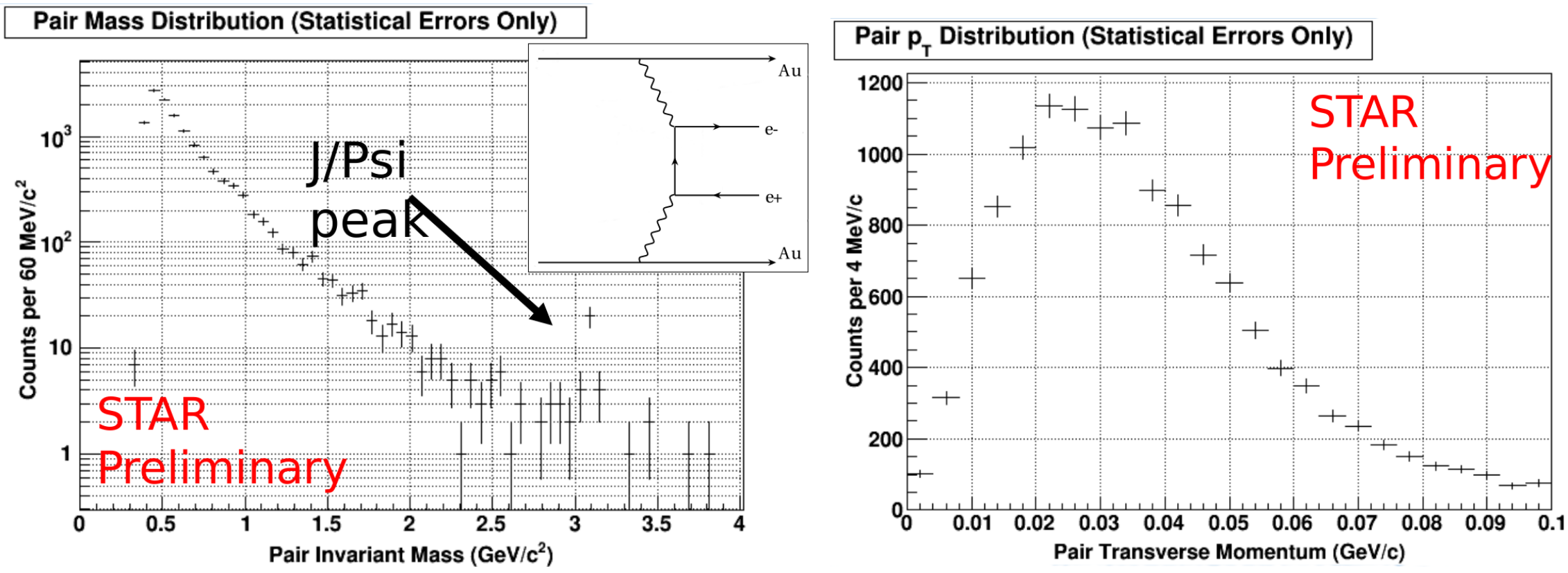

**Fig. 7:** Uncorrected $e^+e^-$ invariant mass distribution (left) and transverse momentum (right) of selected electron-positron candidates. The diagram in the inset shows $\gamma\gamma \to e^+e^-$ production in Au+Au.

# Quarkonium-photoproduction prospects at a fixed-target experiment at the LHC (AFTER@LHC)

*L. Massacrier*[1], *J.P. Lansberg*[1], *L. Szymanowski*[2], *and J. Wagner*[2]
[1] IPNO, Univ. Paris-Sud, CNRS/IN2P3, Université Paris-Saclay, 91406 Orsay, France
[2] National Centre for Nuclear Research (NCBJ), Hoza 69,00-681, Warsaw, Poland

**Abstract**
We report on the potentialities offered by a fixed-target experiment at the LHC using the proton and ion LHC beams (AFTER@LHC project) regarding the study of $J/\psi$ exclusive-photoproduction in pA and AA collisions. The foreseen usage of polarised targets (hydrogen, deuteron, helium) gives access to measurements of the Single-Transverse-Spin Asymmetries of this exclusive process, thereby allowing the access to the helicity-flip Generalised Parton Distribution $E_g$. We detail the expected yields of photoproduced $J/\psi$ in proton-hydrogen and lead-hydrogen collisions and discuss the statistical uncertainties on the asymmetry measurement for one year of data taking at the LHC.

## 1 The AFTER@LHC project

AFTER@LHC is a proposal for a multi-purpose fixed-target programme using the multi-TeV proton or heavy ion beams of the LHC to perform the most energetic fixed-target collisions ever performed so far [1]. If used in a fixed-target mode, the LHC beams offer the possibility to study with high precision pH or pA collisions at $\sqrt{s_{NN}}$ = 115 GeV and PbA collisions at $\sqrt{s_{NN}}$ = 72 GeV, where $A$ is the atomic mass of the target. The fixed-target mode offers several unique assets compared to the collider mode: outstanding luminosities can be reached thanks to the high density of the target; standard detectors can access the far backward center-of-mass (c.m.) region thanks to the boost between the colliding-nucleon c.m. system and the laboratory system (this region remains completely uncharted with hard reactions); the target species can easily be changed; polarised target can be used for spin studies. The physics opportunities offered by a fixed-target programme at LHC have been published in [2–17] and are summarised below in three main objectives.

First, whereas the need for precise measurements of the partonic structure of nucleons and nuclei at small momentum fraction $x$ is usually highlighted as a strong motivation for new large-scale experimental facilities, the structure of nucleon and nuclei at high $x$ is probably as poorly known with long-standing puzzles such as the EMC effect [18] in the nuclei or a possible non-perturbative source of charm or beauty quarks in the proton which would carry a relevant fraction of its momentum [19]. With an extensive coverage of the backward region corresponding to high $x$ in the target, AFTER@LHC is very well placed for performing this physics with a hadron beam.

The second objective of AFTER@LHC is the search and characterisation of the Quark-Gluon Plasma (QGP), a deconfined state of nuclear matter, which was prevailing in the universe few microseconds after the Big Bang. QGP is expected to be formed when the surrounding hadronic matter is extremely compressed or heated. These conditions can be achieved in ultra-relativistic Heavy-Ion collisions (HI). AFTER@LHC with a c.m. energy of 72 GeV provides a complementary coverage to the RHIC- and SPS- based experiments, in the region of high temperatures and low baryon-chemical potentials, where the QGP is expected to be produced. AFTER@LHC will provide crucial information about the phase transition by: (i) scanning the longitudinal extension of the hot medium, (ii) colliding systems of different sizes, (iii) analysing the centrality dependence of these collisions. Together they should provide a measurement of the temperature dependence of the system viscosity both as a QGP or a hadron gas. Additionally, measurements of production of various quarkonia states in HI collisions can provide insight into thermodynamic properties of the QGP. Their sequential suppression was predicted to occur in the deconfined partonic medium due to Debye screening of the quark-antiquark potential [20]. However, other effects (Cold Nuclear Matter effects (CNM), feed-down, secondary production via coalescence...) can also alter the observed yields. AFTER@LHC will provide a complete set of quarkonia measurements (together with open heavy flavours) allowing one to access the temperature of the formed medium in AA collisions, and cold nuclear matter effects in pA (AA) collisions. Thanks to the large statistics expected, a golden probe will be the measurement of $\Upsilon$(nS) production in pp, pA and AA collisions, allowing one to calibrate the quarkonium thermometer and to search for the phase transition by looking at $\Upsilon$(nS) suppression (and other observables like charged hadron $v_2$) as a function of rapidity and the system size.

Finally, despite decades of efforts, the internal structure of the nucleon and the distribution and dynamics of its constituents are still largely unknown. One of the most significant issues is our limited understanding of the spin structure of the nucleon, especially how its elementary constituents (quarks and gluons) bind into a spin-half object. Among others, Single Transverse Spin Asymmetries (STSA) in different hard-scattering processes are powerful observables to further explore the dynamics of the partons inside hadrons [21]. Thanks to the large yields expected with AFTER@LHC, STSA of heavy-flavours and quarkonia –which are currently poorly known– could be measured with high accuracy, if a polarised target can be installed [16].

We show here that AFTER@LHC can also rely on quarkonium exclusive-photoproduction processes to explore the three-dimensional tomography of hadrons via Generalised Parton Distributions (GPDs) [22]. Photoproduction is indeed accessible in Ultra-Peripheral Collisions (UPCs) and the quarkonium mass presumably sets the hard scale to use collinear factorisation in terms of (gluon) GPDs, which are directly related to the total angular momentum carried by quarks and gluons. In fact, exclusive $J/\psi$ production [23] drew much attention in the recent years due to the fact that it is sensitive, even at leading order, to gluon GPDs. Beside, with the addition of a polarised hydrogen[1] target, AFTER@LHC opens a unique possibility to study STSA of such process, which is sensitive to yet unknown GPD $E_g$ [24]. In this contribution, we report on such studies through the collisions of proton and lead beams onto a polarised hydrogen target at AFTER@LHC energies.

## 2 Possible technical implementations at the LHC and projected luminosities

Several promising technical solutions exist in order to perform fixed-target collisions at the LHC. One can either use an internal (solid or gaseous) target coupled to an already existing LHC detector or build a new beam line together with a new detector. The first solution can be achieved in a shorter time scale, at limited cost and civil engineering. Moreover the fixed-target programme can be simultaneously run with the current LHC collider experiments, without affecting the LHC performances.

The direct injection of noble gases in the LHC beam pipe is currently being used by the LHCb collaboration with the SMOG device [25]. However, this system has some limitations, in particular: (i) the gas density achieved inside the Vertex Locator of LHCb is small (of the order of $10^{-7}$ mbar); (ii) there is no possibility to inject polarised gas; (iii) there is no dedicated pumping system close to the target; (iv) the data taking time has so far been limited to few days in a row. The use of more complex systems with higher density gaseous and polarised target inside one of the existing LHC experiment is under study. For instance, an hydrogen gas jet is currently used at RHIC to measure the proton beam polarisation [26]. The H-jet system consists of a polarised free atomic beam source cooled down to 80K, providing an hydrogen inlet flux of 1.3 $\times 10^7$ H/s. With such a device, the gas density can be increased by about one order of magnitude with respect to the SMOG device and probably be continuously run. Another promising alternative solution is the use of an openable storage cell placed inside the LHC beam pipe. Such a system was developed for the HERMES experiment [27, 28]. Polarised hydrogen, deuteron and Helium-3 at densities about 200 times larger than the ones of the H-jet system can be injected, as well as heavier unpolarised gases.

[1]Measurements with deuteron and helium targets are also considered.

Fixed-target collisions can also be obtained with a solid target (wire, foil) interacting with the LHC beams. There are two ways of doing so: either thanks to a system which permit to move directly the target inside the beam halo [29]; or by using a bent crystal (see work by UA9 collaboration [30]) upstream of an existing experiment ($\sim$ 100 m) to deflect the beam halo onto the fixed target. In both cases, the target (or an assembly of several targets) can be placed at a few centimetres from the nominal interaction point, allowing one to fully exploit the performances of an existing detector. The usage of the bent crystal offers the additional advantage to better control the flux[2] of particles sent on the target, and therefore the luminosity determination. Most probably such simple solid targets could not be polarised. The spin physics part of the AFTER@LHC programme could naturally not be conducted with such an option.

Table 1 summarises the target areal density, the beam flux intercepting the target, the expected instantaneous and yearly integrated luminosities for pH and PbH collisions, for the possible technical solutions described above. Luminosities as large as 10 fb$^{-1}$ can be collected in pH collisions in one LHC year with a storage cell. In PbH collisions, similar luminosities ($\sim$ 100 nb$^{-1}$) can be reached with a gas-jet or a storage cell since the gas density has to be levelled in order to avoid a too large beam consumption[3]. We stress that these are annual numbers and they can be accumulated over different runs.

**Table 1:** Target areal density $\theta_{\rm target}$, beam flux intercepting the target, expected instantaneous and yearly integrated luminosities for pH and PbH collisions, for the possible technical solutions described in the text. The solid target is considered to be 5 mm thick along the beam direction. The LHC year is considered to last $10^7$s for the proton beam and $10^6$s for the lead beam.

| | $\theta_{target}$ (cm$^{-2}$) | | Beam flux (s$^{-1}$) | | $\mathcal{L}$ (cm$^{-2}$ s$^{-1}$) | | $\mathcal{L}_{\rm int}$ (fb$^{-1}$) | (nb$^{-1}$) |
|---|---|---|---|---|---|---|---|---|
| Technical solutions | pH | PbH | pH | PbH | pH | Pb | pH | PbH |
| Gas-jet | $1.2 \times 10^{12}$ | $2.54 \times 10^{14}$ | $3.63 \times 10^{18}$ | $4.66 \times 10^{14}$ | $4.4 \times 10^{30}$ | $1.2 \times 10^{29}$ | $\sim 0.05$ | $\sim 100$ |
| Storage Cell | $2.5 \times 10^{14}$ | $2.54 \times 10^{14}$ | $3.63 \times 10^{18}$ | $4.66 \times 10^{14}$ | $9.1 \times 10^{32}$ | $1.2 \times 10^{29}$ | $\sim 10$ | $\sim 100$ |
| Bent Crystal + Solid target | $2.6 \times 10^{22}$ | $2.6 \times 10^{22}$ | $5 \times 10^8$ | $10^5$ | $1.3 \times 10^{31}$ | $2.6 \times 10^{27}$ | $\sim 0.15$ | $\sim 3$ |

## 3 Prospects for $J/\psi$ photoproduction studies with AFTER@LHC

Let us now quickly summarise our feasibility study: 100 000 photoproduced $J/\psi$ have been generated in the dimuon decay channel, using STARLIGHT Monte Carlo (MC) generator [31–34] in pH$^\uparrow$ ($\sqrt{s}$ = 115 GeV) and Pb+H$^\uparrow$ collisions ($\sqrt{s_{NN}}$ = 72 GeV). In pH$^\uparrow$ collisions, both protons can be photon emitters, while in Pb+H$^\uparrow$ only the Pb nuclei was considered as photon emitter (dominant contribution). The $J/\psi$ photoproduction cross sections given by STARLIGHT MC are summarised in Tab. 2. In order to mimic an LHCb-like forward detector, the following kinematical cuts have been applied at the single muon level: $2 < \eta^{\mu}_{\rm lab} < 5$ and $p^{\mu}_{\rm T} > 0.4$ GeV/c. Figure 1 shows the rapidity-differential (left) and $p_{\rm T}$-differential (right) cross sections of the photoproduced $J/\psi$ in the dimuon decay channel in pH$^\uparrow$ collisions, generated with STARLIGHT MC generator. The blue curves have been produced without applying any kinematical cuts, while the red curves are produced by applying the cuts: $2 < \eta^{\mu}_{\rm lab} < 5$ and $p^{\mu}_{\rm T} > 0.4$ GeV/c at the single muon level. On the left panel is also indicated the photon-proton

[2] The deflected halo beam flux is considered to be about $5 \times 10^8$ p/s and $10^5$ Pb/s.
[3] Assuming a total cross section $\sigma_{\rm PbH}$ = 3 barn, 15% of the beam is used over a fill.

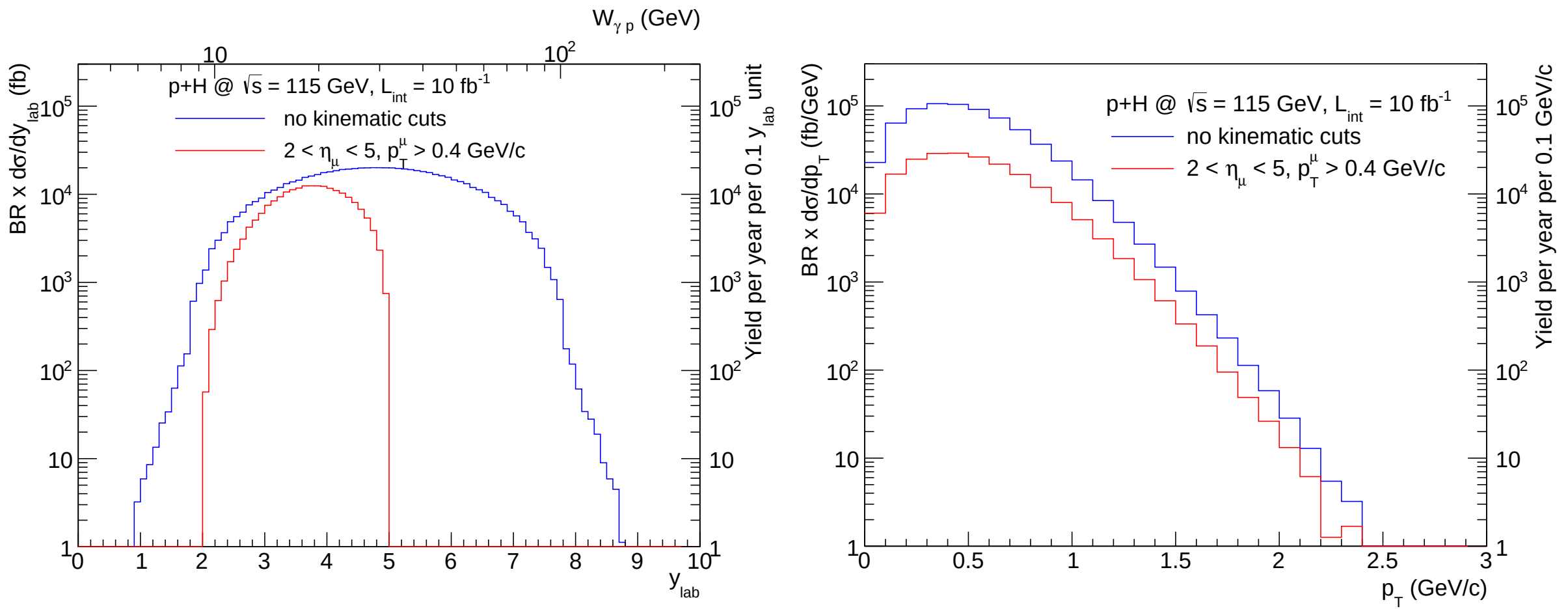

**Fig. 1:** $y_{\rm lab}$- (left) and $p_{\rm T}$-differential (right) cross sections of the photoproduced $J/\psi$ in pH collisions generated with STARLIGHT MC. On the vertical axis is also shown the photoproduction $J/\psi$ yield per year per 0.1 $y_{\rm lab}$ unit (left), and per 0.1 GeV/c unit (right). On the left panel is indicated as well the photon-proton c.m. energy ($W_{\gamma p}$). The blue curves have been produced without applying kinematics cut, while the red curves are produced by applying the $\eta$ and $p_{\rm T}$ cuts described in the text.

c.m. energy ($W_{\gamma p}$), calculated as:

$$W_{\gamma p} = \sqrt{M^2_{J/\psi} + M^2_p + 2 \times M_p \times M_{J/\psi} \times \cosh(y_{\rm lab})}, \tag{1}$$

with $M_{J/\psi}$ and $M_p$, being respectively the $J/\psi$ and proton masses, and $y_{\rm lab}$ the $J/\psi$ rapidity in the laboratory frame. On the vertical axis is shown as well the photoproduction $J/\psi$ yield per year per 0.1 $y_{\rm lab}$ unit (left) and per 0.1 GeV/c unit (right). An integrated yearly luminosity of $\mathcal{L}_{\rm int}$ = 10 fb$^{-1}$, corresponding to the storage cell solution, has been assumed in pH$^{\uparrow}$ collisions in order to calculate the $J/\psi$ yearly photoproduction yield. About 200 000 photoproduced $J/\psi$ are expected to be detected in an LHCb-like acceptance per year with AFTER@LHC[4]. Similarly to Fig. 1, Fig. 2 shows the rapidity-differential (left) and $p_{\rm T}$-differential (right) cross sections of the photoproduced $J/\psi$ in the dimuon decay channel in PbH$^{\uparrow}$ collisions, generated with STARLIGHT MC generator. The collection of an integrated luminosity of 100 nb$^{-1}$ per year is expected at AFTER@LHC with the storage cell option. This would result in about 1 000 photoproduced $J/\psi$ per year emitted in an LHCb-like acceptance.

**Table 2:** Summary table of $J/\psi$ photoproduction cross sections from STARLIGHT MC generator.

| collision type | photon-emitter | $\sigma^{tot}_{J/\psi}$ (pb) | $\sigma_{J/\psi\to\mu^+\mu^-}$ (pb) | $\sigma_{J/\psi\to\mu^+\mu^-}$ ($2 < \eta^{\mu}_{\rm lab} < 5$, $p^{\mu}_{\rm T}$ > 0.4 GeV/c) (pb) |
|---|---|---|---|---|
| pH$^{\uparrow}$ | proton (sum of 2 contributions) | $1.18\times10^3$ | 70.10 | 20.64 |
| Pb+H$^{\uparrow}$ | Pb | $276.77\times10^3$ | $16.50\times10^3$ | $9.81\times10^3$ |

In a forthcoming publication [35], we will report on the evaluation of the uncertainty on the STSA, $A_N$, from the photoproduced-$J/\psi$ yields obtained with STARLIGHT MC and the expected modulation for $E_g$ following [24]. Indeed, $A_N$, the amplitude of the spin-correlated azimuthal modulation of the produced particles, is defined as $A_N = \frac{1}{P}\frac{N^{\uparrow}-N^{\downarrow}}{N^{\uparrow}+N^{\downarrow}}$, where $N^{\uparrow}$ ($N^{\downarrow}$) are the photoproduced-$J/\psi$ yields for an up (down) target-polarisation orientation, and p is the effective polarisation of the target. The statistical uncertainty $u_{A_N}$ on $A_N$ can then be derived using $u_{A_N} = \frac{2}{P(N^{\uparrow}+N^{\downarrow})^2}\sqrt{N^{\downarrow 2}u^{\uparrow 2} + N^{\uparrow 2}u^{\downarrow 2}}$ with $u^{\uparrow}$ ($u \downarrow$) the relative uncertainties on the $J/\psi$ yields with up (down) polarisation orientation. Dividing the sample into two $J/\psi$ $p_{\rm T}$ ranges relevant for GPDs extraction: 0.4 $< p_{\rm T} <$ 0.6 GeV/c and 0.6 $< p_{\rm T} <$ 0.8 GeV/c, the expected statistical precision on $A_N$ is already expected to be better than 10% given the integrated yearly yield of 200 000 photoproduced $J/\psi$ in $pH^{\uparrow}$ collisions at AFTER@LHC. This will also allow for an extraction of $\mathrm{A}_N$ as a function Feynman-$x$, $x_{\mathrm{F}}$ [5].

[4]The detector efficiency has been assumed to be 100%.

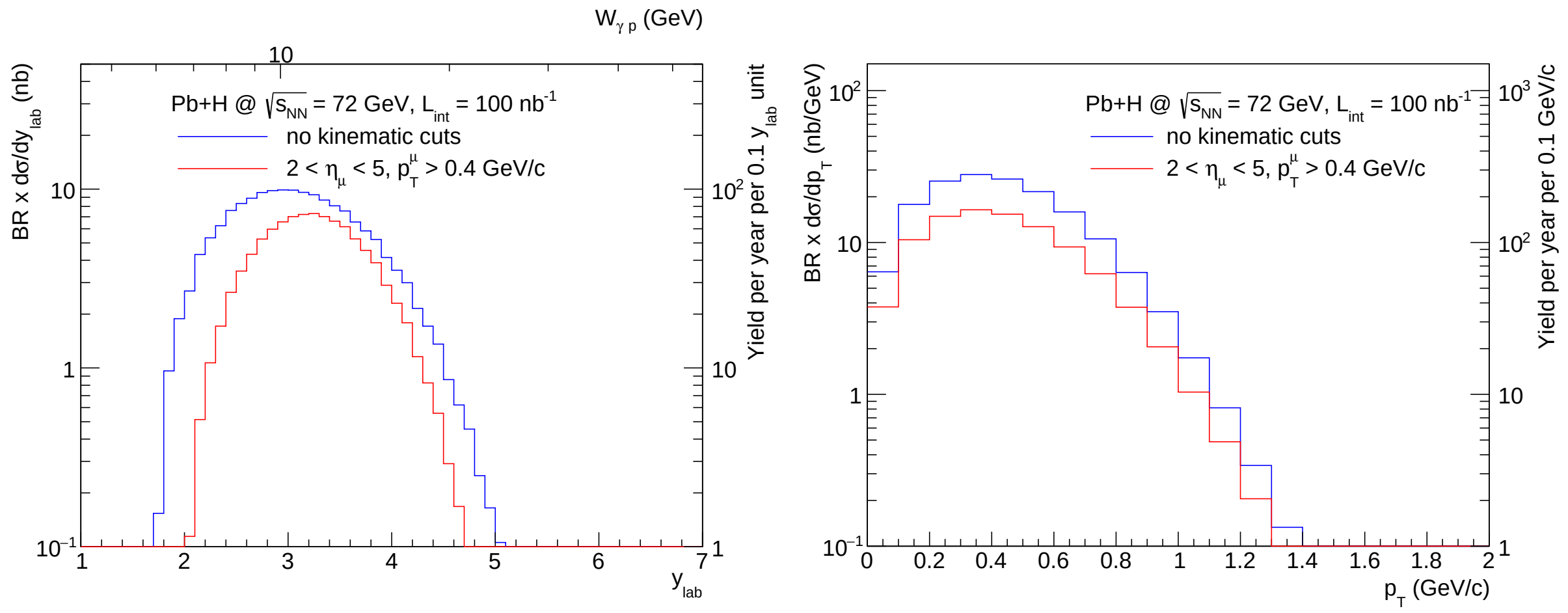

**Fig. 2:** Rapidity-differential (left) and $p_{\mathrm{T}}$-differential (right) cross sections of the photoproduced $J/\psi$ in the laboratory frame in PbH collisions generated with STARLIGHT MC. On the vertical axis is also shown the photoproduction $J/\psi$ yield per year per 0.1 $\mathrm{y_{lab}}$ unit (left), and per 0.1 GeV/c unit (right). On the left panel is indicated as well the photon-proton c.m. energy ($\mathrm{W}_{\gamma p}$). The blue curves have been produced without applying kinematics cut, while the red curves are produced by applying the $\eta$ and $p_{\mathrm{T}}$ cuts described in the text.

## 4 Conclusion

We have presented projections for $J/\psi$ photoproduction measurements in polarised $\mathrm{pH}^{\uparrow}$ and $\mathrm{Pb{+}H}^{\uparrow}$ collisions after one year of data taking at AFTER@LHC energies and assuming a storage cell technology coupled to an LHCb-like forward detector. In $\mathrm{pH}^{\uparrow}$ collisions, a yearly yield of about 200 000 photoproduced $J/\psi$ is expected, allowing one to reach a very competitive statistical accuracy on the $A_N$ measurement differential in $x_{\mathrm{F}}$. A non-zero asymmetry would be the first signature of a non-zero GPD $E_g$ for gluons.

## Acknowledgements

We thank D. Kikola, S. Klein, A. Metz, J. Nystrand and B. Trzeciak for useful discussions. This work is partly supported by the COPIN-IN2P3 Agreement, the grant of National Science Center, Poland, No. 2015/17/B/ST2/01838, by French-Polish scientific agreement POLONIUM, by the French P2IO Excellence Laboratory and by the French CNRS-IN2P3 (project TMD@NLO).

[5] $x_{\mathrm{F}} = 2 \times M_{J/\psi} \times \sinh(y_{\mathrm{cms}}/\sqrt{s})$, with $y_{cms}$ being the $J/\psi$ rapidity in the c.m. system frame and $\sqrt{s}$ the c.m. energy

# Exclusive production at HERA

*O. Lukina*
M.V. Lomonosov Moscow State University, Skobeltsyn Institute of Nuclear Physics, Moscow, Russia

**Abstract**
Recent studies of exclusive processes in $ep$ scattering at HERA are presented. These include the results on exclusive dijets production and the measurement of the production ratio $\psi(2S)/\psi(1S)$ in diffractive deep inelastic scattering, the first HERA measurement of $\rho^0$ meson photoproduction with a leading neutron.

## 1 Introduction

The world only $ep$ collider HERA operated at DESY, Hamburg, with electrons or positrons at 27.5 GeV and protons at 820 or 920 GeV during the years 1992 to 2007. Two collider experiments, H1 and ZEUS, collected data corresponding to an integrated luminosity of 0.5 fb$^{-1}$ each. The high resolution multi-purpose detectors H1 and ZEUS equipped with additional forward sub-detectors allow for detailed analyses of exclusive processes, reviewed in this talk. These processes are clean experimentally, kinematic variables are fully reconstructed by measuring scattered electron and vector meson decay products or jets. Forward baryons ( protons and neutrons), which carry a large fraction of momentum of the incoming proton are detected with low acceptance by forward detectors.

The HERA facility can be viewed as a $\gamma^* p$ collider, which allows measurements within a single experiment to be made over a wide range of the $\gamma p$ centre-of-mass energy, $W$, up to its maximum value, set by the centre-of-mass energy of the electron-proton system $\sqrt{s} = 310$ GeV. The photon virtuality, $Q^2$, ranged up to several thousand GeV$^2$. The reaction phase space can be divided into photoproduction ($\gamma p$), $Q^2 \approx 0$ GeV$^2$, and deep inelastic scattering ( DIS), where $Q^2 > 1$ GeV$^2$. To describe the processes are used four momentum transfer squared at the proton vertex, $t$, Bjorken scaling variable, $x_{Bj}$, which is a fraction of proton's momentum carried by struck quark, and also the fractional loss of proton longitudinal momentum, $x_{IP}$, and a fraction of Pomeron momentum 'seen' by photon, $\beta = x/x_{IP}$. At high energies as at the HERA collider the $ep$ interaction is in fact an interaction of a virtual photon, emitted from the electron with the incident proton. Such processes is usually thought to proceed in two steps: interaction of the virtual photon with partons in the proton and the fragmentation of the intermediate partonic state into final hadrons. When the special configuration of the intermediate partonic state is small the former interaction is $hard$ which implies possibility of the interaction within the perturbative quantum chromodynamics (pQCD), whereas the latter one is typical $soft$ hadronic process, usually described by the phenomenological models. The precision of HERA data allows studies of exclusive processes in their transition from $soft$ to $hard$ interactions as well as in the pQCD domain.

## 2 Production of exclusive dijets in diffractive deep inlelastic scattering

ZEUS Collaboration has reported on the first measurement of exclusive dijet production in high energy electron-proton scattering $e + p \to e' + p' + jet1 + jet2$ with only dijet, electron and proton in the final state Ref. [1]. The process can be viewed as an interaction between the virtual photon, $\gamma^*$, and the proton, which is mediated by the exchange of a colourless object, Pomeron, $IP$, as it shown in Fig. 1. The production of exclusive dijets in DIS is sensitive to the nature of the object exchanged, therefore this measurement allows of different assumptions about the nature of diffractive exchange to be tested. The measurement is based on data collected with the ZEUS detector in 2003-2007 data-taking period and corresponds to an integrated luminosity of 372 pb$^{-1}$. A clean sample of DIS events with a well-reconstructed electron was selected. Events were accepted if a virtuality of exchanged photon $Q^2 > 25$ GeV$^2$ and photon-proton center of mass energy $90 < W < 250$ GeV$^2$. Diffractive events, characterised by the presence of a large rapidity gap (LRG) between proton beam direction and the hadronic final state, were selected, if the fraction of the proton momentum carried by the diffractive exchange, $x_{IP} < 0.01$ and $M_X > 5$ GeV, where $M_X$ denotes the invariant mass of the hadronic state recoiling against the leading proton. The data were analysed as a function of $\beta$.

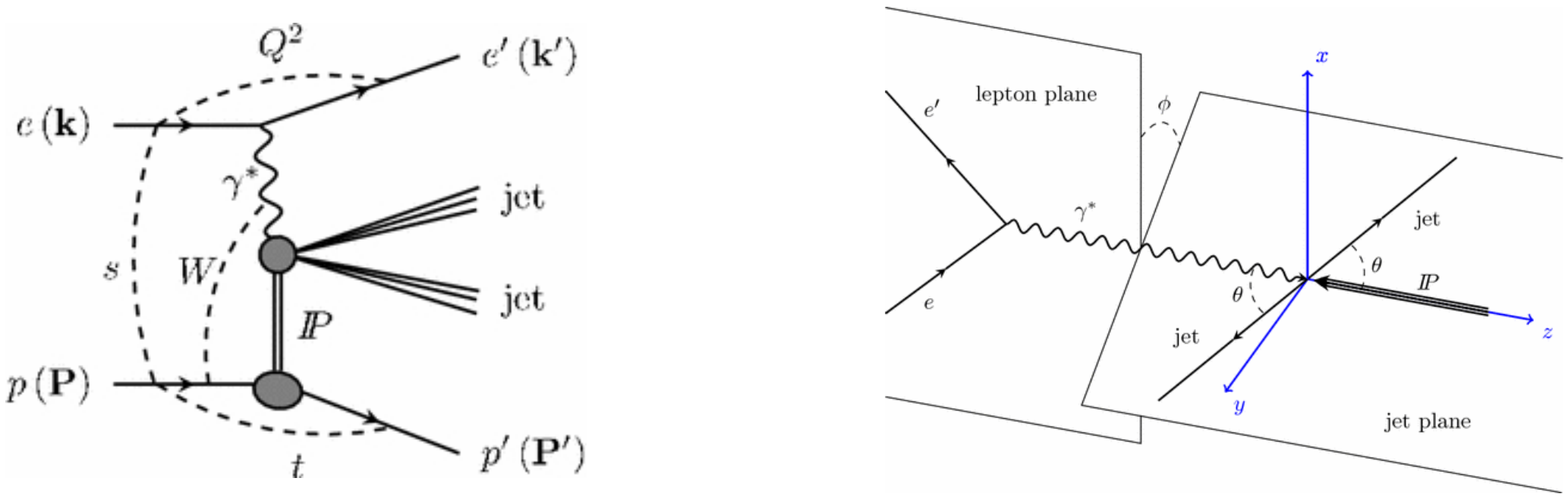

**Fig. 1:** (left) Schematic view of the diffractive production of exclusive dijets; (right) Definition of planes and angles in $\gamma^* - IP$ centre-of-mass frame. The angle $\phi$ is an angle between these two planes.

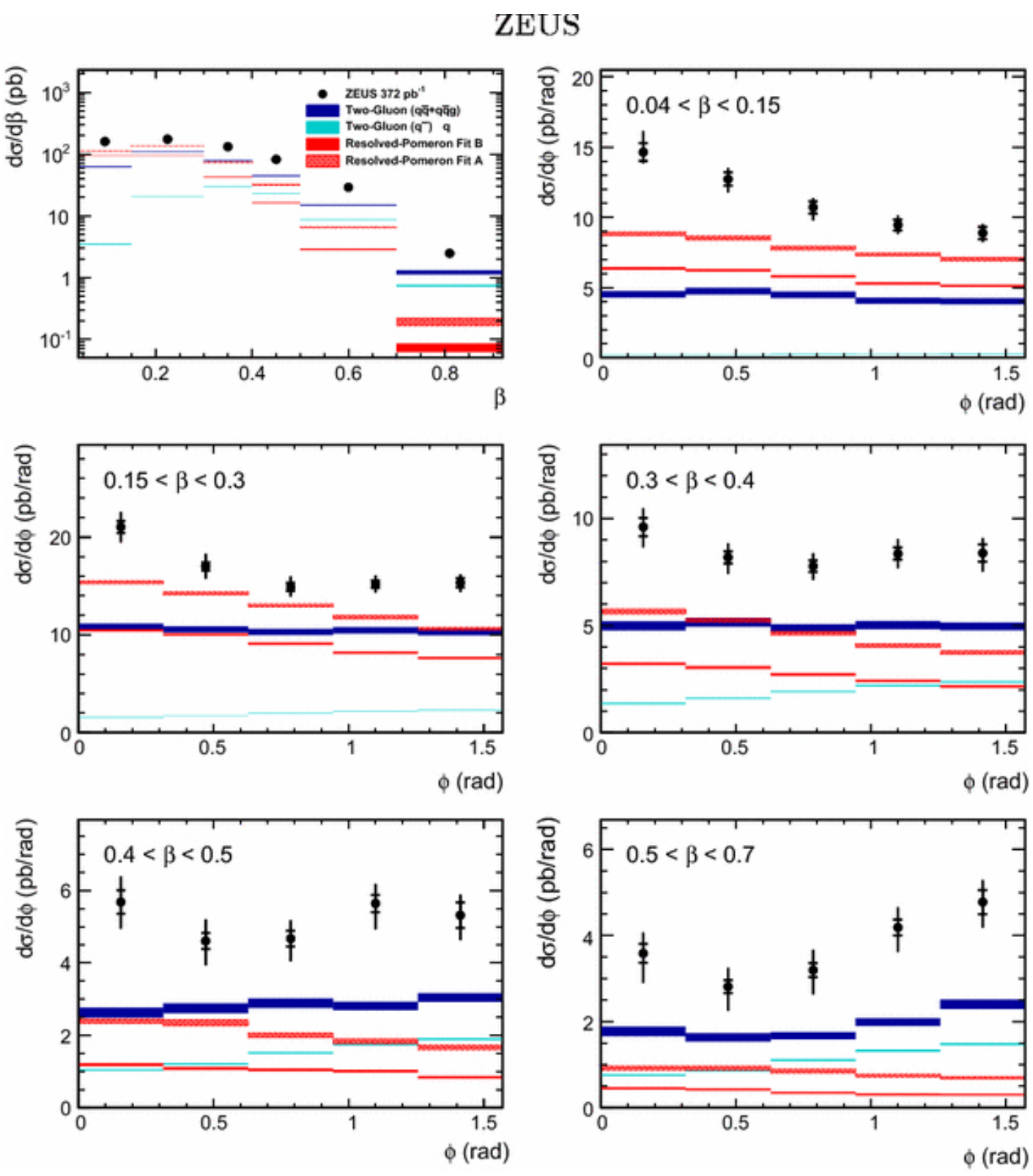

**Fig. 2:** Differential cross sections $d\sigma/d\beta$ ( in log scale) and $d\sigma/d\phi$ in bins of $\beta$ ( in linear scale) in comparison to model predictions.

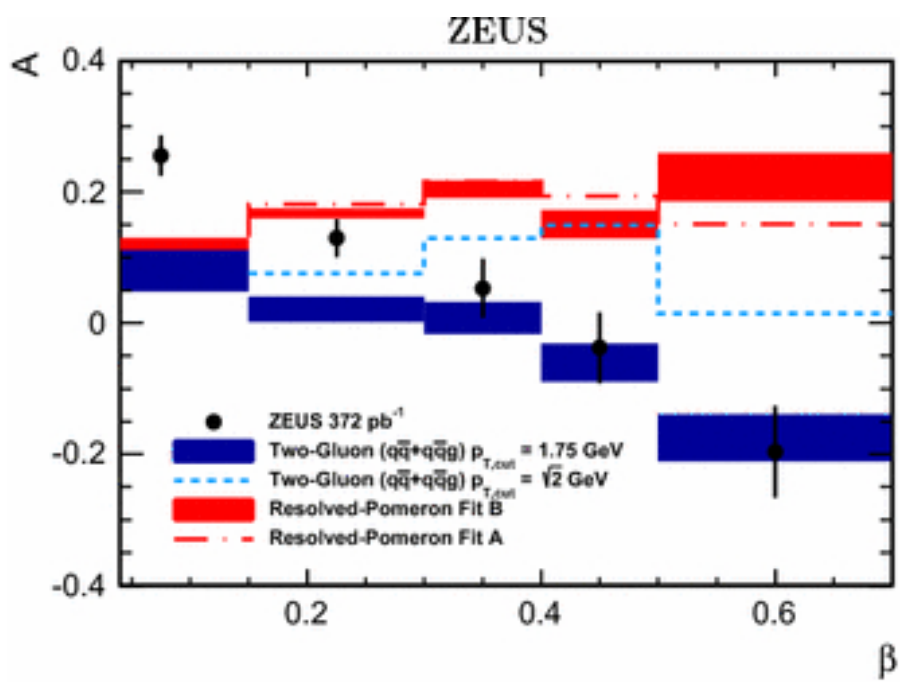

**Fig. 3:** The shape parameter A as a function of $\beta$ in comparison to the values of $A$ obtained from distributions predicted by the Resolved-Pomeron model and the Two-Gluon-Exchange model. The bands in model predictions represent statistical uncertainties.

Jets were found in $\gamma^* - IP$ centre-of-mass frame with $z$ axis along virtual photon momentum and $y$ axis along direction defined by the cross product of virtual photon and scattered lepton momenta, as it shown in Fig. 1. The kt-cluster algorithm known as the Durham jet algorithm Ref. [2] used allows the association of the individual hadrons with a unique jet on an event-by-event basis.

The differential cross sections d$\sigma$/d$\beta$ and d$\sigma$/d$\phi$, as shown in Fig. 2, were compared to MC predictions for the Resolved-Pomeron model Ref. [3] and the Two-Gluon-Exchange model Ref. [4, 5], as implemented in RapGap Monte Carlo generator. The measured absolute cross sections are larger than those of theoretical expectations. Both models predict different shapes in the azimuthal angle $\phi$ between lepton and jet planes. The shape of the $\phi$ distributions were parametrized in different intervals of $\beta$ with the function 1+ Acos $2\phi$ as motivated by theory. The $\phi$ distributions show a significant feature: when going from small to large values of $\beta$, the shape varies and the slope of the angular distribution changes sign. The variation of the shape was quantified by fitting a function to the $\phi$ distributions. The fitted function is predicted by theoretical calculations to be proportional to 1+ A cos $2\phi$ and reproduces the data behaviour. The cross section d$\sigma$/d$\phi$, normalised to the integrated cross section, is compared to predictions of the models in Fig. 3. The Two-Gluon-Exchange model predicts reasonably well the measured value of A for $\beta$ > 0.3, whereas the Resolved-Pomeron model exhibits a different trend. The Resolved-Pomeron model predicts a negative slope and fails to describe the data, while the Two-Gluon-Exchange model predicts a positive slope, which is consistent with the data. In terms of absolute normalisation, both the Resolved-Pomeron and the Two-Gluon-Exchange models are below the data.

## 3 Measurement of the cross-section ratio $\sigma_{\psi(2S)}/\sigma_{J/\psi(1S)}$

The ratio of the cross sections of the reactions $\gamma^* p \to \psi(2S) + Y$ and $\gamma^* p \to J/\psi(1S) + Y$, where Y denoted either a proton or a low mass proton-dissociation system, was measured with the ZEUS detector in the kinematic range 2 < $Q^2$ < 80 GeV$^2$, 30 < $W$ < 210 GeV and $|t|$ < 1 GeV$^2$ Ref. [6]. These two charmonium states have the same quark content, different radial distributions of the wave functions and their mass difference is small compared to the HERA centre-of-mass energy. Therefore, the measurement of the ratio of their electro-production cross sections allows perturbative QCD predictions of wave function dependence of the $c\bar{c}$-proton cross section to be tested. A supression of the $\psi(2S)$ cross section relative to the $J/\psi(1S)$ is expected, as the $\psi(2S)$ wave function has a radial node close to the typical transverse separation of the virtual $c\bar{c}$ pair. The measurement was based on all available data sample amounting to 468 pb$^{-1}$, which consists of data from 1996-2000 and 2002-2007 running periods. Events were selected with no activity in the central ZEUS detector in addition to signals from the scattered electron and the decay products of the studied mesons. The sample contained exclusive and a small fraction of proton-dissociative events with diffractive masses $M_Y$ < 4 GeV which was assumed to cancel in the cross section ratio. The decay channels were: $J/\psi(1S)$, $\psi(2S) \to \mu^+\mu^-$, and $\psi(2S) \to J/\psi(1S)\pi^+\pi^-$ with the subsequent decay $J/\psi(1S) \to \mu^+\mu^-$. The studied process was simulated with the DIFFVM MC package. The GRAPE package was used for simulating exclusive and Bethe-Heitler dimuon production. The final sample contains $\sim$ 2500 $J/\psi(1S)$ and $\sim$ 190 $\psi(2S)$ events. After correcting for the detector acceptance, efficiency and the branching ratios, the cross section ratio $R = \sigma_{\psi(2S)}/\sigma_{J/\psi(1S)}$ was determined in bins of $Q^2$, $W$ and $|t|$. No dependence of the ratio on $W$ and $|t|$ was found. For the $Q^2$ dependence of the ratio $R$ a positive slope is observed. In Fig. 4 the results on $Q^2$ dependence are shown together with previous H1 measurements Ref. [7, 8]. The H1 Collaboration has found a value of $R = 0.150 \pm 0.035$ at photoproduction regime ($Q^2 \approx 0$). The HERA data behaviour is consistent with many of the models, which qualitatively reproduce the rise of $R$ with $Q^2$ although some of the models are excluded.

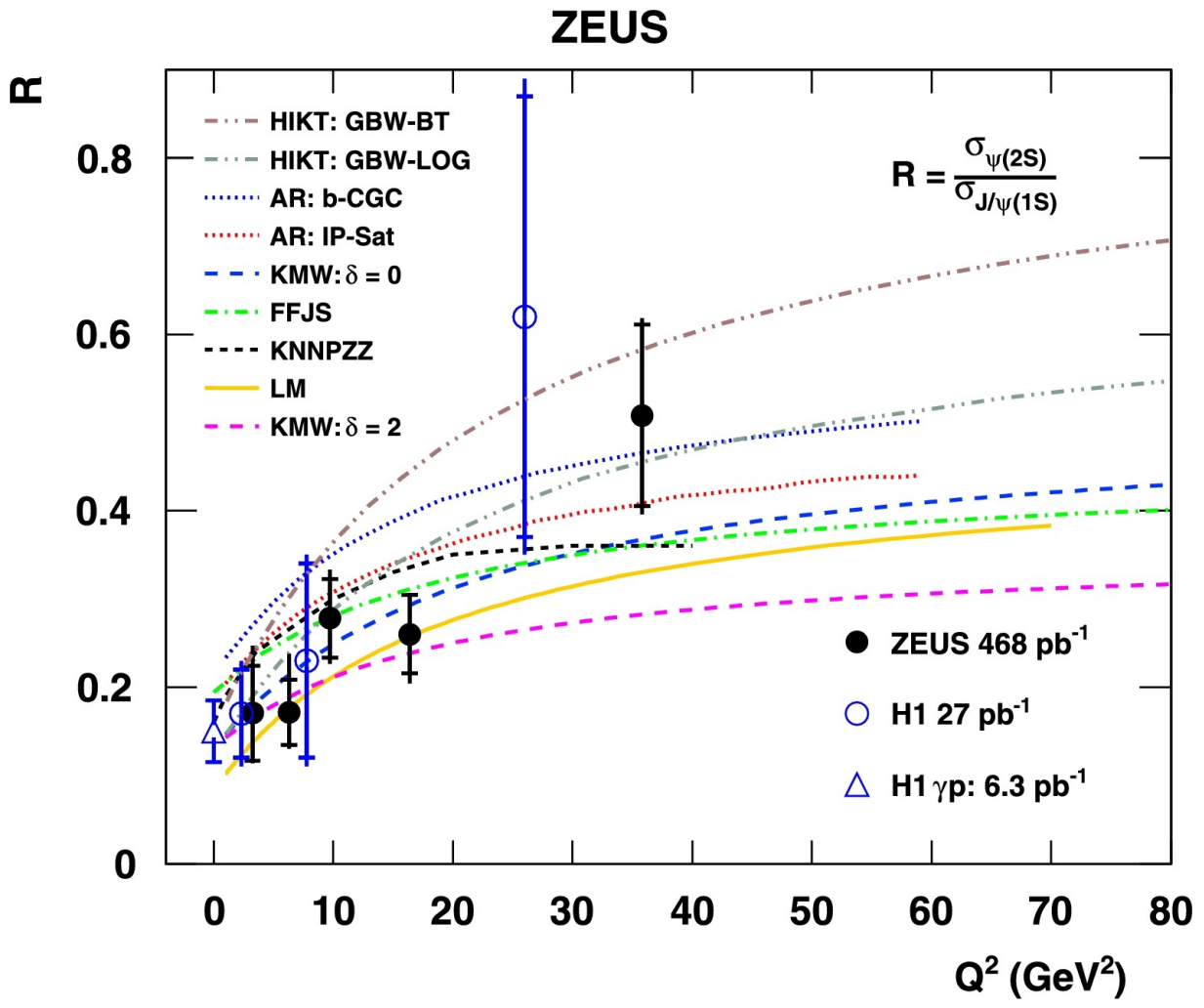

**Fig. 4:** The measured cross section ratio $R = \sigma_{\psi(2S)}/\sigma_{J/\psi(1S)}$ as a function of photon virtuality $Q^2$ in comparison to theoretical predictions.

## 4 Exclusive $\rho^0$ meson photoproduction with a leading neutron

The H1 experiment has reported of the measurement of exclusive photoproduction of $\rho^0$ mesons associated with leading neutrons, $\gamma p \to \rho^0 n \pi^+$, Ref. [9] with the aim to investigate exclusive $\rho^0$ production on virtual pions in the photoproduction and to extract for the first time experimentally the quasi-elastic $\gamma p \to \rho^0 \pi$ cross section. Here, for the $\rho$ meson photoproduction at a soft scale, Reggy phenomenology is most appropriate to describe the reaction. Figure 5 shows a set of Regge diagrams contributing to the signal (a,b,c) and to background (d) for this process. The pion exchange at the proton vertex is followed by elastic scattering of the pion on the virtual photon emitted from the beam electron, $\gamma p \to \rho^0 \pi^+$. This one-pion-exchange diagram (OPE) dominates at small $t \to 0$, where graphs (b,c) contributing to the scattering amplitude with opposite signs largely cancel. The data sample corresponds to an inte-grated luminosity of 1.16 pb$^{-1}$ and was collected at $\sqrt{s_{ep}}$ = 319 GeV using a special low multiplicity trigger. Exclusive events are selected containing only two oppositely charged pions from the $\rho^0$ decay, the leading neutron and nothing else above noise level in detector. This ensure the exclusivity and limits the dissociative background to the range $M_Y < 1.6$ GeV. The pion from the proton vertex is emitted under very small angle with respect to the proton beam and escapes detection. Signal events (Fig. 5a) are modelled by the POMPYT Monte Carlo generator, in which virtual pion is produced at the proton vertex according to one of the available pion flux parametrizations and then the quasi-elastic scattering process, $\gamma p \to \rho^0 \pi^+$, is generated. Diffractive $\rho^0$ production with proton dissociation into a system containing a neutron (Fig. 5d) contributes as background and is modelled by the DIFFVM MC generator. This background is subtracted from the data and its fraction in the final data sample is estimated to be $0.34 \pm 0.05$.

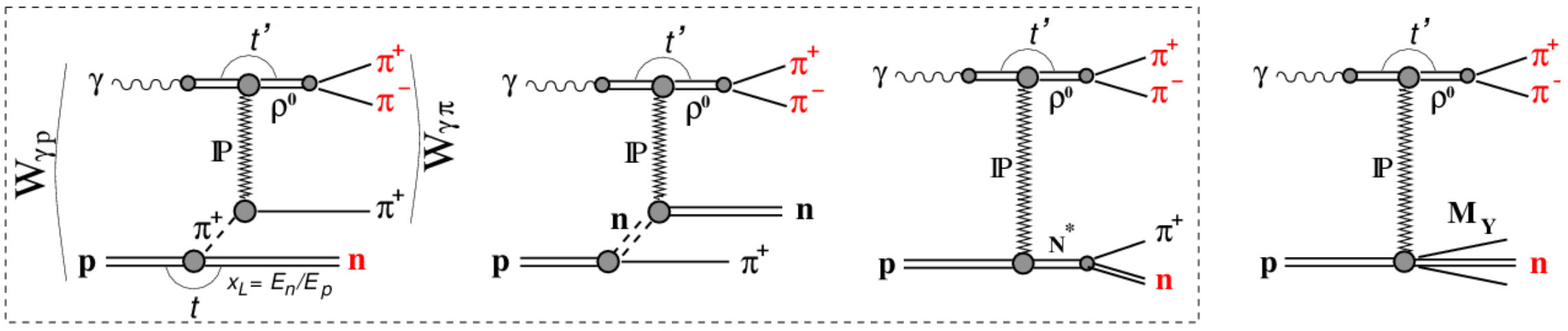

**Fig. 5:** Diagrams for processes contributing to exclusive photoproduction of $\rho^0$ meson associated with leading neutron. Three processes: pion exchange (a), neutron exchange (b) and a direct pole(c), The diffractive $\rho^0$ production with proton dissociation (d) is considered as background.

The cross section of the exclusive reaction $ep \to e\rho^0 n\pi^+$ was measured and converted into $\gamma p$ cross section using the effective photon flux of the Vector Dominance model. The integrated $\gamma p$ cross section in the kinematic range 20 < $W_{\gamma p}$ < 100 GeV, 0.35 < $x_L$ < 0.95 and $\theta$ < 0.75 mrad is measured as $\sigma_{\gamma p}$ = 310 ± 6(stat) ± 45(sys) nb. The differential cross section d$\sigma_{\gamma p}$/d$x_L$ compared to the predictions based on different models for the pion flux is shown in Fig. 6. The shape of the $x_L$ distribution is well reproduced by the pion flux parametrization. The double differential cross section d$^2\sigma_{\gamma p}$/d$x_L$d$p^2_{T,n}$ was measured and fitted with single exponential function and presented in Fig. 6. Steep falling at very high values of $x_L$ is observed. The exponential slope parameter of the leading neutron $b_n$ determined from the $p_{T_n}$ dependence compared to models with various pion flux is shown in Fig. 6. While the shape of the $x_L$ distribution is well reproduced by most of the pion flux approximations, the $x_L$ dependence of the $p_T$ slope of the leading neutron is not described by any of the existing models. The energy dependence of the reaction $\gamma p \to \rho^0 n\pi^+$ is presented in Fig. 7. The cross section drops with $W_{\gamma p}$ in contrast to the POMPYT Monte Carlo expectation, where the whole energy dependence is driven by Pomeron exchange alone. A Regge motivated power law fit $\sigma \propto W^\delta$ with $\delta$ = - 0.26 ± 0.06(stat) ± 0.07(sys) describes the data. Earlier HERA measurements Ref. [10, 11] have shown that the cross section of the reaction $\gamma p \to \rho^0 p$ has another trend and prefers to increase with $W$ with the Pomeron trajectory $\delta \approx$ 0.08. The pion flux models compartible with the data in the shape of $x_L$ distribution are used to extract the photon-pion cross section from d$\sigma$/d$x_L$ in the OPE approximation. The results are presented in Fig. 7. Using the pion flux parametrisation of the Holtmann model Ref. [12, 13] the elastic $\gamma\pi$ cross section is determined at the average energy <$W_{\gamma\pi}$> ∼ 24 GeV: $\sigma(\gamma\pi^+ \to \rho^0\pi^+)$ = 2.33 ± 0.34(exp)$^{+0.47}_{-0.40}$(model) $\mu$b. The estimated cross section ratio for the elastic photoproduction of $\rho^0$ meson on the pion and on the proton, is $r_{el} = \sigma^{\gamma\pi}_{el}/\sigma^{\gamma p}_{el}$ = 0.25 ± 0.06. A similar ratio, but for the total cross sections at <$W$> = 107 GeV, has been estimated by the ZEUS collaboration as $r_{tot}$ = $\sigma^{\gamma\pi}_{tot}/\sigma^{\gamma p}_{tot}$ 0.32 ± 0.03 Ref. [14]. Both measured ratios are significantly smaller the additive quark model predictions. This may be attributed to rescattering, or absorption corrections, which are expected to play an essential role in $soft$ peripheral process. For the studied reaction $\gamma p \to \rho^0 n\pi^+$ this would imply an absorption factor of $K_{abs}$= 0.44 ±0.11.

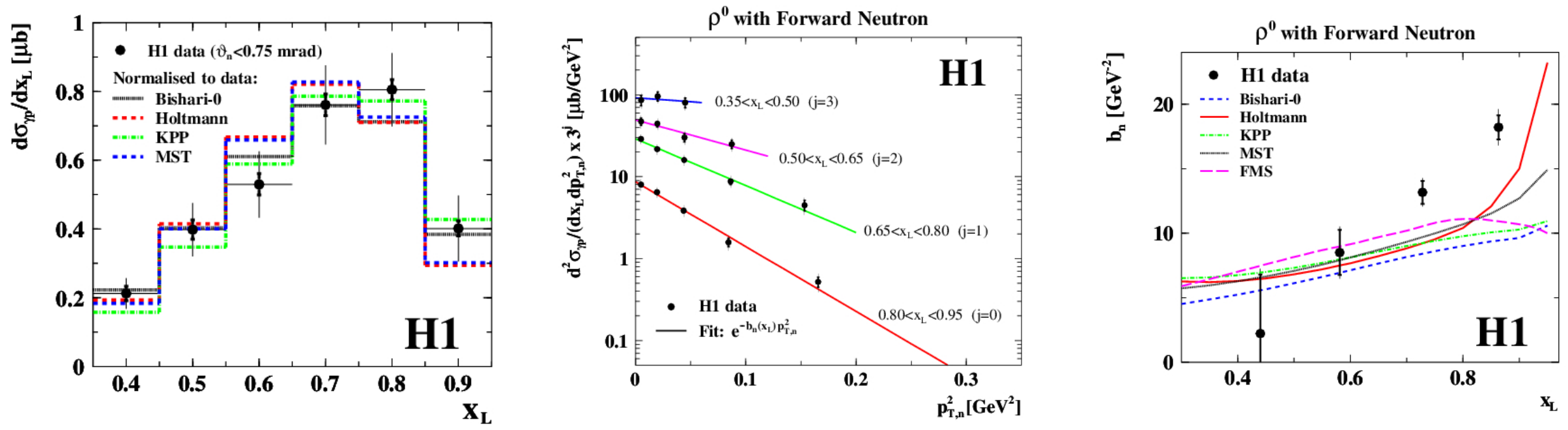

**Fig. 6:** $(left)$ Differential cross section d$\sigma_{\gamma p}$/d$x_L$ compared to the predictions based on different models for the pion flux. $(middle)$ Double differential cross section of neutrons d$^2\sigma_{\gamma p}$/d$x_L$d$p^2_{T,n}$ in the range 20 < $W_{\gamma p}$ < 100 GeV fitted with a single exponential function. $(right)$ The exponential slope fitted through the $p^2_T$ dependence of the leading neutron as a function of $x_L$ compared to the expectations of several parametrisations of the pion flux within OPE model.

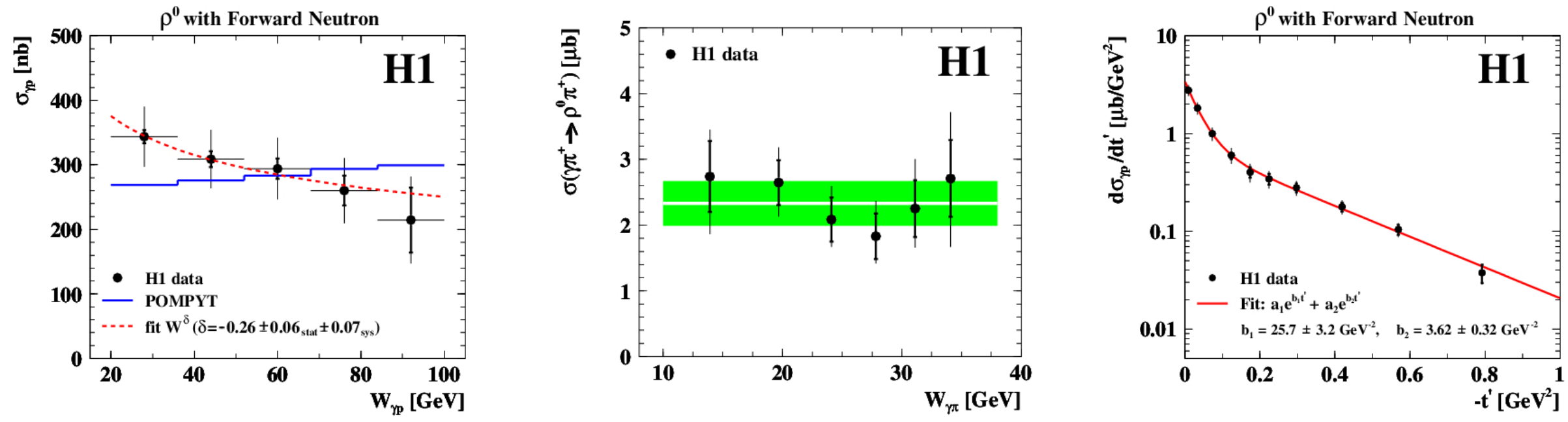

**Fig. 7:** $(left)$ Cross section of the reaction $\gamma p \to \rho^0 n\pi^+$ as a function of $W_{\gamma p}$ . $(middle)$ Elastic cross section $\sigma^{\gamma\pi}_{el} \equiv \sigma(\gamma\pi^+ \to \rho^0\pi^+)$, extracted in the OPE approximation as a function of the photon-pion energy $W_{\gamma\pi}$. $(right)$ Differential cross section d$\sigma_{\gamma p}$ / d$t'$ of $\rho^0$ meson fitted with the sum of two exponential functions.

The cross section as a function of the four-momentum transfer squared of the $\rho^0$ meson, $t'$, is shown in Fig. 7. The $t'$ distribution is fitted with the sum of two exponential functions with different slopes for the low-$t'$ and the high-$t'$ regions, which is typical for double peripheral exclusive reactions. In DPP approach this is due to exchange of two Regge trajectories, Pomeron and pion.

## 5 Summary

A brief review of recent HERA measurements of exclusive processes has been presented. The production of exclusive dijets in deep inelastic $ep$ scattering was measured for the first time and compared to predictions from models based on different assumptions about the nature of diffractive exchange. The measured cross section ratio of the charmonium states, $\psi(2S)$ and $J/\psi(1S)$, in exclusive deep inelastic electroproduction using full HERA data sample, confirms the expectations of QCD-inspired models of vector-meson production. Photoproduction of exclusive $\rho^0$ meson associated with a leading neutron measured first time at HERA which allowed the diffractive production of $\gamma\pi$ scattering to be studied. High accuracy of the presented measurements of the exclusive state production has providing new details to test for QCD theory and experiments at the LHC.

## Acknowledgements

Many thanks to all colleagues in ZEUS and H1 for providing the material in this report.

# Prompt photons at HERA

*P. J. Bussey*
School of Physics and Astronomy
University of Glasgow
Glasgow, United Kingdom, G12 8QQ
for the ZEUS and H1 Collaborations

**Abstract**

The ZEUS detector at HERA has been used to measure the photoproduction of isolated photons in diffractive photoproduction events and deep inelastic scattering (DIS). In the diffractive analysis, cross sections were compared to predictions from the RAPGAP Monte Carlo simulation. First evidence is seen for the direct-photon direct-Pomeron interaction. Cross sections in DIS were evaluated for a number of kinematic two-particle variables and compared to predictions from the AFG and BLZ theoretical models.

## 1 Experimental method: common features

Results are presented here for the production of isolated hard photons ("prompt" photons) in $ep$ interactions at HERA. Two analyses are described: one of the diffractive photoproduction of prompt photons, and one of their production in deep inelastic scattering (DIS) processes. The measurements are based on data samples corresponding to integrated luminosities of 82 and 374 $\mathrm{pb}^{-1}$, taken during the years 1998–2000 and 2004–2007 respectively with the ZEUS detector at HERA, and referred to as HERA-I and HERA-II data respectively. During these periods, HERA ran with an electron or positron beam energy of 27.5 GeV and a proton beam energy of $E_p = 920$ GeV. The samples include $e^+p$ and $e^-p$ data[1].

Charged particles were measured in the ZEUS central tracking detector and a silicon micro vertex detector, which operated in a solenoidal field of 1.43 T. ZEUS used a uranium-scintillator calorimeter, of which the barrel electromagnetic calorimeter (BEMC) had cells with a pointing geometry aimed at the nominal interaction point. Its fine granularity allowed the use of shower-shape distributions in the measurement of outgoing high-energy photons.

Monte Carlo (MC) event samples were employed to evaluate the detector acceptance and event-reconstruction efficiency, and to provide signal and background distributions. RAPGAP 3.2 [1] was used to generate the diffractive process $pe \to pe\gamma X$ for direct and resolved incoming virtual photons at low $Q^2$, incident on resolved Pomerons modelled according to the approach of Ingelman and Schlein. The diffractive proton PDF set H1 DPDF Fit B (2006) was used and, for the resolved photon, the PDF set SASGAM 2D. The program PYTHIA 6.416 was used to generate direct and resolved prompt-photon photoproduction processes for background calculations. In the DIS analysis, PYTHIA was used, and also DJANGOH interfaced with HERACLES to generate events with initial- and final-state photon radiation from the electron.

The measured photons are accompanied by backgrounds from neutral mesons in hadronic jets, in particular $\pi^0$ and $\eta$, where the meson decay products can create an energy cluster in the BCAL that passes the selection criteria for a photon. In the diffractive analysis these were modelled using RAPGAP events with a dijet final state in which one of the jets resembled a photon candidate. Further diffractive backgrounds came from DIS events and Bethe-Heitler events that produced a photon-electron final state. These were simulated using DJANGOH 6 interfaced with HERACLES, and with GRAPE-COMPTON. In the DIS analysis, DJANGOH dijet events were used to evaluate the mesonic backgrounds.

[1]Hereafter "electron" refers to both electrons and positrons unless otherwise stated.

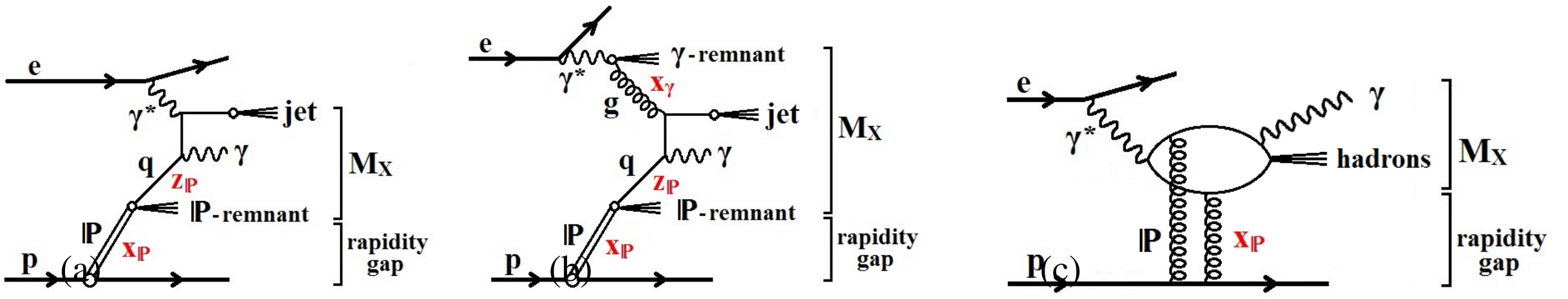

**Fig. 1:** Examples of the diffractive production of a prompt photon and a jet in $ep$ scattering of a resolved Pomeron from (a) direct (b) resolved photons, and (c) a direct Pomeron process.

The basic event selection and reconstruction were performed as previously [2]. A three-level trigger system was used to select events online. The event analysis made use of energy-flow objects (EFO's), constructed from clusters of calorimeter cells, associated with tracks when appropriate, and also unassociated tracks. Photon candidates were identified as EFO's with no associated track, and with at least 90% of the reconstructed energy measured in the BEMC.

Jets were reconstructed using all the EFO's in the event, including photon candidates, by means of the $k_T$ clustering algorithm in the $E$-scheme in the longitudinally invariant inclusive mode with the radius parameter set to 1.0. To reduce the fragmentation contribution and the background from the decay of neutral mesons within jets, the photon candidate was required to be isolated from other hadronic activity. This was imposed by requiring that the photon-candidate EFO had at least 90% of the total energy of the reconstructed jet of which it formed a part.

Events were selected with the following kinematic conditions. A photon candidate was required with pseudorapidity, $\eta^\gamma$, in the range $-0.7 < \eta^\gamma < 0.9$. and with reconstructed transverse energy in the range $5 < E_T^\gamma < 15$ GeV for the diffractive analysis or $E_T^\gamma > 4$ GeV for the DIS,. The hadronic jet, when used, was required to have pseudorapidity, $\eta^{\rm jet}$, range $-1.5 < \eta^{\rm jet} < 1.8$. and transverse energy $4 < E_T^{\rm jet} < 35\,$GeV for the diffractive analysis or $E_T^{\rm jet} > 2.5\,$GeV for the DIS. Background events arising from neutral meson decays were subtracted statistically, following the approach taken in previous ZEUS analyses.

It is possible for a photon to be radiated within a jet. Such processes are hard to model within QCD and are suppressed by requiring that the outgoing photon be isolated. Photon isolation was imposed such that at least 90% of the energy of the jet-like object containing the photon originated from the photon.

### 1.1 Kinematic quantities

When a beam proton, with energy $E_p$, radiates a Pomeron or equivalent, the fraction of the proton energy taken by the radiated Pomeron is given to a good approximation by the variable $x_{I\!P}^{\rm meas} = (E^{\rm all} + p_Z^{\rm all})/2E_p$, where the suffix "all" refers to all final-state particles or detector-measured objects apart from the forward proton and its possible dissociation products, and excluding the scattered electron. In "direct" photoproduction processes (fig. 1(a)), the entire energy of the incoming photon is absorbed by the target, while in "resolved" processes (fig. 1(b)), the incoming photon's hadronic structure provides a quark or gluon that interacts with a parton from the target. These two classes of process are unambiguously defined only at lowest order, but may be partially distinguished in events containing a high-$E_T$ photon and a jet by means of $x_\gamma^{\rm meas} = (E^\gamma + E^{\rm jet} - p_Z^\gamma - p_Z^{\rm jet})/(E^{\rm all} - p_Z^{\rm all})$, which measures the fraction of the incoming photon energy that is given to the photon and the jet. The quantities $E^\gamma$ and $E^{\rm jet}$ denote the energies of the photon and the jet, respectively, and $p_Z$ denotes the corresponding longitudinal momenta. The presence of direct processes generates a prominent peak in the cross section at high $x_\gamma^{\rm meas}$ values. Similarly, direct and resolved Pomeron processes may be defined. The fraction of the Pomeron energy that is taken by the outgoing photon and jet is given to a good approximation by: $z_{I\!P}^{\rm meas} = (E^\gamma + E^{\rm jet} + p_Z^\gamma + p_Z^{\rm jet})/(E^{\rm all} + p_Z^{\rm all})$. Figure 1(c) illustrates a possible event type in which the Pomeron is emulated by two exchanged gluons and there is no Pomeron remnant.

## 2 Diffractive analysis

Diffractive hadronic interactions involve the exchange of a colour-neutral object known as the "Pomeron". Diffractive scattering off protons may be initiated by a second incoming hadron, or by a real or virtual photon. At the HERA *ep* collider, diffractive processes have been studied both in photoproduction and in deep inelastic scattering, the photoproduction processes consisting of those in which the exchanged virtual photon is quasi-real. The diffractive process is characterized by a forward nucleon, followed by a "gap" in rapidity in which little or no energetic scattering is found until the central region of the process where the hard final state is detected and measured. The present measurements follow an earlier study by H1 [3] of inclusive diffractive high energy prompt photons as a function of transverse momentum. Analyses of isolated photons in non-diffractive photoproduction have also been made by the ZEUS and H1 collaborations [2, 4].

In the measurements presented here, a hard prompt photon is detected in the central region of the ZEUS detector and may be accompanied by a jet [5]. Such processes, while rare, are interesting for several reasons. An outgoing photon must be radiated from a charged partonic object, namely a quark, and therefore demonstrates the presence of a quark content in the Pomeron or of scattering in which both the Pomeron and incident photon couple to quarks. In general, they allow QCD-based models to be tested in order to improve our understanding of a type of process which is important at high energies.

### 2.1 Diffractive selections

To select diffractive events further conditions were applied, the first of which was that the maximum pseudorapidity for EFO's with energy above 0.4 GeV satisfied $\eta_{\rm max} < 2.5$. A second diffractive condition was that $x_{I\!P}^{\rm meas} < 0.03$. A selection on the Jacquet–Blondel variable, $0.2 < y_{\rm JB} < 0.7$, removed DIS events. For the HERA-I data sample, the energy in the Forward Plug Calorimeter was required to be less than 1 GeV. This calorimeter was not present in the HERA-II running. A serious background consisted of Bethe-Heitler events containing a high-$E_T$ photon and electron in the final state. These and remaining DIS events were efficiently removed by rejecting events with an identified electron and less than six EFO's in the detector.

The results are potentially affected by proton-dissociation processes, in which the products of the proton dissociation pass undetected inside the central aperture of the Forward Calorimeter. These were not modelled in the version of RAPGAP that was used. In other ZEUS diffractive analyses, they were estimated to be up to 40% for the HERA-II data and up to 16% for the HERA-I data. The HERA-II data sample was used in the main analysis described here to give the best estimation of the shapes of the cross-section distributions, which were assumed not to depend on the presence of proton dissociation. The HERA-I data were analysed similarly, with the addition of the selection on the additional Forward Plug Calorimeter, and were used to evaluate the total cross section within the selected parameter range. Studies of the event shapes using PYTHIA indicated that the fraction of non-diffractive events in the sample lies between 0% and 10%. This was included in the systematic uncertainties, which were otherwise dominated by the uncertainties in the calorimeter and jet calibration.

### 2.2 Results

Differential cross sections were calculated for the diffractive production of an isolated photon, inclusive and with at least one accompanying jet. The kinematic region was defined by $Q^2 < 1$ GeV$^2$, $0.2 < y <$

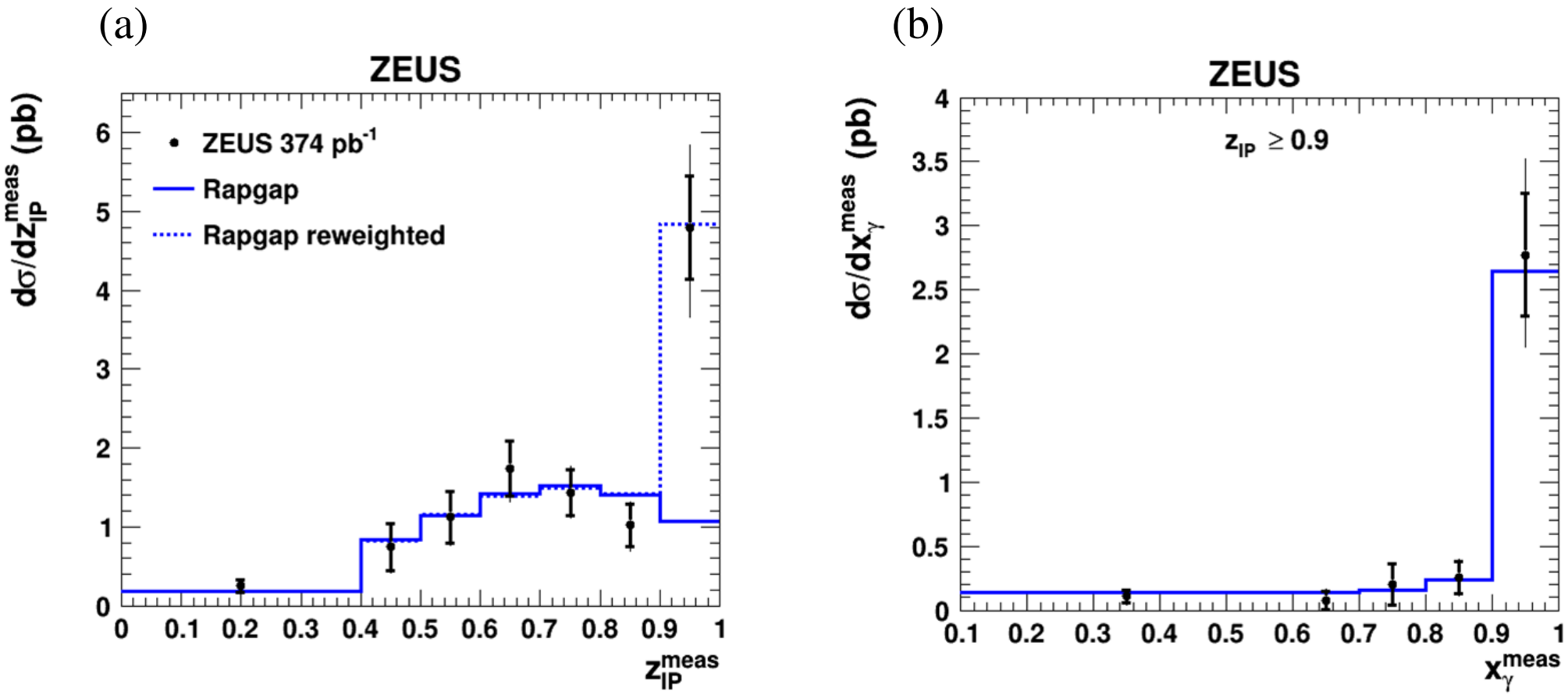

**Fig. 2:** Differential cross sections for events containing an isolated photon accompanied by a jet: (a) as a function of $z_{\mathbb{P}}^{\rm meas}$; (b) as a function of $x_\gamma^{\rm meas}$ for events with $z_{\mathbb{P}}^{\rm meas} > 0.9$ at the hadron level. Comparisons are made to normalized predictions from RAPGAP, with and without reweighting. Thick error bars are statistical and thin are statistical combined with systematic.

$0.7$, $-0.7 < \eta^\gamma < 0.9$, $5 < E_T^\gamma < 15$ GeV, $4 < E_T^{\rm jet} < 35$ GeV and $-1.5 < \eta^{\rm jet} < 1.8$. The diffractive condition consisted in requirements that $x_{\mathbb{P}}^{\rm meas} < 0.03$ and $\eta^{\rm max} < 2.5$. These quantities were evaluated at the hadron level in the laboratory frame, matching the cuts used at the detector level, and the jets were formed according to the $k_T$ clustering algorithm with the radius parameter set to 1.0.

The results here are inclusive of proton dissociation processes. The total cross section within the observed parameter ranges was found to be $0.68 \pm 0.14^{+0.06}_{-0.07}$ pb, to be compared to the RAPGAP prediction of 0.68 pb with no proton dissociation and no resolved-suppression factor. It is found that a high fraction of the events with an observed isolated high-$E_T$ photon are accompanied by at least one jet.

Figure 2(a) shows the differential cross section for $z_{\mathbb{P}}^{\rm meas}$, measured using the HERA-II data set. At the upper end of the distribution, a peak is seen which is not described by RAPGAP and which gives evidence for the presence of direct-Pomeron processes. The solid RAPGAP histogram is normalized to the region $z_{\mathbb{P}}^{\rm meas} < 0.9$, while the dashed histogram includes a reweighting factor and is normalized to the entire data distribution. Figure 2(b) shows the $x_\gamma^{\rm meas}$ cross section for events with $z_{\mathbb{P}}^{\rm meas} > 0.9$ and shows that these events are dominated by a direct-photon component.

Figure 3 shows cross sections for a selection of other variables. The RAPGAP histograms are normalized to the data and describe the shapes of the measured quantities well. Cross sections in $E_T^{\rm jet}$ above 15 GeV are omitted from fig. 3(c) owing to limited statistics, but this kinematic region is included in the other cross-section measurements. Figure 3(e) shows the difference in azimuth between the photon and the jet and demonstrates the back-to-back nature of these events. Figure 3(f) shows that the $x_{\mathbb{P}}^{\rm meas}$ distribution is well-contained within the selection criterion that was applied.

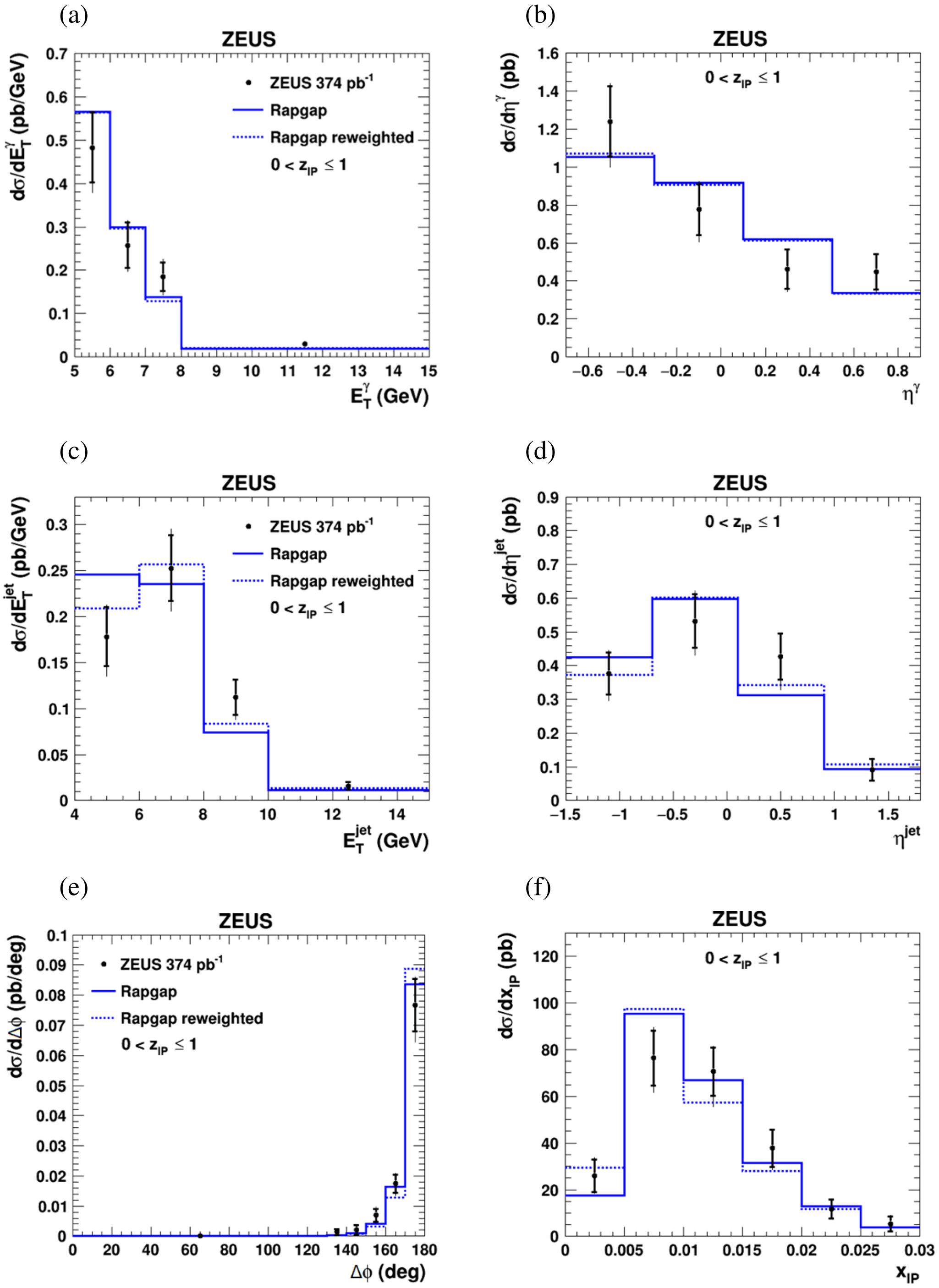

**Fig. 3:** Differential cross sections for events containing an isolated photon accompanied by a jet as functions of: (a) $E_T^{\gamma}$, (b) $\eta^{\gamma}$, (c) $E_T^{\rm jet}$, (d) $\eta^{\rm jet}$, (e) $\Delta\phi$, and (f) $x_{I\!P}^{\rm meas}$.

## 3 DIS analysis

### 3.1 Event selections

The DIS analysis used events taken HERA-II only. In each event, a scattered electron was required to be detected with energy above 10 GeV and with polar angle greater than 140° so that the electron was well measured in the ZEUS Rear calorimeter. The value of $Q^2$ reconstructed from the electron had to be in the range 10–350 GeV$^2$.

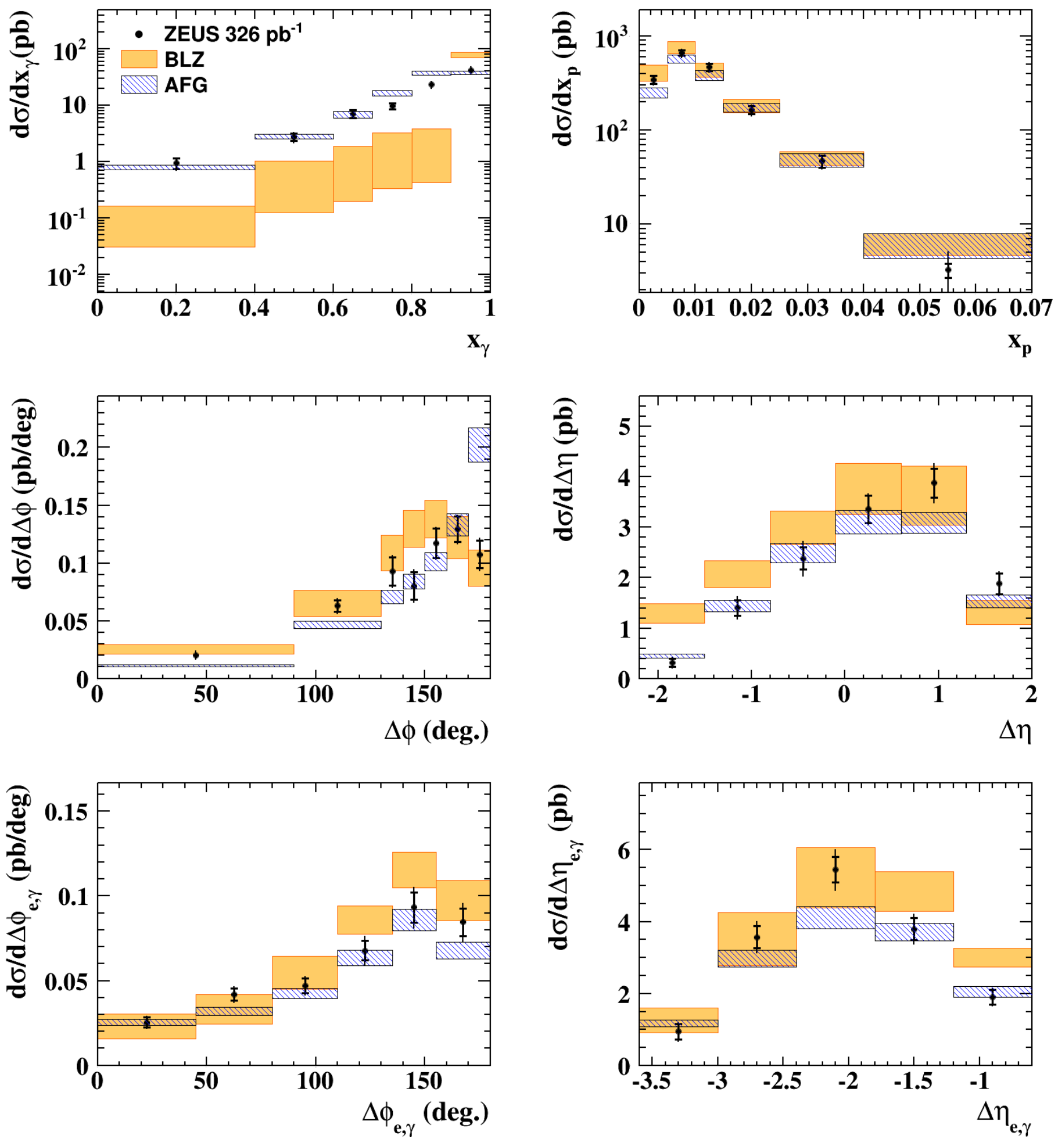

**Fig. 4:** Differential cross sections for DIS events containing an isolated photon accompanied by a jet as functions of: $x_\gamma^{\rm meas}$ (top left), $x_p^{\rm obs}$ (top right), $\Delta\phi$ between photon and jet (middle left), $\Delta\eta$ between photon and jet (middle right), $\Delta\phi_{e,\gamma}*$ (bottom left), $\Delta\eta_{e.\gamma}$ (bottom right).

### 3.2 Results

Cross sections for single-particle variables have been published in an earlier paper [6]. The variables presented in Fig. 4 depend on the event as a whole, or on pairs of final-state objects, and are compared to predictions from the theoretical models of AFG [7] and BLZ [8]. These theoretical predictions have corrections applied to them to allow for the conversion of the calculated parton-level final states into the hadronized final states that are corrected. These corrections were calculated using PYTHIA, with no attempt at this stage made to model thePYTHIA to the shape of the calculations at the parton level. Thus, the BLZ prediction displays a delta-function shape in $x_\gamma^{\rm meas}$ since no higher-order effects were included.

Further improvements could be made by a reweighting of the PYTHIA in these corrections.

Both models describe the data reasonably well within the theoretical uncertainties, which are based on the QCD scale uncertainty, with the exception of the variable $x_\gamma^{\mathrm{meas}}$. For this variable, further correction to the theory would be required to model migration effects downward from the final bin. On the whole, AFG performs slightly better than BLZ.

# Proton tomography by Deep Virtual Compton Scattering

*M. Guidal*
Institut de Physique Nucléaire d'Orsay, CNRS-IN2P3, Université Paris-Sud, Université Paris-Saclay, 91406 Orsay, France.

**Abstract**
We present some recent developments in the field of Generalized Parton Distribution and Deep Virtual Compton Scattering, namely the first extraction of the quark momentum-dependent proton charge radius from data.

These past 20 years, Deep Virtual Compton Scattering (DVCS) has proven to be a very advantageous and effective process to probe the internal quark and gluon structure of the nucleon. DVCS consists in the high-energy exclusive lepto-production of a real photon on a hadronic target, i.e. the $\ell N \to \ell N \gamma$ reaction for a target nucleon $N$. Beam energies at the level of the GeV and higher are in order, so as to probe distances of the order of the fermi and lower. By virtue of a QCD factorization theorem, the DVCS process allows one to access the structure functions of the nucleon called the Generalized Parton Distributions (GPDs). These functions, currently largely unknown, contain, inter alia, informations on the correlation between the spatial and momentum distributions of quarks (and gluons) inside the nucleon, on their angular momentum contribution to the spin of the nucleon, on the pressure distributions inside the nucleon, etc. We refer the reader to Refs. [1–4] for the original articles on GPDs and to Refs. [5–10] for reviews of the domain.

We present here some recent developments in the field. We show a first quasi-model-independent measurement of the proton charge radius as a function of the quarks' momentum fraction. This is often refered to as proton tomography.

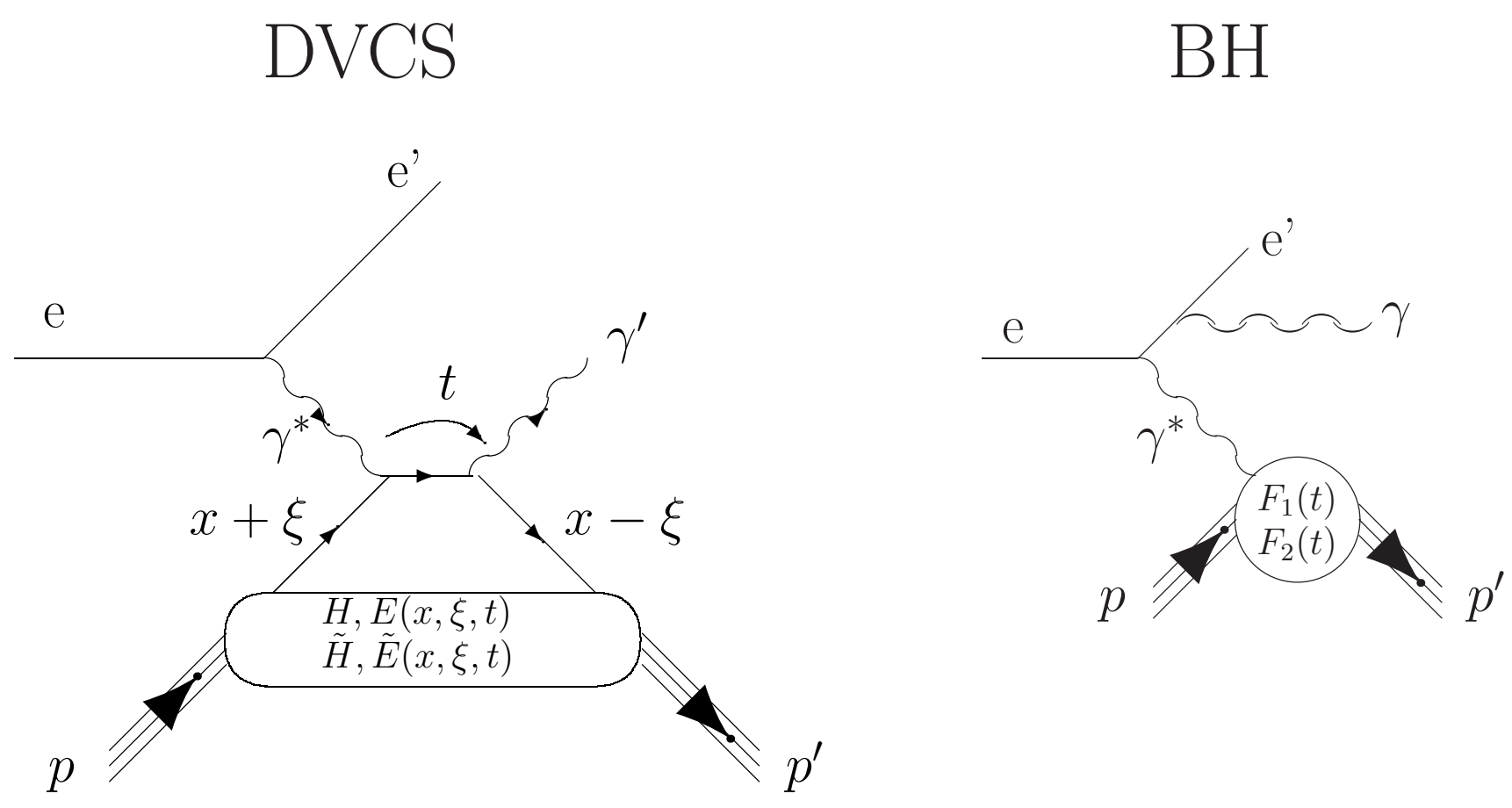

**Fig. 1:** Left: the DVCS process on the proton. Right: the BH process.

In the QCD leading-twist framework, in which this work takes place, there are four quark helicity-conserving GPDs, $H$, $E$, $\tilde{H}$ and $\tilde{E}$ contributing to the DVCS process (Fig. 1-left). They reflect the four independent helicity-spin transitions between the initial and final quark-nucleon systems. The dominant GPD $H$ represents for instance the contribution of unpolarized quarks in an unpolarized nucleon.

In the framework where QCD evolution effects are neglected, the GPDs are functions of three variables: $x$, $\xi$ and $t$. The quantity $x+\xi$ ($x-\xi$) represents the longitudinal momentum fraction of the initial (final) quark w.r.t. the average nucleon momentum, and the variable $t$ is the squared momentum transfer to the nucleon. This latter variable $t$ is actually the conjugate variable of the localization of the quark in the transverse position space (impact parameter $b_\perp$), in a frame where the nucleon goes to the speed of light in a given direction [11–13]. Thus, an intuitive interpretation of GPDs is that they describe the probability amplitude of hitting a quark in the nucleon with longitudinal momentum fraction $x+\xi$ and putting it back with a different longitudinal momentum fraction $x-\xi$ at a given transverse distance $b_\perp$ in the nucleon, relative to the transverse center of mass.

Extracting the GPDs from DVCS data is a very challenging problem because:

- The four GPDs need to be disentangled. The way to do so is to measure a series of observables for the $\ell N \to \ell N\gamma$ reaction, such as unpolarized cross sections, single or double beam and/or target spin asymmetries, charge asymmmetries,... Each observable is indeed in general dominantly sensitive to a given GPD (or a specific combination of GPDs).
- GPDs appear in the DVCS amplitude in the form of integrals over $x$. This is due to the loop in the DVCS diagram of Fig. 1-left, which generates convolution terms such as:

$$\int_{-1}^{+1} dx \frac{GPD(x,\xi,t)}{x-\xi+i\epsilon} + ..., \tag{1}$$

  where the denominator arises from the quark propagator. Thus, only $\xi$ and $t$ are experimentally accessible: $\xi$ is related to $x_B$, the standard Bjorken variable of Deep Inelastic Scattering, via $\xi = \frac{x_B}{2-x_B}$ and can thus be measured by detecting the scattered lepton kinematics; $t$ is measured by detecting the recoil nucleon or the final photon.
- As a consequence of this convolution, by virtue of the residue theorem, the maximum informations that can be extracted from the experimental data at a given $(\xi,t)$ point are quantities of the form $H(\pm\xi,\xi,t)$ when measuring an observable sensitive to the imaginary part of the DVCS amplitude, and $\int_{-1}^{+1} dx \frac{H(\mp x,\xi,t)}{x\pm\xi}$ when measuring an observable sensitive to the real part of the DVCS amplitude. In this work, we call these (real) quantities Compton Form Factors (CFFs). Since there are 4 GPDs, there are 8 CFFs.
- Another concern is that the DVCS process is not the only one contributing to the $\ell N \to \ell N\gamma$ reaction. There is also the Bethe-Heitler (BH) process in which the final state photon is radiated by the incoming or scattered lepton (see Fig. 1-right) and not by the nucleon itself like in DVCS. The BH contribution, which is quite precisely calculable, shall thus be taken into account, at the amplitude level, when extracting GPDs from experiment.

Extracting GPD information from DVCS data involves thus specialized and dedicated algorithms to adress all these issues. Several techniques have been proposed and developped these past years [10, 14–25] to extract the CFFs from different observables, with more or less model-dependency. In these short proceedings, we focus here on the fitting approach pioneered in Ref. [14] which consists in taking, at a fixed ($\xi$, $t$) kinematics, the 8 CFFs as free parameters, varying them in a systematic way in a conservatively bounded 8-fold hyperspace and, knowing the well-established BH and DVCS leading-twist amplitudes, finding the 8-CFF set which minimizes the difference between the theoretical calculation and the data. With this technique, the particular CFF $H_{Im}(\xi,t) \equiv H^q(\xi,\xi,t) - H^q(-\xi,\xi,t)$ could be extracted from several sets of polarized and unpolarized DVCS observables on the proton from the CLAS and Hall A Jefferson Lab experiments [26–29]. Fig. 2 shows such extraction of $H_{Im}$ for several ($\xi$, $t$) bins, at different $Q^2$ values, where $Q^2$ is the squared electron momentum transfer (we recall that, in the framework in which this work is done, CFFs don't depend on $Q^2$) .

Although error bars, which are systematic in nature, are rather large, one can rather clearly distinguish the general behavior where there is an increase of the $t$-slope and of the amplitude as $t \to 0$ of $H_{Im}$ as $\xi$ decreases. Qualitatively, this reflects respectively the increase of the transverse size of the proton (since $t$ is the conjugate variable of $b_\perp$) and of the quarks' density as smaller and smaller longitudinal quark momentum fractions are probed. In order to be quantitative and truely connect $H_{Im}$ to a charge proton radius, a specific procedure, detailed in Ref. [25], has to be applied. It involves:

- An extrapolation of $H_{Im}$ to $\xi = 0$, i.e. $H(\pm\xi, \xi, t)$ to $H(\pm\xi, 0, t)$,
- The connection of the singlet (quark + antiquark) to the non-singlet (quark - antiquark) contribution to which the proton radius is related. This step and the previous one carry some model-dependency, which is ultimately translated into an error bar (which is in general much lower than the uncerttainty associated to the $H_{Im}$ fitting extraction from the data)
- A Fourier transform to shift from the momentum space variable $t$ to the impact parameter space variable $b_\perp$. This latter step can be done analytically if a simple parametrization of $H_{Im}$ is used as in Ref. [25].

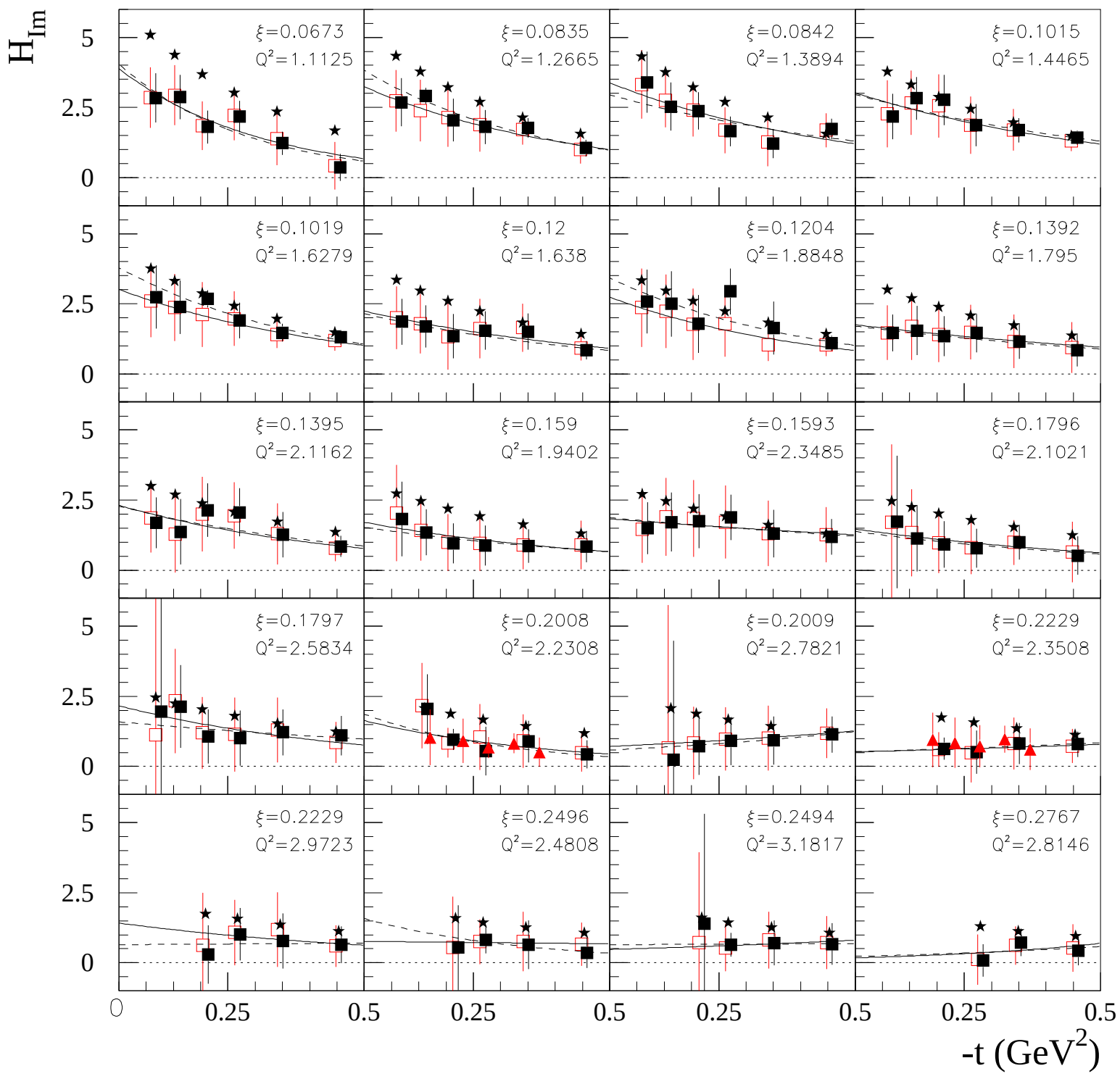

**Fig. 2:** The $H_{Im}$ CFF as a function of $t$ for 20 CLAS ($\xi$, $Q^2$) bins. The meaning of the different symbols and the precise definition of $H_{Im}$ can be found in Ref. [25], where the figure is taken from.

The resulting $x$-momentum-dependence of the proton transverse charge radius is displayed in Fig. 3. The upper plot of Fig. 4 shows a 3-dimensional representation of the fit of Fig. 3. The bottom plot is an artistic view of the tomographic quark content of the proton, with the charge radius and the density of the quarks increasing as smaller and smaller quark momentum fractions are probed.

In summary, ithese proceedings, we have given a very brief overview of one important outcome of the GPD physics, namely the extraction the $x$-dependence of the proton charged radius, for the first time from DVCS data. Several new DVCS experiments are planned with the JLab upgrade at 12 GeV in the short future, which should point to important new advances coming down in the field of nucleon structure.

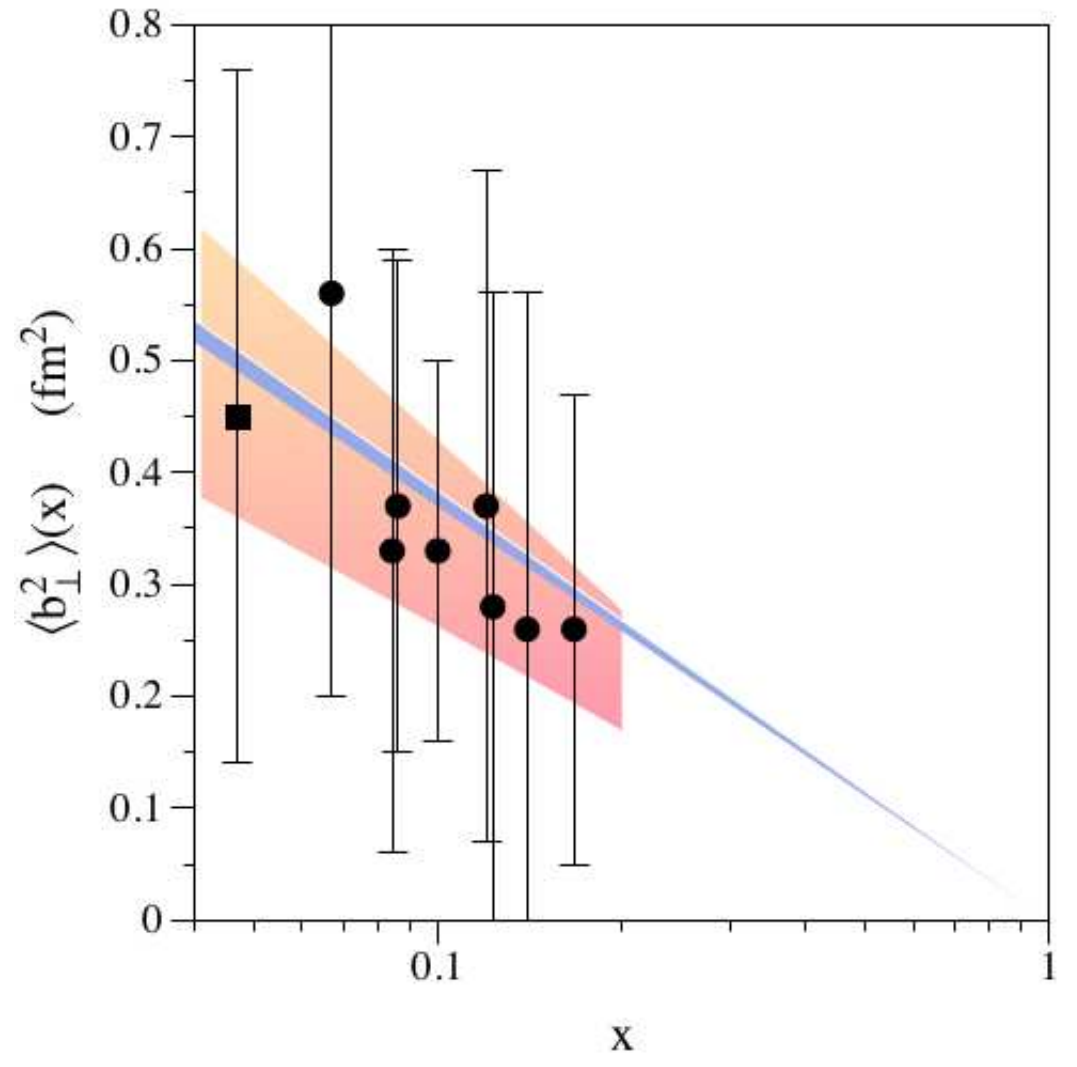

**Fig. 3:** $x$-dependence of the proton charge radius. The definition of the error bars and of the bands can be found in Ref. [25].

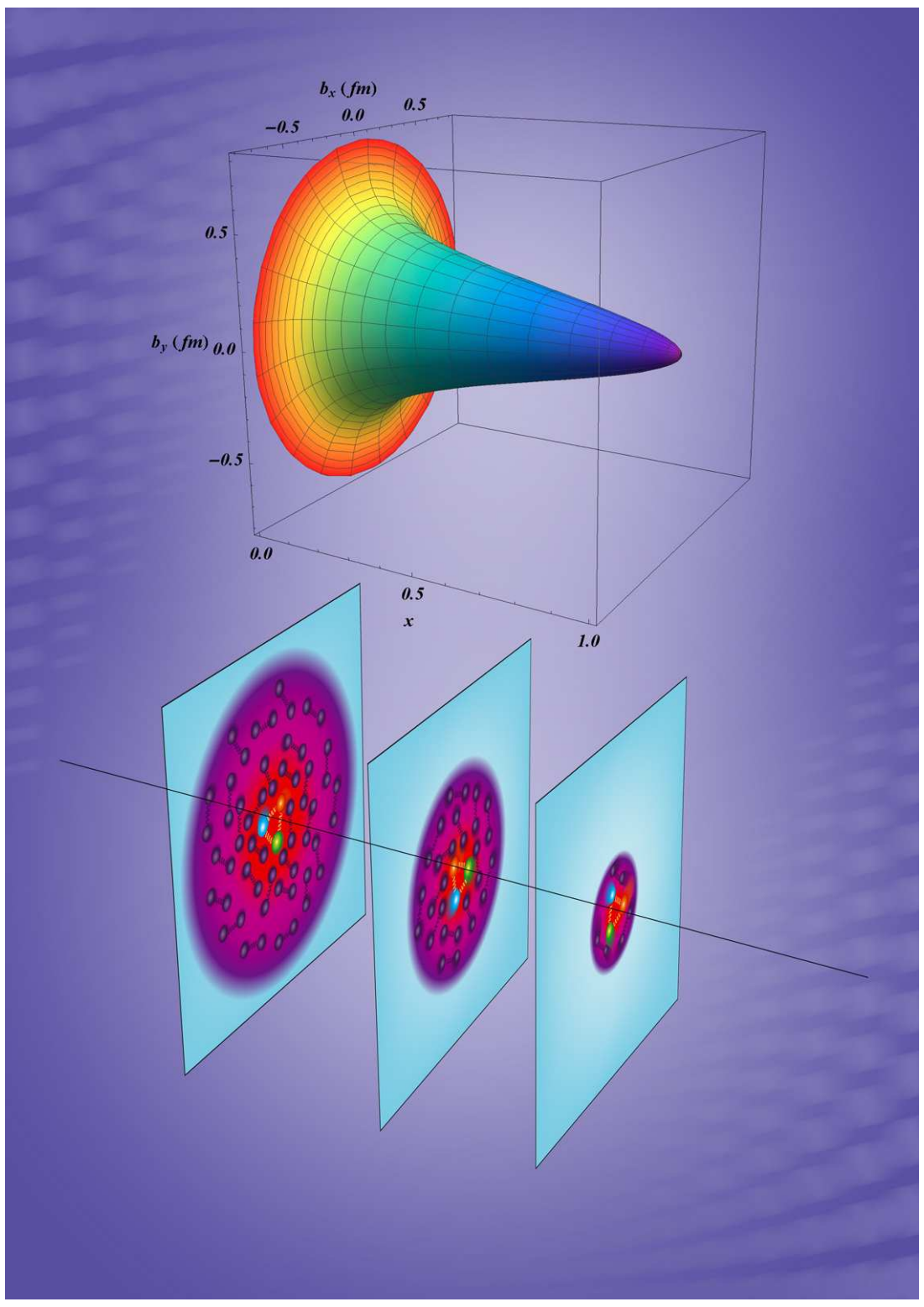

**Fig. 4:** Top panel: three-dimensional representation of the quark-momentum dependence of the proton's transverse charge radius. Bottom panel: artistic illustration of the corresponding rising quark density and transverse extent as a function of their longitudinal momentum.

# Photoproduction at COMPASS

*A. Guskov on behalf of the COMPASS collaboration*
Joint Institute for Nuclear Research, Dubna, Russia

**Abstract**
COMPASS is a multipurpose fixed target experiment at CERN using muon and hadron beams of high intensity for study of hadron structure and hadron spectroscopy. The precision test of the chiral perturbation theory predictions using charged pion scattering off a virtual photon with small momentum transfer is one of the main points of the COMPASS physics programme. The important results for the charged pion polarizability, radiative widths of $a_2(1320)$ and $\pi_2(1670)$ mesons and the cross section dynamics for the reactions $\gamma^*\pi^- \to 3\pi$ are obtained recently. The new results and perspectives for search for photo(lepto-)production of exotic charmonium-like states are also reported.

## 1 Test of chiral perturbation theory predictions in meson-photon scattering

The Chiral Perturbation Theory ($\chi$PT) is one of the most successful effective field theories of strong interaction at low energies. It is based on the low-momentum expansion of the QCD lagrangian. The triplet of $\pi$-mesons in the limit $m_u, m_d \to 0$ or the octet of pseudoscalar mesons ($\pi$, K, $\eta$) under the SU(3) symmetry assumption that $m_u, m_d, m_s \to 0$ are the Goldstone bosons. The $\pi(K)\gamma$ interaction together with the $\pi\pi(KK)$ scattering is one of the main instruments for control of applicability of the $\chi$PT.

Stringent test of $\chi$PT predictions in the pion-photon scattering with different final states is one of the main points of the COMPASS physics programme [1, 2]. The Primakoff reaction, a scattering of a beam pion off a quasi-real photon of the nuclear Coulomb field, is used for that. The Primakoff cross section $\sigma_{\pi Z}$ can be connected to the $\pi\gamma$ cross section using the equivalent-photon approximation:

$$\frac{d\sigma_{\pi Z}}{ds\, dQ^2\, d\Phi_n} = \frac{Z^2\alpha}{\pi(s-m_\pi^2)}\, F^2(Q^2)\, \frac{Q^2 - Q^2_{\rm min}}{Q^4}\, \frac{d\sigma_{\pi\gamma}}{d\Phi_n}. \quad (1)$$

Here, the cross section for the process $\pi$Z$\to XZ$ is factorized into the quasi-real photon density provided by the nucleus of charge $Z$ and $\sigma_{\pi\gamma\to X}$ the cross section for the embedded $\pi\gamma \to X$ reaction. The function $F(Q^2)$ is the electromagnetic form factor of the nucleus and $d\Phi_n$ is the $n$-particle phase-space element of the final-state system $X$. The minimum value of the negative 4-momentum transfer squared, $Q^2 = -(p^\mu_{\rm beam} - p^\mu_X)^2$, is $Q^2_{\rm min} = (s - m_\pi^2)^2/(4E^2_{\rm beam})$ for a given final-state mass $m_X = \sqrt{s}$. For scattering of the negative pions of $E_{\rm beam}$=190 GeV off a nuclear target at COMPASS the typical values of $Q^2_{\rm min}$ are about 1 ( MeV/$c$)$^2$.

### 1.1 Pion polarizability

Pion electric $\alpha_\pi$ and magnetic $\beta_\pi$ polarizabilities characterize the pion interacting as a complex QCD system with external electromagnetic fields. They are fundamental parameters of pion physics and can be probed in the $\pi\gamma$ Compton scattering. In the two-loop approximation, for the charged pion polarizabilities, $\chi$PT predicts $\alpha_\pi - \beta_\pi = (5.7 \pm 1.0) \times 10^{-4}$ fm$^3$ and $\alpha_\pi + \beta_\pi = 0.16 \times 10^{-4}$ fm$^3$ [3].

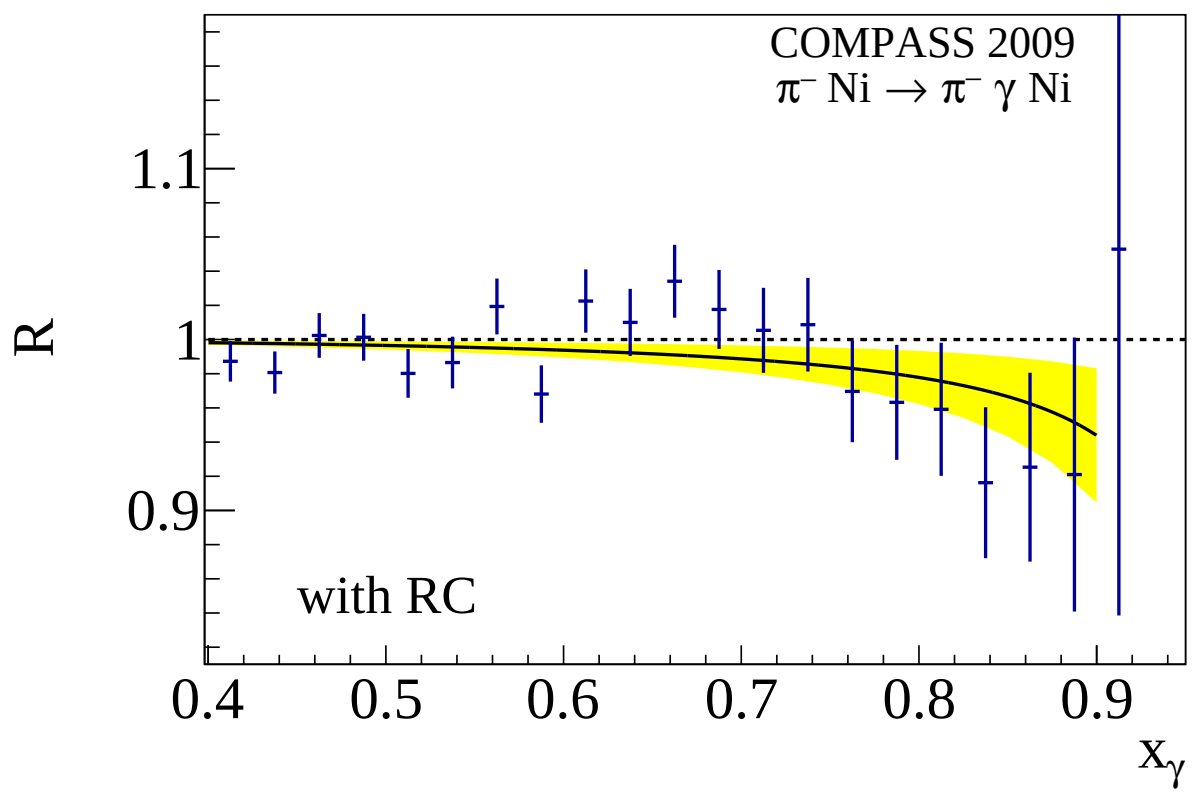

**Fig. 1:** The measured ratio $R_\pi$.

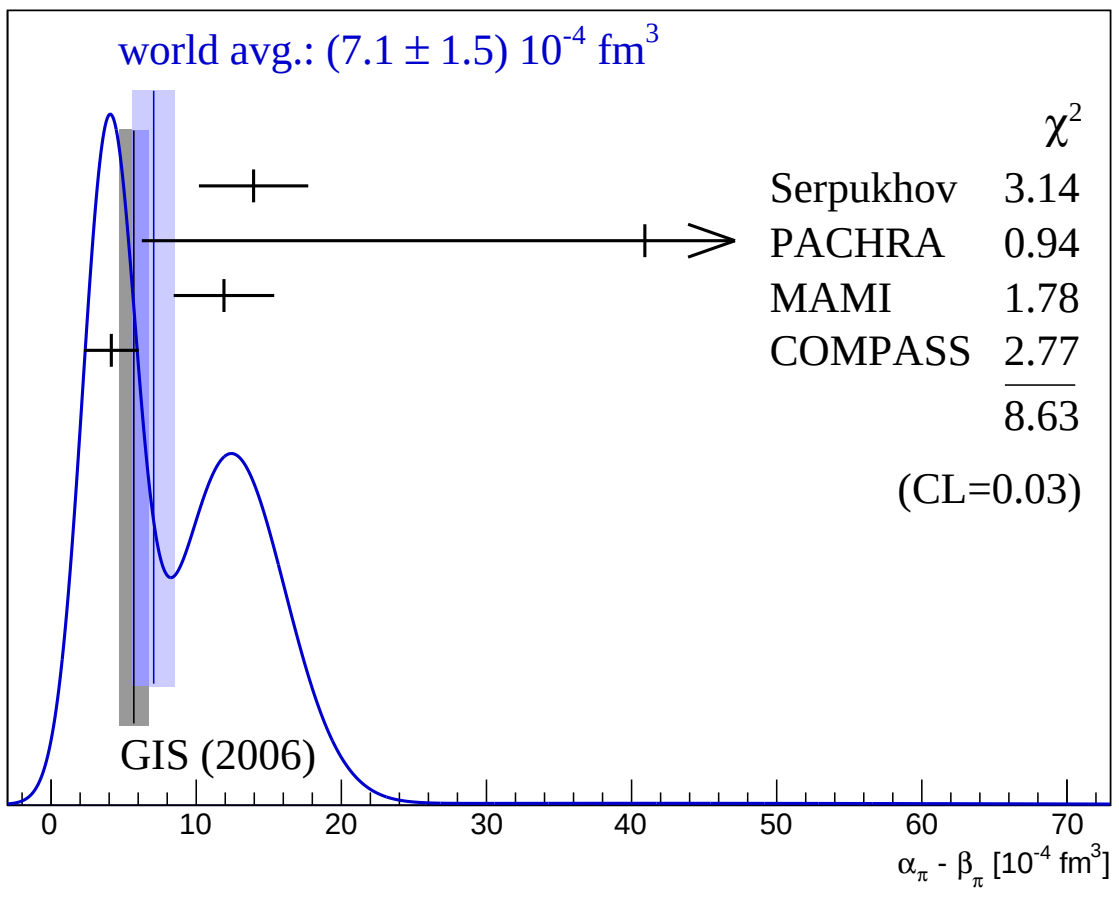

**Fig. 2:** $\alpha_\pi - \beta_\pi$ measured assuming $\alpha_\pi + \beta_\pi = 0$ in the dedicated experiments including COMPASS.

The first measurement of the pion polarizabilities was performed via the Primakoff scattering by the SIGMA-AYAKS Collaboration [4].

For the pion polarizability measurement at COMPASS the reaction $\pi^-\mathrm{Ni} \to \pi^-\mathrm{Ni}\gamma$ was used. A muon beam with similar parameters was used to study and control systematic effects. About 63 000 exclusive $\pi^-\gamma$ events in the kinematic range $p_T > 40$ MeV/c, $m_{\pi\gamma} < 3.5\ m_\pi$, $Q^2 < 1.5 \times 10^{-3}$ GeV$^2/c^2$ and $0.4 < x_\gamma (= E_\gamma/E_{beam}) < 0.9$ were used for pion polarizability extraction under the assumption $\alpha_\pi + \beta_\pi = 0$. Here $m_{\pi\gamma}$ is the mass of the final $\pi\gamma$ state, $p_T$ is the transverse momentum of the scattered pion, $E_\gamma$ is the energy of the produced photon in the laboratory system. The polarizability is determined from the $x_\gamma$ dependence of the ratio $R_\pi$ of the measured cross section $\sigma_{\pi Z\to\pi Z\gamma}$ to the calculated one for the point-like pion

$$R_\pi = \left(\frac{d\sigma_{\pi Z\to\pi Z\gamma}}{dx_\gamma}\right) \Bigg/ \left(\frac{d\sigma^0_{\pi Z\to\pi Z\gamma}}{dx_\gamma}\right) = 1 - \frac{3}{2}\cdot\frac{m_\pi^3}{\alpha}\cdot\frac{x_\gamma^2}{1-x_\gamma}\alpha_\pi. \tag{2}$$

This ratio is presented in Fig. 1 from which we obtained the result $\alpha_\pi = (2.0 \pm 0.6_{\rm stat} \pm 0.7_{\rm syst}) \times 10^{-4}$ fm$^3$, compatible with the expectation from $\chi$PT [3,5]. The COMPASS result together with results of the previous dedicated measurements is presented in Fig 2. The details of the analysis can be found in Ref. [6]. The uncertainty of presented the result is still by a factor two larger than the accuracy of the $\chi$PT prediction. For this reason COMPASS took data for a full year in 2012. Analysis of these data is still ongoing. As the result, the improved accuracy of $\alpha_\pi$ measurement under assumption $\alpha_\pi + \beta_\pi = 0$ and independent determination of $\alpha_\pi$ and $\beta_\pi$ are expected.

### 1.2 Kaon polarizability

Since the kaon is a more compact and rigid object than the pion, it would be natural to expect smaller values for kaon polarizabilities. The prediction of the $\chi$PT states that for the charged kaon the polarizability $\alpha_K$ is $(0.64 \pm 0.10) \times 10^{-4}$ fm$^3$ under the assumption $\alpha_K + \beta_K = 0$ [7]. While the prediction of the quark confinement model is rather different: $\alpha_K = 2.3 \times 10^{-4}$ fm$^3$, $\alpha_K + \beta_K = 1 \times 10^{-4}$ fm$^3$ [8]. As for the experimental results, only the upper limit $\alpha_K < 200 \times 10^{-4}$ fm$^3$ (CL = 90%) has been established from the analysis of X-rays spectra of kaonic atoms [9].

COMPASS can measure the kaon polarizability with the same technique adopted for pions, using the 2.4% kaon contamination in the pion beam and the two Cherenkov detectors (CEDARs) to identify the beam particle, but:

1. the cross section is an order of magnitude smaller;

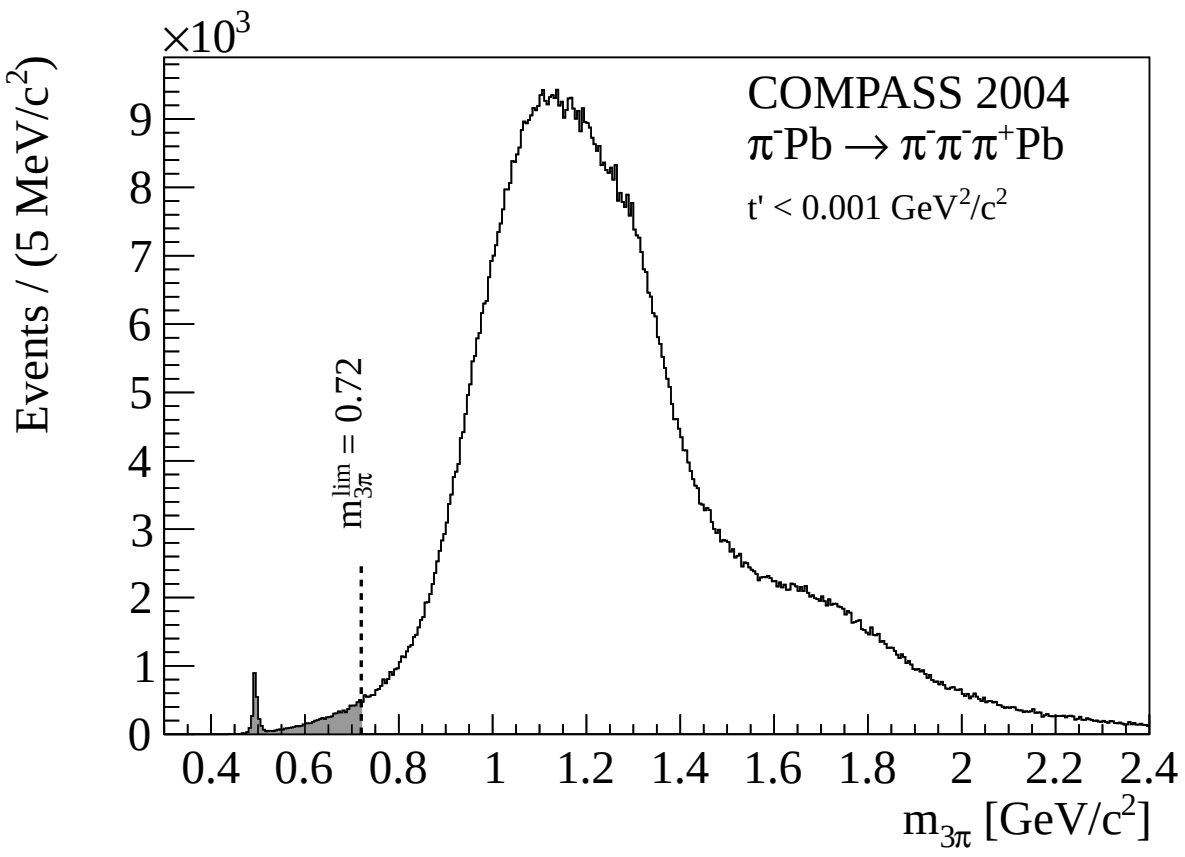

**Fig. 3:** The invariant mass spectrum of the $\pi^-\pi^+\pi^-$ final-state for events with low momentum transfer (below $10^{-3}$ (GeV/$c$)$^2$).

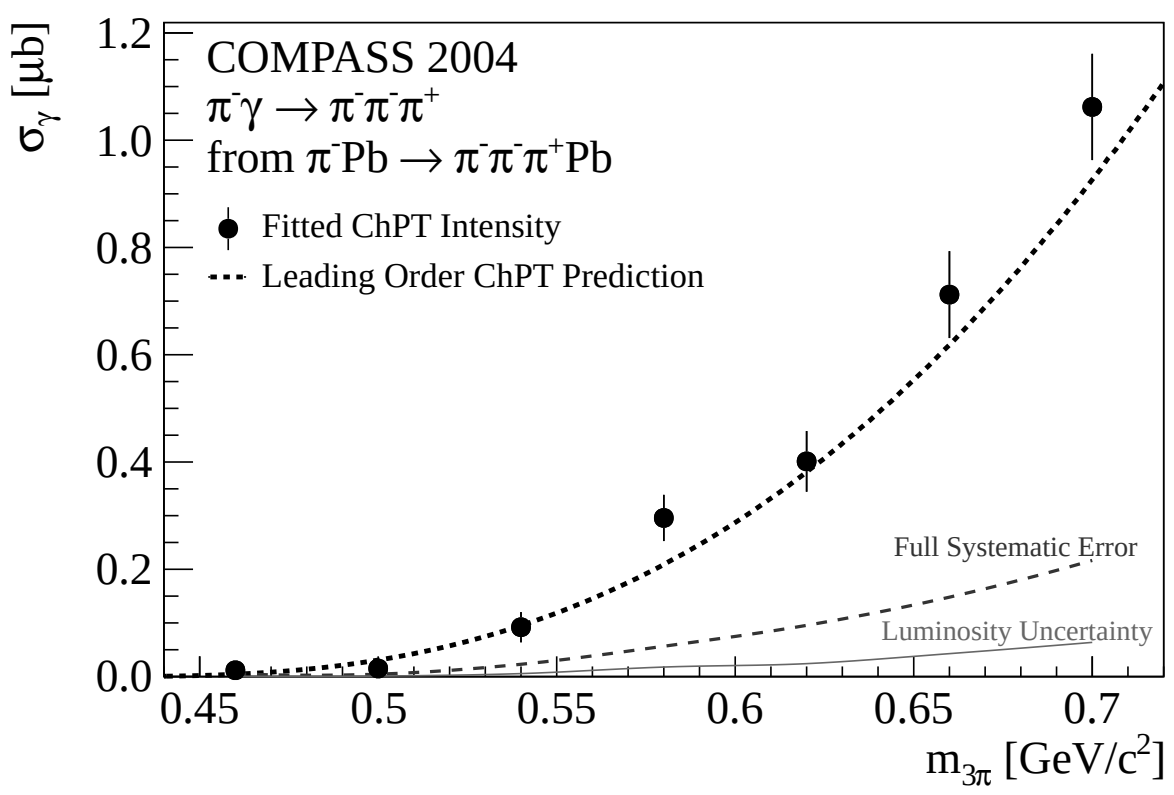

**Fig. 4:** The cross section for reaction $\pi^-\gamma \to \pi^-\pi^-\pi^+$ as a function of the total collision energy $\sqrt{s} = m_{3\pi}$.

2. the purity of particle identification provided by CEDARs is not enough;
3. the kinematic gap between the threshold and the first resonance K*(892) is much smaller than for the pion, that also limits the statistics potentially available for the analysis.

Possibility to measure the kaon polarizability using a dedicated RF-separated hadron beam enriched by kaons [10] is under investigation.

### 1.3 Chiral dynamics in $\pi^-\gamma \to 3\pi$ reactions and radiative width of $a_2(1320)$ and $\pi_2(1670)$ mesons

The cross sections of the reactions $\gamma\pi^- \to \pi^-\pi^+\pi^-$ and $\gamma\pi^- \to \pi^-\pi^0\pi^0$ are governed by the chiral $\pi\pi$ interaction [11]. The first reaction was studied at COMPASS via Primakoff scattering of the negative pion beam off a lead target. The invariant mass spectrum of the $\pi^-\pi^+\pi^-$ final-state for events with momentum transfer below $10^{-3}$ (GeV/$c$)$^2$ is shown in Fig. 3. The mass range from the threshold up to 5m$_\pi$ was used for the cross section determination. A partial wave analysis was performed to extract the intensity of the chiral contribution. In order to perform an absolute normalization the integrated beam flux was determined with good precision by using the decay $K^- \to \pi^-\pi^+\pi^-$ of beam kaons which fraction in the COMPASS hadron beam is precisely known. The result obtained for the cross section (Fig. 4) is in agreement with the tree-level expectation from the $\chi PT$ [12].

The same data were used to study the photoproduction of the resonances $a_2(1320)$ and $\pi_2(1670)$. Since two production mechanisms are possible: Primakoff production via virtual photon exchange and diffractive production via pomeron exchange, the partial wave analysis procedure was used in order to estimate the electromagnetic contribution. The radiative widths obtained are: $\Gamma_{\pi\gamma}(a_2(1320)) = (358 \pm 6_{\text{stat}} \pm 42_{\text{syst}})$ keV and $\Gamma_{\pi\gamma}(\pi_2(1670)) = (118 \pm 11_{\text{stat}} \pm 27_{\text{syst}})$ keV. Radiative corrections to the $\pi^-$Pb cross section are the main contribution to the systematics. The detailed description of the analysis procedure can be found in Ref. [13]. The data for the process $\gamma\pi^- \to \pi^-\pi^0\pi^0$ collected with a nickel target are under analysis.

### 1.4 Chiral anomaly in $\pi^-\gamma \to \pi^-\pi^0$ process

The reaction $\pi^-\gamma \to \pi^-\pi^0$ allows one to determine the chiral anomaly amplitude $F_{3\pi}$, for which the chiral theory makes an accurate prediction already at the leading order by relating $F_{3\pi}$ to the neutral pion decay constant $F_\pi$:

$$F_{3\pi} = \frac{eN_c}{12\pi^2 F_\pi^3} = 9.78 \pm 0.05 \text{ GeV}^{-3}, \tag{3}$$

where $e$ is the electric charge and $N_c$ is the number of colors. The reaction has already been examined in the Primakoff scattering at the SIGMA spectrometer [14], where in the relevant region of $s < 10\ m_\pi^2$ only about 200 events were found, and in the $\pi e$ scattering [15]. The corresponding results for $F_{3\pi}$ are $(10.7 \pm 1.2)$ GeV$^{-3}$ and $(9.6 \pm 1.1)$ GeV$^{-3}$, respectively, and have much lower precision than theoretical predictions.

Acceptance of the COMPASS setup covers large range of the $\pi^-\pi^0$ invariant mass, in particular including the $\rho(770)$-meson peak. Using the approach based on the dispersive relations the data with $\pi^-\pi^0$ mass up to 1 GeV/$c^2$ can be used for extraction of the $F_{3\pi}$ constant [16]. The corresponding analysis is in progress.

## 2 Photoproduction of exotic charmonia

In the last years a lot of new charmonium-like hadrons, so-called the XYZ states, at masses above 3.8 GeV/$c^2$ were discovered. Several interpretations of the new states do exist: pure quarkonia, tetraquarks, hadronic molecules, hybrid mesons with a gluon content, etc. But at the moment many basic parameters of the XYZ states have not been determined yet. New experimental input is required to distinguish between the models that provide different interpretations of the nature of exotic charmonia. COMPASS has a unique possibility to contribute to XYZ physics by investigating photo(lepto-)production of these states. The experimental data obtained for positive muons of 160 GeV/$c$ (2002–2010) or 200 GeV/$c$ momentum (2011) scattering off solid $^6$LiD (2002–2004) or $NH_3$ targets (2006–2011) were used for looking for exotic charmoia production.

### 2.1 Exclusive photoproduction of $X(3872)$

The exotic hadron $X(3872)$ was discovered by the Belle collaboration in 2003 [17]. Its mass is 3871.69 $\pm$ 0.17 MeV/$c^2$ that is very close to the $D^0\bar{D}^{*0}$ threshold. The decay width of this state was not determined yet, only an upper limit for the natural width $\Gamma_{X(3872)}$ of about 1.2 MeV/$c^2$ (CL=90%) exists. The quantum numbers $J^{PC}$ of the $X(3872)$ were determined by LHCb to be $1^{++}$ [18, 19]. Approximately equal probabilities to decay into $J/\psi 3\pi$ and $J/\psi 2\pi$ final states indicate large isospin symmetry breaking.

Photoproduction of the $X(3872)$ at COMPASS was observed in the exclusive reaction $\mu^+ N \rightarrow \mu^+ N' X(3872)\pi^\pm \rightarrow \mu^+ N' J/\psi \pi^+\pi^-\pi^\pm$. The invariant mass spectrum of the $J\psi\pi^+\pi^-$ subsystem is shown in Fig. 5 (two entries per event). It demonstrates two peaks corresponding to production and decay of the $\psi(2S)$ and $X(3872)$ states. Statistical significance of the $X(3872)$ signal depends on the applied selection criteria and varies from 4.5$\sigma$ to 6$\sigma$. The cross section of the reaction $\gamma N \rightarrow N' X(3872)\pi^\pm$ multiplied by the branching fraction for the decay $X(3872) \rightarrow J/\psi\pi^+\pi^-$ was found to be 71 $\pm$ $28_{\text{stat}}$ $\pm$ $39_{\text{syst}}$ pb in the covered kinematic range with a mean value of $\sqrt{s_{\gamma N}} = 14$ GeV. It was also found that the shape of the invariant mass distribution for $\pi^+\pi^-$ produced form the $X(3872)$ decay looks very different from previous results obtained by Belle, CDF, CMS and ATLAS. It could be an indication that the $X(3872)$ object could contain a component with quantum numbers different from $1^{++}$. More detailed information can be found at Ref. [20].

### 2.2 Photoproduction of $Z_c^\pm(3900)$

The $Z_c^\pm(3900)$ state discovered via its decay into $J/\psi\pi^\pm$ by BESIII [21] and Belle [22] is one of the most promising tetraquark candidate. At COMPASS the search for $Z_c^\pm(3900)$ was performed in the exclusive reaction $\mu^+ N \rightarrow \mu^+ N' Z_c^\pm(3900) \rightarrow \mu^+ N' J/\psi\pi^\pm$. The $J/\psi\pi^\pm$ mass spectrum for exclusive events (see Fig. 6) does not exhibit any statistically significant resonant structure around the nominal mass of the $Z_c^\pm(3900)$. Therefore an upper limit was determined for the product of the cross section of the $\gamma N \rightarrow N' Z_c^\pm(3900)$ process and the relative $Z_c^\pm(3900) \rightarrow J/\psi\pi^\pm$ decay probability to be 52 pb (CL=90%) [23]. An upper limit for the partial width of the $Z_c^\pm(3900) \to J/\psi\pi^\pm$ decay was also established basing on the production model described in [24].

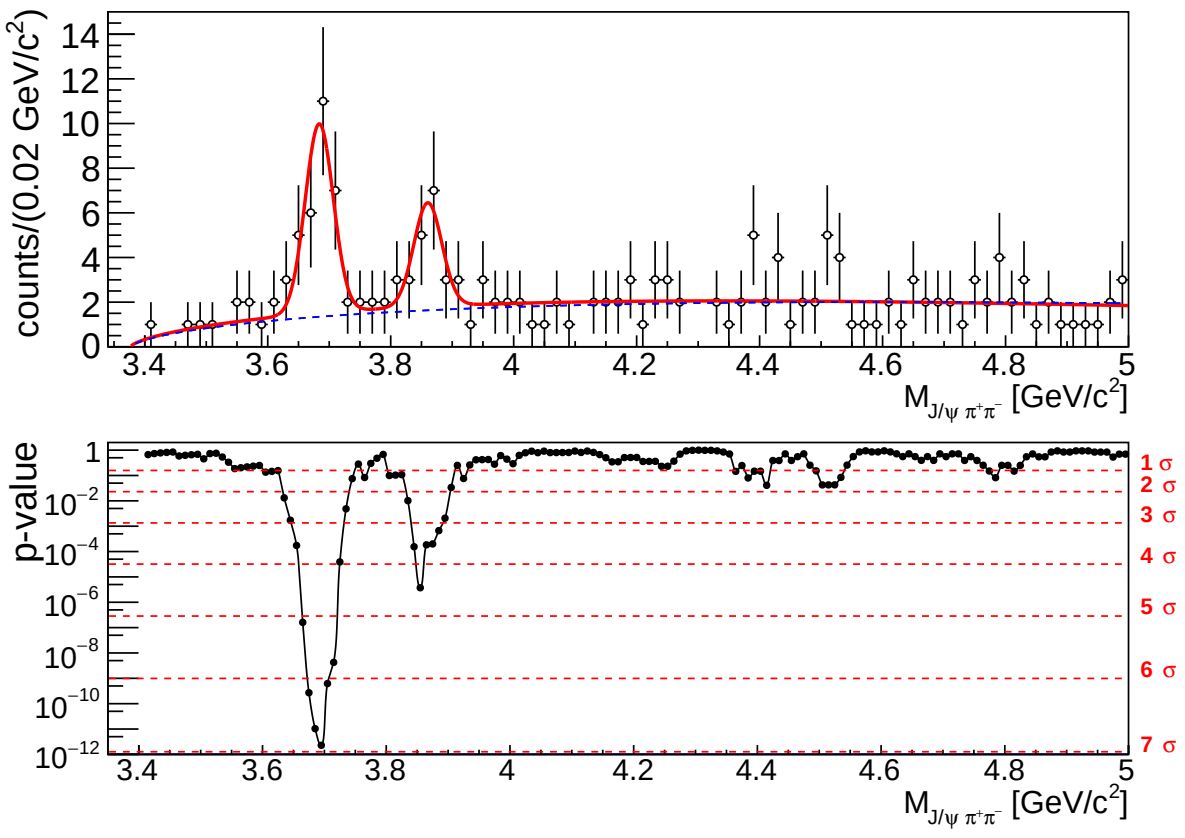

**Fig. 5:** The $J/\psi\pi^+\pi^-$ invariant mass distribution for the $J/\psi\pi^+\pi^-\pi^\pm$ final state (top). The statistical significance of the signal (bottom).

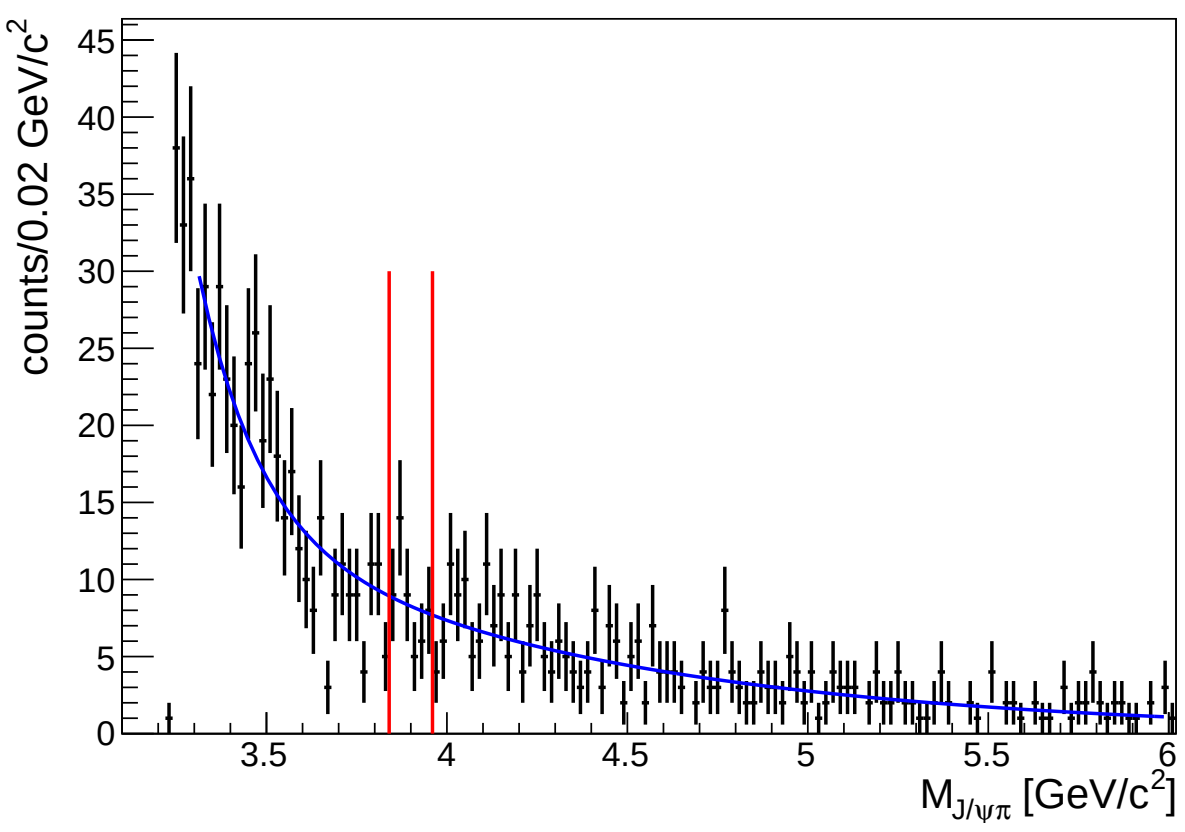

**Fig. 6:** The mass spectrum of the $J\psi\pi^\pm$ subsystem [23]. The searching range is shown by the vertical lines while the curve represents the background fitting.

The $J/\psi\pi^\pm$ mass spectrum measured by COMPASS was used in [25] to estimate the production rate of the $Z_c^\pm(4200)$, another exotic state also observed by Belle [26].

### 2.3 New possibilities

Upgrade of the COMPASS setup related with the data taking in 2016–2017 within the framework of the GPD program [2] provides new opportunities to search for direct production of exotic charmonium-like states. A new, 2.5 m long liquid hydrogen target ($\sim 0.27X_0$) is much more transparent for photons than the $^6$LiD and $NH_3$ targets used before. The target is surrounded by a 4 m long recoil proton detector which can be used to reconstruct and identify recoil protons via time-of-flight and energy loss measurements. The existing system of two electromagnetic calorimeters is extended by installation of the new large-aperture calorimeter. With the new calorimetry system one can expect much better selection of exclusive events. Searching for production of the neutral $Z_c^0(3900)$, discovered by BES-III, decaying into $J/\psi\pi^0$ will be possible. The final states decaying to the $\chi_{c0,1,2}$ -mesons could also be studied.

In more detail the question of the study of exotic charmonia at COMPASS is discussed at Ref. [27].

## 3 Conclusions

COMPASS is a unique apparatus to test of the chiral perturbation theory predictions in meson-photon scattering. Large data sets were collected for various final states. The important results for the charged pion polarizability, chiral dynamics of the $\gamma\pi^- \to \pi^-\pi^+\pi^-$ cross section and the radiative widths of the $a_2(1320)$ and $\pi_2(1670)$ mesons are published. More results are expected.

Photo(lepto-)production of exotic charmonia is a new direction in physics of the XYZ states started by COMPASS. The $X3872$ meson became the first exotic charmonium-like state observed in photo-production. The search results for exclusive production of the $Z_c^\pm(3900)$ state in the charge-exchange reaction are also reported.

# Photoproduction prospects at the EIC

*Heikki Mäntysaari*
Physics Department, Brookhaven National Laboratory, Upton, NY 11973, USA

**Abstract**
We discuss how photoproduction and electroproduction processes will provide a precise tool to study nuclear structure in the small-$x$ region in the future Electron-Ion Collider. In particular, we emphasize how exclusive vector meson production at the EIC can be used to accurately map the gluonic density profile of nuclei, including its event-by-event fluctuations. In addition, we show how measurements of the total diffractive cross section will be sensitive to saturation effects.

## 1 Introduction

The proposed Electron-Ion Collider (EIC) in the United States will be the first collider ever to study the inner structure of both protons and nuclei at high energy [1]. At the EIC, the electron beam probes protons and nuclei in deep inelastic scattering (DIS) processes. These events experimentally and theoretically cleaner compared to for example proton-nucleus collisions, making it possible to study the physics of strong interactions in unprecedented accuracy.

Some key parts of the wide physics program of the EIC are laid out in the White Paper [1] and in the EIC science case [2] review. Recently, as the machine design evolves, the center-of-mass energy required by some of the most important measurements has been studied in detail [3]. In this talk, the possibility to probe the saturated gluonic matter at small Bjorken-$x$ in diffractive events is discussed.

HERA measurements of the proton structure [4] resulted in an accurate picture of the proton in terms of its partonic content as a function of the longitudinal momentum fraction carried by the quarks and gluons. One striking observation has been the rapid rise of the gluon distribution towards small $x$. This raise would eventually violate unitarity, and non-linear effects must limit the growth of the gluon density at very small $x$. These non-linear *saturation* phenomena are included in the Color Glass Condensate (CGC) effective theory of QCD valid in the small-$x$ region [5]. Observing signatures of the saturated dense gluonic matter, described by CGC, is one of the key tasks for the EIC.

In diffractive processes a system of particles (or just e.g. a vector meson) is produced such that there is a large rapidity gap between the system and the beam remnants. In perturbative QCD at leading order, this requires two gluons to be exchanged with the target, as there can not be an exchange of net color charge. Thus, the cross section for such a process is proportional to the target gluon density *squared*, making diffractive scattering a sensitive probe of the gluonic structure of the nucleus. In addition, in exclusive processes where only one particle is produced, it is possible to study the spatial distribution of quarks and gluons in the proton and nuclei as the transverse momentum is Fourier conjugate to the impact parameter.

## 2 Accessing saturation region at the EIC

The Electron-Ion Collider will collide electrons with protons and a variety of different nuclei. In the electron-proton scattering, the proposed maximum center-of-mass energies are $\sqrt{s} = 141$ GeV, which is the region already covered by HERA. The question whether HERA saw saturation or not is not completely settled. However, the scale at which the non-linear saturation phenomena become important, $Q_s^2$ (at given $x$), scales like $A^{1/3}$ as there are $\sim A^{1/3}$ overlapping nucleons. Thus, even at smaller photon-nucleon center of mass energies than in electron-proton mode, it is possible to access the region where saturation scale is large and in the perturbative region by using heavy nuclei. In the electron-nucleus mode, the proposed maximum energy for the EIC is $\sqrt{s} = 40 \dots 90$ GeV. The saturation scales accessed in electron-gold collisions at the EIC, compared to the HERA electron-proton scattering, are illustrated in Fig. 1. As saturation scales probed at the EIC can be up to 2 GeV$^2$, it is expected that the saturation effects are clearly visible.

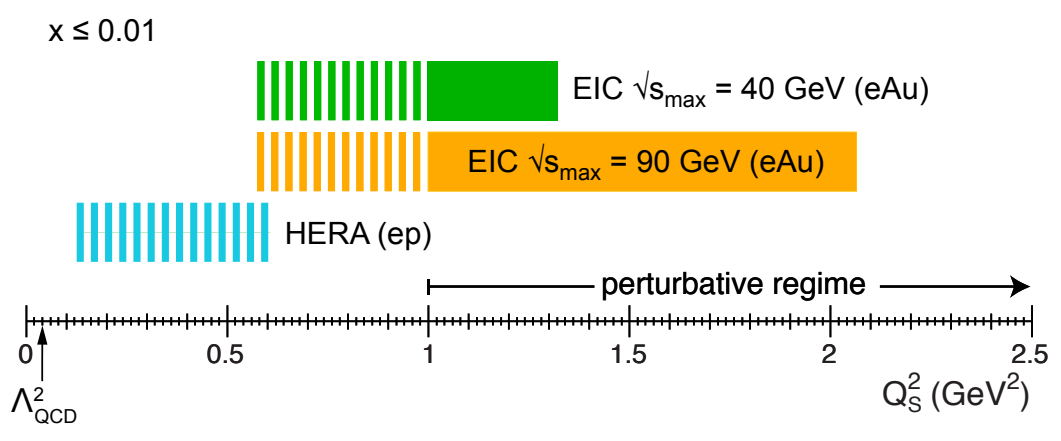

**Fig. 1:** Saturation scales $Q_s^2$ reached at the EIC in electron-nucleus collisions, compared to the ones accessed at HERA in electron-proton scattering. Figure from Ref. [3].

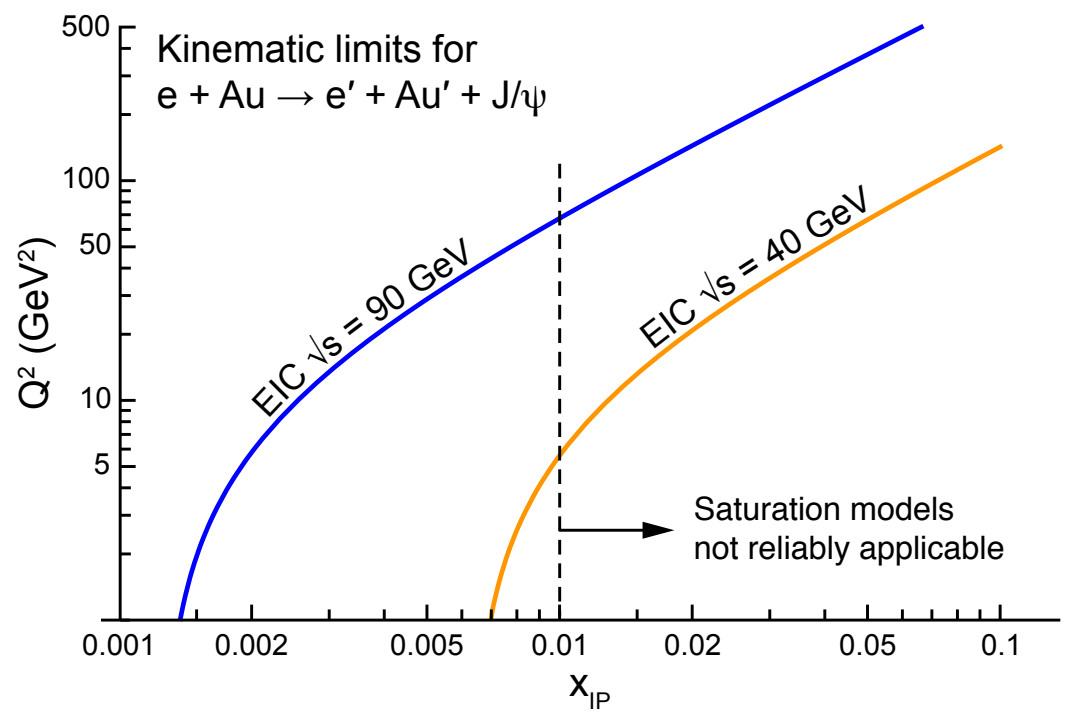

**Fig. 2:** Kinematical coverage for the exclusive $J/\Psi$ production at the EIC. Figure from Ref. [3].

## 3 Diffractive scattering

In diffractive scattering process the incoming lepton emits a virtual photon (virtuality $Q^2$) which interacts with the target hadron (momentum $P$) and forms the final state system $X$ after gaining longitudinal momentum $x_{\mathbb{P}}P$ from the target. Here $\mathbb{P}$ stands for the interpretation that a color neutral object "*pomeron*" is exchanged between the photon and the target. In exclusive processes, the system $X$ is just a single particle, usually a vector meson like $J/\Psi$ or $\rho$. The advantage of $J/\Psi$ is that its large mass allows perturbative treatment also at low $Q^2$, but it is not too heavy to lie outside the kinematical reach of an EIC even at small $x$, as demonstrated in Fig. 2.

A convenient way to describe the scattering process is to look at the scattering in the frame where the photon is very energetic, in the so called dipole picture. In this frame, the scattering process goes like in Fig. 3. First, the virtual photon fluctuates into a quark-antiquark dipole long before the scattering takes place. This $\gamma^* \to q\bar{q}$ splitting is described by the virtual photon wave function $\Psi$. The produced dipole then scatters elastically off the target (dipole scattering given by the dipole-target cross section $\sigma_{\text{dip}}$), and forms the final state $X$. In case of vector meson production, the $q\bar{q} \to$ vector meson transition is described by the vector meson wave function $\Psi_V$. In case of inclusive diffraction, where the final state is specified by the mass of the system $M_X^2$, one has to also take into account $q\bar{q}g$ Fock states of the virtual photon wave function [6].

Let us first study exclusive vector meson production (for exclusive photon production, or Deeply Virtual Compton Scattering, see Ref. [7]). These events can be divided into two categories. The total diffractive cross section is proportional to the squared diffractive scattering amplitude $A$, averaged over all the possible target configurations $\langle|A|^2\rangle$. On the other hand, if we require that the target remains in the same quantum state despite the scattering, the cross section becomes (see Refs. [8,9])

$$\frac{d\sigma^{\gamma^* A \to VA}}{dt} = \frac{1}{16\pi} |\langle A \rangle|^2 . \tag{1}$$

Subtracting this contribution from the total diffractive cross section we obtain the cross section for the process in which the target is left in a different quantum state after the scattering, meaning that it breaks up. This is called incoherent diffraction, and the cross section is

$$\frac{\mathrm{d}\sigma^{\gamma^* A\to VA}}{\mathrm{d}t} = \frac{1}{16\pi}\left(\langle|A|^2\rangle - |\langle A\rangle|^2\right). \tag{2}$$

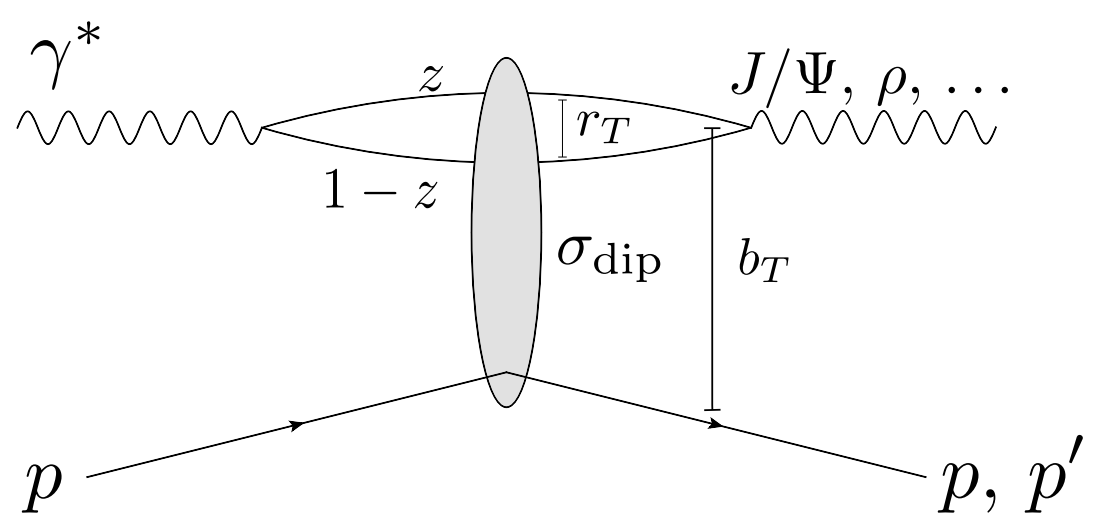

**Fig. 3:** Diffractive scattering in dipole picture.

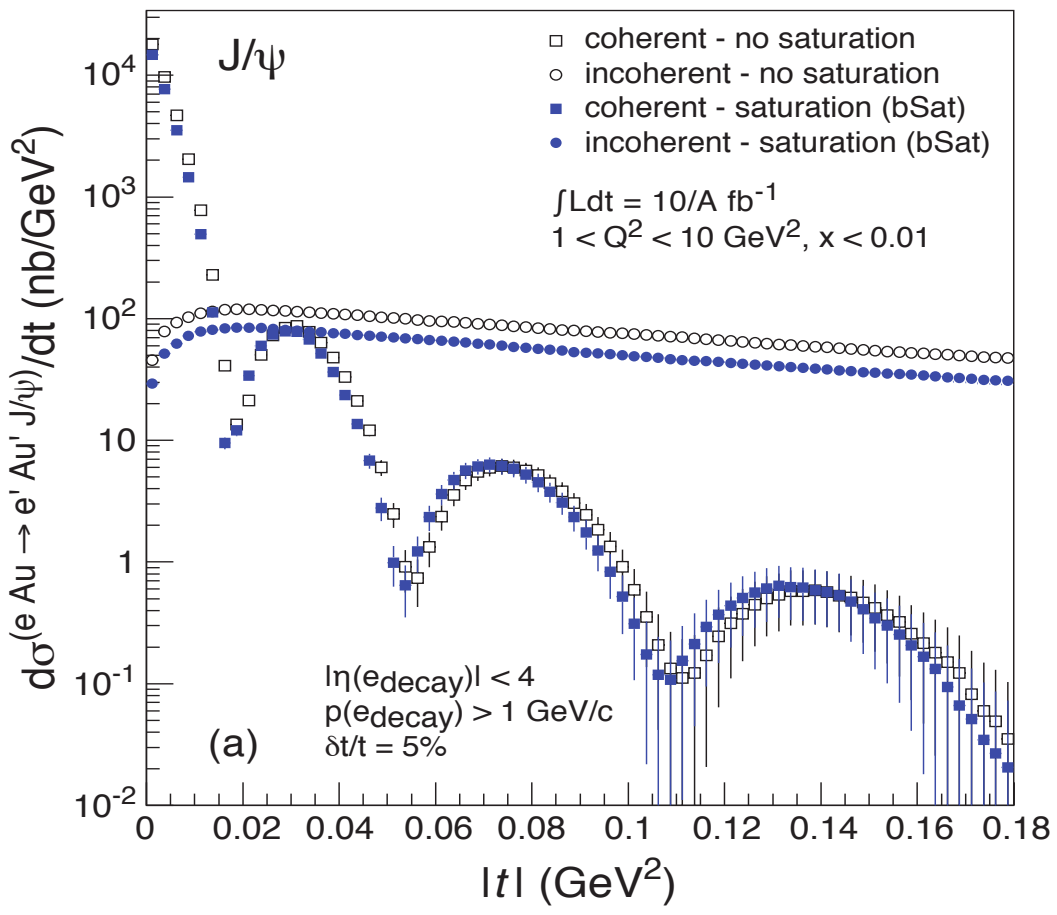

**Fig. 4:** Simulated coherent and incoherent diffractive $J/\Psi$ production at the EIC from Ref. [13].

Because the incoherent cross section is the variance of the scattering amplitude, it measures the amount of fluctuations in the target wave function in the impact parameter space (see e.g. Refs [10–12] and references therein).

The diffractive scattering amplitude in the dipole picture can be written as [14]

$$A_{T,L}^{\gamma^* p\to Vp}(x_{\mathbb{P}}, Q^2, \Delta) = i\int \mathrm{d}^2 r_T \int \mathrm{d}^2 b_T \int \frac{\mathrm{d}z}{4\pi}(\Psi^*\Psi_V)_{T,L}(Q^2, r_T, z) \times e^{-i[b_T-(1-z)r_T]\cdot\Delta}\,\frac{\mathrm{d}\sigma_{\rm dip}(r_T, b_T, x_{\mathbb{P}})}{\mathrm{d}^2 b_T}. \tag{3}$$

Here $\Delta$ is the transverse momentum of the produced vector meson. The overlap between the virtual photon and the vector meson wave functions is $\Psi^*\Psi_V$, given the virtual photon virtuality $Q^2$, dipole size $r_T$ and the longitudinal momentum fraction of the photon carried by the quark $z$.

The EIC capabilities to measure exclusive vector meson production were simulated in Ref. [13], and the simulation results for $J/\Psi$ production are shown in Fig. 4 as a function of $-t \approx \Delta^2$. Due to the large mass of the $J/\Psi$, small dipole sizes dominate the process and the saturation effects are not especially large (but somewhat enhanced in the incoherent process, see also Ref. [15]) unlike in case of lighter meson production. As the coherent cross section is obtained by Fourier transforming the dipole amplitude from coordinate space to momentum space (see Eq. (3)), Fourier transforming the measured cross section back to coordinate space gives the nuclear density profile in the coordinate space. The projected accuracy of the EIC is shown to be good enough to reconstruct the nuclear density profile accurately, as demonstrated in Ref. [13]. From the incoherent cross section one can extract the amount of fluctuations and study at which distance scales these fluctuations take place, as demonstrated in Refs. [12, 16].

If instead of $J/\Psi$ one studies the production of lighter mesons like $\rho$ or $\phi$, then the dominant contribution comes from larger dipoles that make these processes more sensitive to saturation effects. However, due to the small mass, there is no large scale which makes perturbative calculations unreliable at low $Q^2$, and the EIC capabilities to get to higher $Q^2$ are necessary to fully take the advantage of the possibility to measure exclusive production of different particle species.

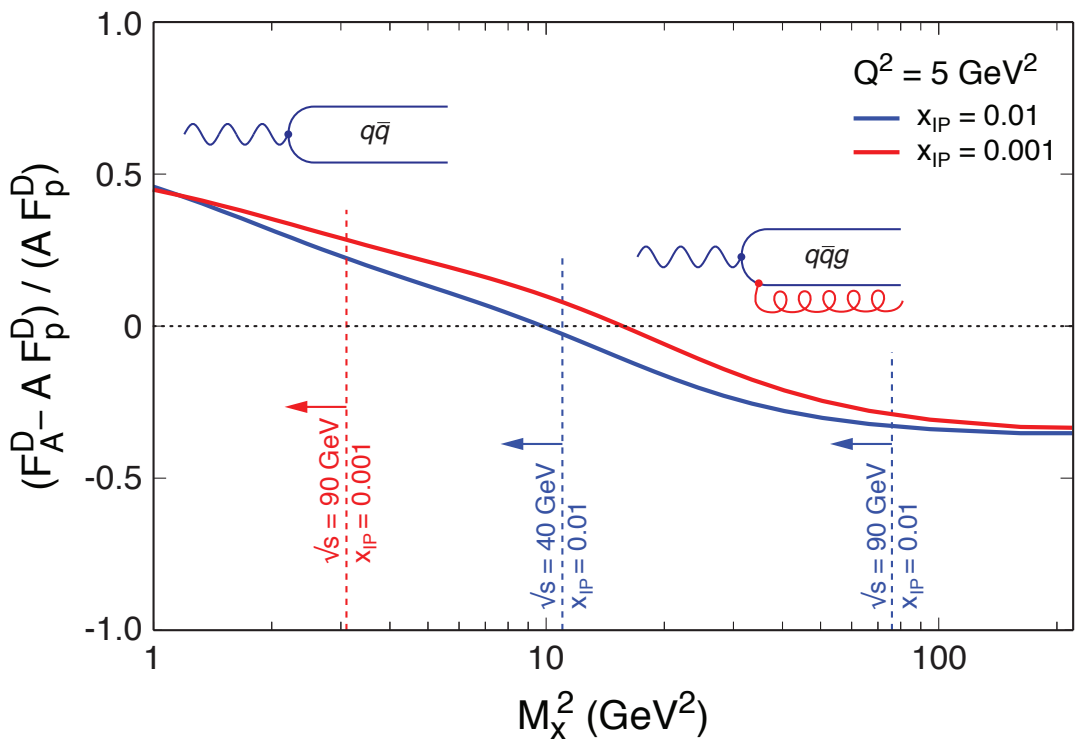

**Fig. 5:** Nuclear modification to the diffractive structure function as a function of the produced system mass. Dashed lines show kinematical limits for the two different EIC energies. Figure from Ref. [3].

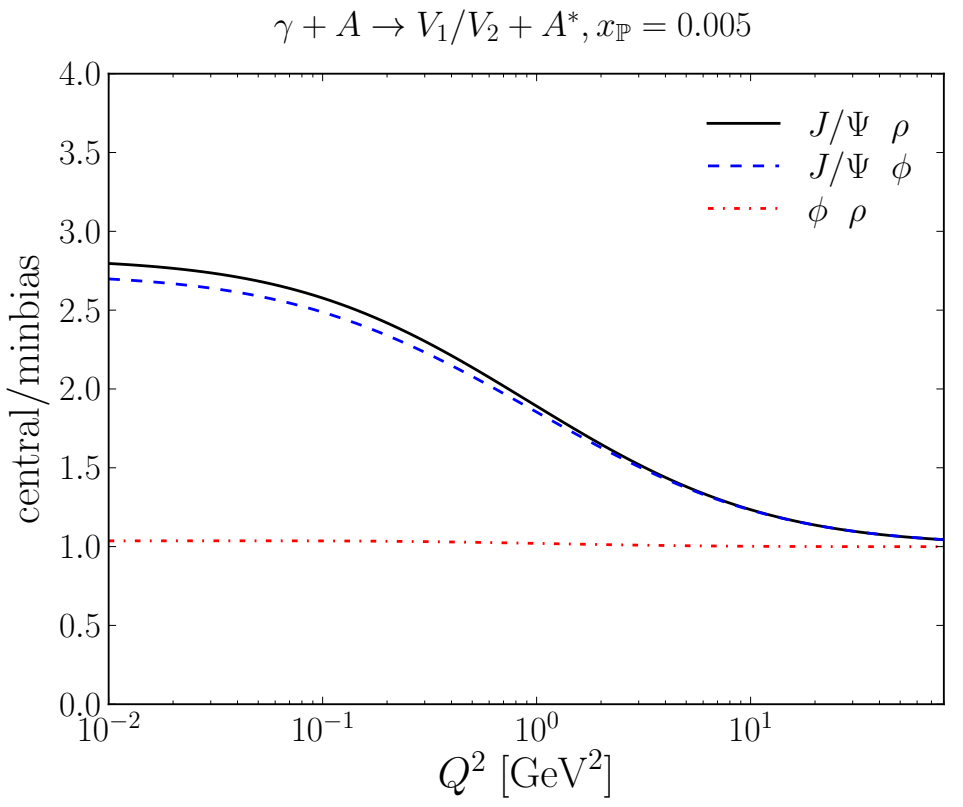

**Fig. 6:** Ratio of diffractive vector meson production cross sections in central and minimum bias events. Figure from Ref. [17].

In inclusive diffraction, one can study the scattering process as a function of the invariant mass of the produced system $M_X^2$. The advantage is that this makes it possible to probe both $q\bar{q}$ dipole-nucleus scattering (low $M_X^2$) and $q\bar{q}g$ dipole-nucleus scattering (large $M_X^2$) [6]. The saturation picture calculations have a characteristic prediction for the nuclear effects in inclusive diffraction. First, the total cross section for the production of small invariant mass systems is enhanced in the nucleus relative to a proton. On the other hand, because the strong color field of the nucleus absorbs $q\bar{q}g$ dipole more strongly compared to the proton, the diffractive cross section at large $M_X$ is suppressed in the nucleus relative to proton [6]. This is demonstrated in Fig. 5, where the nuclear modification to the diffractive structure function $F^D$ is shown (which is proportional to the total diffractive cross section). The suppression of $q\bar{q}g$ dipoles at large invariant masses is expected to be visible at the EIC energies [3].

## 4 Accessing rare configurations

In addition to measurements of average partonic content of the proton and nuclei, and the event-by-event fluctuations of the density profile, it is also possible to study rare partonic configurations in diffractive processes. One possible way to probe the structure of the nuclei in the region where gluon densities are maximally large (very large saturation scale $Q_s^2$) is to study diffractive scattering as a function of centrality. In incoherent diffractive process, the relatively large $|t|$ ensures that the momentum transfer is localized in the nucleon-size area, and one can define centrality classes.

This idea was developed in Ref. [17], where the centrality is related to the number of the so called *ballistic* protons. In incoherent diffraction, one nucleon receives a relatively large $p_T$ kick, and it propagates out of the large nucleus. On its way out, it can scatter off the other nucleons. The more central the photon-nucleus interaction is, the nucleon kicks on average more nucleons out of the nucleus. These "ballistic" neutrons end up in the zero degree calorimeters and can not be distinguished from "evaporation" neutrons that are a result of the nucleus decaying from the excited state long after the scattering process. The ballistic protons, on the other hand, have different transverse momentum than the evaporation protons which follow a thermal spectrum in the rest frame of the nucleus. It was shown in Ref. [17] that the ballistic protons end up in roman pot detectors where their multiplicity is proportional to the centrality.

In the most central events, the saturation scale is significantly larger than on average in the minimum bias events. Thus, one can expect to see a strong suppression in the light vector meson production cross section compared to $J/\Psi$ production in the events where the proton multiplicity in the Roman pot is maximal. This is demonstrated in Fig. 6, where the ratio of production cross sections for mesons $V_1$ and $V_2$ in central events (defined as $|b_T| = 0$ fm), compared to the minimum bias events (integrated over all $b_T$), is shown. Namely, one calculates

$$\frac{\sigma^{\gamma^* A \to V_1 A^*} \Big/ \sigma^{\gamma^* A \to V_2 A^*} \Big|_{\text{central}}}{\sigma^{\gamma^* A \to V_1 A^*} \Big/ \sigma^{\gamma^* A \to V_2 A^*} \Big|_{\text{min. bias}}}. \tag{4}$$

As shown in Ref. [17], in case of $J\Psi$-light meson ratio, the double ratio approaches $Q^4_{s,\text{central}}/Q^4_{s,\text{min. bias}}$ at low $Q^2$, which makes this observable especially sensitive probe of large gluon distributions in the center of the nucleus.

## 5 Summary

Exclusive and inclusive diffractive processes are powerful tools to probe the small-$x$ structure of the nuclei in the future Electron Ion Collider. The advantages are the especially good sensitivity to gluon densities (to the first approximation, cross section is proportional to the *squared* gluon distribution), and the possibility to access the impact parameter profile of various targets. The possibility to simultaneously study, in addition to average quantities, the event-by-event fluctuations and rare partonic configurations makes diffractive processes even more interesting probes of the small-$x$ dynamics.

## Acknowledgments

This work is supported under DOE Contract No. DE-SC0012704. Useful discussions with T. Ullrich and everyone in the BNL EIC Task Force are gratefully acknowledged.

# Photon measurements in proton and nucleus collisions at PHENIX

*Norbert Novitzky for the PHENIX collaboration*
Stony Brook University, Stony Brook, USA

**Abstract**

Direct photons provide an excellent probe in studying both the proton and nucleus collisions. The PHENIX measurements of the direct photon-hadron and $\pi^0$ -hadron correlations in $p+p$ collisions searches for the breakdown of the QCD factorization. The measurements of the fragmentation functions in $\mathrm{Au}+\mathrm{Au}$ collisions provide an excellent tool to understand the dynamics of the energy loss mechanism in the heavy ion collisions. Furthermore, we summarize the results on the direct photon measurements at low-$p_\mathrm{T}$ in order to study the thermal radiation in $\mathrm{Au}+\mathrm{Au}$ and $\mathrm{Cu}+\mathrm{Cu}$ collisions at different collision energies.

## 1 Introduction

The Relativistic Heavy Ion Collider (RHIC) provides a unique capability of colliding polarized protons and light or heavy ion beams. The PHENIX detector records many different particles emerging from RHIC collisions, including photons, electrons, muons, and hadrons. Direct photons are defined as all photons that arise from processes during the collision, rather than from decays of final state hadrons. The biggest challenge in the measurement of direct photons is to distinguish them from the large background of decay photons.

In this paper we summarize the recent PHENIX measurements using direct photons. The first part is focusing on the data collected from $p+p$ collisions at $\sqrt{s} = 510$ GeV, which investigates the possible breakdown of the QCD factorization. In the second part we introduce the new results from the highly asymmetric collisions such as $p$+Al, $p$+Au and $d$+Au at $\sqrt{s_{NN}} = 200$ GeV. In the third part we probe the fragmentation function modification in $\mathrm{Au}+\mathrm{Au}$ collisions and also the thermal photon production from the QGP phase.

## 2 Proton collisions

There has been much recent activity devoted to the study of parton transverse momentum in high energy hadronic collisions. Observables that are sensitive to parton transverse momentum can potentially provide new insight into the structure of hadrons. The theoretical framework that has been developed [1,2] to describe parton dynamics in hadrons involves transverse-momentum-dependent (TMD) parton distribution functions (PDFs) and fragmentation functions (FFs). TMD-factorization should be contrasted with the more common collinear factorization theorems, applicable to cases where observables are not sensitive to intrinsic transverse parton momentum.

In $p+p$ collisions the factorization breaking has been predicted at small transverse momentum scale where the non-perturbative objects become correlated [3]. To have sensitivity to possible factorization breaking and modified TMD evolution effects, a particular observable must be sensitive to a small scale on the order of $\Lambda_{\mathrm{QCD}}$ and measured over a range of hard scales. The back-to-back dihadron correlation provides a unique probe which is sensitive to the initial and final state transverse momenta [4]. We define the out-of-plane momentum component $p_{out} = p_{\mathrm{T}}{}^{\mathrm{assoc}} \sin \Delta\phi$, where $p_{\mathrm{T}}{}^{\mathrm{assoc}}$ is the transverse momentum of the associated particle and $\Delta\phi$ is the azimuthal angle between the trigger and associated particle. The $p_{out}$ quantifies the acoplanarity of the two-particle pair, and as such it is related to the initial- and final-state $k_{\mathrm{T}}$ and $j_{\mathrm{T}}$ [4].

In the PHENIX detector we study the both the direct photon-hadron and $\pi^0$ -hadron correlation function [4]. For the direct photon correlations, we use the statistical subtraction of the decay photon background and also apply an isolation cut to further remove contributions from the fragmentation photons. Figure 1 shows the $p_{out}$ distribution for direct photon and $\pi^0$ triggers. Only the away side hadrons are used in the $p_{out}$ distribution in the $2\pi/3 < \Delta\phi < 4\pi/3$ azimuthal region with respect to the trigger particle. The distributions are fit with a Gaussian function in the small $p_{out}$ region $[-1.1, 1.1]$ and with a Kaplan function ($a(1 + p_{out}^2/b)^{-c}$) in the whole range. The power law behavior is generated from hard gluon radiation in the initial state or final state, whereas the Gaussian behavior is generated from the soft $k_{\mathrm{T}}$ and $j_{\mathrm{T}}$ and is demonstrated in the nearly back-to-back hadrons that are produced around $p_{out} \approx 0$.

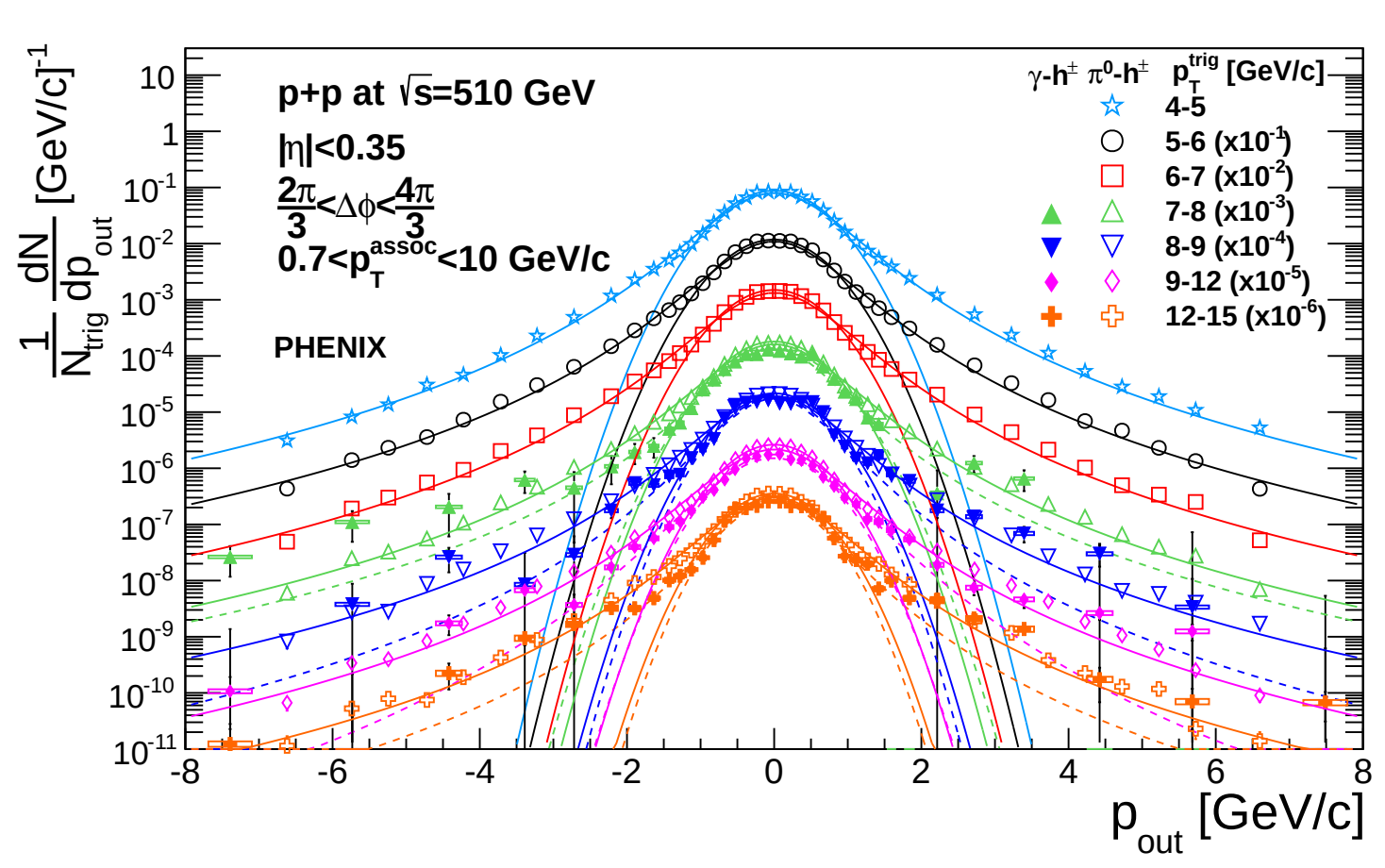

**Fig. 1:** Per trigger yields of associated charged hadrons as a function of $p_{out}$ [4]. The solid markers correspond to direct $\gamma - h^{\pm}$ and the open markers to the $\pi^0 - h^{\pm}$ correlations. The $\pi^0$- and direct photon-triggered distributions are fit with a Gaussian function at small $p_{out}$ and Kaplan function in full range. The 9% total uncertainty on the normalization of the charged hadron yields is not shown.

The left panel in Figure 2 shows the extracted widths from the Gaussian fit for the direct photon- and $\pi^0$ -hadron correlations as a function of the $p_{\mathrm{T}}{}^{\mathrm{trig}}$. The systematic uncertainties were calculated by adjusting the Gaussian fit range by +/- 0.15 GeV/c and they are combined with the statistical uncertainties in quadrature. The extracted widths in both the direct photon-hadron and $\pi^0$ -hadron correlations are decreasing towards larger $p_{\mathrm{T}}{}^{\mathrm{trig}}$ values. The decreasing trend is opposite from the one expected from TMD factorization [5], which predicts an increasing trend and has been shown in phenomenological analyses of Drell-Yan and SIDIS data [6]. PYTHIA [7] was used to compare the widths, and the decreasing trend was reproduced by the simulations. This is surprising since PYTHIA does not consider explicit factorization breaking. However, it does include the initial and final-state interactions which could be indirectly describing the effects of factorization breaking.

## 3 Nucleus collisions

The primary focus on the nucleus-nucleus collisions at RHIC collider is to create and study the Quark Gluon Plasma (QGP) [8]. The QGP is a fundamentally new state of matter where the temperature is so high that the quarks and gluons are deconfined. Photons are an excellent probe of this hot and dense, strongly interacting matter as they do not participate in the strong interaction and they leave the matter without further modification. This in fact means that the photons carry the information from the time of their creation. Experimentally, we measure the space-time integrated photon emission.

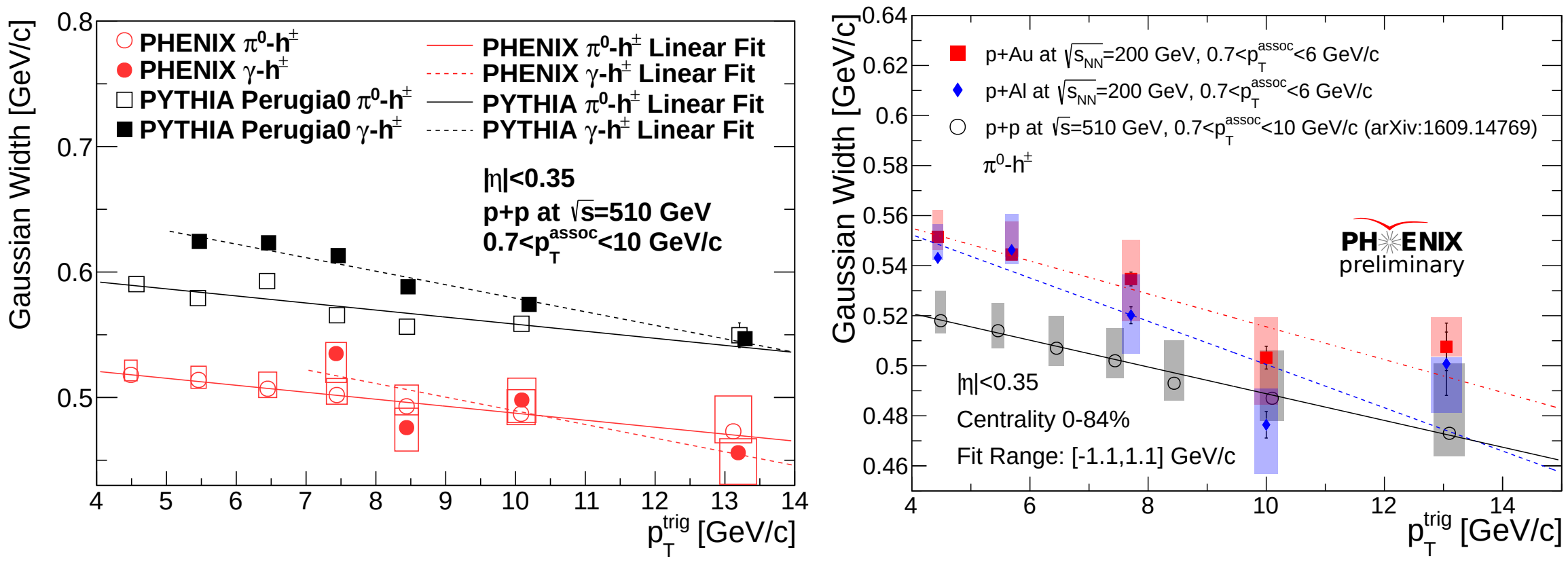

**Fig. 2:** Left panel: The results of Gaussian widths if $p_{\rm out}$ distributions for $\pi^0$ and direct photons from the data and PYTHIA as a function of $p_{\rm T}^{\rm trig}$. PYTHIA produces similar trend as the data, although we obtain a $10-15\%$ higher values for each $p_{\rm T}^{\rm trig}$. Right panel: The results of Gaussian widths of $p_{\rm out}$ distributions for $\pi^0$ as a function of $p_{\rm T}^{\rm trig}$ from $p+$ Au and $p+$ Al collisions at $\sqrt{s_{NN}} = 200$ GeV and from $p+p$ collisions at $\sqrt{s} = 510$ GeV.

We recognize several sources which contribute to the overall direct photon production. The prompt photons are created in the initial parton scattering. As they are not affected by the final state effect, they provide an excellent probe to study the initial state effects in nuclear collisions. The other significant source of direct photons are the thermal photons, which come from the thermal radiation of the Quark Gluon Plasma and the Hadron Gas. In addition we expect more sources of direct photons from jet-medium interactions.

### 3.1 Factorization breaking

The QCD factorization breaking has been also investigated in highly asymmetric proton-nucleus collisions, namely in $p$+Au and $p$+Al collisions at $\sqrt{s_{NN}} = 200$ GeV. Similarly, as in $p+p$ collisions we can study the Gaussian width in these collisions. The right panel in Figure 2 shows the comparison of the Gaussian width with the previous $p+p$ result at $\sqrt{s} = 510$ GeV (note that the $p+p$ collision is at different collision energy). At high-${p_{\rm T}}^{\rm trig}$ the widths from the nucleus collisions are consistent with the $p+p$ data. On other hand, at lower $p_{\rm T}$, there seem to be an enhancement of particles observed around pT 5GeV/c, which could be attributed to the 'Cronin'-effect [9]. The slopes in nucleus collisions are also decreasing towards higher $p_{\rm T}$ values. The data suggest a slightly larger slopes than that in the $p+p$ collisions, but it is not conclusive with the current precision of the data.

### 3.2 Fragmentation Function

It has been established that partons from hard scattering in heavy ion collisions experience energy loss while propagating through the hot and dense medium [10]. The phenomena were observed via the leading hadron and jet suppression also known as "jet-quenching". The parton energy loss can be studied via the jet fragmentation function [11]. The fragmentation function is defined as $D(z) = \frac{1}{N_{\rm jet}}\frac{dN(z)}{dz}$, where $z = p^h/p^{\rm jet}$; $p^{\rm jet}$ is the jet momentum and $p^h$ is the momentum of the hadron jet fragment. As the direct photons are not modified by the medium, they can provide a precise measurement of the momentum of the initial parton or jet, such ${p_{\rm T}}^{\gamma} \approx {p_{\rm T}}^{\rm jet}$. This is an approximation, as the initial transverse momentum of the parton, $k_{\rm T}$, is not taken into account.

The extraction of a sample triggered purely by direct photons is complicated by the presence of photons from meson decays (dominantly from $\pi^0$ decays), which must be removed from the inclusive photon-hadron correlations. In Au+Au, a statistical subtraction determines the direct (i.e. non-decay) photon-hadron correlations from the measured inclusive photon-hadron correlations [11]. The integration in the away side peak is done in the range of $|\Delta\phi - \pi| < \pi/2$, first for the yield per inclusive photon ($Y_{\rm inc}$), then for the estimated decay photon yield $Y_{\rm dec}$. The associated yield per direct photon is $Y_{\rm dir} = (R_\gamma Y_{\rm inc} - Y_{\rm dec})/(R_\gamma - 1)$, where $R_\gamma$ is the ratio of inclusive photons to decay photons. Inclusive photon-hadron correlations are determined from the distribution of photon-hadron pairs as a function of their azimuthal angular separation ($\Delta\phi$).

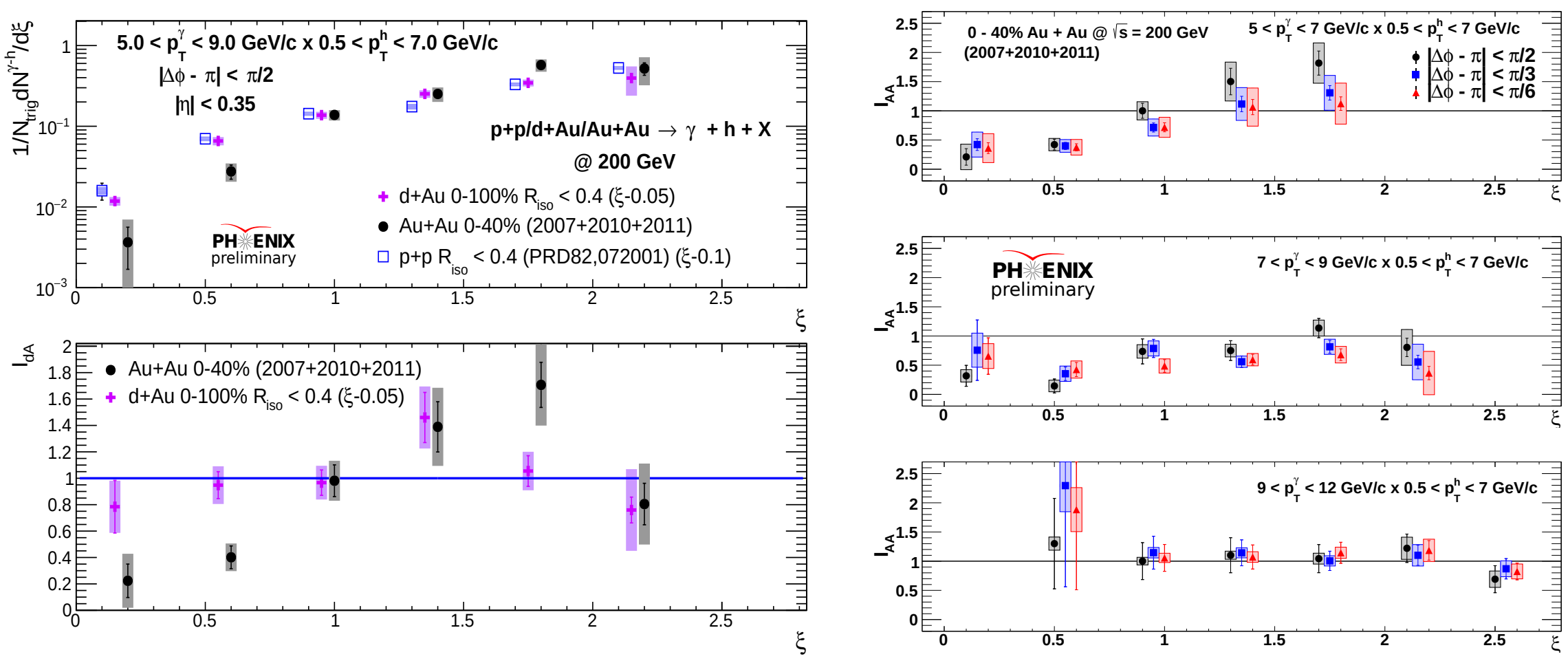

**Fig. 3:** Left panels: The upper panel shows the per trigger yield of charged hadrons associated with direct photons as a function of $\xi = ln(1/z_T)$ in $p+p$, $d+$Au and Au+Au collisions at $\sqrt{s_{NN}} = 200$ GeV. The lower panel is showing the ratio of fragmentation function ($I_{AA}$) between the nucleus-nucleus and $p+p$ collisions. Right panels: The ratios of $I_{AA}$ for the away side charged particles in $0.5 < p_{\rm T} < 7$ GeV$/c$ with three different $p_{\rm T}^{\rm trig}$ ranges. The different colors represent different integration ranges of the away side correlation function.

In order to study the jet fragmentation function, $D(z)$, associated hadron yields are determined as a function of $z_{\rm T} = p_{\rm T}{}^h/p_{\rm T}{}^\gamma$. To focus on the low $z_{\rm T}$ region, one can express the fragmentation function as a function of the variable, $\xi = ln(1/z_{\rm T})$. Left top panel in Figure 3 shows the the conditional, or per trigger, associated yield, extracted after subtraction of photon-hadron pairs from the bulk underlying event as a function of $\xi$ in Au + Au, $d$+Au and $p+p$ collisions at $\sqrt{s_{NN}} = 200$ GeV. To study the jet fragmentation modification function, we take the ratio between the nucleus-nucleus and $p+p$ collisions ($I_{AA} = Y^{A+A}/Y^{p+p}$) and is shown on the bottom panel in the Figure 3 for $d$+Au and Au + Au collisions. The $I_{AA}$ ratio shows no modification of the fragmentation for the $d$+Au collisions, while in Au + Au collisions there is a large suppression observed at low-$\xi$ and an enhancement at high-$\xi$.

Furthermore, the new data from the Au + Au collisions allow us to study the fragmentation modification in different photon trigger ($p_{\rm T}{}^\gamma$) bins, while using the same associated hadron transverse momenta ($p_{\rm T}{}^h$). The right panels in Figure 3 show the three different $I_{AA}$ ratios, the largest modification is visible in the $5 < p_{\rm T}{}^\gamma < 7$ GeV$/c$ bin, while it is consistent with unity at all $\xi$ values in the largest selected trigger bin, $9 < p_{\rm T}{}^\gamma < 12$ GeV$/c$. In addition, we can also study where the missing energy goes by changing the away side integration range between $\pi/2$ and $\pi/6$ shown in different colors in Figure 3. While in the low-$\xi$ region the different integration regions are very consistent, at high-$\xi$ the largest enhancement is in the largest integration range $|\Delta\phi - \pi| < \pi/2$, which suggests that the lost energy of the outgoing parton is scattered in the larger angles.

### 3.3 Thermal Photons

Analogous to the black body radiation, a formation of the hot and dense matter in the heavy ion collisions would result in emittance of thermal radiation in form of photons and di-electron pairs. The temperature of the medium would directly correlate to the rate of the emission, such that as the temperature cools down, the thermal yield emission rate would slow down. However, while the medium is cooling down it is also rapidly expanding which results in a larger blue shift of the emitted photons in the later stage of the medium evolution [12]. The thermal photon yield was predicted to be dominant in the low-$p_{\mathrm{T}}$ region. However, the measurement of direct photons for $1 < p_{\mathrm{T}} < 5$ GeV$/c$ is notoriously difficult due to a large background from hadronic decay photons. In fact, direct photons would contribute only $\approx 10\%$ above the hadronic background.

Figure 4 shows the measured direct photon yield in $\mathrm{Au}+\mathrm{Au}$ and $p+p$ collisions at $\sqrt{s_{NN}} = 200$ GeV [13]. To compare the two different collisions, we fit the $p+p$ data with the modified Hagedorn function and scale it with the number of binary collisions to $\mathrm{Au}+\mathrm{Au}$ data. The scaled fit describes the high-$p_{\mathrm{T}}$ region of the $\mathrm{Au}+\mathrm{Au}$ collisions where the pQCD processes are the dominant source of the photons. At low-$p_{\mathrm{T}}$ we observed a large excess of the direct photon production in comparison to the scaled $p+p$ collisions.

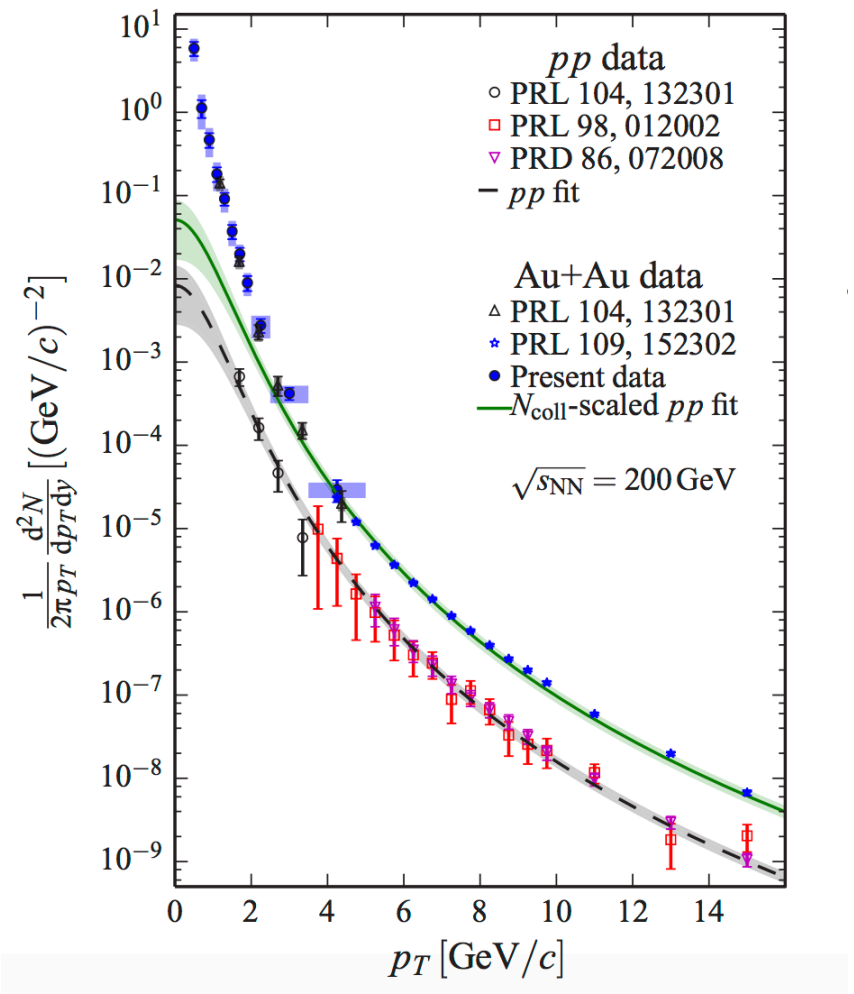

**Fig. 4:** Invariant yield of direct photons in Au+Au and $p+p$ collisions at $\sqrt{s_{NN}} = 200$ GeV.

The excess of these photons is attributed to thermal photon production from the medium. The excess of the data points can be described with an exponential function with the inverse slope of $T_{\mathrm{eff}} \approx 242 \pm 28(\mathrm{stat}) \pm 7(\mathrm{syst})$ MeV. In case of the static medium the inverse slope would be directly proportional to the average temperature of the medium, yet we now know that it is more the space-time averaged temperature. The temperature in the models should be calculated during the entire cooling down of the medium and in addition also have to include the blue shift from the expansion.

In most theoretical models the thermal photons come from binary processes, therefore their number from a unit volume should be proportional to the square of the number of constituents. The bulk properties on other hand would be proportional to the number of constituents. In Figure 5 shows the integrated yield of direct photons as a function of the number of participant nuclei ($N_{\mathrm{part}}$) in $\mathrm{Au}+\mathrm{Au}$ collisions at $\sqrt{s_{NN}} = 39-200$ GeV and $\mathrm{Cu}+\mathrm{Cu}$ collisions at $\sqrt{s_{NN}} = 200$ GeV. We observe that all different collision systems and energies scale on a single curve with the exponent of $\alpha = 1.35 \pm 0.09$. The observed exponent is larger than unity, which suggests that the photons are coming from a thermal source. Although the exponent is smaller than two, that can be attributed to the expanding medium and volume.

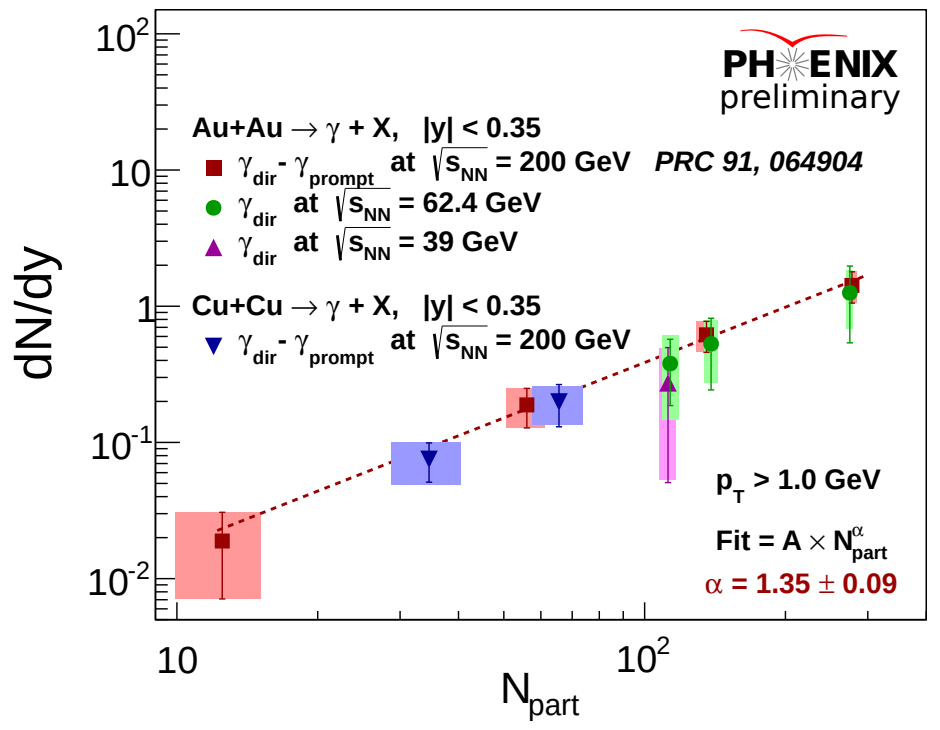

**Fig. 5:** The integrated yield of direct photons as a function of number of participants in Au+Au collisions at $\sqrt{s_{NN}} = 39 - 200$ GeV and Cu+Cu at $\sqrt{s_{NN}} = 200$ GeV.

## 4 Summary

We presented the new direct photon-hadron and $\pi^0$ -hadron correlation results from $p + p$ collisions at $\sqrt{s} = 510$ GeV and minimum bias $p$+Au and $p$+Al collisions at $\sqrt{s_{NN}} = 200$ GeV. The measured Gaussian widths of the $p_{out}$ distributions decrease as the $p_{\mathrm{T}}{}^{\mathrm{trig}}$ increases, which is contradicting the expectation from TMD factorization. We also presented the new measurement on the modification of jet fragmentations in Au + Au collisions at $\sqrt{s} = 200$ GeV. These results suggests that the medium enhances production of soft particles preferentially at large angles in comparison in $p + p$ collisions. Lastly, the presented enhancement of the direct photons at low-$p_{\mathrm{T}}$ shows a scaling property as a function of the charged hadron multiplicity independent of collision system and energy.

# Prompt photon production with POWHEG

*M. Klasen*
Institut für Theoretische Physik, Westfälische Wilhelms-Universität Münster, Wilhelm-Klemm-Straße 9, D-48149 Münster, Germany

**Abstract**
We present a calculation of prompt photon and associated photon-jet production at next-to-leading order that is consistently matched to parton showers with POWHEG. Specific issues that appear in photon radiation are discussed. Numerical results are compared to pp collision data at RHIC energies and shown to describe the data better than inclusive next-to-leading order calculations or those with leading order Monte Carlo generators like PYTHIA alone.

## 1 Introduction

Theoretical calculations of prompt photon production at hadron colliders have a long-standing tradition. Their importance derived originally from understanding perturbative QCD [1], e.g. the fractional quark charges or renormalisation group effects, whereas photon pair production is currently an important background in one of the Higgs-boson discovery channels at the LHC [2, 3]. In heavy-ion collisions, thermal photons are an important probe to determine the effective temperature of the created Quark-Gluon Plasma (QGP) through their characteristic exponential transverse momentum ($p_T^\gamma$) spectrum [4].

Leading-order (LO) calculations, supplemented by parton showers (PS) and hadron fragmentation and implemented in Monte Carlo generators like PYTHIA 8 [5], provide detailed information on the final state and are indispensable tools in the experimental analyses. Inclusive next-to-leading order (NLO) calculations like JETPHOX [6] employ, in contrast, inclusive fragmentation functions (FFs) like BFG II [7] and have a smaller theoretical scale uncertainty. With NLO+PS Monte Carlo methods like POWHEG [8] it is possible to combine the advantages of both approaches. We have recently applied this method to prompt photon and associated photon-jet production [9]. We review the theoretical approach in Sec. 2 and then demonstrate its phenomenological advantages by applying it to PHENIX data from RHIC [10] in Sec. 3. Our conclusions are given in Sec. 4.

## 2 Theoretical approach

The POWHEG method requires first the recalculation of Born processes with spin and colour correlations, in this case for $q\bar{q} \to \gamma g$ and the QCD Compton process $qg \to \gamma q$. Next, the virtual corrections must be recomputed and their ultraviolet and infrared divergences consistently renormalised and subtracted, respectively. The real emission amplitudes must not be subtracted, since POWHEG does so automatically. While the hardest radiation is thus generated first, subsequent emissions are produced by the parton shower implemented, e.g., in PYTHIA [5], leading always to detailed events with positive weights. This method applies not only to QCD radiation, but can be generalised to QED radiation. One must then check that fragmentation photons like those produced in $e^+e^-$ collisions at LEP are as well described with PSs as with inclusive FFs [11].

There are, however, a number of specific issues for photons. First, QED radiation is suppressed with respect to QCD radiation by the smaller coupling, colour factors and multiplicities, so that it must be artificially enhanced. Second, the hard scale is not necessarily the $p_T^\gamma$ of the observed photon, but may be the $p_T$ of an underlying QCD parton. Third, the radiation process $q \to q\gamma$ must be correctly symmetrised, i.e. also include the process $q \to \gamma q$. Finally, inclusive photon production $pp \to \gamma + X$ has a collinear divergence at low $p_T^\gamma$, which must be carefully regularised so that sensitivity to the low-$p_T^\gamma$ region important for heavy-ion collisions and independence on the regularisation method are simultaneously maintained.

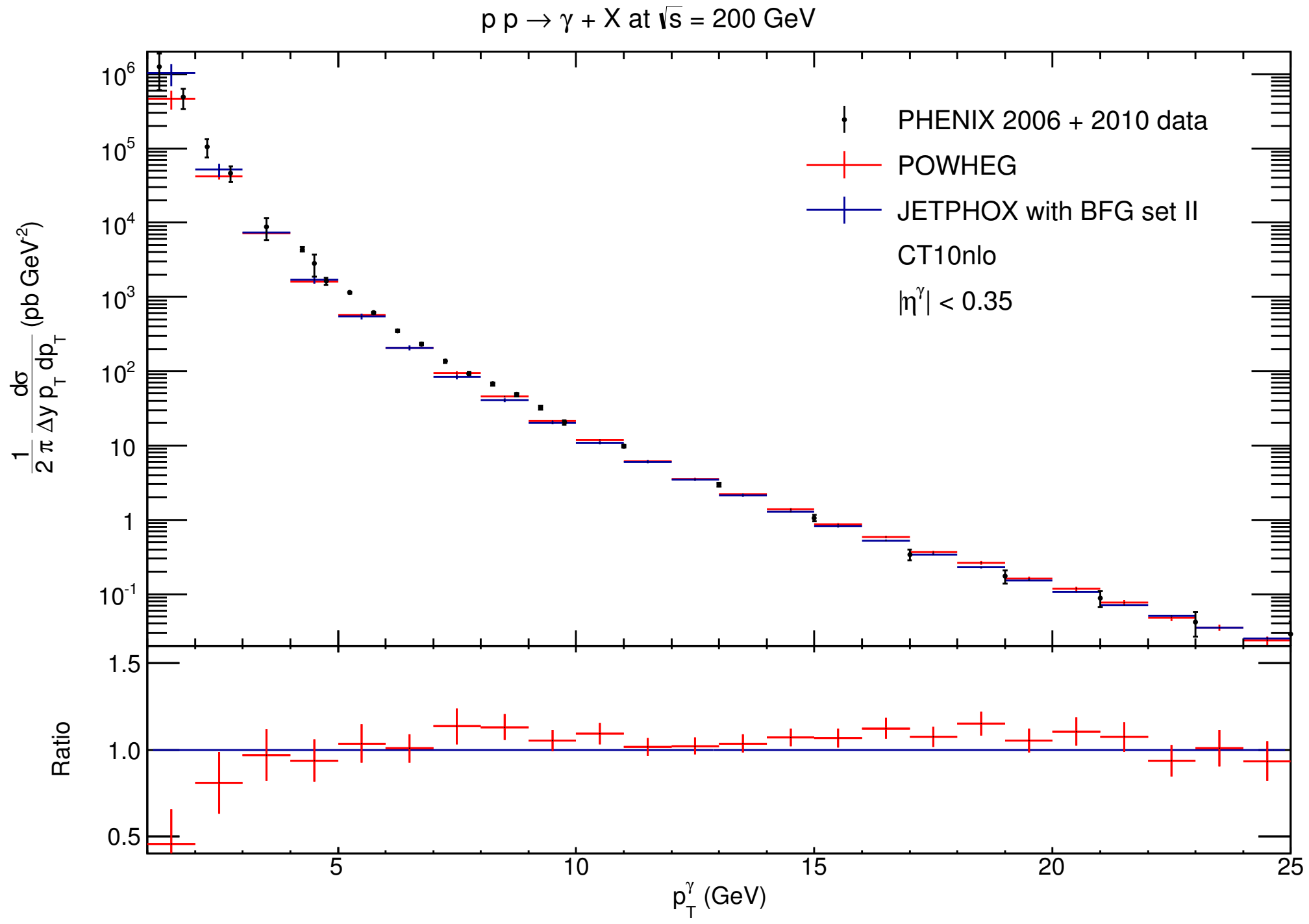

**Fig. 1:** Inclusive photon production in pp collisions at RHIC with a centre-of-mass energy of 200 GeV. PHENIX data (black) are compared with NLO+PS predictions with POWHEG+PYTHIA (red) and NLO predictions with JETPHOX (blue).

## 3 Prompt photons and photon-jet correlations at RHIC

We now compare the implementation of prompt photons in POWHEG to data obtained by the PHENIX collaboration in pp collisions at RHIC [10]. These data represent an important baseline for studies of the QGP produced in heavy-ion collisions and should be described by prompt (direct and fragmentation) contributions alone.

In Fig. 1 the inclusive photon distribution in transverse momentum is shown at the RHIC centre-of-mass energy of $\sqrt{s} = 200$ GeV. As expected for this inclusive quantity, both the NLO calculation with JETPHOX (blue) and the NLO+PS calculation with POWHEG+PYTHIA (red) describe the data up to 25 GeV. The two theoretical predictions differ only at very low $p_T^\gamma$, where fragmentation contributions dominate. There, the POWHEG+PYTHIA predictions are lower than those with JETPHOX and seem to agree better with the PHENIX data, although both are consistent with the data within error bars.

A quantity that is more sensitive to the treatment of the photon fragmentation process is the fraction of isolated photons, defined by a hadronic energy fraction of less than 10% of the energy of the photon in a cone of radius $R = \sqrt{(\Delta\eta)^2 + (\Delta\phi)^2} \leq 0.5$ around it. The comparison with JETPHOX in Fig. 13 of the experimental publication led to the conclusion that neither BFG II nor GRV [12] photon FFs described this fraction correctly, even after accounting for statistical, systematic, and theoretical scale uncertainties [10].

As one can see in Fig. 2, the fraction of isolated photons predicted by POWHEG+PYTHIA describes the data very well, apart from a few fluctuations due to limited Monte Carlo statistics. The fraction rises from about 60% at low $p_T^\gamma$, where a substantial fraction of photons does not survive the isolation cut due to collinear parton radiation, to almost unity for intermediate and large $p_T^\gamma$. There, photons are rather produced back-to-back with a recoiling jet and have little near-side hadronic energy.

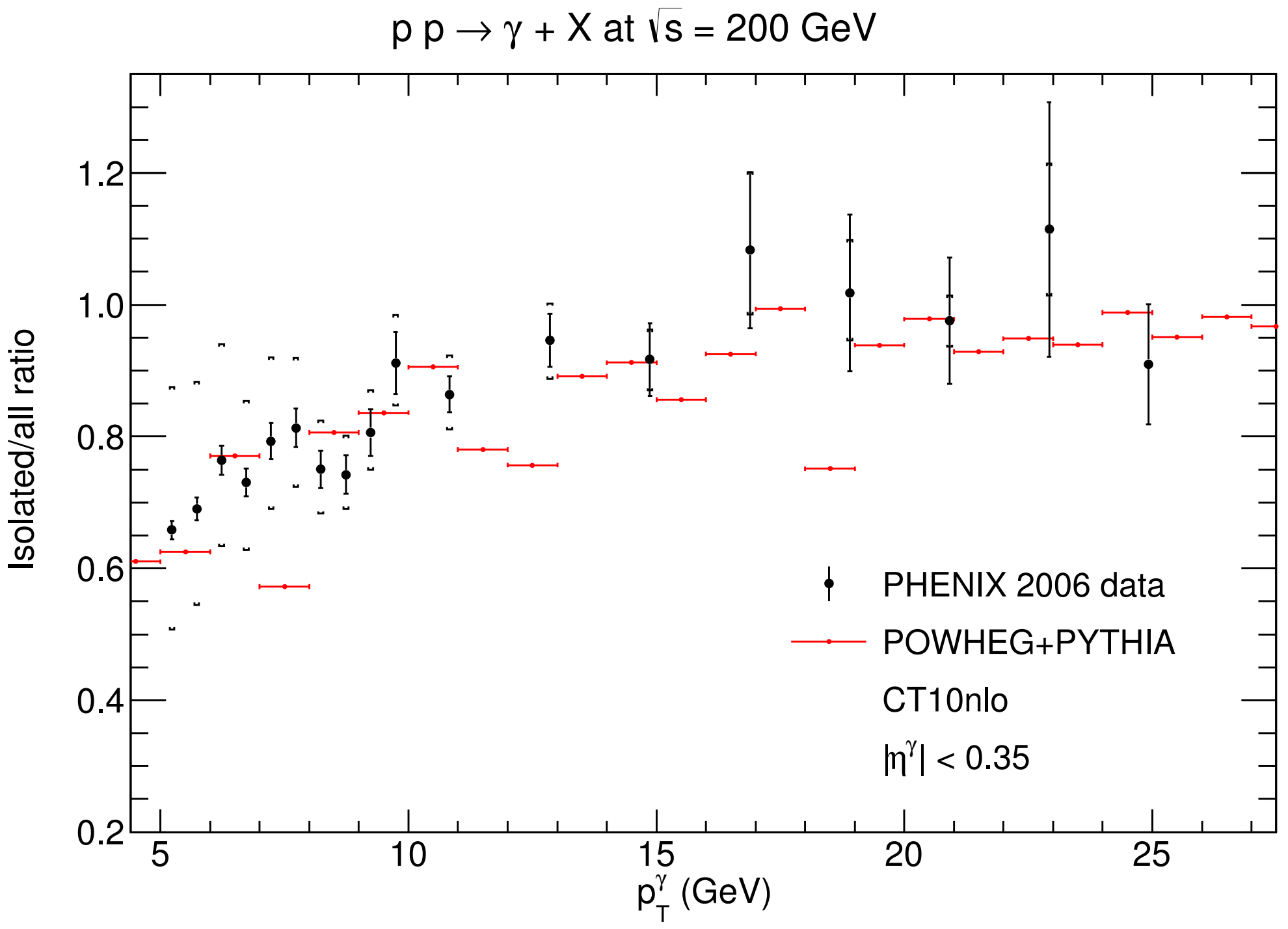

**Fig. 2:** Isolated photon production in pp collisions at RHIC with a centre-of-mass energy of 200 GeV. PHENIX data (black) are compared with NLO+PS predictions with POWHEG+PYTHIA (red).

When both the photon and the recoiling jet are experimentally measured, the two may be correlated either in transverse momentum or in azimuthal angle $\Delta\phi$. Both quantities are sensitive to modifications of the hadronic jet due to rescatterings in the QGP, but also to higher-order perturbative QCD effects [13]. Fig. 3 shows the distribution in azimuthal angle, subtracted for decay photons assuming a Zero-Yield at Minimum (ZYAM) and normalised to the total number of trigger photons. The near- ($\Delta\phi \sim 0$) and away-side ($\Delta\phi \sim \pi$) regions are clearly visible, but only the latter is correctly described by PYTHIA alone, while both are reproduced with POWHEG+PYTHIA. As expected and as we have shown in our original paper [9], the near-side region is dominated by fragmentation photons, which requires a proper matching of NLO and PS contributions.

## 4 Conclusion

To conclude, we have reviewed our recent implementation of prompt photon production in POWHEG. Our calculations now allow for predictions of prompt photon and photon-jet associated production at hadron colliders with reduced theoretical scale uncertainties and sufficient detail of the produced final state. We have successfully applied our calculations to PHENIX data from RHIC. Applications to the LHC will be presented in a forthcoming publication.

## Acknowledgements

We thank T. Jezo and F. König for their collaboration and the organisers of the conference for the kind invitation. This work is supported by the BMBF under contract 05H15PMCCA and by the DFG through the Research Training Network 2149 “Strong and weak interactions - from hadrons to dark matter”.

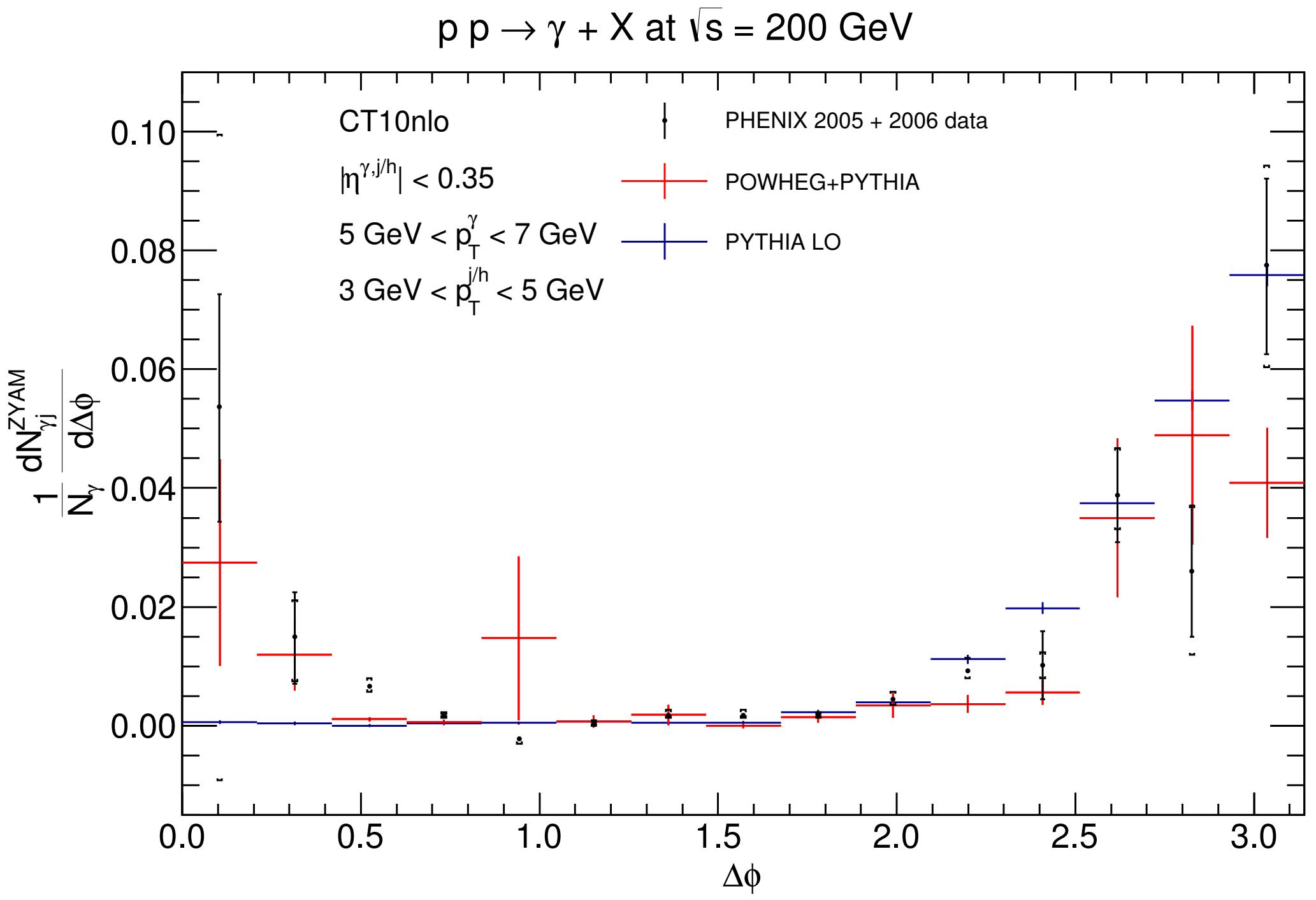

**Fig. 3:** Associated photon+jet production in pp collisions at RHIC with a centre-of-mass energy of 200 GeV. PHENIX data (black) are compared with NLO+PS predictions with POWHEG+PYTHIA (red) and LO+PS predictions with PYTHIA (blue).

# Recent developments on parton-to-photon fragmentation functions

*Tom Kaufmann* [1], *Asmita Mukherjee* [2], *Werner Vogelsang* [1]
[1] Institute Theoretical Physics, Tübingen Univ., Auf der Morgenstelle 14, 72076 Tübingen, Germany
[2] Department of Physics, Indian Institute of Technology Bombay, Powai, Mumbai 400076, India

**Abstract**
We report on some recent developments concerning parton-to-photon fragmentation functions. We briefly summarize the main theoretical concepts on which the currently available extractions of these fragmentation functions rely: evolution and the vector meson dominance model. We present comparisons of the available sets. We argue that the single-inclusive photon production process $pp \to \gamma X$ possesses only little sensitivity to the fragmentation contribution. Instead, the semi-inclusive process $pp \to (\text{jet}\,\gamma)X$, for which the photon is observed inside a fully reconstructed jet and is part of the jet, is shown to offer much potential for providing valuable new information.

## 1 State of the Art

The production of photons with high transverse momentum $p_T$ in hadronic collisions is of fundamental importance in today's particle and nuclear physics. Foremost, it may serve as a tool for determining the gluon distributions of the scattering hadrons, thanks to the presence and dominance of the leading order (LO) Compton subprocess $qg \to \gamma q$ [1,2]. Photons also provide sensitive probes of the medium produced in collisions of heavy ions, being able to traverse and escape the medium with little attenuation. Finally, photon signals play an important role in studies of physics within and beyond the Standard Model, with the process $pp \to \gamma\gamma X$ through production and decay of a Higgs boson and the indication of a 750 GeV diphoton excess seen by the ATLAS and CMS collaborations at the LHC in 2015 [3] arguably being among the most well-known examples. Although the presumed "bump" in the 2015 data disappeared with more statistics in a 2016 update [4], the enormous interest by the community (hundreds of papers tried to give a possible explanation of this excess [5]) demonstrated once again the importance of photon signals in the search for "new physics" beyond the Standard Model of particle physics.

As was discussed a long time ago [6], in perturbative high-$p_T$ processes there are two production mechanisms for photons. The photon may be directly produced in the hard scattering process through its "pointlike" QED coupling to a quark. Such contributions are usually referred to as "direct". On the other hand, photons may also emerge in jet fragmentation, when a quark, antiquark, or gluon emerging from a QCD hard-scattering process fragments into a photon plus a number of hadrons. The need for introducing such a "photon fragmentation" contribution is physically motivated by the fact that the photon may result, for example, from conversion of a high-$p_T$ $\rho$ meson. Furthermore, at higher orders, the perturbative direct component contains divergences from configurations where the photon and a (massless) final-state quark become collinear. These are long-distance contributions that naturally signify the need for non-perturbative fragmentation functions into which they can be absorbed. Schematically, photon production cross sections may be written in a factorized form as

$$d\sigma^\gamma = d\hat{\sigma}^\gamma + \sum_c d\hat{\sigma}^c \otimes D_c^\gamma \,, \tag{1}$$

where the sum runs over all partons (quarks, antiquarks and gluons). Note, that for processes with initial state hadrons, the partonic cross sections in Eq. (1) have to be further convoluted with the respective

PDFs. In general, the parton-to-photon fragmentation functions depend on the longitudinal momentum fraction $z$ which is transferred to the photon and on the factorization scale $\mu$: $D_c^\gamma \equiv D_c^\gamma(z,\mu^2)$. The non-perturbative functions $D_c^\gamma$ are assumed to be universal and thus may in principle be extracted in the same manner as the parton-to-hadron FFs $D_c^h$ via fits to experimental data. While there has been much progress on parton-to-hadron FFs in the last years [7], there have been no new extractions of parton-to-photon FFs for about two decades, and our knowledge of these functions has remained relatively poor. This is mostly due to the fact that the largest data set comes from *single-inclusive* photon data in hadronic collisions, for which the dominant contribution comes from the direct (i.e. non-fragmentation) part. For the reaction $e^+e^- \to \gamma X$ fragmentation yields the dominant contribution; however, here only a very sparse data set exists, and the amount of photon data and their precision does not match that of hadron production data. The two most recent sets of photon FFs available are the "Glück-Reya-Vogt" (GRV) set [8] and the "Bourhis-Fontannaz-Guillet" (BFG) set [9]. The latter is based on the somewhat older "Aurenche-Chiappetta-Fontannaz-Guillet-Pilon" (ACFGP) set [10]. We note, that the BFG set actually consists of two sets of FFs, which mainly differ in the gluon-to-photon fragmentation function.

Besides universality, another key ingredient to extractions of FFs is evolution. The presence of the direct part affects the evolution equations for photonic distributions. Following [8, 9], the DGLAP-like evolution equations may be written as

$$\frac{d}{d\log\mu^2} D_i^\gamma(z,\mu^2) = \sum_j P_{ji}(z,\mu^2) \otimes D_j^\gamma(z,\mu^2)\,, \tag{2}$$

where $i,j$ run over all parton flavors including the photon itself, i.e. $i,j \in \{q_{i,j}, \bar{q}_{i,j}, g, \gamma\}$, so that we also have a photon-to-photon FF $D_\gamma^\gamma$ and photon splitting functions. The symbol $\otimes$ denotes the standard convolution integral. The evolution kernels, also called time-like splitting functions, are double series in the electromagnetic coupling $\alpha$ and the strong coupling $\alpha_s$,

$$P_{ij}(z,\mu^2) = \sum_{m,n} \left(\frac{\alpha(\mu^2)}{2\pi}\right)^m \left(\frac{\alpha_s(\mu^2)}{2\pi}\right)^n P_{ij}^{(m,n)}(z)\,. \tag{3}$$

Usually, only the leading order in the electromagnetic coupling is considered, so that $D_\gamma^\gamma(z,\mu^2) = \delta(1-z)$. Furthermore, the running of $\alpha$ is neglected. The evolution equations then reduce to the frequently used inhomogeneous evolution equations

$$\frac{d}{d\log\mu^2} D_i^\gamma(z,\mu^2) = k_i^\gamma(z,\mu^2) + \sum_j P_{ji}(z,\mu^2) \otimes D_j^\gamma(z,\mu^2)\,, \tag{4}$$

where now just $i,j \in \{q_{i,j}, \bar{q}_{i,j}, g\}$. The inhomogeneous term is given by

$$k_i^\gamma(z,\mu^2) = \frac{\alpha}{2\pi} \sum_n \left(\frac{\alpha_s(\mu^2)}{2\pi}\right)^n P_{\gamma i}^{(1,n)}(z)\,. \tag{5}$$

Like for hadron FFs, these evolution equations are solved most conveniently in Mellin-$N$ space where the convolutions turn into simple products. Furthermore, they usually are decomposed into the standard singlet and non-singlet combinations, see, for instance, [8,9]. The full solution of the evolution equations (4) is given by the sum of a particular solution to the inhomogeneous equation, which can be computed in perturbation theory, and a general solution to the homogeneous equation, which contains the non-perturbative component. Schematically, we have

$$D_i^\gamma = D_i^{\gamma,\text{inhom}} + D_i^{\gamma,\text{hom}}\,. \tag{6}$$

While the first part in Eq. (6) is perturbative, the second non-perturbative part has to be extracted from experiment or modeled. Ideally, one would prefer to extract this part from a clean reference process without contamination of other non-perturbative functions (like PDFs), i.e. from single-inclusive annihilation (SIA) $e^+e^- \to \gamma X$. However, only a very limited amount of data are available for this process, which furthermore have rather large uncertainties [11]. In view of this, the two most recent extractions of FFs have resorted to the vector meson dominance (VMD) model, for which one assumes that the fragmentation process is dominated by conversion of vector mesons. Thus, the ansatz

$$D_i^{\gamma,\mathrm{hom}}(z,\mu_0^2) = \alpha \sum_{v=\rho,\omega,\phi,\ldots} C_v D_i^v(z,\mu_0^2) \tag{7}$$

is used at the initial scale for the evolution, along with a vanishing inhomogeneous piece at the initial scale. Here, the sum runs over all vector mesons under consideration and the $D_i^v$ are the fragmentation functions into the corresponding vector mesons. As the FFs for the latter are rather poorly known as well, the GRV set adopts pion FFs for them instead. The BFG set uses $\rho$ data from ALEPH [12] and HRS [13] to constrain their VMD ansatz. We note that the recent paper [14] presented a detailed Monte-Carlo event generator study of the inclusive hadro-production of photons and vector mesons. By adding the $p_T$ spectra of the light vector mesons $(\rho,\omega,\phi)$, weighted with $\alpha$ divided by the individual vector-meson decay constants, one obtains a fairly good description of inclusive photon data in the low-$p_T$ region, which is the kinematic regime where the fragmentation process dominates over the direct part [15]. This observation supports the VMD ansatz for the nonperturbative part of photon FFs.

It is interesting to compare the different FF sets that are available [1, 15]. In the left and middle panels of Fig. 1 we show $zD_i^\gamma(z,\mu^2 = (20\,\mathrm{GeV})^2)$ for $i = u,d,s,g$, as given by the GRV and the two BFG sets. While the quark FFs are rather similar, the gluon FF is quite different in each of the three sets. SIA data would not be expected to help discriminate among the various $D_i^g$ as the gluon FF enters only at NLO or via evolution. In order to see whether single-inclusive photon production in hadronic collisions, $pp \to \gamma X$, is more promising, we compare in the right panel of Fig. 1 the theoretical cross section at NLO (using the code of Ref. [16]) with data from PHENIX [17]. As can be seen, the different FF sets yield very similar results. Compared to the experimental uncertainties, the difference in the FF sets is negligible. Even though the gluon FF is very different and channels with *initial* gluons dominate, the inclusive photon production is apparently not really sensitive to the gluon FF. We note that this observation has been made in previous literature for various collider and fixed target settings; see, for instance, [1, 15, 18]. We stress that the fact that the presently available sets of photon FFs yield similar cross section predictions does not imply that the fragmentation contribution to photon production is under satisfactory control. For instance, there is arguably a much larger uncertainty in the $u$-quark FF than suggested by the curves shown in Fig. 1.

These observations also have ramifications for photon signals in $pA$ and, especially, $AA$ collisions. For the latter, photons are used in studies of the quark-gluon plasma (QGP). While photons produced directly in the collision will traverse the medium with only little attenuation, the fragmentation photons will originate from partons that suffered energy loss in the medium. To assess this effect properly, good understanding of the “vacuum” photon fragmentation functions is essential.

We finally note that in $pp$ collider experiments usually a photon isolation cut is introduced in order to suppress the large background from $\pi^0 \to \gamma\gamma$ decay. The idea is to center a cone around the final state photon and to demand that the hadronic energy fraction inside this cone be less than a certain amount $\epsilon$. Such isolation cuts also suppress the photon fragmentation contribution [2, 19], since they confine the fragmentation contribution to large values of $z$. This, however, introduces further uncertainties since the FFs are completely unconstrained in the region of large $z$ and since this region is much harder to treat theoretically. For instance, large logarithmic $\log(1-z)$ contributions arise here in the evolution kernels and partonic cross sections [20]. This especially affects the inhomogeneous part which does not vanish for $z \to 1$ [8]. Moreover, the isolation procedure also introduces logarithmic contributions in the energy fraction $\log\epsilon$ [20] and the cone opening [21, 22]. Any experimental observable that can provide direct information on photon fragmentation at high $z$ will provide valuable insights into these questions.

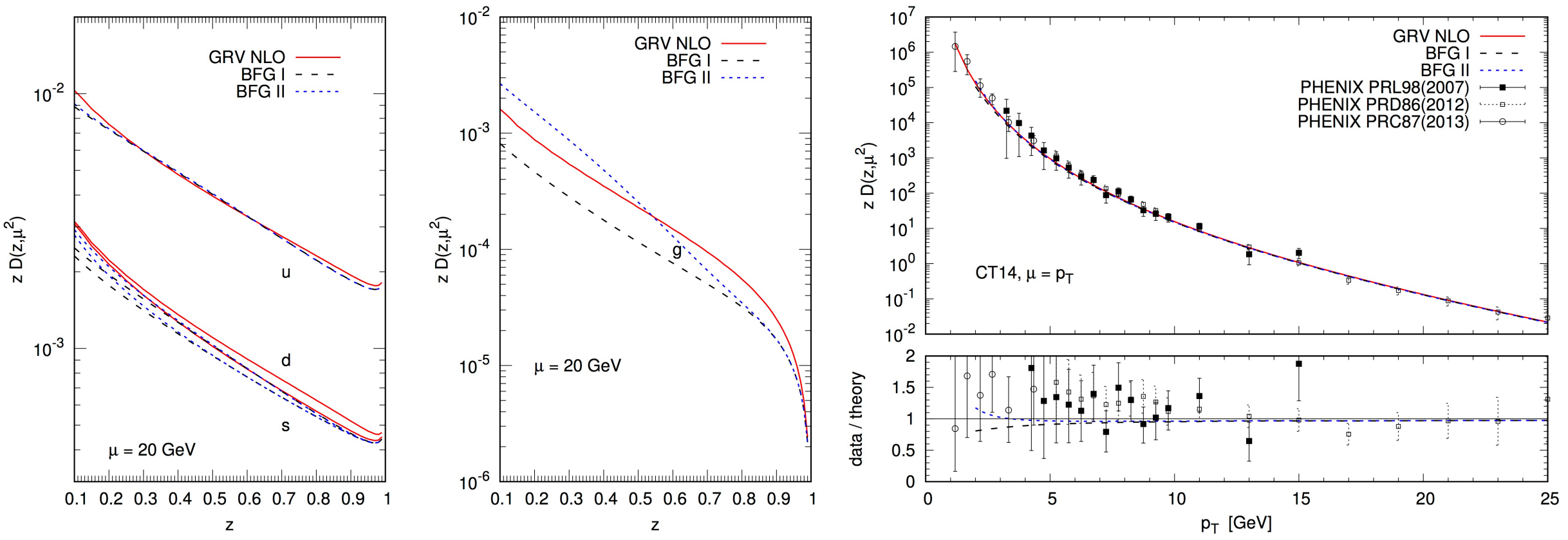

**Fig. 1:** Comparison of the different FF parametrizations [8,9]. In the left and middle panel we show $zD_i^\gamma(z,\mu^2 = (20\,\text{GeV})^2)$ for $i = u,d,s$ and $i = g$, respectively. In the right panel we compare to PHENIX data [17] for $pp \to \gamma X$, where the lower part shows the ratio of the data and the theoretical NLO predictions for the two BFG sets with respect to the one based on GRV FFs.

## 2 Photon-in-jet production

As we have seen, neither SIA nor inclusive photon production in $pp$ collisions are able to provide detailed access to the fragmentation contribution. It is thus important to identify new observables that are able to yield new and complementary information on photon FFs. One such observable was introduced in Ref. [23], where the process $pp \to (\text{jet}\,\gamma)X$ was proposed, for which a photon is observed in the final state inside a fully reconstructed jet, as part of the jet. Previous related studies [24] for the case of final-state hadrons had already established that such "fragmentation-inside-jets" observables may be reliably computed using factorization and perturbative-QCD techniques and may indeed give information on FFs. Recently, an extraction of $D^*$-meson FFs has been performed [25], including, for the first time, $D^*$-in-jet data. It was shown that the in-jet data actually are able to give valuable constraints on, especially, the gluon FF. We now briefly present some of our results in [23]. The cross section is calculated differential in the transverse momentum and rapidity of the jet, $p_T^{\text{jet}}$ and $\eta^{\text{jet}}$, respectively, and the photon-jet momentum correlation variable

$$z_\gamma \equiv \frac{p_T^\gamma}{p_T^{\text{jet}}}\,. \tag{8}$$

The partonic cross sections are calculated analytically in the framework of the "narrow jet approximation" (NJA). In the NJA, the jet is assumed to be relatively narrow, in the sense that the jet parameter $R$ (we have in mind the widely used anti-$k_T$ jet algorithm [26] here) is rather small, $R \ll 1$. Thus, contributions of the order $\mathcal{O}(R^2)$ are neglected throughout the calculation. It was shown in Refs. [27] for inclusive jet production that the NJA works well out to rather large values of $R \lesssim 0.7$. We stress that the observable we have in mind is different from the "away-side" photon-jet correlations considered in Ref. [28] and provides a kinematically simpler and more direct access to the $D_c^\gamma$. The main asset of the process $pp \to (\text{jet}\,\gamma)X$ is that at LO the cross section is *directly proportional* to the FFs probed at $z = z_\gamma$:

$$\left.\frac{d\sigma^{pp\to(\text{jet}\,\gamma)X}}{dp_T^{\text{jet}}\,d\eta^{\text{jet}}\,dz_\gamma}\right|_{\text{LO}} \propto \sum_{\substack{a,b,c\in\\ \{q,\bar{q},g,\gamma\}}} f_a \otimes f_b \otimes d\hat{\sigma}_{ab}^{c,\text{LO}} \times \left[\delta(1-z_\gamma)\delta_{c\gamma} + D_c^\gamma(z_\gamma,\mu^2)(1-\delta_{c\gamma})\right]\,. \tag{9}$$

The first part in the squared brackets of Eq. (9) is the direct part while the second one is the fragmentation contribution. By demanding $z_\gamma < 1$ to ensure that we have a hadronic jet around the photon, the direct part does not contribute at LO and only the fragmentation contribution remains. The various FFs are weighted by appropriate combinations of PDFs and partonic cross sections, which may be regarded as "effective charges". The structure of the cross section hence becomes similar to that for $e^+e^- \to \gamma X$, but with the essential difference that also gluon-to-photon fragmentation contributes at the lowest order. We finally note, that we have presented a detailed study of the $\pi^0 \to \gamma\gamma$ background for processes involving final state photons in Ref. [23]. We have addressed the two main background sources, i.e. when the two decay photons become collinear or when one of the two photons falls below the energy detection threshold.

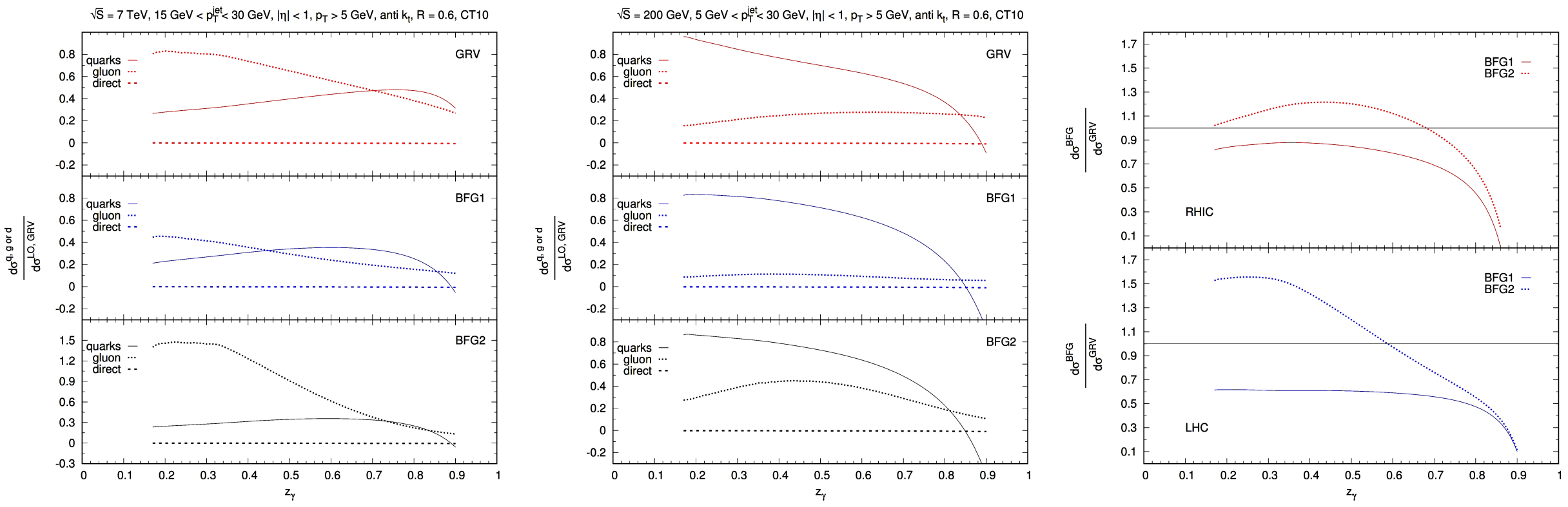

**Fig. 2:** The contribution of different subprocesses (gluon fragmentation, quark fragmentation and direct part) for kinematical setups corresponding to LHC (left panel) and RHIC (middle panel). In the right panel we show the ratio of cross sections computed with the BFG sets with respect to that for GRV.

In Fig. 2 we demonstrate the potential of the process $pp \to (\text{jet}\,\gamma)X$ for providing valuable and new information on the photon FFs. We show, for LHC and RHIC kinematics, the relative contributions of quark FFs, the gluon FF and the direct part for the three different FF parameterizations. For better readability, we have normalized all results to the LO cross section computed with the LO GRV fragmentation set. We fist observe that the direct part is very small, being a pure NLO contribution. Furthermore, we see that the gluon contribution is much larger for LHC compared to RHIC. This due to the large size of the contributions by initial gluons at high center of mass energies. In the right panel of Fig. 2 we compare the full NLO cross sections for the different FF sets. We show the ratios of the cross sections computed with the BFG1 and BFG2 sets, relative to the cross section for the GRV FFs. The potential of this process becomes visible, especially for the LHC setup where we find differences among the cross sections of up to $50\%$. This is in stark contrast to what we saw for $pp \to \gamma X$ (see the lower part of the right panel in Fig. 1), where over a large range in $p_T$ the cross sections for the different FF sets differed by less than $10\%$. We are hence optimistic that experimental data for $pp \to (\text{jet}\,\gamma)X$ would allow one to distinguish between the different FF sets.

## 3 Conclusions

We have presented the current state of the art for parton-to-photon fragmentation functions. While precise knowledge of these functions is important for predictions for all observables with observed final-state photons, only little is actually known about them so far. The single-inclusive process $pp \to \gamma X$ has a dominant direct contribution, so that it is rather insensitive to the fragmentation one. For the SIA reaction $e^+e^- \to \gamma X$, on the other hand, only a very sparse data set exists and it has no sensitivity to gluon fragmentation. In contrast to this, $pp \to (\text{jet}\,\gamma)X$, for which a photon is observed in the final state inside a fully reconstructed jet and is part of the jet, may provide direct and clean access to the parton-to-photon fragmentation functions, including the gluon one. We thus encourage experimental efforts to perform dedicated analyses of this process.

## Acknowledgements

The work of T.K. is supported by the Bundesministerium für Bildung und Forschung (BMBF) under grant no. 05P15VTCA1. A.M. thanks the Alexander von Humboldt Foundation, Germany, for support through a Fellowship for Experienced Researchers.

# Electroweak corrections at the LHC: the contribution of photons

*Davide Pagani*
Technische Universität München, James-Franck-Str. 1, D-85748 Garching, Germany

**Abstract**
In this talk I review the photon contribution to the electroweak corrections for some of the most relevant processes at the LHC, namely, the Drell–Yan, $VV$, $t\bar{t}$ and dijet production. In the discussion I focus on two dominant effects: photon radiation from final-state light particles and the impact of the photon PDF.

## 1 Introduction

In order to match the current and especially the future precision in the measurements of Standard Model (SM) processes at the LHC, both higher-order QCD and electroweak (EW) corrections have to be taken into account in theory predictions. Whenever NNLO QCD corrections are relevant, NLO EW corrections cannot be neglected, since they are expected to have similar sizes: $\alpha_s^2 \sim \alpha$. Moreover, in the boosted regime EW corrections are enhanced by the so-called Sudakov logarithms, with leading terms that are negative and proportional to $\log^2\left(\frac{Q^2}{m_W^2}\right)$, being $Q$ the typical scale of the boosted regime considered. Although Sudakov effects are of purely weak origin, so they do not depend on photon (QED) effects, they can contribute in the same phase-space regions where QED corrections are large and positive, leading to cancellations. This is particular relevant in our discussion since nowadays QED and purely weak corrections are typically computed together within complete NLO EW calculations. Moreover, in the recent years, completely automated NLO EW calculations have been performed for several processes.

In the following I will discuss photon effects from the EW corrections to the cross section of the Drell–Yan, $VV$, $t\bar{t}$ and dijet production processes. In particular I will focus on two dominant effects:

– the impact of photon Final-State-Radiation (FSR) from light particles (typically leptons) in sufficiently exclusive observables,
– the impact of photon-initiated processes, which thus depend on the photon PDF.

When a lepton is present in a final state, FSR induces large corrections due to the collinear enhancement in the $\ell \to \ell\gamma$ splitting. The recombination of photons with leptons (dressed leptons), which is mandatory for the case of electrons only, reduces the impact of these effects, introducing a dependence on the radius $R$ of the recombination. Unless it is explicitly specified (bare muons), we will refer always to dressed leptons.

Regarding the photon PDF it is important to keep in mind two aspects. First, the dependence on the photon PDF is entering at different perturbative orders, depending on the processes. While for neutral-current Drell–Yan and $WW$ production it appears already at LO, for, e.g., $WZ$, $ZZ$, $HV$ production it appears only in the NLO EW corrections. Moreover, for processes such as $t\bar{t}$ and dijet production, it appears and give its dominant contribution at $\mathcal{O}(\alpha_s\alpha)$, i.e., a tree-level contribution that is suppressed w.r.t. the dominant LO, which is of $\mathcal{O}(\alpha_s^2)$. Second, many NLO EW calculations have been performed in the recent years by using NNPDF2.3QED [1] and NNPDF3.0QED [2] distributions, which at large Bjorken-$x$ have large central values and especially very large uncertainties. On the contrary, very recently, a new method for the determination of the photon PDF has been proposed and included in the set LUXQED PDF set [3, 4]. This new photon PDF determination has a much smaller uncertainty and also a considerably smaller central value for large Bjorken-$x$. Unless differently specified, results reviewed in the following have been calculated with an NNPDF photon distribution. The enhancements due to a specific kinematic effect in photon-induced process are in general present also with different PDF parametrizations (including also MRST2004QED [5] and CTEQ14QED [6]), whereas large effects due to solely the PDF luminosity are strongly suppressed with LUXQED, which is considered at the moment the most accurate one.

We invite the interested readers to directly look into the works cited in the following for the details of the calculation set-ups, such as definitions of cuts and input parameters, which will not be in general described here. A few details on the issue of isolated photons and NLO EW corrections will be given in Sec. 2.4.

## 2 Processes

### 2.1 Drell–Yan: $W$ and $Z$

We start with the case of $pp \to \ell^+\ell^-$, referring to the results that have been presented in Ref. [7]. The invariant mass $m(\ell^+\ell^-)$ distribution receives large positive corrections just below the value $m(\ell^+\ell^-) = m_Z$ due to the FSR. Indeed, the emission of a photon from a lepton reduces the value of $m(\ell^+\ell^-)$ in the events, which mostly populate the region around the $Z$ resonance. This effect is particularly large for bare muons (80 %), but also for dressed leptons (40 %), where quasi-collinear photons are recombined. This leads to the necessity of taking into account the effects due to the multiple emission of photons, which amount to a few percents in the aforementioned phase-space region. At LO also the $\gamma\gamma \to \ell^+\ell^-$ is present and does not include $s$-channel diagrams. Thus, the contribution of the photon PDF is sizable at large values of $m(\ell^+\ell^-)$, but it strongly depends on the PDF parametrization used [8].

In the case of $pp \to \ell\nu_\ell$ production FSR effects are also sizable in the distributions for the transverse mass of the lepton-neutrino pair and $p_T(\ell)$, see, e.g., Ref. [9]. These observables are particularly relevant since they can be exploited for the measurement of $m_W$ at hadron colliders. For this purpose, also the effects of multiple emission of photons have to be taken into account, since permille accuracy is necessary in the theory predictions in order to achieve an accuracy $\sim 10-20$ MeV for the value of $m_W$, as it has been discussed in detail in Ref. [10].

### 2.2 $VV$ production, $V = W, Z$

Considering $VV$ production with stable vector bosons $V$, FSR effects are absent due to presence of only massive particles in the final state. The dominant effect from NLO EW corrections is given by Sudakov logarithms, which, e.g., can reach a relative size of order $\sim -40\%$ for $p_T(V_2) \sim 800$ GeV, where $V_2$ is the softest vector boson, see Ref. [11]. However, this effect can be partially compensated by quark radiation via the $\gamma q \to VVq^{(\prime)}$, which is part of the NLO EW corrections and depends on the photon PDF. Very similarly to the case of the QCD giant $K$-factors these effects arise from configurations displaying $\gamma q \to Vq$ production plus a collinear $q \to Vq^{(\prime)}$ splitting, thus growing for large $p_T(V)$. Moreover, if one of the two vector bosons is a $W$, the initial state photon can couple directly to it, leading to further enhancements. Results computed with the MRST2004QED PDF set are available in Ref. [12]; an almost complete cancellation of Sudakov logarithms and $\gamma q$-induced effects for $WZ$ and $WW$ production is found. The latter process, as already said, includes also $\gamma\gamma$ initial-state contributions, which give large corrections for large values of $m(WW)$, see also Ref. [13].

Similarly to the case of the large QCD corrections, enhancements due to the quark radiation can be avoided by directly vetoing jets. This procedure has been studied in Ref. [14] where one of the leptonic signatures of $WW$ production has been considered, namely the $pp \to \nu_\mu\mu^+e^-\bar{\nu}_e$ process. Off-shell effects and non-resonant contributions are consistently included in the calculation. It is found that indeed the impact of the photon PDF in NLO EW corrections in $p_T(e^-)$ distributions is reduced by imposing a jet veto. On the other hand, with the inclusion of leptonic $W$ decays, FSR effects are relevant, especially

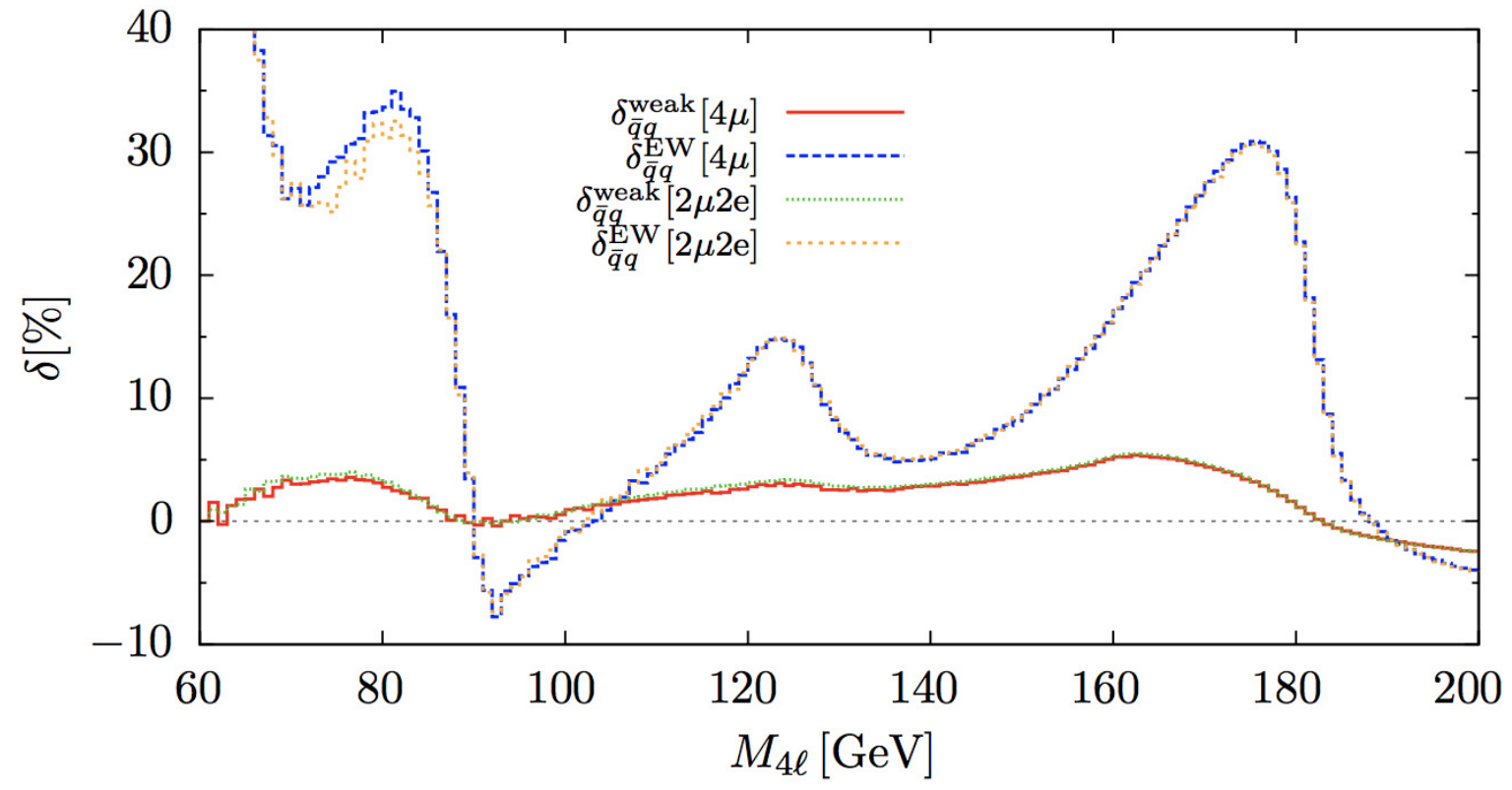

**Fig. 1:** Plot taken from Ref. [15] and adapted for this proceeding. In the text we refer to the blue long-dashed and yellow short-dashed lines.

in the case of bare muons.

FSR effects are particularly important in the case of the leptonic signatures of $ZZ$ production. In Ref. [15] both the $pp \to \mu^+\mu^-e^+e^-$ as well $pp \to \mu^+\mu^-\mu^+\mu^-$ processes have been considered and the distribution of the invariant mass of the four leptons $m(4\ell)$ exhibits different regions where NLO EW corrections are large due to FSR effects, as can be seen from Fig. 1. Similarly to the Drell–Yan case, just below $m(4\ell) = m_Z$ there is an enhancement due to the migration of events from the peak associated to kinematic configurations with only one on-shell $Z$ to smaller values of $m(4\ell)$. Analogously, the same effects is present below $m(4\ell) = 2m_Z$, which corresponds to the threshold for the production of two on-shell $Z$ bosons, and also at $m(4\ell) = m_Z + 2p_{T,\rm min}$, where $p_{T,\rm min}$ =15 GeV is the cut on the transverse momentum of each lepton that has been used in the calculation.

Additional studies have been performed in Ref. [16] for again the $pp \to \nu_\mu\mu^+e^-\bar{\nu}_e$ signature as well as the $pp \to e^+e^-\nu\bar{\nu}$ one, which can emerge from $WW$, $ZZ$ and $Z\gamma^*$ production. The impact of photon PDFs from different distributions is analysed and found to be in general non-negligible. Moreover, first attempts towards the matching of NLO EW corrections and photon showers are discussed. In particular, different approximations for the multiple emission of photons and their accuracy in reproducing hard radiation is scrutinised; reasonably accurate results are found.

NLO EW corrections have been calculated also for $WWW$ production [17]. The contribution of $\gamma q$ initial states is in general large, but its value strongly depends on the PDF set used in the calculation and it is reduced by imposing a jet veto.

### 2.3 Top quark pair production

Similarly to the $VV$ case, with stable top quarks FSR effects are absent. The relevance of the photon PDF for $t\bar{t}$ production has been discussed for the first time in Ref. [18] and carefully analysed in Ref. [19]. The LO cross section of $t\bar{t}$ production is $\mathcal{O}(\alpha_s^2)$; no photon-induced channel is present at this order. However, tree-level $\gamma g \to t\bar{t}$ contributions are present and contribute at $\mathcal{O}(\alpha_s\alpha)$, which we denote as LO EW. Moreover also $\gamma q \to t\bar{t}q$ contributions are present at $\mathcal{O}(\alpha_s^2\alpha)$, that is in the NLO EW corrections.

In Ref. [19] the contribution of the photon PDF has been found to be large for NNPDF2.3QED, especially at large $p_T(t)$ and $m(t\bar{t})$, with a sizable dependence on the definition of the factorization scale. The main contribution arises from the LO EW term, while the NLO EW part is in general small. Also at large values of the top-quark and top-quark-pair rapidity the impact of the photon PDF has been found to be non negligible and potentially measurable via normalized distributions for these variables. However, in Ref. [20], where a detailed study of $t\bar{t}$ distributions at NNLO QCD and NLO EW accuracy has been performed, it has been shown that photon-induced effects are negligible when LUXQED PDF set is used. The reason is due to the fact that no special kinematic enhancement is present in the $g\gamma$-initiated contribution and for this reason its size is completely PDF dependent, in particular very small for LUXQED as can be seen in Fig. 2.

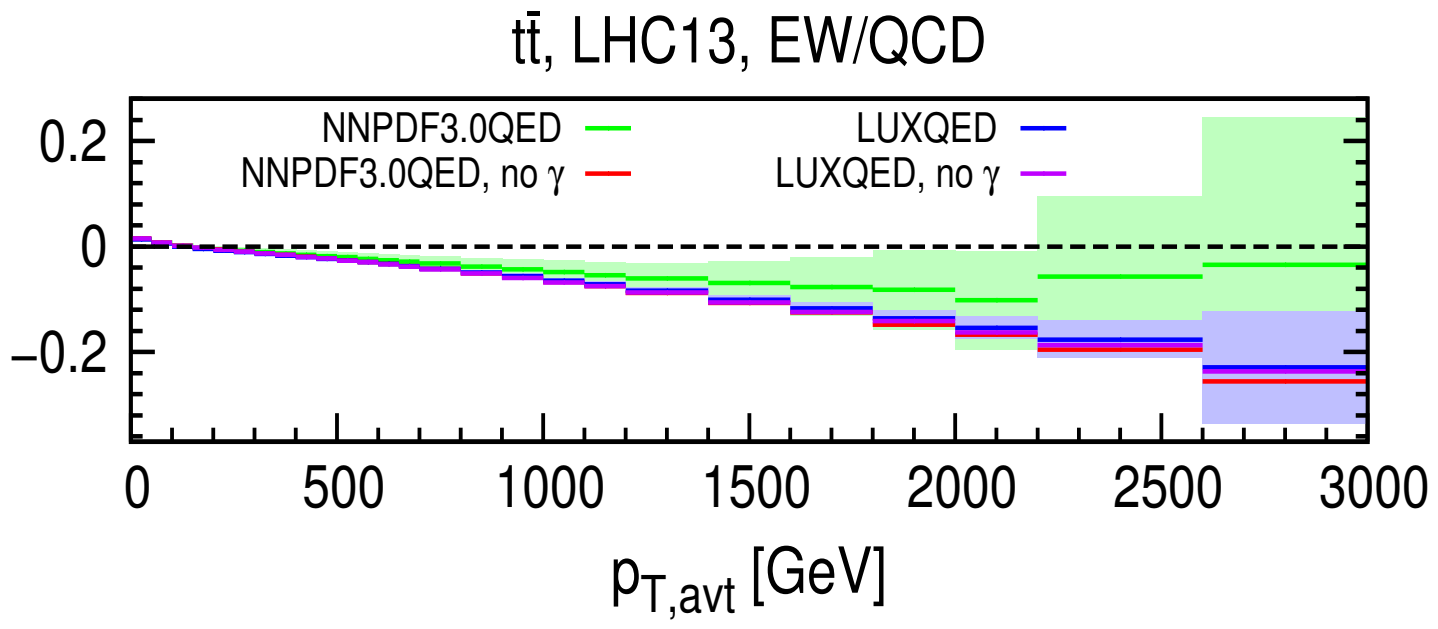

**Fig. 2:** Plot taken from Ref. [20]. The relative impact of EW corrections (LO EW, NLO EW and all the other fixed-order subdominant contributions) for the average transverse momentum of the top quarks at the 13 TeV LHC are displayed for four different cases. NNPDF3.0QED and LUXQED, both including or not the contribution from the photon PDF. Bands refer to the PDF uncertainties.

In Ref. [21], one of the dilepton signatures of $t\bar{t}$ production has been analysed, namely the $pp \to e^+\nu_e\mu^-\bar{\nu}_\mu b\bar{b}$ process. Similarly to what has already been discussed for the $Z$ resonance in Secs. 2.1 and 2.2, the reconstructed top mass $m(e^+\nu_e b)$ receives large corrections in the region below $m(e^+\nu_e b) = m_t$, due to FSR effects. Also, non-negligible photon-induced effects (with NNPDF photon density) have been observed for $m^2(e^+b) > m_t^2 - m_W^2$, a phase-space region that is not allowed with an on-shell top quark.

### 2.4 Dijet production

The first calculation including photon effects from NLO EW corrections to dijet production has been performed in Ref. [22]. As expected, photon PDF effects are important for large values of the inclusive $p_T(j)$, at least with the NNPDF parametrization. On the contrary, the impact of NLO EW corrections is in general modest. However, a few theoretical issues concerning jet and photon discrimination within the calculation of EW corrections have been addressed.

In this calculation not only the LO and NLO EW orders have been considered but also all the tree-level induced $\mathcal{O}(\alpha_s^{(2-i)}\alpha^i)$ with $0 \le i \le 2$ contributions and all the NLO contributions of $\mathcal{O}(\alpha_s^{(3-i)}\alpha^i)$ with $0 \le i \le 3$. At this level of accuracy, the only straightforward way for obtaining infrared-safe observables is the usage of democratic jets, i.e., the inclusion of photons as well leptons in the jet definitions. Nevertheless, one may want to identify the contributions of a processes with a single or two isolated photons to the dijet calculation with democratic-jet definition. While $\mathcal{O}(\alpha_s\alpha)$ and $\mathcal{O}(\alpha_s^2\alpha)$ term for single isolated-photon and $\mathcal{O}(\alpha^2)$ and $\mathcal{O}(\alpha_s\alpha^2)$ for double isolated photon can be calculated via the usage of Frixione isolation [23], subleading terms necessarily involve the usage of fragmentation functions, including the poorly known photon-to-photon one. Nevertheless, we evaluated the aformentioned contributions for which fragmentation functions are not necessary and we concluded that they represent a negligible fraction of the cross section with democratic jets, pointing to the fact that the definition of jets including photons is completely legitimate and in fact closer to the one used in experimental analyses.

# Mixed QCD-EW corrections to Drell–Yan processes

*Stefan Dittmaier*[1], *Alexander Huss*[2], *and Christian Schwinn*[3]
[1] Albert-Ludwigs-Universität Freiburg, Physikalisches Institut, D-79104 Freiburg, Germany
[2] Institute for Theoretical Physics, ETH, CH-8093 Zürich, Switzerland
[3] Institute for Theoretical Particle Physics and Cosmology, RWTH Aachen University, D-52056 Aachen, Germany

**Abstract**
We review the status of NNLO QCD–electroweak corrections to W- and Z-boson production at hadron colliders. We outline the application of the pole approximation to compute the dominant corrections, which arise from the combination of the QCD corrections to the production with electroweak corrections to the decay of the W/Z boson. We compare these results to simpler approximations based on naive products of NLO QCD and electroweak correction factors or leading-logarithmic approximations for QED final-state radiation.

## 1 Introduction

The Drell–Yan (DY)-like production of W and Z bosons, $\mathrm{pp/p\bar{p}} \to V \to \mathrm{l_1\bar{l}_2} + X$, is one of the most important class of "standard-candle" processes at hadron colliders and allows for precision tests of the Standard Model, as highlighted by the first LHC measurement of the W-boson mass $M_\mathrm{W}$ with an accuracy of 19 MeV [1], which may be reduced by a factor of two in the future. First LHC measurements of the forward-backward asymmetry in Z-production have appeared [2] that allow to extract the effective weak mixing angle, where the ultimate precision of the LHC might be competitive with that of LEP.

The sophisticated level of the current theoretical description of the DY-like production of W or Z bosons is reviewed in Ref. [3], where further references can be found. QCD corrections at next-to-next-to-leading-order (NNLO) accuracy are combined with higher-order soft-gluon effects through analytic resummation or matching to QCD parton showers. Electroweak (EW) corrections at NLO are supplemented with leading multi-photon final-state corrections and universal higher-order weak corrections. Compared to QCD corrections, the EW corrections lead to new features such as loop diagrams connecting initial and final states, the need for a consistent treatment of the finite vector-boson width, and the appearance of photon-induced processes. The NLO QCD and EW corrections are shown in Fig. 1 for the distributions of the transverse mass of the lepton pair ($M_\mathrm{T,\nu l}$), and the lepton transverse momentum ($p_\mathrm{T,l}$) in W production and the lepton-invariant-mass ($M_\mathrm{ll}$) and $p_\mathrm{T,l}$ for Z production. The EW corrections significantly distort the distributions near the Jacobian peaks at $M_\mathrm{T,\nu l} \approx M_\mathrm{W}$ and $p_\mathrm{T,l} \approx M_\mathrm{V}/2$ and lead to large corrections in the invariant-mass spectrum in Z production due to photonic final-state corrections shifting the reconstructed value of $M_\mathrm{ll}$ away from the resonance $M_\mathrm{ll} = M_\mathrm{Z}$ to lower values. While the QCD corrections are moderate for the $M_\mathrm{T,\nu l}$ and $M_\mathrm{ll}$ distributions, they become extremely large for the $p_\mathrm{T,l}$ distributions above threshold due to the recoil of the vector boson against the real emission of a jet, and require all-order soft-gluon resummation for a consistent description.

The next challenges to improve fixed-order predictions for the DY processes are given by the N$^3$LO QCD corrections and the mixed NNLO QCD–EW corrections of $\mathcal{O}(\alpha_\mathrm{s}\alpha)$. In this contribution we provide an overview of the relevance and the current status of the $\mathcal{O}(\alpha_\mathrm{s}\alpha)$ corrections and outline our recent computation of the dominant corrections at this order for resonant vector-boson production using the so-called *pole approximation* [4, 5]. We compare these results to naive multiplicative combinations of NLO QCD and EW corrections, and to leading-logarithmic approximations of photon radiation.

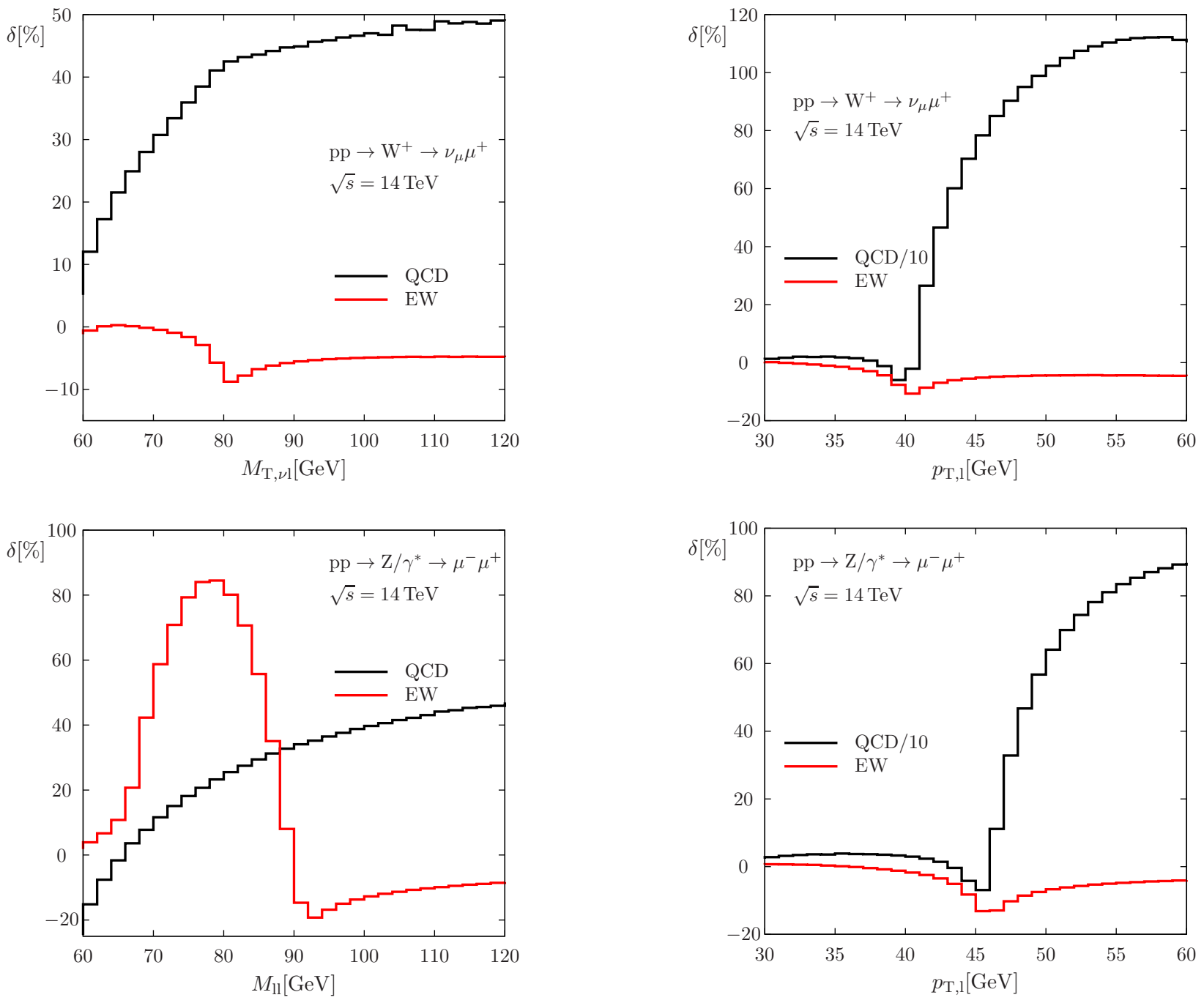

**Fig. 1:** NLO QCD and EW corrections relative to LO. Above: distributions in the transverse-mass (left) and transverse-lepton-momentum (right) for $W^+$ production at the LHC. Below: distributions in the invariant-mass (left) and transverse-lepton-momentum (right) for Z production at the LHC. (Taken from Ref. [4].)

## 2 Mixed QCD–EW corrections

The mixed QCD–EW NNLO corrections are expected to be particularly relevant in two regimes: First, at large invariant masses of the lepton pair the EW corrections are enhanced by so-called Sudakov logarithms, and the size of the $\mathcal{O}(\alpha_s\alpha)$ effects can be estimated to exceed the scale uncertainty of the NNLO QCD result [6]. On the other hand, observables for precision measurements dominated by the vector-boson resonance can show a percent-level sensitivity to the $\mathcal{O}(\alpha_s\alpha)$ corrections, resulting e.g. in an effect on the $M_W$ determination of about 15 MeV [5, 7]. Therefore these corrections must be brought under theoretical control to match the precision goals of the LHC. Efforts are being made towards a full NNLO computation of the $\mathcal{O}(\alpha_s\alpha)$ corrections, which involves complicated multi-scale two-loop integrals [8] and requires a method for the cancellation of infra-red singularities to combine the two-loop corrections with the $\mathcal{O}(\alpha)$ EW corrections to $W/Z+\text{jet}$ production, the $\mathcal{O}(\alpha_s)$ QCD corrections to $W/Z+\gamma$ production (see references in Ref. [5]), and the double-real corrections [9]. Awaiting the completion of these computations, the impact of the $\mathcal{O}(\alpha_s\alpha)$ corrections can be estimated by a naive multiplicative combination of the NLO QCD and EW corrections. In a more sophisticated approach, the fixed-order NLO QCD and EW corrections are matched to a QCD parton shower and a generator for final-state photon radiation (FSR) so that the virtual NLO corrections and the first emitted photon or gluon are treated exactly, while further emissions are generated in the collinear approximation. A careful treatment of the vector-boson resonance is required in order not to introduce spurious effects at $\mathcal{O}(\alpha_s\alpha)$ [7, 10].

### 2.1 Dominant mixed QCD–EW corrections in the pole approximation

A well-established method for the calculation of the EW corrections to precision observables dominated by the production of a resonant W or Z boson is provided by the pole approximation (PA). The PA is based on a systematic expansion of the cross section about the pole of the gauge-boson resonance and splits the corrections into factorizable and non-factorizable contributions. The former can be separately attributed to the production and decay of the gauge boson, while the latter link the production and decay subprocesses by the exchange of soft photons. The application of the PA to the NLO EW corrections shows agreement with the known full result up to fractions of $1\%$ near the resonance [4, 11]. Motivated by this quality of the PA at NLO, in Refs. [4, 5] we have extended this method to the calculation of the $\mathcal{O}(\alpha_s\alpha)$ corrections in the resonance region, which are classified into the four types of contributions shown in Fig. 2 for the case of the double-virtual corrections:[1]

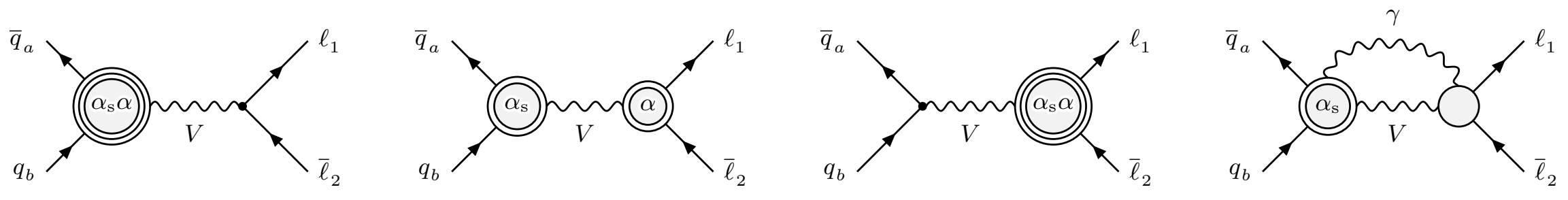

**Fig. 2:** Contributions to the mixed NNLO QCD–EW corrections in the PA illustrated by generic two-loop amplitudes: factorizable corrections of initial–initial (i-i), initial–final (i-f), and final–final type (f-f), and non-factorizable corrections (nf). Simple circles symbolize tree structures, double (triple) circles one-loop (two-loop) corrections.

(i-i) The initial–initial factorizable corrections are given by two-loop $\mathcal{O}(\alpha_s\alpha)$ corrections to on-shell W/Z production and the corresponding one-loop real–virtual and tree-level double-real contributions. Results for individual ingredients are known, but a consistent combination using a subtraction scheme for infrared singularities at $\mathcal{O}(\alpha_s\alpha)$ has not been performed yet.

(i-f) The factorizable initial–final corrections consist of the $\mathcal{O}(\alpha_s)$ corrections to W/Z production combined with the $\mathcal{O}(\alpha)$ corrections to the leptonic W/Z decay and provide the numerically dominant contribution. The main results of their computation [5] are presented below.

(f-f) Factorizable final–final corrections arise from the $\mathcal{O}(\alpha_s\alpha)$ counterterms of the lepton–W/Z-vertices. They yield a relative correction below $0.1\%$ [5], so that they are phenomenologically negligible.

(nf) The non-factorizable $\mathcal{O}(\alpha_s\alpha)$ corrections are given by soft-photon corrections connecting the initial state, the intermediate vector boson, and the final-state leptons, combined with QCD corrections to W/Z-boson production. Their numerical effect is below $0.1\%$ [4], so that for phenomenological purposes the $\mathcal{O}(\alpha_s\alpha)$ corrections can be factorized into terms associated with initial-state and/or final-state corrections and their combination.

The (i-i)-contributions are the only currently missing $\mathcal{O}(\alpha_s\alpha)$ corrections within the PA. Results of the PA at $\mathcal{O}(\alpha)$ show that observables such as the $M_{\mathrm{T},\nu l}$ distribution for W production or the $M_{\mathrm{ll}}$ distributions for Z production are extremely insensitive to photonic initial-state radiation (ISR) [4] and also do not receive overwhelmingly large QCD corrections. Therefore we do not expect significant initial–initial NNLO $\mathcal{O}(\alpha_s\alpha)$ corrections to such distributions. On the other hand, we expect class (i-f) to capture the dominant $\mathcal{O}(\alpha_s\alpha)$ effects, since it combines two types of corrections that are sizeable at NLO and deform the shape of differential distributions. Therefore our default prediction for $\sigma^{\mathrm{NNLO}_{s\otimes ew}}$ is given by the sum of the factorizable (i-f) corrections, $\Delta\sigma^{\mathrm{NNLO}_{s\otimes ew}}_{\mathrm{prod}\times\mathrm{dec}}$, and the full NLO QCD and EW corrections, $\Delta\sigma^{\mathrm{NLO}_s} + \Delta\sigma^{\mathrm{NLO}_{ew}}$. All contributions are consistently evaluated with NLO PDFs.

Figure 3 shows the numerical results for the relative $\mathcal{O}(\alpha_s\alpha)$ corrections for the $M_{\mathrm{T},\nu l}$ and the $p_{\mathrm{T,l}}$ distributions for $\mathrm{W}^+$ production at the LHC. For Z production, the results for the $M_{\mathrm{ll}}$ distribution and a transverse-lepton-momentum ($p_{\mathrm{T,l^+}}$) distribution are displayed in Figure 4. We consider isolated ("bare") muons using the setup and input parameters of Ref. [5]. The corresponding corrections for "dressed leptons" recombined with collinear photons are similar, but typically smaller by a factor of two [5]. The figures show results for the following approximations:

[1] For each class of contributions apart from the (f-f) corrections, also the associated real–virtual and double-real corrections have to be computed, obtained by replacing one or both of the labels $\alpha$ and $\alpha_s$ in the blobs in Fig. 2 by a real photon or gluon, respectively, and including corresponding crossed partonic channels, e.g. with quark–gluon initial states.

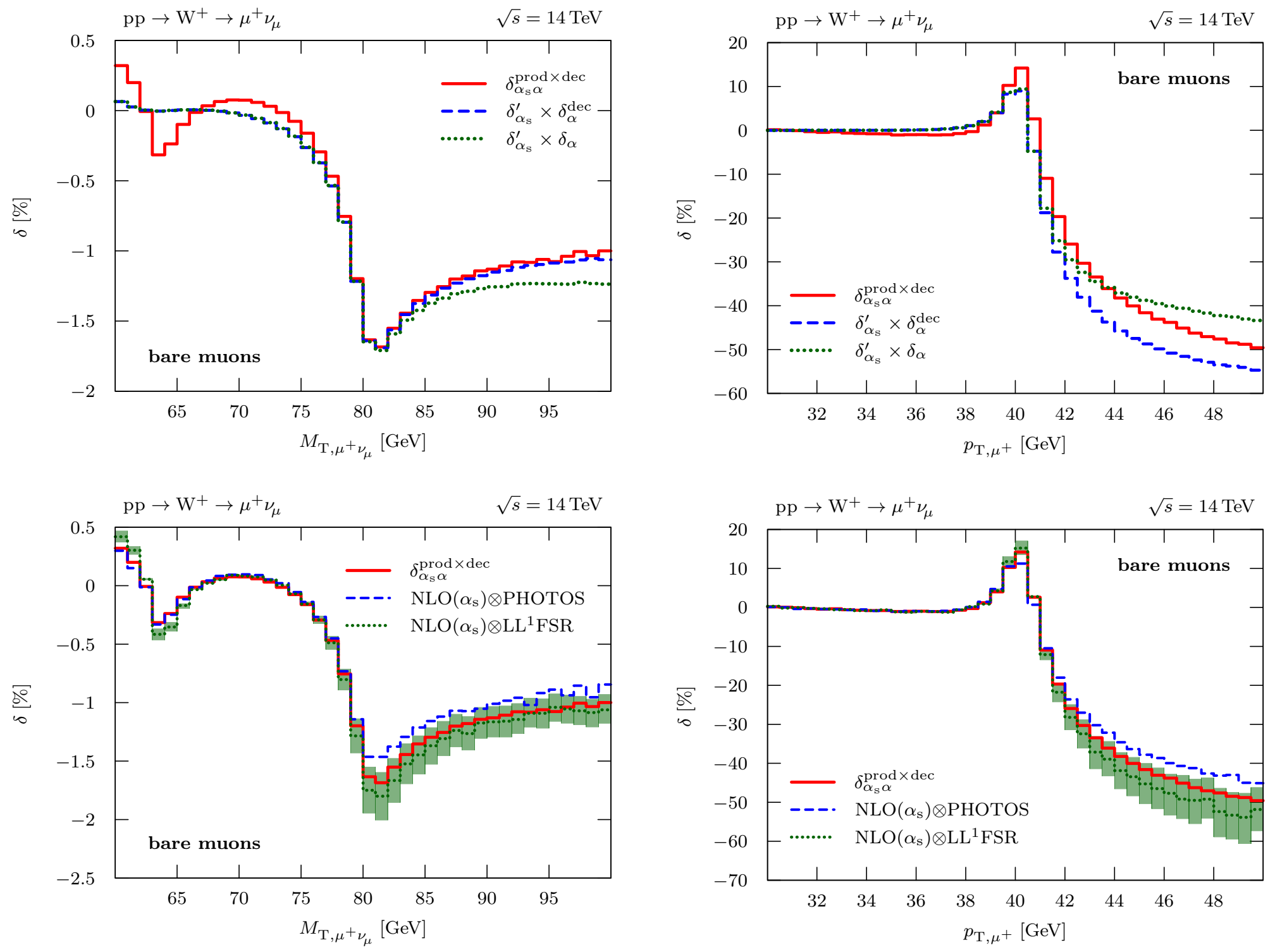

**Fig. 3:** Relative corrections of $\mathcal{O}(\alpha_s\alpha)$ induced by factorizable initial-state QCD and final-state EW contributions for the $M_{T,\nu l}$ (left) and $p_{T,l}$ (right) distributions for $W^+$ production at the LHC. The default prediction $\delta^{\mathrm{prod}\times\mathrm{dec}}_{\alpha_s\alpha}$ is compared to the naive products of the NLO correction factors $\delta'_{\alpha_s}$ and $\delta_\alpha$ (upper plots) and the combination of initial-state QCD with photonic FSR from PHOTOS or the structure-function ($\mathrm{LL}^1\mathrm{FSR}$) approach (lower plots). (Taken from Ref. [5].)

**Pole approximation:** Our default prediction of the (i-f) corrections, $\delta^{\mathrm{prod}\times\mathrm{dec}}_{\alpha_s\alpha} \equiv \Delta\sigma^{\mathrm{NNLO}_{s\otimes\mathrm{ew}}}_{\mathrm{prod}\times\mathrm{dec}}/\sigma^{\mathrm{LO}}$.

**Naive products:** The product $\delta'_{\alpha_s}\delta_\alpha$ of the QCD and EW correction factors,

$$\delta'_{\alpha_s} \equiv \Delta\sigma^{\mathrm{NLO}_s}/\sigma^{\mathrm{LO}}, \qquad \delta_\alpha \equiv \Delta\sigma^{\mathrm{NLO}_{\mathrm{ew}}}/\sigma^0. \tag{1}$$

Note that the LO prediction $\sigma^{\mathrm{LO}}$ ($\sigma^0$) is evaluated with LO (NLO) PDFs. The NLO EW corrections are defined in two different versions: based on the full $\mathcal{O}(\alpha)$ correction ($\delta_\alpha$), and on the dominant EW final-state correction of the PA ($\delta^{\mathrm{dec}}_\alpha$).

**LL-FSR:** The full NLO QCD corrections to W/Z production are combined with a leading-logarithmic (LL) approximation for FSR obtained using structure-functions [12] or PHOTOS [13]. To obtain the strict $\mathcal{O}(\alpha_s\alpha)$ corrections, only a single photon emission is generated in the LL approximation.

The difference of the two naive product versions gives an error-estimate of the PA, in particular of the missing (i-i) corrections. Deviations between our default prediction $\delta^{\mathrm{prod}\times\mathrm{dec}}_{\alpha_s\alpha}$ and the product approximations can be attributed to the double-real corrections, which do not simply factorize into a product of two NLO factors due to the interplay of recoil effects from jet and photon emission [5]. In contrast, both FSR approximations take this effect properly into account, but neglect subdominant finite contributions.

For the $M_{T,\nu l}$ distribution for $W^+$ production (left plots in Figure 3), the $\mathcal{O}(\alpha_s\alpha)$ corrections amount to $\approx -1.7\%$ around the resonance, which is about an order of magnitude smaller than the NLO EW corrections. Both variants of the naive product provide a good approximation to the full result in the region around and below the Jacobian peak, which is dominated by resonant W production. This is consistent with the insensitivity of $M_{T,\nu l}$ to photonic ISR already seen at NLO [4]. For larger $M_{T,\nu l}$, the product $\delta'_{\alpha_s}\delta_\alpha$ using the full NLO EW correction factor deviates from the other curves, which signals that effects beyond the PA become more important, but remain at the per-mille level for $M_{\mathrm{T},\nu l} \lesssim 90$ GeV. The two FSR approximations show good agreement and improve over the naive product approximations. For the structure-function approach (denoted by LL$^1$FSR), the intrinsic uncertainty of the LL approximation is illustrated by the band width resulting from varying the QED scale $Q$ within the range $M_V/2 < Q < 2M_V$ for $V = \mathrm{W, Z}$.

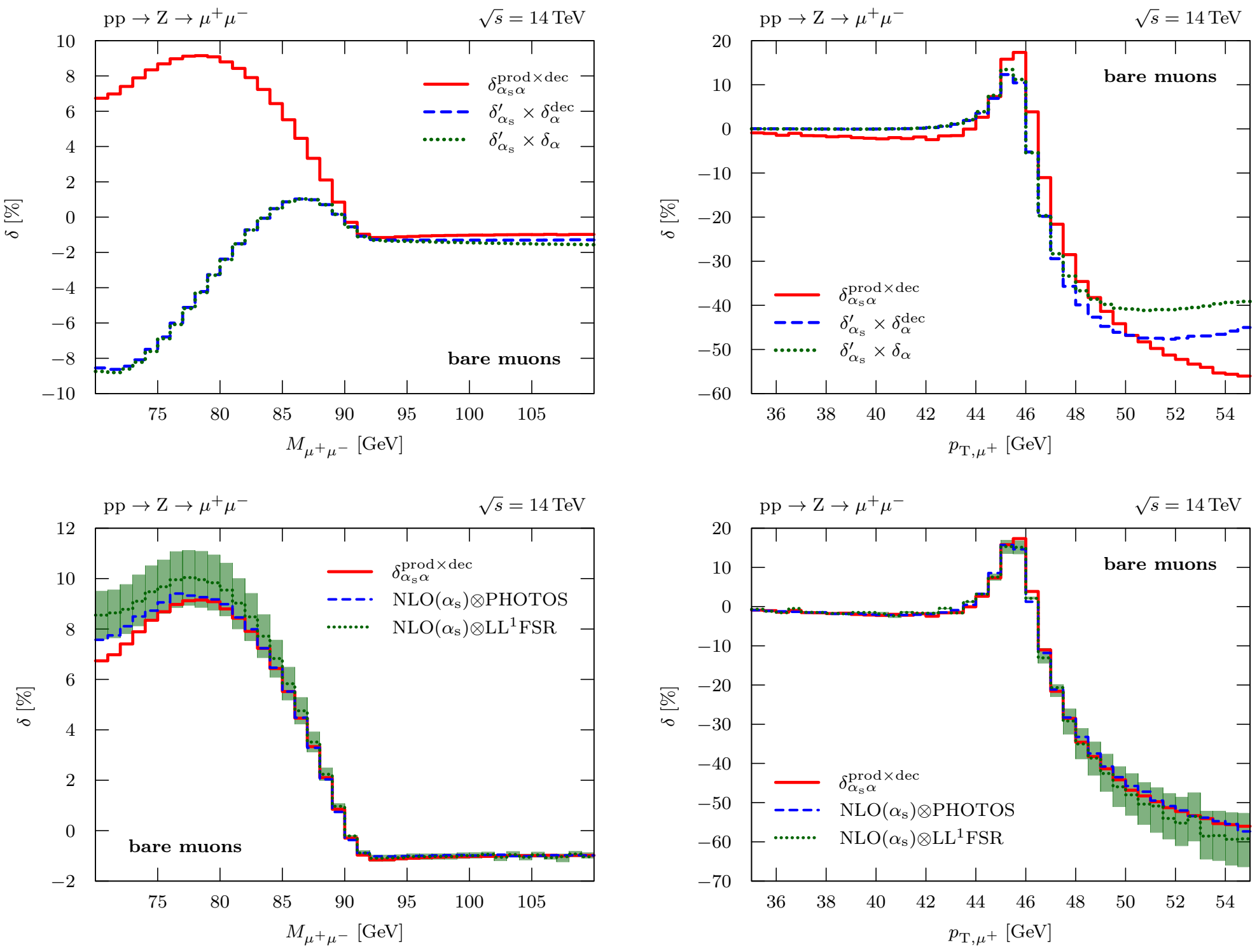

**Fig. 4:** As Fig. 3 but for the lepton-invariant-mass distribution (left) and a transverse-lepton-momentum distribution (right) for Z production at the LHC. (Taken from Ref. [5].)

The corrections to the $p_{\mathrm{T,l}}$ distributions (right plots in Figures 3 and 4) are small far below the Jacobian peak, but rise to about 15% (20%) on the Jacobian peak at $p_{\mathrm{T,l}} \approx M_\mathrm{V}/2$ for the case of the $\mathrm{W}^+$ boson (Z boson) and then drop to $-50\%$. As in the NLO QCD results of Fig. 1, the large corrections above the Jacobian peak arise since recoil due to real QCD radiation shifts events with resonant W/Z bosons above the Jacobian peak. This effect implies a larger impact of the double-real emission corrections, which are not captured correctly by the naive product ansatz that deviates from the full result $\delta^{\mathrm{prod}\times\mathrm{dec}}_{\alpha_s\alpha}$ by 5–10% at the Jacobian peak, where the PA is expected to be the most accurate. The differences of the two versions of the naive products furthermore indicates the potential impact of the missing $\mathcal{O}(\alpha_s\alpha)$ (i-i) corrections.[2] The description of the $p_{\mathrm{T,l}}$ distributions is improved by the combination of the full NLO QCD corrections with LL photon emission, but some differences remain for W-production.

In the $M_{\mathrm{ll}}$ distribution for Z production (left plots in Figure 4), corrections up to 10% are observed below the resonance. This is consistent with the large NLO EW corrections from photonic final-state radiation (FSR) seen in Figure 1. The naive products $\delta'_{\alpha_s}\delta^{(\mathrm{dec})}_\alpha$ approximate the full (i-f) corrections $\delta^{\mathrm{prod}\times\mathrm{dec}}_{\alpha_s\alpha}$ reasonably well for $M_{\mathrm{ll}} \geq M_\mathrm{Z}$ but have the wrong sign already slightly below the resonance. The reason for this failure is that the appropriate QCD correction factor for the events that are shifted below the resonance by photonic FSR is given by its value at the resonance, whereas the naive product ansatz simply multiplies the corrections locally. In contrast, the two FSR approximations model the $M_{ll}$ distribution correctly within their uncertainty.

[2] These deviations should be interpreted with care, since the peak region $p_{\mathrm{T,l}} \approx M_\mathrm{V}/2$ corresponds to the kinematic onset for $V$ + jet production where fixed-order predictions break down and QCD resummation is required for a proper description.

## 3 Conclusions

The precision-physics program in Drell–Yan-like W- and Z-boson production at the LHC requires a further increase in the accuracy of the theoretical predictions, where the mixed QCD–electroweak corrections of $\mathcal{O}(\alpha_s\alpha)$ represent the largest component of fixed-order radiative corrections after the well established NNLO QCD and NLO electroweak corrections. In this contribution, we have reviewed the construction of the pole approximation for evaluating the $\mathcal{O}(\alpha_s\alpha)$ corrections to Drell–Yan processes [4], and summarized our numerical results [5] for the dominant factorizable corrections, which arise from the combination of sizeable QCD corrections to the production with large EW corrections to the decay subprocesses. Naive product approximations fail to capture these corrections in distributions that are sensitive to QCD initial-state radiation and therefore require a correct treatment of the double-real-emission part of the NNLO corrections. Naive products also fail to capture observables that are strongly affected by a redistribution of events due to final-state real-emission corrections, such as the invariant-mass distribution in Z production. A combination of the NLO QCD corrections and a collinear approximation of real-photon emission through a QED structure-function approach or a QED parton shower such as PHOTOS provides a significantly better agreement with our results. In order to reduce ambiguities due to the leading-logarithmic accuracy of these approaches, a consistent matching to the full NLO EW correction is mandatory, as emphasized recently also in Ref. [7].

## Acknowledgements

This project is supported by the German Research Foundation (DFG) via grant DI 784/2-1 and the German Federal Ministry for Education and Research (BMBF). A.H. is supported via the Swiss National Foundation (SNF) under contract CRSII2-160814 and C.S. is supported by the Heisenberg Programme of the DFG.

# Photon+$V$ measurements in ATLAS

*D.V. Krasnopevtsev on behalf of ATLAS collaboration*
National Research Nuclear University MEPhI, Moscow, Russia

**Abstract**
ATLAS measurements of multi-boson production processes involving isolated photons in proton–proton collisions at 8 TeV are summarized. Standard Model cross sections are measured with high precision and are compared to theoretical predictions. No deviations from Standard Model predictions are observed and limits are placed on parameters used to describe anomalous triple and quartic gauge boson couplings.

## 1 Introduction

Photon+$V$ ($V = W, Z$) studies are used to test the electroweak sector of the Standard Model (SM) with high accuracy using multi-boson production cross sections. These measurements allow to probe the $SU(2)_L \times U(1)_Y$ gauge symmetry of the electroweak theory that determines the structure and self-couplings of the vector bosons and to search for signs of new physics using anomalous triple and quartic gauge-boson coupling (aTGC and aQGC) studies. Moreover precision photon+$V$ measurements affect tuning of Monte-Carlo (MC), which describes some backgrounds in SM and beyond the SM studies.

The ATLAS detector [1] uses electromagnetic calorimeter (EM) and inner detector (ID) tracking system to reconstruct photons with high efficiencies. Both photons that do or do not convert to electron-positron pairs are reconstructed in ATLAS. The EM calorimeter is composed of three sampling layers longitudinal in shower depth. The fine $\eta$ granularity of the strips in the first layer is sufficient to provide discrimination between single photon showers and two overlapping showers coming from the decays of neutral hadrons in jets. The jet suppression is about $10^4$ during the first operation period of Large Hadron Collider (LHC) [2].

The measurements in this paper use 20.3 fb$^{-1}$ of proton–proton collisions collected with the ATLAS detector during LHC operation at a center-of-mass energy of 8 TeV. Standard Model cross sections for $V\gamma$ production in ATLAS were measured with high accuracy already with 7 TeV data. Proton-proton collisions at 8 TeV allowed to improve these measurements and provided an opportunity to study rare processes of $Z\gamma$ scattering and triboson productions.

## 2 Standard Model Photon+$V$ measurements

### 2.1 $Z\gamma(\gamma)$ production

$Z\gamma(\gamma)$ production includes electroweak (EWK) and quantum chromodynamics (QCD) components. QCD component is studied using charged and neutral $Z$ decays with initial and final state radiation photons: $Z/\gamma^* \to \ell^+\ell^-\gamma(\gamma)$ (where $\ell$ = e or $\mu$) and $Z \to \nu\bar{\nu}\gamma(\gamma)$. The measurements are compared to SM predictions obtained with a parton-shower MC simulation and with two higher-order perturbative parton-level calculations at next-to-leading order (NLO) [3] and next-to-next-to-leading order (NNLO) [4].

The $Z/\gamma^*$ decays to charged leptons are selected using high transverse momentum ($p_\mathrm{T}$) electrons or muons triggers and photons with transverse energy $E_\mathrm{T} > 15$ GeV. Events with $Z$ boson decays to neutrinos are selected using high $E_\mathrm{T}$ photon triggers: $E_\mathrm{T} > 130$ GeV for the single photon channel and $E_\mathrm{T} > 22$ GeV for the di-photon channel. In this analysis events with any number of jets are considered (inclusive case) as well as events with veto on jets with $p_\mathrm{T} > 30$ GeV (exclusive case). Selected $\ell^+\ell^-\gamma(\gamma)$ event candidates must contain exactly one pair of same-flavor, opposite-charge isolated leptons (electrons or muons) with invariant di-lepton mass ($m_{\ell^+\ell^-}$) greater than 40 GeV and at least one or two isolated photons. $Z\gamma(\gamma)$ events with neutrino $Z$ decay are selected by considering events with $E_{\mathrm{T}}^{\mathrm{miss}} > 100$ GeV (110 GeV for the di-photon channel) and at least one or two isolated photons. $E_{\mathrm{T}}^{\mathrm{miss}}$ is the absolute value of the vector of momentum imbalance in the transverse plane and is used to identify neutrinos in this study. The separation between the reconstructed photon (di-photon system) direction and missing energy vector in the transverse plane is required since in signal events the $Z$ boson should recoil against the photon(s). To suppress $W(\gamma)$+jets and $W\gamma(\gamma)$ backgrounds, events containing a muon or an electron are rejected.

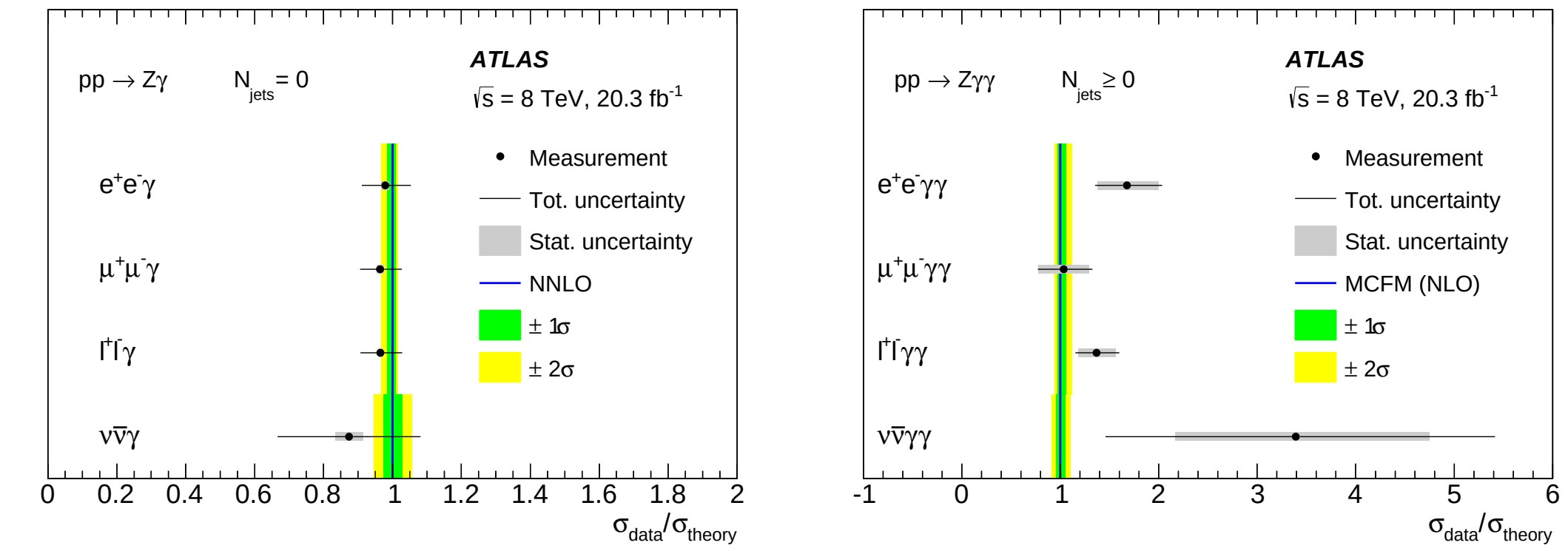

(a) (b)

**Fig. 1:** Comparison between the measured cross sections and the NNLO theory predictions in the exclusive region for the $pp \to Z\gamma$ (left) and NLO theory predictions in the inclusive region for the $pp \to Z\gamma\gamma$ (right) [5].

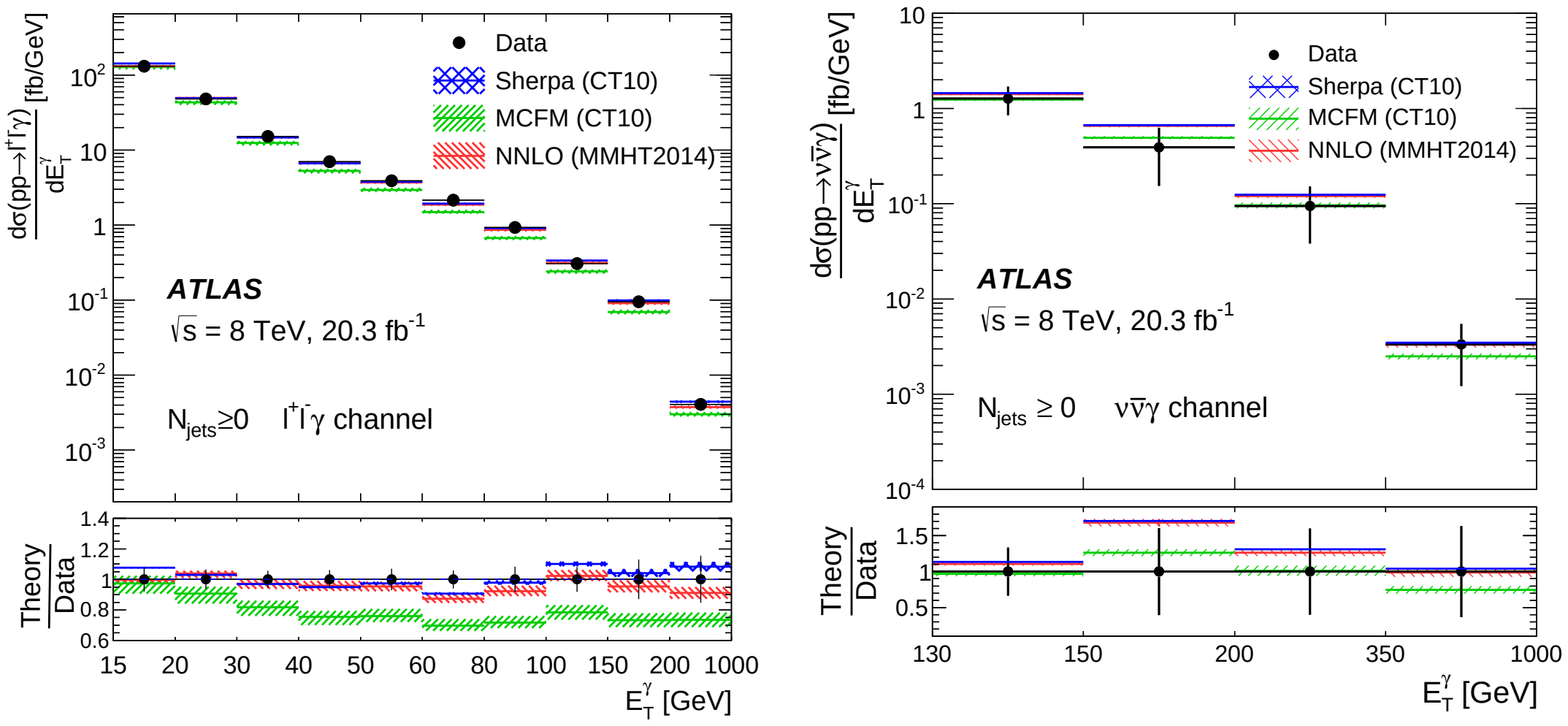

(a) (b)

**Fig. 2:** The measured (points with error bars) and expected differential cross sections as a function of $E_{\mathrm{T}}^{\gamma}$ for the pp→ $\ell^+\ell^-\gamma$ process (left) and the pp→ $\nu\bar{\nu}\gamma$ process (right) in the inclusive fiducial regions. The lower plots show the ratios of various theory predictions to data (shaded bands) [5].

The backgrounds in $Z\gamma(\gamma)$ signal regions are dominated by events in which jets and electrons are misidentified as photons. In neutrino channels additionally there is contribution from events where jet energy is not totally measured causing deviations in missing energy value. The largest background contributions are determined using data-driven techniques, smaller are obtained with a MC simulation.

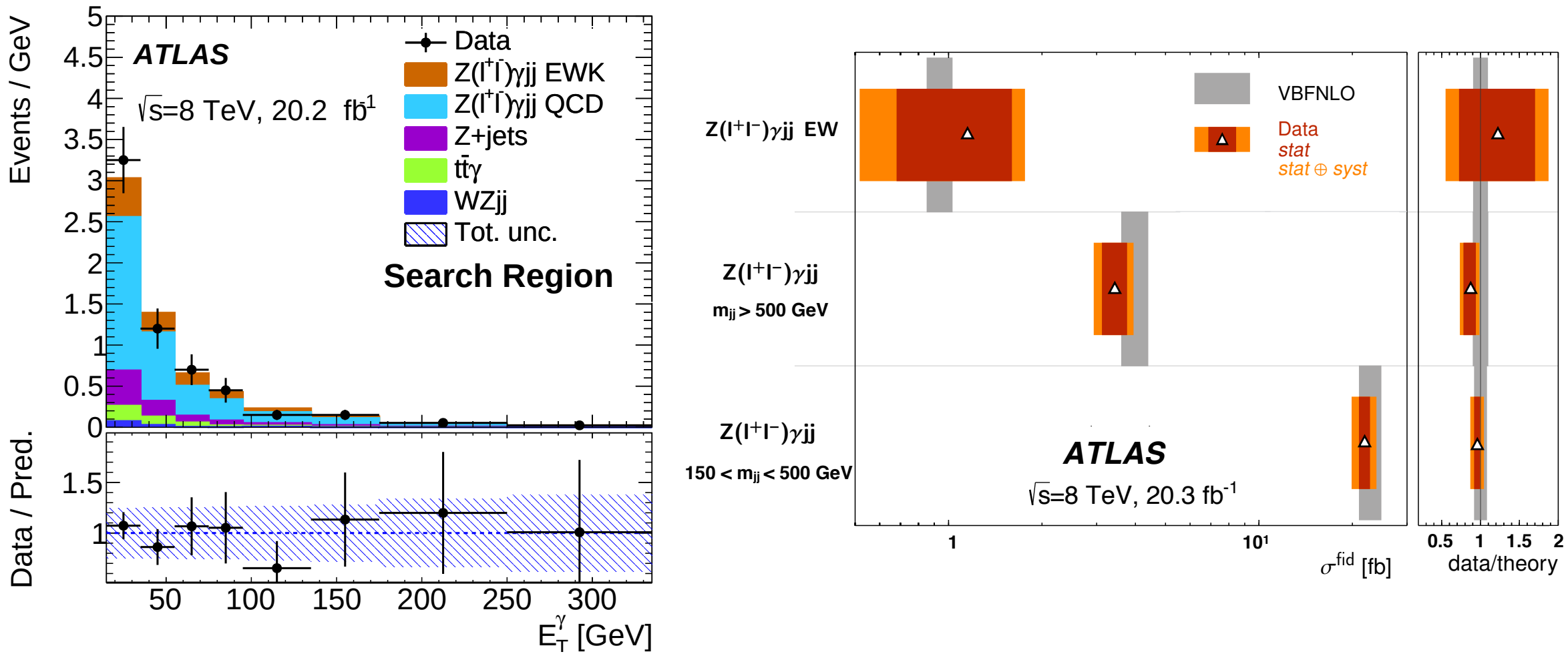

(a) (b)

**Fig. 3:** $Z\gamma$ centrality - (a) and comparison between the measured cross sections and NLO theory predictions for different control regions - (b) for $Z(\ell\ell)\gamma$ channel in $Z\gamma$ VBS analysis [6].

The measured cross sections are compared to the SM predictions. There is generally a good agreement between the measured cross sections in the $Z\gamma$ channels and those predicted by the SM. NNLO calculation of the inclusive cross section for the $Z(\ell^+\ell^-)\gamma$ channel is in a better agreement with the measurement than the NLO calculation. The results for both charged and neutrino $Z\gamma\gamma$ channels are compared to the NLO MCFM predictions. Although the observed number of events is not large, the combined $\ell^+\ell^-\gamma\gamma$ results show 5 standard deviations over non-signal hypothesis. Figure 1 shows the level of agreement between the experimental results and the theory for all channels.

Differential cross sections measurements are performed in $Z\gamma$ channels to remove measurement inefficiencies and resolution effects from the observed distributions. Figure 2 shows the measured and the predicted by the SM $E_T^\gamma$ distributions for $pp \to \ell^+\ell^-\gamma$ and $pp \to \nu\bar{\nu}\gamma$ events with the inclusive selection. Better agreement is found with NNLO predictions.

### 2.2 $Z\gamma$ scattering

EWK processes in $Z\gamma$ production are characterized by jets with wide rapidity separation, large di-jet invariant mass and production with high centrality. Vector boson scattering (VBS) belongs to EWK component and can be selected by presence of two high energetic jets in addition to bosons. Therefore only events with more than 1 jet are considered in this analysis. Jets are considered if they have $p_T > 30$ GeV and $|\eta| < 4.5$. Jet candidates are rejected if they are not well separated from the previously selected leptons and photons. Both charged and neutral decays of $Z$ are studied. $Z(\ell\ell)\gamma$ channel is used for cross section measurements and search for aQGCs, while $Z(\nu\nu)\gamma$ is used only in the search for aQGC. Figure 3(a) shows the distribution of centrality for observed $Z\gamma$ events in the search region for charged channel. Data (full dots) demonstrates reasonable agreement with background and signal predictions.

Signal events in charged channel should contain two same flavour and opposite sign leptons with $p_T^{e,\mu} > 25$ GeV, isolated photon with $E_T^\gamma > 15$, $m_{\ell\ell} > 40$ GeV and $m_{\ell\ell\gamma} + m_{\ell\ell\gamma} > 182$ GeV. Two regions are studied in the analysis. The search region with maximum VBS contribution and invariant di-jet mass ($m_{jj}$) greater than 500 GeV is aimed to measure EWK and EWK+QCD cross sections. The control region with maximum QCD contribution and $150 < m_{jj} < 500$ GeV is used to normalize QCD contribution in the search region and to perform EWK+QCD cross sections measurements. Figure 3(b) summarizes all cross section measurements made in this analysis. The left part of the plot shows exact values, while the right side illustrates data to theory ratio. The significance of the observed EWK production signal is $2\sigma$.

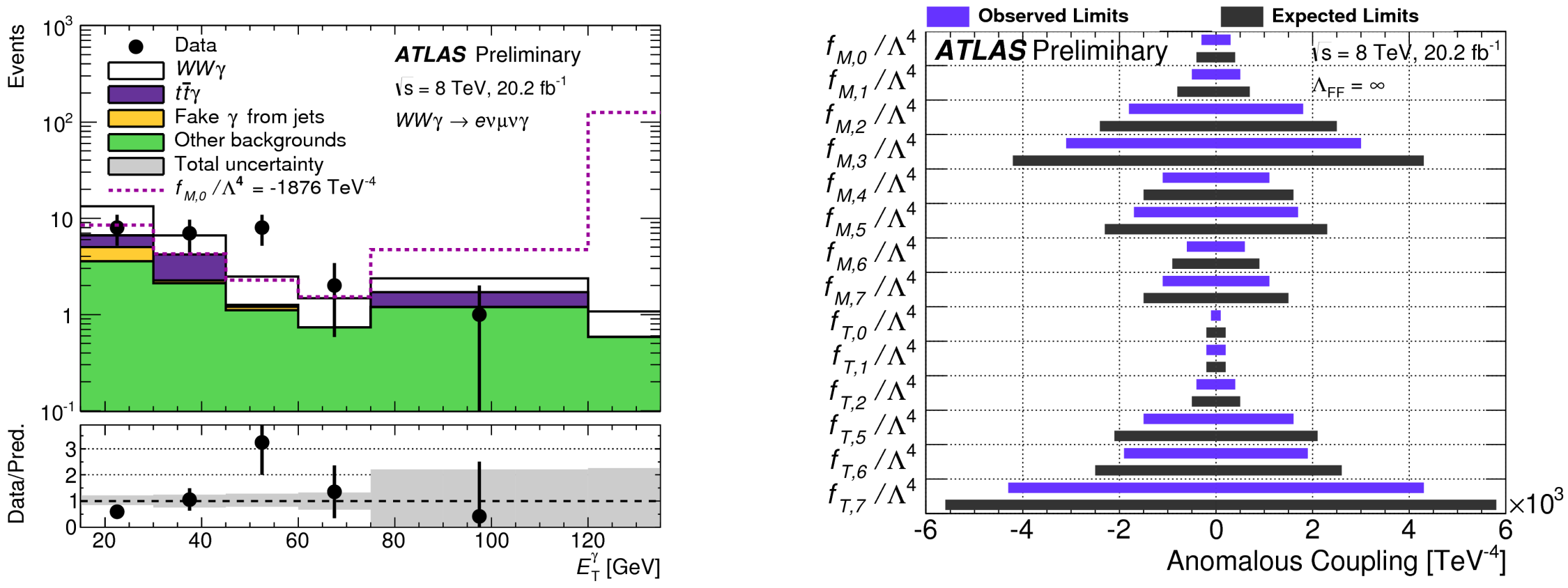

(a) (b)

**Fig. 4:** Transverse energy distributions for photons in $e\nu\mu\nu\gamma$ channel (left) and the observed from $WV\gamma$ analysis ununitarized limits for the studied aQGC parameters (right) [7]. The dashed purple line on left plot shows contribution from aQGC in case $f_{\mathrm{M0}}/\Lambda^4$ parameter is set to be -1876 TeV$^{-4}$.

| | | Observed limit [fb] | Expected limit [fb] | $\sigma_{\mathrm{theo}}$ [fb] |
|---|---|---|---|---|
| Fully leptonic | $e\nu\mu\nu\gamma$ | 3.7 | $2.1^{+0.9}_{-0.6}$ | 2.0 |
| Semileptonic | $e\nu jj\gamma$ | 10 | $16^{+6}_{-4}$ | 2.4 |
| | $\mu\nu jj\gamma$ | 8 | $10^{+4}_{-3}$ | 2.2 |
| | $\ell\nu jj\gamma$ | 6 | $8.4^{+3.4}_{-2.4}$ | 2.3 |

**Fig. 5:** Observed and expected cross-section upper limits at 95% CL for the different $WV\gamma$ final states. The expected cross-section limits are computed assuming no signal is present. The last column shows the theory prediction computed with the VBFNLO [7].

### 2.3 $WV\gamma$ production

$WV\gamma$ production is studied using fully leptonic $WW\gamma$ and and semi-leptonic $WV\gamma$ channels. Bosons in these channels originate from quartic or triple (with initial radiation photon) vertices or come from radiative processes.

$WW\gamma$ events are studied solely in the $e\nu\mu\nu\gamma$ final state since others fully-leptonic channels have low sensitivity because of large $Z\gamma$ backgrounds. The $e\nu\mu\nu\gamma$ event candidates are selected by considering events with $E_{\mathrm{T}}^{\mathrm{miss}} > 15$ GeV, one isolated photon with $E_{\mathrm{T}}^{\gamma} > 15$ GeV, one electron and one muon with $p_{\mathrm{T}} > 20$ GeV and $m_{\ell\ell} > 50$ GeV. Events with zero jets ($p_{\mathrm{T}}^{jet} > 25$ GeV) are only considered. Figure 4(a) shows transverse energy distribution for photon. The measured cross section has significance $1.4\sigma$ and is an agreement with NLO predictions by VBFNLO within statistical uncertainty.

Semi-leptonic $WV\gamma$ events with one leptonically decaying $W$ boson and one hadronically decaying $W$ or $Z$ boson are studied. Signal events are required to have $E_{\mathrm{T}}^{\mathrm{miss}} > 30$ GeV, one isolated photon with $E_{\mathrm{T}}^{\gamma} > 15$ GeV, one electron or one muon with $p_{\mathrm{T}} > 25$ GeV and $m_{\mathrm{T}} > 30$ GeV. Events with at least two jets (not b-jets) and $70 < m_{jj} < 100$ GeV are considered. Figure 5 shows upper exclusion limits on production cross-section for all studied channels. The observed limits are between 1.8 and 4.1 times larger than the SM cross-section.

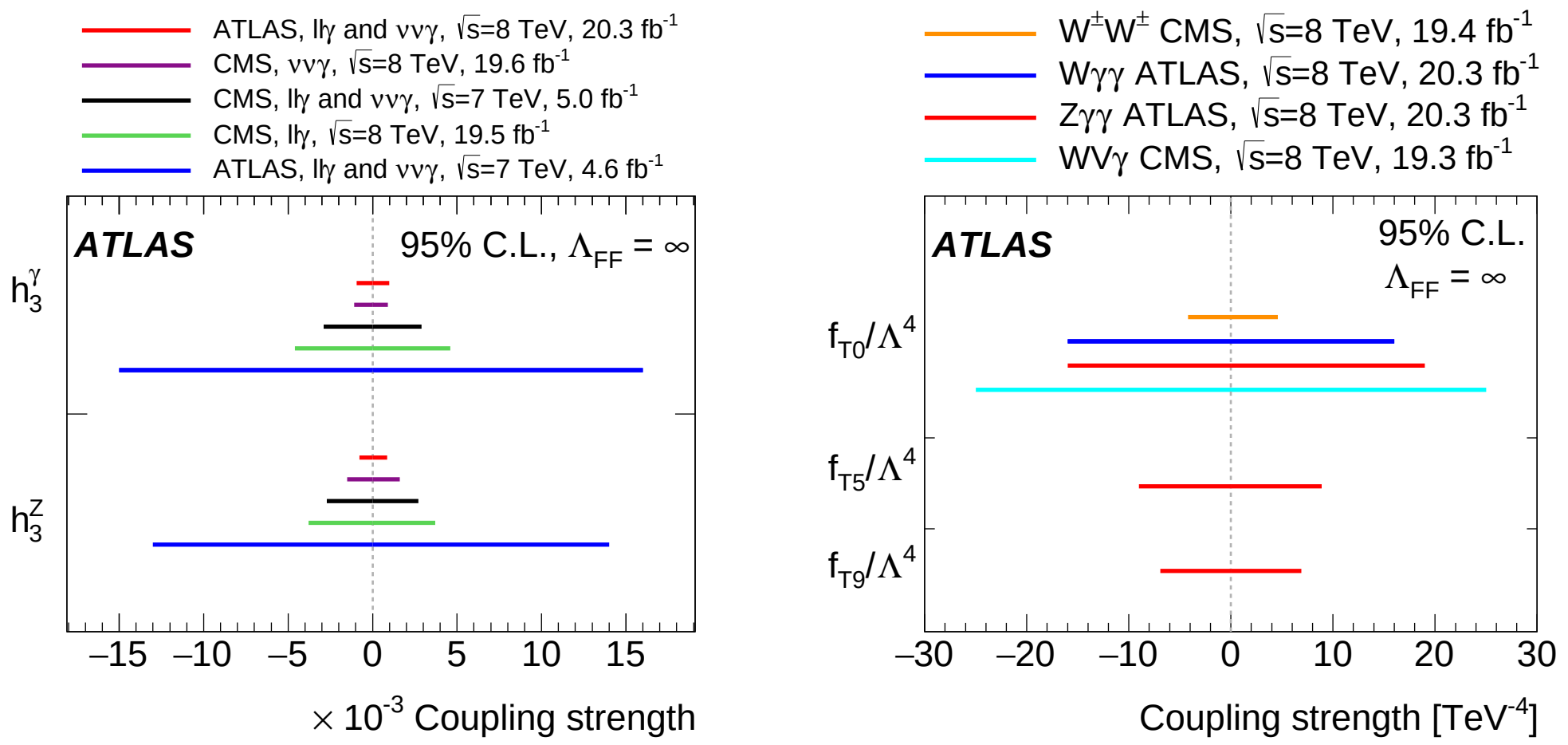

(a) (b)

**Fig. 6:** Comparison of the observed from $Z\gamma$ analysis ununitarized limits for the aTGC parameters ($h_3^\gamma$ and $h_3^Z$) with previous ATLAS and CMS results (left), and comparison of the observed from $Z\gamma\gamma$ analysis ununitarized limits for the aQGC parameters ($f_{\mathrm{T}0}/\Lambda^4$, $f_{\mathrm{T}5}/\Lambda^4$ and $f_{\mathrm{T}9}/\Lambda^4$) with previous ATLAS and CMS results (right) [5].

| | Limits 95% CL | Measured [TeV$^{-4}$] | Expected [TeV$^{-4}$] |
|---|---|---|---|
| ATLAS $Z(\to \ell\bar{\ell}/\nu\bar{\nu})\gamma$-EWK | $f_{T9}/\Lambda^4$ | $[-3.9, 3.9]$ | $[-2.7, 2.8]$ |
| | $f_{T8}/\Lambda^4$ | $[-1.8, 1.8]$ | $[-1.3, 1.3]$ |
| | $f_{T0}/\Lambda^4$ | $[-3.4, 2.9]$ | $[-3.0, 2.3]$ |
| | $f_{M0}/\Lambda^4$ | $[-76, 69]$ | $[-66, 58]$ |
| | $f_{M1}/\Lambda^4$ | $[-147, 150]$ | $[-123, 126]$ |
| | $f_{M2}/\Lambda^4$ | $[-27, 27]$ | $[-23, 23]$ |
| | $f_{M3}/\Lambda^4$ | $[-52, 52]$ | $[-43, 43]$ |
| CMS $Z(\to \ell\ell)\gamma$-EWK | $f_{T9}/\Lambda^4$ | $[-4.0, 4.0]$ | $[-6.0, 6.0]$ |
| | $f_{T8}/\Lambda^4$ | $[-1.8, 1.8]$ | $[-2.7, 2.7]$ |
| | $f_{T0}/\Lambda^4$ | $[-3.8, 3.4]$ | $[-5.1, 5.1]$ |
| | $f_{M0}/\Lambda^4$ | $[-71, 75]$ | $[-109, 111]$ |
| | $f_{M1}/\Lambda^4$ | $[-190, 182]$ | $[-281, 280]$ |
| | $f_{M2}/\Lambda^4$ | $[-32, 31]$ | $[-47, 47]$ |
| | $f_{M3}/\Lambda^4$ | $[-58, 59]$ | $[-87, 87]$ |
| CMS $W(\to \ell\nu)\gamma$-EWK | $f_{T0}/\Lambda^4$ | $[-5.4, 5.6]$ | $[-3.2, 3.4]$ |
| | $f_{M0}/\Lambda^4$ | $[-77, 74]$ | $[-47, 44]$ |
| | $f_{M1}/\Lambda^4$ | $[-125, 129]$ | $[-72, 79]$ |
| | $f_{M2}/\Lambda^4$ | $[-26, 26]$ | $[-16, 15]$ |
| | $f_{M3}/\Lambda^4$ | $[-43, 44]$ | $[-25, 27]$ |

**Fig. 7:** Comparison of the observed ununitarized limits for the aQGC parameters in the Z$\gamma$ VBS analysis with CMS results [6].

### 2.4 Anomalous triple and quartic gauge couplings

Vector-boson self-interactions are completely fixed by the model's $SU(2)_L \times U(1)_Y$ gauge structure. Their observation is thus a crucial test of the model. Anomalous coupling parameters can parameterize possible deviations from the SM at high $E_{\mathrm{T}}$ [8, 9].

Anomalous TGC in $Z\gamma$ channels, which vanish in the SM at tree level, are studied using vertex approach. The dominant contribution from aTGC is expected for region with high-$E_{\mathrm{T}}$ photons therefore the study is done in region with $E_T^\gamma > 250$ GeV for charged channels and with $E_{\mathrm{T}}^\gamma > 400$ GeV for neutrino channel. Having found no significant deviations from SM predictions, the data are used to set limits on anomalous triple couplings of photons to $Z$ bosons. Figure 6(a) shows the experimental ununitarized limits for studied aTGC parameters and comparison with CMS and previous ATLAS results.

Other analyses described in Section 2.1 used data to search for aQGC. An effective field theory with higher-dimensional operators is adopted to parameterize these anomalous couplings. Figure 4(b) summarizes expected and observed ununitarized limits for all studied parameters in $WV\gamma$ study. In this analysis first exclusion limits on the coupling parameters $f_{\mathrm{M4,M5}}$ and $f_{\mathrm{T6,T7}}$ are obtained. Neutral quartic couplings are not present in SM. Signal yields from $Z\gamma\gamma$ channels is used to set limits on parameters $f_{\mathrm{M2,M3}}$ and $f_{\mathrm{T0,T5,T9}}$. To maximize aQGC sensitivity in $Z\gamma\gamma$ channel similarly to $Z\gamma$ a region with high invariant mass of two photons ($m_{\gamma\gamma}$) is considered: $m_{\gamma\gamma}>$ 200 GeV for charged channel and $m_{\gamma\gamma}>$ 300 GeV for neutrino channel. Figure 6(b) shows corresponding ununitarized limits together with results obtained by other experiments. Limits from VBS analysis are shown on figure 7 and compared to CMS results. The neutrino channel provides best expected limits for all studied operators. Combination with charged channels improves results by 5-10%. ATLAS and CMS limits for aQGC parameters obtained using $Z\gamma$ scattering process are of the same order.

## 3 Conclusion

ATLAS provides great opportunities to study $V\gamma(\gamma)$ productions in high energy proton-proton collisions with previously unattainable accuracy. All measurements reported here demonstrate an agreement with the theory predictions. No evidence for physics beyond the SM is found in anomalous boson triple and quartic couplings at the level of statistics obtained at LHC during the first data-taking period.

## Acknowledgments

This work was performed within the framework of Nuclear Physics and Engineering Institute and supported by MEPhI Academic Excellence Project (contract 02.a03.21.0005 of 27.08.2013)

# Latest results on Higgs boson $\rightarrow \gamma\gamma$ in the CMS experiment

*Hamed BAKHSHIANSOHI*[*]
Centre for Cosmology, Particle Physics and Phenomenology (CP3), Université Catholique de Louvain, B-1348 Louvain-la-Neuve, Belgium

**Abstract**
The photonic decay of the Higgs boson is one of the most prominent decay modes for the observation and measurement of the Higgs boson properties, although its branching fraction is as low as 0.002. The CMS collaboration has analyzed 35.9fb$^{-1}$ of proton-proton collision data delivered by the LHC at a center of mass energy of 13 TeV. The inclusive results of the search in different Higgs boson production channels and the differential fiducial cross section measurement results are presented in this talk.

## 1 Introduction

The Higgs boson is the last discovered elementary particle predicted by the standard model of particles (SM). Since it was discovered by the CMS [1] and ATLAS experiments in 2012 [2, 3], many studies have been performed to understand its interaction with other elementary particles. So far all the measurements have been consistent with the standard model predictions.

As the photon is a massless particle, it does not couple to Higgs boson directly. The Higgs boson decays to two photons via a loop involving top-quarks or W-bosons and hence the branching fraction of the $H \rightarrow \gamma\gamma$ is as low as 0.002. Having a clean final state with an invariant mass peak that can be reconstructed with very high precision makes this rare decay of the Higgs boson very important.

The CMS experiment recorded 35.9 fb$^{-1}$ of proton-proton collision data, to be used for physics analysis, delivered by the LHC at a center of mass energy of 13 TeV in 2016. This data has been analyzed to measure the cross section for different Higgs boson production mechanisms, where the Higgs boson decays to two photons. Results have been documented in Ref.[4]. A dedicated analysis was also performed to measure the differential and fiducial production cross section of the Higgs boson in the photonic decay mode and the results have been documented in Ref.[5]. Both of these analyses are reviewed in this report.

## 2 Event reconstruction

Within the superconducting solenoid which is the central feature of the CMS experiment, silicon pixel and strip trackers, a lead tungstate crystal electromagnetic calorimeter (ECAL), and a brass and scintillator hadron calorimeter (HCAL), each composed of a barrel and two endcap sections, are located. The trajectory of charged particles are measured by the silicon pixel and strip tracker. The ECAL extends up to $|\eta| < 1.48$ in the barrel, while the endcaps cover the region $1.48 < |\eta| < 3.0$. ECAL crystal arrays projecting radially outwards from the nominal interaction point, with a slight offset.

The global event reconstruction algorithm, known as particle-flow event reconstruction [6], reconstructs and identifies each individual particle using information from the various elements of the CMS detector. The energies of photons is directly obtained from the ECAL measurement. While the energy of electrons, muons, and hadrons are determined from a combination of the tracker and calorimeter information. Hadronic jets are clustered from these reconstructed particles using the anti-$k_T$ algorithm [7], with a size parameter of 0.4. Jets including B-hadrons are tagged using the information of the displaced decay vertex, using the combined secondary vertex (CSV algorithm) [8]. The missing transverse momentum vector (MET) is taken as negative vector sum of all reconstructed particle candidate transverse momenta.

[*]on behalf of the CMS collaboration

### 2.1 Photon reconstruction

Photons are identified as ECAL energy clusters not linked to the extrapolation of any charged-particle trajectory to the ECAL. The clustering algorithm has been optimized to recollect the total energy of the photon, including conversions in the material upstream of the calorimeter. Clustering starts based on local energy peaks above a given threshold, as the "seeds". Then they are grown by aggregating crystals with at least one side in common with a clustered crystal and with an energy in excess of a given threshold. Finally, clusters are dynamically merged into "superclusters", to allow good energy containment, accounting for geometrical variations of the detector along $\eta$, and optimizing robustness against pileup.

The energy of photons is computed from the sum of the energy of the ECAL reconstructed hits, calibrated and corrected for several detector effects [9]. The energy is corrected to contain the electromagnetic showers in the clustered crystals and the energy losses of converted photons. A multivariate regression technique is used to compute the correction. It allows estimating simultaneously the energy of the photon and its median uncertainty.

The photon candidates used in this analysis are required to satisfy preselection criteria similar to, but slightly more stringent than, the trigger requirements. To reject ECAL energy deposits incompatible with a single isolated electromagnetic shower, such as those coming from neutral mesons, a selection on shower shape variables is applied. To reject hadrons, the ratio of energy in HCAL cells behind the supercluster to the energy in the supercluster is checked. If the supercluster matches to an electron track with no missing hits in the innermost tracker layers, the photon candidate is vetoed. To discard photons within jets, the photon is requested to be isolated in tracker and calorimeters.

The efficiency of all preselection criteria, except the electron veto requirement, is measured with a tag and probe technique using $Z \to$ee events. The efficiency for photons to satisfy the electron veto requirement is measured with $Z \to \mu\mu\gamma$ events, in which the photon is produced by final-state radiation and provides a sample of prompt photons with purity higher than 99%. Simulation is corrected to data to consider the differences in efficiencies of simulation and data.

A Boosted Decision Tree (BDT) is trained to separate prompt photons from photon candidates satisfying the preselection requirements. The photon identification BDT is trained using simulated $\gamma$+jets events where prompt photons are considered as signal and non-prompt photons as background. Shower shape and isolation variables are used as the input variables. Fig. 1 (Left) shows the output of the identification BDT score.

### 2.2 Diphoton reconstruction

Determination of the vertex from which the diphoton originated has a large impact on the diphoton mass resolution. A BDT is trained using observables related to tracks recoiling against the diphoton system. In case of conversion in the tracker material, the track information is also helpful in vertex assignment. The probability that the assigned vertex is within 1cm of the diphoton interaction point is then estimated using a second BDT.

Events with two preselected photon candidates with $p_T^{\gamma 1} > \mathrm{m}_{\gamma\gamma}/3$ and $p_T^{\gamma 2} > \mathrm{m}_{\gamma\gamma}/4$, in the mass range $100 < \mathrm{m}_{\gamma\gamma} < 180$ GeV are selected.

A dedicated diphoton BDT is trained to evaluate the diphoton mass resolution per event. Higher values of the diphoton BDT show that the event has two photons with signal-like kinematics, good mass resolution, and high photon identification BDT score. The input variables to the classifier are:

- $p_T^{\gamma}/\mathrm{m}_{\gamma\gamma}$ for each photon;
- the pseudorapidity of the two photons;
- the cosine of the angle between the two photons in the transverse plane;
- photon ID BDT scores for both photons;
- two per-event relative mass resolution estimates, one under the hypothesis that the mass has been reconstructed using the correct primary vertex, and the other under the hypothesis that the mass has been reconstructed using an incorrect vertex;
- the per-event probability estimate that the correct primary vertex has been assigned to the diphoton.

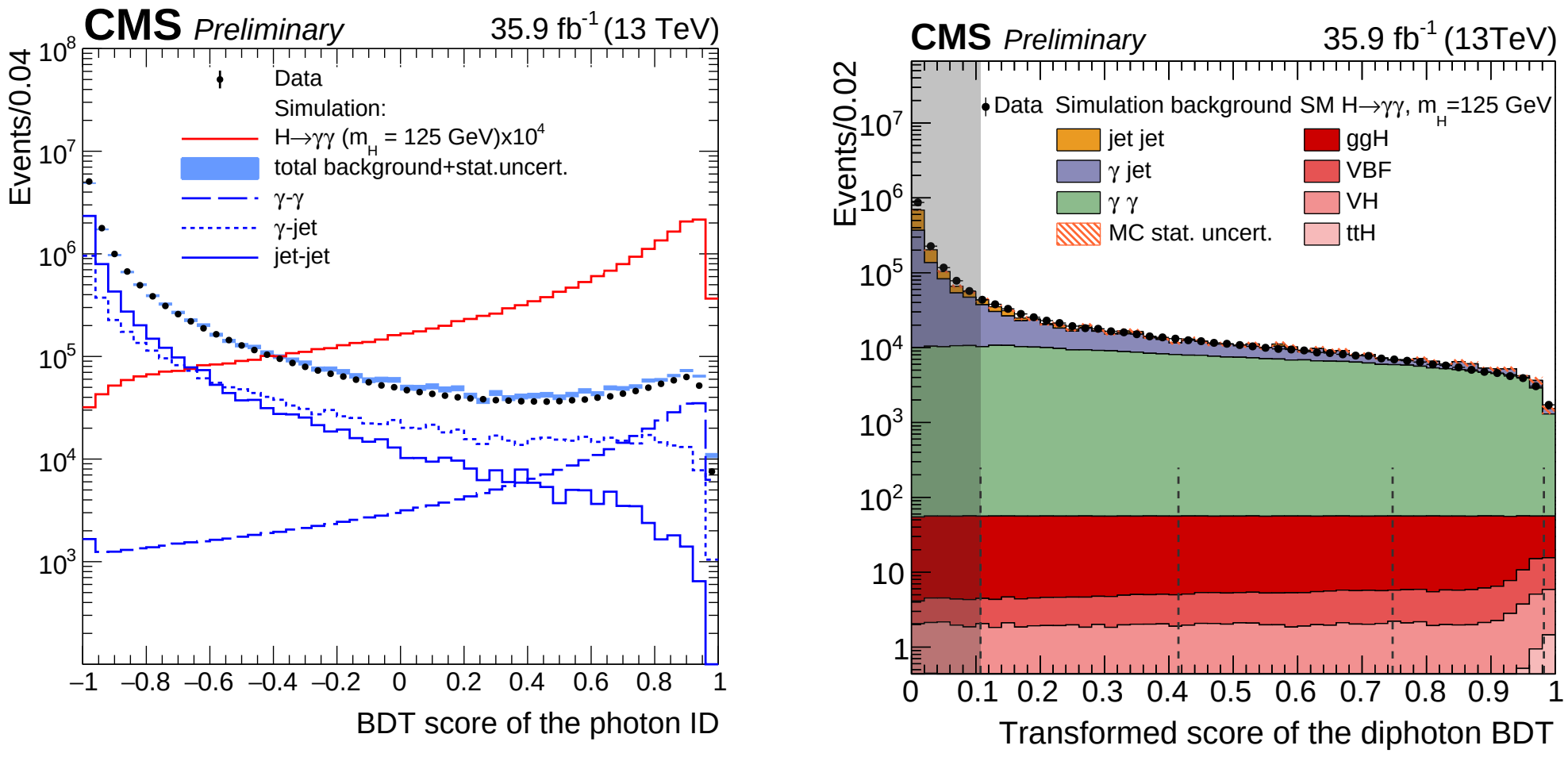

**Fig. 1:** (Left) Photon identification BDT score in the 13 TeV data set (points), and for simulated background events (blue histogram). The red histogram corresponds to simulated Higgs boson signal events. (Right) Transformed score of the diphoton multivariate classifier for events with two photons satisfying the preselection requirements in data (points), simulated signal (red shades), and simulated background (coloured histograms). Both signal and background are stacked together. The grey shade indicates events discarded from the analysis (see Ref.[4]).

The distribution of the diphoton BDT score is shown in Fig. 1(Right).

## 3 Categorization

To increase the sensitivity of the analysis, events are classified according to the production mechanisms and their mass resolution and predicted signal-to-background ratio. Higgs boson production mechanisms considered in this analysis are gluon fusion (ggH), vector boson fusion (VBF), production associated with a vector boson (VH) or with a top quark pair ($\mathrm{t\bar{t}H}$).

In total 14 exclusive categories are defined. These categories and their rankings are shown in Fig 2. A dedicated BDT is trained to separate $\mathrm{t\bar{t}H}$ events, based on the jet information. The cut on the BDT output is optimized together with the cut on the diphoton BDT score to maximize the expected sensitivity to this production mechanism. Events with one lepton, either electron or $\mu$ and at least 2 jets, one of which is tagged as b-jet are tagged under the $\mathrm{t\bar{t}H}$ Leptonic category. Events with no lepton and at least 3 jets that have at least one b-tagged jet are tagged as hadronic decays of $\mathrm{t\bar{t}H}$.

There are 5 VH categories, where V stands for either Z-boson or W-boson. Events with two leptons with invariant mass consistent with the Z-boson mass are tagged as ZH Leptonic event. The next two VH categories are dedicated to events with 1 lepton, among which events with MET>45 GeV and less than 3 jets are good candidates for leptonic decays of WH events. The VH-Leptonic loose category contains events with MET<45 GeV, to include ZH events with one missing lepton. The VH(MET) category is defined to tag ZH events, where Z decays to neutrinos, or WH events when the lepton of the W-boson decay is missing. Events with MET>85 GeV, where $\Delta\phi(\text{MET}, \gamma\gamma)$>2.4 are tagged in VH(MET) category. To include the hadronic decays of vector bosons, events with more than 2 jets when the invariant mass of one of the pair of jets is between 60 GeV and 120 GeV are tagged as VH(Hadronic) category.

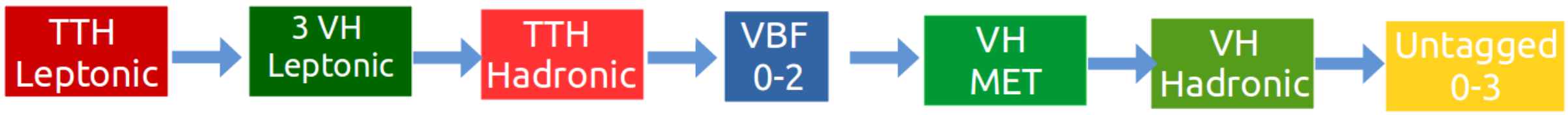

**Fig. 2:** Schematic view of event categorization and the order of exclusive categories.

Having two forward jets is the distinct feature of the VBF production. Events with two forward jets where the invariant mass of them is greater than 250 GeV are selected. Forward information is also used to train a BDT. Events are categorized in 3 categories based on the BDT output to obtain the best expected significance.

All the remaining untagged events are divided into 4 categories according to their mass resolution.

## 4 Results

In each category, Higgs boson events appear as a peak on a falling background of non-Higgs events. The shape of the falling background is modeled fitting on data, while the signal shape is obtained by fitting on simulation.

### 4.1 Background Model

The discrete profiling method [10] is used to describe the background. Exponentials, Bernstein polynomials, Laurent series and power law function families are considered. The F-test [11] technique is used to determine the maximum order for each family of functions to be used.

### 4.2 Signal Shape Modeling

Given that the distribution of $m_{\gamma\gamma}$ depends significantly on the correct vertex assignment of the candidate diphoton, distributions with the correct vertex and wrong vertex are fitted separately when constructing the signal model. For each process, category, and vertex scenario, the $m_{\gamma\gamma}$ distributions are fitted using a sum of at most five Gaussian functions.

Parameter values of the signal shape for each process, category, and vertex scenario are found as a function of the Higgs boson mass in the range from 120 to 130 GeV.

The efficiency times acceptance of the signal model as a function of $m_H$ for all categories combined is shown in Fig. 3.

### 4.3 Yields and inclusive results

The expected signal yield and the composition of different production mechanisms in each category and also the ratio of signal events is shown in Fig. 4.

A fit is performed to obtain the signal strength for each production in all categories. The results are shown in Fig. 5. The expected sensitivity for different production modes are compatible with the observation.

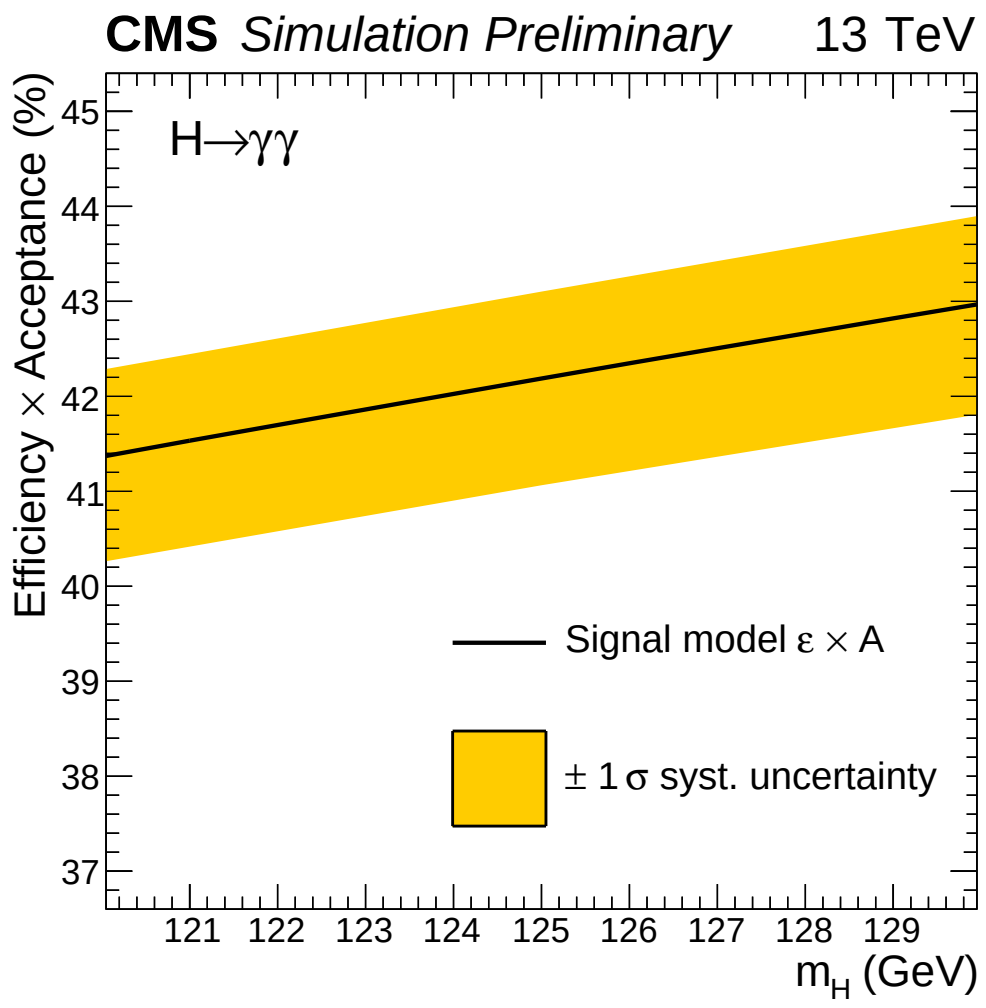

**Fig. 3:** The efficiency $\times$ acceptance of the signal model as a function of $m_H$ for all categories combined. The black line represents the yield from the signal model. The yellow band indicates the effect of the systematic uncertainties for trigger, photon identification and selection, photon energy scale and modelings of the photon energy resolution, and vertex identification (see Ref.[4]).

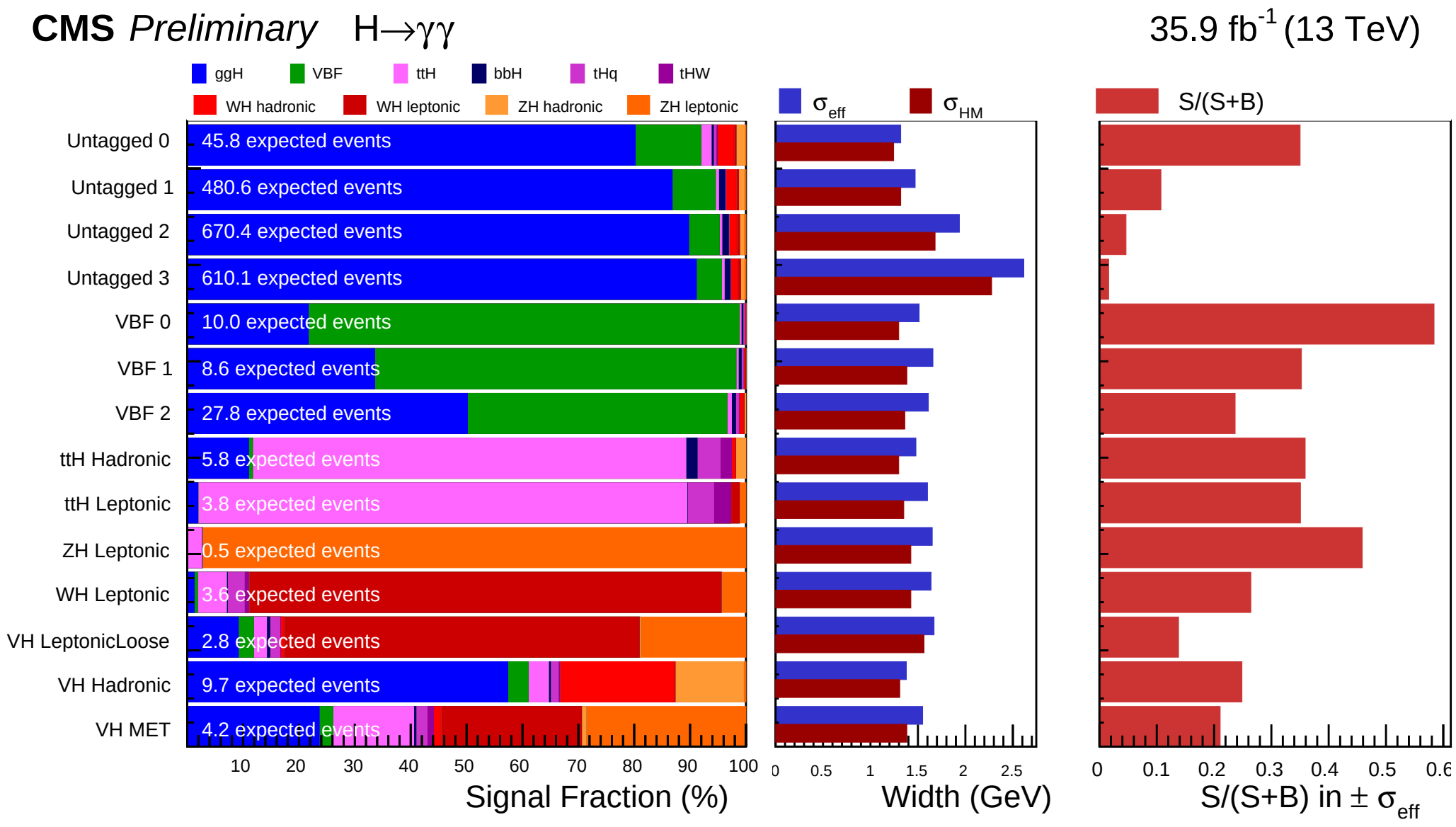

**Fig. 4:** Expected composition of signal events per production mechanism in different categories. The ratio of the number of signal events. (S) to the number of signal plus background events (S+B) is plotted on the right hand side (see Ref.[4]).

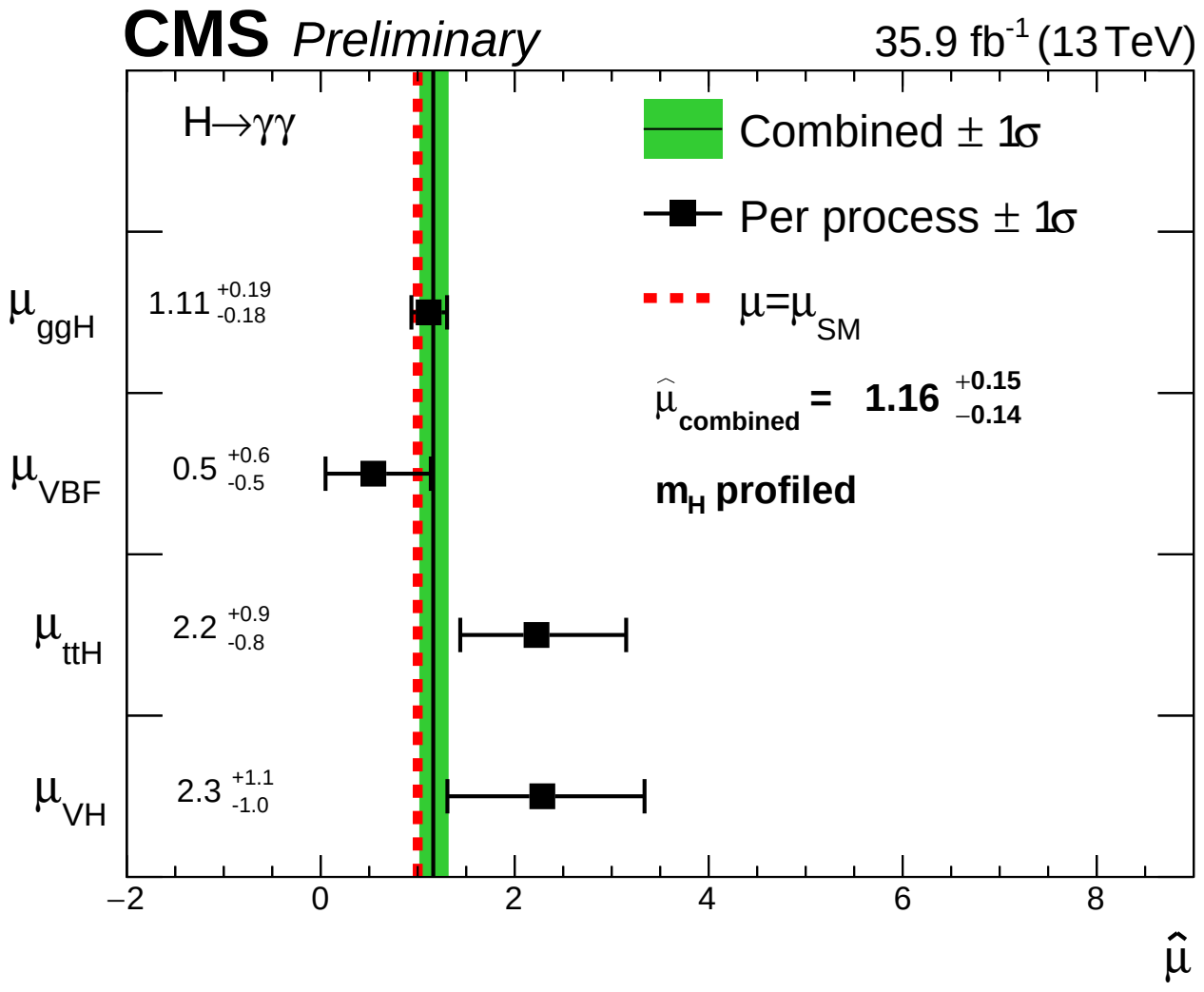

**Fig. 5:** Signal strength modifiers measured for each process (black points), with $\mathrm{m_H}$ profiled, compared to the overall signal strength (green band) and to the SM expectation (dashed red line), (see Ref.[4]).

### 4.4 Differential and Fiducial results

A dedicated analysis is performed to measure the cross section of the Higgs boson production vs. different properties of Higgs boson and event. Diphoton events are categorized based on an uncorrelated mass resolution estimator defined as

$$\mathrm{y}(\sigma_\mathrm{m}/\mathrm{m}|\mathrm{m}) = \int_0^{\sigma_\mathrm{m}/\mathrm{m}} f(\sigma_\mathrm{m}/\mathrm{m}'|\mathrm{m}) d\sigma_\mathrm{m}/\mathrm{m}'$$

where the conditional distribution of $\sigma_\mathrm{m}/\mathrm{m}$, $f(\sigma_\mathrm{m}/\mathrm{m}|\mathrm{m})$, is constructed by sorting the values of $\sigma_\mathrm{m}/\mathrm{m}$ in bins of $\mathrm{m}$. The number of categories was optimized and found to be 3 in order to saturate the maximum achievable sensitivity.

Total and differential cross sections are unfolded to particle-level using a likelihood fit. Total cross section in fiducial phase space with $p_T^{\gamma 1} = \mathrm{m}_{\gamma\gamma}/3$, $p_T^{\gamma 2} = \mathrm{m}_{\gamma\gamma}/4$ and $|\eta^\gamma|$<2.5 is measured and found to be $84^{+13}_{-12}$ fb which is consistent with the theory prediction.

The differential cross sections as a function of the diphoton transverse momentum and the jet multiplicity are reported in Figure 6, together with the corresponding theoretical predictions. Two sets of predictions are shown in each plot. For the first, shown in orange, MADGRAPH_aMC@NLO was used to simulate all the Higgs boson production processes. The second, shown in green, was obtained using POWHEG-generated ggH events, while taking other production mechanisms from MADGRAPH_aMC@NLO. The plots show in blue the sum of the contributions from VBF, VH and ttH (labeled HX).

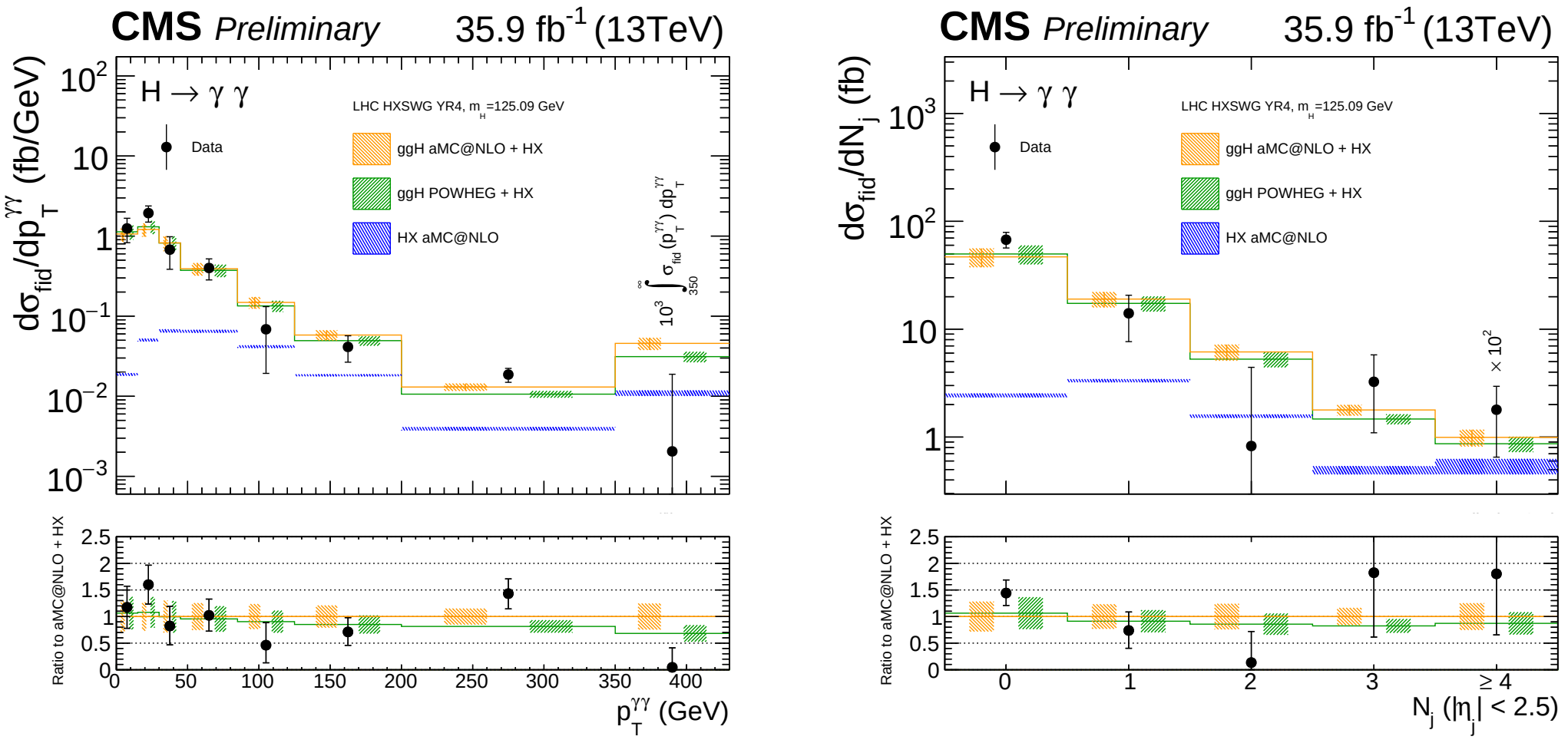

**Fig. 6:** Measured $H \rightarrow \gamma\gamma$ differential cross-section (black points) for (Left) $p_{T,\gamma\gamma}$, (Right) $N_{jets}$. The measurements are compared to the theoretical predictions, combining the Higgs boson cross sections and branching fraction as in the LHC Higgs Cross Section Working Group [12] with two different generators for the gluon-gluon fusion process: MADGRAPH_aMC@NLO (in orange) and POWHEG (in green) (see Ref.[5]).

# Rare radiative decays at LHCb

*Albert Puig Navarro, on behalf of the LHCb collaboration*
Universität Zürich, Zürich, Switzerland

**Abstract**
Radiative $b$-hadron decays are sensitive probes of New Physics through the study of branching fractions, $CP$ asymmetries and measurements of the polarization of the photon emitted in the decay. During Run I of the LHC, the LHCb experiment has collected large samples of radiative $b$-hadron decays. An overview of the LHCb measurements, including results on the time dependence of $B_s^0 \to \phi\gamma$ decays, is presented here. These results help constrain the size of right-handed currents in extensions of the Standard Model.

## 1 Introduction

Rare $b \to s\gamma$ flavour-changing neutral-current transitions are forbidden at tree level in the Standard Model (SM) and, as a consequence, are very sensitive to new physics (NP) effects arising from the exchange of new heavy particles in electroweak penguin diagrams. In such cases, the SM predicts that the emitted photon is predominantly left-handed since the recoil $s$ quark that couples to a $W$ boson is left-handed. However, in several NP models, such as the left-right symmetric model [1–4] or the minimal supersymmetric model (MSSM) [5], the photon can acquire a significant right-handed component. Although effects coming from NP are strongly constrained by measurements of inclusive radiative decays, there is still room for contributions that would enhance the right-handed photon polarization component.

The main challenge for the study of $b$ decays with a final state photon at the LHCb experiment [6] is the fact that their mass resolution is significantly worse (a factor 4–5) than that for decays to only charged particles due to the dominance of the photon energy resolution. This issue, combined with the large levels of background expected from a $pp$ machine, makes the identification of signal remarkably difficult. Despite this fact, the LHCb collaboration has a wide program covering $b$ decays with a final state photon, studying not only the polarization of the photon but also other quantities which could be affected by NP effects. This contribution presents the main measurements of rare radiative $B$ decays at LHCb.

## 2 $B^0 \to K^{*0}\gamma$ and $B_s^0 \to \phi\gamma$

While inclusive decays are theoretically cleaner than exclusive ones—which suffer from large uncertainties due to, e.g., form factors—they are more challenging experimentally, especially in the LHC context. It is possible, however, to find cleaner, form-factor free observables that can be at the same time predicted accurately enough and measured with good precision. Examples of such observables are the ratio of branching fractions of $B^0 \to K^{*0}\gamma$ and $B_s^0 \to \phi\gamma$ decays, with a SM prediction of $1.0 \pm 0.2$ [7], or the direct $CP$ asymmetry in $B^0 \to K^{*0}\gamma$ decays, predicted to be $(-0.61 \pm 0.43)\%$ [8].

Benefitting from large samples of $B^0 \to K^{*0}\gamma$ and $B_s^0 \to \phi\gamma$ decays corresponding to 1 fb$^{-1}$ of $pp$ collisions at $\sqrt{s} = 7$ TeV, shown in Fig. 1, the LHCb collaboration measured the ratio of their branching fractions to be [9]

$$\frac{\mathcal{B}(B^0 \to K^{*0}\gamma)}{\mathcal{B}(B_s^0 \to \phi\gamma)} = 1.23 \pm 0.06\,(\text{stat}) \pm 0.04\,(\text{syst}) \pm 0.10\,(f_s/f_d), \quad (1)$$

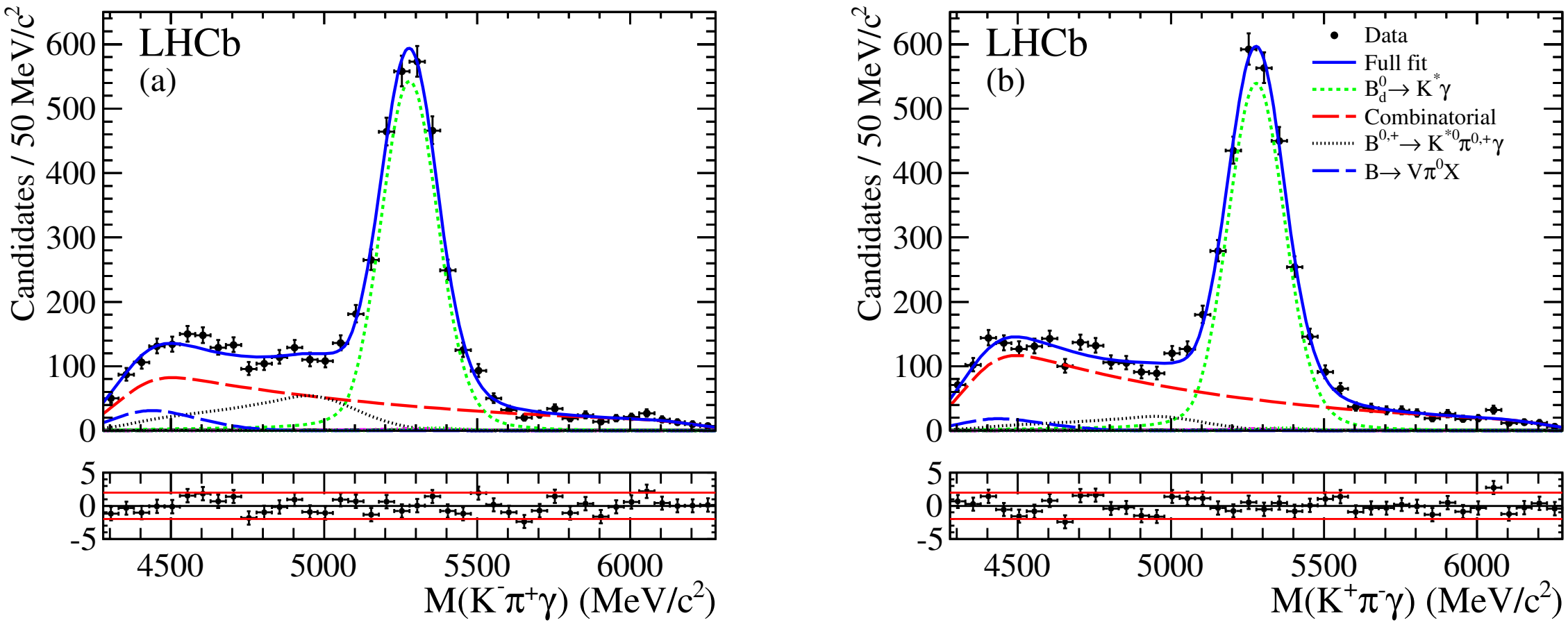

**Fig. 1:** Invariant mass distributions of the (a) $B^0 \to K^{*0}\gamma$ and (b) $B_s^0 \to \phi\gamma$ decay candidates. The fit result is overlaid as a solid blue line, with the signal component represented as a dashed green line.

which is the most precise measurement to date and is in agreement with the SM prediction. Using the world average value of $\mathcal{B}(B^0 \to K^{*0}\gamma) = (4.5 \pm 0.15) \times 10^{-5}$ [10], the branching fraction of $B_s^0 \to \phi\gamma$ was found to be

$$\mathcal{B}(B_s^0 \to \phi\gamma) = (3.5 \pm 0.4) \times 10^{-5}. \tag{2}$$

The same dataset was used to measure the direct $CP$ asymmetry of the $B^0 \to K^{*0}\gamma$ decay, defined as

$$\mathcal{A}_{CP} = \frac{\Gamma(\overline{B}^0 \to \overline{K}^{*0}\gamma) - \Gamma(B^0 \to K^{*0}\gamma)}{\Gamma(\overline{B}^0 \to \overline{K}^{*0}\gamma) + \Gamma(B^0 \to K^{*0}\gamma)}, \tag{3}$$

to be [9]

$$\mathcal{A}_{CP} = (0.8 \pm 1.7\,(\text{stat}) \pm 0.9\,(\text{syst}))\%, \tag{4}$$

well in agreement with the SM prediction.

Once these two results are updated with the full dataset collected by LHCb, the measurements are expected to reach the systematic limitation.

## 3 Search for $B_{(s)} \to J/\psi\,\gamma$ decays

Radiative $B$ decays can also be used to test different approaches to the treatment of QCD calculations in theoretical predictions. In particular, the branching fraction predictions for $B_{(s)} \to J/\psi\gamma$ decays, dominated by the $W$ boson exchange diagram shown in Fig. 2, vary one order of magnitude between the QCD factorization and the perturbative QCD approaches, going from $\sim 2 \times 10^{-7}$ [11] to $5 \times 10^{-6}$ [12], respectively.

Despite not being $b \to s\gamma$ transitions, $B^0 \to J/\psi\gamma$ and $B_s^0 \to J/\psi\gamma$ present similar challenges to other radiative decays, i.e., the large contamination from low-mass backgrounds due to the wide mass resolution and the difficulty to separate $\pi^0$ from photons at high transverse energies. In order to distinguish the signal from the large contamination from decays such as $B_{(s)} \to J/\psi\eta(\to \gamma\gamma)$, $B^0 \to J/\psi\pi^0$, $B^0 \to J/\psi K_S^0(\to \pi^0\pi^0)$, and $B^+ \to J/\psi\rho^+(\to \pi^0\pi^+)$, only photons reconstructed as a electron-positron pairs where used; the fact that only the $J/\psi$ meson and one photon are reconstructed was then used to separate these backgrounds from the signal.

Using 1 fb$^{-1}$ of luminosity at $\sqrt{s} = 7$ TeV collected by the LHCb experiment, no significant signal was observed and an upper limit on the branching fractions was set to

$$\mathcal{B}(B^0 \to J/\psi\gamma) < 7.3 \times 10^{-6} \text{ at } 90\% \text{ CL}, \tag{5}$$

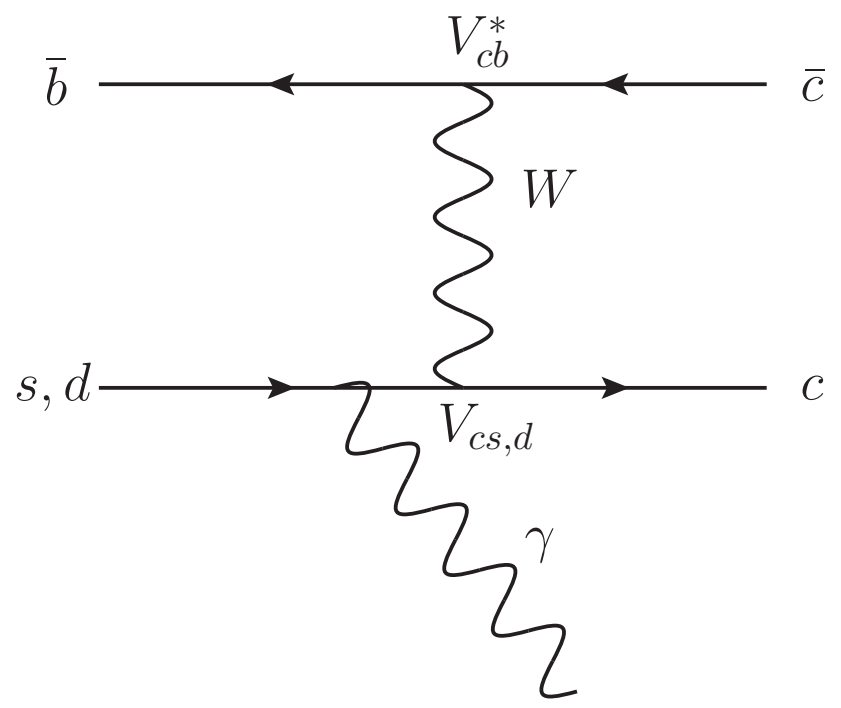

**Fig. 2:** Feynman diagram of the leading contribution to $B_{(s)} \to J/\psi\gamma$ decays, where one quark radiates a photon.

$$\mathcal{B}(B_s^0 \to J/\psi\gamma) < 1.5 \times 10^{-6} \text{ at } 90\% \text{ CL}. \tag{6}$$

The $B^0 \to J/\psi\gamma$ branching fractions is in agreement and competitive with the previous measurement by the BaBar collaboration [13], while the first limit on $B_s^0 \to J/\psi\gamma$ is close to the sensitivity of the prediction based on perturbative QCD.

## 4 $B_s^0 \to \phi\gamma$ lifetime measurement

The time-dependent decay rate of $B_s^0$ and $\overline{B}_s^0$ mesons decaying into a common final state containing a photon, such as $B_s^0 \to \phi\gamma$, is proportional to

$$e^{-\Gamma_s t}\left\{\cosh\left(\frac{\Delta\Gamma_s}{2}\right) - \mathcal{A}^{\Delta}\sinh\left(\frac{\Delta\Gamma_s}{2}\right) \pm \mathcal{C}\cos\left(\Delta m_s t\right) \mp \mathcal{S}\sin\left(\Delta m_s t\right)\right\}, \tag{7}$$

where $\Delta\Gamma_s$ and $\Delta m_s$ are the width and mass differences between the light and heavy $B_s^0$ mass eigenstates and $\Gamma_s$ is the mean decay width; the $\mathcal{A}^{\Delta}$, $\mathcal{C}$ and $\mathcal{S}$ coefficients are functions of the photon polarization. When the initial flavor of the $B_s^0$ meson is unknown, only the $\mathcal{A}^{\Delta}$, predicted to be $\mathcal{A}^{\Delta} = 0.047^{+0.029}_{-0.025}$ in the SM [14], is accessible, and can be measured through the study of the $B_s^0 \to \phi\gamma$ effective lifetime.

Making use of the full statistics collected in the LHC Run I by the LHCb experiment, corresponding to $1\ \text{fb}^{-1}$ and $2\ \text{fb}^{-1}$ of luminosity collected at center-of-mass energies of $7$ and $8\ \text{TeV}$, respectively, the LHCb collaboration performed the first measurement of the $\mathcal{A}^{\Delta}$ parameter [15]. A value of

$$\mathcal{A}^{\Delta} = -0.98^{+0.46}_{-0.52}\,(\text{stat})^{+0.23}_{-0.20}\,(\text{syst}), \tag{8}$$

was found through an unbinned simultaneous fit of the $B_s^0 \to \phi\gamma$ and $B^0 \to K^{*0}\gamma$ background-subtracted decay-time distributions, shown in Fig. 3, where the latter is used as a control sample. This result is compatible with the SM prediction within two standard deviations.

## 5 Angular analysis of $B^+ \to K^+\pi^-\pi^+\gamma$

Another way to access the photon polarization is through the decays of $B$ mesons to a photon and a resonance that decays to three particles; information about the polarization of the photon can then be obtained from its direction with respect to the normal to the plane defined by the momenta of the three final-state hadrons in their centre-of-mass frame ($\tilde{\theta}$) [16, 17].

In general, the differential decay rate of $\overline{B} \to P_1P_2P_3\gamma$ going through a single resonance can be written using the helicity amplitude $\mathcal{J}_\mu$ as

$$\frac{\mathrm{d}\Gamma(\overline{B} \to \overline{K}_{\text{res}}\gamma \to P_1P_2P_3\gamma)}{\mathrm{d}s\,\mathrm{d}s_{13}\,\mathrm{d}s_{23}\,\mathrm{d}\cos\tilde{\theta}} \propto |\vec{\mathcal{J}}|^2(1+\cos^2\tilde{\theta}) + \lambda_\gamma\, 2\,\mathrm{Im}\left[\vec{n}\cdot(\vec{\mathcal{J}}\times\vec{\mathcal{J}}^*)\right]\cos\tilde{\theta}\,, \tag{9}$$

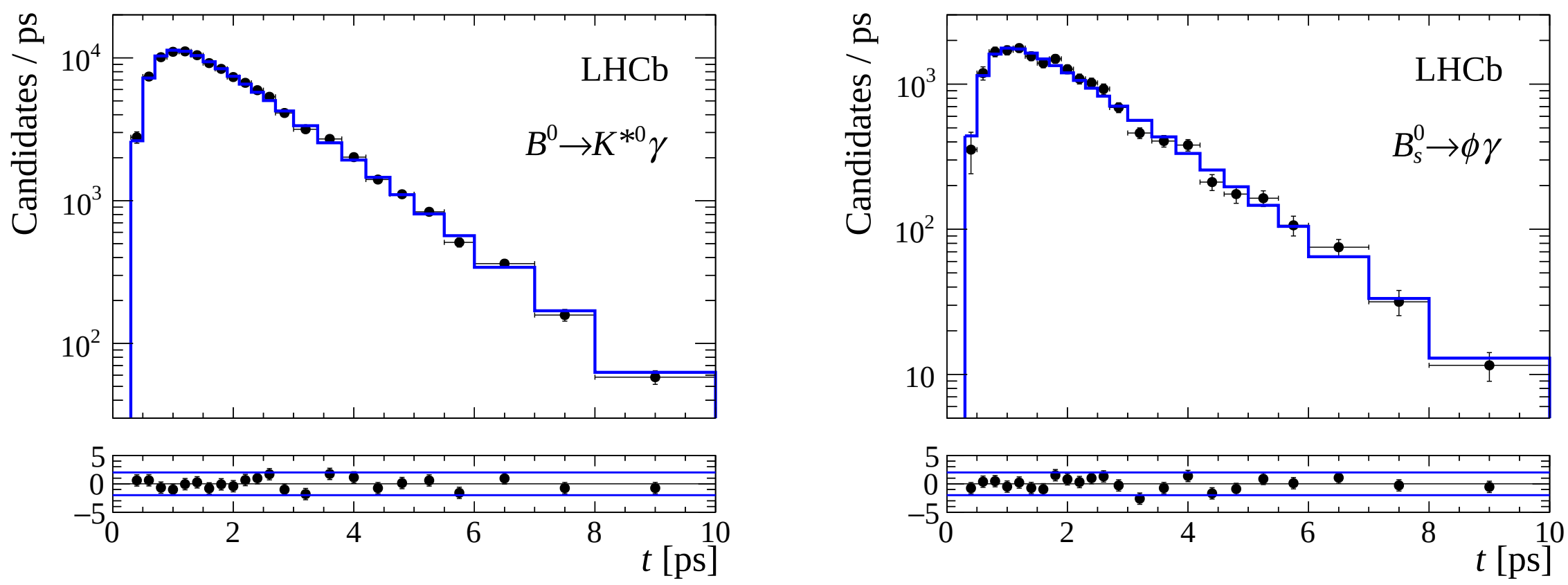

**Fig. 3:** Background-subtracted decay-time distributions of the (left) $B^0 \to K^{*0}\gamma$ and (right) $B_s^0 \to \phi\gamma$ decays. The fit projection is overlaid as a solid blue line.

where $s$ is the invariant mass of the $P_1P_2P_3$ system and $s_{ij}$ the invariant mass of the $P_iP_j$ system. In the case of overlapping intermediate resonances, one needs to consider their interference and Eq. 9 is not valid, leading to more complex dependencies on $\cos\tilde{\theta}$. However, even in the case of multiple resonances one can write the differential decay rate as a function of the even and odd powers of $\cos\tilde{\theta}$ and coefficients $a_j(s_{13}, s_{23})$:

$$\frac{\mathrm{d}\Gamma(\overline{B} \to \overline{K}_{\text{res}}\gamma \to P_1P_2P_3\gamma)}{\mathrm{d}s\,\mathrm{d}s_{13}\,\mathrm{d}s_{23}\,\mathrm{d}\cos\tilde{\theta}} \propto \sum_{j \text{ even}} a_j(s_{13}, s_{23}) \cos^j\tilde{\theta} + \lambda_\gamma \sum_{j \text{ odd}} a_j(s_{13}, s_{23}) \cos^j\tilde{\theta}. \quad (10)$$

The structure of the decay rate can be exploited to study the photon polarization by constructing the up-down asymmetry

$$\mathcal{A}_{\text{ud}} \equiv \frac{\int_0^1 \mathrm{d}\cos\tilde{\theta} \frac{\mathrm{d}\Gamma}{\mathrm{d}\cos\tilde{\theta}} - \int_{-1}^0 \mathrm{d}\cos\tilde{\theta} \frac{\mathrm{d}\Gamma}{\mathrm{d}\cos\tilde{\theta}}}{\int_{-1}^1 \mathrm{d}\cos\tilde{\theta} \frac{\mathrm{d}\Gamma}{\mathrm{d}\cos\tilde{\theta}}} = C\lambda_\gamma\,, \quad (11)$$

where the constant $C$ takes into account the integral over the Dalitz plot, $a_j(s_{13}, s_{23})$, and the angle $\cos\tilde{\theta}$. If $\mathcal{J}$ is known, $C$ can be calculated and this asymmetry allows the determination of the photon polarization. However, in the case of $B^+ \to K^+\pi^-\pi^+\gamma$ the different resonances in the $K^+\pi^-\pi^+$ spectrum, shown in Fig. 4, cannot be easily separated and therefore the measurement needs to be performed inclusively.

The LHCb collaboration has studied the $\cos\tilde{\theta}$ angular distribution—including the up-down asymmetry in $B^+ \to K^+\pi^-\pi^+\gamma$ decays using 1 $\text{fb}^{-1}$ of data collected at $\sqrt{s} = 7$ TeV [18]. The measurement was performed in bins of the $K^+\pi^-\pi^+$ mass, as defined in Fig. 4, in order to separate as much as possible physics effects coming from the dominant resonances.

While the measurements do not allow to determine the photon polarization, the combined significance of the observed up-down asymmetries (shown in Fig. 5) with respect to the no-polarization scenario in which the up-down asymmetry is expected to be zero in every mass interval, was determined to be $5.2\,\sigma$. This is the first observation of non-zero photon polarization.

## 6 Conclusions

The LHCb collaboration has fulfilled its core radiative $B$ decays programme with data collected in the Run I of the LHC, with most of the presented results being either world's best or very competitive with previous measurements. The great potential of the LHCb experiment for the study of radiative $B$ decays is further showcased through its influence on the global fits of the $C_7^{(\prime)}$ Wilson coefficients (see for example Ref. [19]): thanks to measurements such as the $B_s^0 \to \phi\gamma$ effective lifetime and the angular distribution of the $B^0 \to K^{*0}e^+e^-$ decay [20], combined with other results from the $B$-factories, the presence of large NP effects $C_7^{(\prime)}$ has almost been ruled out.

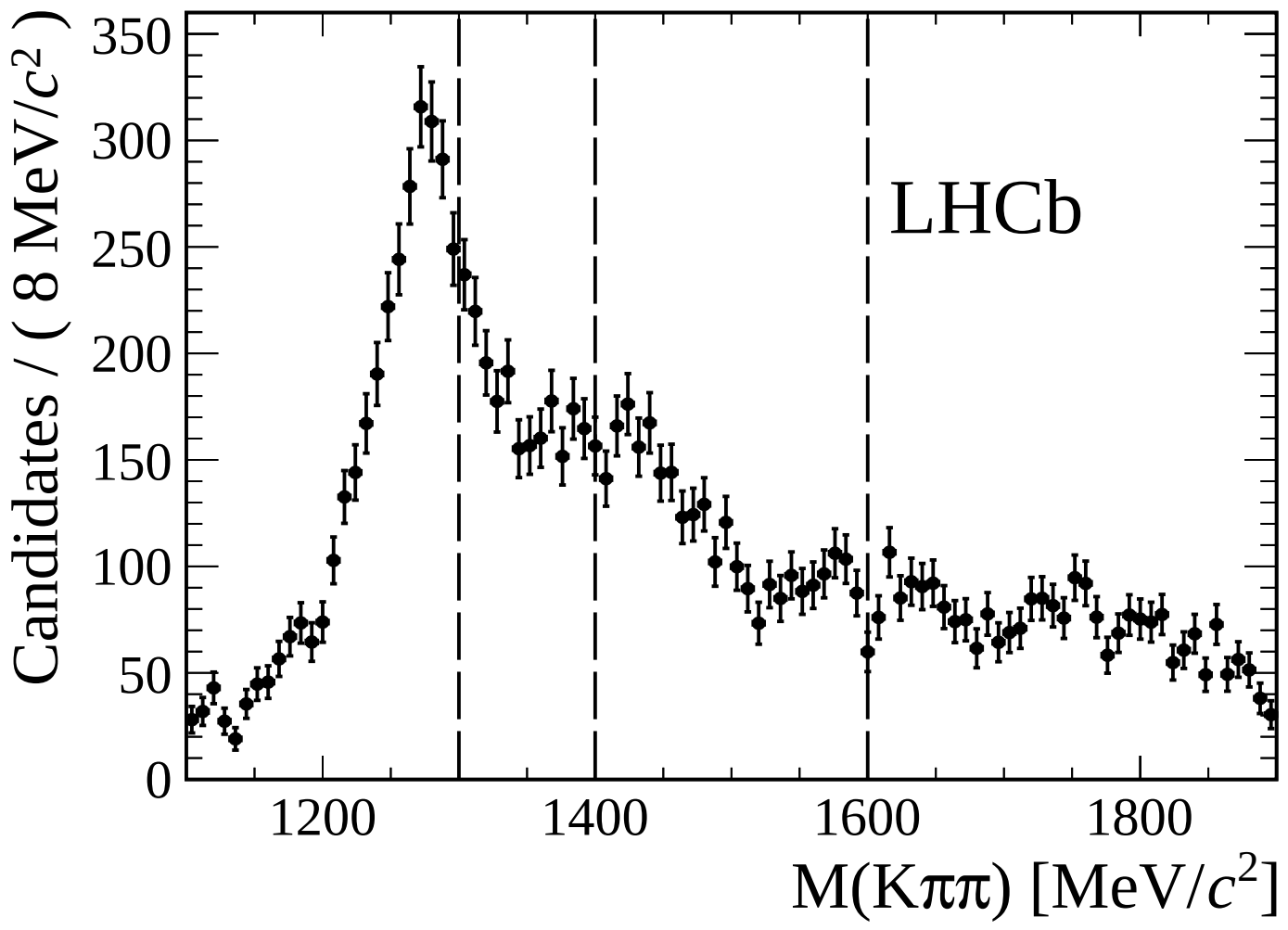

**Fig. 4:** Background-subtracted $K^+\pi^-\pi^+$ mass distribution of $B^+ \to K^+\pi^-\pi^+\gamma$ candidates. The vertical dashed lines indicate the bin edges used in the angular analysis.

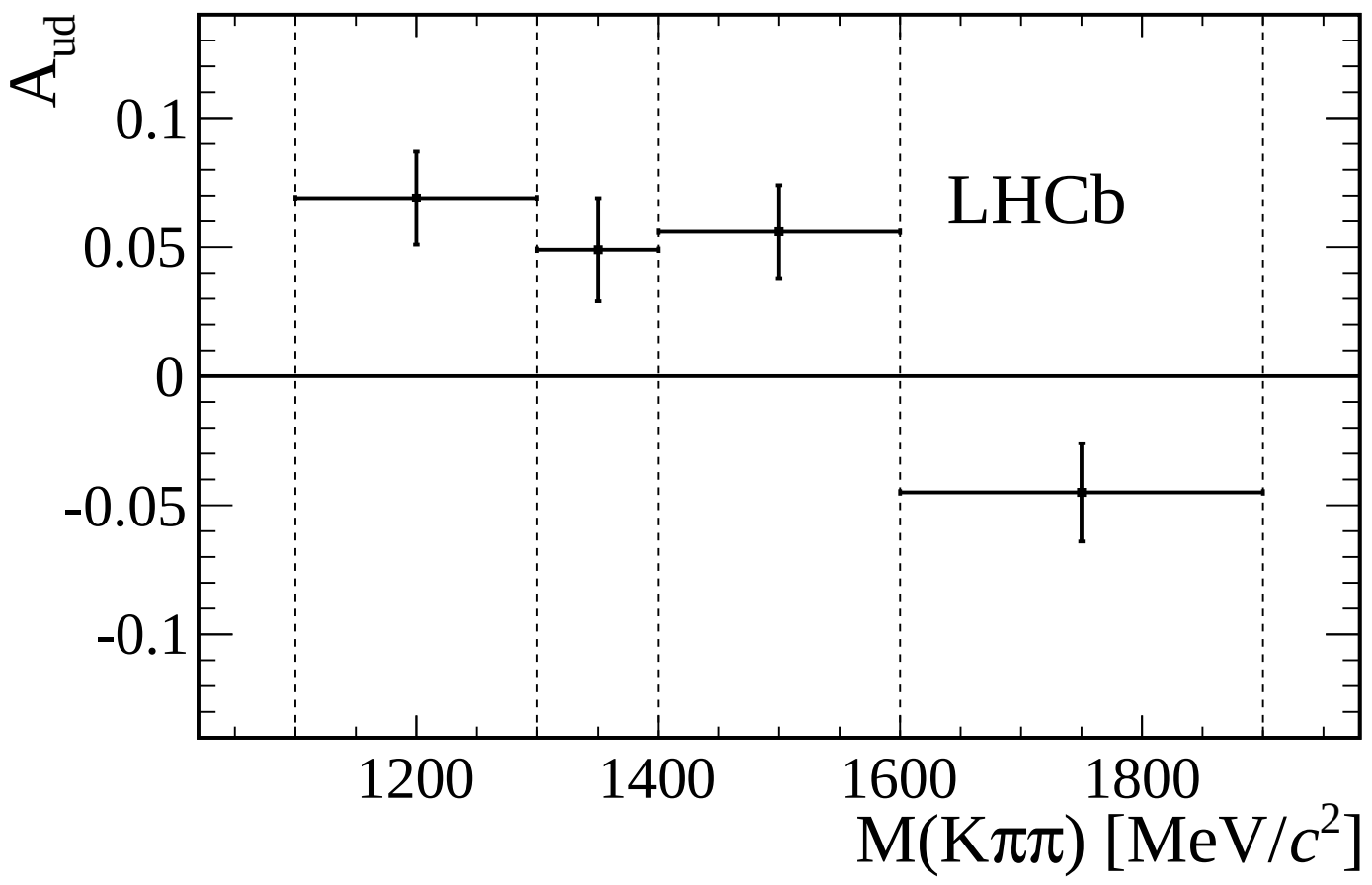

**Fig. 5:** Up-down asymmetry in bins of $K^+\pi^-\pi^+$ mass measured in $B^+ \to K^+\pi^-\pi^+\gamma$ decays. The vertical dashed lines indicate the bin edges used in the angular analysis.

The update of the results presented here with data taken in the Run II of the LHC, along with the addition of more complex measurements, such as the $B$-tagged $B_s^0 \to \phi\gamma$ effective lifetime, $b$ radiative baryon decays, the full amplitude analysis of $B^+ \to K^+\pi^-\pi^+\gamma$ or $b \to d\gamma$ transitions, will improve constraints on NP and will allow to reach an unprecedented knowledge of the polarization of the photon.

# The Gamma Factory proposal for CERN

*M.W. Krasny and the Gamma Factory Study Group* *
LPNHE, University Paris VI et VII and CNRS–IN2P3, Paris, France

**Abstract**

This contribution discusses the possibility of broadening the present CERN research programme by a new component making use of a novel concept of the light source. The proposed, Partially Stripped Ion beam driven, light source is the backbone of the Gamma Factory initiative. It could be realised at CERN by using the infrastructure of the already existing accelerators. It could push the intensity limits of the presently operating light-sources by at least 7 orders of magnitude, reaching the flux of the order of $10^{17}$ photons/s, in the particularly interesting $\gamma$-ray energy domain of $1 \leq E_{photon} \leq 400$ MeV. This domain is out of reach for the FEL-based light sources based on sub-TeV energy-range electron beams. The unprecedented-intensity, energy-tuned, quasi-monochromatic gamma beams, together with the gamma-beams-driven secondary beams of polarised positrons, polarised muons, neutrinos, neutrons and radioactive ions would constitute the basic research tools of the proposed Gamma Factory. A broad spectrum of new opportunities, in a vast domain of uncharted fundamental and applied physics territories, could be opened by the Gamma Factory research programme.

## 1 Existing and future MeV-range light sources

The light sources in the discussed MeV energy range have already been constructed and operating in several countries: HI$\gamma$S-USA, LEPS-Japan, LADON-Italy, ROKK-1-Russia, GRAAL-France and LEGS-USA. The leading future project entering the construction phase is the European Union project ELI-NP. The ELI-NP facility is expected to produce the flux of $10^{13}$ photons/s with the maximal energy of 20 MeV. The highest photon flux which has been achieved so far is $10^{10}$ photons/s.

All the above facilities generate, or are expected to generate, the photon beams by the process of the inverse Compton scattering of the laser photons on the highly relativistic electron beams. Since the cross section of the inverse Compton process is small, in the $\mathcal{O}$(1 barn) range, in order to achieve the quoted above fluxes, the laser system and the energy recovery linac technologies have to be pushed to their technological limits.

*Current members of the Gamma Factory group: E.G. Bessonov, Lebedev Physical Institute, Moscow, Russia; D. Budker, Helmholtz Institute, Johannes Gutenberg University, Mainz, Germany; K. Cassou, I. Chaikovska, R. Chehab, K. Dupraz, A. Martens, F. Zomer, LAL Orsay, France; C.Curatolo, L. Serafini Department of Physics, INFN-Milan and University of Milan, Milan, Italy ; O. Dadoun, M. W. Krasny, LPNHE, University Paris VI et VII and CNRS–IN2P3, Paris, France; P. Czodrowski, J. Jowett, Reyes Alemany Fernandez, M. Kowalska, M. Lamont, D. Manglunki, A. Petrenko, F. Zimmermann, CERN, Geneva, Switzerland; W. Placzek, Jagellonian University, Krakow, Poland; Y. K. Wu, FEL Laboratory, Duke University, Durham, USA; M. S. Zolotorev Center for Beam Physics, LBNL, Berkeley, USA.

## 2 The Gamma Factory proposal for CERN – the leap into the ultra-high gamma-source intensity

The idea underlying the Gamma Factory proposal is to use Partially Stripped Ion (PSI) beams, instead of electron beams, as the drivers of its light source[1]. The PSI beams are the beams of ions carrying one, or more, electrons which have not been stripped along the way from the ion source to the final PSI beam storage ring. In the process of the resonant absorption of the laser photons by the PSI beam, followed by a spontaneous atomic-transition emissions of secondary photons, the initial laser-photon frequency is boosted by a factor of up to $4 \times \gamma_L^2$, where $\gamma_L$ is the Lorenz factor of the partially stripped ion beam. Therefore, the light source in the energy range of $1 \leq E_{photon} \leq 400$ MeV must be driven by the high-$\gamma_L$, LHC-stored, PSI beams. CERN is a unique place in the world where such a light source could be realized.

The cross-section for the resonant absorption of laser photons by the atomic systems is in the **giga-barn range**, while the cross-section for the point-like electrons is in the **barn range**. As a consequence the PSI-beam-driven light source intensity could be higher than those of the electron-beam-driven ones by a large factor. For the light source working in the regime of multiple photon emissions by each of the beam ions, the photon beam intensity is expected to be limited no longer by the laser light intensity but by the available RF power of the ring in which partially stripped ions are stored. For example, the flux of up to $10^{17}$ photons/s could be achieved for photon energies in the 10 MeV region already with the present, U= 16 MV, circumferential voltage of the LHC cavities. This photon flux is by a factor of $10^7$ higher than that of the highest-intensity electron-beam-driven light source, HI$\gamma$S@Durham, operating in the same energy regime.

If photon beams carrying more than $\mathcal{O}$(100 kW) of beam-power can be safely handled, and if the present circumferential voltage could be increased (at LEP2 the corresponding value was 3560 MV), even higher fluxes could be generated.

## 3 Acceleration, storage and use of the PSI beams at CERN

The first steps to understand the storage stability of the PSI beams were already made at BNL. The $^{77+}$Au beam with two unstripped electrons was successfully circulating in the AGS ring at BNL and, more recently, in its RHIC ring [2]. These tests may be considered as a starting point for further exploratory tests which could be carried out initially at the CERN SPS and, if successful, at the LHC.

If stable PSI beams could be produced and stored they would not only drive the photon source but could also be used for the following two unconventional applications.

Firstly, they would allow the LHC to operate as an **electron-proton(ion) collider** [3]. The LHC experiments could simply record collisions of electrons, brought to LHC experiment's interaction points "on the shoulders" of the ion-carriers, with the counter-propagating proton(ion) beam.

Secondly, they may turn out to be efficient driver beams for the hadron beam driven plasma-wakefield acceleration [4] of a witness beam. This is because the PSI bunches, contrary to the proton bunches, could be very efficiently cooled by the Doppler laser cooling techniques, allowing to compress their bunch sizes. A profit could thus be made from the fact that the maximal achievable plasma electric field acceleration gradient increases quadratically with the decreasing bunch length of the driver beam.

Thirdly, they could provide new possibilities for precision electroweak measurements in hydrogen-like, high-Z atoms for indirect searches of new, Beyond the Standard Model (BSM), effects.

It remains to be stressed that a large fraction of the beam cooling and beam manipulation techniques exploiting the internal degrees of freedom of the beam particles, which have been mastered over the three decades by the atomic physics community, could be directly applied to the high energy PSI beams.

[1] For the discussion of the light sources based on PSI beams see e.g. [1] and the references quoted therein

## 4 The photon collision schemes and secondary beams of the Gamma Factory

The high intensity and high brilliance gamma beam could be used to realize, for the first time, a **photon-photon collider at CERN**: (1) in the range of CM energies of 1 - 100 KeV, for collisions of the gamma beam with the laser photons, and (2) in the energy range of 1 - 800 MeV, for the gamma beam collisions with the counter propagating, twin gamma beam.

The gamma beam could also collide with the LHC proton and fully stripped ion beams. The CM energy range of the corresponding **photon-proton** and **photon-nucleus colliders** would be 4 - 60 GeV.

Finally, the gamma beam could be extracted from the LHC and used to produce high intensity secondary beams of:

- **Polarized electrons and positrons with the expected intensity which could reach $10^{17}$ positrons/s**. Such an intensity would be three orders of magnitude higher than that of the KEK positron source and largely satisfy the source requirements for both the ILC and CLIC colliders, and even that of a future high luminosity ep (eA) collider project based on the energy recovery linac.
- **Polarized muon and the tertiary neutrino beams**. The intensity of the Gamma Factory polarized muon beams could be sizably higher than that of the Paul Scherrer Institute's "$\pi$E5" muon beam. If accelerated, they could be used to produce high intensity neutrino beams. Thanks to the initial muon polarization the muon-neutrino (muon-anti-neutrino) beams could be uncontaminated by the electron-neutrino (electron-anti-neutrino) contributions. The neutrino and antineutrino bunches could be separated with 100 % efficiency on the bases of their timing. In addition, their fluxes could be predicted to a very high accuracy, providing an optimal neutrino-beam configuration for the high systematic precision measurements e.g. of the CP-violating phase in the neutrino CKM matrix. To reach high muon (neutrino) intensities two paths could be envisaged. In the first one, based on the conversion of the high energy gamma beam into muon pairs, the present circumferential voltage of the LHC would have to be upgraded and a specialized design of the gamma conversion targets would have to be made. An alternative scheme would be to tune the gamma beam energy to a significantly lower energy – just above the electron-positron pair production threshold, reducing thus both the circumferential voltage and the beam power strains. The positron bunches, produced by such a low energy gamma beam, would need to be accelerated in the dedicated positron ring to the energy exceeding the muon pair production threshold in collisions with the stationary target electrons, $E_e \sim (2m_\mu^2)/(m_e)$. The intensity of the muon beam produced in such a scheme could be increased by replacing the single-pass collisions of the positron beam by the multipass collisions [5]. For both the above two types of muon beams the product of the beam longitudinal and transverse emittances could be at least four orders of magnitude smaller than that for the pion-decay-originated muon source.
- **Neutrons with the expected intensity reaching $10^{15}$ neutrons/s (first generation neutrons) and radioactive, neutron-rich ions with the intensity reaching $10^{14}$ ions/s**. Preliminary estimates show that the intensity of the Gamma Factory beams of neutrons and radioactive ions could approach those of the European projects under construction like ESS (and FAIR) and the planned EURISOL facility. The Gamma Factory beams may turn out to be more effective in terms of their power consumption efficiency since almost 10 % of the LHC RF power could be converted into the power of the neutron and radioactive ion beams if the energy of the photon beam is tuned to the Giant Dipole Resonance (GDR) region of the target nuclei.

New Gamma Factory beam lines of unprecedented intensities and its high luminosity photon-photon, photon-proton and photon-nucleus collision interaction points could attract new scientific communities to CERN. This could lead to a diversification of the CERN-based scientific programme.

## 5 Expected highlights of the Gamma Factory research programme

The physics research domains which could be explored by this proposal include: fundamental QED measurements (for example, for the first time ever, the elastic light-light scattering could be observed with the rate of $\approx$1000 events/s, providing the high-precision QED test); dark matter searches (mainly via the dark photon and neutron portals); investigation of basic symmetries of the Universe (neutron dipole moment, neutron-antineutron oscillations, forbidden muon decays); studies of color confinement; nuclear photonics; physics of neutron-rich radioactive beams, physics with energy-tagged neutron beam and the vast domain of the atomic physics of muonic and electronic atoms.

The Gamma Factory's high brilliance beams of polarized positron and muons may help in addressing at CERN the research programme of: (1) **a TeV-energy-scale muon collider**; (2) **neutrino factory**, (3) **a lepton-hadron collider**, and (4) **the Deep Inelastic Scattering (DIS) fixed target programme**.

The CERN Gamma Factory project could open a wide spectrum of industrial and medical applications in the following domains: muon catalyzed cold fusion; gamma-beam catalyzed hot fusion; Accelerator Driven System (ADS) and Energy Amplifier (EA) research; nondestructive assay and segregation of nuclear waste; transmutation of nuclear waste; material studies of thick objects and production of ions for Positron Emission Tomography (PET) and for the selective cancer-cell therapy with alpha emitters.

## 6 The way forward

The presented above research option may turn out not only to be scientifically attractive but also cost-effective because it proposes to re-use, in a novel manner, the existing CERN accelerator infrastructure. It may be considered as complementary to the present hadron-collision programme and could be performed at any stage of the LHC life-time.

In order to prove that such a future option is not only a conceptually attractive but also a viable one, two initial exploratory paths have been initiated very recently.

The goal of the first one is to perform a detailed validation of the achievable performance figures of the Gamma Factory initiative for each branch of its application domains, to build up the physics case for its research programme and, most importantly, to attract a wide community to this initiative.

The goal of the second one is to prove experimentally the concepts underlying this proposal. Most of the feasibility tests can be performed at the SPS and organised such that the ongoing CERN research programme is hardly affected. Its initial target is to understand the stability of the PSI beams in the CERN storage rings. The following beam tests are already scheduled (or being discussed):

- test runs with the $^{129}_{54}Xe(+39)$ (P-like) ions in the SPS (2017)
- test runs with the $^{208}_{82}Pb(+54)$ (Ni-like) and $^{208}_{82}Pb(+80)$ (He-like) (or $^{208}_{82}Pb(+81)$ (H-like)) ions in the SPS (2018)
- test runs and a short physics run with $^{208}_{82}Pb(+80)$ (He-like) (or $^{208}_{82}Pb(+81)$ (H-like)) ions in the LHC (2018)

# Axion-like Particles from Primakov production in beam-dumps

*Babette Döbrich*
EP Department, CERN, 1211 Geneva 23

**Abstract**
We discuss searches for Axion-like Particles which are coupled predominantly to photons from proton- or electron beam-dumps. In particular, we scrutinize the present state of exclusions from SLAC 141 in the mass range of ∼10-30 MeV

## 1 ALPs in the MeV–GeV mass range

Whilst we have good reason to believe in the existence of particles beyond the Standard Model, there are many well-motivated but also very different places to look for them: In the sub-eV range, e.g. the QCD axion, which is produced non-thermally in the early universe, is an excellent candidate for Dark Matter. At masses around $\sim$ 1 GeV, WIMPs can reveal themselves by scattering in ultra-low background detectors or be seen indirectly at the LHC, see e.g. [1]. The (so-far) non observation of new particles suggests to probe those energies also under a different view-point: The particles that could be Dark Matter may not couple directly to the Standard Model but through a weakly coupled 'portal', for which different possibilities and motivations exist, see [2] for a review.

In the following, we review some physics results and prospects for a pseudo-scalar Axion-like Particles (ALP) somewhat below the GeV scale. Such an ALP can act as portal [3] and can in specific cases be even useful to solve the strong CP problem [4].

If such an ALP $a$ is predominantly coupled to photons we can write the interaction as:

$$\mathcal{L} = \frac{1}{2}\partial^\mu a\, \partial_\mu a - \frac{1}{2} m_a^2\, a^2 - \frac{1}{4}\, g_{a\gamma} a\, F^{\mu\nu} \tilde{F}_{\mu\nu} \,, \tag{1}$$

where $g_{a\gamma}$ denotes the photon-ALP coupling. A powerful method to search for them is in proton- or electron beam-dumps. In such experiments, ALPs can be created from Primakov production. The advantage in ALP production from coherent proton- or electron-scattering is that the production cross-section can be reliably computed. Particularly the transverse momenta of the ALP which are crucial to be known for an accurate estimate of the ALP acceptance in a beam-dump setup. Additionally, ALP Primakov production in the dump can be boosted with heavier target nuclei. In [5] details of computing the differential cross-sections (with respect to energy and transverse angle) of such ALPs have been worked out. Sensitivity estimates for past and future proton-beam dumps have been provided and compared to the literature on electron-beam dump results.

In the following, let us use the opportunity to revisit the electron beam-dump limit put on ALPs from SLAC 141. This is particularly interesting as upcoming and planned experiments can partially overlap with the parameter space excluded by SLAC 141.

## 2 Limits on ALPs from beam-dumps

Figure 1 (adapted from [5] w.r.t. the SLAC 141 region, as explained in the text below) summarizes presently published exclusions on ALPs in the MeV–GeV mass range. The shape of the regions excluded by individual experiments can be understood as follows: The ALP of Eq. 1 has a lifetime

$$\tau = 64\pi/(g_{a\gamma}^2 m_a^3) \,. \tag{2}$$

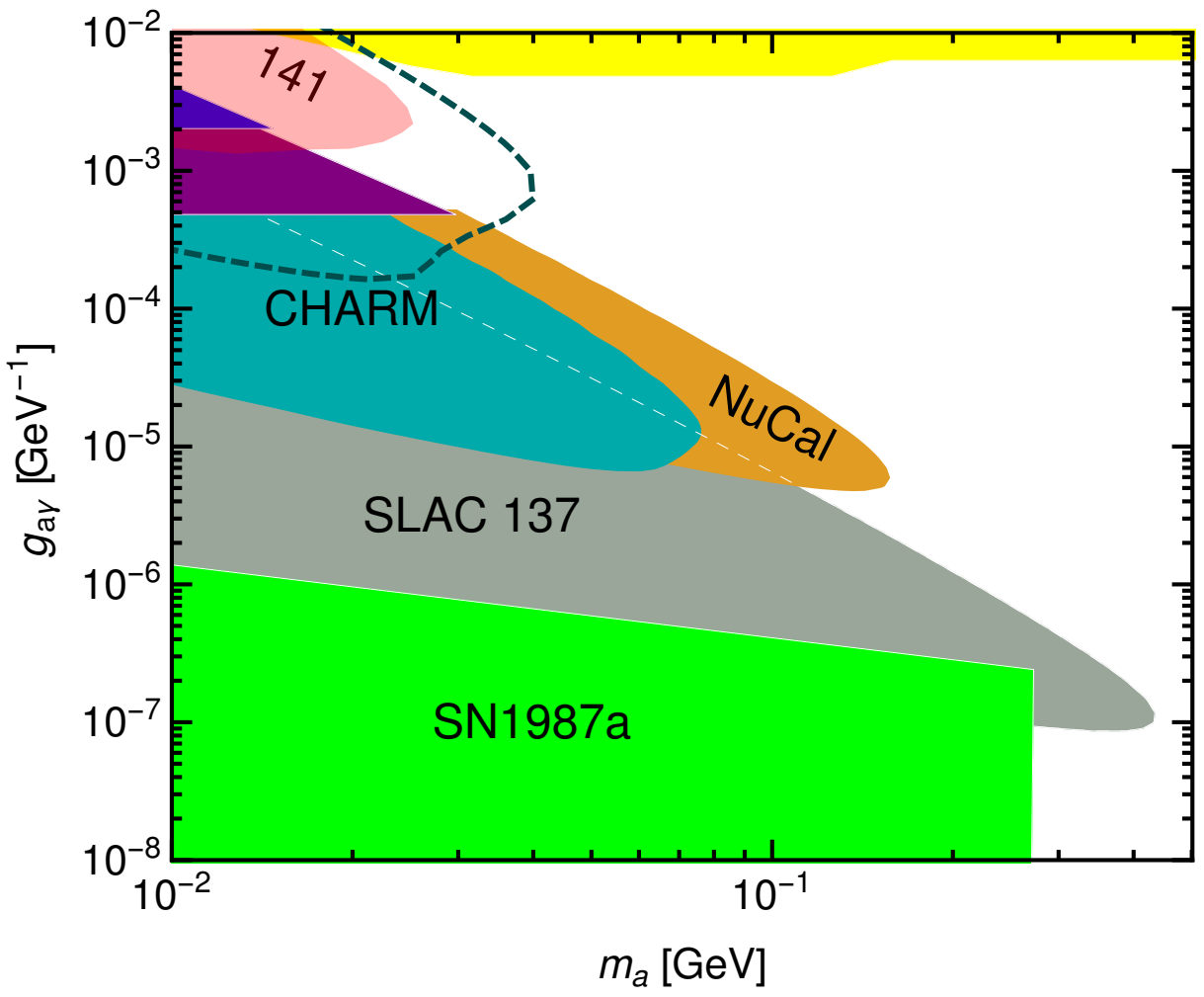

**Fig. 1:** Compilation of limits on ALPs coupled to photons. All limits are shown at 90% C.L. . The compilation is the same as in [5] except for the limits set by SLAC 141: The dark-green dashed line shows its exclusion inferred from a negative result of a $e^+, e^-$ search. The rose region shows exclusion based on an analysis by the collaboration of the same data set.

The requirement that ALPs decay after the beginning of the fiducial region but before its end yield the characteristic nose-shape of the beam-dump exclusion regions. Let us describe these regions in the following.

In Figure 1, 'CHARM' and 'NuCal' are exclusions set from $\sim 2.4 \times 10^{18}$ and $\sim 1.7 \times 10^{18}$ protons on target, respectively. Note, that albeit NuCal had a lower beam energy of 70 GeV than CHARM (400 GeV), the NuCal detector was much closer (64m) to the production point than CHARM (480m) which allows the search for ALPs of much shorter lifetimes. On the other hand, 'SLAC137' is based on $2 \times 10^{20}$ electrons on an aluminum target 200 m upstream of the detector. Details of the limit compilation, references to the experiment papers and a description of the other exclusion regions can be found in [5].

In previous versions of Figure 1, limits from SLAC 141 on ALPs have been shown based on reported negative search results [6] of ALPs decaying to $e^+, e^-$. The experiment used an incident beam of 9 GeV and the spectrometer, located 35m downstream the dump, looked for positrons of energies of minimum 4.5 GeV. This data taking included also the regular insertion of a photon converter such that it had sensitivity to di-photon final states. In the following, we point to a less-known analysis of this data for ALP $\to \gamma, \gamma$ and therewith update the status of exclusion shown previously.

The dark green, dashed line in Figure 1 is the SLAC 141 limit on $g_{a\gamma}$ shown first in a review article of a 2011 community workshop [7], referencing the SLAC 141 publication [6]. The publication [6] reports on a search for light, pseudo-scalar bosons, predominantly coupled to $e^+, e^-$. Based on $\sim 2 \times 10^{15}$ electrons on a tungsten target they excluded such ALPs up to masses of $\simeq$ 20MeV. Part of this data was taken with a photon converter in place as pointed out in the letter.

The rose-shaded labeled '141' in Fig. 1 presents the exclusion limit from Figure 3 of [8]. Ref. [8] is based on the same SLAC 141 data and presents a limit for an ALP coupled to two photons. To translate the results presented in [8] into Figure 1, we have employed a conversion of the ALP lifetime provided in Eq. 2.

As is apparent, the limit on $g_{a\gamma}$ based on the results of [6] seem to be much stronger than the limits reported from the di-photon search, albeit both limits are referring to the same data. The reason for this discrepancy can possibly be understood as follows: As mentioned above, the search for ALPs coupled to two photons was made possible by the insertion of a photon converter. To infer an exclusion limit on ALPs coupled predominantly to photons from the information in [6], however requires the knowledge of the spectrum of the charged particles from the photon converter: Since the cut of 4.5 GeV on the secondary $e^+$ is so close to the beam energy of the dumped beam, the yield is very sensitive to the required minimum ALP energy. It is therefore a non-trivial task to infer limits $g_{a\gamma}$ solely based on information stated in the $e^+, e^-$ search.

The publication of [8] is –to our knowledge– not easily found online, but it is the relevant reference for limits on $g_{a\gamma}$ from SLAC 141. We conclude that for future experiments it can be worth searching the 'white region' below the dash-dotted, green line of Figure 1.

## 3 Future searches for ALP at MeV–GeV

Current and future searches have the prospect of searching the 'blank space' of Figure 1: The fixed target experiment NA62 [9] is currently taking data to measure $K^+ \to \pi^+ \nu\bar{\nu}$, but is in principle able to run in beam-dump mode: Removing the target and 'closing' the (copper) collimator can open up significant discovery potential for ALPs [5]. For a discovery of ALPs, at minimum an analysis for which both photons from the decay are detected, is needed. If tracking of the photons is not possible, searching for any 'Dark Particle' decaying into two photons can be challenging since the decay point cannot be determined when the mass of the decaying particle is unknown. However, it *is* possible to reliably compute the angular distributions of ALPs created from coherent Primakov production. It is useful to exploit that information to single out a potential ALP-signal.

In the near future, depending on the model, e.g. Belle2 [10], the proposed SHiP experiment [11] and the LHC [12,13] have excellent potential of discovering ALPs coupled to photons in that mass range.

In summary, checking for the existence of weakly coupled particles at the MeV–GeV scale deserves dedicated effort. Compiling accurate existing constraints is a crucial input for new searches.

## Acknowledgments

I wish to thank M. W. Krasny for bringing the SLAC 141 results presented at EPS 1987 to our attention. I am grateful to Felix Kahlhoefer for helpful comments on this proceedings article. Also, I would like to thank the organizers of PHOTON 2017 for bringing this great workshop to CERN.

# Status of the light-by-light scattering and the $(g-2)_\mu$

*P. Masjuan*[1] *and P. Sanchez-Puertas*[2]
[1]Grup de Física Teòrica, Departament de Física, Universitat Autònoma de Barcelona, and Institut de Física d'Altes Energies (IFAE), The Barcelona Institute of Science and Technology (BIST), Campus UAB, E-08193 Bellaterra (Barcelona), Spain
[2] Faculty of Mathematics and Physics, Institute of Particle and Nuclear Physics, Charles University in Prague, V Holešovičkách 2, Praha 8, Czech Republic

**Abstract**
In this talk we review the recent progress on the numerical determination of the Hadronic Light-by-Light contribution to the anomalous magnetic moment of the muon and we discuss the role of experimental data on the accuracy of its determination. Special emphasis on the main contribution, the pseudoscalar piece, is made. Gathering recent progress in the light-by-light scattering contribution we consider $a_\mu^{\rm HLBL} = (12.1 \pm 3.0) \times 10^{-10}$ as a good summary of the state-of-the-art calculations which still claims for a $4\sigma$ deviation between theory and experiment for the $(g-2)_\mu$.

**Keywords**
Anomalous magnetic moment of the muon; light-by-light scattering contribution

## 1 Introduction

The anomalous magnetic moment of the muon $(g-2)_\mu$ is one of the most accurately measured quantities in particle physics, and as such is a very promising signal of new physics if a deviation from its prediction in the Standard Model is found.

The present experimental value for $a_\mu = (g-2)_\mu/2$, is given by $a_\mu^{\rm EXP} = 11659209.1(6.3) \times 10^{-10}$, as an average of $a_{\mu^+} = 11659204(7.8) \times 10^{-10}$ and $a_{\mu^-} = 11659215(8.5) \times 10^{-10}$ [1, 2]. Since statistical errors are the largest source of uncertainties, a new measurement with a precision of $1.6 \times 10^{-10}$ is been pursuit at FNAL [3] and JPARC [4], using different experimental techniques.

At the level of the experimental accuracy, the QED contributions has been completed up to the fifth order $\mathcal{O}(\alpha_{em}^5)$, giving the QED contribution $11658471.885(4) \times 10^{-10}$ [5], using the Rydberg constant and the ratio $m_{Rb}/m_e$ as inputs [2]. Also electroweak (EW) and hadronic contributions in terms of the hadronic vacuum polarization (HVP) and the hadronic light-by-light scattering (HLBL) are necessary. The latter represents the main uncertainty in the Standard Model. The common estimates for QED, HVP, HLBL, and EW corrections are collected in Table 1. In this talk, we will update the HLBL contribution.

| Contribution | Result in $10^{-10}$ units | Ref. |
|---|---|---|
| QED (leptons) | $11658471.885 \pm 0.004$ | [5] |
| HVP (leading order) | $690.8 \pm 4.7$ | [6] |
| HVP$^{\rm NLO+NNLO}$ | $-8.7 \pm 0.1$ | [6, 7] |
| HLBL | $11.6 \pm 4.0$ | [8] |
| EW | $15.4 \pm 0.1$ | [9] |
| Total | $11659179.1 \pm 6.2$ | |

**Table 1:** Standard Model contributions to $(g-2)_\mu$.

For the HLBL, two reference numbers can be found in the literature. The one quoted in Table 1 $a_\mu^{\rm HLBL} = (11.6 \pm 4.0) \times 10^{-10}$ [8] but also $(10.5 \pm 2.5) \times 10^{-10}$ [10]. They both imply a discrepancy $\Delta a_\mu = a_\mu^{\rm EXP} - a_\mu^{\rm SM} = (30.0 \pm 8.8) \times 10^{-10}$ of about $3.5\sigma$. The overall HLBL contribution is twice the order of the present experimental error and a third of $\Delta a_\mu$. The striking situation then comes when the foreseen experiments (precision of $1.6 \times 10^{-10}$) would imply the HLBL being a $6\sigma$ effect. On the light of such numbers we really need to understand the HLBL values and their errors since the goal is a HLBL within 10% errors.

The progress on the field is captured in at least three recent dedicated workshops on $(g-2)_\mu$ [11–13] and a newly created $(g-2)_\mu$ theory initiative (https://indico.fnal.gov/conferenceDisplay.py?confId=1379

The results here described update those reported in Ref. [14].

**Table 2:** The HLBL and its different contributions from different references and methods, representing the progress on the field and the variety of approaches considered. † indicates used from a previous calculation. Units of $10^{-11}$.

| Group | HLBL | $\pi, K$ loop | PS | higher spin | quark loop | method |
|---|---|---|---|---|---|---|
| BPP [15] | $+83(32)$ | $-19(13)$ | $+85(13)$ | $-4(3)$ | $+21(3)$ | ENJL, ’95 ’96 ’02 |
| HKS [16] | $+90(15)$ | $-5(8)$ | $+83(6)$ | $+1.7(1.7)$ | $+10(11)$ | LHS, ’95 ’96 ’02 |
| KN [17] | $+80(40)$ | | $+83(12)$ | | | Large $N_c$+$\chi$PT, ’02 |
| MV [18] | $+136(25)$ | $0(10)$ | $+114(10)$ | $+22(5)$ | $0$ | Large $N_c$+$\chi$PT, ’04 |
| JN [8] | $+116(40)$ | $-19(13)\dagger$ | $+99(16)$ | $+15(7)$ | $+21(3)\dagger$ | Large $N_c$+$\chi$PT, ’09 |
| PdRV [10] | $+105(26)$ | $-19(19)$ | $+114(13)$ | $+8(12)$ | $0$ | Average, ’09 |
| HK [19] | $+107$ | | $+107$ | | | Hologr. QCD, ’09 |
| DRZ [20] | $+59(9)$ | | $+59(9)$ | | | Non-local q.m., ’11 |
| EMS [21–23] | | | $+90(7)$ | | | Padé-data driven,’13 |
| EMS [23, 24] | | | $+88(4)$ | | | Large $N_c$ , ’13 |
| GLCR [25] | | | $+105(5)$ | | | Large $N_c$+$\chi$PT, ’14 |
| J [26] | | | | $+8(3)_{\rm axial}$ | | Large $N_c$+$\chi$PT, ’15 |
| BR [27] | | $-20(5)_{\pi\,\rm only}$ | | | | Large $N_c$+$\chi$PT, ’16 |
| MS [28] | | | $+94(5)$ | | | Padé-data driven,’17 |
| CHPS [29] | | $-24(1)_{\pi\,\rm only}$ | | | | Disp Rel, ’17 |

## 2 Dissection of the HLBL and potential issues

The HLBL cannot be directly related to any measurable cross section and requires knowledge of QCD at all energy scales. Since this is not known yet, one needs to rely on hadronic models to compute it. Such models introduce systematic errors which are difficult to quantify. Using the large-$N_c$ and the chiral counting, de Rafael proposed [30] to split the HLBL into a set of different contributions: pseudo-scalar exchange (PS, dominant [8, 10]), charged pion and kaon loops, quark loop, and higher-spin exchanges (see Table 2, notice the units of $10^{-11}$). The large-$N_c$ approach however has at least two shortcomings: firstly, it is difficult to use experimental data in a large-$N_c$ world. Secondly, calculations carried out in the large-$N_c$ limit demand an infinite set of resonances. As such sum is not known, one truncates the spectral function in a resonance saturation scheme, the Minimal Hadronic Approximation (MHA) [31]. The resonance masses used in each calculation are then taken as the physical ones from PDG [2] instead of the corresponding masses in the large-$N_c$ limit. Both problems might lead to large systematic errors not included so far [21, 23, 24, 28, 32]. Results obtained under such assumptions are quoted as Large $N_c$+$\chi$PT in the last column of Table 2.

Actually, most of the results in the literature follow de Rafael’s proposal (see Refs. [8, 10, 15, 18, 19, 21, 23, 25, 27, 29, 33–40], including full and partial contributions to $a_\mu^{\rm HLbL}$) finding values for $a_\mu^{\rm HLbL}$ between basically $6 \times 10^{-10}$ and up to almost $14 \times 10^{-10}$.

Such range almost reaches ballpark estimates based on the Laporta and Remiddi (LR) [41] analytical result for the heavy quark contribution to the LBL. The idea in such ballparks is to extend the perturbative result to hadronic scales low enough for accounting at once for the whole HLBL. The free parameter is the quark mass $m_q$. The recent estimates using such methodology [22, 42–45] found $m_q \sim 0.150 - 0.250$ GeV after comparing the particular model with the HVP. The value for the HLBL is higher than those shown in Table 2, around $a_\mu^{\rm HLBL} = 12 - 17 \times 10^{-10}$, which seems to indicate that the subleading pieces of the standard calculations are not so negligible.

As we said, the Jegerlehner and Nyffeler review [8] together with the *Glasgow consensus* written by Prades, de Rafael, and Vainshtein [10] represent, in our opinion, the two reference numbers. They agree well since they only differ by few subtleties. For the main contribution, the pseudoscalar, one needs a model for the pseudoscalar Transition Form Factor (TFF). They both used the model from Knecht and Nyffeler [17] based on MHA, but differ on how to implement the high-energy QCD constrains coming from the VVA Green's function. In practice, this translates on whether the piece contains a pion pole or a pion exchange. The former would imply that the exchange of heavier pseudoscalar resonances (6th column in Table 2) is effectively included in PS [18], while the latter demands its inclusion. The other difference is whether the errors are summed linearly [8] or in quadrature [10]. All in all, even though the QCD features for the HLbL are well understood [8, 10], the details of the particular calculations are important to get the numerical result to the final required precision. Considering the drawback drawn here, we think we need more calculations, closer to experimental data if possible.

Dispersive approaches [29, 46] relies on the splitting of the former tensor into several pieces according to low-energy QCD, which most relevant intermediates states are selected according to their masses [30, 47]; see Refs. [29] for recent advances. Up to now, only a subleading piece has been computed, cf. Table 2. An advantage we see in this approach is that by decomposing the HLBL tensor in partial waves, a single contribution may incorporate pieces that were separated so far, avoiding potential double counting. The example is the $\gamma\gamma \to \pi\pi$ which includes the two-pion channel, the pion loop, and scalar and tensor contributions. A complete and model-independent treatment would require coupled channel formalism, not developed so far, and a matching to the high-energy region yet to be included. So by now, the calculations are not yet complete, and not yet ready to be added to the rest of contributions.

Finally, for the first time, there have been different proposals to perform a first principles evaluation by using lattice QCD [48]. They studied a non-perturbative treatment of QED which later on was checked against the perturbative simulation. With that spirit, they considered that a QCD+QED simulation could deal with the non-perturbative effects of QCD for the HLBL. Whereas yet incomplete and with some progress still required, promising advances have been reported already [48].

## 3 The role of the new experimental data on the HLBL

The main obstacle when using experimental data is the lack of them, specially on the doubly virtual TFF [49]. Fortunately, data on the TFF when one of the photons is real is available from different collaborations, not only for $\pi^0$ but also for $\eta$ and $\eta'$. It is common to factorize the TFF, i.e., $F_{P\gamma^*\gamma^*}(Q_1^2, Q_2^2) = F_{P\gamma^*\gamma}(Q_1^2, 0) \times F_{P\gamma\gamma^*}(0, Q_2^2)$, and describe it based on a rational function. One includes a modification of its numerator due to the high-energy QCD constraints. Although the high-energy region of the model is not very important, it still contributes around 20%. More important is the double virtuality, especially if one uses the same TFF model (as it should) for predicting the $\pi^0 \to e^+e^-$ decay. Current models cannot accommodate its experimental value (see [50]). The worrisome fact is that modifying the model parameters to match such decay and going back to the HLBL, would result in a dramatic decrease of the HLBL value [50].

While the HLBL requires knowledge at all energies, it is condensed in the $Q^2$ region from 0 to 2 GeV$^2$, in particular above around $0.5$ GeV$^2$. Therefore a good description of TFF in such region is very important. Such data are not yet available, but any model should reproduce the available one. That is why the authors of [21–23, 28], in contrast to other approaches, did not used data directly but the low-energy parameters (LEP) of the Taylor expansion for the TFF and reconstructed it *via* the use of Padé approximants. As demonstrated in Ref. [28], the pseudoscalar TFF driving the PS contributions to the HLBL is Stieltjes functions for which the convergence of the Padé approximants sequence is guaranteed in advanced. As such, a comparison between two consecutive elements in this sequence estimates the systematic error yield by the method. In other words, Padé approximants for the TFFs take full advantage of analyticity and unitary of these functions to correctly extrapolate low- and high-energy regions. The LEPs certainly know about all the data at all energies and as such incorporates all our experimental knowledge at once. This procedure implies a model-independent result together with a well-defined way to ascribe a systematic error. It is the first procedure that can be considered an *approximation*, in contrast to the *assumptions* considered in other approaches. The LEPs were obtained in [21] for the $\pi^0$, in [51] for the $\eta$-TFF and in [52] for the $\eta'$-TFF, taking into account the $\eta - \eta'$ mixing [52, 53] and the determinations of the double virtual $\pi^0$ [50] and $\eta, \eta'$ [54] TFFs. Ref. [28] collects the most updated results for the space- and time-like TFF together with $\gamma\gamma$ decays from 13 different collaborations, and yields the most updated and precise pseudoscalar contribution to the HLBL. The HLBL value from such approach is quoted in Table 2 under EMS and under MS after the double virtual $\pi^0, \eta, \eta'$-TFF were extracted from pseudoscalar decays into a lepton pair [50, 54].

The new pseudoscalar-pole contribution obtained with the Padé method yields $a_\mu^{\mathrm{HLBL,PS}} = (9.4 \pm 0.5) \times 10^{-10}$ [28], which agrees very well with the *old reference* numbers but with an error reduced by a factor of 3. Adding to this quantity the $\pi$ loop $(2.0 \pm 0.5) \times 10^{-10}$ from [27], the axial contribution $(+0.8 \pm 0.3) \times 10^{-10}$ from [26], the scalar contribution $(-0.7 \pm 0.7) \times 10^{-10}$ and the quark loop from [15], and the NLO estimate $(+0.3 \pm 0.2) \times 10^{-10}$ from [55], the HLBL reads:

$$a_\mu^{\mathrm{HLBL}} = (9.9 \pm 1.1) \times 10^{-10} \qquad \text{(errors in quadrature)} \tag{1}$$

$$a_\mu^{\mathrm{HLBL}} = (9.9 \pm 2.5) \times 10^{-10} \qquad \text{(errors linearly added)}\,. \tag{2}$$

The discussion not yet settled is whether one should consider a pseudoscalar-pole or a pseudoscalar-exchange contribution, which is the main difference between the two *old reference* results, see the recent discussion in Ref. [28]. If instead of a pseudsocalar pole, one would consider the pseudoscalar exchange, Ref. [28] tells us that the result MS from Table 2 would grow up to $a_\mu^{\mathrm{HLBL,PS}} = (13.5 \pm 1.1) \times 10^{-10}$. Then, summing up the rest of the contributions (without the quark loop which is effectively included in the PS exchange), the HLBL reads:

$$a_\mu^{\mathrm{HLBL}} = (12.1 \pm 1.5) \times 10^{-10} \qquad \text{(errors in quadrature)} \tag{3}$$

$$a_\mu^{\mathrm{HLBL}} = (12.1 \pm 3.0) \times 10^{-10} \qquad \text{(errors linearly added)}\,. \tag{4}$$

While still marginally compatible, the results from Eqs. (1) and (3) indicate that the role of the dismissed pieces in the standard pseudoscalar-pole framework is as important as subleading contributions which are much larger than the desired global 10% precision. This fact requires, of course, further calculations and focus on missing pieces now that the dominant one is well under control.

In conclusion, the new experimental data used to update the pseudoscalar contribution in Ref. [28] seem to reveal larger contributions from pseudsocalar mesons, meaning that the modeling of the TFF is more important than expected. Also, systematic errors due to both chiral and large-$N_c$ limits are important and difficult to evaluate, but PAs can help. Lattice QCD seems promising but only in the long run. Dispersion relations are useful for subleading terms and at low energies but a consensus will be needed in order to combine such results with those from other contributions are promising. On top of this, the ballpark predictions coincide on drawing scenarios with larger values, indicating in our opinion the need to better understand the process.

## Acknowledgements

P.M is supported by CICYTFEDER-FPA2014-55613-P, 2014-SGR-1450, the CERCA Program/Generalitat de Catalunya, and the Secretaria d'Universitats i Recerca del Departament d'Empresa i Coneixement de la Generalitat de Catalunya. P.S.P. is supported by the Czech Science Foundation (grant no. GACR 15-18080S).

# Exploring pseudo-Nambu-Goldstone bosons in the sub-eV to 10 keV mass range with stimulated photon collider

*Kensuke Homma* *
Graduate School of Science, Hiroshima University, Kagamiyama, Higashi-Hiroshima, Hiroshima 739-8526, Japan

**Abstract**
Generic laboratory searches for pseudo-Nambu-Goldstone bosons (pNGBs) in weak-coupling domains beyond the GUT scale are indispensable for understanding the dark components in the universe. We introduce a novel approach to search for resonantly produced pNGBs coupled to two photons in the mass range from sub-eV to 10 keV by colliding plural laser fields to induce the rare scattering process.

## 1 Introduction

Astronomical observations suggest that the universe is occupied by something invisible: dark matter and dark energy via gravitational effects. The Higgs boson has been recently discovered and we are approaching to the completion of the standard model of elemental particles. However, the standard model particles can explain only 5% of the energy density of the universe. What can we do if we want to make something dark visible? Shining intense light on a dark spot in the vacuum is a quite primitive but natural way to see something unseen. The method we introduce in this paper is essentially based on this quit simple concept. If we assume that something dark can be regarded as a kind of particle or *dark field* which respects the basic principle of particle physics, we may generically express the interaction Lagrangian between a standard model particle and a dark field by introducing two independent parameters: the weak coupling strength and the mass of the exchanged dark field. Among possible standard model fields, we choose photons as the probe fields.

A crucial parameter for generic searches is the energy scale, more specifically the center of mass system (CMS) energy of a photon-photon collision. For instance, the Higgs boson is an example of a scalar resonance state at the highest energy scale, 126 GeV which can decay into two photons. The second example is the neutral pion corresponding to a kind of pseudoscalar resonance state at 135 MeV. This particle can also decay into two photons. We now know at least two energy scales in nature where spin-zero neutral resonance states coupling to two photons surely exist. Since these particles have coupling to two photons, in principle, they can be directly created by photon-photon collisions via the inverse processes.

Spontaneous symmetry breaking can be one of the most robust guiding principles for general discussions of dark components in the universe. Whenever a continuous global symmetry is broken, a massless boson may appear as Nambu-Goldstone-Boson (NGB). In nature, however, an NGB emerges as a pseudo-NGB (pNGB) with a finite mass. The neutral pion is such an example of a pNGB via chiral symmetry breaking. Even though a pNGB is close to being massless, its decay into lighter particles such as photons is kinematically allowed. There are several theoretical models that predict low-mass pNGBs coupling to two photons such as dilatons [1], axions [2], and string-theory-based axion-like particles [3],

*On behalf of SAPPHIRES collaboration

which are relevant to dark components of the universe if the coupling to matter fields is reasonably weak. However, a quantitative prediction on the physical mass of a pNGB in such models is commonly difficult. Indeed, string theories predict pNGBs to be homogeneously distributed on a log scale in the mass range possibly up to $10^8$ eV [3]. Therefore, laboratory tests are indispensable for investigating the physical mass as widely as possible in the lower mass range. The above two evidences of spin-zero boson decaying into two photons encourage us to explore similar types of fields via the two-photon coupling in very different energy scales in general. Especially, in order to produce these resonance states at an lower center-of-mass energy, lower energy colliding beams with massless particles, that is, photon colliders have essential roles.

We have previously advocated a novel method [4] for stimulating $\gamma\gamma \to \phi \to \gamma\gamma$ scattering via an s-channel resonant pNGB exchange by utilizing the coherent nature of laser fields. We have first considered a quasi-parallel colliding system (QPS). This colliding system allows us to reach a low CMS energy in the sub-eV range via the small incident angle even if we use a laser field with its photon energy of 1eV. We also have considered an asymmetric-energy head-on collision system (ACS) [5] to access relatively higher CMS energies in order to discuss a possibility to test an unidentified emission line, $\omega \sim 3.5$ keV, in the photon energy spectra from a single galaxy and galaxy clusters [6, 7] (the arguments are still actively ongoing [8]) with an interpretation of a pNGB decaying into two photons [9]. This motivated us to extend the same method up to 10 keV by combining different types of coherent and incoherent light sources in ACS.

In this paper, we review the basic concept of the proposed method and discuss the future prospect toward direct laboratory searches for pNGBs in the mass range from sub-eV to 10 keV as candidates of the dark components of the universe.

## 2 Generic photon-photon interactions

We assume photon-photon scattering with the following effective Lagrangian in Eq. (1) [4],

$$-L_\phi = gM^{-1}\frac{1}{4}F_{\mu\nu}F^{\mu\nu}\phi, \tag{1}$$

$$-L_\sigma = gM^{-1}\frac{1}{4}F_{\mu\nu}\tilde{F}^{\mu\nu}\sigma, \tag{2}$$

where an effective coupling $g/M$ between two photons and a scalar $\phi$ or pseudoscalar $\sigma$ field is introduced. If we are based on the invisible axion scenario [10], a dark field satisfying the dimensional constant $M = 10^{11} - 10^{16}$ GeV and the mass $m$ = meV - $\mu$eV can be cold dark matter candidates. If $M$ corresponds to the Planck mass $M_P \sim 10^{18}$ GeV, the interaction is as weak as that of gravity and this case would have a great relevance to explain dark energy if $m \geq$ neV [11].

We may discuss possibilities to exchange scalar and pseudoscalar type of fields by requiring combinations of photon polarization in the initial and final states [4]. The virtue of laser experiments is in the capability of specifying photon spin states both in the initial and final states in the two body photon-photon interaction. This allows us to distinguish types of exchanged fields in general.

## 3 Concept of stimulated laser colliders

The novelty of the proposed method is in the following two dominant enhancement mechanisms. The first mechanism is the creation of a resonance state via laser-laser collisions by tuning the CMS energy at a pNGB mass, which is the same approach as that in charged particle colliders. The second mechanism is to stimulate the scattering process by adding another background laser field. This feature has never been utilized in high-energy particle colliders, because controllable coherent fields are not available at higher energy scales above 10 keV. We will explain these two mechanisms in the following subsections in detail.

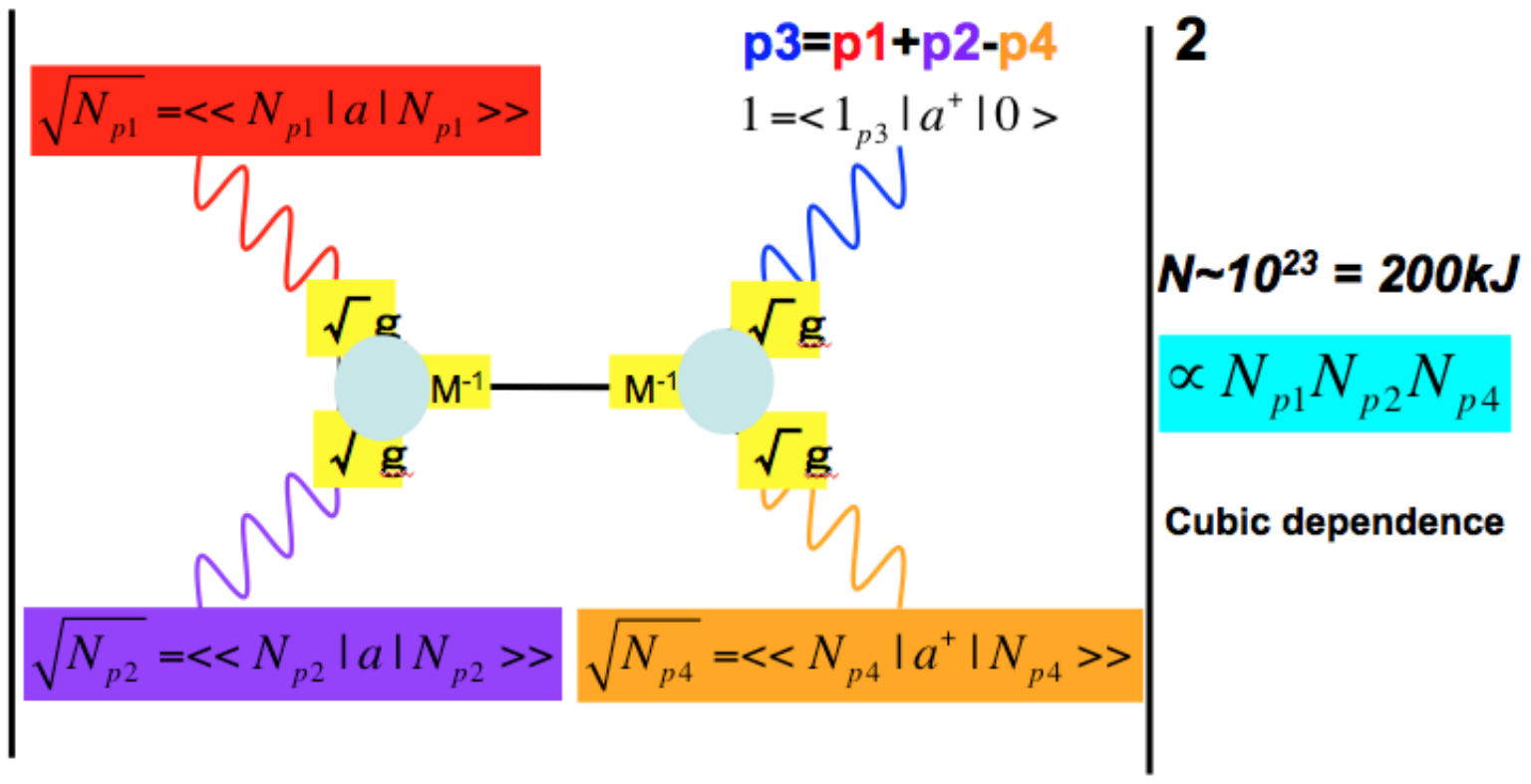

**Fig. 1:** Concept of stimulated photon collider.

### 3.1 Inclusion of a resonance state within an energy uncertainty in photon-photon collisions

A CMS energy, $E_{cms}$, can be generically expressed as

$$E_{cms} = 2\omega \sin\vartheta, \tag{3}$$

where $\vartheta$ is defined as a half incident angle between two incoming photons and $\omega$ is the beam energy in units of $\hbar = c = 1$. This relation indicates two experimental knobs to adjust $E_{cms}$. With $\vartheta = \pi/2$, we can realize a CMS collision. A QPS is realized with a small incident angle by focusing a single laser beam, where $E_{cms}$ can be lowered by keeping $\omega$ constant. We also consider an asymmetric-energy collision in the head-on geometry in order to relatively increase $E_{cms}$ [5]. Both QPS and ACS correspond to Lorentz boosted systems of CMS. A QPS is realized when a CMS is boosted with respect to the perpendicular direction of head-on collision axis, while a ACS is realized when a CMS is boosted in parallel to the head-on collision axis. Owing to these boosted effects, energies of the final state photons are different from any of incident photon energies. Therefore, frequency shifted photons can be clear signatures of photon-photon scatterings if QPS or ACS are realized as laboratory frames.

We then aim at the direct production of a resonance state via s-channel Feynman amplitude in the photon-photon collisions. The square of the scattering amplitude $A$ proportional to the interaction rate can be expressed as Breit-Wigner resonance function [4]

$$|A|^2 = (4\pi)^2 \frac{W^2}{\chi^2(\vartheta) + W^2}, \tag{4}$$

where $\chi$ and the width $W$ are defined as $\chi(\vartheta) \equiv \omega^2 - \omega_r^2(\vartheta)$ and $W \equiv (\omega_r^2/16\pi)(g^2 m/M)^2$, respectively. The energy $\omega_r$ satisfying the resonance condition can be defined as ${\omega_r}^2 \equiv m^2/2(1-\cos 2\vartheta_r)$ [4]. If $\chi^2(\vartheta_r) = 0$ is satisfied, $|A|^2$ approaches to $(4\pi)^2$. This feature is independent of any $W$ in mathematics. However, if $M = M_P$, the width $W$ becomes extremely small. This implies that the resonance width is too narrow to directly hit the peak of the Breit-Wigner function by any experimental ability. How can we overcome this situation? In a QPS realized in a focused laser field, the incident angles or incident momenta of laser photons become uncertain maximally at the diffraction limit due to the uncertainty principle. This implies that $|A|^2$ must be averaged over the possible uncertainty on $E_{cms}$. This unavoidable integration over the possible angular uncertainty results in $W \propto 1/M^2$ dependence of $|A|^2$ compared to the $W^2 \propto 1/M^4$ dependence when no resonance is contained in the energy uncertainty, that is, when $\chi^2(\vartheta) \gg W^2$. In ACS too, a similar uncertainty is expected. We have proposed a ACS with high-intensity pulse lasers [5], where the energy uncertainty is caused by the shortness of the pulse duration time via the uncertainty principle again. In both cases the inclusion of a resonance peak enhances the interaction rate by the huge gain factor of $M^2$.

### 3.2 Stimulated scattering by coherent laser fields

In order to reach the sensitivity to the gravitational coupling strength, however, the inclusion of a resonance state within the uncertainty on $E_{cms}$ is still short. We thus need an additional enhancement mechanism. We then consider the stimulation of the Feynman amplitude by replacing the vacuum state $|0\rangle$ with the quantum coherent state $|N \gg$ [4]. A laser field is represented by the quantum coherent state which corresponds to a superposition of different photon number states, characterized by the averaged number of photons $N$ [12]

$$|N \gg \equiv \exp\left(-N/2\right) \sum_{n=0}^{\infty} \frac{N^{n/2}}{\sqrt{n!}} |n>, \tag{5}$$

where $|n>$ is the normalized state of $n$ photons

$$|n>= \frac{1}{\sqrt{n!}} \left(a^{\dagger}\right)^{n} |0>, \tag{6}$$

with $a^{\dagger}$ and $a$ the creation and the annihilation operators of photons specified with momentum and polarization, respectively. The coherent state satisfies the normalization condition

$$\ll N|N \gg = 1. \tag{7}$$

We can derive following properties of coherent states $|N \gg$ and $\ll N|$:

$$a|N \gg = \sqrt{N}|N \gg \text{ and } \ll N|a^{\dagger} = \sqrt{N} \ll N| \tag{8}$$

from the familiar relations

$$a^{\dagger}|n\rangle = \sqrt{n+1}|n+1\rangle \text{ and } a|n+1\rangle = \sqrt{n+1}|n\rangle. \tag{9}$$

The property in Eq. (8) gives the expectation value of the annihilation and creation operators to coherent states

$$\ll N|a|N \gg = \sqrt{N} \text{ and } \ll N|a^{\dagger}|N \gg = \sqrt{N}. \tag{10}$$

Figure 1 illustrates how the enhancement of the sensitivity arises due to the coherent laser fields. In the production vertex, two incident photons must annihilate from the incident lasers with the momentum $p_1$ and $p_2$, respectively. The expectation values associated with the individual photon legs correspond to the first of Eq. (10). And then if an additional coherent laser field with the momentum $p_4$ is supplied in advance, the expectation value to create a final state photon $p_4$ in the sea of the inducing laser field corresponds to the second of Eq. (10). The overall enhancement factor on the interaction rate to have a signal photon with the momentum $p_3$ is then expressed as

$$(\sqrt{N_{p1}}\sqrt{N_{p2}}\sqrt{1_{p3}}\sqrt{N_{p4}})^2 = N_{p1}N_{p2}N_{p4}, \tag{11}$$

where $N_i$ indicate the average numbers of photons with momenta $p_i$. Because $N_{p_i}$ has no limitation due to the bosonic nature of photons, we can expect a huge enhancement factor by the cubic dependence on the photon numbers. This is in contrast to conventional charged particle colliders where the dependence on the number of particles is quadratic and also there is a physical limitation due to the space charge effect. Compared to the upper number of charged particles, typically $10^{11}$ particles per collision bunch in conventional colliders, Mega Joule laser, for instance, can provide 10 times of Avogadro's number of visible photons per pulse. The cubic of this number results in a enormous enhancement factor on the interaction rates. Therefore, the stimulated photon collider can provide an extremely high sensitivity to feeble couplings compared to that of particle colliders whose prime missions is, of course, to discover new heavy particles even though sacrificing the quadratic luminosity factor.

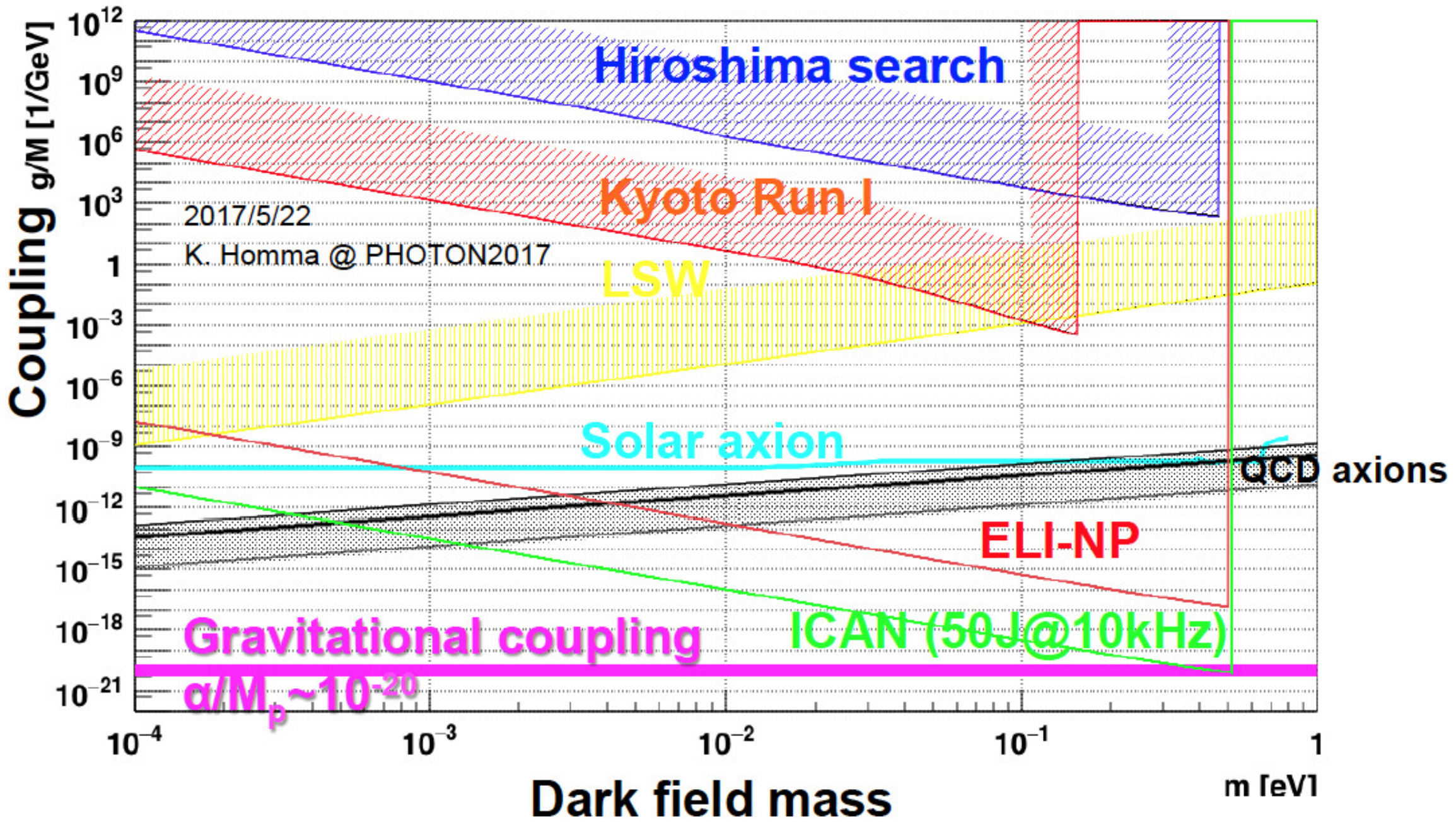

**Fig. 2:** Sensitivity to the mass–coupling domains in QPS.

## 4 Future prospect

The first pilot search based on this method has been already performed in [13]. We then have attempted to apply the same method to higher intensity lasers where background signals from the atomic process in residual gas components appeared and they were successfully suppressed in the experiment [14]. We are now in preparation for further upgrades toward the Extreme-Light-Infrastructure project [15] by forming an international collaboration SAPPHIRES (Search for Axion-like Particle via optical Parametric effects with High-Intensity laseRs in Empty Space) [16, 17] based on the concept introduced here. Figure 2 shows the prospect of sensitivity by searches in QPS where the search in Hiroshima [13], the search in Kyoto [14], and the prospect at the Romanian Extreme-Light-Infrastructure site (ELI-NP) are shown. Figure 3 shows the prospect of sensitivity by searches in ACS. The details of the curves are explained in [5]. In both cases, we foresee that the coupling sensitivities can reach the weakness beyond the GUT scale, $M \sim 10^{16}$ GeV, within the currently available laser technology.

## Acknowledgements

K. Homma acknowledges the support of the Collaborative Research Program of the Institute for Chemical Research, Kyoto University (Grants Nos. 2016–68 and 2017–67) and the Grants-in-Aid for Scientific Research Nos. 15K13487, 16H01100, and 17H02897 from MEXT of Japan.

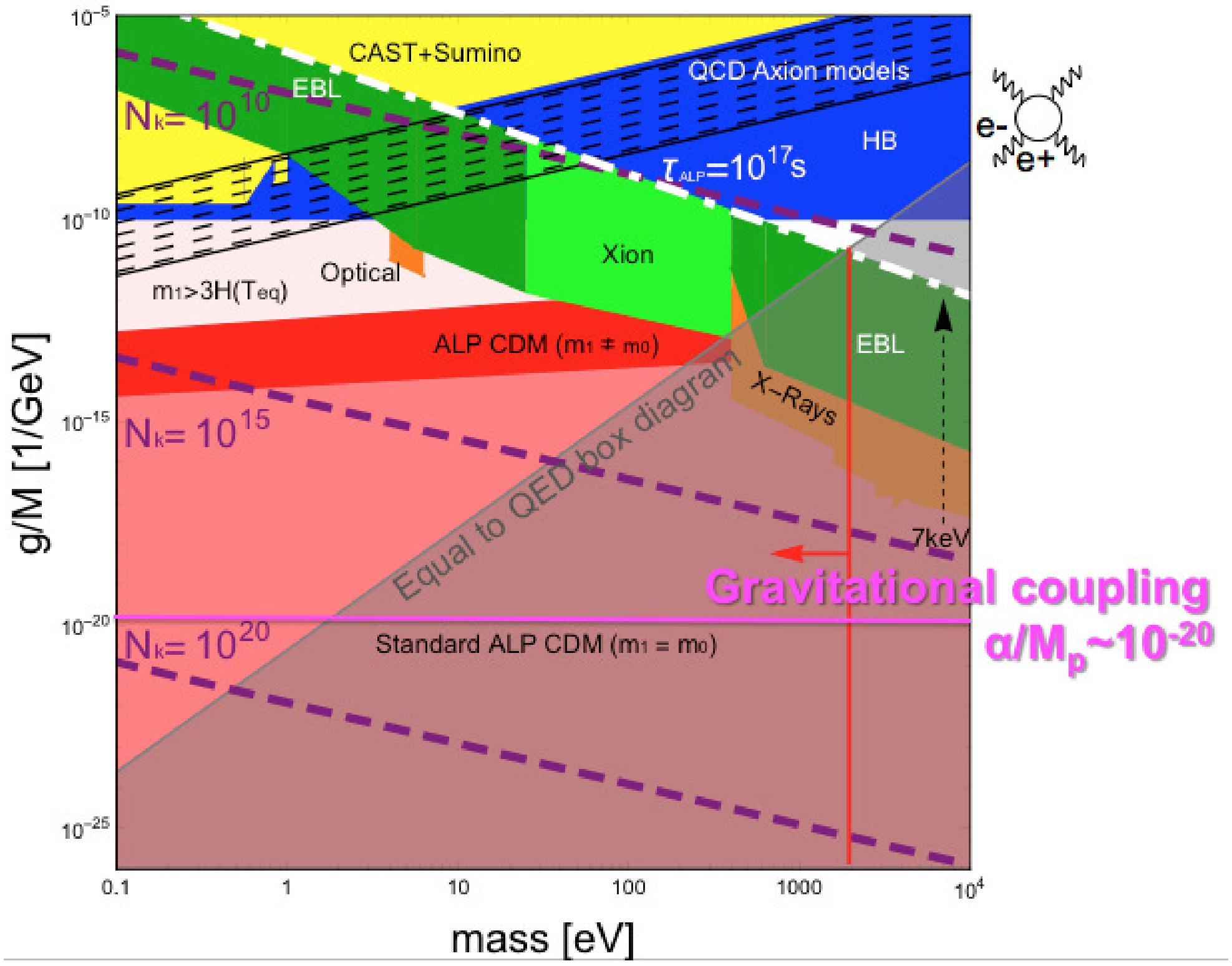

**Fig. 3:** Sensitivity to the mass–coupling domains in ACS.

# Photoproduction with a mini-jet model and Cosmic Ray showers

*F. Cornet* [1], *C. Garcia-Canal* [2], *A. Grau* [1], *G. Pancheri* [3] *, *S. J. Sciutto* [2]
[1] Departamento de Física Teórica y del Cosmos, Universidad de Granada, E-18071 Granada, Spain
[2] Dept. de Física, Univ. Nacional de La Plata, IFLP, CONICET, C.C. 67, 1900 La Plata, Argentina
[3] INFN Frascati National Laboratories, Frascati 00044, Italy

**Abstract**
We present post-LHC updates of estimates of the total photo-production cross section in a mini-jet model with infrared soft gluon resummation, and apply the model to study Cosmic Ray shower development, comparing the results with those obtained from other existing models.

## 1 The total photo-production cross section

We address again the question of how different models for photo-production affect the description of the development of photon-initiated cosmic ray showers and, consequently, how much the estimated photon composition of the showers depends from the models used in the simulation. In this way, we update a previous publication [1]. The simulation of the shower development is performed using the AIRES MC [2], linked to the hadronic models QGSJET-II-04 [3], QGSJET in the following, and EPOS-LHC 3.40 [4], EPOS in the following, that have recently been updated to take into account LHC results. The study includes two different photo-production models:

– the Block and Halzen (BH) model for total cross sections at very high energies [5] presently in AIRES as default model,

– the extension to photoproduction [6] of the so-called Bloch-Nordsieck (BN) model for total hadronic cross sections [7, 8], recently implemented in AIRES.

### 1.1 The Bloch-Nordsieck (BN) model for hadronic processes

This model for the total hadronic cross section is based on a perturbative QCD (pQCD) calculation of mini-jets as being at the origin of the observed rise [9] of the total cross section with energy. For a fixed minimum transverse momentum of the scattering partons, called $p_{tmin} \gtrsim 1$ GeV, the low-x behavior of the PDFs leads to a very strong increase of the minijet integrated cross sections as the c.m. energy increases, shown in the left panel of Figure 1 for different LO Parton Density Functions (PDFs): Glück, Reya and Schienbein (GRS) [10] for the photon, Glück, Reya and Vogt (GRV) [11], and Martin, Roberts, Stirling and Thorne (MRST) [12] for the proton. In the BN model, to be described shortly, infrared gluon resummation tames the fast rise of mini-jet cross sections through soft gluon emissions and can lead to saturation through a phenomenological ansatz for resummation of $k_t \simeq 0$ gluons, which we outline below. Thus the sudden rise around $\sqrt{s} \gtrsim 10 - 20$ GeV morphs into the experimentally observed gentle asymptotic behaviour, which satisfies the Froissart-Martin bound, i.e. $\sigma_{total} \lesssim [\log s]^2$. To this effect, mini-jet collisions are embedded into the eikonal formulation for the total cross section, which, for $pp$ scattering, reads as

$$\sigma_{total} = 2 \int d^2\mathbf{b}[1 - e^{-[\bar{n}_{soft}(b,s)+\bar{n}_{hard}(b,s)]/2}] \quad (1)$$

*Research affiliate with MIT CTP, Cambridge, MA, USA

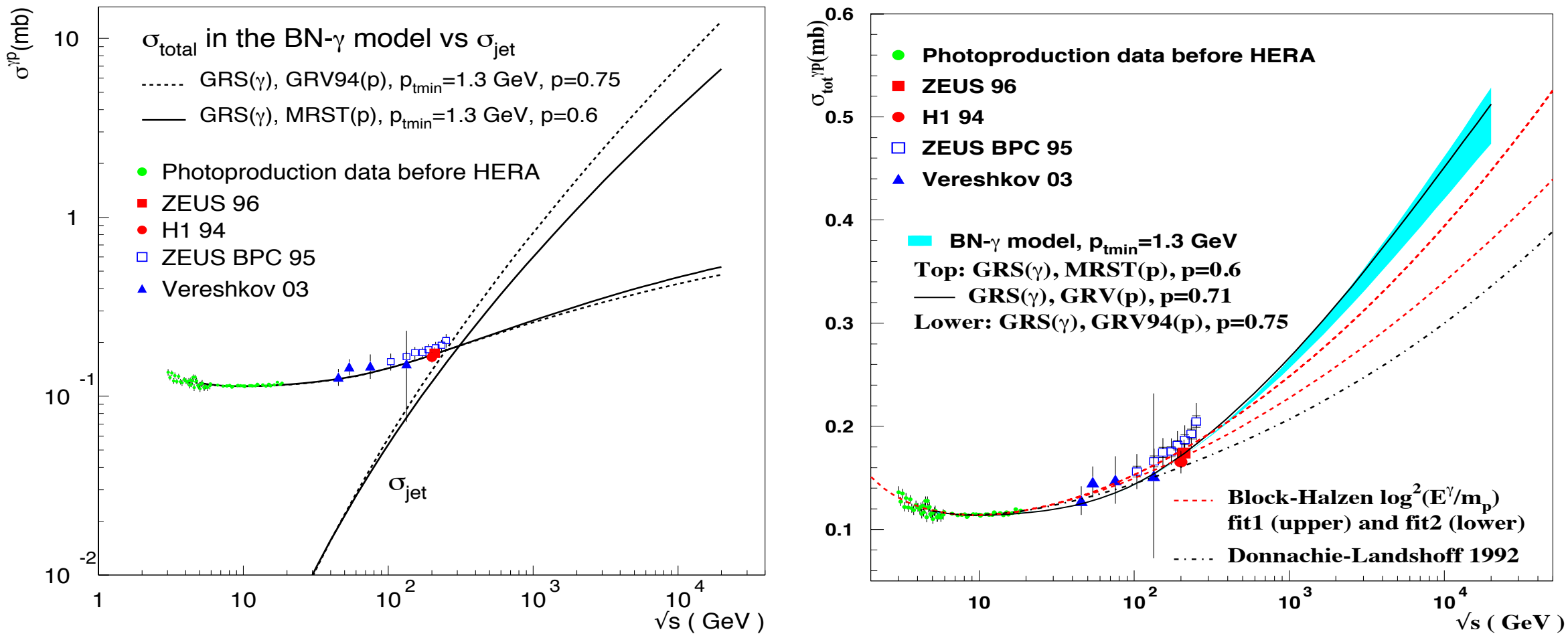

**Fig. 1:** Left panel: Mini-jet integrated cross sections for different PDFs in $\gamma p$ scattering, in comparison with $\gamma p$ total cross section data and the BN model described in the text. Right panel: Total $\gamma p$ cross section calculated for different PDFs with the BN model, called here BN-$\gamma$, and compared with the model of Ref. [5], with lower red dashed curve being the default in AIRES, and the Donnachie and Landshoff description of Ref. [13].

Here $\bar{n}_{hard}(b,s)$ is to be calculated through the LO, DGLAP evolved QCD jet cross section, while our choice for $\bar{n}_{soft}$ is to parametrize it by normalizing $\sigma_{total}$ at low energy, i.e. before pQCD mini-jet production takes up a major role. Phenomenology suggests that this quantity is either a constant or decreases with energy. On the other hand, the main point of the model used here is that $\bar{n}_{hard}(b,s)$ should be fully estimated by means of a pQCD calculation, with saturation effects due to an All Order Resummation procedure, which includes the infrared region.

Once pQCD can be applied, a complete description requires not only the calculation of hard parton-parton scattering but also soft gluon effects accompanying the collision. If, in Eq. (1), we write

$$\bar{n}_{hard}(b,s) = A_{hard}(b,s)\sigma_{jet}(s;p_{tmin},PDFs) \tag{2}$$

then $A_{hard}(b,s)$ will include soft gluon resummation effects, and thus account for the cut-off in impact parameter space, required for satisfaction of the Froissart bound, i.e. the saturation effects.

Our model for resummation of soft gluons is based on a semi-classical approach in which one does not count individual gluons, hence no branching or angular ordering is involved, and follows the Bloch and Nordsieck observation [14] that soft photons emission follows a Poisson distribution and only an infinite number of them can give a finite cross section. The procedure to apply this result to QCD was first discussed in [15] and recently outlined in Ref [16]. Labelling as $A_{BN}$ the impact parameter distribution of partons to use in Eq. (2), the resulting expression is the following:

$$A_{BN}(b,s;p,p_{tmin}) = \frac{e^{-h(b,s;p,p_{tin})}}{\int d^2\mathbf{b}e^{-h(b,s;p,p_{tmin})}} \equiv \tag{3}$$

$$\mathcal{N}(s,p_{tmin})\int d^2\mathbf{K_t}e^{-i\mathbf{K}_t\cdot\mathbf{b}}\frac{d^2P_{soft-resum}(\mathbf{K_t},s;p_{tmin})}{d^2\mathbf{K_t}} \tag{4}$$

$$\tag{5}$$

with

$$h(b,s;p,p_{tmin}) = \frac{8}{3\pi^2}\int_0^{q_{max}} d^2\mathbf{k}_t[1-e^{i\mathbf{k}_t\cdot\mathbf{b}}]\alpha_s(k_t^2)\frac{\ln(2q_{max}/k_t)}{k_t^2} \tag{6}$$

$$q_{max}(s;p_{tmin},PDF) = \frac{\sqrt{s}}{2}\frac{\sum_{i,j}\int\frac{dx_1}{x_1}f_{i/a}(x_1)\int\frac{dx_2}{x_2}f_{j/b}(x_2)\sqrt{x_1x_2}\int_{z_{min}}^1 dz(1-z)}{\sum_{i,j}\int\frac{dx_1}{x_1}f_{i/a}(x_1)\int\frac{dx_2}{x_2}f_{j/b}(x_2)\int_{z_{min}}^1(dz)} \tag{7}$$

where the integration in Eq. (6) makes use of an ansatz of maximal singularity for the behavior of the coupling of infrared gluons to the source current, namely $\alpha_s(k_t \to 0) \propto (k_t^2/\Lambda^2)^{-p}$ with $p < 1/2 < 1$ [15], and $q_{max}$ is calculated from the kinematics of single gluon emission [17].

In Reference [6] the model described above was applied to photon processes, using photon PDFs, and a suitable parametrization for the probability $P_{had}$ that a photon behaves like a hadron, following the simple, but effective model proposed in Ref. [18]. In [6] the total photo-production cross section was then calculated as follows:

$$\sigma_{total}^{\gamma p} = 2P_{had} \int d^2\mathbf{b}[1 - e^{-[2/3\ \bar{n}_{soft}^{pp}(b,s) + \bar{n}_{hard}^{\gamma p}(b,s)]/2}] \tag{8}$$

with

$$\bar{n}_{hard}^{\gamma p}(b,s) = \frac{A_{BN}^{\gamma p}(b,s;p,p_{tmin},PDF)\sigma_{jet}^{\gamma p}(s;p_{tmin},PDF)}{P_{had}}. \tag{9}$$

The probability $P_{had}$ can be extracted from Vector Meson Dominance models, and adjusting it to the normalization of data at low energy, we propose $P_{had} = 1/240$.

After determining the best set of parameters $\{p, p_{tmin}\}$ compatible with existing $\gamma p$ data, we show in the right hand panel of Figure 1 our present calculation [1] for the photo-production total cross section, updated from [6] with more recent sets of proton PDFs, and compare it with the results of the models from Refs. [5] and [13].

## 2 Shower development with post LHC AIRES simulations

We have performed simulations of extended air showers using the AIRES system [2] linked to the packages QGSJET-II-04 [3] and EPOS 3.40 [4] for processing high energy hadronic interactions. The versions used for both hadronic models are those optimized taking into account the recent results of LHC experiments (for details see, for example, references [4, 19]).

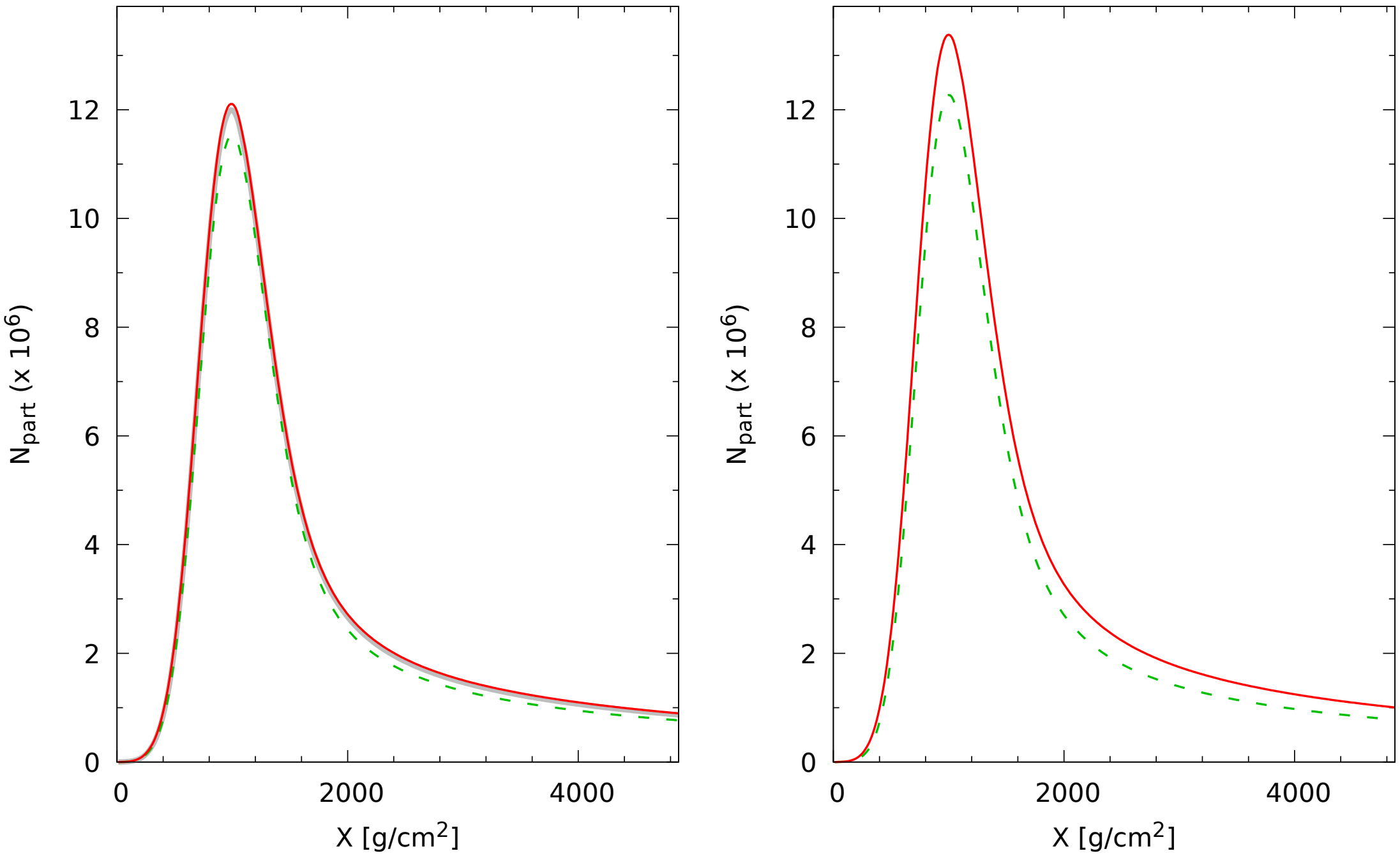

**Fig. 2:** Longitudinal development of muons for $10^{19}$ eV photon showers inclined 80 degrees. The solid (dashed) lines correspond to simulations with the BN-$\gamma$ (BH) model for photonuclear cross section, and processing high energy hadronic interactions with the QGSJET-II-04 (left) and EPOS (right) models.

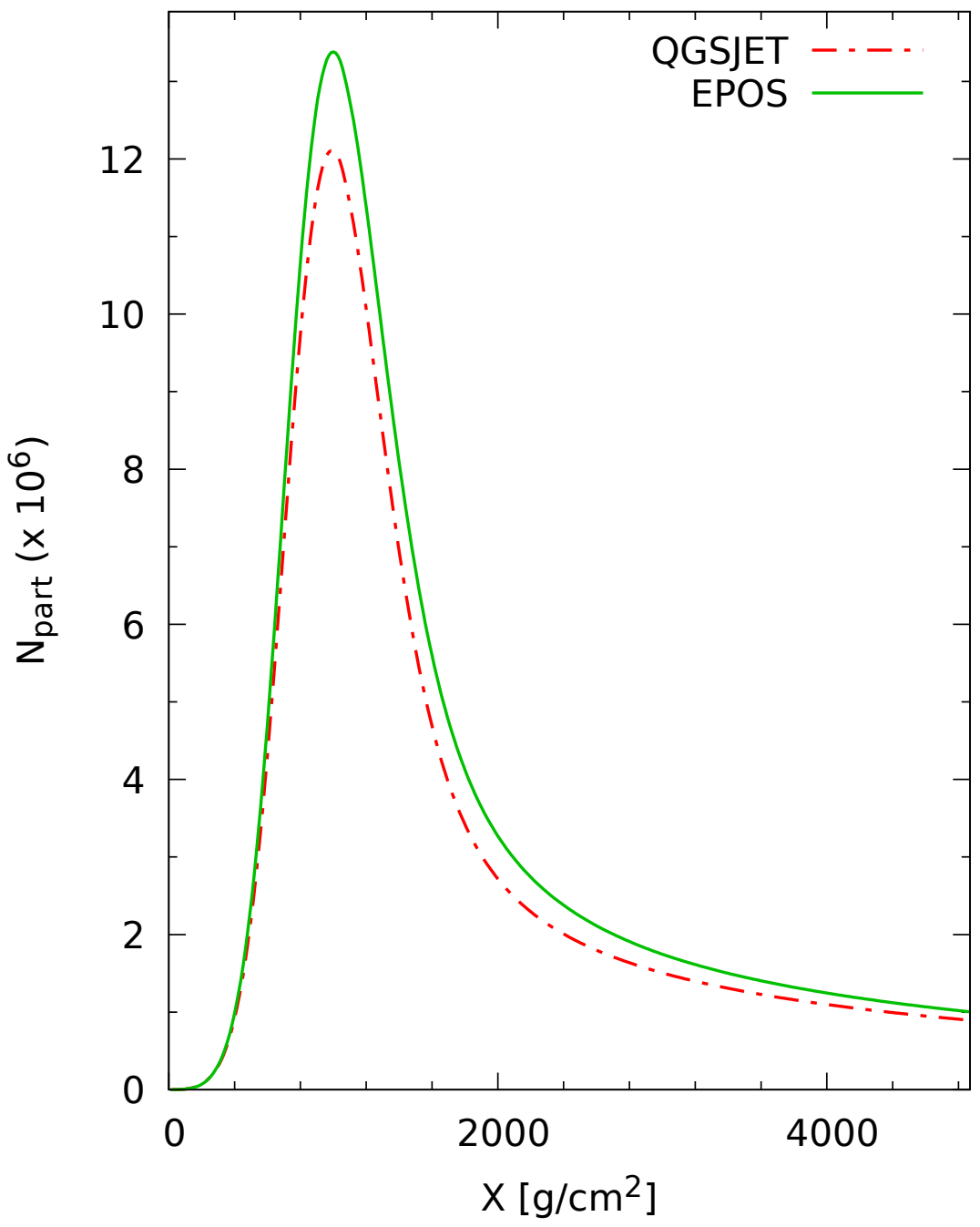

**Fig. 3:** Longitudinal development of muons for $10^{19}$ eV photon showers inclined 80 degrees from QGSJET-II-04 or EPOS 3.40 hadronic packages in the case of the BN-$\gamma$ model for $\gamma p$.

With each of the mentioned hadronic models we have run two sets of simulations, namely, (1) using the cross sections for photonuclear reactions at energies greater than 200 GeV that are provided with the currently public version of AIRES which correspond to the lower of the two BH fits in the right panel of Figure 1 (model BH introduced in Section 1); and (2) replacing those cross sections by the ones corresponding to the present model [6] (model BN-$\gamma$ introduced in Subsection 1.1).

Following the lines of our previous work [1], we report here on the results for the very representative case of $10^{19}$ eV gamma-initiated showers. As already pointed out in [1], at this primary energy, geomagnetic conversion [20] is not frequent, thus allowing photons to initiate normally the atmospheric shower development.

In Figures 2 and 3 the results for the longitudinal development of the mean number of muons is displayed, while in Figure 4 we present the results for the cases of pions and kaons.

For all the secondary particles considered, and for both hadronic models QGSJET and EPOS, the production of muons and hadrons is larger in the case of the simulations using the BN-$\gamma$ model for photon cross sections, in comparison with the corresponding BH model results, as expected (see discussion in reference [1]). This shows up clearly in Figure 2 where the plots corresponding to the present model (solid red lines) are located always over those corresponding to the BH model (dashed green lines).

It is very important to notice that the number of hadrons produced during the shower development depends noticeably on the package used to process the hadronic collisions. Despite the fact that both hadronic packages have been tuned to best reproduce measurements performed at the LHC collider [4, 19], it is evident from the plots presented here that differences between models still persist. For example, the longitudinal profiles obtained using EPOS contain larger number of hadrons when compared with the corresponding ones for the case of QGSJET. Using model BN-$\gamma$ for photon cross sections (red solid lines in Figure 4), the number of pions at the point of maximum development is 11% larger for EPOS with respect to QGSJET. In the case of kaons such figure is only 1%, and for protons (neutrons) (not plotted) at the point of maximum development is 11% (8%) larger for EPOS-LHC with respect to QGSJET. As shower muons are generated after the decay of unstable hadrons, a similar increase can be seen for the maximum number of muons (Figure 2) where EPOS overpasses the prediction of QGSJET-II in 6%, as exemplified in Figure 3 for the case of muons when using model BN-$\gamma$ for photon cross section.

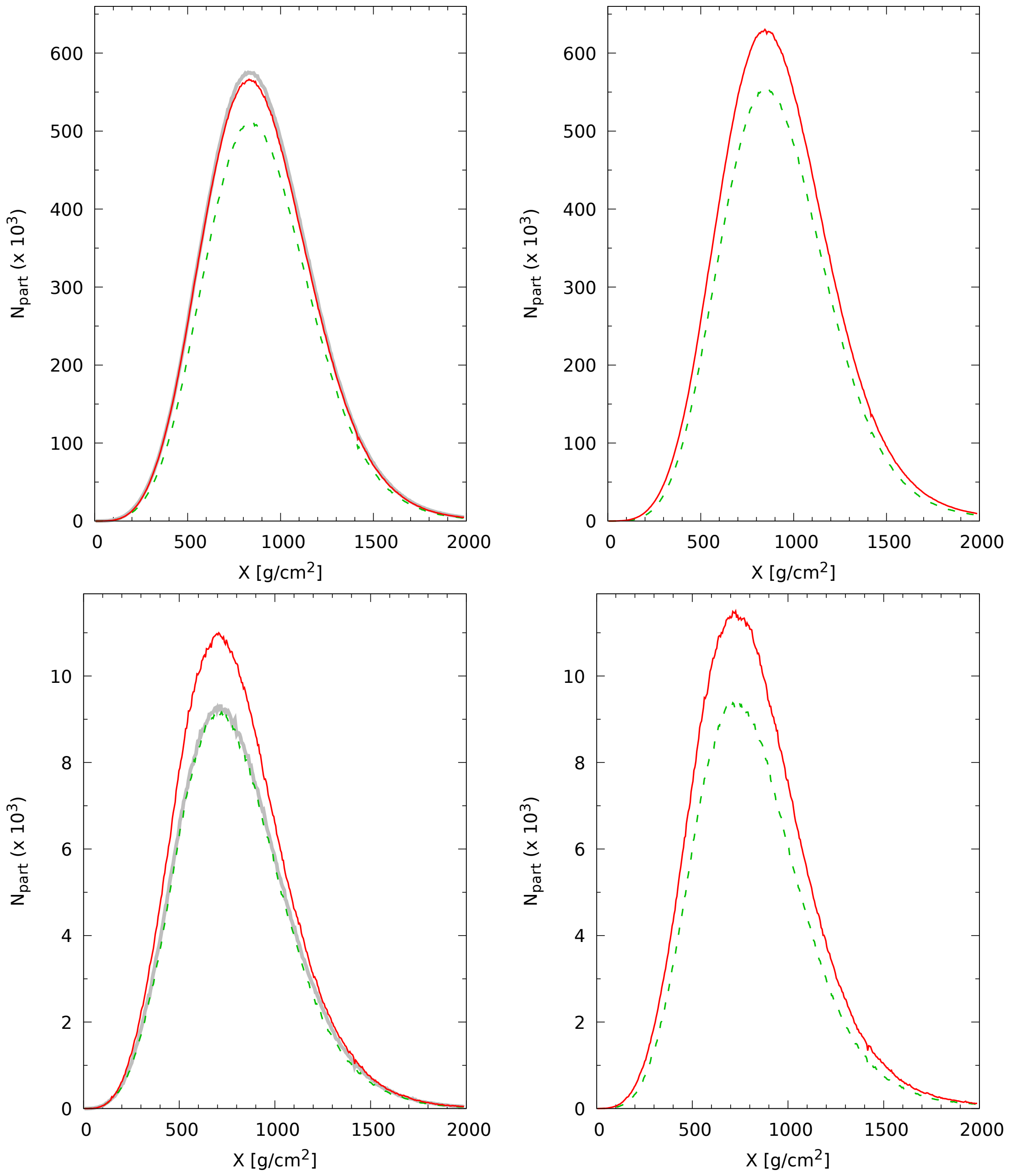

**Fig. 4:** Longitudinal development of pions (upper row) and kaons (lower row) for $10^{19}$ eV photon showers inclined 80 degrees. The solid (dashed) lines correspond to simulations with the present BN-$\gamma$ (BH) model for photonuclear cross section, and processing high energy hadronic interactions with the QGSJET-II-04 (left) and EPOS (right) models. The grey line corresponds to similar simulations performed using the BN-$\gamma$ model for photonuclear cross section and the (pre-LHC) QGSJET-II-03 model (see figure 6 of reference [1])

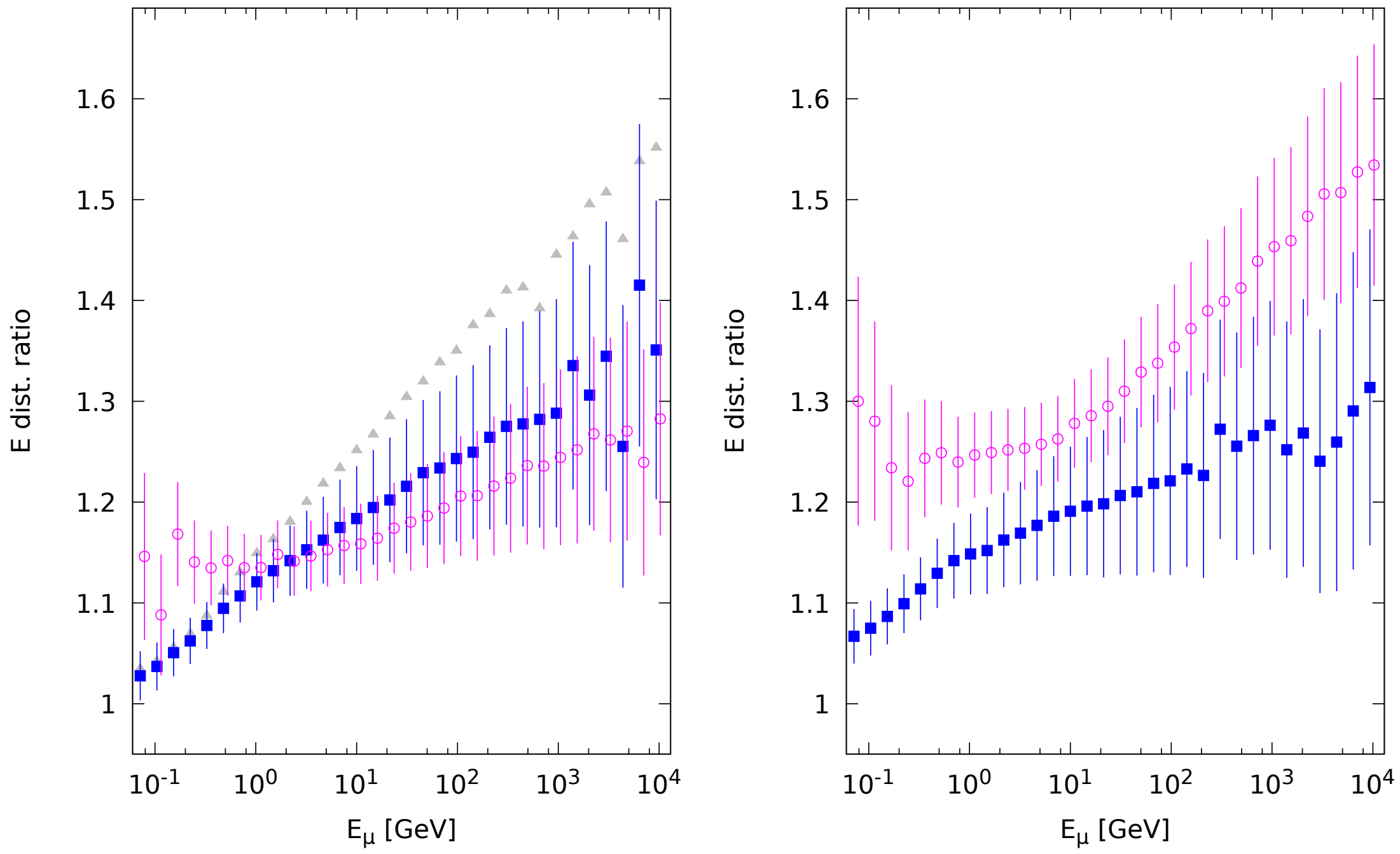

**Fig. 5:** Ratio between ground muon energy distributions obtained with the present (BN-$\gamma$) and old (BH) models, for $10^{19}$ eV photon showers, and simulating the hadronic simulations with QGSJET-II-04 (left) and EPOS (right). The solid squares (open circles) correspond to a shower inclination of 45 (80) deg. Error bars are calculated by propagation of the individual RMS statistical errors of each of the distributions. The abscissas of the 80 deg data set have been shifted by 10% to improve error bar visibility. The grey triangles correspond to similar simulations performed using the present BN-$\gamma$ model for photonuclear cross section and the (pre-LHC) QGSJET-II-03 model for showers inclined 45 degrees (see figure 10 of reference [1]).

Another point that is important to check is the difference between pre and post-LHC versions of each model. In Figure 4 we have included (grey lines in left column plots) the results corresponding to the present model for photon cross sections, but simulated using QGSJET-II-03 (pre-LHC version of QGSJET). These grey lines correspond to the respective solid red lines plotted in figures 5 and 6 of reference [1]. It can be observed that there are virtually no differences between the pre and post-LHC models in all the cases, except, remarkably, in the case of kaons (Figure 4 lower left plot), where the post-LHC model QGSJET-II-04 predicts a noticeably larger number of kaons.

Another quantity that we have included in our study is the ratio of energy distributions of muons reaching ground level (see the discussion on this quantity at reference [1]). Such ratios are plotted in Figure 5 as functions of the muon energy for the representative cases of 45 (solid squares) and 80 (open circles) degrees of inclination. The increased number of high energy muons resulting after the simulations using the BN-$\gamma$ model for photonuclear cross sections shows up clearly. In the case of QGSJET (left) the increase of the number of muons is not much dependent on the inclination of the shower, and is noticeably smaller compared with the results corresponding to QGSJET-II-03 (grey triangles), already studied in reference [1]. In the case of EPOS-LHC (right) the results are somewhat different with respect to the QGSJET ones. In this case the very inclined showers (80 degrees) present the largest increase rate, reaching nearly 50% for the most energetic muons, while for the showers inclined 45 degrees, the distribution ratios are about 25% for the most energetic muons.

The differences between the results obtained in our analysis using different hadronic models, indicate that the simulation of hadronic collisions at the highest energies continue to be an open issue, more than 30 years after the first simulations were reported, and despite all the experimental data that have been collected since then.

## Acknowledgements

G.P. gratefully acknowledges hospitality at the MIT Center for Theoretical Physics. This work was partially supported by CONICET and ANPCyT, Argentina. A.G. acknowledges partial support by the Ministerio de Economía y Competitividad (Spain) under grant number FPA2016-78220-C3-3-P, and by Consejería de Economía, Innovación, Ciencia y Empleo, Junta de Andalucía (Spain)(Grants FQM 101 and FQM 6552). F. C. also acknowledges support by the Ministerio de Economía y Competitividad (Spain) under grant number FPA 2016-78220-C3-1-P, and by Consejería de Economía, Innovación, Ciencia y Empleo, Junta de Andalucía (Spain)(Grants FQM 330 and FQM 6552).

# Very forward photon production in proton-proton collisions at $\sqrt{s}$ = 13 TeV measured with the LHCf experiment

*A. Tiberio, on behalf of the LHCf Collaboration*
University of Florence and INFN, Florence, Italy

**Abstract**
The latest physics results of LHCf photon analysis are reported in this paper. The inclusive energy spectrum of photons produced in proton-proton collisions at $\sqrt{s}$ = 13 TeV are shown. The results are compared with the prediction of hadronic interaction models commonly used in air-shower simulations of ultra-high-energy cosmic rays. Even if experimental data lie between the models, there is not a single model consistent with data in all the energy range. The better overall prediction is given by EPOS-LHC hadronic interaction model. The electromagnetic energy flow as a function of pseudorapidity is also presented, showing a good agreement of EPOS-LHC and SIBYLL 2.3 models with data.

## 1 Introduction

The LHC-forward experiment (LHCf) has measured neutral particles production in a very forward region in proton-proton and proton-lead collisions at the Large Hadron Collider. The main purpose of LHCf is to improve hadronic interaction models of Monte Carlo (MC) simulations used in cosmic rays indirect measurements. Highest energy cosmic rays can only be detected from secondary particles which are produced in the interaction of the primary particle with nuclei of the atmosphere, the so-called air showers. Studying the development of air showers, it is possible to reconstruct the type and kinematic parameters of primary particles. To reproduce the development of air showers, MC simulations with accurate hadronic interaction models are needed. Since the energy flow of secondary particles is concentrated in the forward direction, measurements of particle production in the high pseudorapidity region (i.e. small angles) are very important. The energy flow as a function of pseudorapidity is shown in Fig. 1, where also the coverage of LHCf experiment is reported. In the very forward region soft QCD interactions dominates and MC simulations of air showers utilize phenomenological models base on Gribov-Regge theory [1, 2]. Therefore, inputs from experimental data are crucial for the tuning of that models. The LHC accelerator gives the possibility to study a wide range of collision energies, from 0.9 TeV to 13 TeV in the center of mass frame, which corresponds to an energy range in the laboratory frame from $10^{14}$ eV to $0.9 \times 10^{17}$ eV. This energy range covers the "knee" region of cosmic rays spectrum, which occurs around $10^{15}$ eV. Data collected by central detectors with 7 TeV collisions were already used for the tuning of hadronic interaction models widely used in air shower simulations (EPOS-LHC [3], QGSJET II-04 [4] and SIBYLL 2.3 [5]). However, discrepancies between observed data and MC simulations were observed also with these models [6].

## 2 The detector

LHCf is composed of two independent detectors, called *Arm1* and *Arm2*. Arm1 is located 140 meters away from ATLAS interaction point (IP1) in the IP8 direction, while Arm2 is placed 140 meters away from IP1 in the opposite direction (toward IP2). Detectors are placed inside Target Neutral Absorber (TAN), where the beam pipe turns into two separates tubes. Since charged particles are deviated by the D1 dipole magnet (which bends colliding beams into the two separate beam pipes), only neutral particles, mainly photons and neutrons, reach the detector.

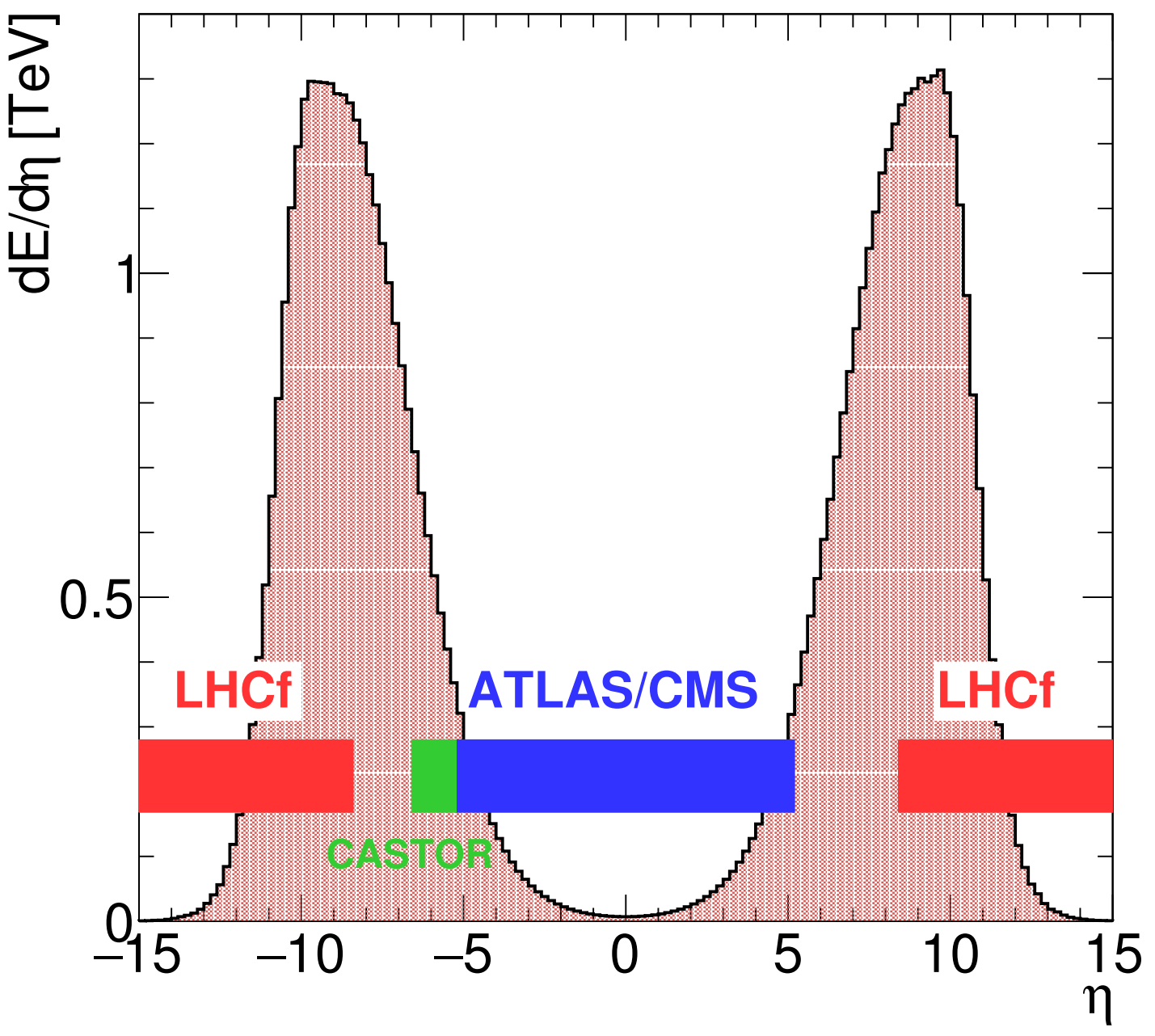

**Fig. 1:** Pseudorapidity dependence of the energy flow of secondary particles produced in a proton-proton collision at $\sqrt{s}$ = 13 TeV. The coverage of several LHC experiments is also shown.

Each detector is made of two sampling and imaging calorimeters (called *towers* hereafter). Each tower is composed of 16 tungsten layers and 16 scintillator layers to measure the energy deposit and it also contains 4 position sensitive layers. During 0.9 TeV, 2.76 TeV and 7 TeV operations at the LHC, plastic scintillators (EJ-260) were used. Arm1 detector used scintillating fiber (SciFi) to measure position, while Arm2 used silicon microstrip detectors. For 13 TeV operation both detectors were upgraded: all the plastic scintillators were replaced by $Gd_2SiO_5$ (GSO) scintillators because of their radiation hardness; also the Arm1 SciFi were replaced by GSO bars. In Arm2 the signal of silicon detectors was reduced using a new bonding scheme of the microstrips to avoid saturation of readout electronics due to the higher energy deposit expected at $\sqrt{s} = 13$ TeV.

Transverse cross sections of towers are $20 \times 20$ mm$^2$ and $40 \times 40$ mm$^2$ for Arm1 and $25 \times 25$ mm$^2$ and $32 \times 32$ mm$^2$ for Arm2. Longitudinal dimension of towers is of 44 radiation lengths, which correspond to 1.6 nuclear interaction lengths. Energy resolution is better than 2% for photons above 200 GeV and of about 40% for neutrons. Position resolution for photons is 200 $\mu$m and 40 $\mu$m for Arm1 and Arm2, respectively, while position resolution for neutrons is of about 1 mm. Smaller tower of each detector is placed on the beam center and covers the pseudo-rapidity range $\eta > 9.6$, while larger tower covers the pseudo-rapidity range $8.4 < \eta < 9.4$. More detailed descriptions of detector performance are reported elsewhere [7–10].

## 3 Photon analysis results from $\sqrt{s}$ = 13 TeV run

Results for inclusive photon energy spectrum in p-p collisions at $\sqrt{s}$ = 900 GeV and 7 TeV have already been published [11, 12]. Proton-proton collisions at $\sqrt{s} = 13$ TeV were produced for the first time in 2015 at LHC. LHCf had a dedicated low-luminosity run from 9th to 13th of June 2015, with an instantaneous luminosity of $0.3 \div 1.6 \times 10^{29}$cm$^{-2}$s$^{-1}$. The data sample used in this analysis was obtained during the LHC Fill #3855 in a 3-hours run started at 22:32 on July 12. The instantaneous luminosity (measured by ATLAS [13]) ranged from 3 to 5 $\times 10^{28}$cm$^{-2}$s$^{-1}$, with 29 colliding bunches, an half crossing angle of 145 $\mu$rad and a $\beta$* of 19 m. The integrated luminosity was 0.191 nb$^{-1}$ for both Arm1 and Arm2, after correcting for data-acquisition live-time. The total inelastic cross section ($\sigma_{inel}$) was extrapolated from TOTEM results at 8 TeV using the best fit of $\sigma_{inel}$ vs $\sqrt{s}$ relation done by the COMPETE collaboration [14, 15].

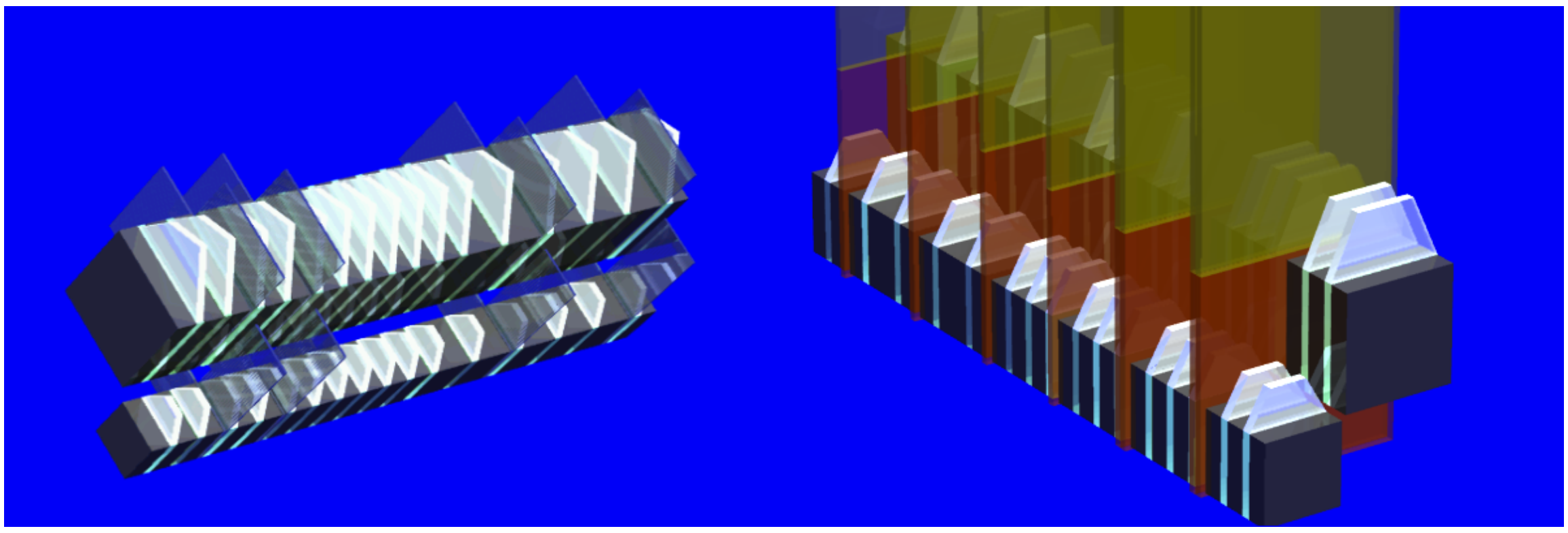

**Fig. 2:** Schematic view of Arm1 (left) and Arm2 (right) detectors. Tungsten absorbers are shown as dark grey layers and GSO scintillators as cyan layers for both detectors. Blue and red layers represent Arm1 GSO bars and Arm2 silicon detectors, respectively.

### 3.1 Photon inclusive energy spectrum

The inclusive energy spectrum of photon produced in p-p collisions at $\sqrt{s}$ = 13 TeV is presented in Fig. 3 for two pseudo-rapidity ranges together with the predictions of DPMJET 3.06 [16, 17], EPOS-LHC, PYTHIA 8.212 [18, 19], QGSJET II-04 and SIBYLL 2.3 hadronic interaction models. The LHCf data lie between MC predictions but there is not an unique model with a good agreement in the whole energy range and in both rapidity regions. In the pseudorapidity range $\eta > 10.94$, QGSJET and EPOS presents a good overall agreement with experimental data; SIBYLL predicts a lower yield of photons, even if it features a shape similar to data; PYTHIA spectrum agrees with data until $\sim$3.5 TeV but become harder at higher energies; DPMJET is generally harder than data. In the pseudorapidity range $8.81 < \eta < 8.99$, EPOS and PYTHIA spectra agree with data until $\sim$3 TeV, while they become harder at higher energies; SIBYLL has a good agreement until $\sim$ 2 TeV, then also it becomes harder than data; QGSJET presents a lower yield of photons, while DPMJET generally predict an harder spectrum than experimental data. The differences between data and models were attributable to a less-than-complete understanding of the soft hadronic interactions implemented in the models as diffractive processes [20, 21].

### 3.2 Electromagnetic energy flow

The energy spectrum gives the possibility to calculate the energy flow by integrating the contribution of each energy bin. The analysis was extended to other three pseudorapidity regions of Arm1 detector: $8.52 < \eta < 8.66$, $8.66 < \eta < 8.81$ and $8.99 < \eta < 9.22$. The contribution of each energy bin to the energy flow in shown in Fig. 4: in the regions between $\eta = 8.52$ and $\eta = 9.22$ the dominant contribution comes from low energy photons, while in $\eta > 10.94$ region also high-energy photons matter. The results are compared with the predicted photon energy flow from EPOS, QGSJET and SIBYLL. As can be seen in Fig. 5, data are consistent with EPOS and SIBYLL models near the peak of energy flow, while QGSJET predicts a much lower energy flow. An higher measured flow with respect to all models is found instead for $\eta > 10.94$. The good agreement of data with EPOS and SIBYLL in the peak region is related to the good agreement of the spectrum predicted by these models with data in the low energy region (as shown in Fig. 3 for $8.81 < \eta < 8.99$), which gives the dominant contribution to the energy

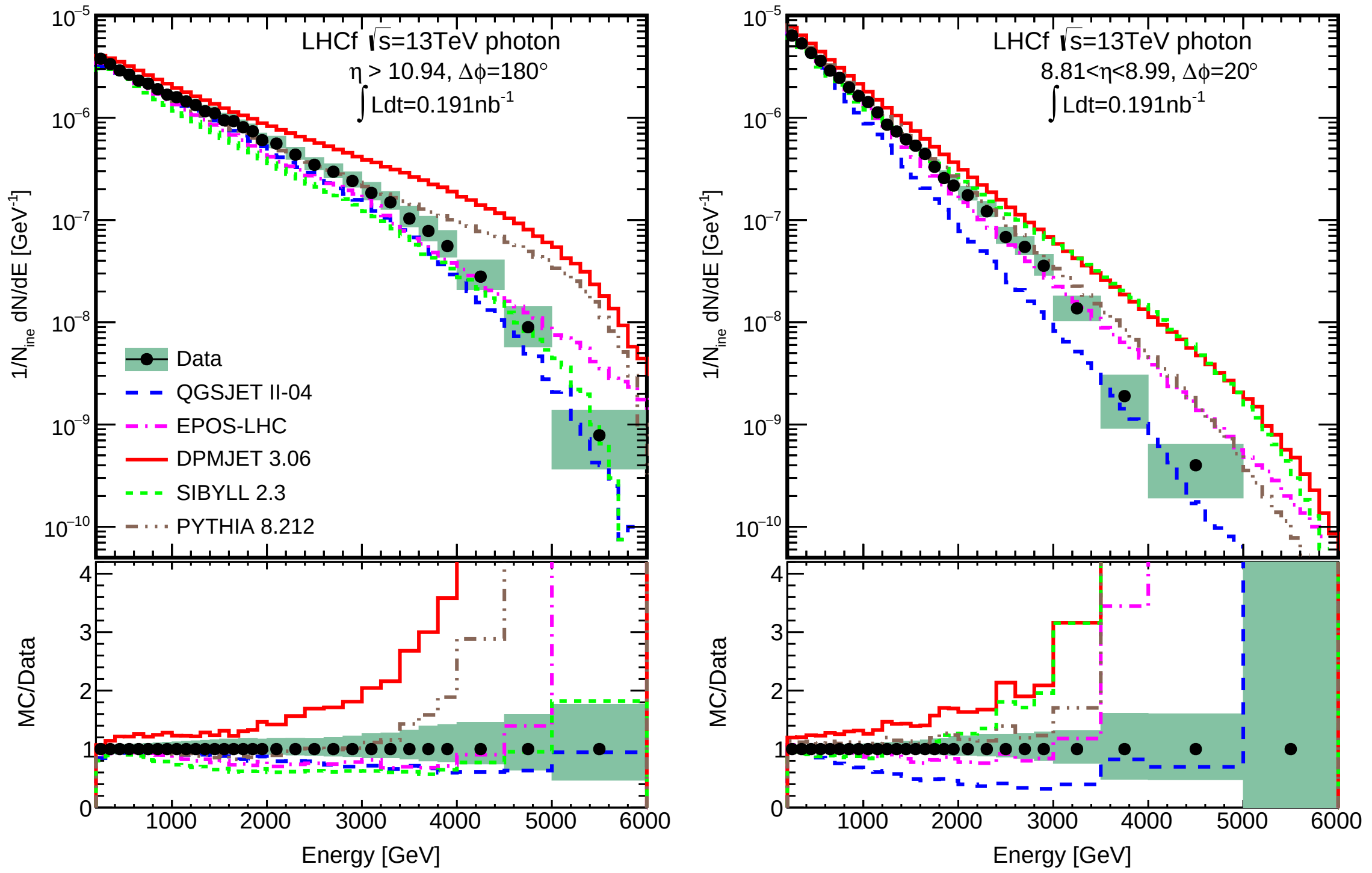

**Fig. 3:** Inclusive photon energy spectrum for pseudorapidity region $\eta > 10.94$ (left) and $8.81 < \eta < 8.99$ (right). Data are represented by black points while MC prediction from several models are represented by coloured histograms. Green shaded area represents statistical+systematic errors of data. Spectra are normalized to the total number of inelastic collisions. Bottom panels show the ratio of MC predictions to data.

flow.

## 4 Summary

LHCf experiment performed measurements on very forward photon production in proton-proton collisions at $\sqrt{s}$ = 13 TeV at the LHC accelerator. These measurements are necessary to calibrate hadronic interaction models used in cosmic rays physics to understand the development of atmospheric showers. Measured photon inclusive energy spectrum indicates that there is not an unique model representing the data in the pseudorapidity regions $\eta > 10.94$ and $8.81 < \eta < 8.99$. However, the measured data lie between the prediction of DPMJET, EPOS, PYTHIA, QGSJET and SIBYLL hadronic interaction models. EPOS-LHC model has an overall better agreement with data than other models; QGSJET II-04 has a good agreement in the $\eta > 10.94$ region; PYTHIA 8.212 is consistent with data in the low energy region (below ~3 TeV); SIBYLL 2.3 shows a good agreement below ~2 TeV in the $8.81 < \eta < 8.99$ region; DPMJET 3.06 predicts an harder spectrum than data in both the rapidity regions. The photon energy flow was also calculated form the energy spectrum. Three more pseudorapidity regions were considered in order to study the $\eta$ dependence of the energy flow. EPOS-LHC and SIBYLL 2.3 show the best agreement with data in the region $8.52 < \eta < 9.22$, while all models predict a lower flow than data in $\eta > 10.94$ region.

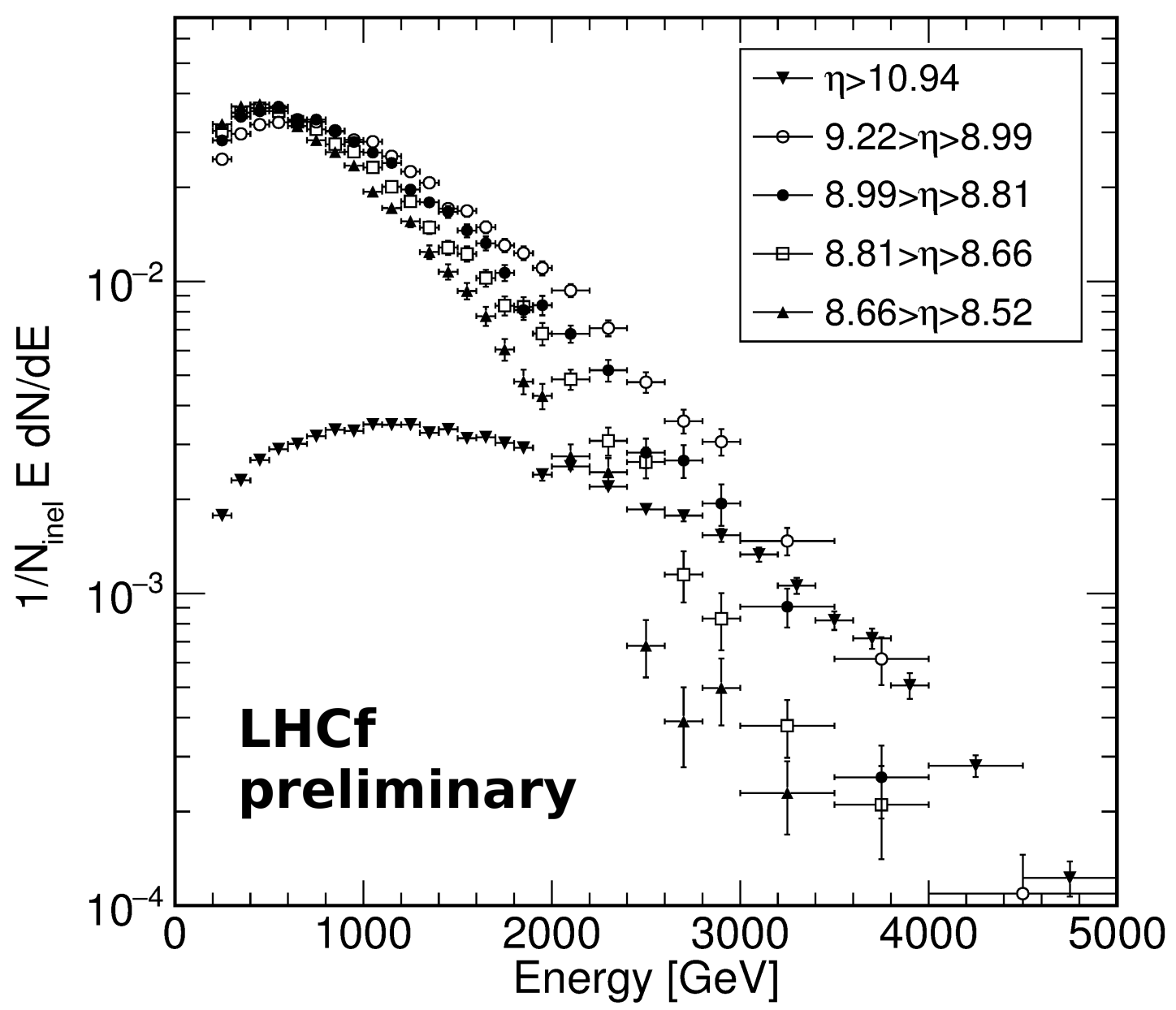

**Fig. 4:** Electromagnetic energy flow contribution of each energy bin for all pseudorapidity regions of Arm1 detector. In $\eta < 9.22$ regions the main contribution comes from low energy photons.

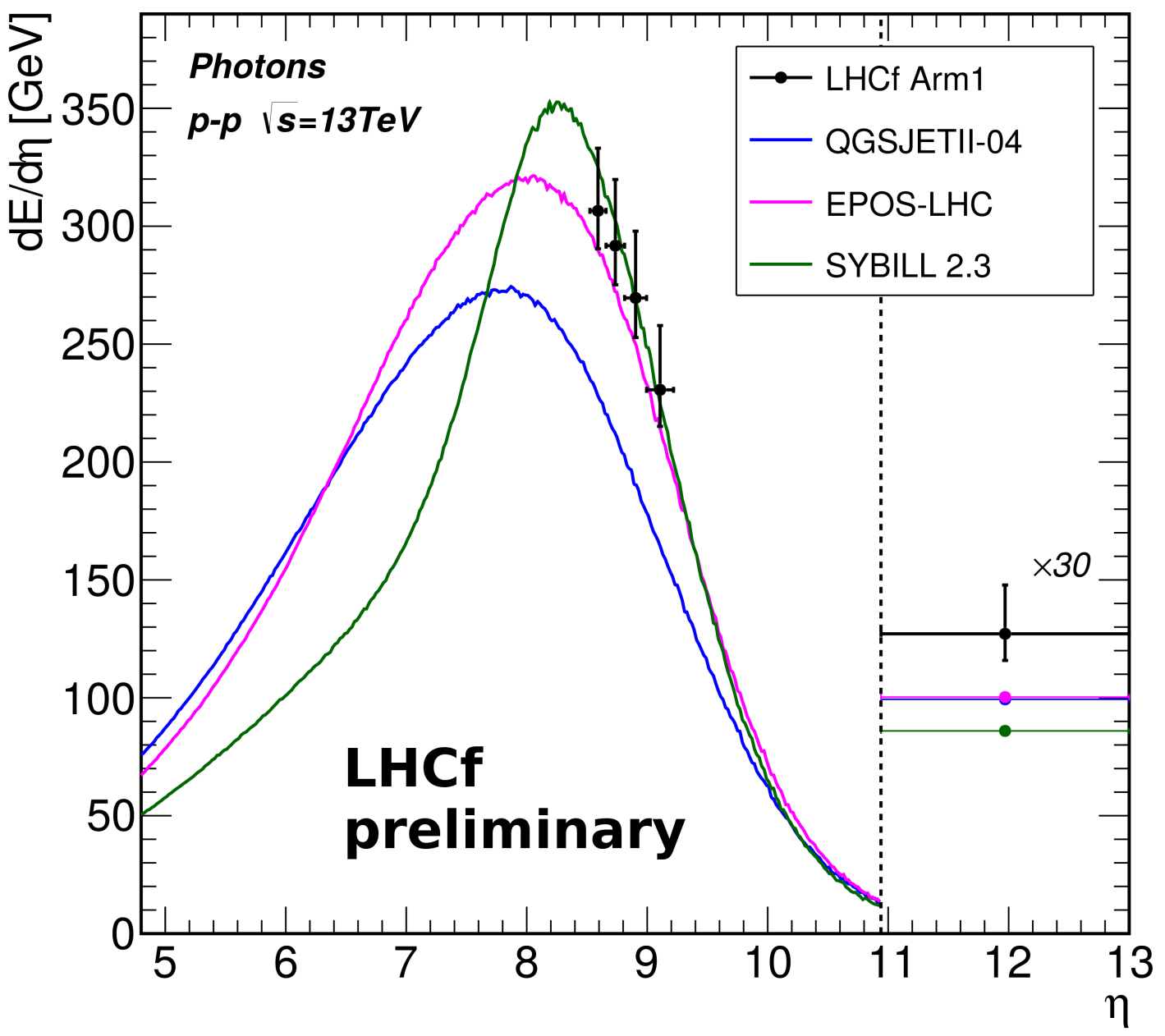

**Fig. 5:** Electromagnetic energy flow as a function of pseudorapidity. Arm1 data are represented by black points while MC prediction from EPOS-LHC, QGSJET II-04 and SIBYLL 2.3 models are represented by coloured histograms. The point above $\eta$ = 10.94 is scaled by a factor of 30 for clarity and it is artificially limited to $\eta$ = 13.

# Search for Ultra-High Energy Photons with the Pierre Auger Observatory

*P. Homola*[1] *for the Pierre Auger Collaboration*[*]
[1] Institute of Nuclear Physics Polish Academy of Sciences, Kraków, Poland
[*] Av. San Martín Norte 304 (5613) Malargüe, Prov. de Mendoza, Argentina

**Abstract**

One of key scientific objectives of the Pierre Auger Observatory is the search for ultra-high energy photons. Such photons could originate either in the interactions of energetic cosmic-ray nuclei with the cosmic microwave background (so-called cosmogenic photons) or in the exotic scenarios, e.g. those assuming a production and decay of some hypothetical super-massive particles. The latter category of models would imply relatively large fluxes of photons with ultra-high energies at Earth, while the former, involving interactions of cosmic-ray nuclei with the microwave background - just the contrary: very small fractions. The investigations on the data collected so far in the Pierre Auger Observatory led to placing very stringent limits to ultra-high energy photon fluxes: below the predictions of the most of the exotic models and nearing the predicted fluxes of the cosmogenic photons. In this paper the status of these investigations and perspectives for further studies are summarized.

## 1 Introduction

So far, photon structure is most efficiently studied with the accelerator instruments, where available energies reach TeV. There is, however, some potential in exploring the cosmogenic photons as well. Despite the cosmic photon fluxes being extremely low in comparison to accelerator data, different physics mechanisms are involved at the production sites and during the subsequent propagation to the Earth. This gives prospects for a complementary study on photons. Also the energies of cosmogenic photons can be higher than in accelerators. Gamma astronomers report on power-law spectra of $\gamma$-rays extending without a cut-off or a spectral break to tens of TeV [1] and plan building instruments capable of detecting photons of energies up to 300 TeV [2]. But these are not the largest photon energies expected on Earth. Within the research concerning ultra-high energy cosmic rays (UHECR), i.e. particles with energies exceeding $10^{18}$ eV, all the prominent scenarios predict photon fluxes reaching the Earth. One distinguishes two major classes of UHECR models: "bottom-up", based on the acceleration and subsequent interaction of nuclei; and "top-down", based on a decay or annihilation of hypothetical supermassive particles (see [3] for a review). The two classes differ significantly in the predicted fractions of photons among UHECR: $< 1\%$ the former and even up to c.a. 50% the latter. Hence, determining the UHECR mass composition, including identification of photons or setting upper limits on their fluxes, is an effort towards distinguishing between the two major classes which should give a hint on photon production and properties at the highest energies. Moreover, it is worth emphasizing that any result on UHE photons, including non-observation, is meaningful for the foundations of physics at the highest energies, allowing constraints on e.g. Lorentz invariance violation (LIV) [4], QED nonlinearities [5], space-time structure [6] or the already mentioned "top-down" scenarios.

The searches performed so far have not confirmed the existence of UHE photons, resulting in setting upper limits to photon fluxes and fractions [7–11]. In this paper we summarize the recent advances in UHE photon search performed based on the data collected by the largest cosmic-ray instrument: the Pierre Auger Observatory [12, 13] resulting in setting the most stringent upper limits. The summary is based on Refs. [14], [15], and [16]. We also briefly sketch an outlook on the further possible efforts, complementary to the current searches, based on the cascade approach.

[*]Full author list: http://www.auger.org/archive/authors_2017_08.html

## 2 Photon signatures

Investigations on cosmic rays of energies above $10^{15}$ eV can be done only indirectly, through the interpretation of properties of extensive air showers (EAS) induced in the atmosphere by primary cosmic particles. The interpretation is made based on interaction models which incorporate cross section data from accelerators within the available energy range and assume extrapolations beyond this range. Obviously, the higher the primary energy under investigation, the higher the risk of being mistaken with the extrapolations being used to interpret the data. Another disadvantage and potential uncertainty is the progressing degeneracy of the information about primary particles and their first interactions in the atmosphere with the rise of subsequent generations of secondary particles. Thus, if there is some mistreatment in the models at the very beginning of an air shower creation and propagation, it might significantly affect the final outcome of the analysis. Keeping in mind all the theoretical assumptions implying the existence of uncertainties, an effort is being made to identify the ultra-high energy primary particles that initiate the largest air showers observed. The effort is based on using simulations to predict air shower properties, characteristic for primaries of different types and confronting these predictions with observations.

EAS initiated by UHE photons should posses two independent properties: significantly delayed development and reduced muon content [17]. The former signature is based on the assumption that at the highest energies the pair production formation length gets elongated so much that the destructive interference of the fields associated with the atoms and particles nearby the primary suppresses the pair production process in the upper atmosphere, thus delaying the first interaction and the consequent air shower development. This is so-called LPM effect [18], a standard in air shower modeling [20]. Here it is important to note that the LPM suppression is sensitive to the possible LIV effects including both increase and reduction of the pair production formation length [21]. While the former would strengthen the UHE photon discrimination power based on delayed air shower development, the latter would do the opposite: reduce or even invert the LPM effect, and hence make photon-induced air showers develop more similarly to those initiated by nuclei. The present state-of the art photon identification methods involving the expectation of a delayed development of an air shower induced by a photon assume the standard LPM effect, without any LIV effects.

The other expected property of photon-induced air showers, the low muon content $N_\mu$, is also founded on conventional physics assumptions concerning the photonuclear cross section $\sigma_{\gamma-air}$. Air shower muons are thought to originate mostly from the charged pion decays, and charged pions originate from hadronic interactions. Therefore the observed $N_\mu$ should correspond to the size of a hadronic component of an air shower. If the interaction initiating an air shower is not hadronic, one would consequently expect that the hadronic component is initiated only by one of the secondary particles, thus its size and the corresponding $N_\mu$ are smaller comparing to an air shower induced by a hadronic interaction. According to the standard extrapolations for a photon of energy $E_\gamma = 10^{19}$ eV, $\sigma_{\gamma-air}$ should be ca. 30 times smaller than the electron pair production cross section, although some existing models allow $\sigma_{\gamma-air}$ to be only ca. 3 times smaller at the highest energies [17]. The energies of the secondary and virtual photons inside air showers are naturally lower than the primary energy and thus the $\sigma_{\gamma-air}$ uncertainty is correspondingly lower. Since in any case one expects that primary photons at the highest energies would pair produce more readily than interact with air molecules, the hadronic component and $N_\mu$ in the corresponding air showers will be in general smaller than in case of hadron-induced showers. The data collected so far by the Pierre Auger Observatory do not reveal muon-poor showers, just the opposite – the muon content seems to exceed the simulated values at the highest energies [22], which might point to a yet unconsidered physical source of uncertainty related to $N_\mu$.

## 3 Diffuse photon flux and hybrid UHE photon limits with multivariate analysis

In this section we follow Ref. [15] to present the upper limits to diffuse UHE photon fluxes obtained using the hybrid detector of the Pierre Auger Observatory. The observables used for this analysis are tightly connected with the key photon signatures introduced in the previous section: $X_{max}$ [g/cm$^2$] – the depth in the atmosphere where an EAS reaches maximum of its development, $N_{stat}$ – the number of triggered stations in an effective EAS footprint composed of the triggered stations, and $S_b = \sum_i^N S_i(R_i/R_0)^b$, where $S_i$ and $R_i$ are the signal and the distance from the shower axis of the $i$-th station, $R_0$ = 1000 m is a reference distance and b = 4 is a constant optimized to have the best separation power between photon and nuclear primaries in the energy region above $10^{18}$ eV. $S_i$ are measured in units of VEM (Vertical Equivalent Muon, i.e. the signal produced by a muon traversing the station vertically). Due to the standard LPM effect $X_{max}$ in an EAS induced by a UHE photon is expected to occur deeper in the atmosphere than in the case of showers initiated by nuclei. On the other hand, the low muon content expected in showers initiated by UHE photons results in a lower trigger capability at larger distances, and consequently smaller $N_{stat}$ and $S_b$ when comparing to the footprints of hadron-induced EAS. The primary type discrimination power of the above three observables is illustrated in Fig. 1. The Boosted

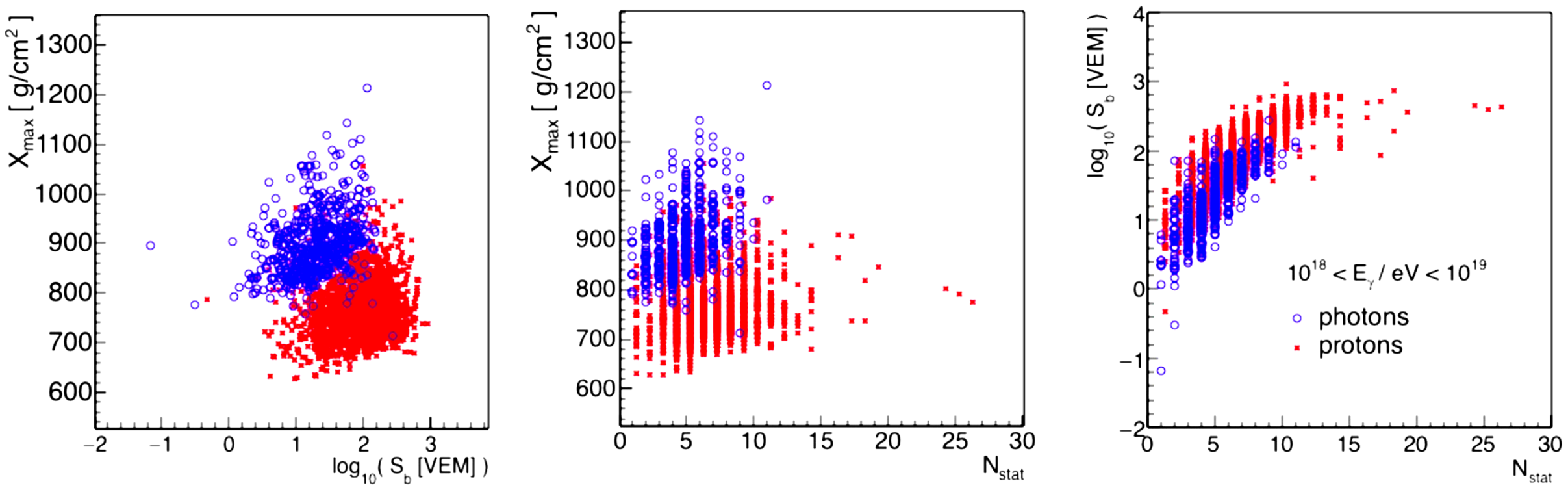

**Fig. 1:** Photon-proton discrimination power of the shower observables used for the UHE photon search in the hybrid detector of the Pierre Auger Observatory. [15]

Decision Tree (BDT) multivariate analysis method has been applied to the available data set using $X_{max}$, $N_{stat}$ and $S_b$. The BDT method has been chosen after examining several variants, as illustrated in the left panel of Fig. 2, leading to the optimal selection cut suitable for $E_\gamma > 10^{18}$ eV (Fig. 2, right panel). The few photon candidates found in this way were checked to be consistent with the proton background which led to determining the new hybrid upper limits to photon fluxes, as seen in Fig. 3. The new limits are more stringent in comparison to previous ones (see Ref. [9]), and the uncertainties of the limits are specified for the first time. The strong constraints on the "top-down" models can be concluded under the assumptions explained in Sections 1 and 2. For the first time also the photon flux predictions from one of the "bottom-up" scenarios ("GZK proton I") are constrained below $E_\gamma$=$10^{19}$ eV. The reader is referred to Ref. [15] for details and references.

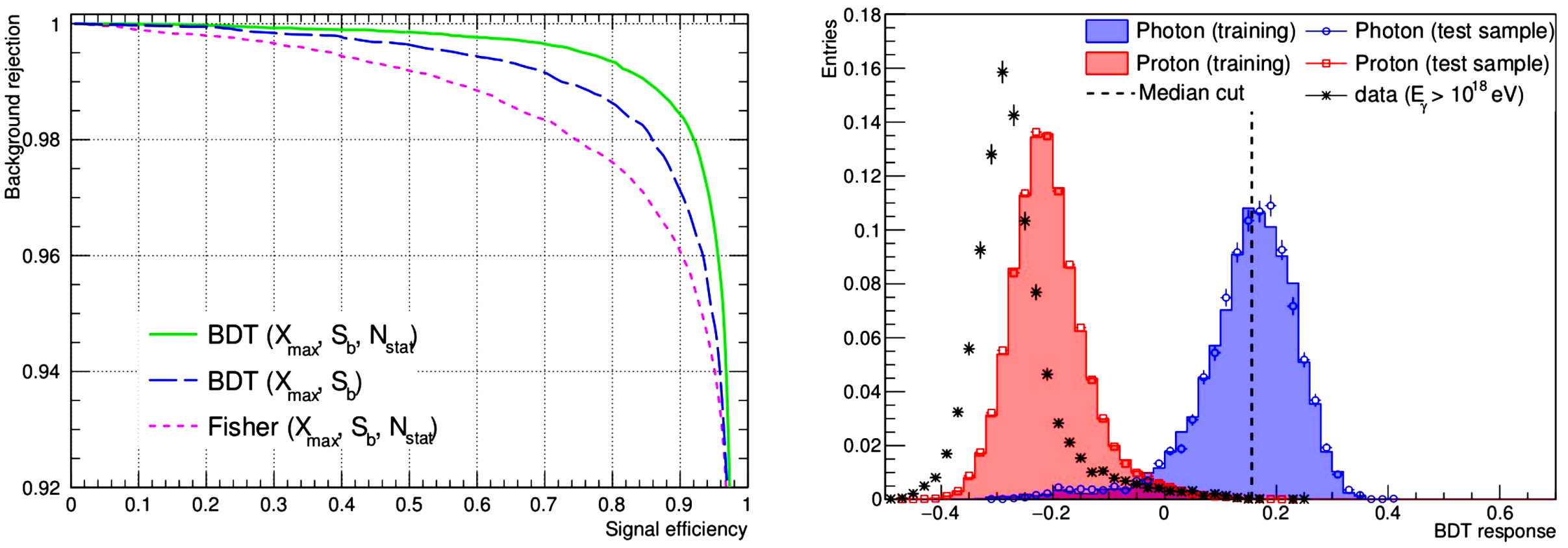

**Fig. 2:** The variants of the multivariate analysis method considered for the diffuse photon flux studies (left) and the selection cut on the response of the optimum variable (right). [15]

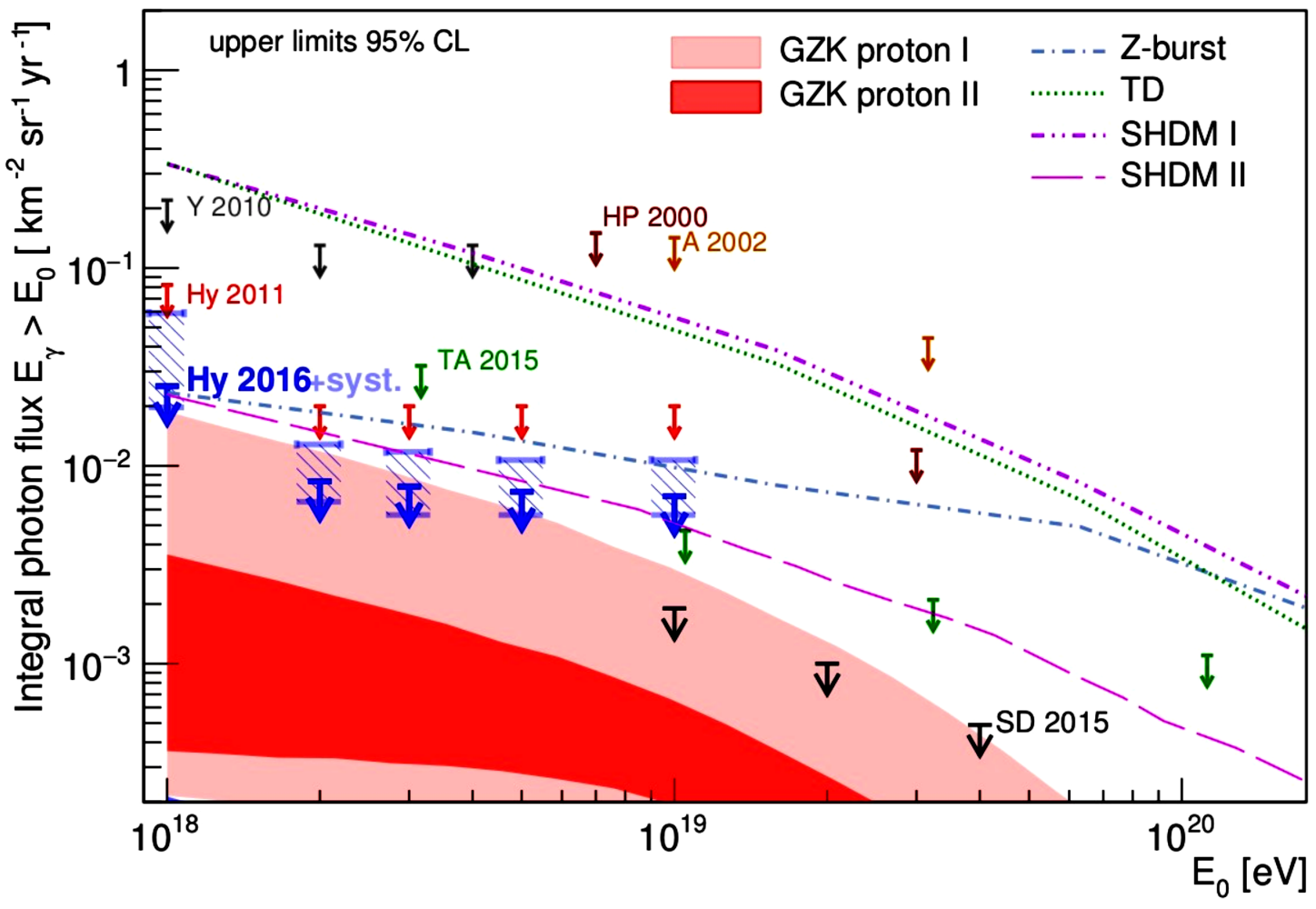

**Fig. 3:** The newest hybrid upper limits to the diffuse photon fluxes based on the Pierre Auger Observatory data (blue arrows) and the corresponding uncertainties (dashed regions around the arrows) compared to the theoretical predictions (check Ref. [15] for references) and to the limits provided by other experiments. [15]

## 4 Directional blind search

Complementing the diffuse photon flux study, the Pierre Auger Observatory performed also the blind search for arrival directions where photon excesses could be observed [14]. While the diffuse flux study aims at identifying all the photon candidates regardless their arrival directions, the directional blind search tries to identify photon candidates grouping directionally, and thus pointing to possible photon sources. The sensitive search for point sources was performed within a declination band from $-85°$ to $+20°$, and in an energy range from $10^{17.3}$ eV to $10^{18.5}$ eV, were the photon fluxes are not precluded by the diffuse photon search. No photon point source has been detected and an upper limit on the photon flux has been derived for every direction. In the study the method to derive a skymap of upper limits to the photon fluxes of point sources (Fig. 4) was specified.

## 5 Targeted search

To reduce the statistical penalty of many trials from that of a blind directional search in Ref. [16] several Galactic and extragalactic candidate objects were grouped in classes and analyzed for a significant photon excess above the background expectation. No evidence for photon emission from candidate sources was found and the corresponding particle and energy flux upper limits were given. These limits significantly constrain predictions of EeV proton emission models from non-transient Galactic and nearby extragalactic sources, as illustrated for the particular case of the Galactic center region (Fig 5).

## 6 Summary and Outlook

In this report the status of the ultra-high energy photon search at the Pierre Auger Observatory is summarized with emphasis on the recent advances, including the searches for the diffuse UHE photon flux, emission from discrete sources and targeted photon search. Although none of the searches performed so far proved an existence of ultra-high energy photons, the diffuse, directional and grouped upper limits to photon fluxes provide valuable astrophysical constraints. The end of the road to discovering UHE photons in the Pierre Auger Observatory has not yet been reached. The new substantial amount of data of increased sensitivity to primary mass will be acquired within the next several years with the AugerPrime upgraded detectors [13] which could lead either to identification of photons or to further constraints on the “bottom-up” predictions.

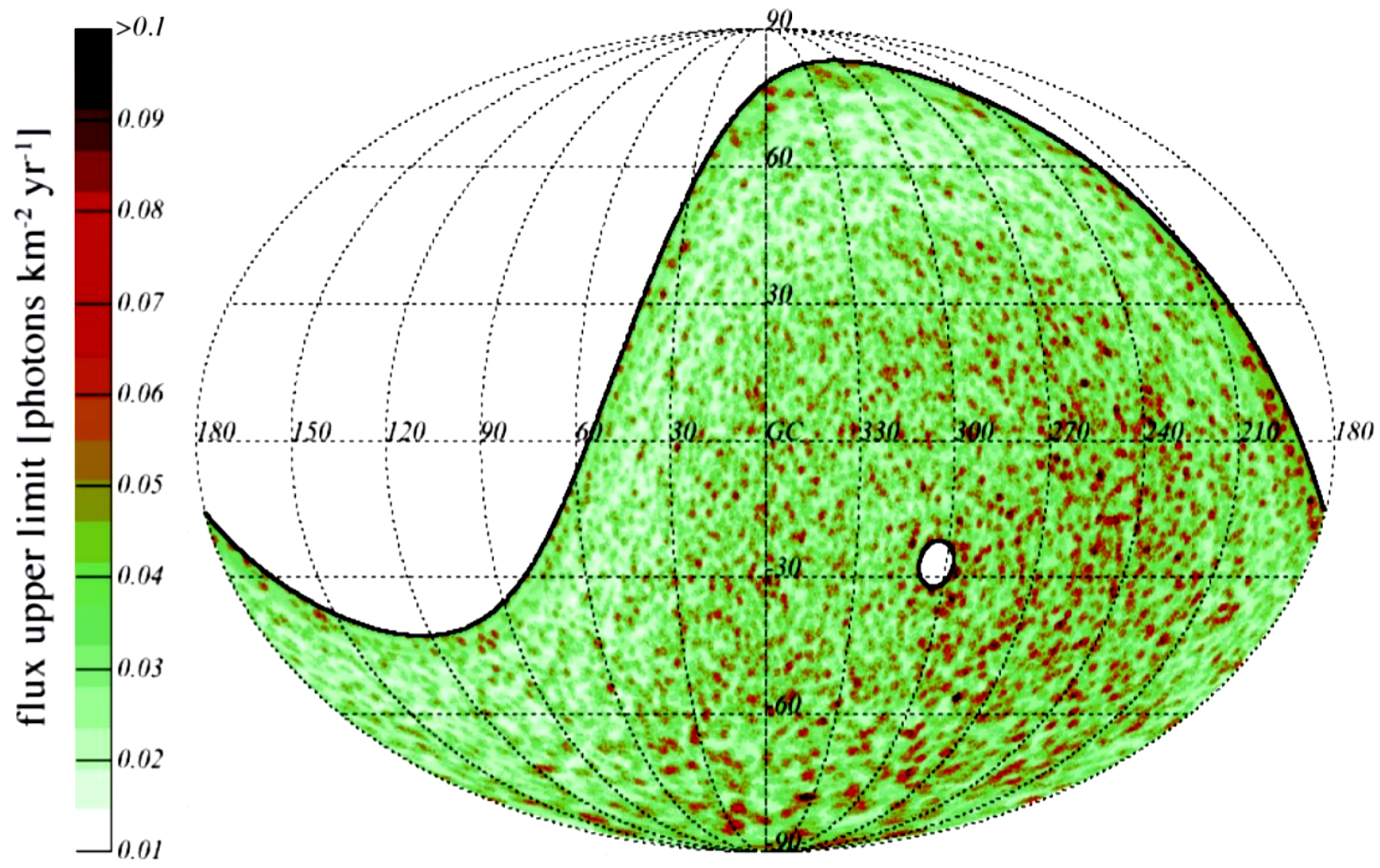

**Fig. 4:** Celestial map of photon flux upper limits in photons km$^2$yr$^1$ illustrated in Galactic coordinates. [14]

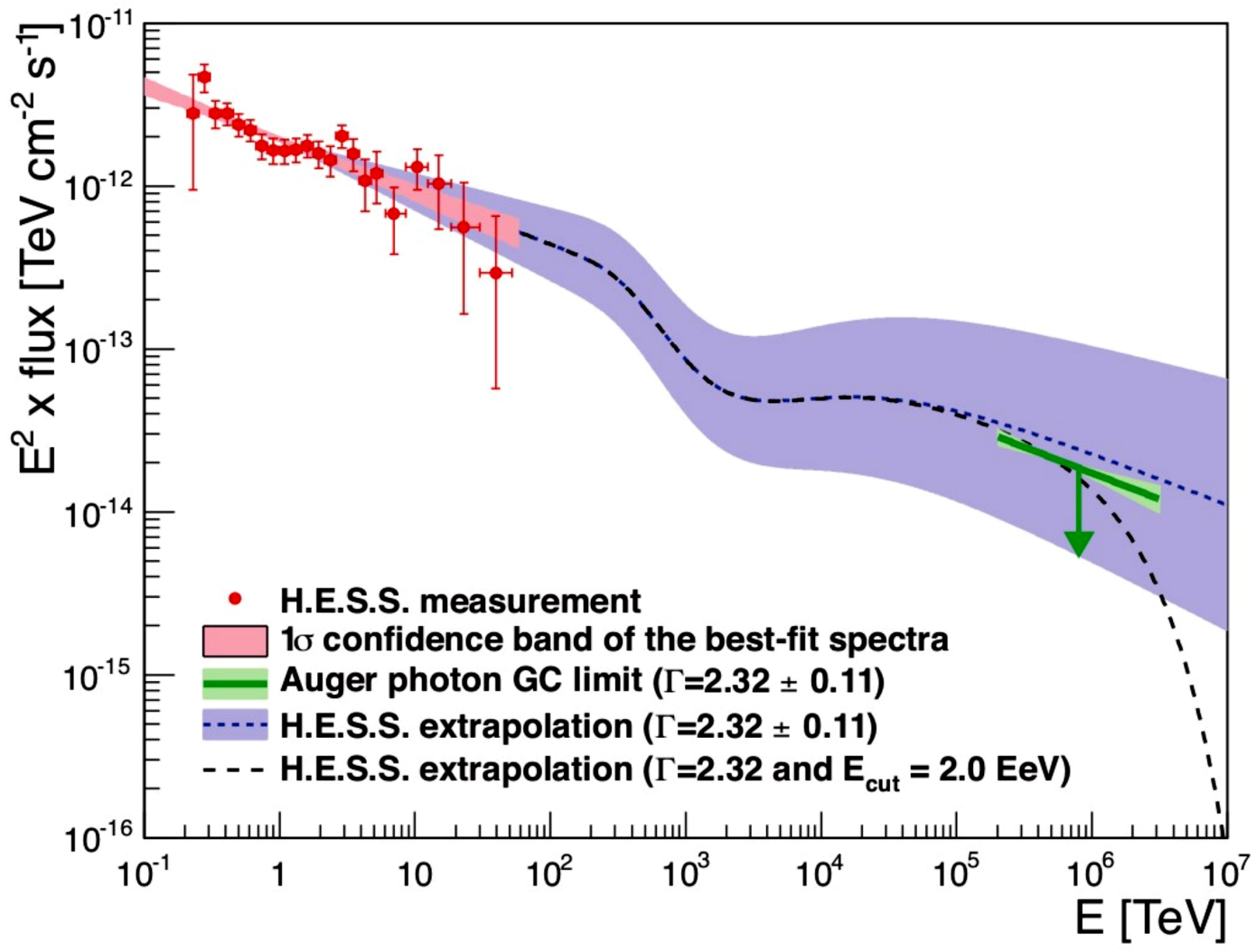

**Fig. 5:** Photon flux as a function of energy from the Galactic center region. Measured data by H.E.S.S. are indicated, as well as the extrapolated photon flux at Earth in the EeV range, given the quoted spectral indices ( [1]; conservatively the extrapolation does not take into account the increase of the $p$-$p$ cross section toward higher energies). The Auger limit is indicated by a green line. A variation of the assumed spectral index by $\pm 0.11$ according to systematics of the H.E.S.S. measurement is denoted by the light green and blue band. A spectral index with cutoff energy $E_{cut}$=2.0$\times 10^6$ TeV is indicated as well. [16]

The UHE photon search can be continued also with alternative methods. UHE photon sensitivity comparable to the Pierre Auger Observatory is possible e.g. with the Cherenkov Telescope Array [23]. Considering $\gamma$-ray detection technique for the UHE photon search implies using an alternative set of observables, unreachable by standard cosmic-ray arrays. The first studies in this direction point to a chance for a very precise identification of a photon primary or photon pre-shower, even event-by-event [24, 25].

Another interesting scenario involving UHE photons and their interactions is based on questioning some of the assumptions underlying the physical interpretation of the presently reported non-observation of UHE photons. In these scenarios (e.g. LIV with rapid photon decay) UHE photons produced at astrophysical distances have very short lifetimes and have a negligible chance to reach Earth [26–29]. On the other hand, the experimental verification of such models would be possible only if there is a chance to observe at least partly the products of UHE photon interactions: extensive cascades of cosmic rays. Such a chance should then be determined for the scenarios leading to UHE photon cascading in order to proceed with the experimental effort. Alternatively one could also think of a global cosmic-ray analysis strategy dedicated to hunting for large scale time correlations independently of the expectations from the existing theoretical models.

**Acknowledgments –** P.H. thanks the Pierre Auger Collaboration, in particular Daniel Kuempel and Marcus Niechciol for help in summarizing the status of the UHE photon search, and Mikhail V. Medvedev, Łukasz Bratek, and David d'Enterria for inspiring discussions about the perspectives of further studies in this direction.

# Search for Extensive Photon Cascades with the Cosmic-Ray Extremely Distributed Observatory

*P. Homola for the CREDO Collaboration**
Institute of Nuclear Physics Polish Academy of Sciences, Kraków, Poland

**Abstract**
Although the photon structure is most efficiently studied at accelerators, there is also a scientifically complementary potential in investigations on photons produced in the outer space. This potential is already being explored with $\gamma$-ray telescopes, ultra-high energy cosmic ray observatories and, since very recently, by the Cosmic-Ray Extremely Distributed Observatory (CREDO). Unlike the former instruments focused on detection of single $\gamma$, CREDO aims at the detection of cascades (ensembles) of photons originating even at astrophysical distances. If at least a part of such a cascade reaches Earth, it might produce a unique pattern composed of a number of air showers observable by an appropriately dense array of standard detectors. If the energies of air showers constituting the pattern are relatively low and if the typical distances between neighbors are large, the ensemble character of the whole phenomenon might remain uncovered, unless the CREDO strategy is implemented.

## 1 Introduction

While the photon structure and it properties in the high energy regime are investigated mainly using the data from man-made accelerators, one should remain aware of the scientific potential of astrophysical studies: gamma-ray astronomy and ultra-high energy cosmic ray (UHECR) research. Within the former field we deal with photons at energies unavailable in terrestrial instruments which adds complementarity to the accelerator photon investigations, despite incomparably low flux of gamma rays reaching the Earth. Considering the photon energies even larger than in gamma-ray astronomy, one enters the realm of UHECR which concerns particles of energies exceeding $10^{18}$ eV, with the few extreme events clearly above $10^{20}$ eV. The existence of such extremely energetic particles have remained a puzzle since decades. Interestingly, the two main classes of scenarios describing production of UHECR: "bottom-up" models based on acceleration of nuclei and "top-down" class postulating stable existence and decay or annihilation of super-massive particles of energies reaching even $10^{23}$ eV, predict that photons should contribute to the UHECR flux [1]. A clear distinction between the two classes is based on the scale of this contribution: in "bottom-up" models one would expect very small fraction of photons in the UHECR flux while in "top-down" scenarios the photon contribution to the observed flux is expected to exceed even 50% at $10^{20}$ eV. The research performed over the last decade by the largest cosmic-ray instruments does not indicate the existence of UHE photons, thus stringent upper limits are placed which under some basic assumptions might allow a severe constraining of the "top-down" class as a whole (see e.g. [2]). The point that we undertake in this paper is based on a trivial note concerning the mentioned "basic assumptions": we do not know the physics at UHE, relaying on extrapolations over many orders of magnitude from the accelerator energy region. Being aware of the fundamental theoretical uncertainties involved in the interpretation of the UHECR data allows considering the variety of logical and observational consequences of taking significantly different theoretical assumptions. In this paper we highlight one of such consequences: since UHE photons are expected to exist and the available evidence does not confirm their existence, we propose considering scenarios in which UHE photons exist but have negligibly little chance to reach Earth due to the interactions during their propagation on the way to us. Such scenarios prompt a purely technical challenge: can one see the products of these interactions - ensembles of cosmic rays? Actually this question needs an answer also within the state-of-the-art set of assumptions: if UHE photons exist, they should interact with the matter and fields during their propagation through the Universe which would lead to the initiation of extremely large cascades composed mainly of photons. And, continuing the logics within the paradigm, we also ask whether under some circumstances the possible sizes of such cascades might compensate a very small flux, as pointed out by the stringent upper limits to UHE photons. In other words we ask whether the scenarios involving UHECR photons can be tested more efficiently with the focus put on possible detection of photon cascades rather than single particles, complementarily to the current state-of-the-art research. The photon cascade approach has been initiated only recently by the CREDO Collaboration [3] and the scope of the addressed issues defines a wide physics program with long-term perspectives rather than a short term project. The basic research channel proposed by CREDO is the experimental verification of the astrophysical models where particle cascades are initiated, with the emphasis on photon ensembles. Such a verification would be possible only if there is a chance to observe at least partly the products of primary particle (e.g. UHE photons) interactions and this chance should be determined for the scenarios to be verified before proceeding with the experimental effort. Complementarily, we also propose another type of investigation which we call "fishing for unexpected physics". This approach is oriented on hunting for clearly non-statistical excesses above the diffuse and random cosmic-ray background, or arrival time correlations of air showers and single muons (or other secondary cosmic rays) in distant detectors, independently of the expectations from theoretical models. In this paper we highlight the CREDO science case and instrumentation and analysis strategies related to these two research channels: testing scenarios and fishing for unexpected.

*Full author list: http://credo.science/publications

## 2 $N_{ATM}>1$: mysterious air shower observations and generalized cosmic-ray research

It seems to be not very well known within the cosmic-ray community, especially among the younger colleagues, that there exist published reports on multi-cosmic-ray events looking like footprints of ensembles of primary cosmic rays correlated in time [4, 5]. The reports discuss a) a burst of air showers at estimated mean energy of $3 \times 10^{15}$ eV lasting 5 minutes [4], and b) unusual simultaneous increase in the cosmic-ray shower rate at two recording stations separated by 250 km [5]. Both observations were taken in two independent experiments in 1981 and 1975, respectively, and were the only events of their kinds seen during the lifetimes of both detecting systems. Other few hints of such possibly correlated cosmic-ray phenomena were seen by some small cosmic-ray experiments dotted around the world, such as a Swiss experiment that deployed four detector systems in Basel, Bern, Geneva and Le Locle, with a total enclosed area of around 5000 km$^2$ [6]. As proposed in Ref. [6], a globally coordinated cosmic-ray detection and analysis effort seems to be in place to verify whether the peculiar air shower observations carry any physical essence or maybe are just artifacts. The proposal concerned building small and cheap detectors which were planned to be installed in high schools at different locations around the globe, then operated and maintained by the high school pupils and staff. The science case addressed the cascading processes initiated by nuclei, mainly the photodisintegration of high-energy cosmic-ray nuclei passing through the vicinity of the Sun, first proposed by N. M. Gerasimova and G. Zatsepin back in the 1950s. The challenge had been undertaken in several research centers across the world where scientists in cooperation with their educational partners established small size experiments and approached a global coordination. Insufficient funding together with deficit of enthusiasm among the participants which grew with the continuing lack of exciting observations gave no scientifically meaningful outcome and led to reducing or closing up the activity in most of the involved high schools.

Now the idea of a global cosmic-ray research is being revoked in an enriched incarnation of CREDO with the following novelties:

1. The enriched CREDO science case includes photon cascades which addresses foundations of physics at the highest energies, allowing constraints on e.g. Lorentz invariance violation (LIV) [7], QED nonlinearities and space-time structure [8], or the "top-down" UHECR scenarios [1], similarly as in the UHE photon search. Furthermore, the generalization of the global approach allowing consideration of photon cascades changes the detection strategy. Photon cascades might contain even millions of particles, comparing to few or at best several in case of nuclear cascading like the Gerasimova-Zatsepin effect. This implies the necessity of implementing novel algorithms and triggers, and at the same time gives promising perspectives for unique, global observable signatures enabling event-by-event identification of cosmic-ray ensembles.
2. CREDO points to the necessity of involving as many detectors as possible, regardless the technical diversity or complexity of the whole network, focusing in the first stage only at the timing of single events and particles, looking for excesses in time windows of different scales. This approach addresses a very wide variety of instruments: satellites, stratospheric balloons, cosmic-ray arrays, fluorescence telescopes, radio air shower detectors, gamma-ray telescopes (see Ref. [9] for the first multi-primary gamma ray study in the CREDO context), neutrino observatories, accelerators, educational arrays, university laboratories, high school detectors, popular pocket-size detectors and finally smartphones equipped with a detection app and educational toys with simplest detectors. Such a global network can serve as a universal scenario tester: a subnetwork of detectors meeting the optimum requirements of a specific scenario can be used and more detectors can be built "on demand" if scientifically justified by the expectations of the tested scenario. At the same time the network can be used as a whole to fish for unexpected physics, i.e. unusual rate excesses or arrival time orders, as highlighted in the Introduction.
3. As the acquisition of the data from "everywhere" recorded with "everything" would generate an enormous stream of information, a sensible analysis and interpretation would automatically require an enormous manpower, including an effort of non-professional but enthusiastic scientific partners. Therefore CREDO takes the public engagement as the key scientific tool, putting the emphasis on the clarity of the key objectives, and easy, intuitive usage of the relevant tools. In parallel we will offer the paths for both education and science careers for all the participants.
4. CREDO puts a particular emphasis on the exploration of an alerting potential of the global cosmic-ray network, addressing not only the astrophysical strategies but also multidisciplinary research involving climate changes or seismic studies [3].

With the novel approach to a global cosmic-ray effort proposed by CREDO it becomes evident that a) widening of the scientific perspectives, b) including as much of the available data as possible, and c) involving as many of the potentially interested and enthusiastic colleagues as possible, increases the chances for scientific discoveries of a fundamental importance. Therefore CREDO postulates a fully open project, with free access to data and open source tools, where both financial and in-kind contributions are welcome. It all offers an unprecedented chance for multidisciplinary research and education based on cosmic ray data which are available everywhere and at negligibly small cost.

Following the Introduction and the above considerations one defines the CREDO mission in the simplest way by admitting more than one cosmic ray particle (including photons) entering the atmosphere simultaneously, where the term "simultaneously" denotes a temporal correlation and the specification "more than one" is equivalent to "ensemble" or to the mathematical expression $N_{ATM}>1$ (see Fig. 1).

## 3 Ensembles of photons: scenarios plus fishing

The "$N_{ATM}>1$" definition brings us to the already mentioned detection channels: A) testing theoretical scenarios and B) hunting for unexpected physics manifestations. Let us illustrate the channel A) with the two examples: exotic and standard. The exotic example is based on noting that there are variants of LIV with critically different predictions concerning the UHE photon fluxes, depending significantly on the assumed alteration of the dispersion relation at the highest energies. E.g. taking the dispersion relation in the shape defined in Ref. [10]:

$$E_\gamma(\vec{k}) = \sqrt{\frac{(1-\kappa)}{(1+\kappa)}}|\vec{k}| \tag{1}$$

one understands that the sign of parameter $\kappa$ changes the UHE photon flux expectations dramatically. If $\kappa$ is positive, the pair production by a primary UHE photon is suppressed, which should lead to increased UHE photon fluxes observed at Earth, in comparison to the implications concluded with using non-altered dispersion relation. In this scenario the non-observation result allows constraining $\kappa$ and therefore also LIV. On the other hand, if $\kappa$ is negative, the lifetime of a UHE photon would be extremely short, even of the order of 1 second, which on astrophysical scale is equivalent to an immediate decay [11–13]. We note that if the latter scenario is real, non-observation of UHE photons at Earth and the subsequent upper limits would be a trivial, inconclusive result. However, even then one still has at hand one yet not checked research option – approaching an observation of products of a UHE photon decay: cosmogenic electromagnetic cascades. Although it is widely assumed that such cascades get completely dissipated before reaching Earth, thus contributing to the diffuse photon flux, there are no precise calculations of the horizon (the distance within which a cascade can reach Earth at least in part, i.e. as an ensemble of a minimum two particles) with different theoretical assumptions. Such calculations within the Standard Model of particles are possible with the currently available tools [14] and the first steps in this direction have already been made, as described in Ref. [15]. In addition, when one takes into consideration physics beyond the Standard Model, either of particles or cosmological – more scenarios allowing observation of cascade-like signatures at the Earth appear verifiable (see e.g. Ref. [13] for a review on concepts relating to potential observation of quantum gravity manifestations). In this context it becomes apparent that a complete study and search for UHE photons should include both an effort towards identification of single UHE particles and a search for products of their decay: ensembles of photons correlated in time, most likely dispersed significantly in space, maybe also in time, with energies spanning even a very wide spectrum. The existence of a logically obvious and experimentally available, although yet not probed UHE photon search direction can be illustrated with considering two extreme cases: obvious detection of a photon ensemble and its obvious extinction. If photons in a cosmic-ray ensemble which reaches Earth travel very closely to each other, both in space and time, they would induce a set of extensive air showers (EAS) which would effectively behave as one big EAS, being a superposition of the smaller ones, detectable with the state-of-the art techniques, e.g. with a giant array of particle counters or with fluorescence telescopes. On the other hand, if the ensemble components are distant one from another on average more that the size of the Earth, then obviously no conclusion about the cascade-like nature of the phenomenon is possible: we see at best one particle which contributes to the diffuse and random cosmic-ray particle background. What is in between of these two "extremes", ensembles of particles (photons) distant one from another on average by less than the size of the Earth, remains to be studied, and, possibly, observed.

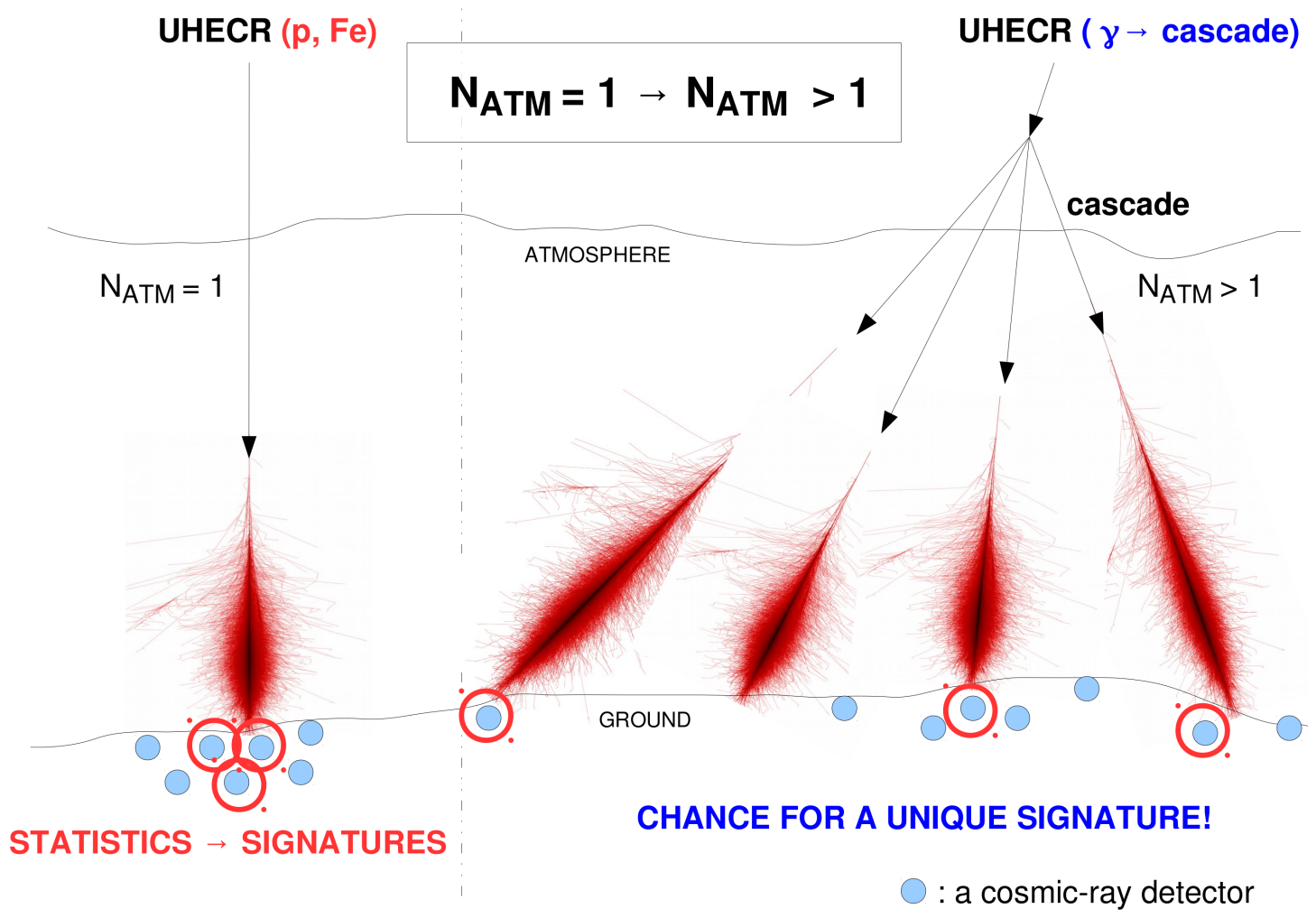

**Fig. 1:** Generalization of cosmic-ray research by admitting ensembles of particles as observation target.

An example of a non-exotic scenario within the channel A is a cascade of photons initiated by a UHE photon primary passing through the vicinity of the Sun and interacting with its magnetic field (see Fig. 2). This phenomenon, known in the literature as the preshower effect [16], is expected within the standard quantum electrodynamics and can be simulated with the available open source tools [17] which are also being used as a standard in the studies involving UHE photon-induced EAS [18]. The expected particle distribution at the top of atmosphere is very much elongated (even 10000 km!) in the West-East direction and super-thin (meters) along the North-South line, promising a unique observable signature built of temporal sequence of arrival times of the secondary cosmic rays on ground, and a very characteristic pattern of the triggering detectors [3]. In the preshower effect, once the primary UHE photon converts into an electron-positron pair, the electrons begin to radiate magnetic bremsstrahlung photons. The further the electrons travel the lower their energies and the larger deflection with respect to the primary direction. This is reflected in the photon distribution on ground: the photons near the core corresponding to the primary direction posses high energies as they were emitted right after the electron-positron pair creation, when the electrons still had energies comparable to the primary and they did not get deflected significantly in the magnetic field of the Sun. The further from the core, the lower photon energies. In the example shown in Fig. 2 the primary photon energy is $10^{19}$ eV and the spectrum of photons at the top of the Earth atmosphere extends from below GeV to above EeV (not the whole spectrum shown!). A feature such as shown in Fig. 2 could be observed with a global cosmic-ray network, or with a single large cosmic-ray observatory, or with a dedicated experiment tuned to the particle densities expected on ground. Testing this scenario, which we call Sun-SPS (SPS for Super Pre-Shower), is one of the first scientific tasks of the CREDO Collaboration. It is worthwhile to mention that the largest observatories are tuned to record EAS with energies typical for the very vicinity of the Sun-SPS core, landing at distances not further than few tens km, while the whole Sun-SPS footprint might be even 3 orders of magnitude longer. It points to the advantage of the global and diversified approach to the available cosmic ray data implemented in CREDO, at least as far as testing the Sun-SPS scenario is considered.

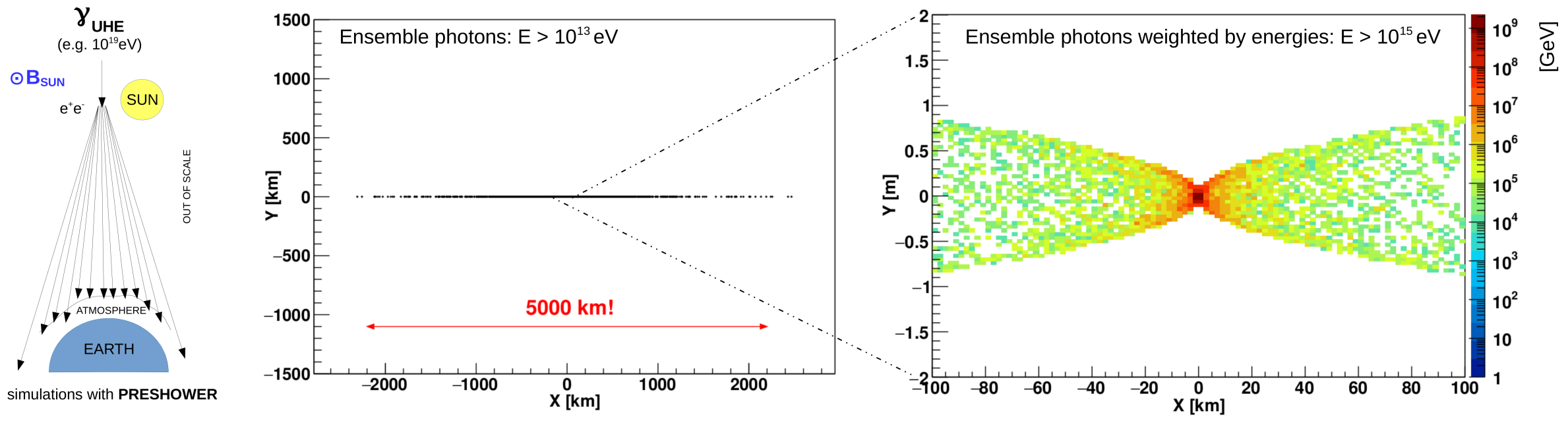

**Fig. 2:** The particle distribution on Earth of an ensemble of photons originated from an interaction of UHE photon of energy $10^{19}$ eV with the Sun magnetic field.

A special attention in CREDO is put to the channel B: fishing for unexpected physics. The idea of the "unexpected physics" trigger based on arrival time correlations and order in distant detecting stations is sketched in Fig. 3. Complementarily to the standard search of neighbor detectors triggered simultaneously by an EAS (clusters in space) one might also look for distant detectors triggered within some predefined time window by an ensemble of cosmic rays (clusters in time). In addition one might expect some order in the arrival times of the particles or events contributing to the cluster. The presence of such a feature would increase statistical significance of the observation.

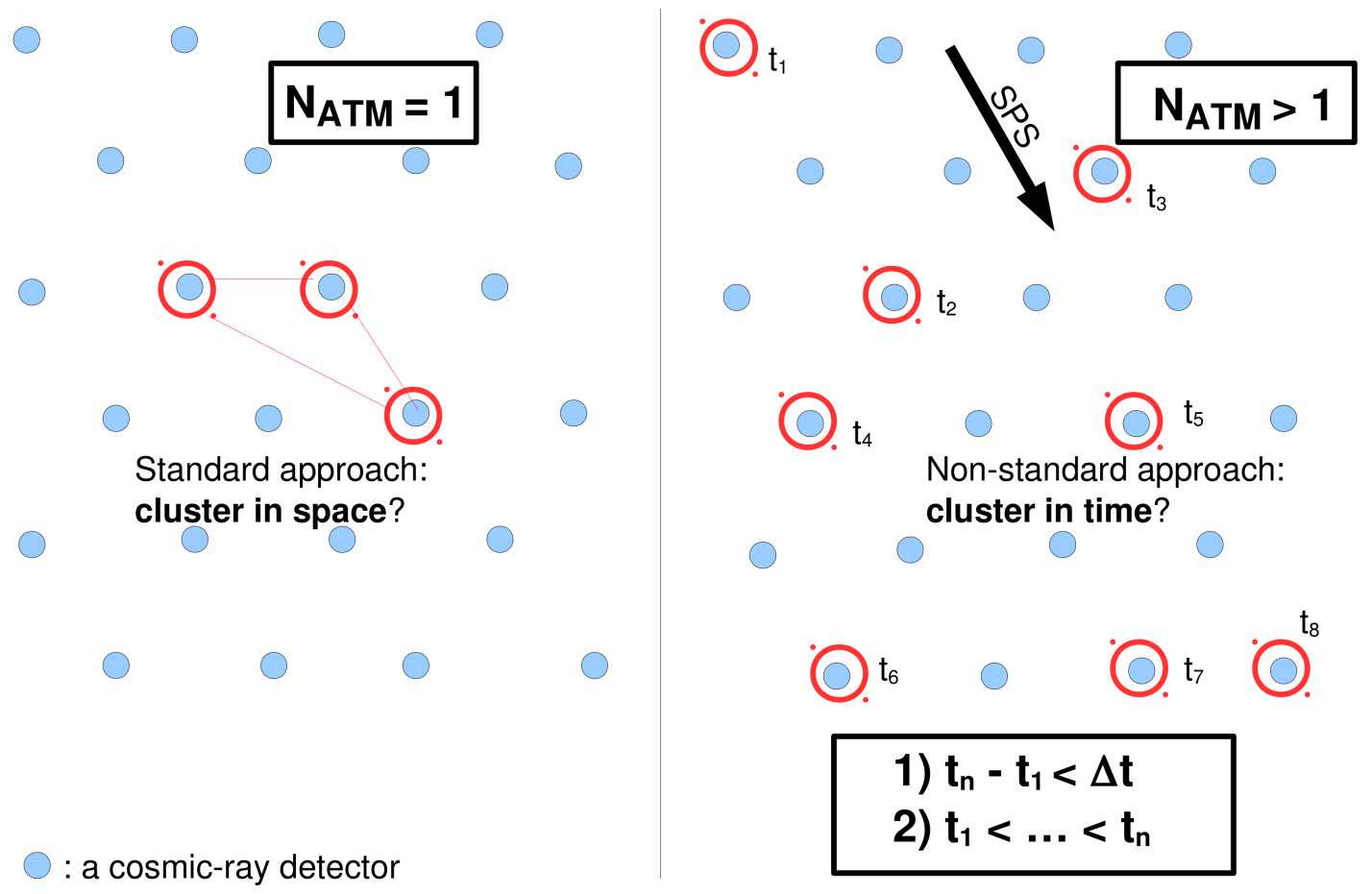

**Fig. 3:** A standard (cluster in space) vs ensemble (cluster in time) trigger in a ground array of a cosmic-ray observatory.

## 4 Observation: public engagement as a scientific tool

As already explained, the scientific success of the CREDO mission is strictly determined by the scale of the project possible to be achieved: total collecting area, geographical distribution of the detecting sites, and availability of manpower. The optimum can be reached by combining the available professional resources and wide public engagement. Apart from social reasons for which the public should be kept informed and even involved in the professional scientific research, it is obvious that public engagement in an exciting scientific project must induce a growth of professional scientific resources bringing profits to the whole science community and to the society as a whole. The key condition for this scientific growth is to show opportunities and paths of individual development and education within the project. In CREDO public engagement is going to be driven by three simple tools that would help to reach both social and scientific objectives of the project. Firstly, a massive participation will be achieved with an open source mobile application which turns a smartphone into a particle detectors. Such applications already exist [19, 20] although they are not yet open, thus not enabling sufficient flexibility required for a society driven software engine. For this reason CREDO opens its own app, to be freely distributed among science enthusiasts across the world with the encouragement to contributing to the development [21]. This of course does not exclude contributions from the users of the other applications to the common worldwide database. Another potentially available channel to involve even the youngest generations of science enthusiasts is related to the educational toys capable of detecting secondary cosmic rays and networked worldwide to help the CREDO mission. The detection of a particle and the link to the community dedicated to reach common and ambitious scientific goals should stimulate the passion, enthusiasm and a desire to get involved deeper in the project, i.e. to get educated.

Both using the smartphone particle detection app or an educational toy will enable passive participation in the CREDO project by collecting the data. The next level of involvement will be the activity within the CREDO community environment. The pilot component of this environment is the CREDO citizen science platform Dark Universe Welcome (DUW) [22] installed on the Zooniverse engine. With the easily understandable analysis format of DUW (see Fig. 4) one will be able to analyze "private" particles in the global context, search for "strange" detection patterns and help to train the "scientific fishing" algorithms. Other social facilities of the CREDO community environment, like e.g. individual and group rankings, will increase the pleasure of doing science and further stimulate the motivation to get involved deeper. The educational and scientific career paths supplementing the popular devices and software will in turn strengthen the stream of creativity and "fresh blood" to power the community of science professionals, that ultimately should be reflected in the increase of the ability of the society as a whole to develop by making scientific discoveries.

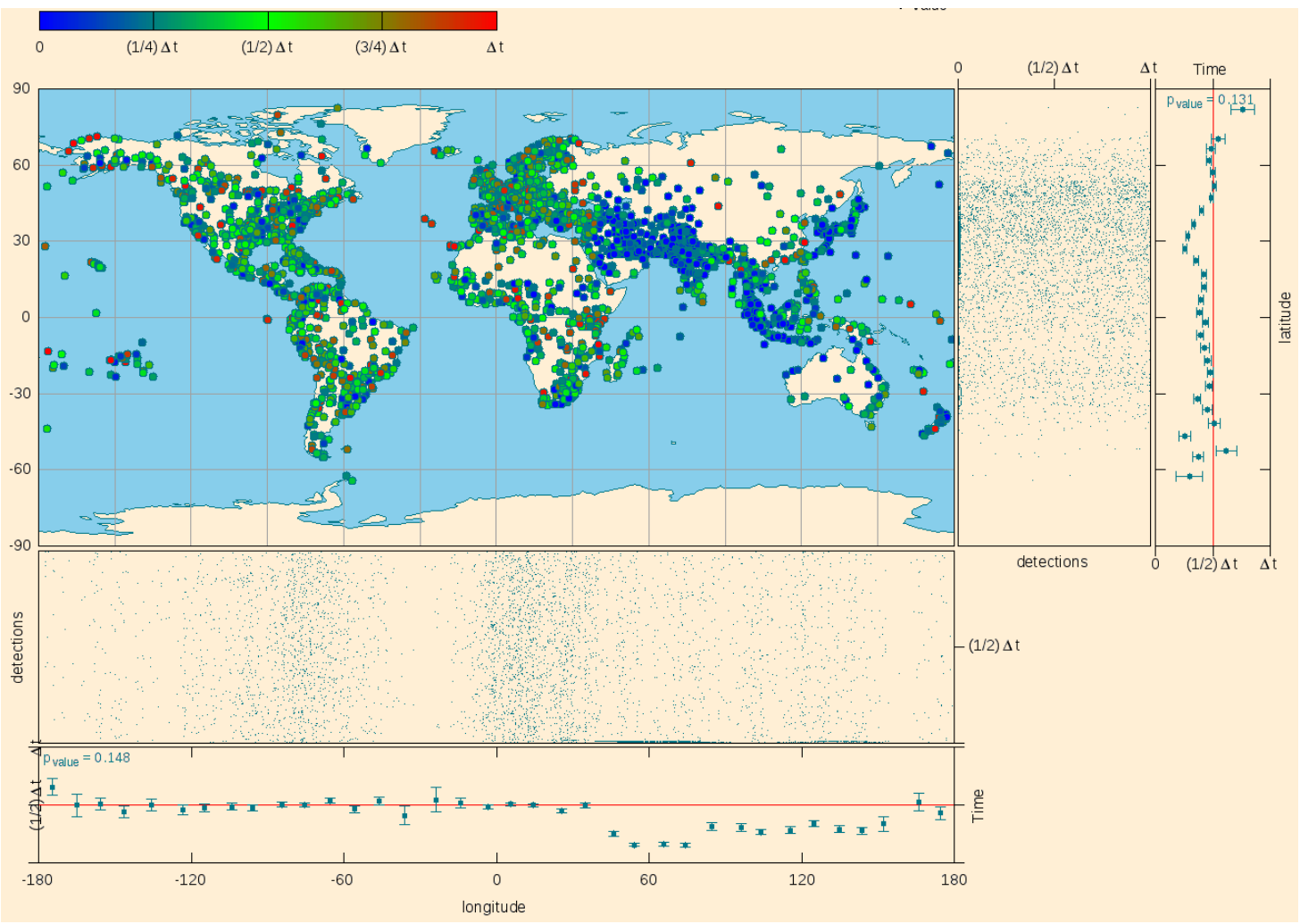

**Fig. 4:** A simple visualization of the "fishing for unexpected" strategy. The average arrival time within a certain temporal an spatial interval should be statistically consistent with the mean of the time interval if the received signal is composed of uncorrelated particles. A significant departure from the mean might be a footprint of an ensemble of correlated cosmic rays.

The third and scientifically the most exciting tool of the public engagement will be the automated procedure to monitor the cosmic-ray data globally which will provide the easily classifiable monitoring images with the largest discovery potential. The prototype of such a monitoring machine, called CREDO Monitor, fed by some of the publicly available cosmic-ray data, has been launched recently and is internally available for the CREDO members [23]. The available data is migrated periodically from the active acquisition sites to the data storage and computing center maintained at ACC Cyfronet AGH-UST [24], then after basic processing (scanning for time-clustering) classifiable global detector patterns (maps as in Fig. 4) are generated and stored on a web server ready for the inspection with a human eye. The receivers of CREDO Monitor will be able to tune the view of the most interesting discovery proposals selected by classifying machines and initiate a collective human-based classification according to the predefined crowdsourcing requirements, and finally open a professional analysis with a variety of algorithms which would actually lead to specifying of statistical significance of the proposed discovery patterns.

The above three pillars of the public engagement in CREDO should attract large number of participants and increase the chances for a scientific success of the project. Importantly, all the contributions to data acquisition and analysis, no matter from scientists or from "just" science enthusiasts, would give the right to claim co-authorship of scientific publications and the share in the possibly accompanying awards. Moreover, it is planned that the contributions will be easily registered and evaluated, leading to the estimate of the share in the project. Such an evaluation system, including e.g. the already mentioned user rankings, would offer a potential to activate additional motivations of the participants: positive competition.

## 5 Summary

We consider cosmic-ray cascades composed of photons correlated in time as a yet not checked channel of information about the Universe and the physics at the highest energies known. If such cascades exist they might have a wide spatial distribution which might make them observable only with a worldwide network of detectors, and keep out of the reach of even the largest cosmic-ray observatories with their state-of-the-art configuration. We introduce the Cosmic-Ray Extremely Distributed Observatory, the infrastructure and physics program tuned to cosmic-ray cascades, with potential impact on ultra-high energy astrophysics, the physics of fundamental particle interactions and cosmology, offering also a multidimensional interdisciplinary opportunities. We implement the CREDO strategy by applying a trivially novel approach to the cosmic-ray data taking - a global and massive approach. Within the CREDO strategy based on the collective and global approach to the available and future cosmic-ray data the chances for detecting and studying the astrophysical cascades by definition exceed the capabilities of even the largest observatories and detectors working independently of each other. Everybody, from theorists to non-experts, both institutions and private persons, are invited and welcome to contribute.

**Acknowledgments –** Research supported in part by PLGrid Infrastructure. The staff at ACC Cyfronet AGH-UST is warmly thanked for their always helpful supercomputing support. The Dark Universe Welcome citizen science experiment was developed with the help of the ASTERICS project (EU Commission Framework Programme Horizon-2020, Research & Innovation action grant no. 653477). PH thanks Andrew Taylor, Marcus Niechciol, Daniel Kuempel, and David d'Enterria for inspiring discussions.